\documentclass[10pt,a4paper,twoside,dvips]{report}
\usepackage[T1]{fontenc}
\usepackage{ifthen}
\usepackage{exscale}
\usepackage{tabularx}
\usepackage[german,american,spanish]{babel}
\usepackage[latin2]{inputenc}
\usepackage[intlimits,fleqn]{amsmath}
\usepackage{amsfonts}
\usepackage{amssymb}
\usepackage{fancyhdr}
\usepackage[numbers,sectionbib]{natbib}
\usepackage{chapterbib}
\usepackage[final]{graphicx}
\usepackage{afterpage}
\usepackage{subfigure}
\usepackage{psfrag}
\usepackage{IEEEtrantools}
\usepackage{wrapfig}
\usepackage{textcomp}
\usepackage{ltxtable}
\usepackage{lscape}
\usepackage{layout}
\usepackage{wasysym}
\usepackage{makeidx}
\usepackage{eurosym}
\usepackage{calc}
\usepackage{url}
\usepackage{mathrsfs}
\usepackage{calligra}
\usepackage{yfonts}[1998/10/03]
\usepackage[
a4paper,
dvips,
pdftitle={Depth of Interaction Enhanced Gamma-Ray Imaging for Medical Applications}
pdfauthor={Christoph W. Lerche}
bookmarks=true,     
bookmarksopen=false,
bookmarksnumbered=true,
colorlinks,
citecolor=OliveGreen,
linkcolor=MidnightBlue,
urlcolor=RoyalPurple
]{hyperref}

\usepackage[dvips,usenames]{color}
\usepackage[dvips,usenames]{pstcol}
\usepackage{pst-node}
\usepackage{pst-tree}
\usepackage{lscape}

\graphicspath{{figures/}}
\input{config/mystyle.sty}
\DeclareMathOperator{\erf}{Erf}

 \newcommand{\half}{\frac{1}{2}\,}

 \newcounter{saveeqn}

\def\Quadrat#1#2{{\vcenter{\hrule height #2
  \hbox{\vrule width #2 height #1 \kern#1
    \vrule width #2}
  \hrule height #2}}}
\def\dAlemb{\mathop{\kern 1pt\hbox{$\Quadrat{8pt}{0.4pt}$} \kern1pt}}
\def\dAlember{\mathop{\kern 1pt\hbox{$\Quadrat{4pt}{0.4pt}$} \kern1pt}}
\def\Chi{{\mathop{\kern 2pt\vcenter{\hbox{$\chi $}}\kern2pt}}}
\def\dslash{\partial\kern-.6em\slash}
\def\kslash{k\kern-.5em\slash}
\def\slh#1{#1\kern-.5em\slash}
\def\fslash#1{#1 \kern-.5em\slash}
\def\fbar#1{#1\kern-.5em\raise6pt\hbox{\footnotesize /}}
\newcommand{\lsim}{\mathrel{\rlap{\lower4pt\hbox{\hskip1pt$\sim$}}
\raise1pt\hbox{$<$}}}

\newcommand{\nn}{\nonumber}        
\newcommand{\ds}{\displaystyle}
\newcommand{\ts}{\textstyle}
\newcommand{\Ss}{\scriptstyle}

\newcommand{\doped}{\raisebox{0.2ex}{:}\:\!}
\newcommand{\mdef}{\;\raisebox{0.1ex}{:}\!=\;}
\newcommand{\stagpm}[2]{\pm\raise.86ex\hbox{$\scriptstyle#1$}\kern-.4em\raise-.3ex\hbox{$\scriptstyle#2$}\,}

\newcommand{\chapterquote}[2]{%
\vspace*{4eX}
\begin{center}
\begin{minipage}{0.75\textwidth}
\hrule
\Large\gothfamily
\vspace{1ex}
#1
\Large\fraklines
\flushright{{\gothfamily #2}}%
\vspace{1ex}
\par
\hrule
\end{minipage} 
\end{center}
\vspace*{4eX}
}

\newcommand{\tincaps}[1]{\mathit{\mbox{\tiny #1}}}
\newcommand{\chapterbib}{\renewcommand{\bibname}{\Large\bibtitle}\bibliographystyle{IEEEtran}\bibliography{mrabbrev,IEEEabrv,bibliography}}
\newcommand{\mycite}[3]{(#1 #2\cite{#1:#3})}
\DeclareMathOperator{\var}{var}
\DeclareMathOperator{\cov}{cov}
\newcommand{\chemform}[1]{$\mathrm{#1}$}
\newcommand{\isotope}[2]{$\mathrm{^{#2}#1}$}
\newcommand{\chform}[1]{$\mathrm{#1}$}
\newcommand{\g}{$\mathrm{\gamma}$}

\newcommand{\mathdegree}{\mbox{\textdegree}}

\author{Christoph W. Lerche}
\title{\sc Depth of Interaction Enhanced Gamma-Ray Imaging for Medical Applications}

\begin{document}

\selectlanguage{american}

 \pagestyle{empty}
 \begin{titlepage}

\selectlanguage{spanish}

\centering
\vspace*{2.5cm}
{\LARGE\parbox{\textwidth}{\centering{\sc%
Universidad De Valencia\\
\vspace*{1ex}
{\large Departamento de F\'{\i}sica At\'{o}mica, Molecular y Nuclear}\\
\vspace*{1ex}
\rule{0.4\linewidth}{1pt}\\
\vspace*{2ex}
Consejo Superior de Investigaciones Cient\'{\i}ficas
}}}

{\vspace{1cm}
\begin{figure}[h]
  \centering
  \includegraphics[height=0.2\textwidth]{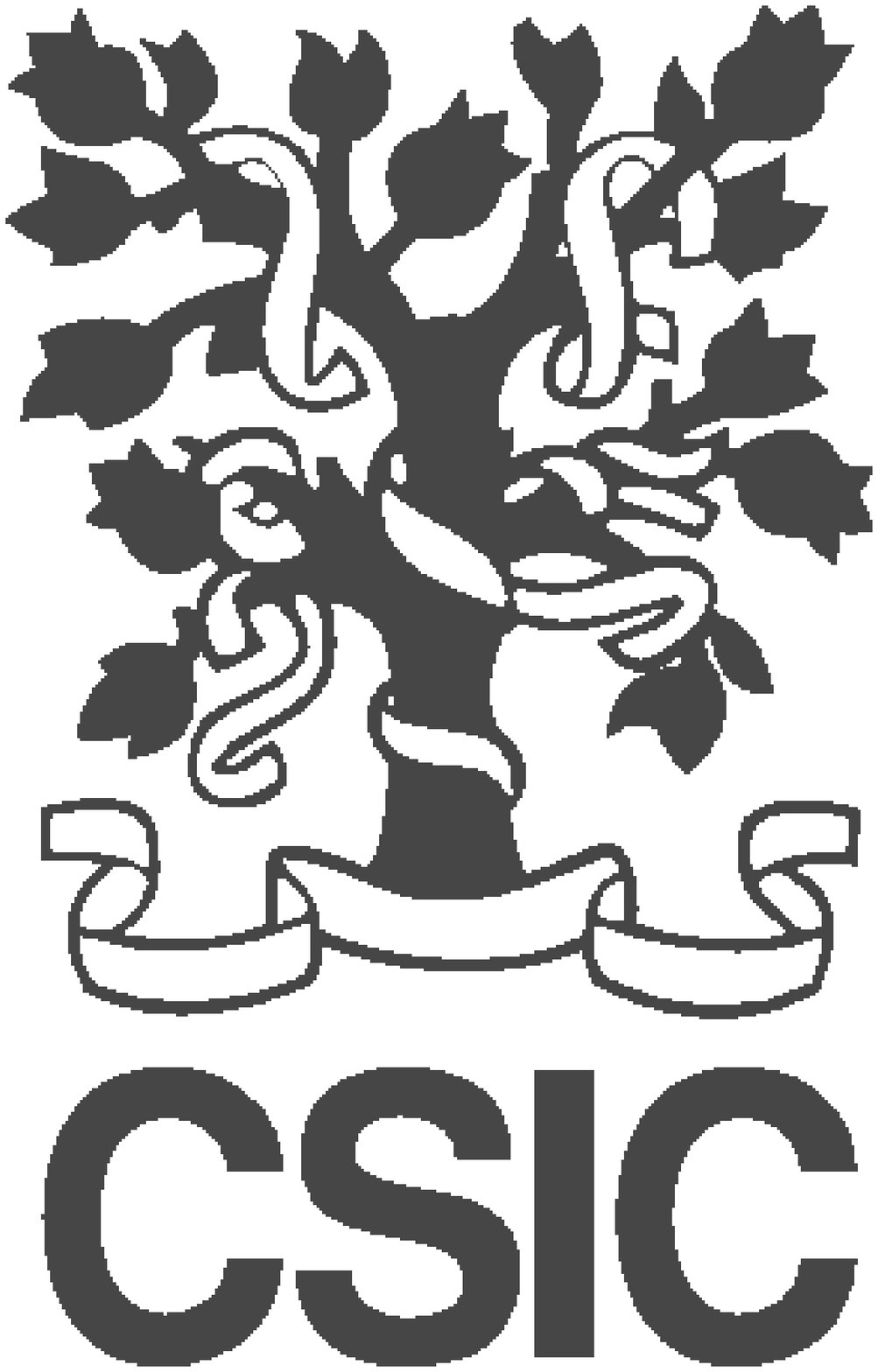}
\hspace*{0.7cm}
  \includegraphics[height=0.2\textwidth]{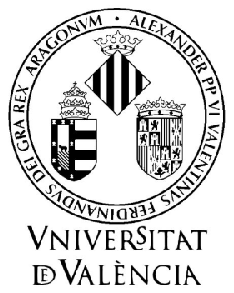}
\end{figure}
\vspace{1cm}}

\selectlanguage{american}
\boldmath
{\huge\parbox{\textwidth}{\centering{\bf%
Depth of Interaction Enhanced\\ 
Gamma-Ray Imaging\\ 
for Medical Applications}}}\\
\unboldmath
\vfill
\selectlanguage{spanish}
\flushright
{\Large\parbox{0.4\textwidth}{\centering {{\sl Christoph Werner
        Lerche}\\Tesis Doctoral\\Junio 2006}}}
\vspace*{0.5cm}
\end{titlepage}
\selectlanguage{american}
\cleardoublepage

 \pagenumbering{roman} 
 \pagestyle{plain}
 \vspace*{0.94\textheight}

\begin{flushright}
\calligra 
\Large
To Judit and Eva
\end{flushright}
\cleardoublepage

\pagestyle{fancy}

\cleardoublepage{}

\vspace*{15eX}

\centerline{\LARGE\bf Abstract}

\vspace*{2eX}

\begin{center}
  \begin{minipage}[t]{0.8\linewidth}\bf
    A novel design for an inexpensive depth of interaction capable detector for 
    gamma rays, suitable for nuclear medical applications, especially Positron
    Emission Tomography, has been developed, studied via simulations, and tested 
    experimentally. The design takes advantage of the strong correlation
    between the width of the scintillation light distribution in continuous
    crystals and the depth of interaction of the gamma-ray. For measuring
    the distribution width, an inexpensive modification of the
    commonly used charge dividing circuits that allows analogue and
    instantaneous computation of the 2nd moment has been developed and
    is presented in this work. This measurement does not affect the
    determination of the centroids of the light distribution. The
    method has been tested with a detector made of a continuous
    LSO-scintillator of dimensions 42x42x10 $\mathbf{mm^3}$ and optically
    coupled to the compact large area position sensitive
    photomultiplier H8500 from Hamamatsu. The mean resolution in all
    non-trivial moments was found to be rather high (smaller than
    5\%). However the direct use of these moments as estimates for the
    three-dimensional photoconversion position turned out to be unsuitable.
    Especially, for gamma-ray impact positions near the edges and
    corners of the scintillation crystal, there is a strong
    interdependence between first and second moments. Nevertheless, it
    could be demonstrated that the measurement of the centroids is not
    affected at all by the simultaneous measurement of the second
    moment. Also it is has been shown that the bare moments can be
    used to reconstruct the true photoconversion position. This is a
    typical inverse problem also known as the truncated moment
    problem. Standard polynomial interpolation in higher dimensions
    has been adopted to reconstruct the impact positions of the
    gamma-rays from the measured moments. For this, a
    parameterization of the signal distribution has been derived in order
    to predict the moment response of the detector for all possible
    gamma-ray impact positions inside the scintillation crystal.
    The starting point is the inverse square law but other important
    effects have been included: refraction and Fresnel transition at
    the crystal-window interface, angular response of the
    photocathode, exponential attenuation of the scintillation light, and
    background from residual diffuse reflections at the black painted
    crystal surfaces. This model has been verified by experiment. For the
    three non-trivial moments, a very good agreement with measurements  
    was observed. When using the reconstructed impact positions,
    the intrinsic mean spatial resolution of the detector was found to be
    $\mathbf{1.9\,mm}$ for the transverse components and
    $\mathbf{3.9\,mm}$ for the depth of interaction. Using directly the 
    bare moments as position estimate, the intrinsic mean spatial
    resolution of the detector was found to be $\mathbf{3.4\,mm}$ and
    $\mathbf{4.9\,mm}$, respectively. The cost for the required detector
    improvements are essentially negligible.
  \end{minipage}
\end{center}
\cleardoublepage{}


\cleardoublepage{}
\selectlanguage{spanish}

\chapter*{Resumen en Castellano}

\selectlanguage{american}
\chapterquote{The pure and simple truth is rarely pure and never simple.}{%
Oscar Wilde, $\star$ 1854 -- $\dagger$ 1900
}
\selectlanguage{spanish}

\section*{Antecedentes, Objetivos y Organización del Trabajo}

\PARstart{E}{n} los últimos ańos, las técnicas de imagen en Medicina
Nuclear han ganado en importancia debido a sus éxitos en el
diagnóstico de oncología, neurología y cardiología. Imágenes
tridimensionales pueden ser obtenidas actualmente por tomografía
computerizada, mediante resonancia magnética nuclear (RMN)
o mediante el empleo de isótopos radioactivos incorporados en una droga o en
un compuesto biológico activo en general. 
La tomografía por emisión de positrones (Positron Emission Tomography
o PET en inglés), gammagrafía y  tomografía por
emisión de un solo fotón (Single Photon Emision Tomography, SPECT) son
las técnicas más usadas en diagnóstico por 
imagen en Medicina Nuclear y  se basan
en la reconstrucción de la distribución de  pequeńas cantidades
de radiofármacos  administrados previamente. Si el radiofármaco
administrado es
específico para un cierto proceso metabólico, el empleo de los medios
diagnósticos permite estudiar, caracterizar y valorar este mismo proceso.
Las imágenes obtenidas son por tanto imágenes funcionales del cuerpo
entero, de órganos o de las células. Por el contrario, imágenes
médicas obtenidas por rayos X, tomografía
computerizada, ecografía o similares aportan información morfológica y
estructural del cuerpo o de los órganos. La RMN es capaz de
proporcionar imágenes estructurales y funcionales, aunque la
RMN funcional requiere la administración de grandes cantidades de
sustancias de contrastes y su sensibilidad es de alrededor de seis ordenes de magnitud inferior
que la de PET, SPECT y gammagrafía.

Los detectores de centelleo han constituido durante ańos los instrumentos
primordiales para la detección de la radiación gamma procedente de los
radiofármacos. Los más simples comprenden un único cristal de centelleo
y un único fotodetector. Para obtener imágenes con dicho detector se
inventó el escáner rectilíneo que aporta la información espacial al
mover el detector sobre el objeto, registrando a la vez su seńal junto con su
posición actual. La primera cámara gamma fue desarrollada por Hal Anger
en 1952 y consistió en un cristal y 7 fotomultiplicadores con una lógica
analógica (lógica de Anger) que calcula las posiciones por suma con
pesos. La configuración de
cámaras gamma actuales se diferencia muy poco de este primer diseńo
aunque los constituyentes modernos de dichas cámaras se han mejorado
significativamente  en los últimos ańos.
Hoy en día hay una amplia gama de cristales centelladores con muy
diferentes propiedades y lo mismo ocurre con los fotodetectores. 
Una mejora muy importante de los últimos ańos es el uso de sistemas de
dínodos especiales para que los fotomultiplicadores sean sensibles a la
posición (Position Sensitive Photomultiplier Tube, PSPMT). Esto hizo
posible el desarrollo de cámaras gamma muy
compactas para su aplicación en la visualización de órganos pequeńos.
La gran mayoría de detectores de rayos-$\gamma$ para todas las
modalidades de Medicina Nuclear son cámaras de este tipo y que se han
especializado para su función eligiendo los componentes
más adecuados. 

Desgraciadamente, los detectores de centelleo para rayos-$\gamma$,
 en general, padecen de un problema común.
Dado que los cristales de centelleo han de ser de un grosor finito
para conseguir parar los rayos-$\gamma$ que se pretenden detectar,
ellos mismos introducen una incertidumbre debido al hecho
de que hasta el día de hoy existen pocas técnicas ya comercializadas para detectar
la profundidad de interacción del rayo-$\gamma$ dentro del cristal
centellador. Como consecuencia, la posición del origen del rayo-$\gamma$
no se calcula correctamente, conduciendo al error de paralaje.
Este error es especialmente importante para la modalidad PET porque
los fotones de aniquilación que se tienen que detectar tienen una
energía alta de $\mathrm{511\,keV}$ y en consecuencia su probabilidad
de ser detectados es relativamente baja. Para detectores de PET con una
eficiencia intrínseca aceptable, centelladores gruesos son necesarios.
Debido a la falta de una componente de la posición de
fotoconversión, el origen de la radiación $\gamma$ no se puede
computar exactamente siempre que la incidencia del
rayo-$\gamma$ no es normal respecto al plano del área sensible del fotodetector.
Este error es especialmente importante para puntos de la región de
interés lejos del centro del detector.

En los últimos ańos se han dedicado muchos esfuerzos a mejorar los
 parámetros claves como eficiencia intrínseca, resolución espacial
y resolución energética. La detección de la profundidad de
interacción es uno de  estos campos de investigación.
Entre los métodos más conocidos para determinar la profundidad de interacción
figura la llamada técnica {\it phoswich}
que usa el hecho de que los tiempos de desintegración (desexcitación) de diferentes 
centelladores se distinguen entre ellos y por lo tanto dan lugar a pulsos de luz de centelleo 
de diferente duración (Seidel et al.\ \cite{Seidel:1999}). Usando dos
materiales de centelleo  diferentes,
se puede determinar en cual de los dos cristales se ha efectuado la foto-conversión
del rayo-$\gamma$. Las desventajas son la necesidad de dos cristales distintos
para cada detector y la electrónica para diferenciar los dos tiempos de 
caída de la seńal. Otra técnica muy usada es el {\it light-sharing}
(Moses and Derenzo \cite{Moses:1994}).
Esta técnica se usa sobre todo con cristales pixelados y requiere
dos fotodetectores de los cuales por lo menos uno ha de aportar la información
espacial. 
Para cada pixel del centellador, se puede 
deducir la profundidad de interacción usando el reparto de
la luz de centelleo entre los dos detectores. 
Uno de los fotodetectores tiene que ser
un detector de semiconductores para no atenuar demasiado los rayos-$\gamma$. A parte de estas dos
técnicas existen otras posibilidades no tan comunes. 
Una gran desventaja de los métodos mencionados es la necesidad de foto-detectores o/y cristales de centelleo
adicionales para realizar la medida de la profundidad de interacción. Debido a
que estos componentes son los más caros de un detector,
estas técnicas encarecerían significativamente su construcción.
Para permitir el amplio uso de métodos
diagnósticos por imagen, tanto en medicina como en la investigación
se requieren técnicas baratas y con razonables prestaciones.
La segunda desventaja de todas las técnicas, a excepción del {\it light-sharing},
es que la resolución de la profundidad de interacción es no-continua
(discreta) y que esta depende del tamańo de los pixels. 

El objetivo principal de este trabajo fue el
desarrollo de un  detector de rayos-$\gamma$ de un coste de fabricación reducido
pero con prestaciones comparables a los de otros detectores actuales.
Con este fin, se emplearon cristales de centelleo continuos y de grandes
dimensiones, ya que de esta forma se puede evitar el costoso proceso de
segmentación de los cristales. Se estima que este proceso encarece
el cristal en un factor 7 debido al material del centellador que
se pierde y también a las rupturas involuntarias. El uso de un fotomultiplicador
sensible a la posición y con un área sensible elevada se fundamenta en
una reducción en los costes de fabricación. Aunque estos dispositivos
son relativamente caros, el precio por unidad de área sensible no es muy elevado. 
Varios estudios anteriores mostraron, que el empleo de cristales
continuos es problemático especialmente para la tomografía
por emisión de positrones. Debido a la elevada energía de los fotones
de aniquilación, los cristales deben tener un grosor también
elevado para asegurar una eficiencia intrínseca de detección
suficientemente alta. Esto introduce variaciones importantes en la
distribución de luz de centelleo que dependen de la profundidad del
impacto de rayo-$\gamma$ y de su posición en el plano del fotocátodo.
Obviamente, la determinación de la posición del impacto es más difícil
que en el caso de cristales pixelados en el que es suficiente
identificar el pixel activo. El empleo de cristales continuos requiere
analizar la distribución de luz y deducir a partir de este análisis
los parámetros de impacto. Por su extremadamente bajo coste, el método más común es 
el algoritmo de centro de gravedad. 
Desgraciadamente, su uso junto con cristales gruesos produce efectos
no lineales y dependientes de la profundidad de interacción cerca de
los bordes de los cristales. Como resultado se perjudica  la
resolución espacial y energética en estas zonas, siempre que la
profundidad  de interacción no pueda ser medida
(Freifelder et al.\ \cite{Freifelder:1993}, Siegel et al.\
\cite{Siegel:1995}, Seidel et al.\ \cite{Seidel:1996}, Joung et al.\ \cite{Joung:2002}).
No obstante, la configuración de un cristal continuo con un
fotomultiplicador sensible a la posición y de área sensible amplia ofrece
la estimación de la profundidad de interacción a partir de la anchura
de la distribución de luz de centelleo detectada
(Kenneth et al.\ \cite{Kenneth:2001}, Antich et al.\ \cite{Antich:2002}). El principal problema consiste en
medir esta anchura de forma rápida y con modificaciones de bajo
coste. La digitalización de cada uno de los segmentos del ánodo
permite su cálculo pero requiere muchos canales electrónicos. 
Si se implementara este método en un PET de animales pequeńos 
compuesto por ocho módulos y cada uno con un PSPMT de 64 canales se
requerirían en total 512 canales electrónicos. Para una versión con PSPMTs
con 256 canales, el número se cuadruplica. Incluso con muy bajos
costes por canal, el coste total para el sistema de
adquisición de datos, almacenamiento y procesamiento sería elevado.

La idea fundamental para la resolución de este problema es una pequeńa
mejora de los circuitos de división de carga que se usan para la
implementación analógica del algoritmo de centro de gravedad y se
explica de la siguiente manera: el cómputo del centroide o del primer
momento de una distribución discreta de cargas se puede realizar con
una cadena de resistencias del mismo valor. Una carga que se inyecta
en una de las interconexiones de la cadena se divide en dos
fracciones. Según la posición donde se inyecta la carga, estas
fracciones tienen diferentes valores siendo su suma
siempre la misma. Si las fracciones de
cargas varían linealmente con la posición, la diferencia de las
cargas totales extraídas de los extremos de la cadena de resistencias
es proporcional a la posición de la inyección, o, lo que es lo mismo, al
centroide. La variación lineal de las cargas se consigue  con
una variación lineal de las resistencias y por lo tanto con
resistencias de igual valor. Esto implica que la
carga inyectada ve la conexión en paralelo de las dos ramas de la
cadena. Como la variación de las resistencias con la posición es
lineal, la variación de la impedancia total vista por la carga es
cuadrática. Teniendo en cuenta que la anchura de la
distribución de luz en cristales continuos esta
correlacionada con la profundidad de interacción y que la
desviación estándar es un buen estimador estadístico para la anchura
de una distribución, la observación de la codificación cuadrática de
los voltajes ofrece un método muy eficaz para la medida de la
profundidad de interacción. 

El punto de partida del presente trabajo se basó en estas
observaciones ya que el desarrollo de circuitos de división de carga,
con capacidad para medir
un momento adicional sin perjudicar la medida de los centroides, puede
proporcionar un diseńo para detectores de rayos-$\gamma$ relativamente
barato pero con prestaciones similares a los de los diseńos basados en
cristales segmentados. No obstante, la mera medida del segundo momento
o de la desviación estándar a partir del segundo momento y de los
centroides no es suficiente para obtener una buena resolución
espacial. Aunque la medida de la profundidad de interacción pueda
ayudar a eliminar el error de paralaje, la resolución espacial
intrínseca de los detectores empeora sustancialmente hacia los bordes
del cristal debido a que los centroides están sometidos a una compresión
no lineal y dependiente de la profundidad de interacción. Este último
hecho posibilita la reconstrucción de la posición
verdadera del impacto del rayo-$\gamma$ a partir de los tres primeros
momentos no triviales. Este problema es un problema inverso típico pero también se
conoce como el problema de los momentos truncados
(Tkachenko et al.\ \cite{Tkachenko:1996}). Se trata de reconstruir la distribución 
a partir de una secuencia incompleta de los momentos de esta. 

Este trabajo esta organizado de la siguiente manera. Tras una
introducción general e histórica a la materia de Medicina Nuclear
en el capitulo~\ref{ch:introduction} se recapitulan en el
capitulo~\ref{ch:detect-comps-for-nuc-img} el diseńo
típico para detectores de rayos-$\gamma$ para esta disciplina, sus limitaciones
más comunes y propuestas de mejoras. El capitulo~\ref{ch:motivation}
resume la motivación para este trabajo. Una gran parte del trabajo se
destinó al estudio de las distribuciones de luz de centelleo
(capitulo~\ref{ch:light-distribution}) y al diseńo y comportamiento
teórico de circuitos de división de carga con capacidad de computar
analógicamente el segundo momento
(capitulo~\ref{ch:enhanced-charge-dividing-circuits}). Cada uno de los
dos capítulos se puede leer con independencia. En los 
capítulos~\ref{ch:experiment} y \ref{ch:position-reconstruction}  
se tratan respectivamente la validación experimental de los resultados de los capítulos
anteriores y un algoritmo para reconstruir la posición
del impacto del rayo-$\gamma$ a partir de las medidas
proporcionadas por las circuitos de división de carga modificadas.
Finalmente, se resumen los resultados más importantes en las
conclusiones (capitulo~\ref{ch:conclusiones-and-outlook}).

El capitulo~\ref{ch:light-distribution} está dedicado al estudio del
reparto de luz de centelleo sobre el área sensible del fotodetector.
Para este fin se optó por el uso de una parametrización analítica de
los efectos supuestamente más importantes. No se usó el método de
las simulaciones Monte Carlo aunque es muy común para estudios
similares. Las razones para esta decisión son la mejor
comprensión de la distribución de luz de centelleo finalmente
detectada y, una vez encontrado un modelo fiable y conforme con las
observaciones, su adopción más sencilla a nuevos diseńos de
detectores. Para llegar a las mismas
conclusiones que permite tal modelo analítico con simulaciones de
Monte Carlo,  muchas horas de simulación y muchas repeticiones con
diferentes parámetros hubieran sido necesarias. 
En todo caso, simulaciones de Monte Carlo incluyen los
mismos efectos físicos conocidos que se incluyeron en la
parametrización usada en este trabajo pero con la ventaja de que el modelo analítico
permite atribuir fácilmente detalles de la distribución a efectos
fundamentales aislados. Por ejemplo, se puede estudiar muy bien con
este modelo el efecto de usar cristales de extensión espacial
finita. Simplemente hay que establecer un modelo para un
cristal de dimensiones finitas y otro con dimensiones infinitas y
comparar los resultados. En el caso de simulaciones de Monte Carlo, ni
siquiera es posible hacer esta comparación. A parte de esto, las
simulaciones se llevan a cabo evento por evento, es decir, fotón por fotón, y
por lo tanto requieren un tiempo elevado de computación. 

En el modelo de la distribución de luz de centelleo se incluyeron lo
siguientes efectos: El punto de partida fue la ley del inverso cuadrado
que describe el reparto de intensidades en superficies esféricas para
fenómenos de radiación. Se ha de tener en cuenta que los fotodetectores en
general (y en particular el que se usa para este
trabajo) disponen de una superficie plana para la detección de los
fotones. Por lo tanto hay que multiplicar por el coseno del ángulo de
incidencia para compensar la diferencia de las áreas irradiadas.
Otro efecto fundamental es la auto-absorción de luz de centelleo por
el mismo cristal centellador. Aunque esta es normalmente muy baja por
razones obvias, puede resultar relevante para posibles caminos
de luz muy largos. La auto-absorción obedece la ley de atenuación
exponencial. Sobre todo
para puntos de observación lejos de la posición de
impacto se reducirá la intensidad detectada de la luz.
Estrictamente, la atenuación exponencial incluye dos efectos:
la absorción y la dispersión elástica de la luz incidente. Esta
última contribución causa un fondo de luz ya que la luz distribuida
puede ser detectada en otro punto de la superficie sensible del
fotodetector. Se supone que esta contribución es muy baja y no se 
incluyó en el modelo. 

El siguiente efecto es debido al interfaz óptico entre el
fotodetector y el cristal de centelleo. Para su protección, los
fotodetectores disponen siempre
de una ventana de entrada hecha de un material transparente para la
radiación que se quiere detectar. Esta ventana es de
un grosor finito y en general su índice de refracción es diferente al
del cristal de centelleo. Muchos de los cristales de centelleo con
aplicación para Medicina Nuclear tienen un índice de refracción muy
elevado y mayor al de la ventana de entrada del fotodetector. En este
interfaz óptico se produce reflexión total cuando
el ángulo de incidencia supera al ángulo crítico. 
Debido a este hecho, el cristal centellador ha de acoplarse al fotodetector
mediante grasa óptica de un índice de refracción intermedio. De otro
modo no se pueden evitar la inclusión de una fina capa de aire que
reduce considerablemente la eficacia de recolección de luz.
La luz de centelleo que pasa a la ventana de entrada se desvía según la ley de
Snell y la amplitud de la misma viene descrita por las ecuaciones de
Fresnel que también reproducen bien la reflexión de una fracción
residual de la luz incidente para ángulos de incidencia menor al ángulo crítico.
La refracción de Snell y la transmisión y reflexión de Fresnel se incluyeron
en el modelo analítico suponiendo además, que la luz de centelleo no
esta polarizada. El siguiente fenómeno que se tuvo en cuenta requiere
especificar que tipo de fotodetector se usa para el diseńo del detector de
rayos-$\gamma$, ya que las propiedades de los mismos pueden resultar muy
diferentes. Como ya se había mencionado arriba, para el presente
trabajo se optó por un fotomultiplicador sensible a posición.
La sensibilidad del fotocátodo del mismo no es constante para
diferentes  ángulos de incidencia. Esto es debido a varias
circunstancias. Una de ellas es la limitación de que
los fotocátodos tienen que tener un grosor muy pequeńo para asegurar que
los fotoelectrones puedan salir del mismo y ser recolectados por el
primer dínodo. Por esta razón, la eficiencia cuántica no es muy
elevada porque una fracción alta de la luz de centelleo es transmitida
sin ser detectada. Los fotones de luz con un ángulo de incidencia elevado
tienen que recorrer una trayectoria más larga dentro del fotocátodo y
su probabilidad de crear un fotoelectron es más elevado. 

Aparte de la luz de centelleo que llega directamente a ser detectada
por el fotomultiplicador también existen contribuciones debidas a
reflexiones en los cinco lados del cristal centellador que no están
acopladas ópticamente al fotodetector. Se conocen numerosos
estudios que demuestran que el acabado de estas superficies es muy importante para la
eficiencia de recolección de luz y la resolución espacial. No
obstante, el método de deducir la profundidad de interacción a partir
de la anchura no permite usar acabados reflectantes sino que requiere la
supresión de esta luz. Por este motivo se cubrieron estos lados con resina epóxica negra. Aunque el
coeficiente de reflexión de este material es muy bajo, el área total
de las superficies cubiertas con este material es elevado, y, como se
verá en el capitulo~\ref{ch:experiment} de los resultados
experimentales, no es despreciable especialmente para
profundidades de interacción cerca del limite superior de los
posibles valores. Ya que los cristales de centelleo no están pulidos sino
cubiertos con resina epóxica negra, dicha reflexión
residual es supuestamente difusa y su comportamiento se 
aproximó con la ley de Lambert. También hay que tener en cuenta
que gran parte de la luz procedente del punto de fotoconversión no es capaz de entrar a la ventana de
entrada debido a la reflexión total. En su lugar, esta luz reflectada
se refleja una segunda vez y de forma difusa en las otras superficies negras.
Esta contribución es igual de importante que la luz que se refleja
directamente en las superficies negras y por lo tanto se incluyó en el
modelo analítico.
Otros efectos como refracción, transmisión de Fresnel o sensibilidad
angular del fotocátodo no se tuvieron en cuenta para la luz de fondo debido a
la reflexión difusa. La distribución de seńal observada en los
segmentos del ánodo del fotodetector es el resultado de la acción conjunta
de todos los efectos descritos y suponiendo como procesos ideales la
recolección de los fotoelectrones por el sistema de dínodos y su
multiplicación.

En el siguiente capítulo~\ref{ch:enhanced-charge-dividing-circuits} se
analizaron  detalladamente las propiedades de diferentes
implementaciones de circuitos de división de carga. También se
mostró como se pueden mejorar estos circuitos para que computen
simultáneamente el segundo momento sin perjudicar a los centroides.
Aparte de la lógica de Anger tradicional existen otras
posibilidades de implementación del algoritmo de centro de gravedad con
redes de resistencias (Siegel et al.\ \cite{Siegel:1996}). Las tres variantes más
comunes muestran una calidad de posicionamiento muy parecido pero
hay importantes diferencias en la cantidad de resistencias
necesarias. Como se ha explicado anteriormente, las
corrientes (o equivalentemente las cargas) inyectadas causan un
potencial codificado cuadráticamente. Esto se debe a la codificación lineal
para la computación de los primeros momentos, los llamados
centroides. Por lo tanto un sumador analógico ya es suficiente para
obtener una única seńal adicional que representa el segundo momento. 
Ya que los componentes necesarios para este sumador son un amplificador
operacional y unas pocas resistencias y condensadores, el coste total
viene únicamente dado por el canal electrónico adicional (por detector
de rayos-$\gamma$) para la digitalización del segundo momento. Si
retomamos el ejemplo de un PET para animales pequeńos construido con 8 detectores
con cristales continuos, el uso del algoritmo de centro de gravedad
analógico reduce el numero total de canales electrónicos necesarios a 32
en vez de 512 para PSPMTs de 64 ánodos o de 2048 para PSPMTs de 256
canales. Además el número de canales electrónicos necesarios no
depende del numero de segmentos de ánodos del tipo de PSPMT
usado. Esto hace el método del centroide muy versátil. Con la mejora
para la medición simultánea de los segundos momentos harán falta 40
canales en vez de 32 lo que no supone ningún problema de realización.

Aunque el comportamiento de las diferentes variantes de los circuitos
de división de carga es muy similar para el centroide, el
comportamiento respecto a la medida del segundo momento muestra
importantes diferencias. Un aspecto es la simetría en el
comportamiento de los circuitos de división de cargas
respecto al intercambio de las coordenadas espaciales $x$ e $y$. La
lógica de Anger original posee esta simetría inherentemente. Sin
embargo, tanto las configuraciones electrónicas de la versión basada en
cadenas proporcionales de resistencias como la de la versión híbrida,
que es una mezcla de las otras dos, rompen esta simetría. Para
restablecer la simetría por completo para los centroides en los dos
casos, algunas resistencias tienen que tener valores determinados que
dependen de la configuración y los valores de las otras.
En el caso del segundo momento se
puede restablecer la simetría sólo para la variante híbrida. Para la
versión basada en cadenas proporcionales de resistencia no es posible
fijar los valores de las resistencias de una manera tal que el
circuito se comporte exactamente igual en la medida de los cuatro
momentos (energía, centroides y segundo momento) para las dos
direcciones espaciales. No obstante, en el caso optimizado, la
disimetría residual para el segundo momento es menor del 1\% y los
otros tres momentos se comportan de forma totalmente simétrica. Para obtener esta
simetría óptima hay que aceptar que el segundo momento contiene
contribuciones de los ordenes $\mathcal{O}(x^4)$, $\mathcal{O}(y^4)$ y
$\mathcal{O}(x^2y^2)$. Este hecho tiene consecuencias cuando se quiera
usar la desviación estándar como estimador de la profundidad de
interacción pero no significa ninguna desventaja para el método
de reconstrucción de la posición que se presentara en el
capitulo~\ref{ch:position-reconstruction}.

Otro aspecto estudiado en el
capitulo~\ref{ch:enhanced-charge-dividing-circuits} es la influencia
de la impedancia de entrada del sumador analógico sobre la medida de los
centroides. Obviamente no se puede permitir que la medida
complementaria reduzca significativamente la calidad de las medidas de
los centroides o la de la energía. Para que este requisito se cumpla
hay que asegurar que el sumador extraiga muy poca corriente de los
circuitos originales. Desgraciadamente no se pueden usar seguidores
de tensión para este fin ya que el consumo medio de tal amplificador
sería de unos $\mathrm{20\,mA}$ que asciende a unos $\mathrm{1.2\,A}$
 para el modulo entero si el
PSPMT tiene 64 segmentos de ánodo. La impedancia de entrada de una
rama del sumador analógico viene dada aproximadamente por la
resistencia de entrada que a su vez determina el peso con que la
seńal correspondiente entra en la suma total. Por lo tanto, 
estas resistencias tienen que tener valores elevados pero no deben
superar cierto límite, ya que valores demasiado altos introducirán ruido
térmico. Como criterio de diseńo se usa el hecho de que las
resistencias reales y comerciales tienen una tolerancia en su valor de
1\%. Carece de sentido calcular los valores de resistencia con
mayor precisión. Este aspecto se tiene que tener en cuenta para 
las tres variantes de los circuitos de división de carga.

El efecto de dispersión de Compton de los rayos-$\gamma$ dentro del
cristal centellador es el objetivo del capitulo~\ref{ch:compton}.
El modelo de la distribución de seńal que se desarrolló en el
capitulo~\ref{ch:light-distribution} es solo válido para eventos que
depositan toda su energía en una sola interacción, es decir, para
fotoconversiones por efecto fotoeléctrico. Especialmente para los
fotones de energía $\mathrm{511\,keV}$ de la
modalidad de PET cabe la posibilidad de que experimenten
varias dispersiones de Compton antes de ser absorbidos por completo. 
Obviamente, sólo la posición de la primera interacción corresponde a
la linea de vuelo correcta del fotón gamma. Con la lógica de Anger y
sus variantes descritas anteriormente no es posible determinar
esta posición. En su lugar, se medirá el centroide de la superposición 
de varias distribuciones procedentes de deposiciones puntuales de
energía, ya que cada interaccíon por efecto Compton depositará una
fracción de la energía inicial del fotón incidente. Esto resultará en
un error de la posición de impacto medida y del segundo momento y reducirá la
resolución espacial transversal y la de la profundidad. 
Para estimar el impacto de dispersión de Compton sobre dicha
resolución se llevó a cabo una simulación de Monte Carlo con el
paquete GEANT 3 (Brun and Carminati \cite{Geant3:1994}). Se simularon las interacciones de
20000 rayos-$\gamma$ de $\mathrm{511keV}$ en un cristal centellador de LSO con
dimensiones $\mathrm{40\times40\times20\,mm^3}$. Como resultado se
pueden resumir las siguientes dos observaciones. La incertidumbre
introducida por este efecto es en la mitad de los casos menor a
$\mathrm{300\,\mu m}$ tanto para las coordenadas paralelas al
fotocátodo como para la componente normal. La otra mitad de los eventos
se reparte en una cola muy larga de baja intensidad atribuyendo sobre
todo ruido de fondo por que las distancias son más grandes que las
resoluciones espaciales medidas obtenidos en los capítulos~\ref{ch:experiment} y
\ref{ch:position-reconstruction}. El otro efecto observado es la
opresión de eventos de dispersión hacia delante, o sea con ángulos de
dispersión cerca de cero grados. Esta opresión se puede explicar de la siguiente
manera. Debido a la formula de Klein-Nishina (Leo \cite{Leo:1994}), la
dispersión de Compton para rayos-$\gamma$ de $\mathrm{511\,keV}$ favorece
fuertemente ángulos de dispersión alrededor de cero grados. Además, para esta
energía, la probabilidad de que un fotón experimente una
interacción de Compton es ya muy reducida y se requiere un grosor
elevado para parar eficientemente dichos rayos-$\gamma$. No obstante, el
cristal simulado es solo de un grosor de ${\mathrm{20\,mm}}$ y un gran número de
fotones pasará el cristal sin ser detectado. Otro número elevado de
fotones experimentará una interacción de Compton con un ángulo de
dispersión muy pequeńo. En consecuencia, la energía del fotón que sale seguirá
siendo muy elevada y la probabilidad de interacción muy baja. Por lo
tanto, todos estos fotones probablemente escapen del cristal
sin ser detectados. Sin embargo, en el caso de que la primera
interacción sea de tipo Compton y con un ángulo de dispersión elevado, la pérdida de
energía del fotón también será elevada y el fotón que salga de esta
interacción tendrá una probabilidad de interacción mucho más
alta. Aparte de esto se moverá más o menos en paralelo al fotocátodo
de forma que aumentará aún más la probabilidad de su detección, ya que la extensión
transversal del cristal simulado es el doble de la extensión normal.
Estos efectos dan lugar a una predisposición hacia ángulos de
dispersión elevados para el subconjunto de eventos detectados.
Probablemente esta es la causa de que el método para medir la
profundidad de interacción que se presenta en este trabajo de lugar a
resultados suficientemente buenos para su aplicación en detectores
reales.

El siguiente capítulo~\ref{ch:experiment} abarca las verificaciones
experimentales de los resultados de los tres anteriores capítulos. 
Para llevar a cabo los experimentos se
usaron dos detectores iguales. Cada uno esta compuesto por un único
cristal del centellador LSO de grandes dimensiones
($\mathrm{42\times42\times10\,mm^3}$) y de un PSPMT del tipo H8500
de la empresa Hamamatsu Photonics Inc.\ (Hamamatsu \cite{data:H8500}). 
Debido a la radiactividad intrínseca del LSO, medidas con fuentes de
actividad menor de $\mathrm{20\,mCi}$ se tienen que llevar a cabo en
coincidencia temporal (Huber et al.\ \cite{Huber:2002}). A parte de
ello, medidas en coincidencia con dos detectores de rayos-$\gamma$ con
resolución espacial permite una colimación electrónica del haz.
El fotomultiplicador H8500 tiene un área sensible de
$\mathrm{49\times49\,mm^2}$ y dispone de 64 segmentos de ánodo.
La seńal de disparo para el módulo se derivó de los últimos dínodos de
los PSPMT. Un discriminador
del tipo {\em leading edge} admitió solo eventos a partir de cierto umbral y 
creó pulsos lógicos de anchura temporal de $\mathrm{\lesssim 5\,ns}$. 
A partir de estas dos seńales se creó la seńal de coincidencia
temporal con una puerta lógica de función booleana AND que sirvió para
dos funciones. Se usó para derivar otro pulso lógico de anchura
temporal de $\mathrm{400\,ns}$ y retrasado por $\mathrm{200\,ns}$. El
flanco de subida de este pulso se usó para iniciar el proceso de integración y
su flanco de bajada la finalizó. El resultado de esta integración de la
corriente es proporcional a la carga total extraída de los
fotomultiplicadores y fue convertida a valor digital. Las operaciones
de restauración de línea base, integración de carga y digitalización se
realizaron con una tarjeta de 12 canales electrónicos
(Zavarzin and Earle \cite{Zavarzin:1999}). 
 La ventana temporal se obtiene como suma directa de las dos anchuras
 de las seńales que proporcionan los discriminadores. Estos se 
 ajustaron a su límite  inferior de $\mathrm{\lesssim5\,ns}$ con lo cual la
 ventana de coincidencia fue de unos $\mathrm{10\,n s}$. 

Una vez digitalizadas las 10 seńales de los dos módulos se transfirieron
al ordenador para la computación de los 4 momentos. Un modulo se usó
como detector de testeo mientras el otro sólo tuvo las funciones de
detector de coincidencia temporal y de la colimación electrónica. 
La distancia total entre los dos módulos era de unos
$\mathrm{\approx25\,cm}$ y la fuente radiactiva (\isotope{Na}{22},
actividad nominal $\mathrm{10\,\mu Ci}$) se colocó entre ellos de
forma centrada y muy cerca (a unos $\mathrm{\approx3\,mm}$ del cristal)
del detector de testeo. De esa manera se pudo colimar el haz
electrónicamente al seleccionar eventos de coincidencia temporal con
una posición de impacto (en el detector de coincidencia) que cayó
dentro de un circulo central de diámetro $\mathrm{12\,mm}$. Por
argumentos geométricos, la región de posiciones en el detector de
testeo tiene que ser un circulo de  diámetro $\mathrm{0.2\,mm}$. 
Mientras el detector de coincidencia y la fuente radiactiva estuvieron
alineados y estacionarios, el detector de testeo estuvo montado encima de
una mesa $x$-$y$ computerizada. Esto permitió
variar la posición del impacto del rayo-$\gamma$ a lo largo del plano del
fotocátodo y se pudieron medir de forma automática los diferentes
momentos en diferentes posiciones.

Dos detalles muy importantes que hay que tener en cuenta son los
siguientes. Primero, la fuente de radiación no fue puntual sino que tuvo
un diámetro de aproximadamente $\mathrm{1\,mm}$. Además, como el
\isotope{Na}{22} decae pro radiación $\beta^+$, estos positrones penetran
hasta dentro la cápsula de resina. Por lo tanto, el diámetro efectivo 
que se obtiene a partir de la radiación de aniquilación es diferente a 
$\mathrm{1\,mm}$. Se estimó por simulaciones Monte Carlo, que
el diámetro efectivo es de unos $\mathrm{0.92\,mm}$. La resolución espacial
que se espera para el detector de rayos-$\gamma$ diseńado en este trabajo
es del mismo orden y por lo tanto se tuvo que corregir mediante el
diámetro efectivo de la fuente radiactiva. El segundo efecto que
juega un papel muy importante es que no se puede prepara el haz de
fotones gamma para que estos interaccionen en una profundidad del
cristal determinada. Mientras las componentes paralelas al plano del
fotocátodo se puede prepara fácilmente posicionando el detector de
testeo en la posición deseada, la profundidad de interacción es una
variable completamente aleatoria. Sin embargo, diferentes detecciones
con diferentes profundidades causan distribuciones de luz de centelleo
con diferentes segundos momentos y por lo tanto se pueden distinguir
estos sucesos si la resolución en la medida de este momento es 
suficientemente alta. Se ha de observar una estadística muy
característica para este momento que refleja una caída exponencial
debido a la absorción de los rayos-$\gamma$ dentro del cristal de LSO,
una resolución intrínseca para el segundo momento y los limites superiores
e inferiores para el momento, ya que el cristal es de un grosor finito y
solo se pueden detectar eventos cuyo segundo momento corresponda a una
profundidad real existente y de acuerdo con las dimensiones del
cristal. Se ha ideado un modelo que describe bien el comportamiento
de dicha distribución y que permite extraer los parámetros clave que
son los limites superiores e inferiores y la resolución intrínseca.

Una vez acabados todos estos preparativos se verificó el modelo
de la distribución de luz establecido en el
capitulo~\ref{ch:experiment}. Para este fin se midieron los cuatro
momentos en $\mathrm{9\times9}$ posiciones distribuidas de forma
centrada, cartesiana y con una distancia de $\mathrm{4.75\,mm}$ entre
ellas. Estas medidas se compararon con las predicciones del modelo
analítico para las mismas posiciones y momentos. Para finalizar este
capitulo, se estimó la resolución en la posición tridimensional del
detector propuesto en el caso de que se usaran estos mismos momentos
como estimador de posición de impacto y energía. Para obtener resoluciones
reales se hubo que corregir por la compresión que introducen los
centroides y por el efecto del diámetro efectivo de la fuente radiactiva.

El capítulo~\ref{ch:position-reconstruction} se motivó por las
observaciones en los
capítulos~\ref{ch:enhanced-charge-dividing-circuits} y
\ref{ch:experiment}. 
Los resultados del
capitulo~\ref{ch:experiment} muestrearon, que los momentos de la
distribución de luz se pueden medir con buena resolución aunque estos
momentos no constituyen estimadores de gran validez para la posición de
impacto real. También se observó en el capitulo anterior, que el
modelo analítico para la distribución de seńal derivada en el
capitulo~\ref{ch:light-distribution} proporciona momentos que
concuerdan muy bien con las medidas. Por lo tanto se tiene un modelo
que es capaz de predecir muy bien el comportamiento de las seńales que
proporciona el detector de rayos-$\gamma$. A parte de esto se tiene para
cada posición tridimensional de impacto y su energía un número total
de cuatro momentos de la distribución. La reconstrucción de la
posición real a partir de estos momentos es un típico problema inverso
y su viabilidad se estudió en el
capitulo~\ref{ch:position-reconstruction}.
El problema también se conoce como problema de momentos
truncados que se ha estudiado intensivamente desde su
descubrimiento (Talenti \cite{Talenti:1987}, Kre\u{\i}n and  Nudel'man
\cite{Krein}, Jones and Opsahl \cite{Jones:1986}). Desgraciadamente,
todos los algoritmos basados en este método requieren una secuencia de más
de 4 momentos para una reconstrucción viable. Por lo tanto se optó por
la interpolación polinómica que se aplicó con éxito para un
problema muy similar (Olcott et al.\ \cite{Olcott:2005}). Este método usa para la
reconstrucción de las posiciones reales una matriz de corrección que se
obtiene a partir de la inversa de Moore-Penrose de los coeficientes
de los polinomios de interpolación de los momentos y sus
correspondientes posiciones de impacto. El funcionamiento correcto de
este método se comprobó usando los datos de medida en todas las
81 posiciones obtenido en el capitulo~\ref{ch:experiment} de forma
cualitativa. Después se midió la resolución espacial del propuesto
detector en los 81 puntos usando esta vez la posición reconstruida en
vez de los momentos y se compararon con los anteriores resultados
usando los momentos como estimadores de posición. Se intentó
también, pero sin éxito, la reconstrucción de la energía verdadera.
Otra vez hubo que corregir por el efecto del diámetro efectivo de la
fuente radiactiva y el de la compresión residual de las posiciones
reconstruidas.

Para terminar el trabajo se resumieron en
capitulo~\ref{ch:conclusiones-and-outlook} las principales conclusiones
de los diferentes capítulos y algunas perspectivas para
investigaciones futuras. Por último se incluyeron los
apéndices~\ref{app:common-radiotracer}-\ref{app:elec-config} que
contienen algunos datos de interés como radiofármacos comunes,
centelladores típicos, resultados complementarios y configuraciones
electrónicas detalladas.

\section*{Discusión de los resultados y conclusiones}

En el presente trabajo  se ha desarrollado un método innovador
para medir la profundidad de interacción de rayos-$\gamma$ en cristales
de centelleo gruesos y continuos. La nueva técnica consiste en estimar este 
parámetro usando la anchura de la distribución de luz de centelleo
en los cristales que es detectada por un fotomultiplicador.
Para su medida rápida y sencilla se ideó una modificación de muy bajo
coste de los circuitos convencionales de división de carga que se usan
con gran frecuencia para la determinación de la posición del impacto
en detectores de rayos-$\gamma$ para la Medicina Nuclear.

Para que la calidad de la imagen médica sea alta respecto a la relación
seńal a ruido, el contraste y la resolución espacial, el detector de
rayos-$\gamma$ tiene que proporcionar información sobre la posición
tridimensional del impacto, especialmente para al modalidad de PET.
Sin esta información, se introduce un error de paralaje para todas las
posiciones fuera del centro y que tiene mayor importancia en zonas de
la región de interés que están lejos de este centro.
Es más, detectores de rayos-$\gamma$ de tipo Anger convencionales aproximan los
componentes transversales de la posición de impacto usando los
centroides, o bien los primeros momentos normalizados, de la
distribución de seńal. Varios grupos han observado que el algoritmo de
centro de gravedad como estimador de posición de impacto produce
errores no-lineales y dependiente de la profundidad de interacción
para cristales gruesos. Esto perjudica la resolución espacial cerca de
los lados y especialmente en las esquinas del detector.

En el capitulo\ref{ch:experiment} se discutieron con detalle los
errores debidos al algoritmo de gravedad. Se reveló que la compresión
de las posiciones es causada por la ruptura de la simetría de la
distribución de seńal por culpa de un cristal de dimensiones
espaciales finitas. Por esta razón, la insuficiente resolución
espacial obtenida con cámaras Anger convencionales no es debida a una
medida de baja resolución de los momentos por los circuitos de división de carga,
sino a que la aproximación de la posición de impacto por estos
momentos no es valida para estas regiones del área sensitiva.
No obstante, se mostró en el capítulo~\ref{ch:experiment} que los
momentos pueden ser medidos con alta resolución. Igualmente se
observó, que la impedancia de entrada de los circuitos de división de
carga están codificados cuadráticamente con la posición y por tanto
producen voltajes con la misma propiedad bajo la inyección de
corrientes procedentes de los fotomultiplicadores. Una configuración
tan simple como un sumador analógico puede ser usado para sumar estos
voltajes y proporcionar una seńal adicional linealmente
correlacionada con el segundo momento. Junto con la observación de
otros grupos (Rogers et al.\ \cite{Rogers:1986}, Kenneth et al.\
\cite{Kenneth:2001}, Antich et al.\ \cite{Antich:2002}) de que la
anchura de la distribución depende fuertemente de la profundidad de
interacción, esto proporciona un método potente para medir la misma
profundidad de interacción.

En el capítulo~\ref{ch:enhanced-charge-dividing-circuits} se
demostró, que todas las versiones conocidas de circuitos de
división de carga pueden ser modificados con un sumador analógico para
medir el segundo momento. También se vio, que las diferencias teóricas
en las cualidades de estas versiones solo varían poco de una a otra.
Esto se observó también experimentalmente para los centroides y la carga total
(Siegel et al.\ \cite{Siegel:1996}) y por lo tanto el criterio para la elección de la
variante del circuito de división de carga puede ser la complejidad de
la red de resistencias. Se dieron también en el
capítulo~\ref{ch:enhanced-charge-dividing-circuits} expresiones
explicitas para la dependencia de los voltajes y la suma de ellos en
función de la posición de la corriente inyectada y  en función
de la configuración del circuito. Comparaciones con simulaciones con
{\sc Spice}, (Simulation Program with Integrated Circuits Emphasis,
  Tietze and Schenk \cite{Tietze}) 
concuerdan muy bien con las predicciones hechas con estas
fórmulas y las diferencias máxima es en todos los casos menor de un 3\%.
Un resultado complementario está relacionado con el comportamiento
de la simetría de los circuitos. La lógica de Anger convencional es
inherentemente simétrica respecto al intercambio de las posiciones $x$
e $y$, pero las otras dos versiones no lo son. Afortunadamente se puede
restaurar esta simetría por completo para los centroides, y, en el
caso de la red híbrida, también para el segundo momento. Para el
circuito basado completamente en cadenas de resistencias, sólo se
puede minimizar la disimetría, aunque se consiguen valores residuales muy
pequeńos de 1\% o menos. Para conseguir esto, hay que
aceptar que se introducirán ordenes de $\mathcal{O}(x^4)$,
$\mathcal{O}(y^4)$ y $\mathcal{O}(x^2y^2)$ en el segundo momento.
Esto sólo supone un problema si se quiere usar la desviación estándar
como estimador para la profundidad de interacción. Para el método
presentado en el capítulo~\ref{ch:position-reconstruction}, estos
ordenes elevados no suponen ninguna complicación adicional. La impedancia de entrada
de los sumadores se tiene que dimensionar de tal forma, que evite la
extracción excesiva de corriente del circuito para los centroides. En
 caso adverso, esto perjudicaría a los mismos lo que no es aceptable. 
La solución ideal sería usar seguidores de tensión, ya
que estos tienen una impedancia de entrada muy elevada. Su uso no es
posible debido a su alto consumo. Esta opción
esta reservada para un futuro diseńo de un circuito ASICs (Application-Specific Integrated
Circuit) y no forma parte del presente trabajo. Por estas razones, los
valores de las resistencias para los sumadores tienen que ser
en general muy elevados. Una indicación adversa al uso de valores
demasiado altos es el ruido térmico. En el presente caso se obtuvieron
 resultados aceptables con valores de resistencias al sumador
que extraen como máximo un 1\% de corriente en cada nodo del circuito de
división. Para la realización del detector de rayos-$\gamma$ se usó 
un circuito basado completamente en cadenas de resistencias y un
sumador con 64 entradas ya que esta versión es
la que más fácilmente se implementa.

En el capítulo~\ref{ch:experiment} se presentaron medidas de los 4
momentos de un detector real. El detector está basado en un cristal de
LSO de dimensiones de $\mathrm{42\times42\times10\,mm^3}$ y un
fotomultiplicador H8500.  Los experimentos muestran que los centroides
no están afectados por la medida del momento adicional. La resolución
media en estos momentos es menor del $\mathrm{5\%}$. También se
observó, que la aproximación de usar estos cuatro momentos como posición de
impacto es inadecuada. Los mismos resultados obtuvieron otros
grupos que investigaban el comportamiento de los centroides sin
medida del segundo momento. El momento trivial representa la energía
del impacto y los momentos no-triviales son los centroides y el
segundo momento. El momento trivial se
ve afectado por efectos y condiciones adicionales y no alcanza la
resolución de los momentos no triviales. Una causa para esto es la
inhomogeneidad del fotocátodo de los fotomultiplicadores. La
eficiencia y la ganancia puede variar de un segmento de ánodo a otro
hasta alcanzar diferencias de un factor 3. Esta falta de uniformidad introduce
una variación de energía detectada adicional e importante. Por otro
lado, el método para medir el segundo momento que se presenta con este
trabajo requiere que todas las superficies que no están acopladas al
fotodetector estén cubiertas de una capa muy absorbente para evitar
reflexiones, ya que estas destruyen por completo la correlación de la
profundidad de interacción con el segundo momento. Obviamente esto
reduce la eficacia de recolección de luz y por lo tanto la resolución
energética. Este efecto es de muy elevada importancia en las esquinas
del detector y las resoluciones energéticas son muy bajas en estas zonas.
En los experimentos se observó una resolución energética
media del $\mathrm{25\%}$ con el valor mínimo en el centro
del $\mathrm{17\%}$ y el valor máximo ($\mathrm{70\%}$) en una de las
esquinas. La degradación de la resolución energética se compone de dos
efectos. Un efecto importante es la variación del total de la luz
detectada por razones
geométricas. Para puntos de fotoconversión muy cerca de una superficie
negra, menos luz es detectada y el máximo del espectro está en canales
más bajos. En los histogramas de energía se superponen muchos eventos con
diferentes posiciones y por lo tanto se obtiene una única distribución
muy ancha debido al movimiento del máximo del fotopico. Este efecto se
puede corregir una vez obtenida la posición real del impacto y
conociendo el comportamiento del momento trivial para todo el volumen
del cristal. El hecho de que no se podía corregir la energía  
como parte del presente trabajo es probablemente debido a una resolución espacial aún no
suficiente para este fin. El otro efecto es el de la variación por
estadística de Poisson. Este efecto no se puede corregir con la
posición aún teniendo 
una resolución muy buena en la misma. El uso de retroreflectores
(Karp and Muehllehner \cite{Karp:1985}, Rogers et al.\
\cite{Rogers:1986}, McElroy et al.\ \cite{McElroy:2002}) puede probablemente mejorar
este aspecto.

El modelo para la distribución de luz se verificó
experimentalmente en el capítulo~\ref{ch:experiment}. Para los tres
momentos no triviales se observó que las predicciones del modelo
reproducen muy bien las medidas de estos momentos. Las desviaciones
siempre estaban por debajo del $\mathrm{11\,\%}$, excepto para el momento
trivial. En este último caso, el modelo no reproduce correctamente
todos los detalles de las medidas. Las predicciones del modelo concuerdan
bien con los momentos medidos para profundidades de interacción cerca
del limite inferior. En el caso opuesto, es decir, para  profundidades
de interacción cerca del limite superior, se producen discrepancias
obvias entre el modelo y las medidas. Estas observaciones se pueden
explicar fácilmente con las aproximaciones que se hicieron para llegar
al modelo para la distribución de luz de fondo. Se suponía que la
contribución total no fuera muy elevada. No obstante, para profundidades
de interacción elevadas, la contribución de luz de fondo a la distribución
total se vuelve muy importante. Esto se verificó con un modelo
alternativo que no disponía de luz de fondo. Sin luz de fondo, el modelo reproduce
las variaciones del momento trivial a lo largo del fotocátodo mucho
peor. Sin embargo, los resultados para los momentos no-triviales se
reprodujeron con una calidad muy similar. Esto se espera, ya que la
normalización de estos momentos elimina de forma eficiente la
dependencia del momento trivial. Razones para los
errores en estos momentos son probablemente la influencia de
dispersión de Compton y sobre todo la precisión mecánica. Aunque la
mesa $x$-$y$ dispone de muy buena precisión, el resto del montaje,
que incluye la fijación de la fuente y de las carcasas de los detectores de
rayos-$\gamma$, no alcanza la misma precisión. Esta última fuente de error ha de minimizarse para
obtener mejores resultados en medidas futuras.

El capítulo~\ref{ch:position-reconstruction} se dedicó a encontrar un
algoritmo para la reconstrucción de la posición de impacto real a
partir de los momentos. Para este fin se usó el modelo de la
distribución de seńal, ya que en el
capítulo~\ref{ch:experiment} se verificó que esta reproduce bien 
los momentos no-triviales. Se usó el modelo para predecir el
comportamiento del detector en 40000 diferentes posiciones de
impacto. La respuesta del detector consiste en los tres momentos
no-triviales y el momento trivial. Los resultados para los dos
centroides y el segundo momento se interpolaron con ordenes 12 para los
componentes transversales y con orden 5 para el componente normal. 
Según el capítulo~\ref{ch:position-reconstruction}, se puede usar la
inversa de Moore-Penrose en conjunto con las 40000 posiciones de
impacto para obtener una matriz del detector que permite la
reconstrucción de la posición de impacto. A continuación, se calcularon
las posiciones a partir de los momentos. La resolución espacial del
detector era en este caso de $\mathrm{1.9\,mm}$ para las dos dimensiones
transversales y de $\mathrm{3.9\,mm}$ para la profundidad de
interacción. Esto presenta una mejora sustancial con respecto a la
resolución del detector obtenido usando los momentos ($\mathrm{3.4\,mm}$
y $\mathrm{4.9\,mm}$) para las mismas coordenadas. Especialmente el
resultado para la resolución en profundidad es muy importante, ya que
existen muy pocos métodos que llegan a esta resolución.
No se consiguió corregir por completo la no-linealidad
de la posición con este método. Con los valores mencionados aquí, se obtuvo una
no-linealidad residual de aproximadamente el $\mathrm{10\%}$. Este error
es del mismo orden que el error del modelo observado en
capítulo~\ref{ch:experiment}. Probablemente, la precisión del modelo
analítico tiene que superar este valor para obtener  mejores linealidades y
resoluciones. La resolución espacial y tridimensional que se
obtiene de momento con el método presentado no es suficiente para
reconstruir la energía real a partir del momento trivial con la
información de los momentos no-lineales.

En este trabajo se ha presentado un método simple y barato para medir
el segundo momento de la distribución. Se ha mostrado, que esta
información adicional se puede utilizar conjuntamente con los centroides
para reconstruir la posición real del impacto. De esta manera se
pueden realizar detectores de rayos-$\gamma$ para Medicina Nuclear que
proporcionan información sobre la profundidad de interacción y que permiten
reducir el error de paralaje. El método presentado es apto para
cualquiera de las modalidades en las que hace falta saber la información de
profundidad y es muy barato. No obstante, el algoritmo
de inversión no es óptimo y requiere futura investigación.

\selectlanguage{american}

\renewcommand{\bibname}{\Large Referencias}
\bibliographystyle{IEEEtran}\bibliography{mrabbrev,IEEEabrv,bibliography}

\cleardoublepage{}


\renewcommand{\headrulewidth}{0.5pt}

\tableofcontents
\cleardoublepage
\pagenumbering{arabic}

 \cleardoublepage
\chapter{Historical Introduction}
\label{ch:introduction}

\chapterquote{%
Fortune knocks but once, but misfortune has much more patience.}{%
Laurence J.\ Peter, $\star$ 1919 -- $\dagger$ 1988
}

\PARstart{G}{amma}-ray imaging covers only a small area of the large 
spectrum of imaging techniques applied to medical diagnostics. Many of these
techniques, {\em e.g.}\ Radiography, Sonography and Nuclear Magnetic
Resonance (NMR), have already been in routine use for many years. Others are
at an early stage of development and far from being widely applied. 
Since 1895,  when the possibility of using X-rays  
for planar transmission imaging was discovered by Wilhelm Conrad
Röntgen at the university of Würzburg (Germany), all techniques have been
under active development to a greater or lesser extent.
In that year, Röntgen observed a green colored fluorescent light generated by a
material located a few feet away from a working cathode-ray tube.
He attributed this effect to a new type of ray that he supposed had
been emitted from the tube and found that the penetrating power of the new
ray also depended on properties of the exposed substances casting 
the object's density distribution into a two dimensional projection. 
One of Röntgen's first experiments with the newly discovered
radiation was a projection image of the hand of his wife Bertha.
An important contribution to radiography diagnostics was made by Carl
Schleussner. He developed the first silver bromide photographic X-ray
films, which made archival storage  of diagnostic results possible and also
lowered the necessary exposure.
Within only a month after the announcement of the discovery, 
several medical radiographs had been built. They were used by surgeons to guide 
them in their work and only a few months later they were used to
locate bullets in wounded soldiers. 

Although radiography was the first medical imaging modality, the
first attempts to see inside the human body without invasive 
operation go back a longer time (Wayand, \cite{Wayand:2004}). 
When in the year 1879 Maximilian Nitze and Josef Leiter introduced 
the first optical system in Vienna using a platinum glow wire as light 
source, they laid the foundations of Endoscopy. Only two years later
and also in Vienna, the surgeon Jan Mikulicz-Redecki demonstrated the first 
Gastroscopy (telescopic inspection of the inside of the gullet,
stomach and duodenum). However, the first commercial semi-flexible 
Gastroscope, designed by Georg Wolf and Rudolph Schindler, did not
appear until 1932. 

Thermography and Electrocardiography are two other examples of medical
imaging modalities that were known before the discovery of X-rays by
W.C.~Röntgen. As for thermography, the knowledge even goes back to
Hippocrates, who first obtained
thermograms of the chest. He proposed covering the patient's thorax
with a piece of thin linen soaked with earth, and observing the
process of drying. At the warmer areas of the thorax, the earth-soaked
cloth dries faster and the pattern of enlargement of the dry areas
represents the temperature distribution \mycite{Otsuka}{{\em et al.}\ }{1997}.
Sir John Herschel rediscovered
thermography in 1840 and created the first thermal image of modern times by
evaporating a thin film of alcohol applied to a carbon-coated surface.
The first detector that was able to measure infrared radiation was
invented in 1880 by Samuel P. Langley, 80 years after the
discovery of this radiation by Sir William Herschel. Herschel measured 
the temperature of light split by a prism and found that the
temperature increased through the colors of the spectrum and
furthermore continued to increase into the non-visible region, today
called infrared. Also, bio-electricity was known long before the
late 19th century. It was first observed by A.L.~Galvani
in 1787, when he exposed a frog muscle to electricity (Zywietz \cite{Zywietz}). The first
measurements of currents and voltages of the frog itself were
possible after 1825, when Nobili {\em et al}.\ constructed sufficiently
sensitive galvanometers \mycite{Mehta}{{\em et al.}\ }{2002}. Eighteen years later,
C.~Matteucci measured electrical currents originating in a resting
heart muscle and Augustuts D.~Waller was the first to record
electric potentials (originating from the beating heart and measured
from the body surface) as a function of time. He used the
capillary electrometer, a device invented and constructed 14 years before 
by G.~Lippmann that visualizes potential differences by changing the
surface tension of a mercury sulfuric acid interface. This was the
first Electrocardiograph. Between 1893 and 1896 George J.~Burch and 
Wilhelm Einthoven strongly improved this method by calibration and
signal correction.

With the beginning of the 20th century new findings piled
up. Investigation and development focused on the improvement of
the technologies known hitherto; natural sciences experienced a boom
leading to numerous new imaging modalities that emerged as a direct 
consequence and also the two world wars strongly fuelled the technologic 
progress. The first practical use of Laparoscopy (endoscopic
exploration of body cavities without natural external access) 
was reported by the internist Hans-Christian Jacobäus, who published
in 1910 the results of endoscopies of the abdominal cavity. 
Nearly at the same time, radiographic imaging was enhanced by using
collimators (E.A.O.~Pasche, 1903) and the employment of high-vacuum
hot-cathode Röntgen-tubes engineered by William D.~Coolidge in
Massachusetts, USA. However, the imaging technique that the clinicians
were mainly interested in was one which was able to isolate in focus some
particular plane in the patient. The superimposition of
three-dimensional objects in a two-dimensional display clearly leads
to relevant structural information loss. That is to say, the aim was
to create sharp images of some particular plane with all other planes
sufficiently blurred out. Nearly simultaneously appeared
{\it Stratigraphy} developed by Allesandro Vallebona, {\it
  planigraphy} by André Edmond Marie Bocage, Bernard Ziedses des
Plantes, Ernst Pohl and Carlo Baese and 
{\it tomography}\footnote{The term tomography is derived from the Greek
  word $\mathrm{\tau o\mu o\varsigma}$ for {\it slice}.} by 
Gustave Grossman. This long list of names shows the increased interest
in section imaging in the 1920s, the more so as the inventors were
working independently from each other \cite{Webb}. At the same time,
other scientists focused their investigation on methods that allow
sharp images of the patient's specific slices using geometric
arrangements of the X-ray source and more than one film. If two films
are used, this is called {\it stereo imaging} and its origins have
been attributed to Elihu Thomson. In 1896, he published  a
description of X-ray stereo images taken from phantoms with metal
objects and mice. An almost contemporary development of X-ray
stereo imaging was put forward in by Imbert and Bertin-Sans in France
and by Czermak at the University of Graz. The use of stereo imaging 
was indicated for measuring distances within solid objects.

A further milestone was reached in 1929 when the Austrian Hans
Berger recorded the first electroencephalogram (EEG) with a
string galvanometer \mycite{Wright}{{\em et al.}\ }{2003}, developed by W.~Einthoven
between 1900 and 1903. With his development, Einthoven wanted to
overcome the slow temporal response and the poor accuracy of the
capillary electro-meter constructed by G.~Lippmann. The importance 
of EEG has to be attributed to the fact that until recently this
modality was the only non-invasive method for recording brain
functions. After the invention of the vacuum tube in 1913, bio-electricity 
could be amplified making the ECGs and EEGs portable. The final
breakthrough of these technologies came with the first implementation of
direct writing instruments by Duchosal and Luthi in 1932 and the use
of cathode-ray tubes by W.~Hollmann and H.E.~Hollmann in
1937. Compared to mechanical recording systems, oscilloscopes based on
cathode-ray tubes are much more suitable for displaying rapidly
varying signals owing to their faster response. Two years after the
first EEG, Dr. Michael Burman published an article
on Myleoscopy (spinal canal Endoscopy). He reports the results from 
the {\em ex-vivo} examinations of eleven vertebral columns
\mycite{Gorchesky}{}{1999}, whilst the first mylescopic exam on an
anesthetized patient was performed by Dr. J.~Lawrence in 1937.

A completely new imaging modality was born in 1916, when P.~Langevin used
ultrasonic waves to locate a submarine that was sunk in shallow water 
\mycite{Tiggelen}{{\em et al.}\ }{2003}. P.~Langevin was a student of Pierre Curie who
analyzed together with his brother Jacques Curie the piezoelectric
qualities of crystals. In 1880, they were successful in producing
ultrasound waves. However, it took sixty-two years until the first
attempt at medical application was made by the Austrian K.~Dussik in
the year 1942. Unfortunately, he tried to take ultrasound images 
of the patient's brain where sonography could not be applied due the 
skull.  Also the foundations of NMR (also called Magnetic Resonance
Imaging -- MRI) were laid in the 1930s. Isidor Rabi 
first described nuclear magnetic resonance in beams in the year 1937. 
But it was not until 1946 that Felix Bloch and independently Edward
Mills Purcell 
observed the same phenomenon in liquids and solids
\mycite{Keevil}{}{2001}. A further major step forward was a paper
from Bloembergen, Purcell and Pound about their observations on
relaxation effects of matter and the influence of motion 
\mycite{Boesch}{}{1999}. Three years later, Arnold reported that
the nuclear magnetic resonance frequency of protons depends on their
chemical environments.

During World War II, much work was concentrated on the sharp
imaging of projectiles within wounded soldiers. This was a period 
of consolidation of the known technologies, of their improvement and
practical implementation. New development was started only for
technologies that give a clear advantage to their owner. This is how 
many new inventions like penicillin, Sonar (Sound Navigation and Ranging), Radar
(Radio Detection and Ranging) and the use of nuclear energy appeard. 
However, after the war, an extensive transfer of technologies towards
other fields of investigation also promoted the science of medical imaging.
A prominent example of this is the first live ultrasonic image taken
by the radiologist D. Howry with the patient
submerged in the water-containing declassified gun turret from a B29
bomber. Reflection of sound waves as the underlying principle for sonographic
imaging at the same time was its major problem. Since the fraction of
reflection at tissue interfaces depends quadratically on the
differences of the acoustical impedance defined by
$Z_{acoustic}=\rho\nu$, where $\rho$ is the tissue density and $\nu$
the speed of sound, the fraction of reflected sound
waves reaches 99.9\% at the skin of the patient when coming from air,
but will be minimized when coming from water.
The necessity of submergence in water avoided its widespread application but was 
required for ultrasound imaging until 1958, when the gynecologist 
I. Donald introduces contact sonography using viscous gel. This method
was immediately accepted by the medical world and is still in use today.
Similarly, the pioneers Inge Edler,
cardiologist, and Hellmuth Herts, physicist used a borrowed and
improved sonar device from a local shipyard to record cardiac echoes
and by this means started the new field of echocardiography.

In 1951 nuclear imaging appeared due to two coinciding 
and breaking events that heralded a new area for medical diagnostics. 
With the January issue of {\it
  Nucleonics}, the invention of the rectilinear scanner from Benedict
Cassen was published \mycite{Wagner}{}{2003}. It consists of a
scintillation counter with a collimator in a radiation shield 
moving slowly back and forth across the region of interest in the patient.
A mechanical or electrical register produces a permanent record from
the detected light pulses of the crystal \mycite{Johns}{{\em et al.}\ }{1983}. 
The second important event was an experiment of Gordon
L.~Brownell and William Sweet carried out at the Massachusetts General
Hospital. They attempted to localize a tumor within a brain probe 
using two facing sodium iodide scintillation detectors \mycite{Nutt}{}{2001}. 
Independently, Wrenn {\em et al.}\ published in the journal {\em Science} studies on how
to use annihilation radiation for localizing brain tumors. Only one
year later, in 1952, Hal Anger reported in the journal {\em Nature} about his
\begin{figure}[t]
  \centering
  \subfigure[][Thyroid image taken with a miniature \g-camera. With
  courtesy of {\sc GEM Imaging S.A.} and the hospital {\sc 9 de Octubre},
  Valencia, Spain.]{%
    \label{subfig:thyoride-scentinella}
    \includegraphics[width=0.4\textwidth]{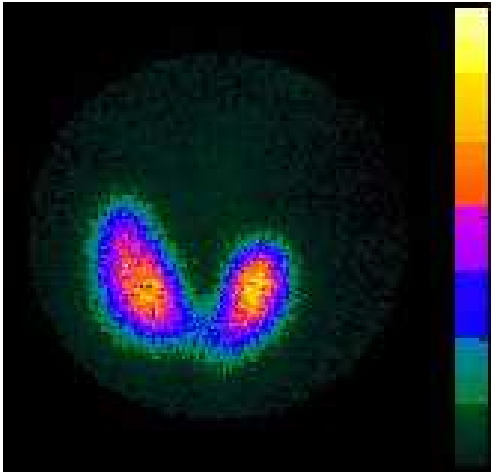}
  }\hspace*{0.1\textwidth}
  \subfigure[][Positron Emission Tomography of the uptake of 
    \isotope{I}{124} in a car\-cin\-o\-gen\-ic mouse. With
  courtesy of {\sc Klinikum} {\sc Rechts} {\sc der} {\sc Isar}, Munich, Germany. ]{%
    \label{subfig:MADPET-rat}
    \raisebox{6eX}{\includegraphics[width=0.4\textwidth]{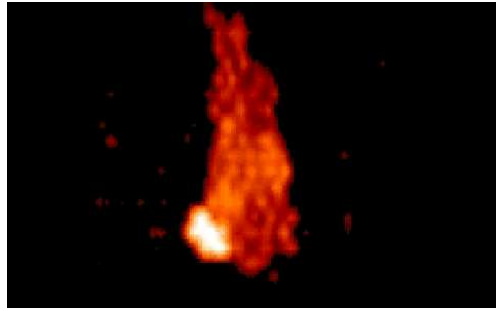}}
  }
  \caption[Two examples of functional imaging]{Two examples of
    functional imaging. Left image: \isotope{Tc}{99m} uptake of the
    thyroid. Right image: small animal positron emission tomography.}
\label{fig:thyprid-img-and-mouse-img}
\end{figure}
first pinhole camera for {\em in vivo} studies of tumors. In this invention, 
gamma photons of the isotope \isotope{I}{131} that passed the pinhole 
collimator excited a large size sodium iodide crystal whose 
scintillation light produced the image on an extensive photographic paper.
Anger further developed his invention and presented its second
scintillation camera in 1957, named after him. Now seven
photomultiplier tubes replaced the photographic film making possible an 
image representation on a cathode ray oscilloscope. However, 
due to the limited number of $\gamma$-photons from the isotope
\isotope{I}{131} the original Anger camera produced only poor
images. 

In 1960, Paul Harper proposed the use of \isotope{Tc}{99m} 
for gamma-scintigraphies. Its physical properties are almost ideal
for the use with Anger-type cameras and further, as advertised by 
Stang and Richards in the same year, can be easily obtained from the
generator \isotope{Mo}{99}. Obviously, the invention of this 
kind of device was only possible after the development of
photocathodes and secondary emission multipliers, called dynodes.
Although the photoelectric effect was discovered in 1887 by Hertz and
afterwards explained by the quantum theory from Albert Einstein, the
first photoelectric tube did not appear until 1913, produced by Elster
and Geiter \mycite{Kume}{}{1994}. In order to achieve higher electron
multiplication, research on secondary emission surfaces was put
forward.

At nearly the same time, very interesting work was going on another
field of science, solid-state electronics. The theoretical tool
of quantum mechanics of the late 1920s with its concept of 
band-structures led to a detailed understanding of solids.
By 1940, Russel Ohl, member of a solid-state working group at 
the Bell-Laboratories was able to prepare $p$- and $n$-type silicon,
and a little later on, even a sample that was of $p$-type at one side
and of $n$-type at the other \mycite{Brinkmann}{}{1997}.
He also found that this sample generated a voltage when it was
irradiated by visible light. During World War II, Radar 
requirements produced a very strong desire to fine-tune solid-state 
materials when it became obvious that shorter wavelength radar, and
thus devices working at higher frequency than the conventional 
vacuum tubes were needed. Finally, William Shockley, John Bardeen
and Walter H.\ Brattain were the first to make a working transistor in
November of the year 1947\footnote{The two physicists, Herbert
  Mataré and Heinrich Welker, from the German radar program independently
  invented a very similar semiconductor device and called it {\em
transitron} (Dormael \cite{Dormael:2004}, Riordan \cite{Riordan:2005}).}. It took only ten
years until Jack Kilby of Texas Instruments developed the first
Integrated Circuit, one of the most important inventions of modern
time for all disciplines of medical imaging.

In the beginning of the 1960s, Kuhl and Edwards focused their work on
image reconstruction for single photon tomography. They successfully
attempted to apply reconstruction techniques to scanners for
radioisotope distributions that were formerly employed in X-ray
section imaging. Furthermore at the beginning of the 1960s, Alexander
Gottschalk began to work with the new Anger-Camera and found that it
could also be used for positron imaging. About the same time, in 1963,
Allan M.~Cormack constructed, together with David Hennage, the first 
experimental X-ray computed tomography (CT) scanner. That this was not
the starting point for its widespread acceptance and application to
medical diagnostics was mainly due to two reasons. First, he applied
his own reconstruction technique to the experimental data with only
partial success and did not discover until 1970 that this mathematical
problem had been solved by J.~Radon in 1917. Secondly, since at that time
there was much interest in positron emission tomography, he designed
his scanner and phantoms adapted to this modality. Thus, the
work of Godfrey N.~Hounsfield on transmission computed tomography 
marks the beginning of this new area in diagnostic imaging. Though
Hounsfield's first experimental scanner used a gamma-ray source, it
needed nine days for data collection, $\mathrm{2\,^1\!/\!_2}$ hours 
for reconstruction and further 2 hours for displaying the digitized
image, the prototype developed for the Atkinson Morly Hospital in 1971
took an image in 18 seconds.

Also in 1971, Raymond Damadian wrote in the journal {\it Science}
about his observation of variations in the relaxation times of NMR
signals obtained from cancerous and normal tissues. This publication
essentially stimulated the medical interest in NMR as diagnostic
technique, although already in 1960 a report from the US National 
Heart Institute advised that NMR experiments might lead to a medical 
imaging modality. Actually, long before 1971, important progress was made
concerning the basic understanding of NMR and how to obtain images
from it. Before the publication of Hahn in 1950, where he described
how decaying NMR-signals could be partially refocussed by a determined
sequence of radio-frequency pulses, the state-of-the-art was
the measurement of a spectrum with continuous irradiation with
radio-frequency. Modern MRI is heavily based onto these so-called
{\it spin-echos}. A further important landmark of NMR was established
by Lauterbur and independently Mansfield in the year 1973. They were
the first ones to propose the use of magnetic field gradients in order to
spatially encode the NMR signals for a subsequent image reconstruction.
Another important result prior to the latter was the observation of 
Ernst and Anderson in 1966 that the free induction decay signal
(studied by Hahn) contained the whole spectrum information. They
introduced the Fourier framework into NMR. Finally, the discovery of
superconductivity and the subsequent development of super-conduction
magnets in the 1970s of the past century marked another important step
in MRI (Coupland \cite{Coupland:1987}, Schrieffer {\em et al.}\ \cite{Schrieffer:1999}).

In emission tomography, one could also observe a time-dense series  
of key developments converging steadily towards the first ring 
tomograph for coincidence imaging as well as the first Single Photon
Emission Computed Tomograph (SPECT). Once again, Hal Anger played a
key role in the development of Single Photon Emission Tomography (SPET).
In his work presented in June of 1967 at the 14th
Meeting of the Society of Nuclear Medicine, he explained how to form
cross-sectional images obtained from a patient rotated in front of a
static gamma-camera using back projection onto a likewise rotating 
film. The necessity of analogue techniques to form section images was
overcome with the appearance of computers after the invention of the
IC. Muellehner and Wetzel were the first in reconstructing projection 
camera data using a computer in 1971, therefore being the first ones 
to use the medical imaging modality that today is called SPECT. 
The reconstruction was done with an IBM 360/30 computer rendering the 
emission information into a 40$\mathrm{\times}$40 pixel
matrix. Simultaneously with this SPECT scanner they proposed
iterative reconstruction algorithms. The technology needed to rotate
the gamma-camera instead of the patient became available in 1977,
when John Keyes got the ``Humongotron'' working. 
In 1955, the first clinical positron scanner had already appeared, 
constructed by Gordon Brownell. However, the images from this device
where rather crude, since it performed only planar scans and estimated
the distribution of the radioactive tracer using the difference in the
average of counting rates of the two detectors. 

The first true
tomographic image (from a dog heart) was scanned with the positron
camera at the Massachusetts General Hospital in the late 1960s by
Brownell and Burmann. After this event many different PET scanners have
been developed. It turned out that the optimal geometry is a circular
structure and that the general tendency of the different generation of
PET scanners in the last decades points towards small-sized segmented
crystal detectors. This is known as Nutt's law, an analogue to More's
law for micro electronic devices. It points out that the number of individual crystal
elements in a positron emission tomograph has doubled every two years for
the past 25 years \mycite{Nutt}{}{2001}. The scintillation
material used for the detector has also a major impact on the 
performance of generations of PET scanners. While in the mid-70s thallium doped
sodium iodide was the only choice for PET, the development of
bismuth-germanate, known as BGO, provided the physicists with a more suitable
crystal for high energy gamma photons. During the last two decades, 
this has been the scintillator of choice due to its significantly greater
stopping power. However, this was at the cost of only 15\% of the
\chemform{NaI\doped Tl}'s scintillation light yield at 511 keV. Therefore, the
announcement of Lutetium Oxyorthosilicate (LSO) by C.~Melcher 1989 resulted
once again in higher image quality. LSO has a relative light yield of 75\%
compared to sodium iodide, is more than six times as fast as BGO and more
than four times as fast as \chemform{NaI\doped Tl}.

 \def\myleftarrow{\includegraphics[height=1cm]{left-arrow}}
 \def\myrightarrow{\includegraphics[height=1cm]{right-arrow}}
 \def\mycenterarrow{\includegraphics[height=1cm]{center-arrow}}
 \def\mysleftarrow{\includegraphics[height=0.6cm]{left-arrow}}
 \def\mysrightarrow{\includegraphics[height=0.6cm]{right-arrow}}
 \def\myscenterarrow{\includegraphics[height=0.6cm]{center-arrow}}

  \makeatletter
  \def\pstree@balancedfit#1#2{%
  \edef\next{\noexpand\pstree@@balancedfit#1\noexpand\@nil#2\noexpand\@nil}%
  \next
  \ifnum\pst@cntg=\z@
  \pstree@max{#1}\pst@cnth
  \else
  \pstree@max{#2}\pst@cnth
  \fi
  \advance\pst@cnth\pst@cnth
  \advance\pst@cnth\psk@thistreesep\relax
  \advance\pst@cnth\pstree@tspace\relax
  \gdef\pstree@tspace{\z@}}
  \def\pstree@@balancedfit#1,#2\@nil#3,#4\@nil{%
  \ifnum#1=\pstree@stop
  \let\next\relax
  \pst@cntg=\@ne
  \else
  \ifnum#3=\pstree@stop
  \let\next\relax
  \pst@cntg=\z@
  \else
  \def\next{\pstree@@balancedfit#2\@nil#4\@nil}%
  \fi
  \fi
  \next}

\newcommand{\MIClassNodeR}[2][]{\Tr[#1]{\psframebox[fillstyle=solid,fillcolor=RedOrange]{#2}}}
\newcommand{\MIClassNodeM}[2][]{\Tr[#1]{\psframebox[fillstyle=solid,fillcolor=Yellow]{#2}}}
\newcommand{\MIClassNodeF}[2][]{\Tr[#1]{\psframebox[fillstyle=solid,fillcolor=CornflowerBlue]{#2}}}
\newcommand{\MIClassNodeS}[2][]{\Tr[#1]{\psframebox[fillstyle=solid,fillcolor=SeaGreen]{#2}}}

\begin{figure}[!ht]
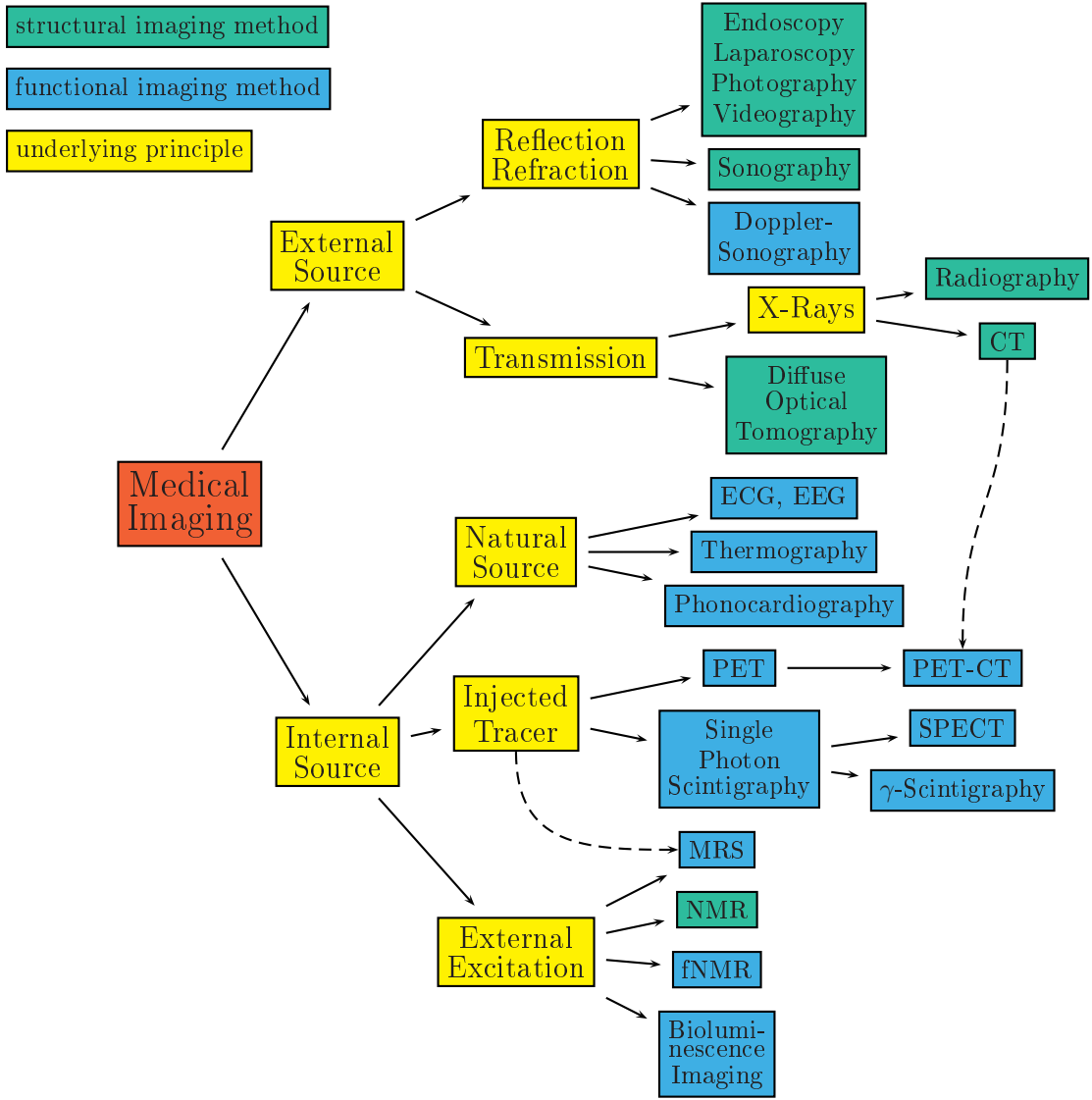

  \vspace*{2eX}
  \centerline{\sc\large The spectrum of medical imaging modalities}
  \vspace*{4eX}
  \begin{flushleft}
    \psframebox[fillstyle=solid,fillcolor=SeaGreen]{structural imaging method}\newline~\newline
    \psframebox[fillstyle=solid,fillcolor=CornflowerBlue]{functional imaging method}\newline~\newline
    \psframebox[fillstyle=solid,fillcolor=Yellow]{underlying principle}
  \end{flushleft}
  \centering
  \vspace*{-17.8eX}
  \psset{arrows=->,treefit=tight}
  \pstree[treemode=R,linestyle=solid,treesep=2ex,nodesep=1ex]{%
    \MIClassNodeR{\shortstack{\Large Medical\\\Large Imaging}}}{%
    \pstree[treesep=1ex,levelsep=20ex]{\MIClassNodeM{\shortstack{\large External\\\large Source}}}{%
      \pstree[treesep=1ex,levelsep=20ex]{\MIClassNodeM{\shortstack{\large Reflection\\\large Refraction}}}{%
        \MIClassNodeS{\shortstack{Endoscopy\\Laparoscopy\\Photography\\Videography}}%
        \MIClassNodeS{Sonography}\MIClassNodeF{\shortstack{Doppler-\\Sonography}}}\
      \pstree[treesep=2ex,levelsep=22ex]{\MIClassNodeM{\large Transmission}}{%
        \pstree[levelsep=18ex]{\MIClassNodeM{\large X-Rays}}%
        {\MIClassNodeS{Radiography}\MIClassNodeS[name=ct]{CT}}%
        \MIClassNodeS{\shortstack{Diffuse\\Optical\\Tomography}}}}
    \pstree[treesep=2ex,levelsep=16ex]{\MIClassNodeM{\shortstack{\large Internal\\\large Source}}}{%
      \pstree[treesep=1ex,levelsep=24ex]{\MIClassNodeM{\shortstack{\large Natural\\\large Source}}}{%
        \MIClassNodeF{ECG, EEG}\MIClassNodeF{Thermography}\MIClassNodeF{Phonocardiography}}%
      \pstree[treesep=2ex,levelsep=20ex]%
      {\MIClassNodeM[name=it]{\shortstack{\large Injected\\\large Tracer}}}%
      {\pstree{\MIClassNodeF{PET}}{\MIClassNodeF[name=pet]{PET-CT}}%
        \pstree{\MIClassNodeF{\shortstack{Single\\Photon\\Scintigraphy}}}%
        {\MIClassNodeF{SPECT}\MIClassNodeF{$\gamma$-Scintigraphy}}}%
      \pstree[treesep=2ex,levelsep=18ex]{\MIClassNodeM{\shortstack{\large External\\\large Excitation}}}%
      {\MIClassNodeF[name=mrs]{MRS}\MIClassNodeS{NMR}\MIClassNodeF{fNMR}%
        \MIClassNodeF{\shortstack{Biolumi-\\nescence\\Imaging}}}
    }}
  \nccurve[linestyle=dashed,ncurv=1,angleA=-90,angleB=180]{->}{it}{mrs}
  \nccurve[linestyle=dashed,ncurv=1,angleA=-90,angleB=90]{->}{ct}{pet}
  \caption[The spectrum of medical imaging modalities]{The scheme is divided into two main branches, depending
    if any kind of radiations/fields has to be applied from the
    outside or not. For the case that the physical process which is
    going to be represented graphically has its origins inside the
    body, there is also the possibility of external excitation. There
    are also methods that combine two ore more principles and
    therefore can be functional as well as structural, {\em e.g.}\ PET-CT,
    which is quiet a new development. Another example is MRS, which
    requires internal tracer molecules as well as external excitation.}
  \label{fig:mispec}
\end{figure}

Further important landmarks in medical imaging are the appearance of
Doppler-sonography in 1979 due to the work of Donald W.\ Baker and many
other scientists and the appearance of Human MR spectroscopy (MRS) in 1980 
and functional NMR (fNMR) in 1990. With MRS one is able to measure the
concentration of biochemical compounds and the spatial variations
in the concentration can be used to form images. (refer to
\mycite{Hendee}{}{1999}). The isotopes used for this
modality are \isotope{P}{31} and \isotope{H}{1} \mycite{Kuijpers}{}{1995}.
Seiji Ogawa from the AT\&T Bell Laboratories was the first who
observed NMR-signal variation induced by the oxygenation level changes
in brains from rats \mycite{Ogawa}{{\em et al.}\ }{1992} and thus initiated
fNMR.
Also, a revitalization of investigation on Diffuse Optical Tomography
could be observed after having been abandoned twice in the
past century. The first attempt was made by Cutler in 1929. He proposed
to detect breast lesions with continuous light, but found that the
necessary intensity would overheat the patient's skin. It was
abandoned for the second time in the year 1990 when a study found too
many false negatives for small breast lesions with a technique
introduced by Gros {\em et al.}\ in 1973, where the breast was
positioned between a visible-light source and the physician's unaided eye.
In 1989, the first helical scan X-ray computed tomography scanner was put in 
operation using the new slip-ring technology of 1988 for continuous
rotation of tube and detectors. 
Helical-scan CT allows one to freely select the increment between
slices to be imaged as a reconstruction parameter. This enables
reconstruction of overlapping slices without increasing the dose.
With the replacement of Xenon gas detectors by solid-state detectors in the
beginning of 1997, the radiation dose applied to the patient could
be significantly reduced without loss in image quality.
And finally, in the past few years Bioluminescent Imaging (BLI) has
emerged \mycite{Dikmen}{{\em et al.}\ }{2005}. It is particularly well suited
for imaging small animals and uses the photochemical reaction between
luciferin and luciferase that depends on Adenine-Tri-Phosphate (ATP) and 
$\mathrm{O_2}$. Therefore only living cells can emit photons.

The whole spectrum of medical imaging technologies has become very
broad due to a far-reaching scientific effort during the last century
and now includes many modalities besides radiography. Furthermore, the
clinician today can choose either modalities for
structural imaging or modalities for functional imaging, where 
the latter provides a relevant complement to conventional methods like
radiography, sonography or X-ray CT. The strict, historical definition 
of functional imaging includes all techniques that use either repeated 
structural scanning of the region of interest (ROI) or other
measurable variables for the representation of the temporal dependence of
specific physiological processes. While this is possible with almost all
the mentioned imaging modalities, nowadays functional imaging mainly
refers to methods which are capable of rendering metabolic processes
of body regions. Likewise, structural imaging makes the structure of
the ROI accessible to the observer. Using the underlying
physical principles and processes of image formation as key property 
allows the more detailed subclassification of the different known
modalities shown in figure~\ref{fig:mispec} \mycite{Deconinck}{}{2003}.  

\section{Gamma-Ray Imaging in Nuclear Medicine}

The field involving the clinical use of non-sealed radionuclides is
referred to as nuclear medicine. The same criterion holds for nuclear
imaging and, although not {\em a-priori} obvious, Magnetic Resonance
Imaging should also be included in this subgroup
 of imaging modalities because it is actually based on the
properties of the nucleus of hydrogen, which is omnipresent as
\chform{H_2O} in biological tissue. It is due the negative
connotations of the word ``nuclear'' in the 1970s that the imaging
technology was marketed as ``Magnetic Resonance Imaging'', while
``Nuclear Magnetic Resonance'' refers to the underlying physical
effect. For nuclear imaging, two important requisites are needed:
the particle detector, able to provide good temporal, spatial and
energetic information, and the radioactive substance to be traced by
the detector. The radionuclides can rarely be administered to the
patient in their simplest chemical forms. Instead, there are many
compounds in biology and medicine that may be of interest because of
their specific biochemical, physiological or metabolic properties
whenever they are prepared in a form that is suitable for use
with the specimen to be examined. Many of these compounds can be marked by 
replacing one of their atomic constituents by a radioactive isotope 
and are called radiopharmaceuticals. After administration to the
patient, the radiopharmaceutical spreads out within the explored
patient due to the normal (or abnormal) vital functions in a way that
is characteristic for a specific metabolic process. In each period of
time, the unstable atoms of a small fraction of the pharmaceutical 
disintegrate giving rise to $\beta^+$- or $\gamma$-radiation.
$\alpha$-radiation is not desired owing to its heavy
molecular-biological impact and furthermore is only generated by
isotopes of very high atomic number, which are generally
toxic. Furthermore, since $\alpha$-particles are strongly ionizing, they would
not leave the explored body and therefore could not be detected by a
external detector. The same holds for $\beta^-$-particles.
After the detection of the radiation by the gamma-ray imager, the
spatial distribution of the radiopharmaceutical is reconstructed from
the detection positions of the photons. If the spatial distribution is
scanned repeatedly, it can be mapped as a function of time.

\begin{figure}
  \centering
  \includegraphics[width=0.35\textwidth]{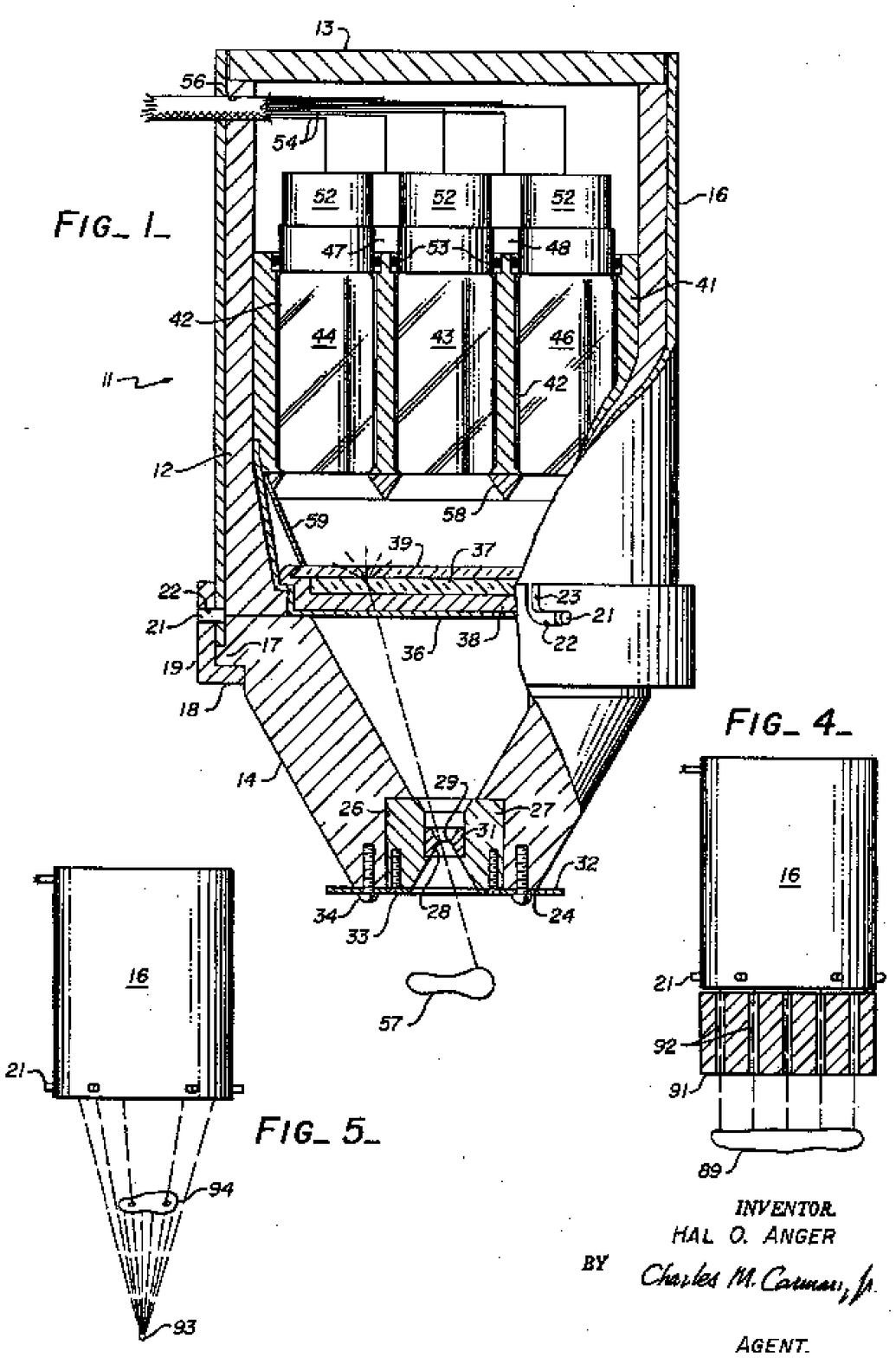}\hspace*{0.1\textwidth}
  \includegraphics[width=0.35\textwidth]{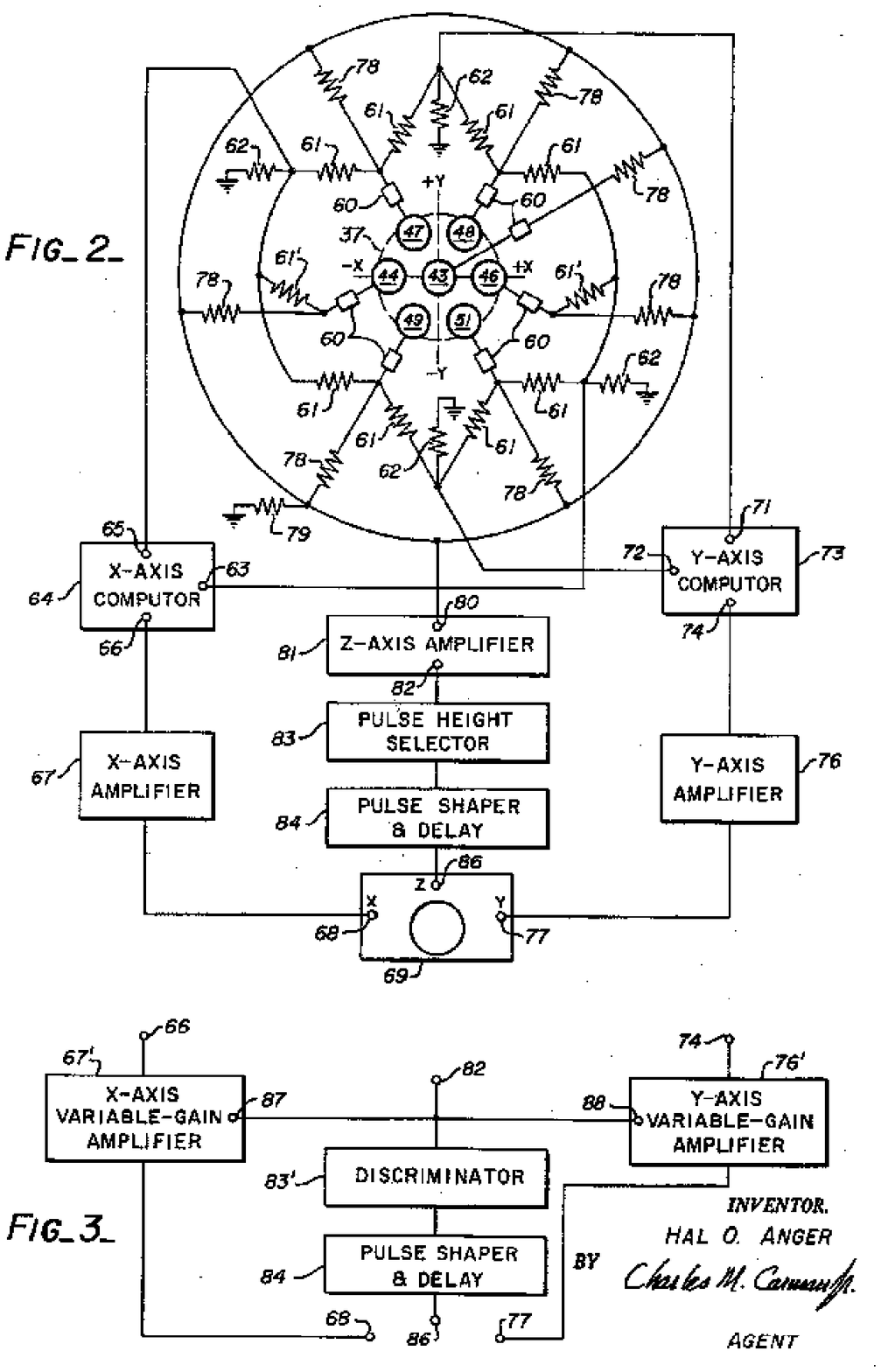}
  \caption[Original proposal of the Anger camera]{Original proposal of an {\em Radiation Image Device}, now
    called {\em Anger-camera} and patented by Hal O.\ Anger in November 1961 (US-Patent number
  3.011.057).}
  \label{fig:anger-patent}
\end{figure}

The decay mode of the radiopharmaceutical normally determines the
gamma-ray detector to be used. In the simplest case, exactly one
$\gamma$ photon is produced. This mode is suitable for
$\gamma$-scintigraphy or SPECT. If the radionuclides disintegrates via
$\beta^+$-decay, it is nearly instantly followed by collinear
$\gamma$-photon pair emerging from the positron-electron annihilation. 
Whilst this radiation can be used for each of the three modalities
PET, SPECT and $\gamma$-scintigraphy, its is particularly well suited
for coincidence imaging. Certain radionuclides emit two gammas of
different energies in cascade (\isotope{Se}{75}, \isotope{In}{111},
\isotope{Cr}{48} and \isotope{K}{43}) or one gamma and a positron 
like \isotope{Fe}{52}, and electron capture is known to give rise 
to X-ray and $\gamma$ radiation (\isotope{I}{125} and \isotope{Hg}{197}).
This opened up the possibility for $\gamma$-$\gamma$ coincidence
tomography, as proposed by Powell and Mohan in 1970 and 1970.
Once again, the radiation resulting from these decay mode can be imaged
with PET, SPECT and $\gamma$-scintigraphy. As a consequence of the 
arguments and examples above, nuclear imaging is essentially based 
on $\gamma$-ray imaging and the energy of the emerging
$\gamma$-photon is a further important parameter that has to be taken
into account for the construction of the imaging detector.
Due to the quantum nature of photons, they cannot be partially stopped
by the detector material. It either comes close enough to an atom to
undergo one of the possible elementary interactions or it will not be
affected at all. The probability for the interaction strongly depends
on the energy $E$ of the $\gamma$-ray, on the effective atomic number
$Z_\mathit{eff}$ of the atom and the density of these atoms (or compounds)
forming the detector material. That is to say, the $\gamma$-imaging
detector has not only to detect the photon, but also has to stop it for
detection. The ability of the detector to stop the $\gamma$-ray is
known as its intrinsic efficiency. This is the reason why Anger-type 
$\gamma$-cameras, optimized for 140 keV radiation from the
tracer \isotope{Tc}{99m} will only produce low quality images when
used for SPECT with 511 keV annihilation radiation. However,
efficiency is not the only parameter of interest and its optimization 
 is very often in conflict with the optimization of others.

PET, SPECT and $\gamma$-scintigraphy are now widely applied to medical
diagnostics and research in the fields of Cardiology, Oncology and Neurology.
They are sometimes referred to as molecular imaging, since
being photon-counting methods, they are actually able to visualize
single molecules. It is that striking sensitivity that makes them very
useful for the evaluation of biological systems that are particularly
sensitive to small quantities of metabolic active substances such as many
receptor systems and intracellular processes. Compared to MRS, which
requires concentrations of paramagnetic tracers of $\mathrm{10^{-6}}$
mol/l in order to change relaxativity, the normally administered
concentrations of \isotope{Tc}{99m} is with $\approx\mathrm{10^{-10}}$
mol/l four orders of magnitudes smaller. Iodinated contrast agents
even need a concentration of $\mathrm{10^{-2}}$ mol/l to achieve
opacification with X-ray CT \mycite{Blankenberg}{{\em et al.}\ }{2002}.
As a result, for the moment only nuclear imaging allows the {\em in vivo}
uninfluenced study of certain specific biochemical, physiological or 
metabolic processes, since the normally administered concentrations
will not cause a pharmalogical effect.

\section{Gamma-Camera (Planar Imaging)}
\label{sec:gamma-cam-planar-imaging}

\begin{figure}
  \centering
  \includegraphics[width=0.8\textwidth]{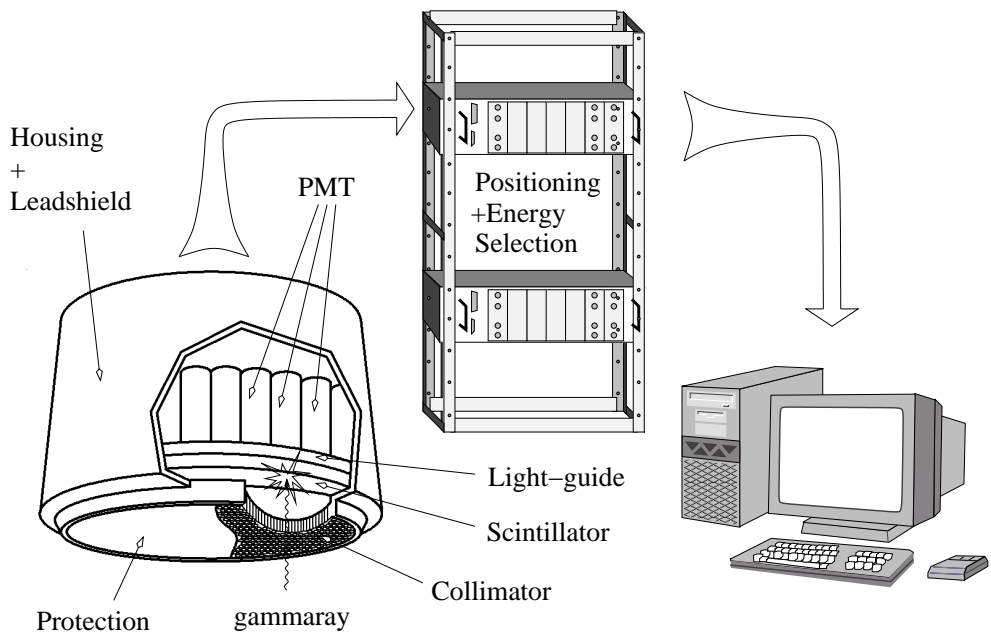}
  \caption[Schematics of the most important components of an Anger type
  scintillation camera]{Schematics of the most important components of an Anger type
  scintillation camera.}
  \label{fig:anger-cam}
\end{figure}

The principle of the Anger-type scintillation camera which was
presented by Hal O.~Anger in 1958 (Anger \cite{Anger:1958}, refer to
figure~\ref{fig:anger-patent}) has changed
little in the past half century and is based on the ``camera oscura'' 
from the early days of photography. While this principle was replaced
by sophisticated optics in photography, it is not possible to construct
lenses for \g-radiation except for some special cases.
Figure~\ref{fig:anger-cam} shows a schematic diagram of an Anger-type gamma-camera.  
The object to be studied is placed in front of
the device, and, after the injection of the radiopharmaceutical, is
continuously emitting $\gamma$-radiation. Although a large fraction of
radiation is lost due to isotropic emissions, absorption by the
collimator or because it is not absorbed by the scintillator, there are
still sufficient $\gamma$-photons that interact with the scintillator. 
The $\gamma$-photons that interact with the scintillation
crystal will originate isotropic scintillation light whose spatial
distribution is projected through the light-guide onto the sensitive
area of the photodetector. This distribution is sampled by the
photodetector array or a position sensitive photodetector and used to compute
electronically the position of the impact within a plane within the
scintillation crystal and normal to the detector axis. In order to
protect the detector from environmental and backscattered radiation,
it has to be fitted into a shield. Modern gamma-cameras occasionally wrap 
the photodetector into \chform{\mu}-metal to protect it from
external magnetic fields. Furthermore, the holes of the collimator have to
be covered to avoid diffuse straylight to be detected by the
photodetector. In general, this is done with an aluminum sheet of
small thickness that does not stop the \g-radiation.

\begin{figure}
  \centering
  \subfigure[t][Parallel hole collimator]{\label{fig:parallel-hole}\raisebox{1.6cm}{
    \includegraphics[width=0.45\textwidth]{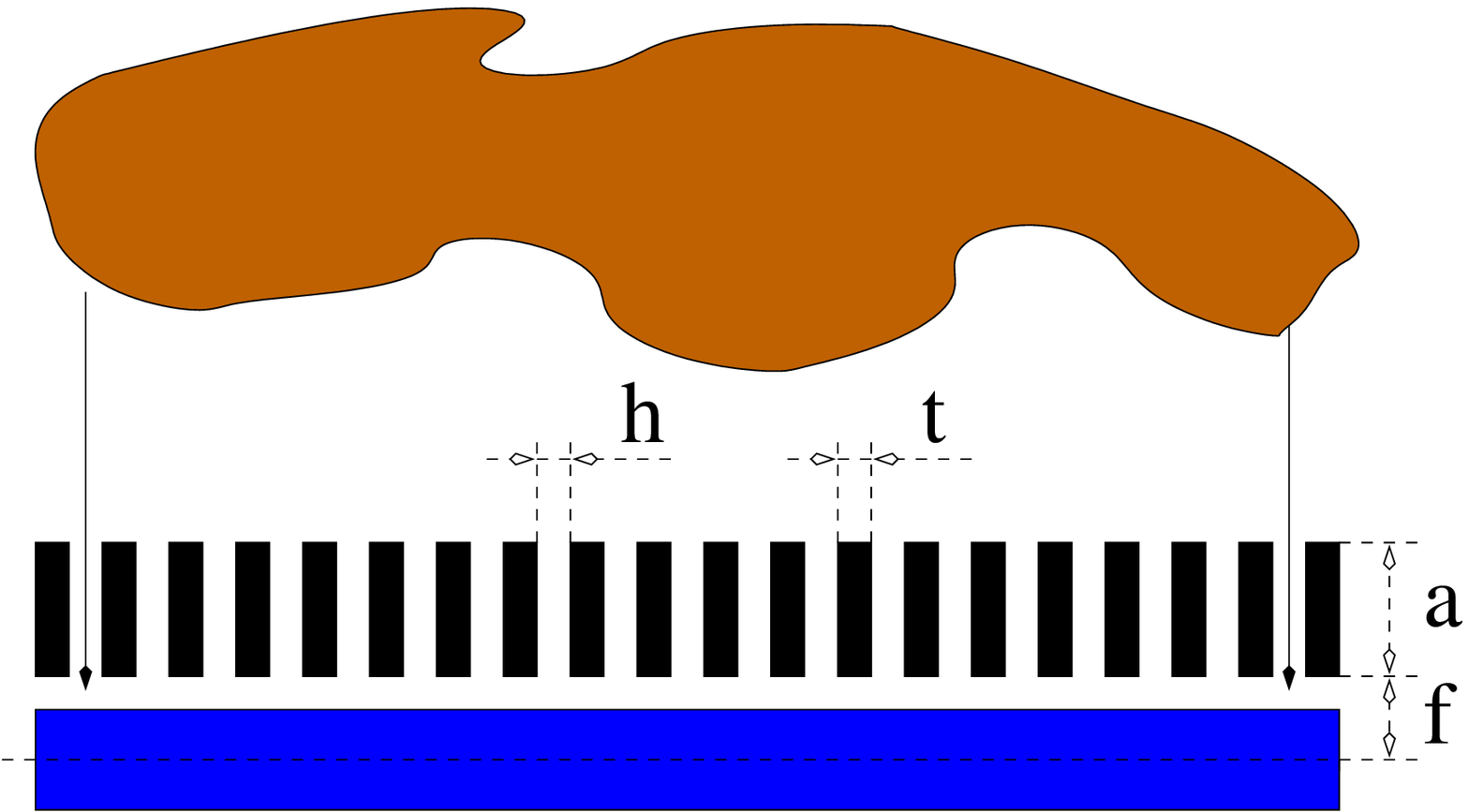}}}
  \subfigure[Pinhole collimator]{\label{fig:pin-hole}%
    \includegraphics[width=0.45\textwidth]{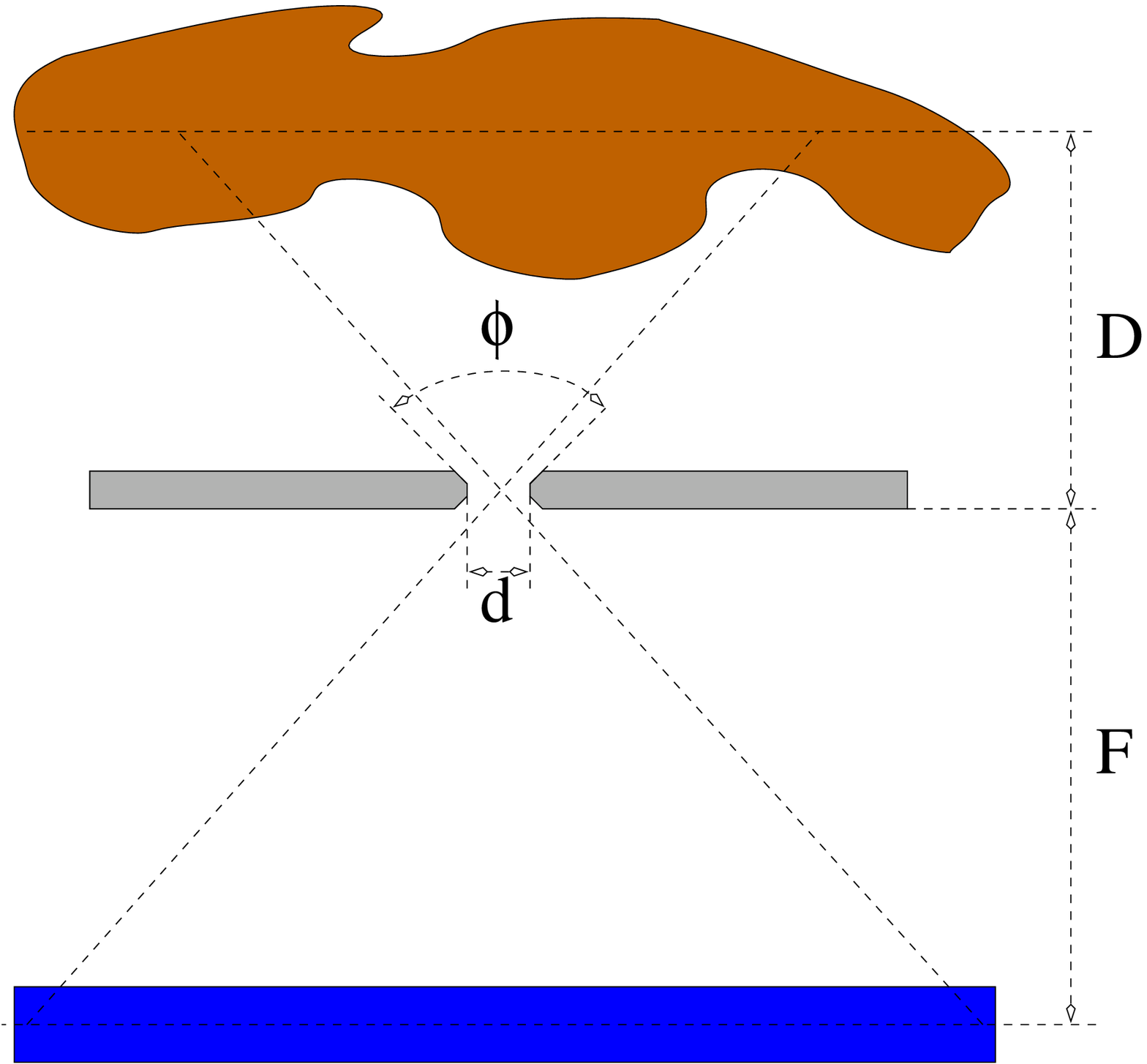}}
  \caption[Diagram showing the most important collimators for
  Anger-type gamma-cameras]{Diagram showing the most important
    collimators for Anger-type gamma-cameras.}
\end{figure}

While this principle is repeated in almost all gamma-cameras, there
exist clear differences from one camera to another depending on the
components used and the applications that they are designed for.
If the object as well as the scintillation crystal is large, 
{\em e.g.}\ for a whole-body scan of a human being, the pinhole collimator leads to an
important loss of sensitivity and the use of a multiple hole collimator
is advised \mycite{Anger}{}{1964}. The main differences of these two collimator
types are their field of view (FOV) and their efficiency. The FOV of
the parallel hole collimator (figure~\ref{fig:parallel-hole}) is just
its spatial extension parallel to the source plane and is independent of 
the distance between object and collimator.

In the case of the pinhole collimator (figure~\ref{fig:pin-hole}), the
FOV depends on the distance $D$ between object and collimator. It can be
approximated using pure geometrical arguments as follows:
\begin{equation}
  \label{eq:fov-pinhole}
  FOV\approx\frac{D}{F}A_\mathit{eff}^\mathit{C}
\end{equation}
where  $A_\mathit{eff}^\mathit{C}$ is the effective area of the
scintillation-crystal photodetector combination and $F$ its distance to the collimator. The
geometric efficiency is given by 
\begin{equation}
  \label{eq:geoeff-pinhole}
  \epsilon_\mathit{geo}^\mathit{pin}=\left(\frac{d_\mathit{eff}}{4D}\right)^2\quad\mbox{with}\quad
d_\mathit{eff}=d\cdot\sqrt{1+\frac{2}{d}\frac{\tan{\Phi/2}}{\mu}\left(1+\frac{\tan{\Phi/2}}{\mu}\right)}
\end{equation}
where the effective pinhole diameter $d_\mathit{eff}$ accounts for effects
due to near-hole collimator penetration of the \g-rays and $\mu$ is
the collimators absorption coefficient for the \g-ray energy of interest. Clearly
the geometric design of the pinhole collimators also contributes to
the total spatial resolution of the camera by
\begin{equation}
  \label{eq:spatres-pinhole}
  \Delta_\mathit{tot}^\mathit{pin}\approx\sqrt{(\Delta_\mathit{geo}^\mathit{pin})^2%
+(\Delta_\mathit{int}^\mathit{pin})^2}%
\quad\mbox{with}\quad\Delta_\mathit{geo}^\mathit{pin}=\frac{D+F}{D}d_\mathit{eff},
\end{equation}
where $\Delta_\mathit{int}^\mathit{pin}$ is the scintillation
detectors intrinsic resolution.
For a circular parallel hole collimator, the geometric efficiency is given by
\begin{equation}
  \label{eq:geoeff-parahole}
  \epsilon_\mathit{geo}^\mathit{para}=\frac{A_\mathit{hole}}{2\pi\left(a-2/\mu\right)^2}%
\frac{N A_\mathit{hole}}{A_\mathit{col}},
\end{equation}
where $A_\mathit{hole}$ denominates the area covered by one aperture,
$A_\mathit{col}$ is the area of the whole collimator, $N$ the number of the
apertures and $a$ the thickness of the collimator.  Its
contribution to the spatial resolution can be approximated by 
\begin{equation}
  \label{eq:spatres-parahole}
  \Delta_\mathit{geo}^\mathit{para}=\frac{h(a-2/\mu+D+f)}{a-2/\mu}
\end{equation}
It is clear from the equations
\ref{eq:fov-pinhole}-\ref{eq:spatres-parahole} that type and quality
of the collimator is of striking importance for FOV, spatial
resolution and detection efficiency. 

\section{Single Photon Emission Computed Tomography}

When the  Anger-type gamma-camera is rotated step-wise around the
patient while acquiring projection images at each step, a tomographic 
image can be reconstructed from a sufficiently large set of
projections. Today this is known as Single Photon Emission Tomography 
or Single Photon Emission Computed Tomography. The spatial resolution
of the resulting images depends on the performance of the  
gamma-camera used and is thus comparable to the resolution of the planar 
gamma-ray imaging modality. However, the contrast is significantly 
increased compared to planar imaging leading to more reliable lesion
detectability. The sensitivity of SPECT is clearly outperformed 
(by approximately two orders of magnitude) by Positron Emission
Tomography and also its spatial resolution falls behind because it
is  strongly affected by the necessity of collimators. 
Nevertheless, SPECT is an imaging technique nowadays widely used in
medical diagnostics and the reason for this is that SPECT devices
currently used for clinical purposes consist of from one to three
Anger-type gamma-cameras. Thus, unlike PET scanners, SPECT scanners are 
multi-modality devices and allow for a higher cost-effectiveness and
broader field of application. Historically, SPECT was introduced into
nuclear medicine facilities by upgrading existing gamma-cameras and
using the same radiopharmaceuticals. For this reason, there was no need
for expensive compact cyclotrons and radiochemistry facilities for
tracer production, although the
low $\gamma$-ray energy give raise to a large fractions of Compton
scattered events and totally absorbed $\gamma$-photons.
It is therefore far from being ideal for tomography of the
human body but remains an interesting imaging modality for small animal
studies, where it is commonly used in experimental oncology
\mycite{Weissleder}{}{2002}.

\section{Positron Emission Tomography}
\label{sec:positron-emiss-tomo}

\begin{figure}
  \centering
  \psfrag{B}{$\mathrm{\beta^+}$}
  \psfrag{g}{$\mathrm{\gamma}$}
  \psfrag{e}{$\mathrm{\e^-}$}
  \psfrag{v}{$\mathrm{\nu}$}
  \psfrag{r}{positron range}
  \psfrag{R}{radiotracer}
  \psfrag{CH2OH}{\tiny $\mathsf{CH_2OH}$}
  \psfrag{C}{\tiny $\mathsf{C}$}
  \psfrag{O}{\tiny $\mathsf{O}$}
  \psfrag{F1}{$\mathsf{^{18}F}$}
  \psfrag{F2}{$\mathsf{^{18}O}$}
  \psfrag{bd}{before decay}
  \psfrag{desintegration}{desintegration}
  \psfrag{ad}{after decay}
  \psfrag{annihilation}{\hspace*{-1em} annihilation}
  \psfrag{Noncolinearity}{\hspace*{-1.4em} non-collinearity}
  \includegraphics[width=0.86\textwidth]{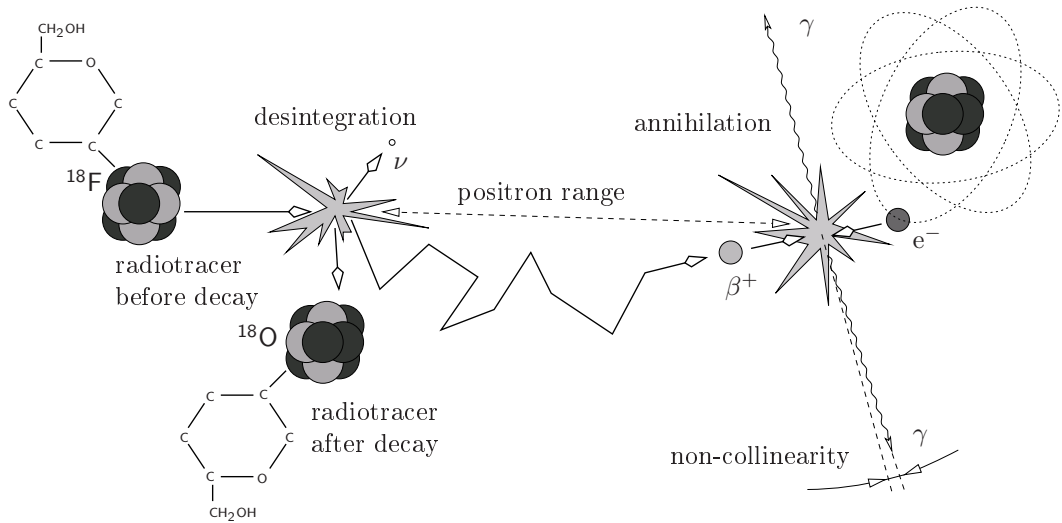}
  \caption[Sketch of important degrading effects in PET]{Sketch of all
    important degrading
    effects in PET, from the decay of the
    radionuclide of the pharmaceutical to the annihilation of the positron.}
  \label{fig:pet-physics-principle}
\end{figure}

Positron Emission Tomography is often referred to as the benchmark
technique for functional imaging. At the same time, it presents the
highest level of resourcing for {\em in vivo} studies among all
inherently contrast agent imaging modalities. In 1950, Gordon
L. Brownell suggested that spatial resolution
and sensitivity of nuclear imaging of brain diseases might be 
improved using annihilation radiation \mycite{Brownell}{}{1999}.
PET is based on the decay of a $\beta^+$ emitting radioisotope tagged
to a derivative of a metabolic active substance. The released positron
has only a very short lifetime after the decay of the radiotracer
and its path ends in an annihilation with one of the millions of
surrounding electrons. As a consequence of energy and momentum
conservation, the annihilation process gives rise to the nearly
collinear \g-photons with an energy of 511 keV 
(refer to figure~\ref{fig:pet-physics-principle}). 

If both \g-photons are detected, the two points in space where
they are detected define a straight line, the line of response (LOR),
which necessarily contains the annihilation point. Since the path of
the \g-rays is known, there is no need for absorptive collimation of
the tracer radiation which is necessary in planar imaging like SPECT.
Consequently, PET offers a wide acceptance angle for detecting the
\g-photons and its sensitivity of approximately $\mathrm{10^{-10}}$ mol
outperforms the sensitivity of SPECT by two orders of magnitude 
\mycite{Cassidy}{{\em et al.}\ }{2005} and, that of a single headed 
Anger-type camera by three orders of magnitude \mycite{Jones}{}{1996}.
In addition, the temporal coincidence detection provides an inherent
and efficient rejection of background radiation and a further advantage
is that the higher energy of 511 keV of the annihilation
radiation entails a much higher probability for the \g-rays to escape
from the examined object, which is not the case for the 140 keV radiation
normally applied with SPECT and \g-scintigraphy. While for a 140 keV
photon the half-layer of soft-tissue (International Commission on
Radiation Units and Measurements, Four-Component soft-tissue) 
is $\mathrm{\approx4.5\, cm}$, this value reaches $\mathrm{\approx7.3\, cm}$
for 511 keV radiation. It is therefore no surprise that, within the 
spectrum of nuclear imaging, PET is the modality {\em par excellence} for 
qualitatively and quantitatively imaging molecular pathways and 
interactions {\em in vivo}.

However, annihilation coincidence detection does not present a miracle
cure for the physical problems of nuclear imaging and also involves 
fundamental problems. First, the higher penetrating power of 511~keV
\g-rays not only is true for the examined object, but also applies to
the photon-detector. One of the fundamental principles of Quantum
Mechanics requires the \g-photon to be stopped and destroyed in order
to detect its path and energy. Due to the greater penetrating power of
the annihilation radiation, 
the detector has to be more massive and normally of larger dimensions.
Another drawback is that both annihilation photons have to be
detected to register a counted event. This directly translates to a
quadratic dependency of the tomograph's detection efficiency on the 
intrinsic detection efficiency of one detector module. Furthermore,
since there are now two \g-rays that have to escape from the examined
object, both of them can undergo Compton scattering
and the probability of misspositioned events is also squared.

\begin{figure}
  \centering
  \psfrag{t}{$\mathrm{\tau(d,\omega)}$}
  \psfrag{A}{$\mathrm{A}$}
  \psfrag{d}{$\mathrm{d}$}
  \psfrag{w}{$\mathrm{\omega}$}
  \psfrag{t1}{$\mathrm{\tau_1(d_1,\omega)}$}
  \psfrag{A1}{$\mathrm{A}_1$}
  \psfrag{d1}{$\mathrm{d_1}$}
  \psfrag{t2}{$\mathrm{\tau_2(d_2,\omega)}$}
  \psfrag{A2}{$\mathrm{A}_2$}
  \psfrag{d2}{$\mathrm{d_2}$}
  \subfigure[t][Single detector efficiency]{\label{fig:pet-geo-eff}\raisebox{0cm}{
      \includegraphics[height=0.126\textwidth]{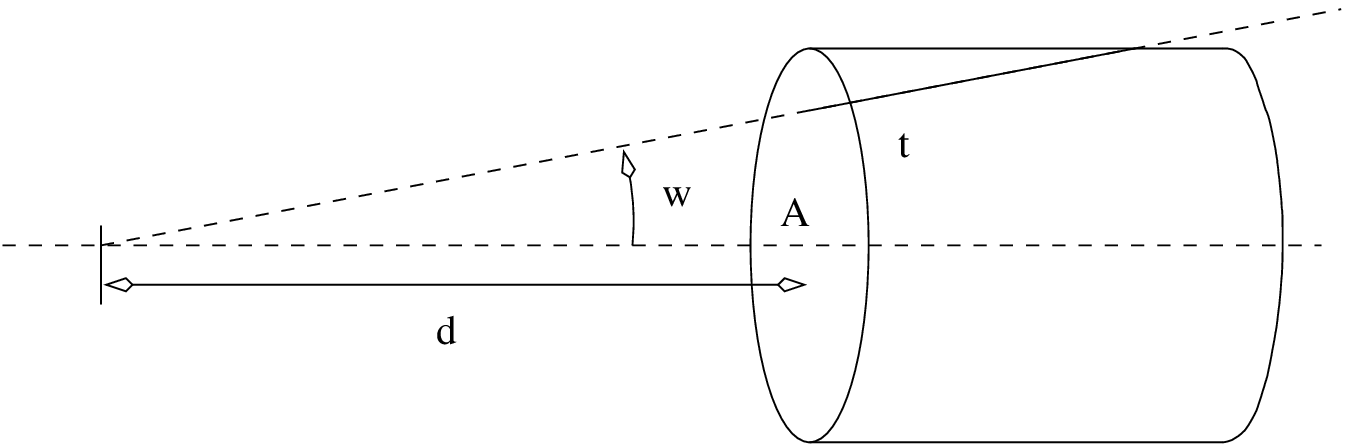}}}
  \subfigure[t][Coincidence detector efficiency]{\label{fig:pet-geo-eff-coin}\raisebox{0cm}{
      \includegraphics[height=0.126\textwidth]{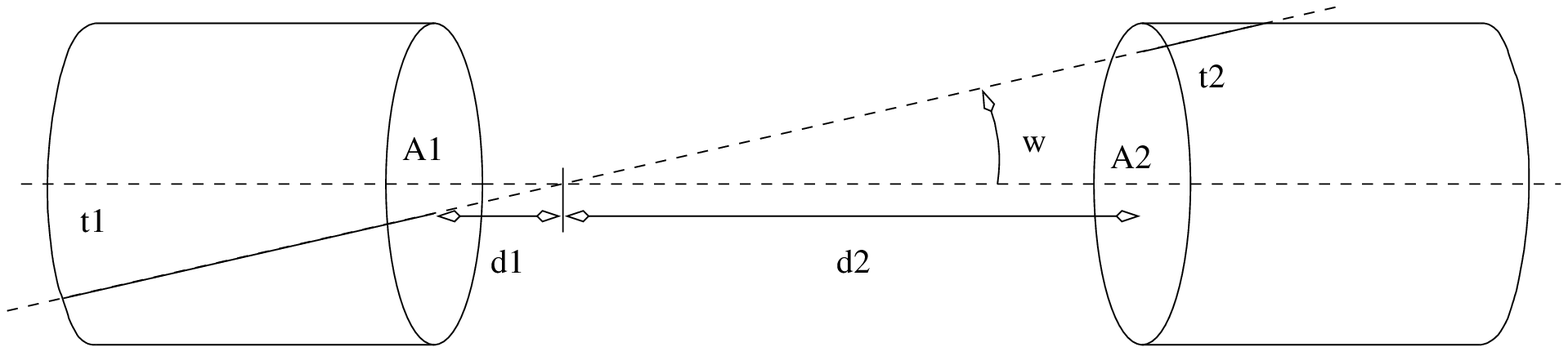}}}
  \caption[Calculation of the detection efficiency of single detector
    and a coincidence detector]{Calculation of the detection
      efficiency of single detector
    and a coincidence detector.}
\end{figure}

The overall sensitivity of PET is determined by the intrinsic
detection efficiency $\mathrm{\epsilon_{int}}$ of the detector system
and the fraction of the solid angle  $\mathrm{4\pi}$ that is covered
by the detector. The latter  is called the geometric efficiency
$\mathrm{\epsilon_{geo}}$ and is normally a
nontrivial function of the detector's and FOV's geometry. Actually,
the distinction between $\mathrm{\epsilon_{geo}}$ and
$\mathrm{\epsilon_{int}}$ is hard to clearly define. This is due
to the fact that one has to stop the impinging \g-ray in order to
detect it, {\em i.e.}\ a non-vanishing detector volume is required and
consequently the detector element's geometry is already involved.
For one single detector element, one can compute its efficiency by 
\begin{equation}
  \label{eq:single-eff-general}
  \epsilon_1(d,\omega)=\frac{1}{A}\int_A \left(1-e^{-\lambda\tau(d,\omega)}\right)
\end{equation}
where $A$ is the area of the detector facing the radiation
source, $d$ the distance of this surface from the source, $\omega$ the
solid angle as shown in figure~\ref{fig:pet-geo-eff}, $\lambda$ the
linear absorption coefficient of the detector material and 
$\tau(d,\omega)$ the length of the path of the \g-particle through the
active detector volume. Likewise one finds for the efficiency of an
annihilation coincidence detector
\begin{equation}
  \label{eq:coin-eff-general}
  \epsilon_\mathit{\gamma-\gamma}=\epsilon_1(d_1,\omega)\cdot\epsilon_2(d_2,\omega)=\frac{1}{A_1A_2}%
  \left[\;\int_{A_1} \left(1-e^{-\lambda\tau_1(d_1,\omega)}\right)\right]%
  \left[\;\int_{A_2} \left(1-e^{-\lambda\tau_2(d_2,\omega)}\right)\right].%
\end{equation}
In order to obtain the total efficiency, equation~(\ref{eq:coin-eff-general})
has to be multiplied by a factor that
represents the fraction of the spherical surface occupied by
$A_1$ and $A_2$. Furthermore, one has to take into
account that the examined object also absorbs \g-photons. 
The angular dependency is, however, highly nontrivial because it also
depends on the density distribution of the examined object, normally an animal
or a human. Approximately, one can write the
coincidence detection efficiency as (Cherry {\em et al.}\ \cite{Cherry})
\begin{equation}
  \label{eq:pet-eff-general}
  \epsilon_\mathit{eff}\simeq\epsilon^2_\mathit{int}\epsilon_\mathit{geo}e^{-\mu T},
\end{equation}
where $\mu$ is the (mean) absorption coefficient of the examined object
and $T$ its total thickness. Instead of $\epsilon_\mathit{geo}$,
 the average geometric efficiency
\begin{equation}
  \label{eq:average-geo-eff}
  \overline{\epsilon_\mathit{geo}}\approx\frac{2A}{3\pi(d_1+d_2)^2}
\end{equation}
is frequently used. This expression is valid when the detector dimensions are
small compared to the distance $(d_1+d_2)$ between both detectors. The
factor 2 in the numerator accounts for the fact that if one \g-photon
is detected, the second photon inherently points towards the second
detector module. The factor 3 in the denominator is the average
coincidence detection efficiency across the sensitive volume at
mid-plane. Commercial state-of-the-art PET scanners typically arrange 
up to 18000 small detectors in circular or (regular) polygonal
arrays. Each detector element operates in multi-coincidence mode with
a very large number of the opposite detector elements. If a
perfect ring-detector is supposed, {\em i.e.}\ a complete ring of very small
detectors of thickness $h$ in axial direction, diameter $D$ with
$h\ll D$ and
vanishing inter-detector spacing of adjacent elements, one obtains for a point
source located at the center of the ring by geometrical
arguments
\begin{equation}
  \label{eq:center-ring-detector-eff}
  \epsilon_\mathit{ring}\approx\frac{h}{D}
\end{equation}
and
\begin{equation}
  \label{eq:mean-ring-detector-eff}
  \overline{\epsilon_\mathit{ring}}\approx\frac{h}{2D}
\end{equation}
for the average geometric efficiency when moving the point-source
axially away from the center plane of the ring towards its
end-planes. Equations \ref{eq:center-ring-detector-eff} and
\ref{eq:mean-ring-detector-eff} are valid for a small volume element
near the center of the ring and also apply to polygonal arrays and
continuous detectors that use the Anger positioning logic.

\begin{figure}
  \centering
  \psfrag{D1}{$\mathrm{D_1}$}
  \psfrag{D2}{$\mathrm{D_2}$}
  \psfrag{dd}{$\mathrm{d'}$}
  \psfrag{d}{$\mathrm{d}$}
  \psfrag{D}{$\mathrm{D}$}
  \psfrag{S}{$\mathrm{S}$}
  \psfrag{A}{A}
  \psfrag{B}{B}
  \psfrag{C}{C}
  \psfrag{P1}{$\mathrm{P_1}$}
  \includegraphics[width=0.65\textwidth]{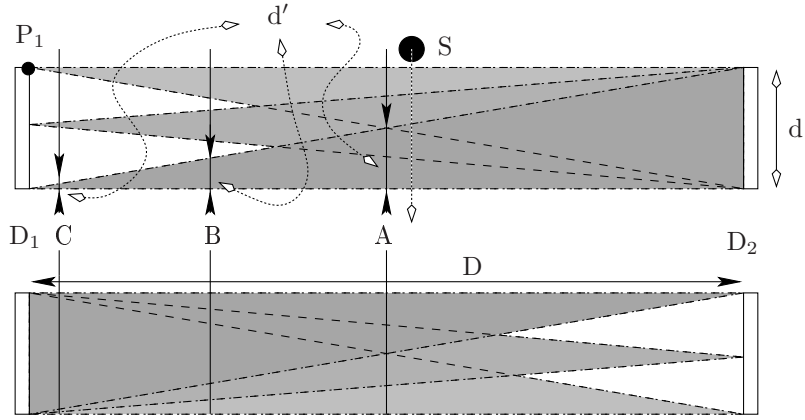}
  \caption[Intrinsic spatial resolution for coincidence detection]{Intrinsic spatial resolution for coincidence detection.}
  \label{fig:pixel-detect-derive}
\end{figure}

The spatial resolution of coincidence detectors depends in a
nontrivial way on the detector module's intrinsic resolution. For the
case of discretized detectors, it depends primarily on the size of the
individual detector elements since there is no possibility to decide
{\em where} within the element the \g-ray interacted. One can derive
the functional dependency of the spatial resolution on the position
along the interconnection line in the following way. Suppose,
an ideal point source is placed at the center between two ideal detectors. The
distance between both detectors is $D$. If discretized
detectors are used, the elements are normally all of equal size
$d$. Annihilation radiation in coincidence detection results in lines
of response that are defined by the two interaction points
within the crystal. Now, the source is moved along a plane between the two
detectors $D_1$ and $D_2$ and parallel to their surfaces through the
field of view of the coincidence detector (refer to
figure~\ref{fig:pixel-detect-derive}). By doing so, one obtains the
point-source response profile of the coincidence detector for that
plane. This profile is the convolution of the geometrical projections
of both detectors onto the plane where the point source is situated and
with the center of dilatation lying within the plane of the opposite detector.
Since the geometric projection depends on the position $x$ along the
optical axes $D$, the spatial resolution also does. At mid-plane (plane A)
of the detector pair, two identical rectangle functions with
width $d'=d/2$ are convolved and the response profile is a triangle
function with a FWHM of $d/2$. As the source moves towards either
detector, the response profile becomes trapezoidal (plane B) and a rectangle
function of width $d'=d$ at the face of either detector (plane C). For
any plane in between, the FWHM of the response profile can be given as
\begin{equation}
  \label{eq:spatial-res-pixel}
  \mathrm{FWHM}(x)=\left(\frac{1}{2}+\frac{|x|}{D}\right)d
\end{equation}
and the mean spatial resolution as
\begin{equation}
  \label{eq:mean-spatial-res-pixel}
  \overline{\delta_\mathit{pix}}=\frac{1}{D}\int_{-\frac{D}{2}}^{\frac{D}{2}}d\left(\frac{1}{2}+\frac{|x|}{D}\right)dx = \frac{3}{4}d.
\end{equation}
\begin{figure}[t]
  \centering
  \psfrag{f1}{\scriptsize FWHM $\mathrm{= d}$}
  \psfrag{f2}{\scriptsize FWHM $\mathrm{= \frac{3}{4}d}$}
  \psfrag{f3}{\scriptsize FWHM $\mathrm{= \frac{1}{2}d}$}
  \psfrag{A}{\small A}
  \psfrag{B}{\small B}
  \psfrag{C}{\small C}
  \psfrag{d}{$\mathrm{d}$}
  \subfigure[t][Discrete detector]{\label{fig:spatial-res-pixel}
    \includegraphics[width=0.8\textwidth]{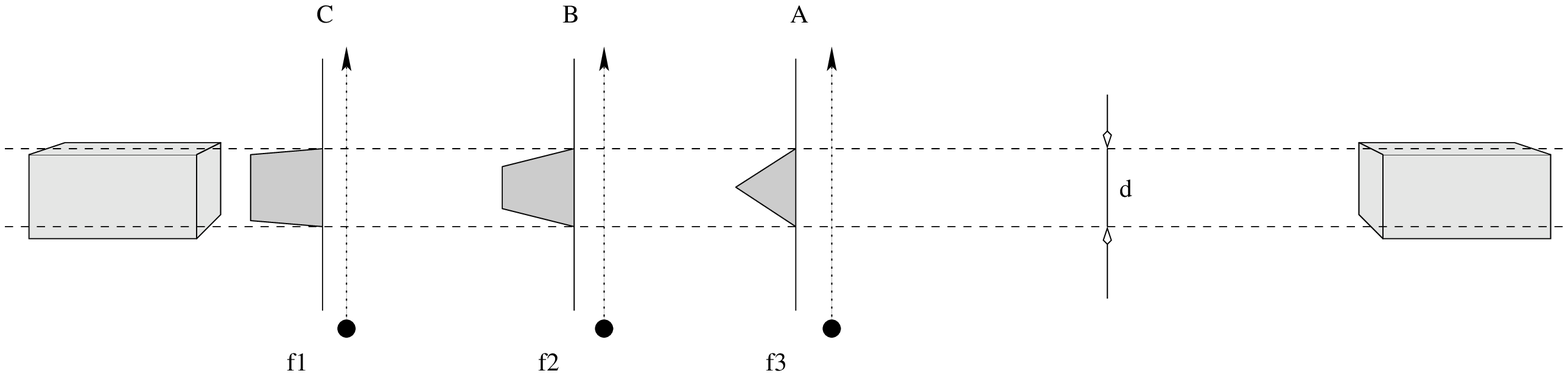}}
  \psfrag{f4}{\scriptsize FWHM $\mathrm{= \delta^i_S}$}
  \psfrag{f5}{\scriptsize FWHM $\mathrm{= \sqrt{\frac{5}{8}}\;\delta^i_S}$}
  \psfrag{f6}{\scriptsize FWHM $\mathrm{= \frac{\delta^i_S}{\sqrt{2}}}$}
  \psfrag{A}{\small A}
  \psfrag{B}{\small B}
  \psfrag{C}{\small C}
  \subfigure[t][Continuous detector]{\label{fig:spatial-res-cont}
    \includegraphics[width=0.8\textwidth]{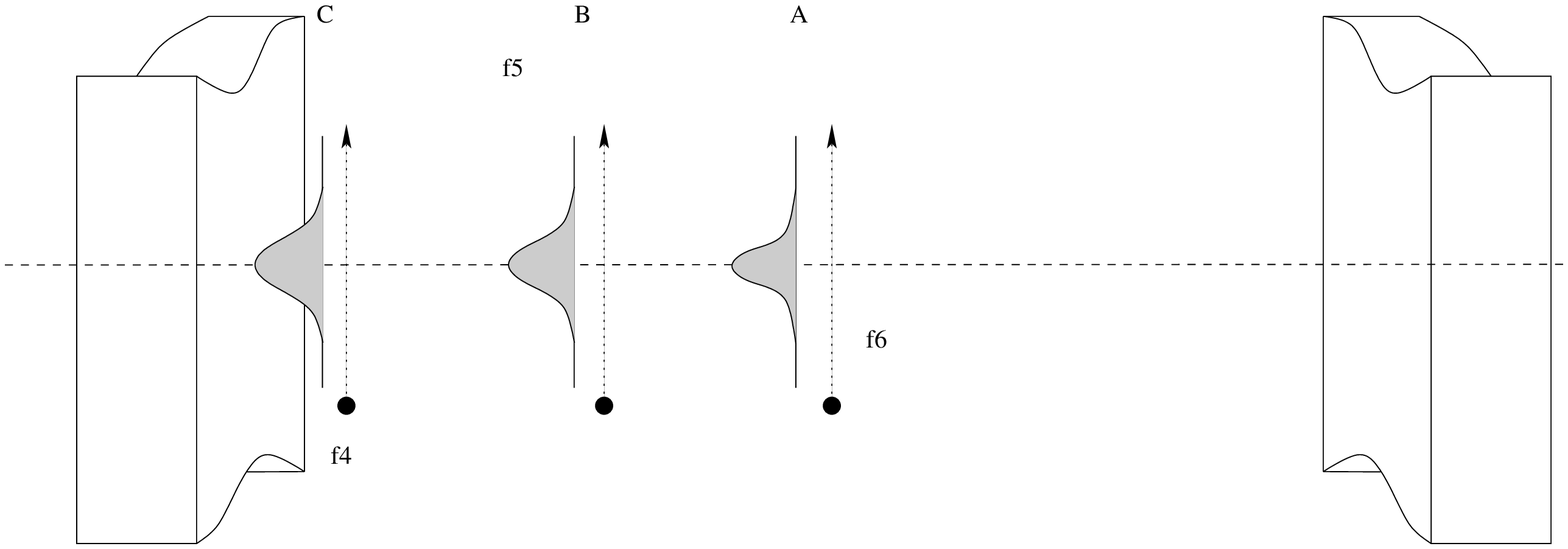}}
  \caption[Intrinsic spatial resolution for coincidence detection]{Intrinsic spatial resolution for coincidence detection.}
\end{figure}
The very same arguments hold for deriving the spatial resolution as
function of the inter-detector position $x$ for the case of continuous
detectors. If one assumes the intrinsic spatial resolution $\delta^i_\mathit{S}$ of both
detectors to be equal and its response profile of Gaussian form, then the convolution
of their projections is also a Gaussian. One obtains the standard
deviation of the resulting Gaussian by the geometric sum of the standard
deviation of the both projected Gaussians. Thus the spatial resolution
along $D$ is given by
\begin{equation}
  \label{eq:spatial-res-cont}
  \mathrm{FWHM}(x)=\frac{\delta^i_\mathit{S}}{\sqrt{2}}\sqrt{1+\left(\frac{2x}{D}\right)^2}
\end{equation}
and the mean spatial resolution by
\begin{equation}
  \label{eq:mean-spatial-res-cont}
  \overline{\delta_\mathit{cont}}=\frac{1}{D}\int_{-\frac{D}{2}}^{\frac{D}{2}}\frac{\delta^i_\mathit{S}}{\sqrt{2}}\sqrt{1+\left(\frac{2x}{D}\right)^2}dx%
  = \left(\frac{1}{2}+\frac{\ln{\big(1+\sqrt{2}\big)}}{2\sqrt{2}}\right)\delta^i_\mathit{S}\approx0.81\delta^i_\mathit{S}.
\end{equation}

\section{PET Designs}
\label{sec:PET-designs}

\begin{figure}
  \centering
  \psfrag{Cuts}{\hspace*{-3em}\footnotesize Saw cuts}
  \psfrag{PMTs}{\hspace*{-2em}\footnotesize four PMTs}
  \psfrag{Scintillator}{\hspace*{-1em}\footnotesize Block of Scintillator}
  \subfigure[t][block detector]{\label{fig:block-detector}\raisebox{0cm}{\hspace*{2em}
      \includegraphics[height=0.13\textheight]{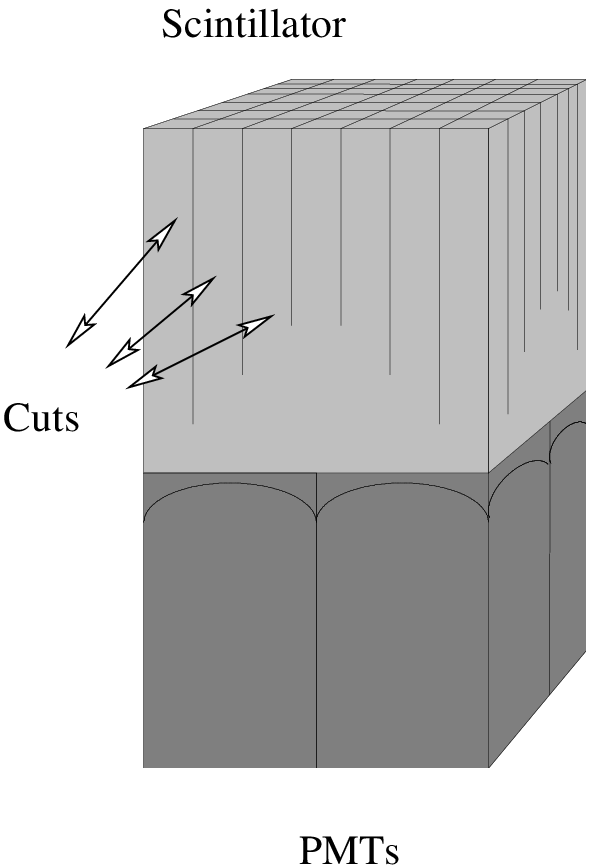}}\hspace*{2em}}\hspace*{0.025\textwidth}
  \subfigure[t][partial ring scanner]{\label{fig:part-ring-scanner}\raisebox{0cm}{\hspace*{2em}
      \includegraphics[height=0.13\textheight]{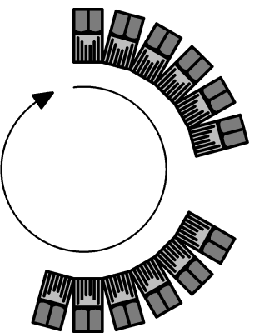}}\hspace*{2em}}\hspace*{0.025\textwidth}
  \subfigure[t][full ring scanner]{\label{fig:full-ring-scanner}\raisebox{0cm}{
      \includegraphics[height=0.13\textheight]{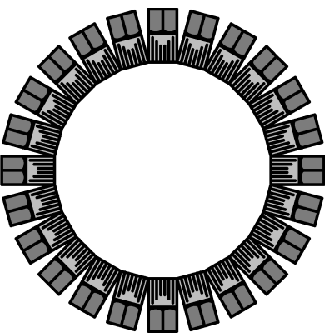}}}\\
  \subfigure[t][planar scanner]{\label{fig:planar-scanner}\raisebox{0cm}{\hspace*{2em}
      \includegraphics[height=0.13\textheight]{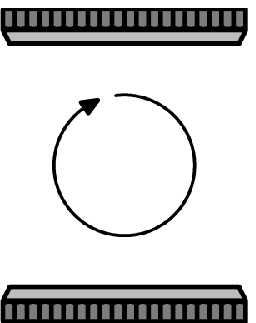}}\hspace*{2em}}\hspace*{0.05\textwidth}
  \subfigure[t][hexagonal scanner]{\label{fig:hexagonal-scanner}\raisebox{0cm}{
      \includegraphics[height=0.13\textheight]{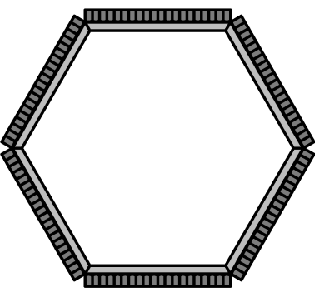}}}
  \caption[Positron Emission Tomography scanner
    geometries]{Positron Emission Tomography scanner
    geometries. Photo-detection elements are dark gray shaded and
    scintillator crystals are light gray shaded.}
\end{figure}

The first PET systems were either two or more Anger-type cameras
operated in temporal coincidence or used individual detector units
consisting of a scintillator block coupled to a photomultiplier tube
and arranged in one or multiple rings around the subject. This
geometry required a huge number of the expensive photomultiplier tubes
and as a consequence there was a large interest in multiplexing schemes
that would reduces the number of PMTs. In 1984, Scanditronix designed 
a PET scanner that used detector modules of one PMT and two scintillators
with different decay time \mycite{Nutt}{}{2001}. The crystal where the \g-ray is stopped
could be identified by measuring the decay time of the current
pulse. Only a few tomographs were produced and the technique was
given the name of {\em phoswich}-detection\footnote{The word phoswich
 is derived from the words {\em \underline{phos}phor} and {\em
   sand\underline{wich}}.}. It now carries the hope of
enabling depth of interaction detection in \g-ray imaging.
Burmham {\em et al.}\ at the Massachusetts General Hospital placed many
individual photodetectors on a circular light-guide. By taking the
ratio of two adjacent PMTs, the position could be estimated just in
the way it is done in Anger-type cameras. Mike Casey and Ronald
Nutt from CTI recognized the advantages of this method and developed
finally the {\em Block detector} that significantly eases the
manufacturing process and {\em eo ipso} reduces costs. Figure
\ref{fig:block-detector} shows the basic configuration. A large piece
of scintillation material is partially segmented into many small
elements. A variation in he depth of the saw cuts from element to
element is used to control the distribution of the scintillation light
to the four single-channel photomultiplier tubes. Two  position
estimates $X$ and $Y$ can be computed from the four signals by simple arithmetic operations:
\begin{align}
  \renewcommand{\arraystretch}{1.6}
  \label{eq:anger-log-block-detect}
  X=\mathit{\frac{PMT^{up}_{right}+PMT^{down}_{right}-PMT^{up}_{left}-PMT^{down}_{left}}%
{PMT^{up}_{left}+PMT^{up}_{right}+PMT^{down}_{left}+PMT^{down}_{right}}}\\[10pt]
  Y=\mathit{\frac{PMT^{up}_{right}+PMT^{up}_{left}-PMT^{down}_{right}-PMT^{down}_{left}}%
{PMT^{up}_{left}+PMT^{up}_{right}+PMT^{down}_{left}+PMT^{down}_{right}}},
\end{align}
which are used afterwards to identify the element where the
scintillation light came from. Typical Block detectors are made from
20 to 30 mm thick BGO or LSO with 4 to 6 mm wide sub-elements. While the
first Block detector used  only 32 crystal elements, up to 144
crystals per PMT can be found in modern state-of-the art PET scanners.

The geometry of dedicated PET-systems is strongly confined by the facts that
temporal coincidence detection is required and that sectors without 
\g-ray detection result in artifacts due to the image reconstruction methods used
 (Natterer and Wübbeling \cite{Natterer}). Therefore, one finds either closed
ring and regular polygon systems that exhibit a clear axial symmetry
or partial ring and polygon systems that are rotated in order to
re-establish the broken symmetry. Figures
\ref{fig:part-ring-scanner}-\ref{fig:hexagonal-scanner} illustrate
some different approaches. Especially in PET scanners with circular
geometry one often finds several stacked full rings in order to
increase the axial FOV. PET scanners with a
polygonal arrangement of panel detectors normally inherently exhibit a
sufficiently large axial FOV. Further advantages of full-ring systems
are a higher efficiency and reduced scanning time. In particular the gain
in efficiency has to be mentioned since it grows with the square of
the active detecting volume when no septa are used. Suppose one has a
system consisting in one single detector ring. If one now adds a second
full ring, one not only gains the events detected by this new ring, but
also those events where one \g-photon is detected by one ring and the
second by the other.

\section{Radioactive Agents and Nuclear Medicine}
\label{sec:radiopharmacos}

The very first study of metabolism using radioactive substances 
transport within living systems dates back to 1923, when de Hevesy and
Paneth investigated the transport of radioactive lead in plants.
Seven years after, Ernest Lawrence {\em et al.}\ constructed the first 
cyclotron \mycite{Bryant}{}{1993} with the aim of bombarding elements
and exploring the nature of
the atomic nucleus \mycite{Wagner}{}{1998}. It took further 4 years
until Curie and Joliot discovered that elements could be activated 
by irradiating them with alpha particles from radium or polonium.
The importance of the radiotracer principle together with the fact
that virtually every element could be activated now became clear, and
 in 1935 Hevesy and Chiewitz were the first to apply this new
principle to a study of \isotope{P}{32} distribution in rats. In
1936, Martin David Kamen came to Berkeley to visit Lawerence's famed 
radiation laboratory. He was finally hired and used the cyclotron to 
produce the radioactive \isotope{C}{11} in order to clarify
the still poorly understood process of photosynthesis. However, the usefulness
of this isotope was limited by its very short half-life of only 21
minutes that forced Kamen and its colleagues to work at high speed
and even to use carrier pigeons to interchange samples between
laboratories. The discovery of \isotope{C}{14} with its half-life of
5730 years was the result of the search to find a more suitable
radioactive carbon isotope. This isotope is very important because it
can be used for determining the age of
fossils as shown by William Libby in 1946. Until the development of
CT and MRI, nuclear medicine physicians focused their attention to 
the imaging of organs, which were insufficiently differentiated by 
X-ray imaging. However, the overwhelming potential of radiotracer
imaging lies in its possibility to mark and trace virtually every 
metabolic active substance allowing a graphical representation of its
distribution within the examined object. Thus  the development 
of radiopharmaceuticals was focussed on biological elements like
\isotope{C}{11}, \isotope{O}{15}, \isotope{F}{18} and  \isotope{N}{13}
that could be generated with cyclotrons. 

One of the first {\em in vivo} studies was the measurement of cerebral blood
flow with \isotope{Kr}{79} and later \isotope{Xe}{133} and
\isotope{I}{131}. With the emergence of the Anger-type camera and 
detectors for positron-emitting radionuclides in 1953, the {\em in vivo} 
localization of tumors  became possible. In 1960, Stang and Richards
from the Brookhaven National Laboratory, USA, announced the new
radionuclide \isotope{Tc}{99m}, obtainable from \isotope{Mo}{99} and
Paul Harper at the University of Chicago realized that it was almost
ideal for nuclear imaging for various reasons: Its decay energy
perfectly matches the requirements of Anger-type cameras. Its physical
and biological half-life was short enough to allow administration of
high doses and it was readily available through its long-living
generator \isotope{Mo}{99}. Owing to its properties, \isotope{Tc}{99m}
became the workhorse of nuclear medical imaging. Decades elapsed
before chemical analogues of interest could be tagged with
radionuclides and even nowadays many radioactive tracer substances in
routine diagnostic imaging are of a very basic chemical structure and
often do not exist naturally in the human body. The renaissance of
the  biological radionuclide began with a study of Ter-Pogossian and
Powers in 1958 on the oxygenation of tumors. They worked with
\isotope{O}{15} and mice with mammary adenocarcinomas. Directly
connected to \chform{O_2}-metabolism one finds the different saccharides as
important energy substrates in living organisms explaining the high
interest in a radiotracer that could visualize this metabolism. 
Of special interest is glucose owing to its importance as an energy pool
for the nervous system. Since an adult human requires at least 
180 grams of glucose every day (Ruhlmann \cite{Ruhlmann}), standard scintigraphy
could not be used due to the high signal level. Another inconvenience
of this molecule was that it could not be labelled with one of the usual
radionuclides known at that time. A first attempt was made in 1977 by
Sokoloff {\em et al.}\ using \isotope{C}{14}-deoxyglucose. Deoxyglucose (DG)
is an analogue of glucose in which the hydroxyl group on the second
carbon has been replaced by hydrogen (Phelps \cite{Phelps}). DG has also been
labeled with the positron-emitting \isotope{C}{11} thereby making it 
accessible to positron emission tomography. The much shorter half-life
of only 20 min.\ allows repeated studies in a single day, but 
unfortunately the DG's blood pool clearance only takes about 40
min. Attempts were also made with \isotope{C}{11}-glucose, that is,
using the radioactive labeled natural substrate glucose. Unlike the
analogue tracer, labeled substrates cannot usually be used to isolate
single reactions steps. In the case of glycolysis, the metabolic
process in which glucose is converted to pyrovate\footnote{Pyruvate is
the compound to start the energy cycle within cells that yield direct 
precursors to ATP or ATP itself.}, ten different chemical reactions
are involved. The \isotope{C}{11}-labeled glucose undergoes a
substantial fraction of these reactions and the following
ATP-metabolism until it is removed from the cell as
\chform{^{11}CO_2} or \isotope{C}{11}-labeled lactate. It is clear that
the radiation from these end-products cannot be distinguished
from radiation coming from the tracer in its original form. 
Further, labeled intermediates could also be taken up by other
processes completely different for glucose. The kinetic data is thus
distorted and less reliable. By substituting the hydroxyl group on
the second carbon of glucose by \isotope{F}{18} one obtains 
[\isotope{F}{18}]-FDG. This radiotracer has a more suitable physical
half-life of 110 min. and furthermore provides the advantage that its
glycolysis breaks down after a few chemical reaction inside the cell.
The idea for this tracer was born at a wine-tasting event where Lou
Sokoloff and Martin Reivich coincided. They commented on their findings
to Al Wolff and Joanna Fowler from the Brookhaven National
Laboratory, where \isotope{F}{18}-labeled FDG  has been synthesized
for the first time by Ido {\em et al.} The
first images using this tracer in the brain were taken with the
Mark-IV scanner by Kuhl {\em et al}.\ and numerous later studies have been
realized with [\isotope{F}{18}]-FDG making it the prime radiotracer
for routine medical PET explorations. Studies that were 
carried out with this radiotracer over decades have demonstrated the
accumulation of [\isotope{F}{18}]-FDG in cancerous tissue and showed 
that PET is very useful for different clinical requirements, such as
detecting primary sites of cancer, differentiating benignity or
malignity of lesions, staging and grading the degree of malignity,
planing and monitoring of the treatment and detection of recurrent
diseases and metastasis.

Nuclear imaging modalities are now established as a powerful
scientific and clinical tool for probing biochemical processes in the
human body. Since the discovery of \isotope{Tc}{99m} and the
development of [\isotope{F}{18}]-FDG, many other more specific
radiotracers have been found and successfully applied. It finally leads
to the emergence of nuclear pharmacy as a specialty practice and
motivated a large number of research groups to search for further
possible radiopharmaceuticals. At the moment, there is already  a very
large spectrum of specific pharmaceuticals which can be used for
visualizing very particular aspects of the metabolism of diseases and the
prospects of finding new ones are promising. Routine clinical
application nowadays embraces above all oncology, nuclear cardiology,
and neurology (Schlyer \cite{Schlyer:2004}, Acampa {\em et al.}\
\cite{Acampa:2000}). Nevertheless, the most commonly used radiotracers are
still very unspecific or do not exist in the human metabolism. Present
and future developments are strongly focused on new radiotracers with
high specificity and with adequate kinetic characteristics for
quantitative analysis. This includes the development of radiotracers
for imaging neurotransmitter systems, amino acid transport, protein
synthesis, DNA synthesis and receptor imaging (Kim and Jackson
\cite{Kim:1999}, Fowler {\em et al.}\ \cite{Fowler:2003}). A list of
commonly mentioned radiotracers together with the medical tests that
they are used for can be found in
Appendix~\ref{app:common-radiotracer}.

\subsection{Requirement for Radiotracers}

Molecular imaging is  very promising  for providing physiological
or biochemical information for individual patients and, as has been
mentioned before, is already
becoming a routine diagnostic technique in medicine. Generally, an
imaging tool for routine medicine has to obtain the results in a short 
time in order to be useful. Functional imaging also requires that
the uncertainty in each measurement is smaller than the reconstructed 
contrast among different physiological states so that a change in
these states can be easily detected. Consequently, radiotracers have
to fulfill some important requirements  in order to be suitable for
molecular imaging, especially if they are applied in human medicine
(Phelps {\em et al.}\ \cite{Phelps}, Hamilton \cite{Hamilton:2004}). First
of all, the tracer must be related predominantly to the process of
concern, whatever this may be. Also it is necessary that the
turnover time of the tracer within the specimen must be within the
time window of the nuclear imaging modality. This second requirement
includes many different aspects of the radiotracer and radionuclide,
the scanner to use and the biological tissue and function that is
imaged. The choice of radionuclide has to be made on the basis of
providing the maximum diagnostic information together with minimum 
radiation dose to the patient. Therefore, the radionuclide must have a
sufficiently long physical half-life $T_\tincaps{PHY}$ in
order to obtain sufficient
statistics for medical images of reasonable quality and at the same
time, they have to be short for a fast lowering of the dose that
 the patient receives. There are two processes that reduce the
radioactivity inside the patient once the radiopharmaceutical has been
administered. Beside the radioactive decay, the radionuclide is
removed from the body physiologically with a biological half-life
$T_\tincaps{BIO}$. The effective half-life is the inverse sum of
$T_\tincaps{BIO}^{-1}$  and $T_\tincaps{PHY}^{-1}$.
The pharmaceutical is also required to be non-toxic in the administered
amounts, to be present in the target tissue in a sufficiently short
time and the uptake in pathology should show differences from normal.
Moreover, many of the radiopharmaceuticals are synthesized {\em in
  situ} and just before they are administered to the patient. This
requires  that the synthesis and the imaging process must be finished
within a time frame that is compatible with the effective half-life of
the radiopharmaceutical. Synthesis and imaging may take few minutes up
to hours.

\subsubsection{Generation}

Radiopharmaceuticals can be  divided into two categories by their
properties whether they are brought ready for use from outside the site
of application or whether they have to be reconstituted {\em in situ}.
The earliest method of producing commercial artificial radionuclides 
was to expose a target material to the neutron flux of nuclear
reactors. However, even in intense neutron fluxes only a very small
fraction of the target nuclei is activated. Since the target and
product nuclei are different isotopes of the same elements (neutrons
carry no charge), they cannot be separated chemically. Hence, the
product is not target- or carrier-free and of low specific activity.
Carrier-free products can be generated by choosing particular targets
whose activation results in short-lived intermediate products. If
the subsequent decay to the desired radionuclide includes a change in
the element, chemical separation is possible. Alternatively,
radionuclides can be obtained by using particle accelerators,
especially compact cyclotrons. Since the accelerated particles are
charged, the product is normally a different element and can be
separated chemically for obtaining carrier-free radionuclides. 

A very important concept for obtaining radionuclides is the production
of a long-lived generator. The principle consists in the creation of an
unstable but long-lived parent radionuclide at the production
facility. This parent decays steadily to produce a short-lived
daughter nuclide which has all properties of the desired radionuclide
and moreover whose chemical characteristics allow for easy separation.
Examples for generators are \isotope{Mo}{99}, \isotope{Rb}{81} and
\isotope{Sn}{113}. They decay to the radionuclides \isotope{Tc^m}{99},
\isotope{Kr^m}{81} and \isotope{In^m}{113} with the half-lives of 67
h, 4.7 h and 115 days respectively.

\chapterbib


  \cleardoublepage{}
\chapter{Motivation and Outline}
\label{ch:motivation}

\vspace*{-1.1eX}

\chapterquote{%
Those who fail to learn the lessons of history are doomed to repeat
them.
}{%
George Santayana, $\star$ 1863 -- $\dagger$ 1952
}

\vspace*{-1.1eX}

\PARstart{N}{uclear} imaging modalities not only have proven very
valuable in the medical disciplines of oncology, cardiology,
neurology, neuropsychology and cognitive neuroscience and psychiatry,
but also play a crucial role in pharmacology. Due to their dedicated
functional imaging techniques, it is not only possible to evaluate in
preclinical trials new drugs very effectively in small animals but also
opens the doorway to a much more sophisticated and function specific
drug development. Moreover, nuclear imaging has significantly enhanced
the ability to study animal models of diseases by enabling the
continuous \mbox{\em in vivo} monitoring of disease development and
rendering many necropsy studies unnecessary. The possibility of transgenic
manipulation of mice emerged in recent years and led to a more
detailed understanding of existing crosslinks between specific genes
and molecular, cellular and organ functions. So-called {\em Knock-out} mice are
now the basis for studying the physiology of the brain, heart and any
other function. The perspectives for molecular imaging become strikingly
clear if one bears in mind that this progress was made with
imaging equipment of rather poor performance. Spatial resolutions above
1 mm, sensitivity of mere 5\% and less for dedicated small animal
PETs together with energy resolutions of about 15\% made many of these
results possible. 
It is against this background that many research groups all over the
world are making a great effort to further improve all performance parameters. 
Unfortunately, many of these parameters constitute design conflicts 
and nearly all conflict with the requirement of low
fabrication cost. Yet, one of the most important prerequisites for
a widespread application of the developed technologies is a low
initial outlay.

A main motivation for this work has been the reduction of fabrication
costs, under the condition that a reasonable detector performance is
achieved. The decision to use continuous, large-sized scintillation
crystals has been taken for this reason. It avoids costly
segmentation of the crystals which involves the loss of a 
substantial amount of material due to involuntary rupture and the saw
cuts and requires either a rather tedious assembling or the
installation of a robot. Likewise, the choice of using large-area
position sensitive photomultipliers together with the center of gravity
algorithm (CoG) is a direct consequence of this goal. The price per unit
sensitive area of these devices is acceptable for use in dedicated
small animal scanners and the analogically implemented CoG algorithm
effectively reduces the number of required
electronic channels. Unfortunately, electronic implementations of the CoG algorithm in its present
form are not able to
provide information about the depth of interaction (DOI) of the incident $\gamma$-ray. This leads to
artifacts that considerably reduce the spatial resolution of
the detector, especially for the PET modality 
where thick scintillation crystals are required for
stopping the incident $\mathrm{511\,keV}$ $\gamma$-ray effectively.

In order to fulfill the precondition of reasonable detector
performance, important enhancements of the analogue CoG algorithm
become necessary. The known fact of the correlation between the width
of the scintillation light distribution and the depth of interaction 
motivated the design of DOI enhanced charge dividing circuits and an
algorithm for recovering the true impact positions from their output
signals. For this purpose, a detailed comprehension of all aspects of light
propagation within the scintillation crystal until its detection is
indispensable and in chapter~\ref{ch:light-distribution}, the distribution of
the scintillation light across the photosensitive surface of the
detector is parameterized. In
chapter~\ref{ch:enhanced-charge-dividing-circuits}, the behavior
of existing charge dividing circuits is studied. It will be shown
that all configurations analogically compute the trivial and the first
non-trivial moment of the applied signal distribution and a novel
modification will be presented that allows the simultaneous
computation of the second non-trivial moment, which relates to the DOI. Chapter~\ref{ch:compton}
briefly summarizes the impact of Compton
scattering on the expected accuracy of the moment measurements.
The results of chapters~\ref{ch:light-distribution} and
\ref{ch:enhanced-charge-dividing-circuits} have been verified by
experiment. How this was done and what results have been obtained is
explained in chapter~\ref{ch:experiment}.
Chapter~\ref{ch:position-reconstruction} presents a standard
method on how the new information provided by the depth-enhanced charge
divider circuits can be used to reconstruct the full, three-dimensional
photoconversion position. The final
chapter~\ref{ch:conclusiones-and-outlook} summarizes the main
conclusions and discusses the future outlook for this method.


  \cleardoublepage
\chapter{Detector Components and Limits}
\label{ch:detect-comps-for-nuc-img}

\chapterquote{%
Every generation laughs at the old fashions, but follows religiously the new.}{%
Henry David Thoreau, $\star$  1817 -- $\dagger$ 1862 
}

\PARstart{T}{he} first radiation detector that was used in medicine was clearly the
photographic film for X-ray transmission radiography. Film-based
radiography is still used extensively in daily medical applications, a
circumstance which is due to the high amount of structural
information, the low cost and complexity and an outstanding spatial
resolution of the photographic film. The overwhelming progress in
atomic and nuclear physics caused an ever-increasing interest in more
sophisticated and accurate particle detectors with constant
improvement in spatial resolution, energy resolution, timing
resolution, etc. In particular, the field of nuclear medicine benefits
from these advances for obvious reasons. 

Radiation detectors can be classified into three large families: Ionization
detectors, scintillation detectors and semiconductor
detectors. Ionization detectors were the first electrical devices that
were developed for radiation detection. Members of this family are the
ionization chamber, the proportional counter, the Geiger-Müller
counter and later the multi-wire proportional chamber, the drift
chamber and the time projection chamber (see for instance Leo
\cite{Leo:1994}). Except for very few cases, such as the HIDAC PET, developed
by Jeavons {\em et al.}\ \cite{Jeavons:1978}, and a more recent proposal of
using lead walled straw tubes (Lacy {\em et al.}\ \cite{Lacy:2001}), none of
these detectors are applied to medical imaging, mainly because it is
difficult to design compact detectors. However, they are
extensively used in experimental high energy physics since there is no
need for compactness in large particle accelerator installations.

The most important detector design for medical imaging is the
scintillation detector. They are undoubtedly the most versatile
particle detectors with a wide range of applications. In 1903, Crooks invented
 the {\em spinthariscope}. This is probably the first use of
scintillators for particle detection and consisted of a \chemform{ZnS} screen
that produced weak scintillation when struck by $\alpha$-particles that
could be observed with the naked eye. This, however, required some
practice and it was tedious to use. After the success of Iams {\em et al.}\
in 1935 to produce the first triode photomultiplier tube, the human
eye was replaced in 1944 by these new devices for the first time by
Curran and Baker. The scintillation could now be counted using
electronic devices. After further development and improvement of the
photomultiplier tubes and the scintillators in the following
decades, scintillation particle detectors are now among the most
reliable and convenient detectors. This, and the fact that nowadays
very compact scintillation detectors can be manufactured, leads to their
wide application in medical imaging. Modern Scintillation detectors
can provide information about the energy and two-dimensional position of the
radiation. Moreover, they are very fast and, depending on the
scintillator, can show very short recovery times compared to other
types of detectors. With certain scintillators it is even possible to
distinguish between different types of particles by means of pulse
shape discrimination.

Semiconductor detectors are the youngest family of particle detectors.
Real development of these devices began in the late 1950's. They
provided the first high-resolution detectors for energy measurement and
were quickly adopted for particle detection and especially gamma
spectroscopy. The use of semiconductors for position measurement is
rather new, however development in this field
has seen huge progress in recent years. This was mainly due to the
adaption of technologies used in micro-electronics and has been further
enhanced by enormous improvements in silicon detector technology, front
end electronics and signal processing. Micro-strip detectors and silicon
drift chambers exhibit excellent properties for use as
high-resolution particle track detectors and are possibly the
detectors for the next generation experiments in high energy physics.
A very important drawback of semiconductor particle detectors arises
from their intrinsic noise. Especially for imaging modalities that are
based on high energy photons, this is an important design
conflict. Due to the high energy of the photons, $\gamma$-ray
detectors need to have an elevated stopping power which is achieved by
sufficiently sensitive volume of high density materials with high
effective atomic number. The density and effective atomic number are
already determined by the choice of the semiconductor and only the
volume remains as a free design parameter. Unfortunately, the noise of
semiconductor devices scales with the volume and very strong
constraints exist. Although the noise can be reduced by active cooling,
this would further increase the cost of the system. Therefore there
are very few and very specific applications that allow the use of
solid state detectors in medical imaging. 

\section{Solid-State Gamma-Ray Detectors}
\label{sec:solid-state-gamma-ray-detect}

The basic operating principle of semiconductor detectors is analogous
to gas ionization chambers. The incident ionizing particle creates a
large number of electron-hole pairs along its trajectory through the
semiconductor material and these charges are collected by an electric
field. Compared to gaseous ionization detectors, there are important
advantages. In the case of semiconductor material, the average energy
that is required to create a single electron-hole pair is 
$\mathrm{\sim 3.6\,eV}$, one order of magnitude smaller than the
average ionization energy of $\mathrm{\sim 30\,eV}$ for gases. 
Furthermore, semiconductors have a higher effective atomic number and density.
Therefore the energy loss per traversed length is very large resulting
in a high stopping power for ionizing particles. As a consequence,
semiconductor particle detectors can be made very compact and the
mechanical rigidity allows for building self-supporting structures. 
Despite the high material density, electrons and holes have a very
high mobility and the charge can be collected in a few nanoseconds.
Additional advantages are the possibility of embedding fixed-space
charges for creating sophisticated field configurations and, as both
detectors and electronics can be built out of silicon, they can be
integrated in a single device (Lutz \cite{Lutz:1995}).
Important disadvantages are the already mentioned intrinsic noise that
often makes intense cooling necessary and the possibility that their
crystalline structure can be damaged by the detected radiation.

\begin{figure}[!t]
  \centering
  \psfrag{Al}{$\mathrm{Al}$}
  \psfrag{p+}{$\mathrm{p}^+$}
  \psfrag{n+}{$\mathrm{n}^+$}
  \psfrag{n}{$\mathrm{n}$}
  \psfrag{Bias}{Bias}
  \psfrag{Signal}{Signal}
  \psfrag{s}{\raisebox{1eX}{\hspace*{-2cm}$\underbrace{\hspace*{3.6cm}}_\mathrm{Sensitive\, Volume}$}}
  \includegraphics[width=0.6\textwidth]{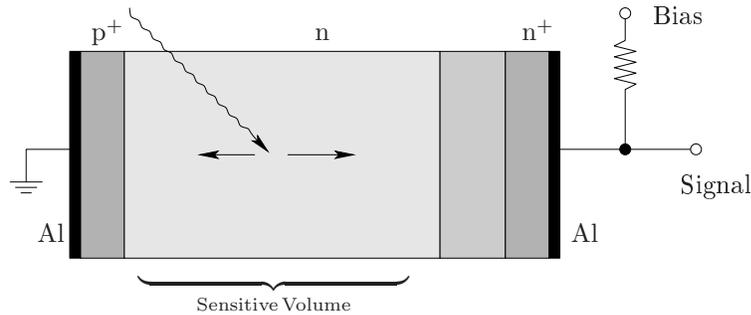}
  \caption[p-n diode junction  detector in reversed bias operation
  mode]{p-n diode junction  detector in reversed bias operation mode.}
  \label{fig:pn-diod-junc}
\end{figure}

A simple $\mathrm{p}$-$\mathrm{n}$ diode junction can already be used
for the detection of light or ionizing radiation. The diode may be
used without the application of an external bias voltage or in the
reverse bias mode. Unbiased detectors are preferred for radiation
level measurements. The example for a $\mathrm{p}$-$\mathrm{n}$
junction in reverse bias mode is shown in
figure~\ref{fig:pn-diod-junc}. This is the common operation mode
for single particle measurements. The diode consists of a very low
doped $\mathrm{n}$ substrate. In order to be able to collect the
charges released by the radiation, metal contacts must be bonded onto
both sides of the diode. Since simple contacts between many metals
and semiconductors lead to depletion zones, highly doped but shallow 
$\mathrm{p}^+$ and $\mathrm{n}^+$ regions are introduced at both
surfaces of the diodes before fitting the contacts. Electron-hole pairs
that are generated inside the sensitive volume are separated by the
electric field and will move towards the electrodes. For signal
isolation purposes, the bias voltage is supplied to the
$\mathrm{p}$-$\mathrm{n}$ diode through a resistor. 

Nearly all existing semiconductor detectors are based on this
principle. More or less sophisticated modifications are made in order to
achieve special required properties, such as position resolution, etc. 
Besides the elemental semiconductors \chemform{Si} and \chemform{Ge}, there are composed
semiconductors consisting of combinations of elements from different
periodic table groups. Well known members of this family are \chemform{GaAs} and
\chemform{InAs}. In recent years, much investigation has been done in this field
and the list of possible candidates is now very long and allows for
choosing semiconductor materials that are most suitable for the
application of interest. For instance, nitrides of the third group have
high tolerance to ionizing radiation, making them suitable for
radiation-hardened electronics. Among this large range, the
II-IV semiconductor Cadmium telluride (\chemform{CdTe}) and the II-VI ternary
alloy semiconductor (\chemform{CdZnTe}) are promising materials for hard X-ray and
$\gamma$-ray detection. The high atomic numbers of the materials give
a reasonable quantum efficiency that is suitable for operation at the
energy range from $\mathrm{10\,KeV}$ to $\mathrm{500\,KeV}$ (Scheiber and
Gaikos \cite{Scheiber:2001}). These materials are already used in
commercial devices for digital radiography, computerized tomography,
small field-of-view $\gamma$-cameras and intra-operative probes for
radio-guided surgery. However, large multi-element detectors made from
these materials are very expensive which is still the main drawback of
this technology.

\section{Scintillation Detectors}

\begin{figure}[!t]
  \centering
  \psfrag{Housing}{Housing}
  \psfrag{Electronics}{Electronics}
  \psfrag{Scintillator}{Scintillator}
  \psfrag{Opical coupling}{Optical coupling}
  \psfrag{Photodetector}{Photodetector}
  \includegraphics[width=0.46\textwidth]{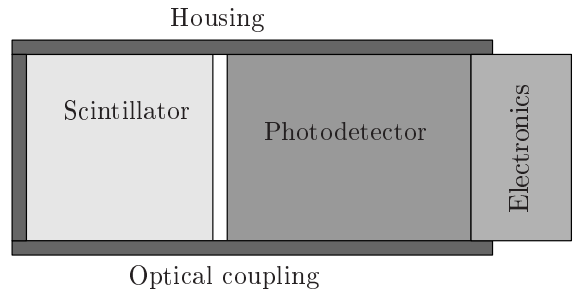}
  \caption[Schematic diagram of the most basic scintillation
  counter]{Schematic diagram of the most basic scintillation
    counter. The housing is required for avoiding the detection of
    straylight. It may also contain lead for reducing the detection of
    background radiation and/or $\mu$-metal for shielding against
    magnetic fields.}
  \label{fig:scint-detect-principle}
\end{figure}

The basic configuration of a scintillation detector is shown in
figure~\ref{fig:scint-detect-principle} and consists of a
scintillator that is optically coupled to a photosensor. 
As radiation interacts with the scintillator by elementary processes
like Compton scattering or photoelectric effect, a {\em knock-on} electron is
produced that then excites or ionizes the atoms and molecules of the
scintillation crystal. The decay of these excited scintillation centers
give rise to the predominantly visible light photons which can be converted
to electrical signals with the photodetector.

\subsection{Scintillators}
\label{subsec:scintillators}

Visual counting of the (at that time) recently discovered X-rays and
also natural radioactivity became possible with the powdered phosphor
materials $\mathrm{CaWO_4}$ and $\mathrm{ZnS}$ at the end of the 19th
century and made possible the previously mentioned invention of the {\em
  spinthariscope}. Since that time, inorganic
scintillators especially have been an
integral part of medical imaging and are still heavily used in this
discipline. This can be explained by their comparatively good detection
efficiency for hard radiation. After the invention of the photomultiplier tube, the final
breakthrough of scintillation detectors was due to the discovery of
the pure and activated alkali halide crystals. In 1948, Hofstadter
introduced the scintillators $\mathrm{NaI\doped Tl}$ and
$\mathrm{CsI\doped Tl}$ for efficient photon and particle
detection (Novotny, \cite{Novotny:2005}). They have been used
extensively for nearly 60 years and are still very commonly
applied. Nevertheless, scintillator development is in full progress,
and the volume of scintillator material required each year for medical
imaging devices is considerable. All in all, a total of 175 tons of
scintillator are produced annually (Moses \cite{Moses:1999}).

Yet, the various nuclear medical imaging modalities differ in
their underlying principles and above all in the energy of the
$\gamma$-rays they have to detect. Therefore, there are different
requirements for scintillation detectors and the scintillators to be
used. An ideal scintillator must have a very large density and
effective atomic number, high luminous efficiency (light yield), short decay time of
the excited states, neither afterglow nor other background, and low
cost. These requirements cannot be met by a single scintillator and many
of them are still sub-optimal for many applications. This is essentially
the reason for further research and development in this field.
Since the {\em all-in-one} scintillator  does not exist, one has to
choose instead the most suitable from all available
materials. Fortunately, the list of available scintillators is already
rather long and the properties of the different materials cover wide
ranges.

For developing a new scintillator material, materials with relatively
high density, $\rho$, and effective atomic number, $Z_\mathit{eff}$, should be
used because the absorption probability by photoelectric effect is
proportional to $\rho Z_\mathit{eff}^{3-4}$ (van Eijk, \cite{vanEijk:2002}).
The material should also be transparent to the scintillation
light. Hence, crystals with a forbidden energy gap, $E_\mathit{gap}$, between valence and
conduction bands must be used and clearly $E_\mathit{gap}$ has to be
large enough to permit the transmission of the light. This is fulfilled
by ionic crystals or crystals with some degree of covalence.
Also good energy resolution is required by many imaging
modalities. Since the photon counting is dominated by Poisson processes,
a large number of scintillation photons is desired and this is achieved by a
low small forbidden gap. This is clearly a design conflict with the
former requirement and a reasonable tradeoff has to be found.
Further aspects are non-proportionality, crystal growth,
photofractions etc. A list (which is not intended to be complete) of
commonly mentioned scintillators together with their key properties
is given in Appendix~\ref{app:com-inorg-scints}.

The special requirements for X-ray scintillators are, in order of
decreasing importance: high light yield, high density, good matching
of the peak emission wavelength, well defined granularity and
homogeneity. All these properties are nearly excellently combined by
$\mathrm{Gd_2O_2S\doped Tb}$ making it the universally used
scintillator for this purpose. 

For X-ray computed tomography, the most important property is low
afterglow, followed in order of decreasing importance, high temporal and
chemical stability and insensitivity to radiation damages, high
density, good matching of the peak wave length and high light yield. 
$\mathrm{CdWO_4}$ and ceramic scintillators are often used here.

The requirements for Gamma-Scintigraphy and SPECT are already
influenced by the higher energy of the imaged radiation. They are, in
order of decreasing importance, high light yield for good energy
and spatial resolution, high density for efficiently stopping the
radiation, low cost because in general large quantities are required,
good matching of the peak emission wavelength and short decay time. 
Almost all these requirements are well matched by the alkali halide crystals
$\mathrm{NaI\doped Tl}$ and $\mathrm{CsI\doped Tl}$. Recently, the new
scintillators $\mathrm{LaCl_3\doped Ce}$ and $\mathrm{LaBr_3\doped
  Ce}$ have been introduced. These materials combine high light yield
and good energy resolution with a response time much smaller than
that of the former two crystals. Since their detection efficiency is
comparable to that of $\mathrm{NaI\doped Tl}$, they may be interesting
alternatives for future detector developments of this modality.

Finally, PET and PEM require above all crystals of a very short
attenuation length, since annihilation radiation is already very
penetrating. Further selection criteria, in order of decreasing
importance, are short decay times of the excited scintillation centers,
low cost because a high sensitive volume is necessary for a
reasonable detection efficiency, high light yield and a good matching
of the peak emission wavelength. Common scintillators for PET imaging
are $\mathrm{NaI\doped Tl}$, \chemform{Bi_4Ge_3O_{12}} (BGO) and
\chemform{Lu_2(SiO_4)O\doped Ce^+} (LSO).
$\mathrm{NaI\doped Tl}$ has long been the standard scintillator for
Gamma-cameras and positron cameras for its high light yield and
because it matches well the sensitivity of photomultiplier tubes.
Moreover, it can be grown easily and large crystal volumes are possible.
With the appearance of BGO in 1973, it was quickly replaced because
BGO has significantly better stopping power and
photofraction. However, light yield and energy resolution are far
below the values obtained with $\mathrm{NaI\doped Tl}$. 
LSO was introduced by Melcher and Schweitzer in
1992 \cite{Melcher:1992}. LSO achieves nearly the same detection
efficiency as BGO having at the same time a much larger light yield and
faster response. While there are different important drawbacks of this
material, for instance afterglow (Rogers {\em et al.}\ \cite{Rogers:2000}),
radioactive background from Lu (Huber {\em et al.}\ \cite{Huber:2002}) and
poor intrinsic energy resolution, for the moment it is the
scintillator of choice for the PET imaging modality. More recently,
other LSO related scintillators have been developed. They are mixed
lutetium silicate (MLS), \chemform{Lu_{1.8}Gd_{0.2}SiO_5\doped Ce^+}
(LGSO) and \chemform{Lu_{2-x}Y_{x}SiO_5\doped Ce^+} (LYSO).
Other scintillators that have been used for PET detectors are
\chemform{Gd_2SiO_5\doped Ce^+} (GSO), \chemform{LuAlO_3\doped Ce}
(LuAP) and \chemform{Lu_{1-x}Y_{x}AlO_3\doped Ce}  (LuYAP).

\subsection{Photodetectors}

All designs of radiation detectors demand a component that
converts the scintillation light into electrical signals. A large
variety of detectors have been constructed during the past years and their
different properties match a wide range of individual application
needs. They can be grouped in vacuum devices, solid-state devices and
a combination of both called hybrid devices and gaseous detectors
(Renker \cite{Renker:2004}).
The latter group has been mainly designed for high-energy physics
experiments and, up to now, there is no application of these devices for
medical imaging in combination with scintillators. They will not be
discussed here. Likewise, the group of hybrid photodetectors has no
major significance for commercial nuclear medical imaging at present
and they are not discussed here either.

All photodetectors have in common that the energy of the light
photons is transferred to electrons and these electrons are detected.
This process is called photoconversion.
The groups of photodetectors differ, however, in which material is
used and how
exactly this energy transfer occurs. Normally, they also provide an
inherent system of amplification that has to be adapted to the 
detector material used and to the kind of photoconversion that takes place. 

A photon that impinges on the surface of any material can liberate an
electron from its atom or even from out of the volume of the material
of which the atom forms a part. This effect is called photoelectric effect and
requires that the photon's energy is higher than the electric
work function of the material\footnote{The electric work function of a
  material is the minimum required energy for liberating an electron
  from the material volume.}. Therefore, the work function defines the
low energy limit of the light spectrum that can be detected when using
a specific material. Typical photocathodes of photomultipliers have their
threshold in the visible red light region.
Semiconductors have a very low work function and thus can be
used for detecting infrared photons. However, semiconductors also
offer a second process for photoconversion. It is sufficient to lift a
electron into the valence band and to avoid, by application of an
electric field in an adequate way, that this electron recombines and can be collected. 
Due to the very low band gap of semiconductors, very little energy is
necessary and a highly efficient photodetector can be realized.

\subsubsection{Vacuum Tubes}

After the discovery of the photoelectric effect by Heinrich Hertz in
1887 and its successful description by Albert Einstein in 1905, all
conditions had been met for the development of the first
photoelectric tube by Elster and Geiter in 1913 (see also Kume
\cite{Kume:1994}). 
The next milestone was the discovery of the compound photocathode called S-1
(\chemform{AgOCs}) which increased the sensitivity by two orders of
magnitude and was the beginning of dedicated photocathode development.

Photoelectric tubes can be significantly enhanced by introducing
secondary emission surfaces as was first mentioned in 1902. The photoelectrons
are accelerated towards an additional electrode where their impact
produces secondary electrons. In its first implementation by the RCA
laboratories in 1936, only
one additional electrode was used, but devices with more than one dynode were
quickly developed. The repeated generation of secondary electrons
results in an effective multiplication of the primary photoelectron and
therefore an amplification of the signal. The implementation is very
simple because the required vacuum tube was already realized with the
simple photoelectric tube.
Since this time, large progress has been made in the development of new
photocathodes and different dynode types.

Modern photomultiplier tubes (PMTs) have a light transmitting window
and a semi-transparent photocathode is normally evaporated directly
onto the inner vacuum side of this window. Nowadays, there are also
different types of windows with different properties such as cutoff
wavelength, thickness and refraction index. Most often borosilicate
glasses and UV-transmitting glasses are applied (Flyckt and Marmonier
\cite{Flyckt:2002}). The most used photocathode materials are the already
mentioned silver-oxygen caesium, antimony-caesium (\chemform{SbCs}),
and the bi- and trialkali compounds \chemform{SbKCs}, \chemform{SbRbCs}
and \chemform{SbNa_2KCs}. Directly behind the photocathode, an input
electron optics focuses all photoelectrons onto the useful area of
the first dynode. Over the past 70 years, different dynode geometries
and materials have been developed. The configuration of the
dynode system has a very high importance for the achieved gain. 
Each single dynode has to be arranged adequately in order to allow
 the secondary electrons they emit to hit the following
dynode. In PMTs with a typical number of 10-16 stages, gains of about
$\mathrm{10^6}$ are achieved. A number of dynode systems have been
designed putting the emphasis on different key requirements. A very
important property provided by {\em mesh dynodes}, {\em foil dynodes}
and {\em micro channel plate dynode} is the low spatial dispersion of
the electron avalanche. This allows position sensitive photomultiplier
tubes (PSPMTs) that are now widely used for different imaging techniques.
{\em Circular cage dynodes} allow for very compact and fast PMTs, but
provide no spatial information.

 \begin{figure}[!t]
   \centering
   \label{fig:bleeder-circ}
   \includegraphics[width=0.8\textwidth]{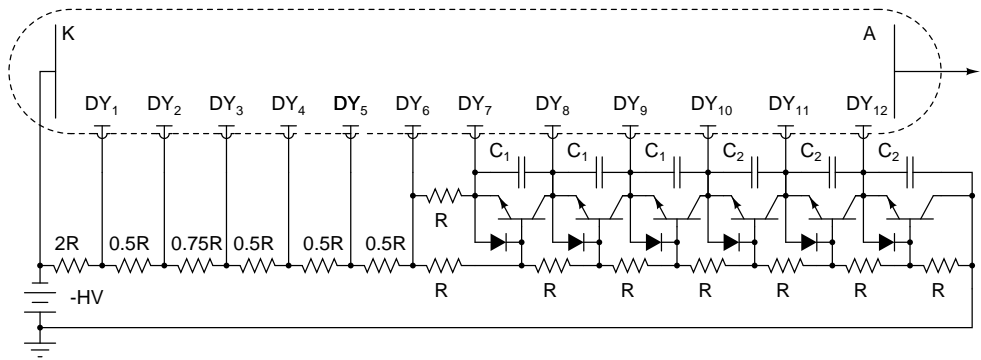}
   \caption[Schematic of a modern high output linearity bleeder circuit]{Schematic of a modern high output linearity bleeder circuit.}
 \end{figure}

Dynode materials must have a satisfactory secondary emission
coefficient and a sufficient rigidity. Typical materials are
\chemform{AgMg}, \chemform{CuBe} and \chemform{NiAl}. Their surfaces
can be easily oxidized and yield \chemform{MgO}, \chemform{BeO} and
\chemform{Al_2O_3} which offer a reasonable secondary emission
coefficient. For operation of the PMT, the dynodes have to be supplied
with high voltages where the potentials of all dynodes must differ for
correct acceleration and multiplication. This can be achieved with
separate high-voltage supplies for each stage or  using 
voltage divider known as {\em bleeder} circuits. Simple resistor chains normally
show an insufficient linearity behavior, especially for high
output currents. For this reasons, sophisticated and active bleeder
circuits have been designed. An example for a high output linearity
circuit is shown in figure~\ref{fig:bleeder-circ}.

The anode or anodes of photomultiplier tubes are the electrodes that
collect all secondary electrons from the last dynode and provide the
output signals. Its geometric configuration has to assure large
collection efficiency, minimized space charge effects and good
impedance matching with the following electronics. In position
sensitive PMTs, pixelated anodes, crossed wire anodes or crossed plate
anodes must be used.

An additional synchronous signal can be derived from the last dynode
of the electron multiplier system. The requirement for such a signal
often arises from the desire to synchronize different measurement
instruments. Especially in the case of position sensitive
photomultiplier tubes, the last dynode signal is of interest. Owing to
the segmentation of the anode system, there is no signal that
provides position independent information about the total amount of
detected light. In order to obtain this information from multi-anode
PMTs, all anode charges have to be summed up. This can be avoided by
sensing one of the dynodes because the electrons for the avalanche are
liberated from the dynodes and the sudden loss of many of them  can
be noted as a variation in the potential. To obtain an amplitude
comparable with that of the anode, usually the last dynode is used for
the derivation of the signal. Care has to be taken not to disturb
the anode signal and an appropriate matching of the impedances must be
assured. Otherwise, the anode signal will be distorted and the dynode signal
differentiated respectively. The properties of this additional signal
allows one to use it for the generation of trigger signals and also for
hardware energy discrimination.

An important characteristic not only for photomultiplier tubes but all
photon detecting devices is its quantum efficiency $\mathrm{QE}$ which
gives the probability that an impinging photon leads to an output
signal. A typical value for PMTs is 25\%.

\subsubsection{Solid-State Photodetectors}

A very important advantage of semiconductor devices over
photomultiplier tubes is their much higher quantum efficiency of about
80\%. This reduces significantly signal noise caused by the Poisson
process of photon detection. They can also be produced in a fully
automated process and therefore can be cheap. It is also possible to
make them very compact because there is no need for mechanical
parts. At the same time, this makes them  very
insensitive to magnetic fields. The detector
itself is only $\mathrm{0.3 mm}$ thick (Renker, \cite{Renker:2004}). 

The silicon PIN diode is an example of a semiconductor detector for
visible light photons.
Its functional principle has already been explained in
section~\ref{sec:solid-state-gamma-ray-detect} about solid state
$\gamma$-ray detectors. It can be produced by standard
semiconductor processes and its operation is simple and reliable.
However, it has no gain at all, making it only suitable for the
integrating detection mode but not for single photon counting
purposes. Arrays of PIN diodes are very easy to produce and the time
response is, at 1-2 $\mathrm{ns}$, quite low.

An important improvement of PIN-photodiodes are avalanche photodiodes (APDs).
Theses devices are designed to allow a very high reverse bias
voltage. For sufficiently high voltages, impinging light photons cause
electron avalanches with achieved gains of between 50 and 200. Even higher
gains are possible but they make the device very sensitive to
ambient instability like temperature drift or residual ripple of the
high voltage power supply. APDs are suitable for single photon
counting and already have been successfully used for small animal
PETs. Very interesting enhancements are large area APDs, (Shah {\em et al.}\
\cite{Shah:2001}) position sensitive avalanche photodiodes (PSAPDs, Shah
{\em et al.}\ \cite{Shah:2002}) and solid-state photomultipliers (SiPMs,
Petroff and Stapelbroek \cite{Petroff:1989}, Saveliev
\cite{Saveliev:2004}).  PSAPDs consist of a large area APD covered
with a thin layer with homogeneous resistivity. This layer performs an
integration with linear weighting of the impinging light. From four
currents extracted at the corners of the resistive layer, the
centroids of the light distribution can be computed. Sensitive areas
up to $\mathrm{28\times28\,mm^2}$ with a planar spatial resolution of
$\mathrm{300\,\mu m}$ are currently possible. Important drawbacks of
all APD based devices are their high intrinsic electronic noise and
the low temperature stability. Silicon photomultipliers consist of a
large array of small APDs operated in Geiger mode. In this way,
respectable gains can be obtained and their timing resolution of
$\mathrm{50\,ps}$ is also excellent. However, these devices are
still under development and there is no commercial application in
medical imaging to this day.

\section{Degrading Factors}

All modalities of nuclear imaging are limited by different physical
effects and uncertainties that impact on their final performance, including
all qualifying parameters like spatial resolution, noise and
efficiency. In all modalities, namely PET/PEM, SPECT and
gamma-scintigraphy, \g-photons are the information carrier and the
detector modules are often of the same type, even though some of the
design parameters differ. The detection of high and medium energy photons
involves, however, a number of difficulties that are mainly due to
non-ideal behavior of the detector components. Finite size of the
detectors leads to a limited geometric efficiency, while the finite
size of a detector pixel limits the spatial resolution. Likewise, none
of the discussed radiation- or photodetectors have a quantum efficiency
of 100\% and all introduce electronic noise. Above all the signal to
noise ratio (SNR) and the spatial resolution of the final image are
degraded by a number of physical effects that are hard to avoid. 
Some aspects, {\em e.g.}\ efficiency of Anger-type cameras and positron
emission tomographs, as well as the spatial resolution degradation
caused by collimators, have been discussed already in
sections~\ref{sec:positron-emiss-tomo} and
\ref{sec:gamma-cam-planar-imaging} respectively and are therefore not
repeated here.

\subsection{Parallax Error and Depth of Interaction}
\label{sec:parallax-and-DOI}

\begin{figure}[!t]
  \centering
  \subfigure[][Historical significance of the parallax error. The
  absolute error $d$ depends on the distances $D$ and $D'$.]{\label{subfig:parallax}
   \psfrag{observation}{{\scriptsize observation}}
   \psfrag{Point}{{\scriptsize point}}
   \psfrag{optic}{{\scriptsize optical}}
   \psfrag{Axis}{{\scriptsize axis}}
   \psfrag{D}{{\scriptsize $D$}}
   \psfrag{DD}{{\scriptsize $D'$}}
   \psfrag{d}{{\scriptsize $d$}}
    \includegraphics[height=0.16\textheight]{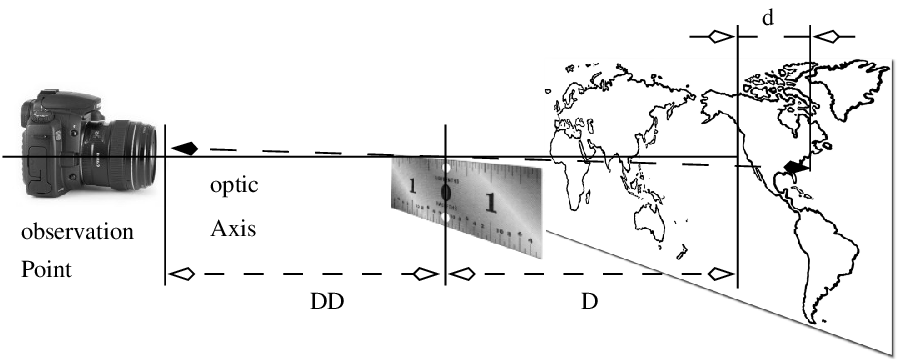}}\hspace*{2em}
  \subfigure[][Manifestation in \g-ray detectors.]{\label{subfig:para-err-schem}
   \psfrag{Photodetector}{{\scriptsize Photodetector}}
   \psfrag{Crystal}{{\scriptsize Crystal}}
   \psfrag{g}{{\scriptsize \g}}
   \psfrag{x=m}{{\white\scriptsize $x$}}
   \psfrag{y=m}{{\white\scriptsize $y$}}
   \psfrag{z=m}{{\white\scriptsize $z=$?}}
  \includegraphics[height=0.16\textheight]{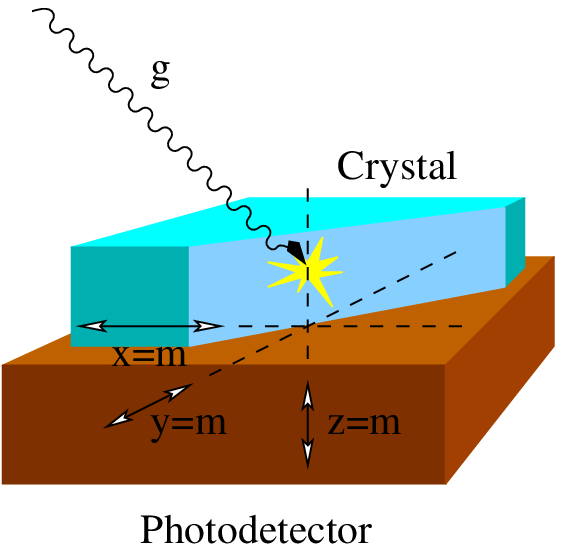}}\\
  \subfigure[][Parallax error for a full ring PET scanner. The
  green (light-gray) lines are true LORs and the red (dark) line is a mis-positioned event.]{\label{fig:pet-parallaxe-circ}
    \includegraphics[width=0.3\textwidth]{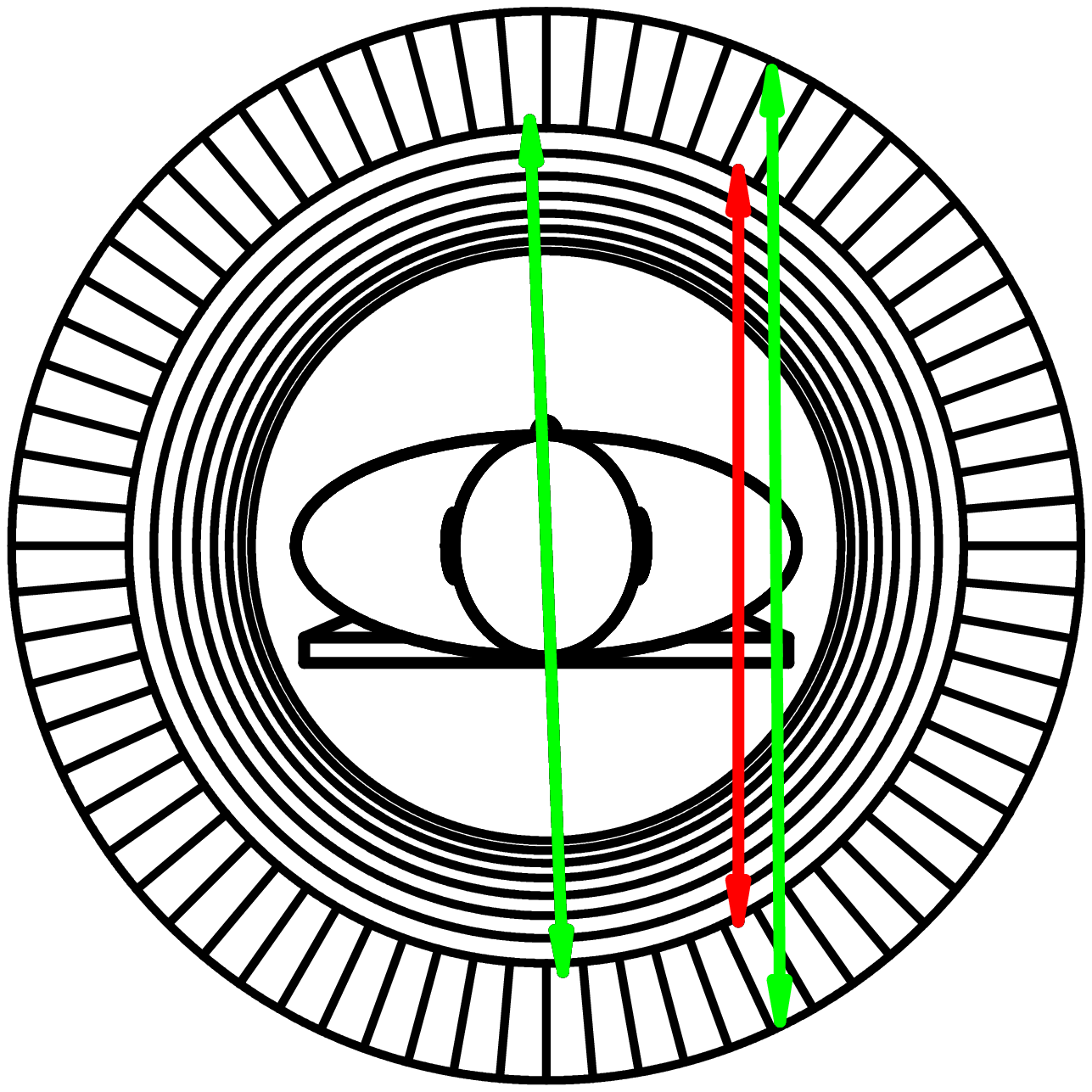}}
  \subfigure[][Parallax error for planar PET scanner. The
  green (light-gray) lines are true LORs and the red (dark) line is a mis-positioned event.]{\label{fig:pet-parallaxe-plan}%
    \includegraphics[width=0.29\textwidth]{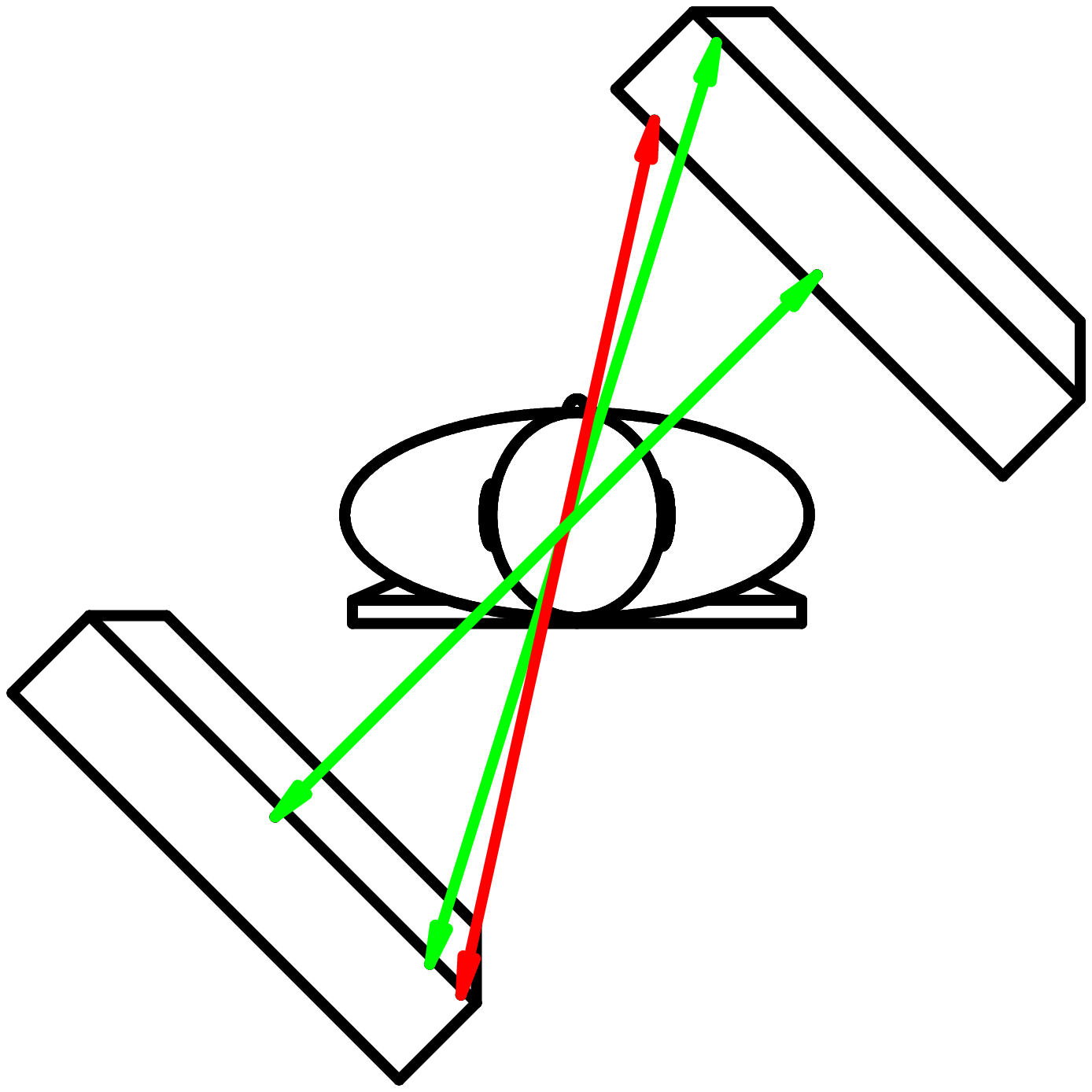}}
  \subfigure[][Parallax error for \g-cameras using pinhole
  collimators. The
  green (light-gray) lines are true lines of flight (LOFs) and the red
  (dark) line is a mis-positioned events.]{\label{fig:gc-parallaxe}%
    \includegraphics[width=0.3\textwidth]{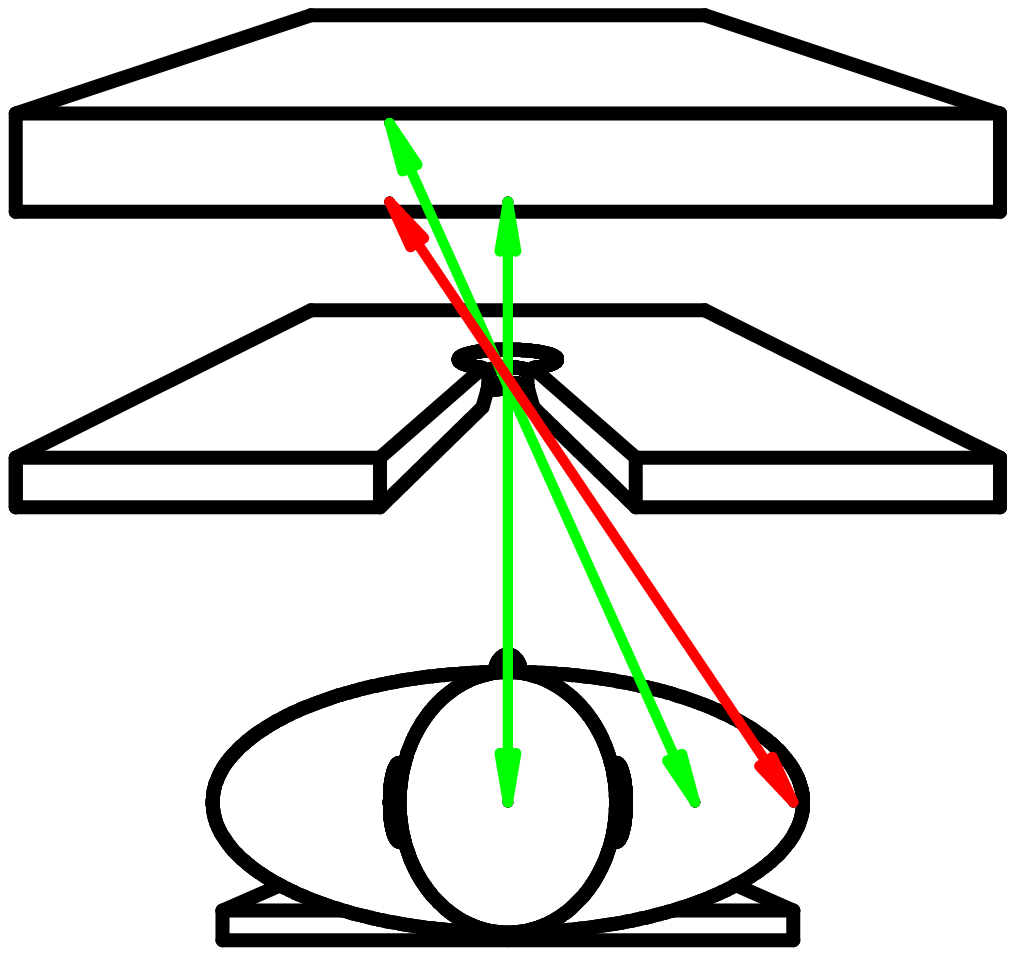}}
  \caption[Formation of the parallax error and its manifestation]{Formation of the parallax
    error. Detectors that are not capable of measuring neither the incidence
  angle nor the $z$-coordinate cannot distinguish between $\gamma$-rays
  impinging parallel to the actual ones. This leads to
  misinterpretation of the $\gamma$-ray's origin. The three figures
  below demonstrate the manifestation of this error for different
  imaging modalities and detector configurations.}
  \label{fig:parallax}
\end{figure}

One of the most important degrading factors for Positron Emission
Tomography is the depth of interaction (DOI) problem. It is caused by
the failure of nearly all \g-ray imaging detectors to provide
full three-dimensional information about the photoconversion position
within the scintillation crystal (or solid-state detector). As a
consequence of its nature, it is inherent to all nuclear imaging
modalities where the \g-rays enter the detector in an off-normal
directions. The lack of DOI information leads to a parallax error in
the determination of both remaining spatial direction. Its existence
is known from other fields, especially from astronomy, where it is
used for determination of stellar distances. Parallax is the Greek
word for alteration and denotes the change of angular position of two
stationary points relative to each other if the observer's position
alters. This is shown in figure~\ref{fig:parallax}. A Parallax error
is introduced in \g-ray imaging whenever the thickness of the absorbing
material is finite, the incidence direction of the \g-ray is oblique
and only the spatial coordinates normal to the thickness of the
detector are measured. Hence, the parallax error is due to the
thickness of the detector and scales with the same. There is no parallax
error for \g-rays whose angle of incidence equals 90\textdegree\,
respect to the plane defined by both measured coordinates. For this
reason, Anger-type cameras for scintigraphy or SPECT that use parallel
hole collimators do not suffer from this error. Contrariwise, the
pinhole collimator introduces parallax errors because it allows
oblique \g-rays. The
thicker the detecting material is, the larger will be the parallax error
because the uncertainty in the depth of interaction increases. On the
other hand one has the requirement of higher mass attenuation for
increasing \g-ray energy. This can be obtained in two ways. The first 
way is the use of material with high attenuation coefficient and the
second method is to increase the thickness. Therefore, the interaction
depth uncertainty poses a more serious problem to positron emission
tomography than to SPECT and to Gamma-scintigraphy with
radiopharmaceuticals of low decay energy.

For the PET-modality, the parallax effect appears for any 
positron source displaced from the geometrical center of the tomograph's field-of-view, or in general
for any $\gamma$-ray impinging non-normal on the crystal (refer to
figures~\ref{fig:pet-parallaxe-circ} and
\ref{fig:pet-parallaxe-plan}). Knowing neither the incidence angle
nor the interaction depth, the incidence direction of the $\gamma$-ray
cannot be unambiguously determined. In positron emission tomography,
the parallax error is also known as radial elongation or radial
astigmatism. For a typical commercial whole-body positron emission
tomograph with $\mathrm{4\,mm}$ wide detectors and a ring diameter of
$\mathrm{80\,cm}$, the lacking DOI information reduces the spatial
resolution by $\mathrm{\approx40\%}$ at a distance of only
$\mathrm{10\,cm}$ from the center of the field of view (Cherry {\em et al.}\
\cite{Cherry}). Since the DOI
effect increases strongly with the distance from this center, PET
scanners are often designed with inner ring diameters much larger than it would be necessary
to fit the patient.

Lack of interaction depth information leads to an additional
positioning error when large-sized
continuous scintillation crystals together with the center of gravity
positioning scheme are used. This error is caused by the incomplete
scintillation light collection near the borders of the detector. It
leads to a strong, depth of interaction dependent image compression
which is the main reason that nearly all research groups abandoned
this particular design of PET detector modules. This particular error will be
discussed in detail in section~\ref{ch:errors-of-cog-and-cdr}.

\subsection{Compton Scattered Events and Randoms}

Another very important error in emission tomography is the presence of
scattered \g-photons. Although it is likewise caused by the
inaccessibility of an important parameter, it is completely different
in nature to the just discussed uncertainty. For all imaging modalities of
nuclear medicine, it is necessary not only to detect \g-photons with
energies between $\mathrm{15\,KeV}$ and $\mathrm{511\,KeV}$, but also
to determine information about their direction and line of flight. 
At these energies, \g-photons are very likely to undergo elementary
processes like Rayleigh scattering, Compton scattering or
photoelectric absorption within the detecting elements and
also inside the explored patient, animal or tissue. Since the \g-rays
are emitted isotropically, actually all surrounding matter, including
detector housing, electronics etc., can cause scattered radiation. 
Since the photoelectric absorption destroys the \g-photons, they
can no longer be detected and therefore only reduce the efficiency but,
except for PET, do not lead to mis-positioned events. In positron
emission tomography, two $\mathrm{511\,keV}$ photons have to be
detected. If one annihilation photon is absorbed before arriving at the
detector element, no coincidence signal is possible and this event is
lost as in the single photon case. However, the necessity of detecting
two \g-photons can lead to the random coincidence detection of two
different annihilation events when only one photon of each
annihilation arrives at the detector. This leads to randomly positioned events.

Coherent
(Rayleigh) scattering involves the interaction with an atom. Hence there
is virtually no change in energy. Moreover, the scattering angles are
usually low; that is, the original direction of the photon is only
changed marginally. Compton (incoherent) scattering is an inelastic
process that is due to interaction with the atomic and molecular
electrons. In the energy range of relevance for nuclear imaging,
especially PET, the
probability of Compton scattering is high in all materials and
tissue involved. Three types with different impact on the final image can be
distinguished: Compton scatter within the explored body/tissue, Compton
scatter within the detector but without detection by the same detector
element (Back scatter) and 
inter- and inner-detector element Compton scatter. The last effect is
of special relevance for the present work and will be discussed in
detail in chapter~\ref{ch:compton}. Back scatter includes large
changes in the direction of the photon. The change in energy depends
on the energy of the photon itself. While 180\textdegree\, scattered
$\mathrm{15\,KeV}$ photons leave the interaction position with
$\mathrm{\approx95\%}$ of their initial energy,  $\mathrm{511\,KeV}$
photons will loose 2/3 of its energy under the same circumstances. 
Hence, in PET, one can effectively filter backscattered events by means
of energy discrimination. Also, the detection efficiency of
backscattering is normally low (Zaidi and Koral \cite{Zaidi:2004}) and can
be further optimized by means of adequate detector design. 
The impact of this scatter type on the final image is therefore
normally low compared to Compton scattering within the region of interest
(ROI). 

\begin{figure}[!t]
  \centering
  \subfigure[][Occurrence of correctly (1) and mis-positioned (2)
  events in single photon imaging modalities.]{\label{subfig:Compton-Pinhole}
    \psfrag{c}{1}
    \psfrag{m}{2}
    \psfrag{ROI}{ROI}
    \psfrag{Collimator}{collimator}
    \psfrag{Detector}{detector}
    \includegraphics[height=0.19\textheight]{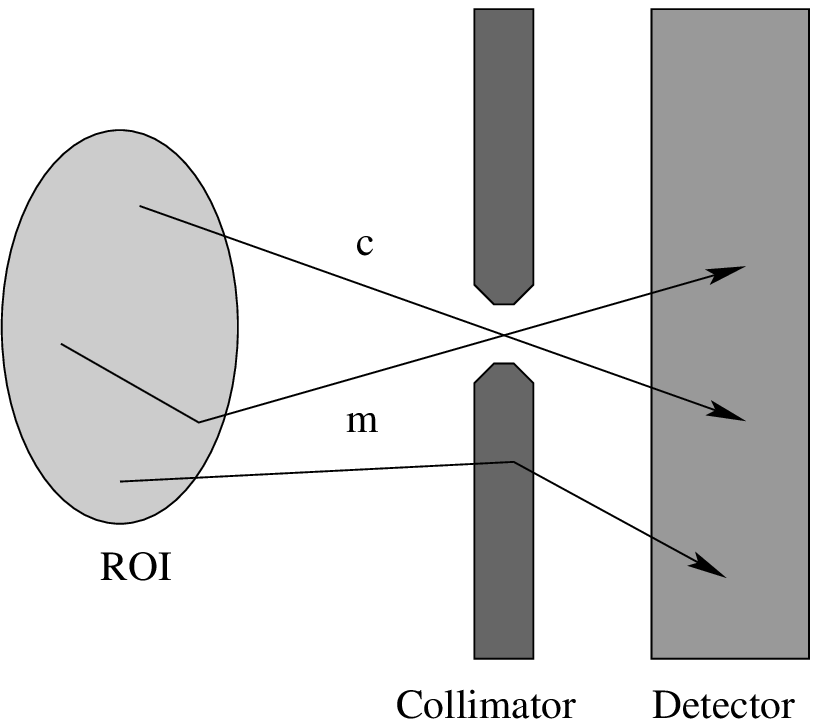}}\hspace*{0.08\textwidth}
  \subfigure[][Occurrence of correctly (1), mis-positioned
  events due to Compton (2) and random coincidences in PET (3).]{\label{subfig:Compton-Random-PET}
    \psfrag{Detector}{detector}
    \psfrag{ROI}{ROI}
    \psfrag{1}{1}
    \psfrag{2}{2}
    \psfrag{3}{3}
    \includegraphics[height=0.19\textheight]{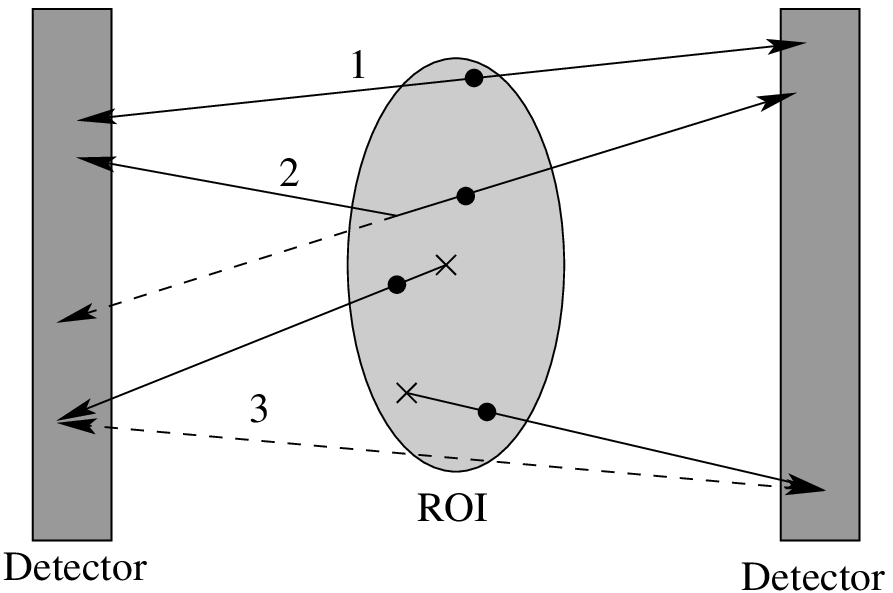}}
  \caption[Schematic diagrams for possible detection of Compton
  scattered events]{Schematic diagrams for possible detection of
    Compton scattered events in single emission tomography (l.h.s.)
    and positron emission tomography (r.h.s.).}
  \label{fig:examples-for-compton-events}
\end{figure}

Figure~\ref{subfig:Compton-Pinhole} shows the different types
of mis-positioned events for single photon emission modalities. Clearly, the presence of collimators in single
photon emission modalities reduces significantly the fraction of detected
\g-photons that were scattered inside the ROI, but leads to additional
mis-positioned events due to scattering within the
collimator. Typical values for fractions of mis-positioned events in this modality
are of the order of 5\% when a \isotope{Ga}{67} based
radiopharmaceutical (energy: $\mathrm{93\,keV}$) is used and
$\mathrm{\approx30\%}$ when a   \isotope{I}{131} based
radiopharmaceutical (energy: $\mathrm{360\,keV}$) is used (Zaidi and
Koral \cite{Zaidi:2004}).

In annihilation coincidence detection, an event is registered whenever
two events are recorded within a specified temporal window. True
coincidences are those events that really arise from the pair of
annihilation photons without intermediate scatter. Erroneous LORs can
arise from a pair of annihilation photons when one of them or both
undergo Compton scattering before being detected (event (2) in
figure~\ref{subfig:Compton-Random-PET}). Also it is possible that two
photons are detected (they may be scattered or not) from different
annihilation events (event (3) in
figure~\ref{subfig:Compton-Random-PET}). It is rather likely that one
of the two photons of an annihilation pair is absorbed inside the ROI or
fails to be detected by the system. The fraction of mis-positioned
events is much larger in coincidence detection modalities because the
\g-radiation is in general not collimated.\footnote{In many commercial
  PET scanners, passive collimation parallel to the plane of the
  drawing~\ref{subfig:Compton-Random-PET} is performed by {\em
    septa}. This is referred to as 2D-mode PET. If the septa are
  removed, it is called 3D-mode PET. Even with septa, all events in
  ~figure~\ref{subfig:Compton-Random-PET} can occur and there is
  actually no collimation for two spatial directions.} Random events
do not exist in single photon emission tomography.

An increase in scattered and random events results in a decreasing
signal to noise ratio (SNR) of the reconstructed image. Even though
the total of true events can be increased by designing detectors with higher
efficiency, this would lead also to higher detecting efficiency for
scattered events and randoms. In general, the different fractions and
the final SNR depend in a complex manner on the geometry of the
scanner, the detector element design and also the explored object (Humm
{\em et al.}\ \cite{Humm:2003}). For
instance, Compton scattered events are of lower importance in small
animal PET. In order to provide a physical measure for characterizing
the influence of scattered events and randoms, the {\em noise
  equivalent count rate} (NEC) was introduced:
\begin{equation}
  \label{eq:nec-def}
  f_\tincaps{NEC}\mdef\frac{f_\tincaps{True}^2}{f_\tincaps{True}+f_\tincaps{Random}+2f_\tincaps{Single}},
\end{equation}
where $f_\tincaps{True}$, $f_\tincaps{Random}$ and
$f_\tincaps{Single}$ are the count rates for true events, random
events and single detected photons. Noise equivalent count exhibits a
maximum when plotted against the administered activity. This is
the activity at which an optimum SNR is expected.

\subsection{Errors Contributed by the Radiopharmaceutical}
\label{subsec:radio-source-errors}

\begin{figure}[!t]
  \centering
  \psfrag{9F18}{\scriptsize ${ }_{\,\;9}^{18}$F}
  \psfrag{8O18}{\scriptsize ${ }_{\,\;8}^{18}$O}
  \psfrag{6C11}{\scriptsize ${ }_{\,\;6}^{11}$C}
  \psfrag{5B11}{\scriptsize ${ }_{\,\;5}^{11}$B}
  \psfrag{7N13}{\scriptsize ${ }_{\,\;7}^{13}$N}
  \psfrag{6C13}{\scriptsize ${ }_{\,\;6}^{13}$C}
  \psfrag{7N15}{\scriptsize ${ }_{\,\;7}^{15}$N}
  \psfrag{8O15}{\scriptsize ${ }_{\,\;8}^{15}$O}
  \psfrag{3/2- }{\scriptsize $3/2^-$}
  \psfrag{1/2- }{\scriptsize $1/2^-$}
  \psfrag{0+}{\scriptsize $0^+$}
  \psfrag{1+ 109.8m }{\scriptsize $1^+\quad 109.8 m$}
  \psfrag{3/2- 20.38m}{\scriptsize $3/2^-\quad 20.38 m$}
  \psfrag{1/2- 122s }{\scriptsize $1/2^-\quad 122s$}
  \psfrag{1/2- 9.96m}{\scriptsize $1/2^-\quad 9.96 m$}
  \psfrag{Qec 1.6555}{\scriptsize $Q_{EC}\quad 1.6555$}
  \psfrag{Qec 1.9821}{\scriptsize $Q_{EC}\quad 1.9821$}
  \psfrag{Qec 2.7539}{\scriptsize $Q_{EC}\quad 2.7539$}
  \psfrag{Qec 2.2205}{\scriptsize $Q_{EC}\quad 2.2205$}
  \psfrag{EC 3.1\%}{\scriptsize $EC\quad 3.1\%$}
  \psfrag{EC 0.24\%}{\scriptsize $EC\quad 0.24\%$}
  \psfrag{EC 0.11\%}{\scriptsize $EC\quad 0.11\%$}
  \psfrag{b+96.9\%}{\scriptsize $\beta^+\quad 96.9\%$}
  \psfrag{b+ 99.76\%}{\scriptsize $\beta^+\quad 99.76\%$}
  \psfrag{b+99.89\%}{\scriptsize $\beta^+\quad 99.89\%$}
  \includegraphics[width=0.25\textwidth]{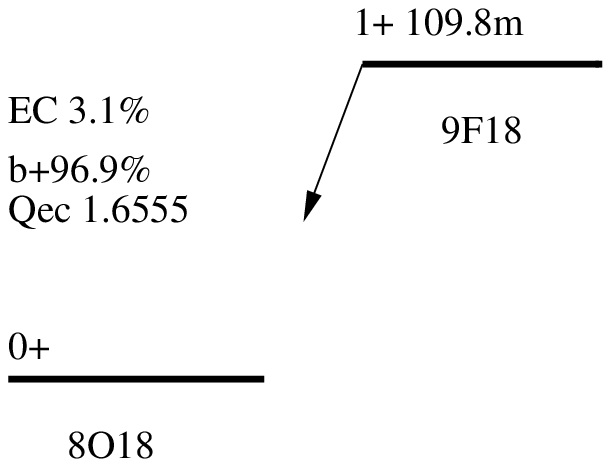}\hspace*{3em}
  \includegraphics[width=0.25\textwidth]{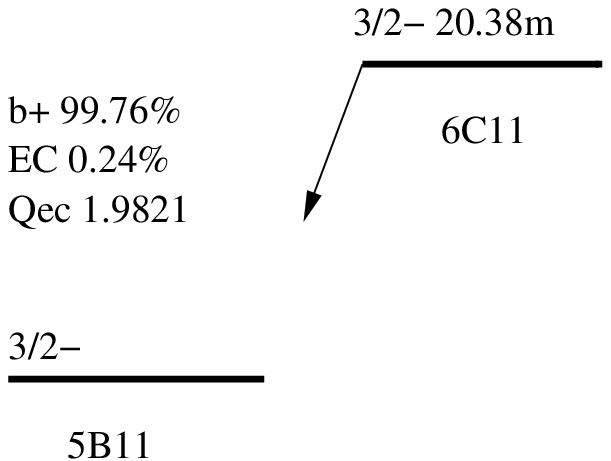}\\
  \includegraphics[width=0.25\textwidth]{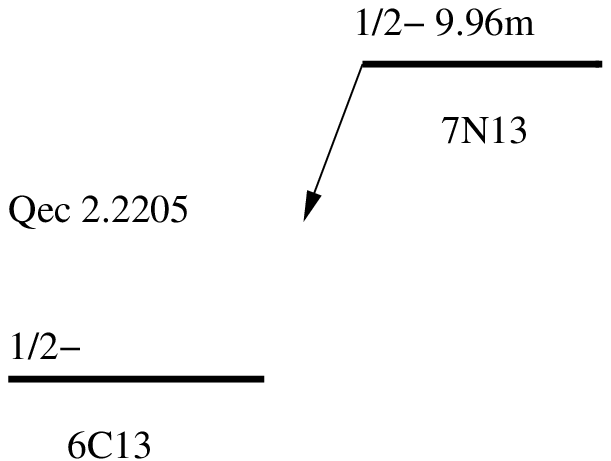}\hspace*{3em}
  \includegraphics[width=0.25\textwidth]{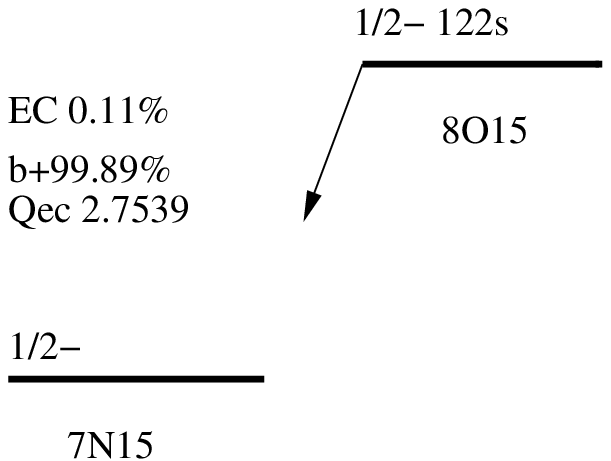}
  \caption[Decay schemes of four often used radionuclides for PET]{Decay schemes together
    with branching ratios, half-lives and $Q$-values
    of four often used radionuclides for PET.}
  \label{fig:isotopes-decay-schemes}
\end{figure}

Beside the limitations mentioned so far, Positron Emission Tomography has
to deal with a further position uncertainty arising from the
nature of the $\beta^+$ decay which was already shown in
figure~\ref{fig:pet-physics-principle}. In particular, this measurement error is
caused by the range of the positron within the explored tissue (Phelps
{\em et al.}\ \cite{Phelps:1975}, Palmer
and Brownell \cite{Palmer:1992}, Levin and Hoffman \cite{Levin:1999}) and the
non-collinearity of the annihilation photons (Beringer and Montgomery
\cite{Beringer:1942}, de Benedetti {\em et al.}\ \cite{DeBenedetti:1950}).

Radionuclides that are used for PET have to be
$\beta^+$-emitters. They are nuclides with a proton excess and can decay
by two processes:
\begin{gather}
\label{eq:beta+decay}
p \longrightarrow n+e^++\nu_e\\
\label{eq:electron-capture}
p + e^- \longrightarrow n+\nu_e.
\end{gather}
The first reaction~(\ref{eq:beta+decay}) describes the ordinary
$\beta^+$-decay. For this mode, decays directly to the ground states
are possible. The second reaction~(\ref{eq:electron-capture}) is the
competing electron capture process.

In most cases, radiopharmaceuticals for PET are based on fluorine
$^{18}$F, oxygen $^{15}$O, carbon $^{11}$C, and nitrogen $^{13}$N. The
transitions of these radionuclides are allowed Gamow-Teller ($^{18}$F)
and allowed Fermi-transitions ($^{15}$O, $^{11}$C, and
$^{13}$N). Therefore, they exhibit the required short half-lives in the
order of minutes as it was discussed in
section~\ref{sec:radiopharmacos}. Decay schemes for these
radionuclides are displayed in figure~\ref{fig:isotopes-decay-schemes}.
Since for $\beta^+$-decay the
positron is always accompanied by a hardly detectable 
neutrino $\nu_e$ that shares the momentum and energy, the energy
spectrum of the positron is continuous.

\begin{figure}[!t]
  \centering
  \subfigure[][\isotope{F}{18}, $\mathrm{FWHM\approx0.1\,mm}$,
  $\mathrm{FWTM\approx1\,mm}$]{
    \label{subfig:flour-pos-range}
    \psfrag{y}{}\psfrag{x}{\hspace*{-2.5em}$\beta^+$-range [mm]}
    \includegraphics[width=0.44\textwidth]{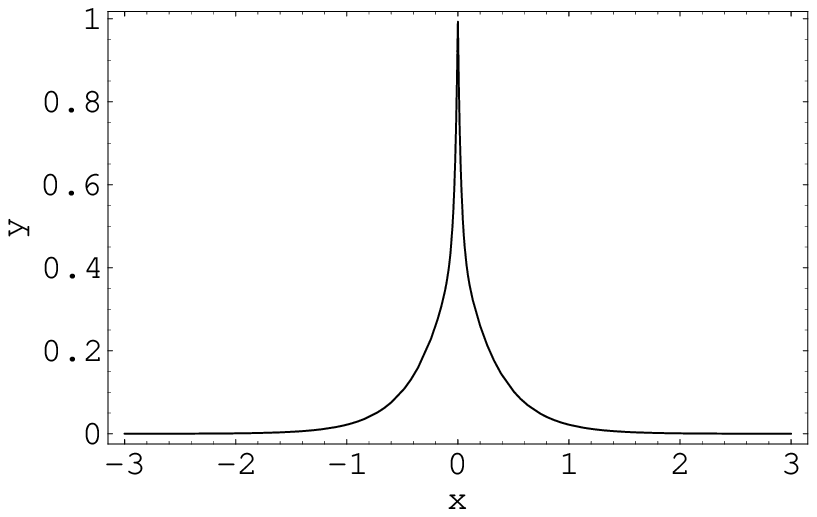}}
  \subfigure[][\isotope{C}{11}, $\mathrm{FWHM\approx0.19\,mm}$,
  $\mathrm{FWTM\approx1.9\,mm}$]{
    \label{subfig:carbon-pos-range}
    \psfrag{y}{}\psfrag{x}{\hspace*{-2.5em}$\beta^+$-range [mm]}
    \includegraphics[width=0.44\textwidth]{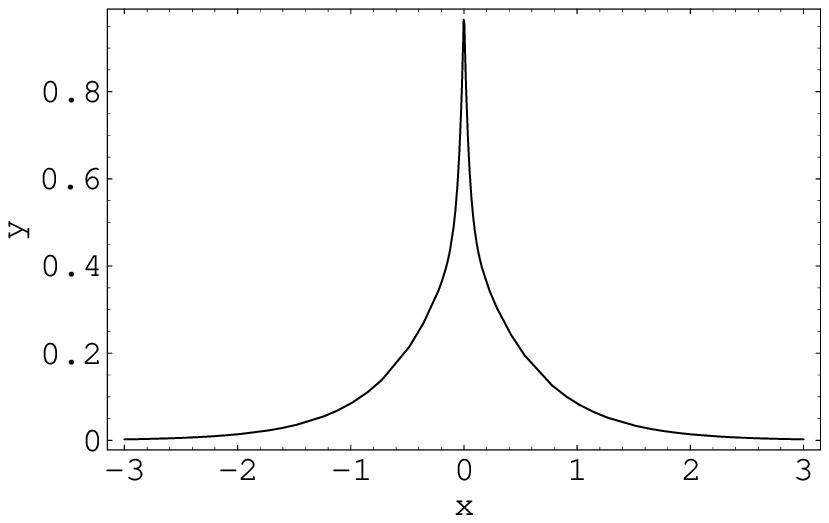}}
  \subfigure[][\isotope{N}{13}, $\mathrm{FWHM\approx0.28\,mm}$,
  $\mathrm{FWTM\approx2.5\,mm}$]{
    \label{subfig:nitro-pos-range}
    \psfrag{y}{}\psfrag{x}{\hspace*{-2.5em}$\beta^+$-range [mm]}
    \includegraphics[width=0.44\textwidth]{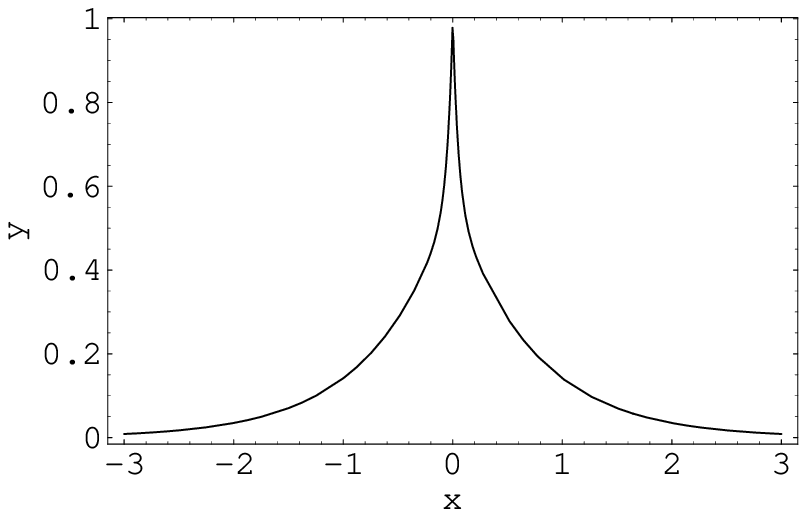}}
  \subfigure[][\isotope{O}{15}, $\mathrm{FWHM\approx0.5\,mm}$,
  $\mathrm{FWTM\approx4.1\,mm}$]{\label{subfig:oxy-pos-range}
    \psfrag{y}{}\psfrag{x}{\hspace*{-2.5em}$\beta^+$-range [mm]}
    \includegraphics[width=0.44\textwidth]{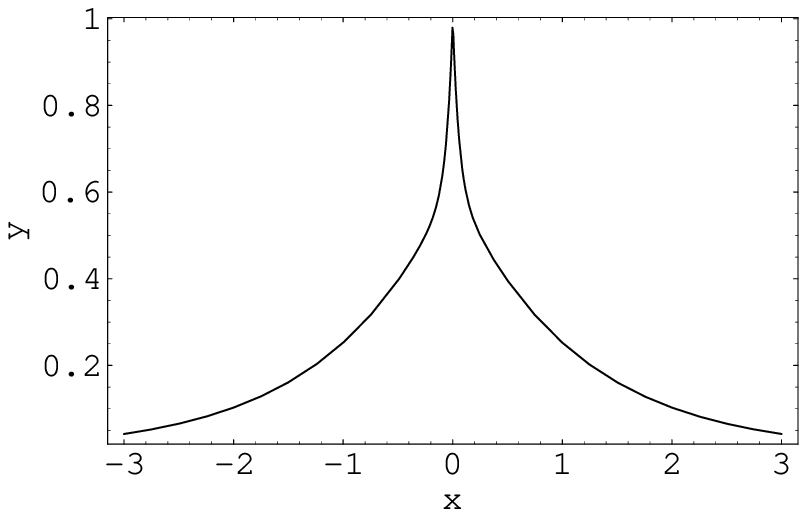}}
  \caption[Radial point spread function for the positron range of
  four radionuclides for PET]{Radial point spread function for the positron range of the
    four radionuclides for PET. Below each figure, FWHM and FWTM
    for each isotope are indicated (Values and functional behavior
    taken from Levin and Hoffman \cite{Levin:1999}).} 
  \label{fig:isotopes-ranges}
\end{figure}

After the decay process, the positron loses energy since it interacts with the tissue until it
finally annihilates with an electron and gives rise to the emission of
two $\mathrm{511\,keV}$ \g-rays propagating in nearly exactly opposite directions. 
The permanent collisions  with the molecules slow the positron down
and the annihilation in general does not occur until the $e^+$ is in thermal
equilibrium with the environment. For this reason, the range of the
positron strongly depends on its initial energy. This process can
be described diffusively (Palmer and Brownell\cite{Palmer:1992}) or
alternatively by the Bethe-Bloch formula (Leo
\cite{Leo:1994}).
Figure~\ref{fig:isotopes-ranges} shows the results for the
distributions of the positron range in water and corresponding FWHM
and FWTM values that were obtained by Levin and Hoffman
\cite{Levin:1999} from Monte Carlo simulations.

Besides the loss of spatial resolution in positron imaging devices caused by the positron's range, there is 
a second equally fundamental effect, which in certain coincidence detector configurations can result in a
more serious loss of resolution. After the thermalisation process of the positron, the kinetic energy of the 
annihilating pair is typically of a few eV which is mainly ascribed to the orbital momentum of the electron. 
In their center-of-mass frame, the photon energies are
$\mathrm{0.511\,MeV}$ and their directions are exactly
opposite. However, having a non-vanishing residual momentum, the two
back-to-back $\gamma$-ray photons formed upon 
annihilation are not seen at exactly $180^\circ$ with respect
to each other in the rest frame of the particle detectors (de Benedetti {\em et al.}\ \cite{DeBenedetti:1950}).
The angular spread leads to an approximately Gaussian-like distribution with the full width of $\mathrm{\sim 0.6^\circ}$
at half maximum (Beringer and Montgomery \cite{Beringer:1942}). For a system with a large detector-detector separation, this
effect can result in a significant positioning error when reconstructing the location of the radiopharmaceutical. 
Both positron range and  photon non-collinearity cannot be measured
with present $\gamma$-ray imaging detectors and represent fundamental limitations to the spatial
resolution of positron emission tomography.

\section{Detector Improvements}

Block detectors as discussed in section~\ref{sec:PET-designs} made
from BGO and LSO are still the most commonly used PET detector designs
for commercial PETs. Typical performance parameters are $\mathrm{80\%}$
detection efficiency, $\mathrm{20\%}$ energy resolution, 
$\mathrm{2\,ns}$ timing resolution and $\mathrm{5\,mm}$ spatial
resolution (Moses, \cite{Moses:2001}). In the case of scintillation
detector modules, the scintillation crystal poses the most important
limit to the performance parameter. One important issue is to increase
the very low total detection efficiency of usually $\mathrm{<1\%}$ for
commercial PET scanner. This can be achieved by increasing the
intrinsic efficiency of the modules or by increasing the total volume
of the scintillation crystal. While the latter possibility is mainly a
questions of costs, the first possibility together with high spatial
resolution constitutes a pair of antagonistic design goals. This is because
a higher intrinsic detection efficiency can only be achieved by
thicker crystals or crystals of higher stopping power than BGO. For
the moment, there is no alternative scintillator with higher stopping
power than BGO and only the first possibility is left, but this would
increase radial elongation as explained in
section~\ref{sec:parallax-and-DOI}. 

The timing resolution of scintillation detectors is mainly determined
by the decay time of the scintillator. The shorter this parameter is,
the higher will be the concentration of the detected light at the very
first moments of the scintillation pulse leading finally to smaller
statistical errors and a better temporal definition of the rising edge
of the pulse. If the timing resolution of the \g-ray detector is
sufficiently high, the possible positions of the positron's
annihilation point can be confined to a fraction of the FOV. This
effectively increases the SNR of the acquired image.
LSO is a very attractive alternative to BGO because it
has a significantly faster decay time but nearly the same stopping
power. Moreover, it has 3-4 times as much light yield compared to BGO
and therefore allows for higher energy resolution and better definition
of the detected position.

Developing PET detector modules with high timing resolution and 
the ability to measure depth of interaction information are two active
fields of research. Notably, techniques of DOI measurement have
mushroomed in recent years.

\subsection{Depth of Interaction Detection}
\label{ch:doi-detectors}

\begin{figure}[!t]
  \centering
  \subfigure[][Light sharing technique for DOI determination.]{\label{subfig:light-sharing-doi}\hspace*{2em}
    \includegraphics[height=0.15\textheight]{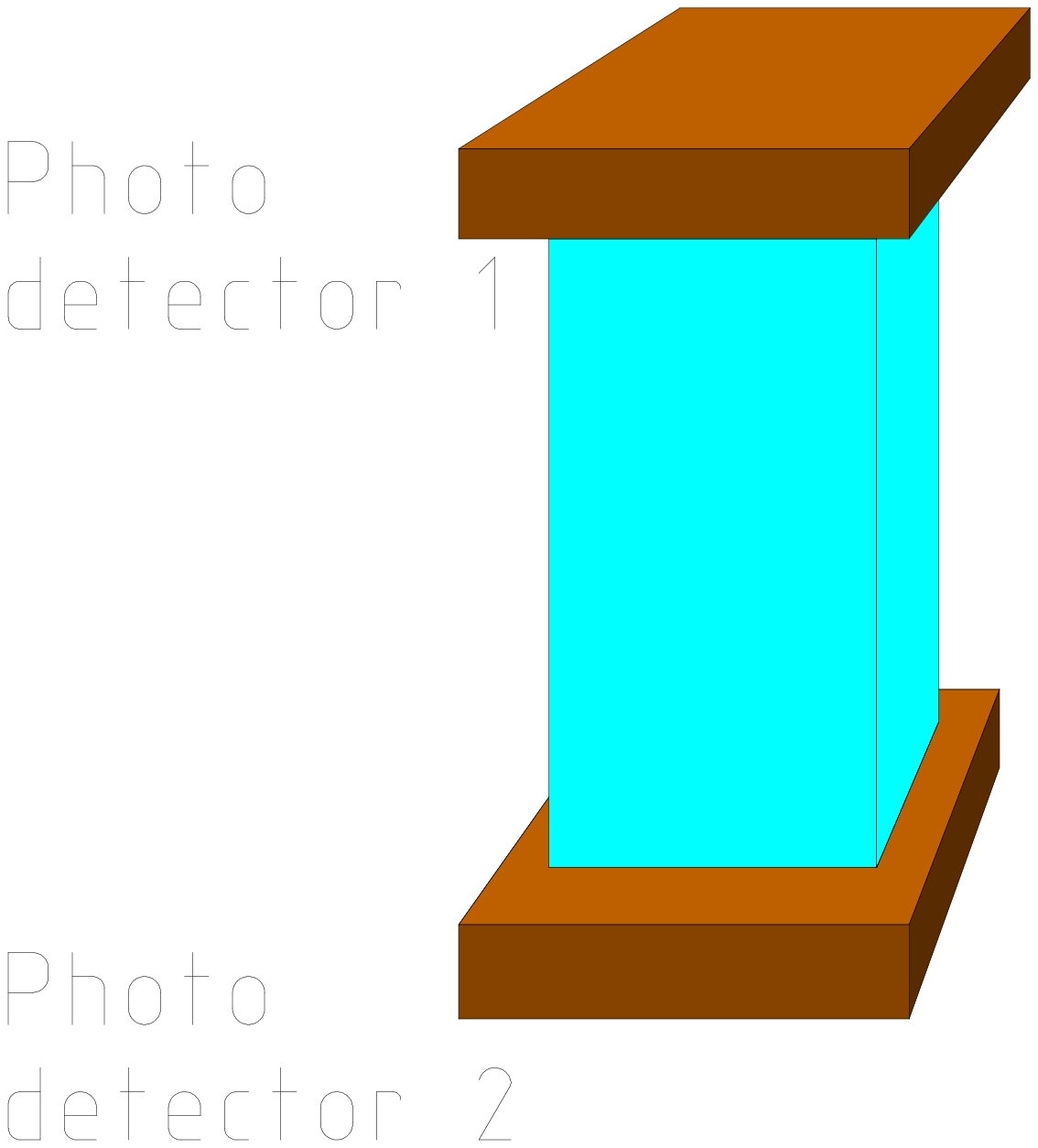}\hspace*{2em}}
  \subfigure[][Phoswich technique for DOI determination.]{\label{subfig:phoswich-doi}\hspace*{2em}
    \includegraphics[height=0.15\textheight]{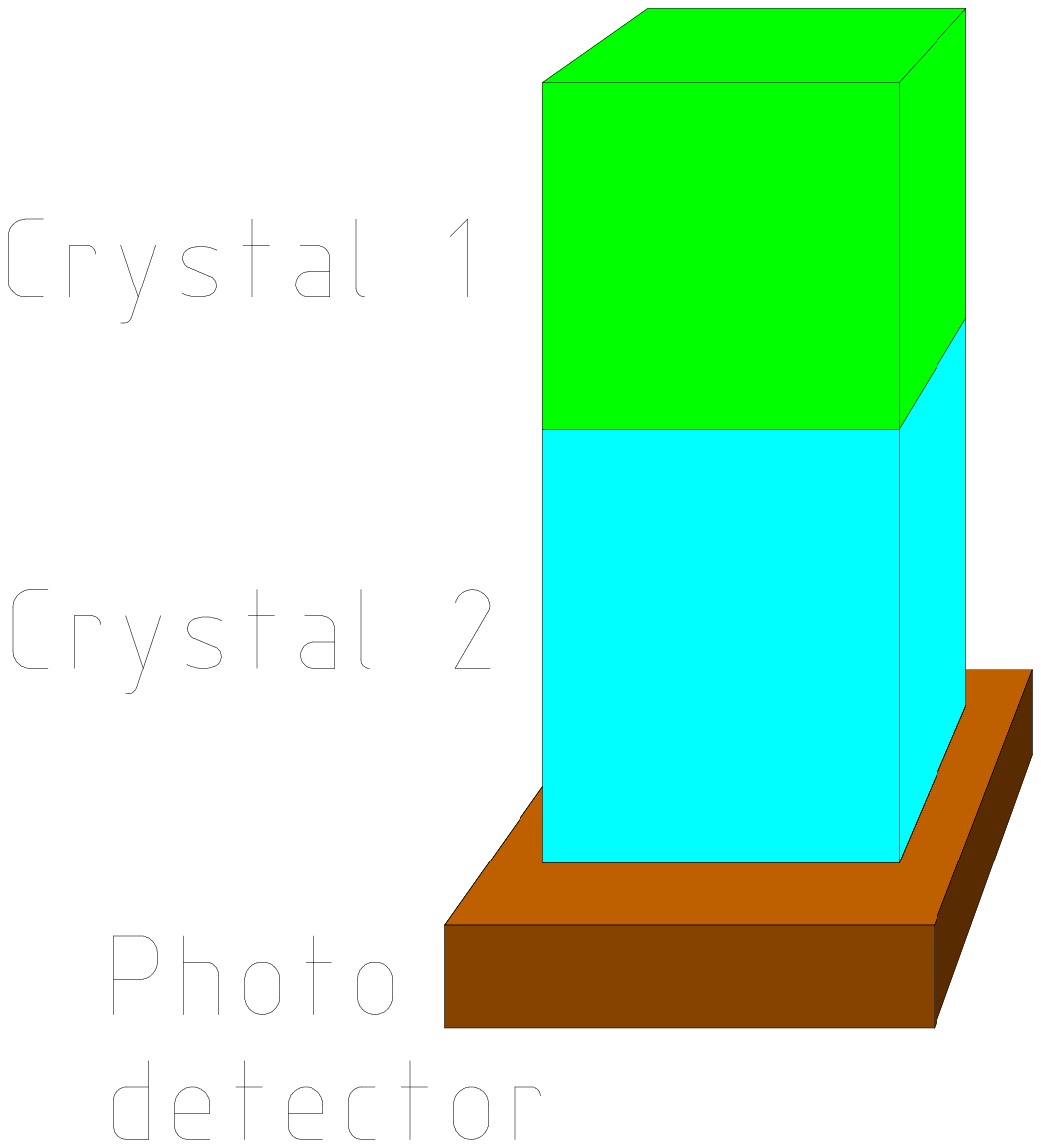}\hspace*{2em}}
  \subfigure[][DOI determination by reduction of light yield in one
  detector segment.]{\label{subfig:absor-band-doi}\hspace*{2em}
    \includegraphics[height=0.15\textheight]{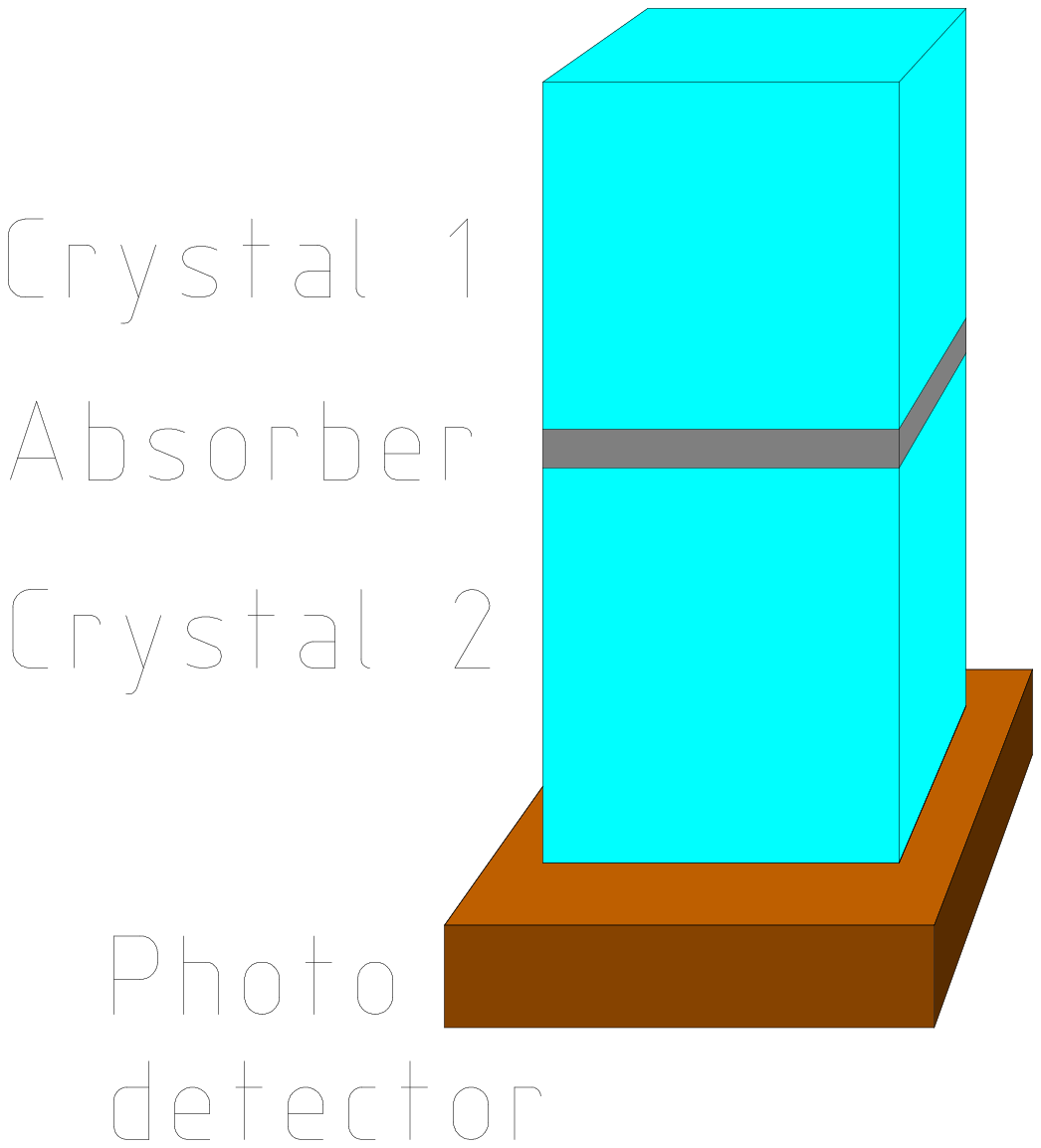}\hspace*{2em}}\\
  \subfigure[][Staggered double layer of crystal pixels for DOI
  determination.]{\label{subfig:staggered-layer-doi-detector}\hspace*{1em}
    \includegraphics[height=0.13\textheight]{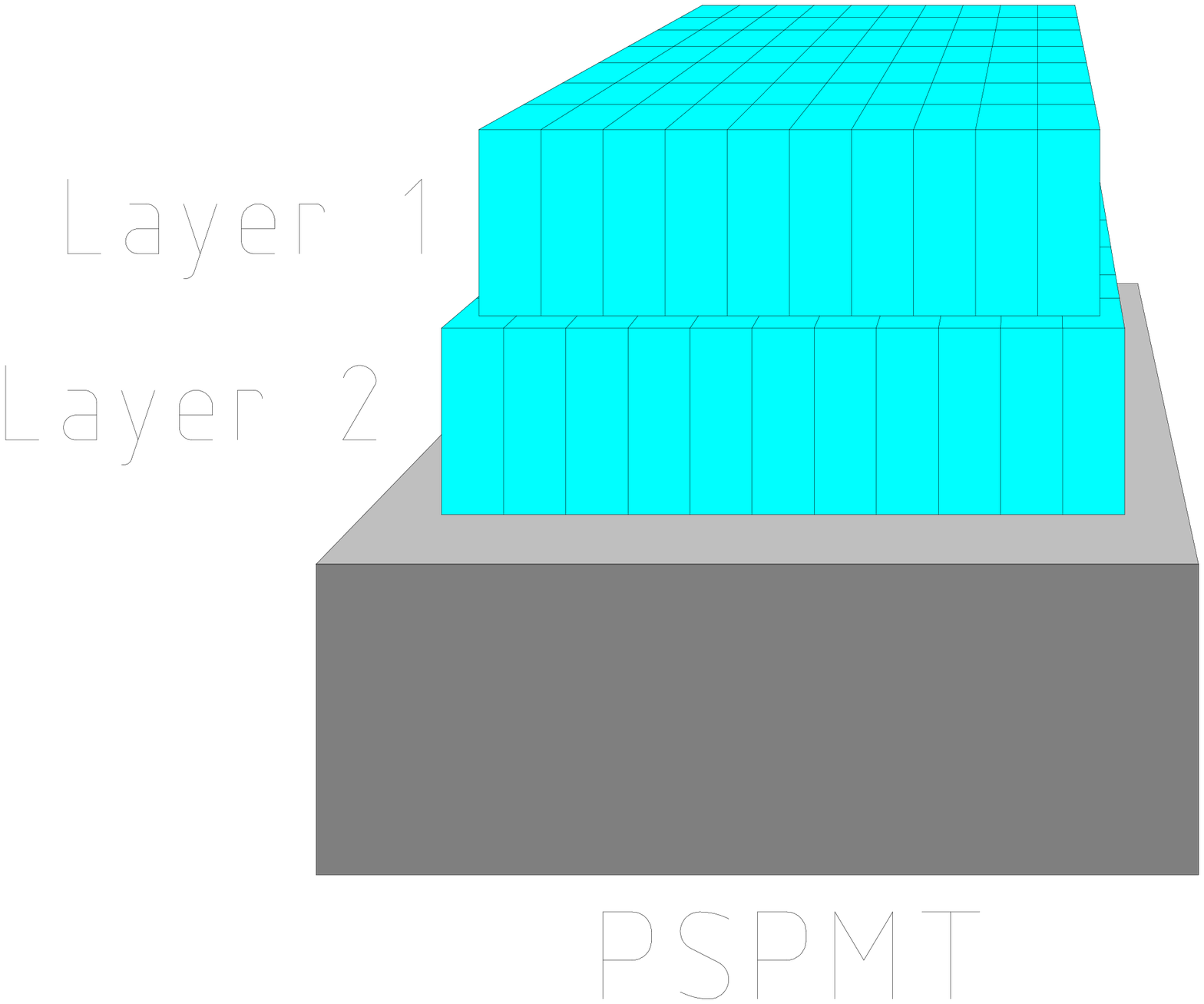}\hspace*{1em}}
  \subfigure[][Multiple pixel layer detector for DOI detection by a
  complex reflection scheme.]{\label{subfig:reflection-encoded-DOI-detector}\hspace*{2em}
    \includegraphics[height=0.13\textheight]{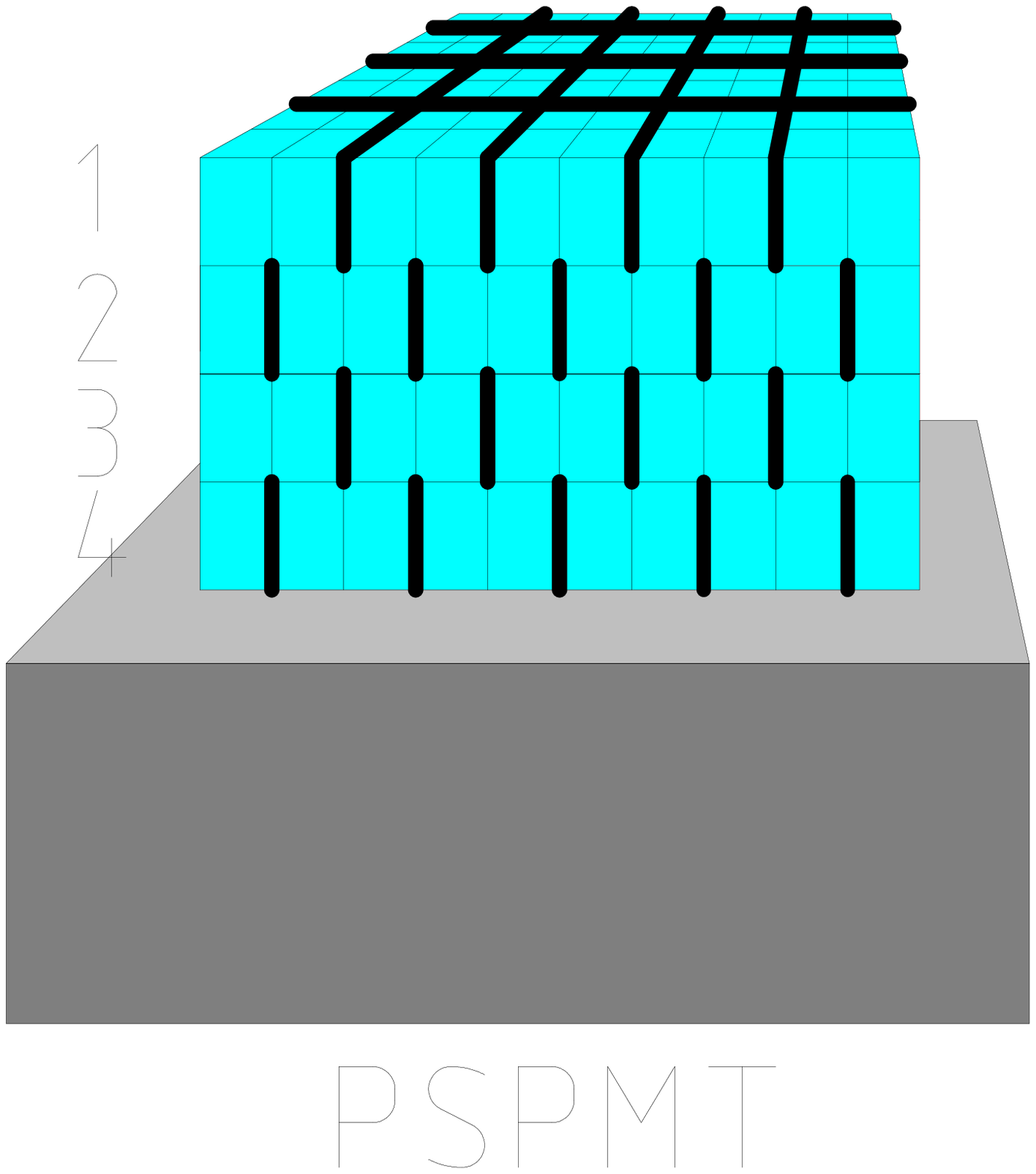}\hspace*{2em}}
  \subfigure[][Fully independent dual layer of pixelated detectors.]{\label{subfig:madpet-detector}
    \includegraphics[height=0.13\textheight]{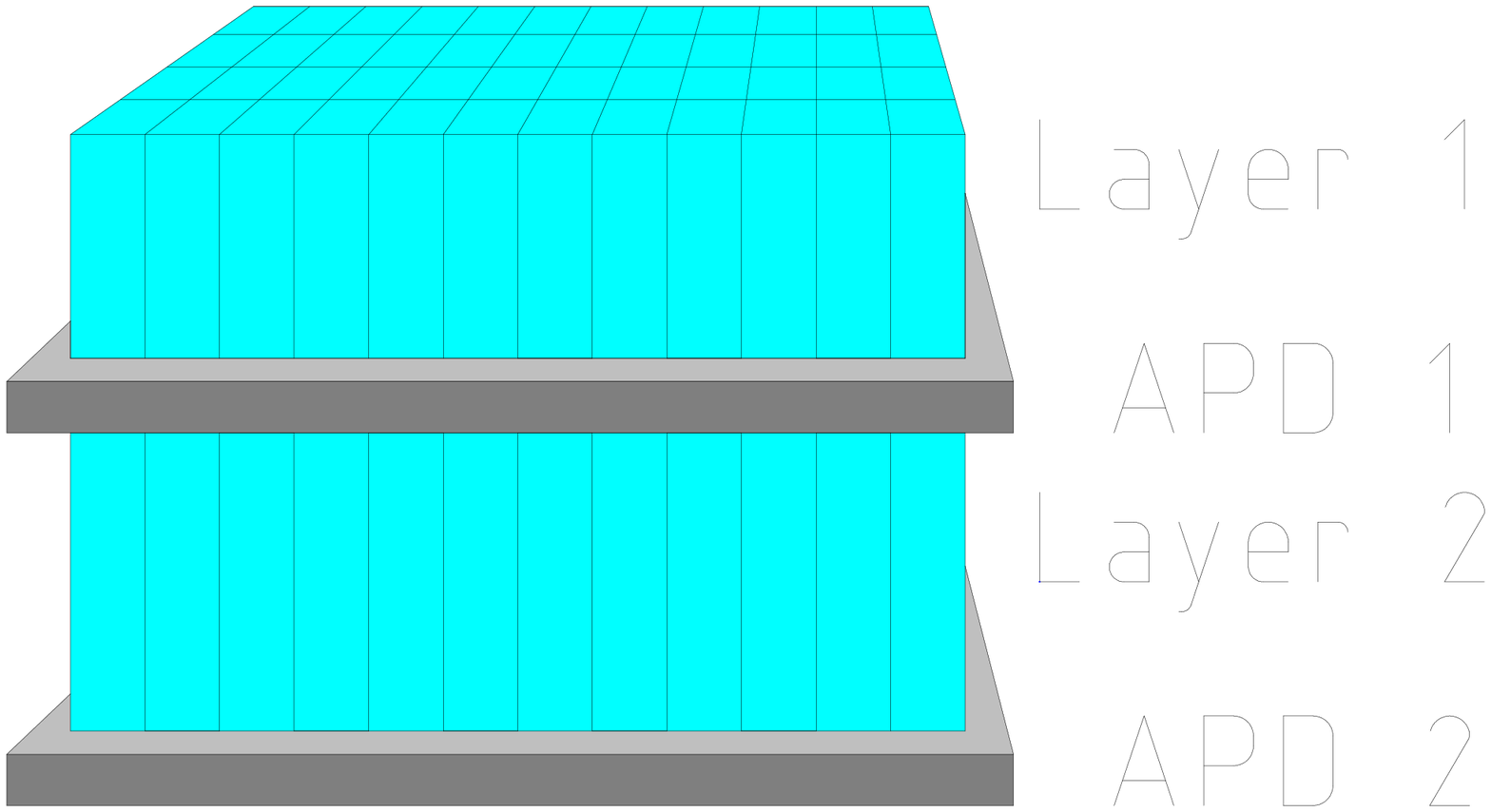}}\\
  \caption[Most common approaches of realizing DOI capable
    $\gamma$-ray detectors]{Most common approaches of realizing DOI capable
    $\gamma$-ray detectors.}
  \label{fig:diff-DOI-detects}
\end{figure}

Detectors that do not provide information about the DOI of the
recorded $\gamma$-ray introduce parallax errors which 
lead to uncertainties in the parameterization of the line of response
(LOR) and, as a consequence, to a nonuniform and non-isotropic spatial
resolution (Hoffman {\em et al.}\ \cite{Hoffman:1989}, Kao {\em et al.}\ \cite{Kao:2000}). One possible strategy to
partially overcome the effect of parallax errors is to restrict the
transaxial field of view (FOV) of PET-systems to a small fraction at
the center of the PET-system's sensitive volume. Unfortunately this
causes other unwanted effects such as loss of efficiency and an 
increasing relevance of photon non-collinearity, besides the fact that
it makes compact PET scanners impossible. For existing $\gamma$-ray
imaging systems, the image quality can be improved potentially if a
sufficiently good estimate of DOI is provided by the
detector. Especially for PET-detectors based on pixelated
scintillators, a number of techniques for designing DOI capable
detectors for $\gamma$-rays in the fields of nuclear medical imaging
have been proposed. One of them consists in measuring the ratio of
scintillation light detected at opposite crystal surfaces using two
photodetectors and is shown in figure~\ref{subfig:light-sharing-doi}
(Moses and Derenzo \cite{Moses:1994}). The DOI information is derived
from the difference in the signal amplitudes of the upper and lower detector.
To avoid additional expensive photodetectors and their associated
electronics, other techniques can be used. Phoswich detector use pulse
shape discrimination for DOI detection. Various layers of crystal
pixels with different scintillation characteristics are stacked on a
single position sensitive photodetector (refer to
figure~\ref{subfig:phoswich-doi}). The DOI information is derived
from measuring the pulse widths for the coincidence events (see
for instance Seidel {\em et al.}\ \cite{Seidel:1999}).
Very similar to this approach is the introduction of absorbing bands
between the different crystal pixel layers as shown in
figure~\ref{subfig:absor-band-doi}. This was also proposed by
different research groups (Bartzakos and Thompson
\cite{Bartzakos:1991} and Rogers {\em et al.}\ \cite{Rogers:1996}).
A staggered double-layer array of crystal needles as displayed 
in figure~\ref{subfig:staggered-layer-doi-detector} was proposed by 
Liu {\em et al.}\ \cite{Liu:2001}. The relative displacement of the pixel
layers allows the identification of the active layer by the detected
transverse position. Similar to this method is the one proposed by
Tsuda {\em et al.}\ \cite{Tsuda:2006} and shown in
figure~\ref{subfig:reflection-encoded-DOI-detector}. By introduction
of vertical reflecting sheets between specific pixels and at different
interfaces in each layer, the measured transverse positions are
displaced by a well defined amount and allow for identification of the
active layer. Completely independent crystal pixels have been proposed
for the MADPET-II dedicated small animal PET scanner by Rafecas et
al.\ \cite{Rafecas:2001a}. This is shown in figure~\ref{subfig:madpet-detector}.
Also combinations of different techniques are possible. However, all these 
approaches imply costly detector modifications to a greater lesser
extent. They need additional photodetectors, smaller crystals or
crystals of a different type. 

Proposals for interaction depth measurements in continuous
scintillation crystals for nuclear imaging were reported by 
Karp and Daube-Witherspoon \cite{Karp:1987}, Matthews {\em et al.}\ \cite{Kenneth:2001}
and Antich {\em et al.}\ \cite{Antich:2002}.
Except for Karp and Daube-Witherspoon, who used a temperature gradient
applied to \chemform{NaI\doped Tl} in order to force a variation of the
signal shape with the interaction depth that could be used for
discrimination, the other groups sampled the scintillation light
distribution and computed the standard deviation from the data set
because of its correlation to the interaction depth.

\subsection{Time-of-Fight PET}

\begin{figure}[!t]
  \centering
  \includegraphics[width=0.4\textwidth]{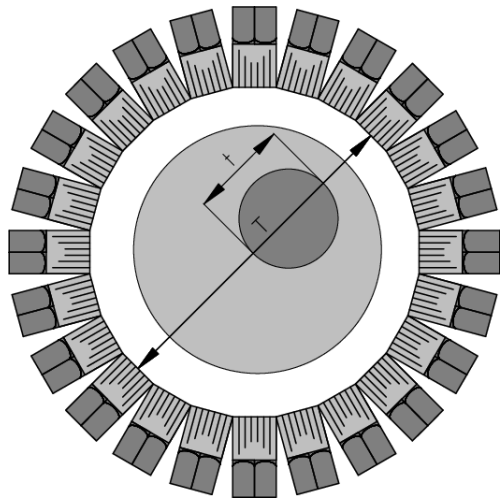}
  \caption[Image reconstruction with TOF capable PET camera]{Image reconstruction with TOF capable PET camera. If the
    time of flight for the LOR T can be measured with sufficient accuracy, the
    possible annihilation position can be restricted to region of
    diameter t and smaller than the FOV (light-gray circular region).}
  \label{fig:tof-pet}
\end{figure}

Commercial PET scanners have an inner ring diameter of $\mathrm{80\,cm}$.
This distance is covered by a \g-ray traveling at the speed of light
in $\mathrm{T\approx2.7\,ns}$. If
the detector elements were able to estimate the time of arrival with
this resolution, the supposed annihilation position can be restricted
to a region of diameter $t$ and smaller than the disk defined by the PET scanner (refer to
figure~\ref{fig:tof-pet}). The higher the timing resolution of the
coincidence system is, the smaller will be the region from which the
annihilation photons can originate. If this information is implemented
in the reconstruction algorithm, the quality of the image can be
improved significantly (Moses \cite{Moses:2003}). One of the parameters
that is improved is the SNR. But also the noise equivalent count rate
is increased, because many random events can be now rejected. For
a timing resolution of $\mathrm{500\,ps}$, the random rate
is reduced by a factor 1.5-2, $f_\tincaps{NEC}$ is increased by a
factor 1.4-1.6 and the noise variance is lowered by a factor of 5.

\chapterbib


  \cleardoublepage{}
\chapter{Parameterization of the Signal Distribution}
\label{ch:light-distribution}

\chapterquote{%
Where there is much light, the shadow is deep.
}{%
Johann Wolfgang von Göthe, $\star$ 1749 -- $\dagger$ 1832
}

\PARstart{I}{n} chapter~\ref{ch:motivation} it has been
pointed out that the use of continuous scintillation crystals
carries several general advantages and that good results can be
obtained with $\gamma$-ray imaging detectors based on this design.
When Positron Emission Tomography emerged as a novel imaging
technology, it was straightforward to take the design from existing
Anger-type cameras and use it for {\em positron cameras}, because they
are just $\gamma$-cameras optimized for the detection of the
annihilation radiation. The key problem arises from the high energy of
this radiation. While $\gamma$-photons from the $\mathrm{^{99m}Tc}$
isotope have an energy of 
approximately $\mathrm{140\;keV}$, the energy of annihilation photons
is more than three times as large. 
The detection of these photons presupposes the photoconversion within
the scintillation crystal but the interaction probability decreases
with increasing  $\gamma$-ray energy. Therefore, the crystals have to be
thicker in order to stop the same number of photons. Since it
is not normally known at which depth the $\gamma$-ray interacts, an uncertainty
is introduced. This may cause serious positioning errors for
\g-rays with non-normal angles of incidence and, depending on which positioning method
is used, depth-dependent  variations in the position measurement. 
Apart from Anger \cite{Anger:1963}, there have been
attempts by several research groups (Karp {\em et al.}\
\cite{Karp:1985}, Rogers {\em et al.}\ \cite{Rogers:1986}, Siegel {\em et al.}\
\cite{Siegel:1995} and Seidel {\em et al.}\ \cite{Seidel:1996}) to develop
$\gamma$-ray detectors for positron emission tomography using
continuous scintillation crystals instead of arrays of small crystal
segments. Except for a few, all groups abandoned this design due to
the strong border artefacts in large-sized continuous scintillation
crystals caused by the center of gravity (CoG) algorithm (see
section~\ref{ch:errors-of-cog-and-cdr}). Clearly, the crucial point is
how the position measurement is implemented. The mentioned resolution
degradation at the crystal edges is inherent in the CoG algorithm,
but not necessarily in other positioning methods. There are 
statistical based position reconstruction schemes (Liu {\em et al.}\
\cite{Liu:1990}, Joung {\em et al.}\ \cite{Joung:2002}) that do not suffer
from this systematic error. However, a sampling of the signal
distribution, {\em i.e.}\ the digitization of all channels, is required
and the border-effects are avoided at the expense of higher computation time
and detector costs. An alternative access to the real
three-dimensional position where the photoconversion took place seems
to be possible by using the width of the signal distribution, since this
parameter is correlated with the interaction depth (Siegel {\em et al.}\
\cite{Siegel:1995}, Matthews {\em et al.}\ \cite{Kenneth:2001}, Takacs et
al.\ \cite{Takacs:2001}, Antich {\em et al.}\ \cite{Antich:2002} and Lerche
{\em et al.}\ \cite{Lerche:2005b}).

In this chapter, a parameterization for the density distributions of
light seen by the photocathode and charge seen by the anode-segments of
the position sensitive photodetector is derived. Although there are
many simulation tools based on the Monte Carlo method, the analytic
model was preferred, because it leads to a better understanding of the
resulting distribution.

\section{Included Contributions and Conventions}
\label{sec:included-contribs}

The signal distribution that is finally accessible for
the reconstruction software emerges as an interplay of a broad 
variety of physical phenomena. The generation starts with the photoconversion of the
incident $\gamma$-ray via a sequence of elementary interactions, {\em e.g.}\
Rayleigh scattering, Compton scattering and the photoelectric effect. However,
only the position of the very first $\gamma$-ray interaction
corresponds to the true line of response (LOR). The photoelectric effect and
Compton scatter events produce so-called {\em knock-on electrons}
that excite the atoms of the crystal material along its path
until being completely stopped. Also, ionization with secondary
electrons of high energy ({\em $\mathit{\delta}$-electrons}) is
possible. The scintillation light arises from the de-excitation of the 
scintillation centers along the different electron paths. Since common
scintillators for PET are normally made of a high density material, these
paths are rather short. One can estimate the range of these electrons
using an empirical formula from Tabata, Ito and Okabe (Tsoulfanidis
\cite{Tsoul:1995}) based on
experiments that measured the range of positrons by extrapolating 
the linear part of the transmission function and defining the
intersection with the background to be the extrapolated range
$R_{ex}$. The largest possible distance of
$\mathrm{\approx 150\mu m}$ is covered by an electron that
emerges from a unique photoelectric interaction of an incident 
$\mathrm{511 keV}$ $\gamma$-ray and that produces no
$\delta$-electrons. However, the majority of produced electrons 
will travel much shorter distances before being stopped and we
assume these elementary interactions to be point-like.

In this chapter, only signal distributions arising from single
photoelectric interactions will be considered. This seems to be a very rough
approximation, since for $\mathrm{511 keV}$ $\gamma$-rays one expects
various Compton scattering interactions as part of the photoconversion even for
dedicated PET scintillators such as LSO. The influence of inner-crystal
Compton scattering on spatial resolution that can be achieved with the
CoG algorithm will be discussed in detail in section~\ref{ch:compton}.
In section~\ref{ch:experiment} it will be shown that even with this
rigorous approximation, very good agreement with the real detector
response is achieved. Signal distributions that arise from various 
interactions  are superpositions
of the  independent single-event distributions. If the
energies deposited at these points are different
(as in the vast majority of the cases), the distribution arising from the point
with maximum energy deposition will dominate. 

Four issues have to be clarified before beginning with the derivation
of the theoretical model. The first is the significance of {\em continuous
  scintillation crystals}. Crystals of almost every possible spatial
dimension can be called continuous as long as one does not reach the
molecular scale. Actually, the definition depends strongly on the scale,
since a crystal of dimensions $\mathrm{2\times2\times2\,mm^3}$ can be
considered continuous, if the region of interest is confined to this
same volume. The volume of interest for the present work is however
defined by the product of the required scintillator thickness and the
sensitive area of the photodetector. Throughout this work I will
refer to a {\em continuous crystal} whenever this volume of interest
is covered by one single crystal. Similarly, the expression {\em
  pixelated crystal} or {\em crystal array} is used, if the same
volume is covered by several independent crystals.
The second issue is the treatment of the crystal surfaces. Generally,
only one of the six scintillator faces is coupled to a
photodetector. For all other faces there exist many possible
treatments that allow an optimization of the detector performance. 
However, the aim of the present work is not the detector optimization,
and therefore, all {\em free} surfaces are assumed to be fine ground
and covered by highly absorptive coatings in order to avoid
reflections while the one that is coupled to the photodetector is
assumed to be polished.

\begin{figure}[!t]
  \centering
  \psfrag{depthofinteraction}{\it\small depth of interaction}
  \psfrag{scintillator}{\it\small scintillator}
  \psfrag{entrancewindow}{\it\small entrance window}
  \psfrag{photosensitivelayer}{\it\small photosensitive layer}
  \psfrag{photoconvpos}{\it\small position of photoconversion}
  \psfrag{interactiondistance}{\it\small interaction distance}
  \includegraphics[width=0.72\textwidth]{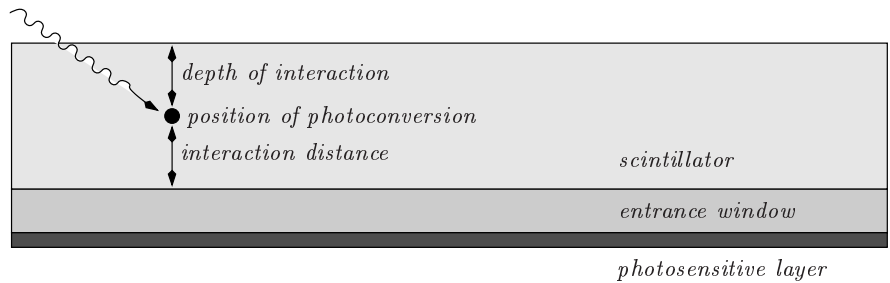}
  \caption[Naming convention for the interaction depth and interaction
  distance]{Naming convention for the interaction depth and interaction distance.}
  \label{fig:naming-conv}
\end{figure}

The third issue concerns the choice of the point of origin. Throughout the
following sections and chapters the reference system is chosen in such
a way that the sensitive area of the photodetector is parallel to the
$x$-$y$-plane and centered at the point of origin  $x_0=y_0=z_0=0$. The $x$- and $y$-spatial
directions are understood to be the {\em transverse} and {\em horizontal}
spatial directions respectively and the $z$-direction the {\em normal} or
{\em vertical} direction. The scintillator 
is placed within 
the half-space with $z>0$ and the detection material of the
photodetector within the half-space with $z<0$. 
The last issue is a pure naming convention. The expression {\em depth
  of interaction} (DOI) refers to the distance between the position of the $\gamma$-ray's
photoconversion and the crystal surface facing to the photodetector
(refer to figure~\ref{fig:naming-conv}). For the derivation below, the
distance between the photoconversion point and the crystal surface
coupled to the photodetector is more intuitive. Throughout this work,
this parameter will be primarily used and it will be called the {\em interaction distance} (ID). 

\subsection{The Inverse Square Law}
\label{sec:inv-square-law}

The starting point for the derivation of the signal distribution is the
inverse square law. It says that the intensity of light radiating
from a point source is inversely proportional to the square of the
distance from the source. This law is founded in strictly geometrical
considerations and therefore applies to all isotropic radiation
phenomena. 
It can be easily derived using the divergence theorem and gives 
\begin{equation}
  \label{eq:general-inv-square-law}
  J(\mathbf{r},\mathbf{r}_\mathrm{c})=\frac{J_\mathrm{c}}{4\pi |\mathbf{r}-\mathbf{r}_\mathrm{c}|^2},
\end{equation}
where $J(\mathbf{r},\mathbf{r}_\mathrm{c})$ is the  amount of
scintillation light at the observation point $\mathbf{r}$, $J_\mathrm{c}$ is the total amount of released
scintillation light and $\mathbf{r}_\mathrm{c}$ is the position of photoconversion.
Equation~(\ref{eq:general-inv-square-law}) shows the geometrical
character of the inverse square law since the denominator is just the
expression for the surface area of a sphere with radius
$|\mathbf{r}-\mathbf{r}_\mathrm{c}|$. Moreover, it
states that the total number of light photons released at the
scintillation center is conserved and does not change as the light
propagates through the crystal.

\subsection{The Cosine Law}

In the previous paragraph, the inverse square was derived for
spherical surfaces. Commercially available photodetectors 
offer always, except for a few special cases, a planar design of the
sensitive area. Only for almost normal incidence of the scintillation
light, the difference between planar and spherical surface element
will be not very important. However, since continuous and large-sized crystals
in the sense of section~\ref{sec:included-contribs} are used in this
study, a large fraction of the scintillation light impinges on the
photodetector at large angles.

\begin{figure}[t]
  \centering
  \psfrag{th}{$\vartheta$}
  \psfrag{d}{$z_c\!-\!z_0$}
  \psfrag{z0}{$z_0$}
  \psfrag{J}{$J_\mathrm{c}$}
  \psfrag{dS}{$d\mathbf{S}$}
  \psfrag{dS1}{$d\mathbf{S}'$}
  \psfrag{r}{$\mathbf{r}$}
  \psfrag{r0}{$\mathbf{r}_\mathrm{c}$}
 \includegraphics[width=0.7\textwidth]{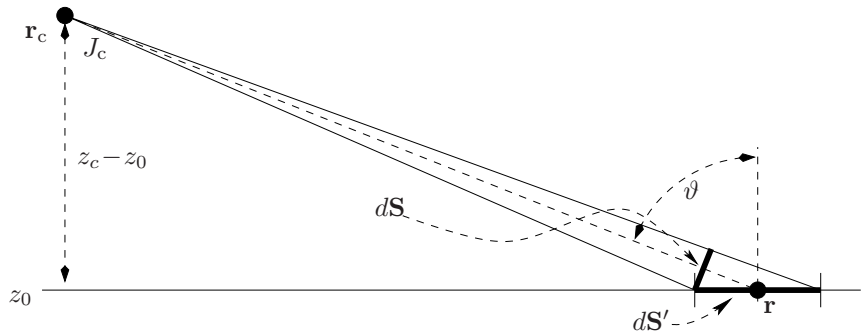}
  \caption[Cosine law for the irradiation of planar surfaces]{\label{fig:cosine-law}Diagram of the planar surface element
    $d\mathbf{S}'$ on the detectors sensitive area is irradiated by
    the source at $\mathbf{r}_\mathrm{c}$ and with the angle $\vartheta$. A photon density
    reduced by the factor $\cos\vartheta$ is observed, because the
    flux of photons through $d\mathbf{S}$ is dispersed over $d\mathbf{S}'$.}
\end{figure}

In the case of scintillation detectors, the position of the light
source is confined to the volume of the scintillation crystal. For
optimum performance, it has to be coupled through light guides as close as
possible to the photodetector. Therefore, the normal distance between
the light source and the photodetector never exceeds the thickness of
the crystal. Furthermore, it is small compared to the planar
extension of the detector. The photon
flux corresponding to the surface element $d\mathbf{S}$ is covered by
the detectors sensitive surface element $d\mathbf{S}'$ as shown in
figure~\ref{fig:cosine-law}. By similarity of the triangles spanned by
$d\mathbf{S}'$ and $d\mathbf{S}$ and $(\mathbf{r}-\mathbf{r}_\mathrm{c})$ and
$(z_c-z_0)$ one deduces that
\begin{equation}
  \label{eq:cosine-law}
  d\mathbf{S}=d\mathbf{S}'\cos\vartheta=d\mathbf{S}'\frac{z_c-z_0}{|\mathbf{r}-\mathbf{r}_\mathrm{c}|}.
\end{equation}
With this result, the inverse square law of equation~(\ref{eq:general-inv-square-law})
becomes 
\begin{equation}
  \label{eq:inv-square-law-2}
  J(\mathbf{r},\mathbf{r}_\mathrm{c})=\frac{J_\mathrm{c}}{4\pi}\frac{(z_c-z_0)}{|\mathbf{r}-\mathbf{r}_\mathrm{c}|^{3/2}}.
\end{equation}
Equation~(\ref{eq:inv-square-law-2}) inherently has the correct
normalization. Integrating $J(\mathbf{r},\mathbf{r}_\mathrm{c})$ over an
infinite plane normal to the $z$-unit vector $\mathbf{\hat{e}}_z$ and at $z_c\neq z_0$, one
obtains $J_\mathrm{c}/2$ which is exactly the expected result since just the
half of all released scintillation light will be collected.

\subsection{Exponential Attenuation}
\label{sec:exp-attenuation}

While the scintillation light travels within the optical components of
the detector, it is attenuated according to the exponential law
\begin{equation}
  \label{eq:exponetial-decrease}
  I(\mathbf{r},\mathbf{r}_\mathrm{c})=I_0 e^{-\alpha(\lambda)|\mathbf{r}-\mathbf{r}_\mathrm{c}|},
\end{equation}
where $\alpha(\lambda)$ is the inverse of the attenuation length at
the wavelength $\lambda$, $I_0$ is the initial intensity and
$I(\mathbf{r},\mathbf{r}_\mathrm{c})$ the attenuated intensity. An important requirement for scintillators 
is a high transparency to its own scintillation light in
order to allow an effective light collection. Therefore, one can
assume that $\alpha(\lambda)^{-1}$ is large compared to the spatial
extensions of the scintillator. Otherwise, the chosen
scintillation material would not be adequate for the application.
Furthermore, $\alpha(\lambda)$ can be approximated by the constant average
value $\bar{\alpha}$, since the emission spectra
of many scintillators are confined to intervals of less than a few
hundred $\mathrm{nm}$ with only small variations of $\alpha(\lambda)$ over this interval.

\subsection{Light Transmission to the Photodetector Window}

The scintillation light that reaches the borders of the crystal is
either absorbed at one of the five absorbing crystal sides or detected
by the photodetector coupled to the transparent side. The photodetector itself
contains a transparent entrance window for protection from
environmental influences. Although this window is optimized with respect to
collection efficiency and transparency, its refraction index in
general does not match the one of the scintillation
crystal. Furthermore, since the refraction index of the entrance
window and the refraction index of the scintillator are much larger
than the refraction index of air, air-gaps between both media have to
be avoided. Optimal optical coupling is achieved with optical gel or
silicone of intermediate refraction index.
Hence, the scintillation light traverses an optical multi-layer system
consisting of several homogeneous media in sequence, {\em e.g.}\ the
scintillation crystal, the optical grease, the entrance window and the
photocathode. There is also a fraction of light that is transmitted to
the the vacuum if photomultiplier tubes
are used for detection.  The behavior of light at the interface
between optical media of different refraction indexes is well
described by Snell's Law and the Fresnel equations. In order to reduce
the complexity of the derived model-distribution, the intermediate
layer of optical fat is not included in the considerations. This
approximation can be made, since the refraction index is chosen to
lie in between the values for the entrance window and the
scintillation crystal, and because the thickness of the additional layer will be
very small. Thus, it will cause only a very small deviation of the
light path.

\subsubsection{Refraction of the Light}
\label{sec:refrac-of-light-path}

\begin{figure}[t]
  \centering
  \psfrag{n1}{$n_1$}
  \psfrag{n2}{$n_2$}
  \psfrag{a}{$a$}
  \psfrag{b}{$b$}
  \psfrag{t}{$t$}
  \psfrag{h}{$h$}
  \psfrag{t1}{$\vartheta_1$}
  \psfrag{t2}{$\vartheta_2$}
  \psfrag{do}{$d$}
  \psfrag{dn}{\hspace*{-0.6ex}$d^\mathit{virt}$}
  \psfrag{rn}{$r$}
  \psfrag{z=zpc}{$z=z_\tincaps{PC}$}
  \psfrag{r0}{$\mathbf{r}_\mathrm{c}$}
  \psfrag{ro}{$r^\mathit{virt}$}
  \includegraphics[width=0.94\textwidth]{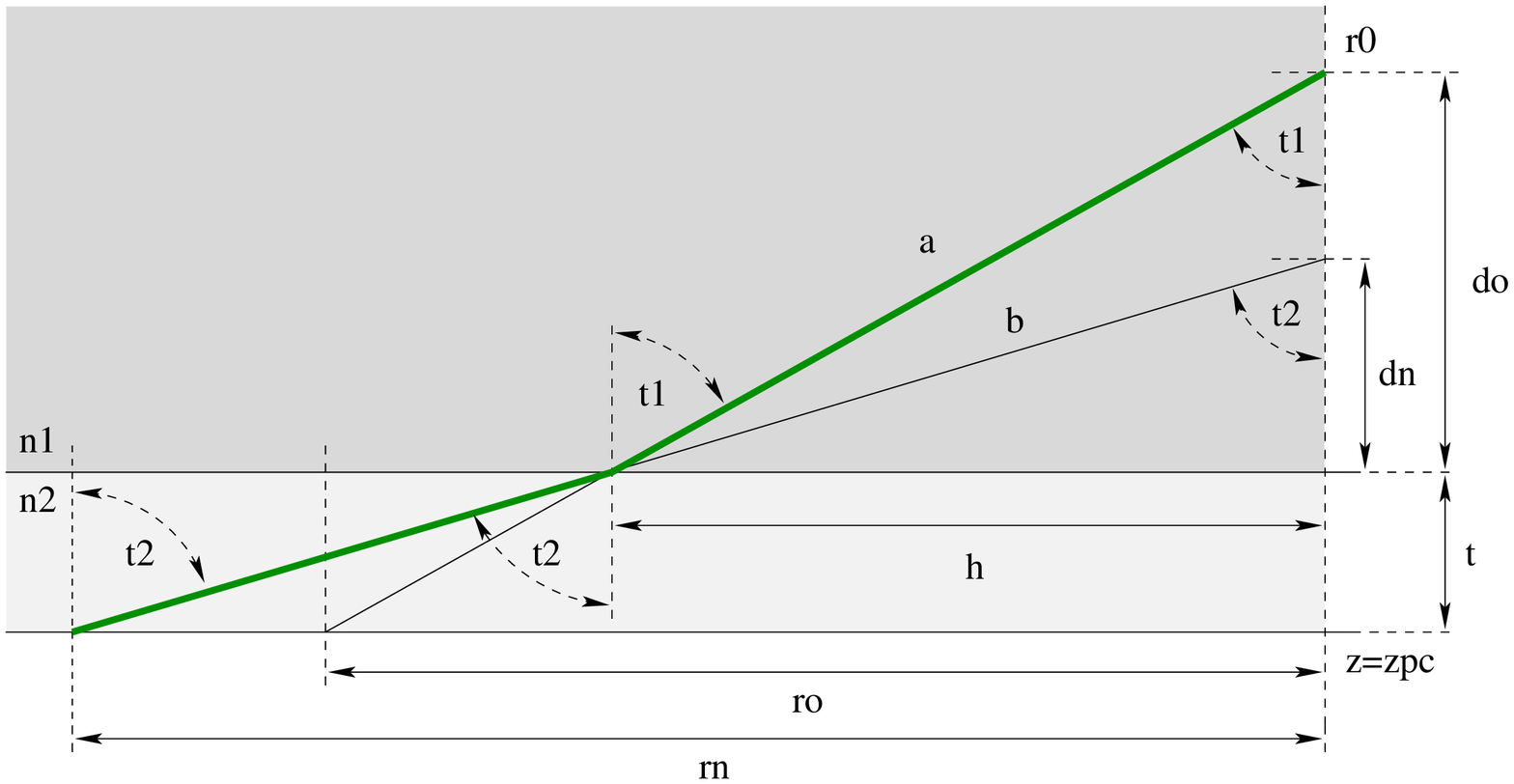}
  \caption[Geometric refraction of the scintillation light
  path]{Geometric refraction of the scintillation light path (thick,
    green line). Due to
    Snell's Law, light detected at a distance $r$ seems
  to be released at the virtual point $(x_c,y_c,z_c^\mathit{virt})=\mathbf{r}_\mathrm{c}^\mathit{virt}$}
  \label{fig:refract-light-path}
\end{figure}

\begin{figure}[!t]
  \centering
  \subfigure[][Virtual interaction distance $d^\mathit{virt}$ seen by
  the photodetector behind the entrance window.]{\label{subfig:dvirt}%
    \psfrag{d1}{\scriptsize$\mathrm{1\,mm}$}
    \psfrag{d10}{$\mathrm{10\,mm}$}
    \psfrag{d}{\hspace*{-2em}$d^\mathit{virt}$, [mm]}
    \psfrag{r}{\hspace*{-2em}$r$, [mm]}
    \includegraphics[width=0.45\textwidth]{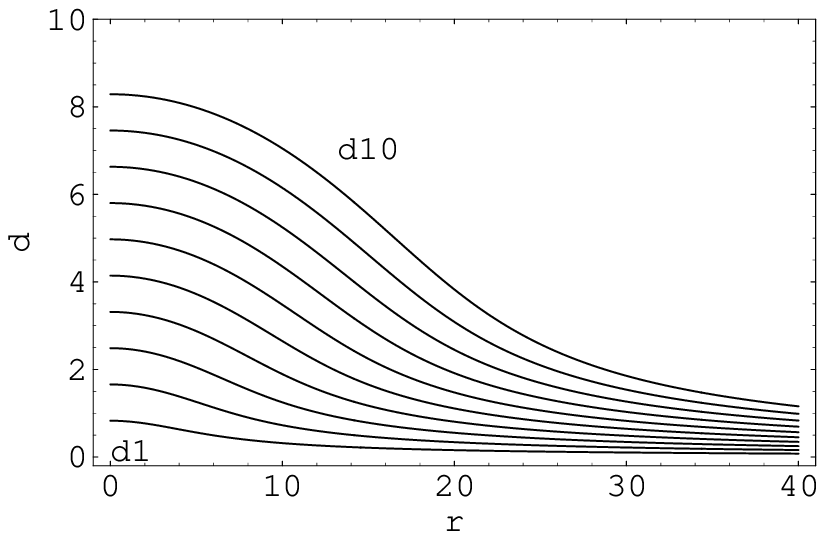}
  }
  \subfigure[][Effectively observed incidence angle $\vartheta_2$ at the sensitive detector-material.]{\label{subfig:virtual_angle}%
    \psfrag{d1}{$\mathrm{1\,mm}$}
    \psfrag{d10}{$\mathrm{10\,mm}$}
    \psfrag{t}{\hspace*{-3em}$\vartheta_2$, [degrees]}
    \psfrag{r}{\hspace*{-2em}$r$, [mm]}
    \includegraphics[width=0.45\textwidth]{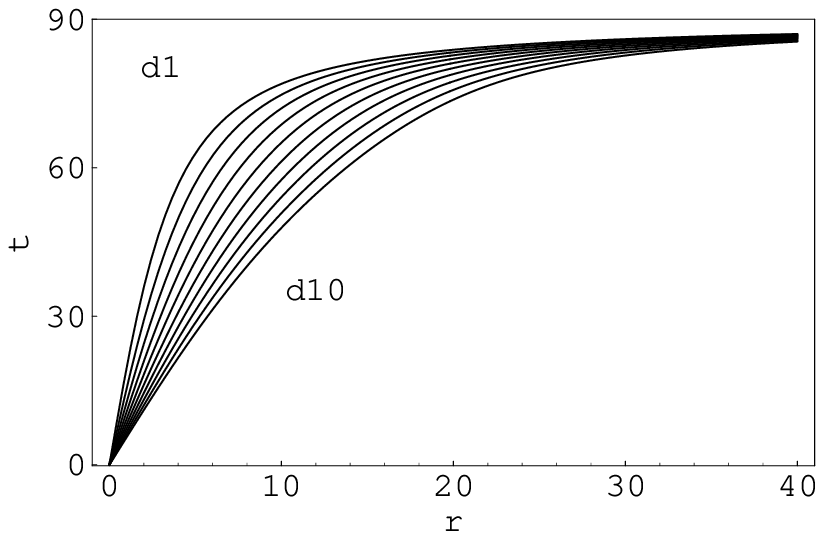}
  }
  \caption[Angle of incidence and virtual depth of the refracted scintillation
    light path]{Angle of incidence and virtual depth of the refracted scintillation
    light path as a function of the distance $r$ and for different
    interaction distances $d$. In each figure ten graphs for the depths
    $\mathrm{d\in\{1,2,\ldots,10\}\,mm}$ are shown.}
  \label{fig:deff-from-r}
\end{figure}

The relationship between the angles of incidence and refraction for a wave
impinging on an interface between two media is given by Snell's law
\begin{equation}
  \label{eq:snells-law}
  \sin\vartheta_2=\kappa\sin\vartheta_1\quad\mbox{with}\quad \kappa=\frac{n_1}{n_2},
\end{equation}
where the angles $\vartheta_1$ and $\vartheta_2$ are taken with respect to the
normal on the interface plane and $n_1$ and $n_2$ are the refraction indices
of the scintillator and the entrance window respectively. The
refraction index of inorganic scintillation crystals for $\gamma$-ray
detection is comparatively high. $n_1$ is normally larger
than $n_2$. This leads to total reflection whenever the angle
$\vartheta_1$ of the incident ray is larger than the critical angle
$\vartheta_c=\arcsin\left(\kappa^{-1}\right)$. As a consequence, only the
fraction $\Phi=(1-\cos\vartheta_c)/2$ of the scintillation light
that is confined to the cone of aperture $2\vartheta_c$ enters the
photodetector and contributes directly to the signal distribution
\mycite{Ambrosio}{{\em et al.}\ }{1999}. The scintillation light outside
this cone is trapped within the crystal due to total reflection and
most of it will be lost by absorption. The light inside the
cone is refracted away from the normal on the interface. This is shown
in figure~\ref{fig:refract-light-path}. 
For a  photodetector, the sensitive material, {\em i.e.}\ the
photocathode or the semiconductor will be located at the plane
$z=z_\tincaps{PC}$ behind the window of thickness $t$. Due to the
refraction, light that is emitted from a scintillation event at the
impact position $\mathrm{r}_0$ will not be detected at the projected
distance $r^\mathit{virt}$ but at $r$. The scintillation light seems
to come from a point at normal distance $d^\mathit{virt}$ from the
interface instead of coming from point $\mathbf{r}_\mathrm{c}$.
One therefore has to compute $d^\mathit{virt}$ as a
function of the true interaction distance $d$, the distance
$r$, the window thickness $t$ and $\kappa$.
This is equivalent to find the roots of the following quartic equation:
\begin{gather}
  \label{eq:light-path-variables}
  0  \stackrel{!}{=}  b^2-d^2_{\!\mathit{virt}}-a^2+d^2,\mbox{
    with }\\
  a  =  \kappa^{-1}b\quad\mbox{and}\quad
  b=\frac{d^\mathit{virt}}{d^\mathit{virt}+t}\sqrt{\left(d^\mathit{virt}+t\right)^2+r^2}\nonumber
\end{gather}
Equation~(\ref{eq:light-path-variables}) has eight mathematical
solutions but only one is of physical interest for the present
problem. It is given by
\begin{gather}
  \label{eq:light-path-solution}
  d^\mathit{virt}  = 
  \frac{1}{2}\left(q-t+\sqrt{2t^2-w-u+\frac{2t(u-t^2+2d^2\kappa^2)}{q}}\right)\mbox{,}\quad\mbox{where}\\
  q  =  \left|\sqrt{w+t^2-u}\right|\mbox{,}\nonumber\\
  w  = 
  \frac{1}{3}\left(u+u^2\left(\frac{v}{2}\right)^{-1/3}+\left(\frac{v}{2}\right)^{1/3}\right)\mbox{,}\nonumber\\
  u  =  t^2-\kappa^2d^2+r^2\left(1-\kappa^2\right)\mbox{,}\nonumber\\
  v  =  2\left(p+\left|\sqrt{p^2-u^6}\right|\right)\quad\mbox{and}\nonumber\\
  p  =  u^3+54\;t^2d^2\kappa^2r^2\left(1-\kappa^2\right)\mbox{.}
\end{gather}
With the aid of $d^\mathit{virt}$ we can define the new, {\em virtual}
photoconversion position
$\mathbf{r}_\mathrm{c}^\mathit{virt}\mdef(x_c,y_c,z_c^\mathit{virt})=(x_c,y_c,z_c-d+d^\mathit{virt})$. This is the
position from which the scintillation light seems to come from.
The dependence of $d^\mathit{virt}$ on the projected distance $r$ from the
impact position $\mathbf{r}_\mathrm{c}$ is shown in figure~\ref{subfig:dvirt}
for $d\in\{1,2,\ldots,10\}\,\mathrm{mm}$. In figure~\ref{subfig:virtual_angle},
the dependence of the angle of incidence $\vartheta_2$ of the light path on
the detecting surface is shown for the same $d$-values. At this point,
important observations can be made. First, although the
effect of total reflection prohibits the transmission of scintillation
light into the photodetectors entrance window for certain angles, the light distribution 
seen at the detecting material will never be truncated. This is
due to the fact that the light within the cone of aperture $2\vartheta_c$
can pass to the medium of refraction index $n_2$ but will be refracted
away from the normal onto the optical interface. For a
light ray of incident angle $\vartheta_c$, the refracted light would be
bent in such a way that it propagates just within the plane of the
interface. For all light rays with incident angles
$\vartheta<\vartheta_c$, the refracted angle will be in the interval $[0,90^\circ[$.
That is to say, the entering light which is confined to the
cone of half-angle $\vartheta_c$ will be redistributed and causes a new
continuous distribution illuminating completely the infinite plane at
$z=z_\tincaps{PC}$. 

Another interesting phenomenon is that this resulting
distribution is narrower than a light distribution without
refraction. This effect was already observed by Tornai {\em et al.}\
\cite{Tornai:1994} and is shown in figure~\ref{subfig:width-samples}
for the simple distribution (\ref{eq:inv-square-law-2}). From the
geometrical arguments in figure~\ref{fig:refract-light-path} one would
intuitively expect that the light distribution observed at the
detection material would be broader. However, the distribution is
biased towards forward angles because $d\vartheta_2/d\vartheta_1$
increases more rapidly at larger $\vartheta_2$ for $n_1>n_2$ and
consequently the photon density decreases with $\vartheta_1$ for larger
distances $r$ (refer to figure~\ref{subfig:photon-density}).
Figures~\ref{subfig:depth-fwhm-rel} and \ref{subfig:lightdist_fwhm-ratio}
show the behavior of the widths of both distributions for different
IDs (and also DOIs). One can see that the FWHM of the
distribution~\ref{eq:inv-square-law-2} depends linearly on the ID and
that the detected width varies only slightly at around 80 \% of the original
width. For completeness, the fraction of detected light as a function
of the critical angle $\vartheta_c$ for total reflection is shown in
figure~\ref{subfig:accepted-light}. Since the energy resolution of
scintillation detectors depends strongly on the light collection efficiency,
a good matching of the refraction indices $n_1$ and $n_2$ has to be
achieved. A low light collection efficiency leads to larger
statistical errors when the scintillation light is converted into
electrical signals and consequently to a lower signal-to-noise ratio (SNR).
This is shown in figure~\ref{subfig:SNR-dependence} for a hypothetical
distribution of 2000 detected photoelectrons.

{\setlength{\captionwidth}{0.8\textwidth}
\begin{figure}[!tp]
  \centering
  \subfigure[][Real detected intensity distribution (solid line) for
  $\kappa=n_\tincaps{LSO}/n_\tincaps{BS}$ compared to an
  inverse square distribution without jump in the  refractive index
  (dashed line).]{\label{subfig:width-samples}%
    \psfrag{i}{\hspace*{-1.5em}$F$ [a.u.]}
    \psfrag{x}{\hspace*{-1.5em}$r$ [mm]}
    \includegraphics[width=0.485\textwidth]{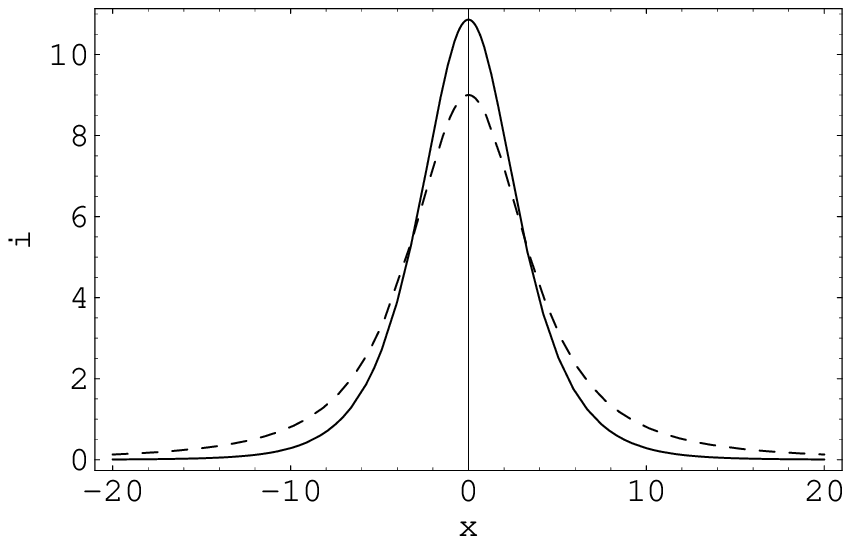}
  }
  \subfigure[][$\mathit{FWHM}$ of both sample distributions as a function of the
  interaction distance $d$.]{\label{subfig:depth-fwhm-rel}%
    \psfrag{fwhm}{\hspace*{-2.5em}$\mathit{FWHM}$ [mm]}
    \psfrag{d}{\hspace*{-1.2em}$d$ [mm]}
    \includegraphics[width=0.485\textwidth]{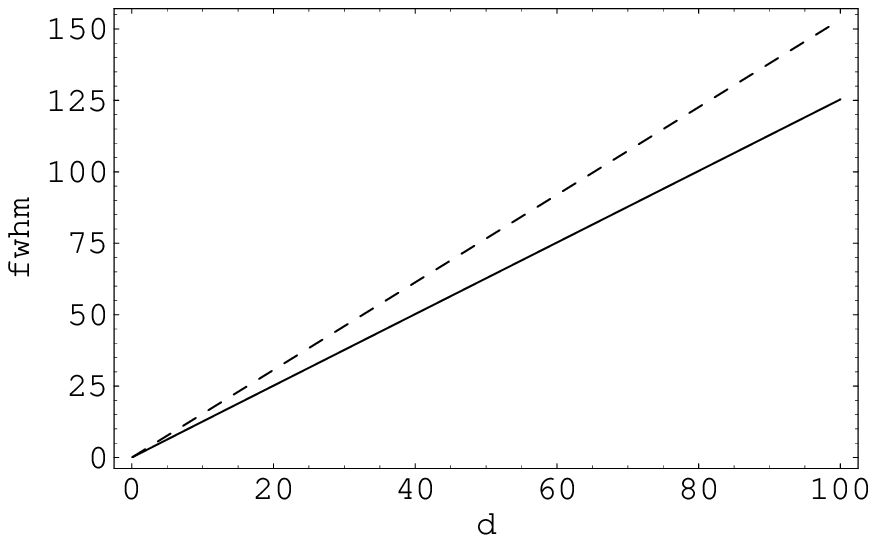}
  }\\\vspace*{-1eX}
  \subfigure[][Ratio of $\mathit{FWHMs}$ of both sample distributions
  in \%.]{\label{subfig:lightdist_fwhm-ratio}%
    \psfrag{r}{\hspace*{-2em}ratio [\%]}
    \psfrag{d}{\hspace*{-1.5em}$z_0$ [mm]}
    \includegraphics[width=0.485\textwidth]{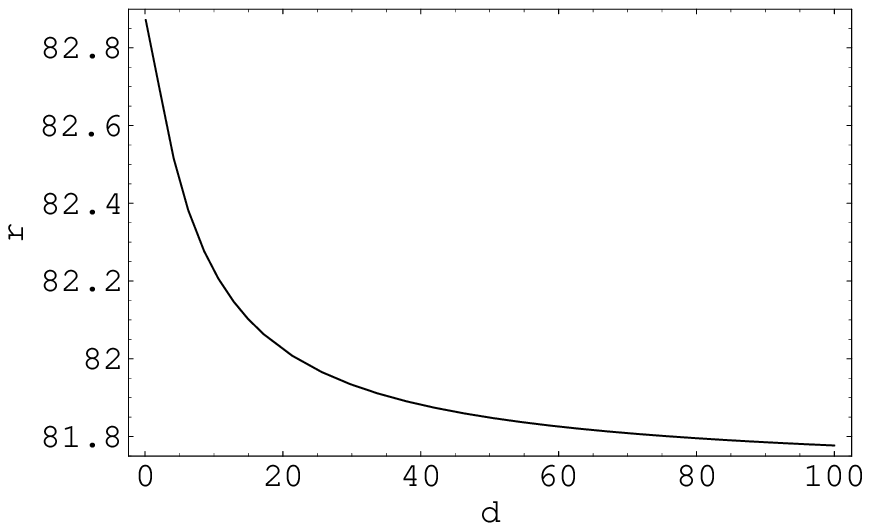}
  }
  \subfigure[][Normalized refracted photon density as a function of the
  refracted angle and for $\kappa\in\{1,1.0,1.1,\ldots,1.4\}$. The
  solid line corresponds to $\kappa=1$ and the dot-dashed line to $\kappa=1.4$.]{%
    \label{subfig:photon-density}%
    \psfrag{normalizeddens}{\hspace*{-1em}$d\vartheta_2/d\vartheta_1$ (normalized)}
    \psfrag{refractedt}{$\vartheta_2$ [degree]}
    \includegraphics[width=0.485\textwidth]{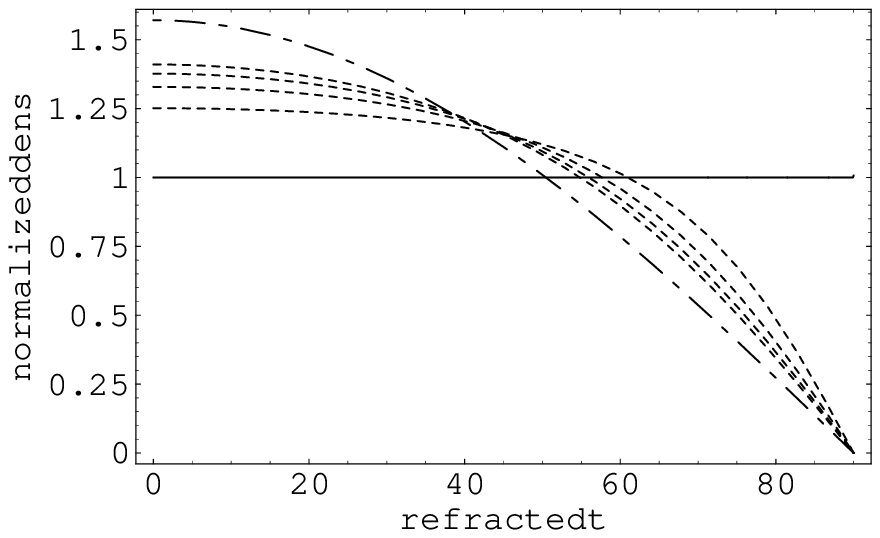}
  }\\\vspace*{-1eX}
  \subfigure[][Light fraction that is transmitted from the scintillator
  through the entrance window to the photocathode as a function of the
  critical angle $\vartheta_c$.]{\label{subfig:accepted-light}%
    \psfrag{t}{\hspace*{-2em}$\vartheta_c$ [degree]}
    \psfrag{F}{$\Phi$}
    \includegraphics[width=0.485\textwidth]{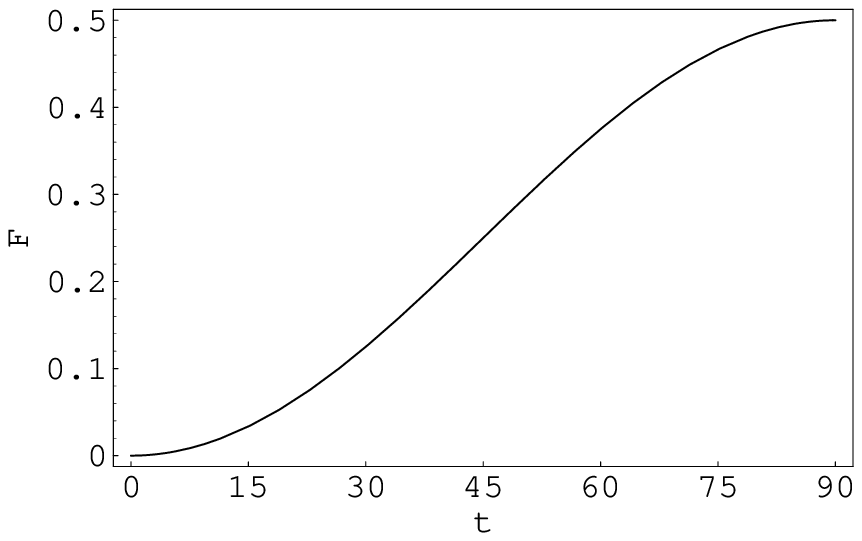}
  }
  \subfigure[][Variation of the signal-to-noise ratio with the
  critical angle $\vartheta_c$ for a supposed distribution of 2000
  detected photoelectrons and Poisson-like statistical errors.]{%
    \label{subfig:SNR-dependence}%
    \psfrag{t}{\hspace*{-2em}$\vartheta_c$ [degree]}
    \psfrag{SNR}{SNR}
    \includegraphics[width=0.485\textwidth]{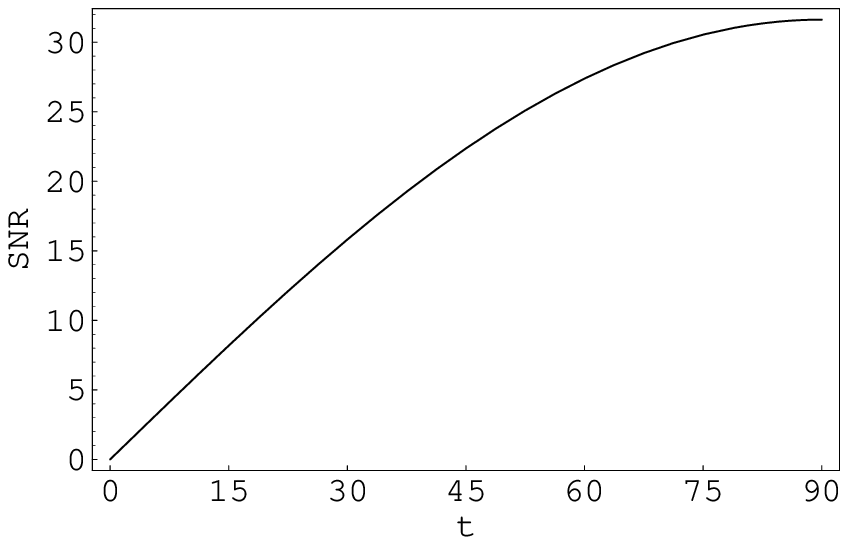}
  }
  \caption[Impact of different refractive indices of
  detector-window/scintillator on signal distributions]{Signal
    distributions for a detector with and without a jump in the
    refractive index. Not only is the overall collected amount of
    scintillation light affected, but one also
    observes an important change in the widths of the signal
    distribution when two optical media of different refractive
    indices are used.}
  \label{fig:width-light-dists}
\end{figure}
}

\subsubsection{Fresnel Transmission}
\label{sec:fresnel-trans}

Apart from the refraction of the scintillation light path, 
reflection at the interface even for angles smaller than the
critical angle $\vartheta_c$ are observed. Actually, Snell's law only
predicts the
relationship between the angles of the incoming and the refracted
light-path. The fractions of the intensities of incident light that is
reflected from the interface and  incident light that is transmitted
to the second medium is described by the Fresnel coefficients for
reflection $\mathcal{R}$ and transmission $\mathcal{T}$.
\begin{gather}
  \label{eq:fresnel-transmit-s-wave}
  \mathcal{T}_s(\vartheta_1,\vartheta_2)  =  1-\mathcal{R}_s(\vartheta_1,\vartheta_2) = 1-\left[\frac{\sin\left(\vartheta_1-\vartheta_2\right)}{\sin\left(\vartheta_1+\vartheta_2\right)}\right]^2\\
  \label{eq:fresnel-transmit-p-wave}
  \mathcal{T}_p(\vartheta_1,\vartheta_2)  =  1-\mathcal{R}_p(\vartheta_1,\vartheta_2) = 1-\left[\frac{\tan\left(\vartheta_1-\vartheta_2\right)}{\tan\left(\vartheta_1+\vartheta_2\right)}\right]^2
\end{gather}
Here the subscripts $s$ and $p$ specify the polarization of the
incoming light. The light is called s-polarized if its electric field
vector is normal to the plane spanned by the incoming light path and
its normal projection onto the interface. If the electric field vector lies
completely within this plane, the light is called p-polarized. 
Alternatively, these polarizations are often called transverse
magnetic mode (TM) and transverse electric mode (TE) respectively. 
Scintillation light is not expected to be polarized along a
specific direction but is supposed to be completely unpolarized.
One therefore has to average over all possible polarizations 
thereby obtaining the fraction of light transmitted to the
photodetectors window
\begin{equation}
  \label{eq:fresnel-transmit}
  \mathcal{T}(\vartheta_1,\vartheta_2)  =  1-\frac{1}{2}\mathcal{R}_p(\vartheta_1,\vartheta_2)-\frac{1}{2}\mathcal{R}_s(\vartheta_1,\vartheta_2)
  = 1-\frac{1}{2}\left[\frac{\sin\left(\vartheta_1-\vartheta_2\right)}{\sin\left(\vartheta_1+\vartheta_2\right)}\right]^2
-\frac{1}{2}\left[\frac{\tan\left(\vartheta_1-\vartheta_2\right)}{\tan\left(\vartheta_1+\vartheta_2\right)}\right]^2,
\end{equation}
where the refracted angle can be expressed as a function of $\mathbf{r}_\mathrm{c}^\mathit{virt}$
\begin{equation}
  \label{eq:refracted-angle}
  \vartheta_2=\tan^{-1}\left(\frac{\sqrt{(x-x_0)^2+(y-y_0)^2}}{z_0^\mathit{virt}+t}\right)
\end{equation}
and the incidence angle $\vartheta_1$ can be obtained by virtue of 
Snell's law~(\ref{eq:snells-law}), once the refracted angle  
$\vartheta_2$ is known.

\subsection{Angular Sensitivity of the Photocathode}
\label{sec:angular-sens}

The physical effects considered so far, {\em e.g.}\ inverse square law,
cosine law and optical transmission into the entrance window, are
 common to many different scintillation detector configurations. Also, the
 photodetector itself has a characteristic influence on the
detected signal distribution because it is not an ideal device. Therefore,
the type of photodetector used has to be specified at this point. For
this work, a position-sensitive photomultiplier tube has been chosen
as photodetector. A photomultiplier converts
light into an electrical signal by photoemission of electrons from the photocathode.
There are two main kinds of photocathodes: semi-transparent cathodes
that are evaporated onto the inner side of the input window and opaque
cathodes that are deposited on a metal electrode inside the tube
(Flyckt {\em et al.}\ \cite{Flyckt:2002}, Kume, \cite{Kume:1994}).
Position-sensitive photomultipliers mostly use semi-transparent
photocathodes because then the cathode can be very large and the entrance
window on which it is deposited can be flat or curved. In this case,
the photoelectrons are emitted from the side opposite the incident
light.

\begin{figure}[!t]
  \centering
  \psfrag{t1}{$\vartheta_2$}
  \psfrag{t2}{$\vartheta_3$}
  \psfrag{entrancewindow}{{\it entrance window}}
  \psfrag{photocathode}{{\it photocathode}}
  \psfrag{vacuum}{{\it vacuum}}
  \psfrag{multiple}{\renewcommand{\baselinestretch}{0.7}\parbox{5em}{\it multiple\\internal\\reflections}}
  \psfrag{n1}{$n_2$}
  \psfrag{n2}{$n_3$}
  \psfrag{n3}{$n_4$}
  \psfrag{in}{\rotatebox{45}{{\it incoming light}}}
  \psfrag{tr}{\rotatebox{49}{\hspace*{-1.7em}{\it transmitted light}}}
  \psfrag{re}{\rotatebox{-47}{{\it reflected light}}}
  \includegraphics[width=0.7\textwidth]{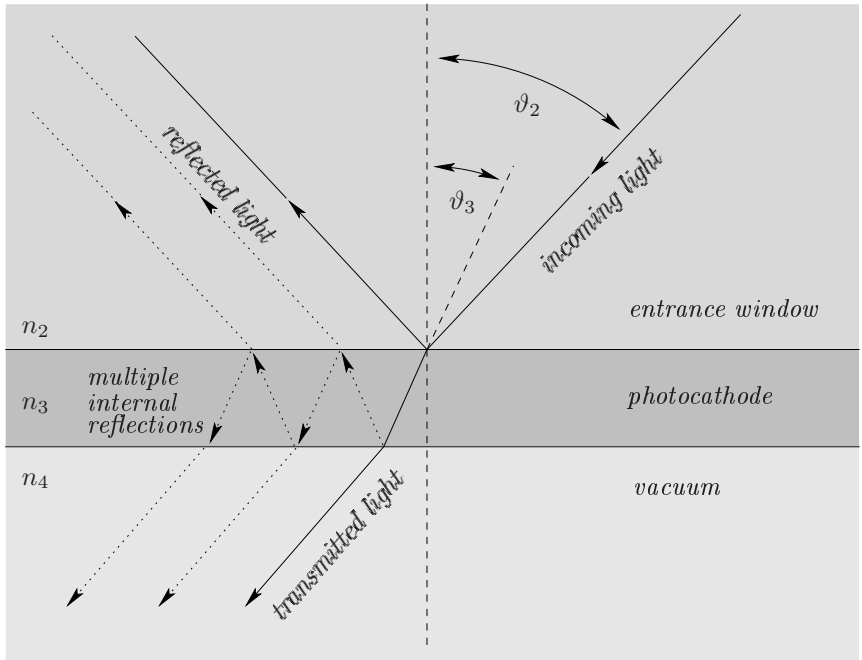}
  \caption[Light transmission and reflection at the
  photocathode]{Light transmission and reflection at the photocathode.
When the scintillation light moves from the entrance window into the
photocathode, only a fraction of this light transfers its energy to
the electrons of the material. The remaining light is either reflected or
transmitted without contributing to the photoemission process. The
fact that $n_2<n_3>n_4$ makes multiple internal total reflections
very probable and leads to a significantly increased absorption for
high incidence angles (cf.\ figures~\ref{subfig:ref-and-trans} and
\ref{subfig:absorbtance}).}
  \label{fig:abs-pc}
\end{figure}

Most photocathodes are made of compound semiconductors consisting of
alkali metals with a low work function\footnote{The (photoelectric) work
  function is the minimum energy required to liberate an electron from
  the surface of the particular metal compound.}. Often, the materials
applied are silver-oxygen-cesium ($\mathrm{AgOCs}$),
antimony-cesium ($\mathrm{SbCs}$), Bi-alkali compounds
($\mathrm{SbKCs}$, $\mathrm{SbRbCs}$ and $\mathrm{SbNaCs}$) and the
tri-alkali compound $\mathrm{SbNa_2KCs}$, besides the less frequent
applied {\em solar-blind} types ($\mathrm{CsI}$ and $\mathrm{CsTe}$)
and semiconductors of negative electron affinity and extended
near-infrared sensitivity ($\mathrm{GaAs}$, $\mathrm{InGaAs}$ and
$\mathrm{InGaAsP}$) (Flyckt {\em et al.}\ \cite{Flyckt:2002}, Kume,
\cite{Kume:1994}). Each of these photocathodes has its typical
spectral response, which is, however, of less importance
for the present derivation of the signal distribution. The
quantum efficiency for photoemission not only depends on the
wavelength of the incident light but also on the angle of incidence
of the light.
The dependence on this angle is called the angular sensitivity. It is
mainly a consequence of reflections and multiple internal reflections
inside the photocathode (refer to figure~\ref{fig:abs-pc}) and is
therefore well described by a standard application of the Fresnel
equations for reflectance and transmittance of multi-layer systems
(Jones \cite{Jones:1976}, Chyba {\em et al.}\ \cite{Chyba:1988} and
Motta {\em et al.}\ \cite{Motta:2005}).

The photocathode can be treated as a continuous medium with complex
refractive index $n_3$.\footnote{Complex refraction indices are used to
  describe absorptive media.} It is deposited onto the photomultiplier's
entrance window of refractive index $n_2$ and faces the PMTs interior
vacuum on the other side ($n_4=1$). Calculating the reflectance and
transmittance for a light beam incident from the entrance window on
the photocathode one obtains
\begin{gather}
  \label{eq:reflectance}
  \mathcal{R}(\vartheta_2)  = %
  \frac{|z_1|^2e^{2\eta\nu}+|z_2|^2e^{-2\eta\nu}+2|z_1||z_2|\cos(\arg{z_2}-\arg{z_1}+2u\eta)}%
    {e^{2\eta\nu}+|z_1|^2|z_2|^2e^{-2\eta\nu}+2|z_1||z_2|\cos(\arg{z_2}+\arg{z_1}+2u\eta)}\\[1em]  \label{eq:transmittance}
  \mathcal{T}(\vartheta_2)  = %
  \frac{\sqrt{(1-n_2^2\sin^2\vartheta_2)}16n_2^2\cos^2(\vartheta_2)(u^2+v^2)G}%
  {n_2\cos\vartheta_2%
    \left[e^{2\eta\nu}+|z_1|^2|z_2|^2e^{-2\eta\nu}+2|z_1||z_2|\cos(\arg{z_2}+\arg{z_1}+2u\eta)\right]}
\end{gather}
where
\begin{gather}
  \label{eq:reflectance-help}
  z_1  =  \frac{n_2\cos\vartheta_2-u-iv}{n_2\cos\vartheta_2+u+iv},\quad
  z_2  = 
  \frac{u+iv-\sqrt{1-n_2^2\sin^2\vartheta_2}}{\sqrt{1-n_2^2\sin^2\vartheta_2}+u+iv}\\
  G  =  \frac{1}%
  {|n_2\cos\vartheta_2+u+iv|^2|\sqrt{1-n_2^2\sin^2\vartheta_2}+u+iv|^2}
\end{gather}
for s-polarized light (TM waves) and 
\begin{gather}
  \label{eq:reflectance-help-2}
  z_1  =  \frac{n_3^2\cos\vartheta_2-n_2(u+iv)}{n_3^2\cos\vartheta_2+n_2(u+iv)},\quad
  z_2  = 
  \frac{u+iv-n_3^2\sqrt{1-n_2^2\sin^2\vartheta_2}}{n_3^2\sqrt{1-n_2^2\sin^2\vartheta_2}+u+iv}\\
  G  =  \frac{|n_3|^4}%
  {|n_3^2\cos\vartheta_2+n_2(u+iv)|^2|n_3^2\sqrt{1-n_2^2\sin^2\vartheta_2}+u+iv|^2}
\end{gather}
for p-polarized light (TE waves). In addition, $\eta=2\pi
t_\tincaps{PC}/\lambda_\tincaps{S}$ and 
\begin{gather}
  \label{eq:art-help}
  u+iv  =  n_3\cos\vartheta_2\\
  n_2\sin\vartheta_2  =  n_3\sin\vartheta_3
\end{gather}
was used, where $t_\tincaps{PC}$ is the thickness of the photocathode
and $\lambda_\tincaps{S}$ the vacuum wavelength of the incident
scintillation light. Equations~(\ref{eq:reflectance}) and
(\ref{eq:transmittance}) have been derived and verified by various research
groups (Jones \cite{Jones:1976}, Chyba {\em et al.}\ \cite{Chyba:1988} and
Motta {\em et al.}\ \cite{Motta:2005}) and agreement with
experimental data was found. 

\begin{figure}[t]
  \centering
  \subfigure[][Reflectance and transmittance for the different
  polarizations: solid line: TM-reflectance; dashed line:
  TE-reflectance; dotted line: TM-transmittance and dash-dotted line:
  TE-transmittance.]{\label{subfig:ref-and-trans}%
    \psfrag{angle}{\hspace*{-1em}$\vartheta_2$ [degree]}
    \psfrag{reflectivtrans}{$\mathcal{R}(\vartheta_2)$ \& $\mathcal{T}(\vartheta_2)$}
    \includegraphics[width=0.49\textwidth]{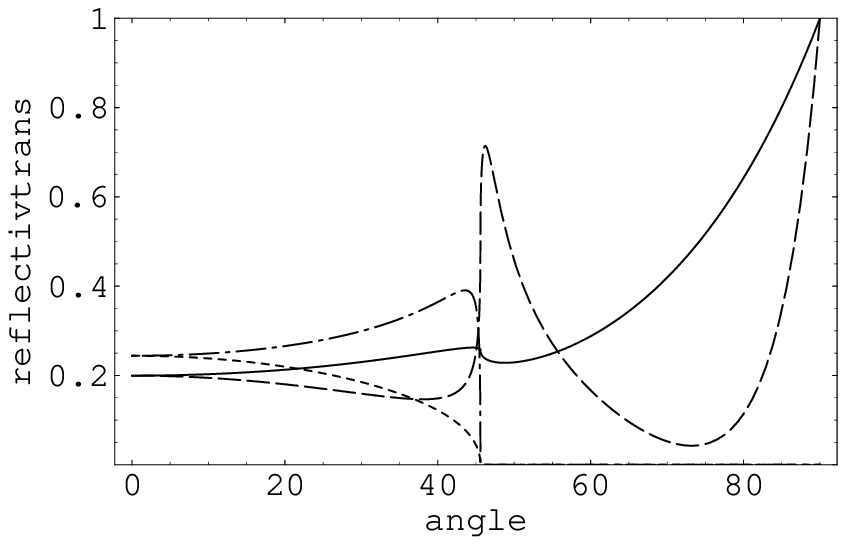}}
  \subfigure[][Absorbance for unpolarized light (solid line), for
  TM-waves (dash-dotted line) and TE-waves (dashed line).]{\label{subfig:absorbtance}%
    \psfrag{angle}{\hspace*{-1em}$\vartheta_2$ [degree]}
    \psfrag{absorbtivity}{\hspace*{2.3em}$\mathcal{A}(\vartheta_2)$}
    \includegraphics[width=0.49\textwidth]{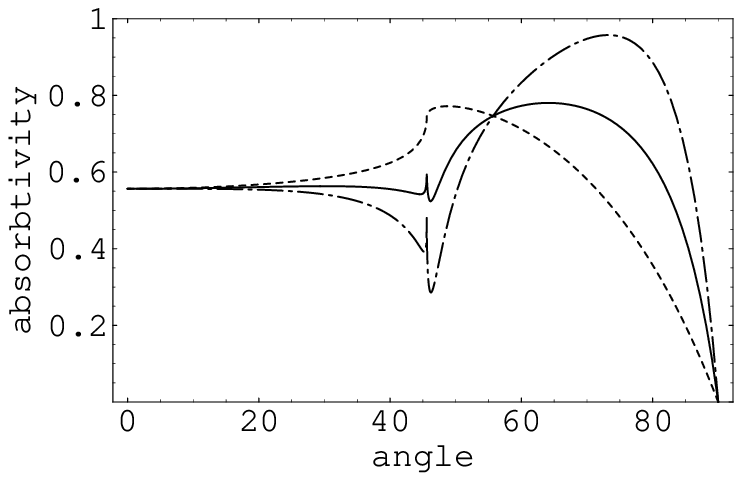}}
  \caption[Reflectance, transmittance and absorbance of the
  photocathode]{Reflectance, transmittance and absorbance of the
    photocathode for both polarizations assuming the following
    parameter values: $\mathrm{n_2=1.4}$, $\mathrm{n_3=2.7+1.5i}$,
    $\mathrm{\lambda_\tincaps{S}=420\,nm}$ and $\mathrm{t_\tincaps{PC}=20\,nm}$. The strong
    variation of all coefficients at $\mathrm{\vartheta_2\approx 46^\circ}$
    are due to total reflection at the photocathode-vacuum interface.}
  \label{fig:pc-absorbtivity}
\end{figure}

Neither the reflectance (\ref{eq:reflectance}) nor the transmittance 
(\ref{eq:transmittance}) are formulae of direct interest for the considered
model for the signal distribution. Except for spurious reflections at
the PMT's dynode system and housing, and photoemissions at the first
dynodes, the transmitted light does not contribute to the signal
distribution. The reflected flux will also be neglected since we
suppose our detector to have absorbing borders and therefore the
major part will not be detectable. The scintillation light that will neither
be reflected nor transmitted must have been absorbed by the
photocathode and one has
\begin{equation}
  \label{eq:pc-absorbtivity}
  \mathcal{A}(\vartheta_2)=1-\mathcal{T}(\vartheta_2)-\mathcal{R}(\vartheta_2).
\end{equation}
Since the light is supposed to be completely unpolarized, one has to
average equation~(\ref{eq:pc-absorbtivity}) over all possible polarizations. 
Figure~\ref{fig:pc-absorbtivity} shows a typical photomultiplier absorbance
(\ref{eq:pc-absorbtivity}) and the different contributions
for TE and TM waves.

\subsection{Background Light}
\label{sec:background-light}

Besides the scintillation light that reaches the photocathode
directly, there are also photons that are scattered or reflected.
The contribution of the scattered photons is expected to be very
small for reasons explained in section~\ref{sec:included-contribs}. 
However, the overall contribution of light reflected at the absorbing
borders can become very important for different reasons. First, it is impossible to
achieve ideal absorbing surfaces. Residual reflection is expected
from all absorbing surfaces of the scintillation crystal. The total
area of these surfaces is always larger than the area that is
coupled to the photodetector. Secondly, if total reflection occurs at the
entrance window, the ratio of reflected light to direct light even becomes more biased towards higher
fractions of reflected light. Apart from this, we have Fresnel
reflections at all optical interfaces.

A consequent inclusion of all possible contributions produced by
 reflection phenomena is clearly impractical. Instead, one has to look for
a simple, computable model that satisfactorily describes the most
important functional dependences of the distribution caused by
reflections. Actually, an exact model is not even required because the
distribution of reflected scintillation light is already a first
order correction to the model. The following approximations are made:
\begin{itemize}
\item Two or more successive reflections are neglected since they will be very
  improbable for black surfaces. 
\item The black surfaces are supposed to exhibit ideal Lambertian
  reflectance. This assumption can be made since the crystal is only
  polished at the side that is coupled to the photodetector. All other
  surfaces
  are fine-ground and covered with a black epoxy resin composite of a
  very low reflection coefficient.
\item Light rays that undergo first a total reflection on the
  interface between the scintillator and the entrance window, and are
  subsequently diffusely reflected at one of the black-painted surfaces,
  are also considered because they are of the same intensity as light
  rays directly reflected at the black sides.
\item Fresnel reflected light from the different optical
  interfaces is not included. This would be already a second-order
  correction.
\item Likewise, corrections discussed in
  sections~\ref{sec:exp-attenuation} to \ref{sec:angular-sens} are not
  applied. 
\item It was shown in section~\ref{sec:refrac-of-light-path} that the
  refraction of the light leads to a narrower detected distribution
  when passing into the entrance window. We define $t_\mathit{eff}$ as
  the effective window thickness that leads to the same average
  distribution width as the true refracted light distribution for the
  considered range of interaction distances. $t_\mathit{eff}$ is used
  to approximate the effect of refraction of the background light.
\end{itemize}

\begin{figure}[!t]
  \centering
  \subfigure[][Diffuse reflection at black coating opposite the
  entrance window.]{\label{subfig:upper-reflect}
    \psfrag{ro}{$\mathbf{r}_\mathrm{c}$}
    \psfrag{r}{$\mathbf{r}$}
    \psfrag{z=a}{\raisebox{-4em}{$z=0$}}
    \psfrag{z=T}{\raisebox{4.4em}{$z=T+t_\mathit{eff}$}}
    \psfrag{gray}{$\gamma$-ray}
    \psfrag{diffuse}{\parbox{4em}{{\scriptsize
          diffuse\vspace*{-1.6ex}\\reflection}}}
    \psfrag{rR}{$\mathbf{r}_\tincaps{R}$}
    \psfrag{t1}{$\vartheta_\tincaps{RE}$}
    \psfrag{t2}{$\vartheta_\tincaps{D}$}
  \includegraphics[width=0.48\textwidth]{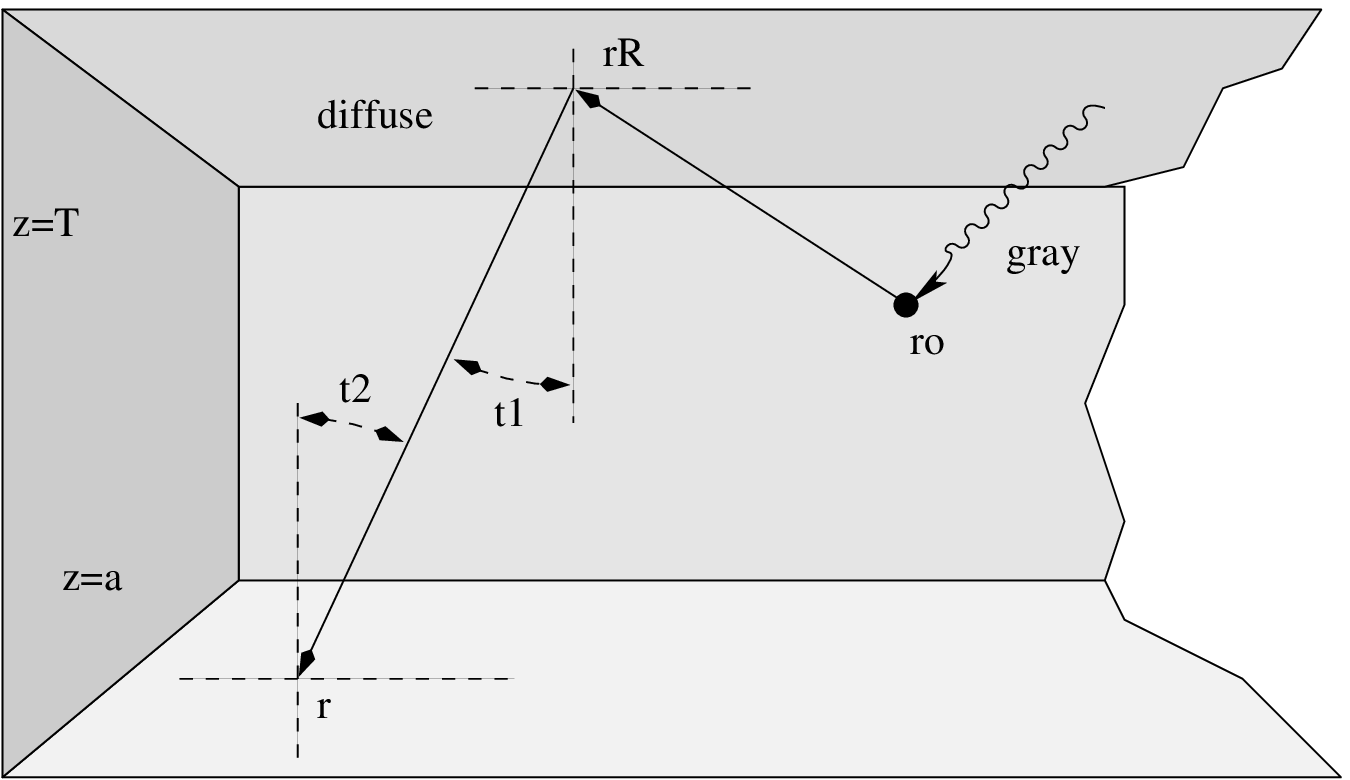}}
  \subfigure[][Total reflection at the scintillator-window interface
  and subsequent diffuse reflection at black coating opposite the
  entrance window.]{\label{subfig:total-upper-reflect}
    \psfrag{ro}{$\mathbf{r}_\mathrm{c}$}
    \psfrag{ro1}{${\mathbf{r}'}_{\!0}$}
    \psfrag{r}{$\mathbf{r}$}
    \psfrag{z=a}{\raisebox{-4em}{$z=0$}}
    \psfrag{z=T}{\raisebox{4.4em}{$z=T+t_\mathit{eff}$}}
    \psfrag{gray}{$\gamma$-ray}
    \psfrag{diffuse}{\parbox{4em}{{\scriptsize
          diffuse\vspace*{-1.6ex}\\reflection}}}
    \psfrag{total}{\parbox{4em}{{\scriptsize
          total\vspace*{-1.6ex}\\reflection}}}
    \psfrag{rR}{$\mathbf{r}_\tincaps{R}$}
    \psfrag{t1}{$\vartheta_\tincaps{RE}$}
    \psfrag{t2}{$\vartheta_\tincaps{D}$}
  \includegraphics[width=0.48\textwidth]{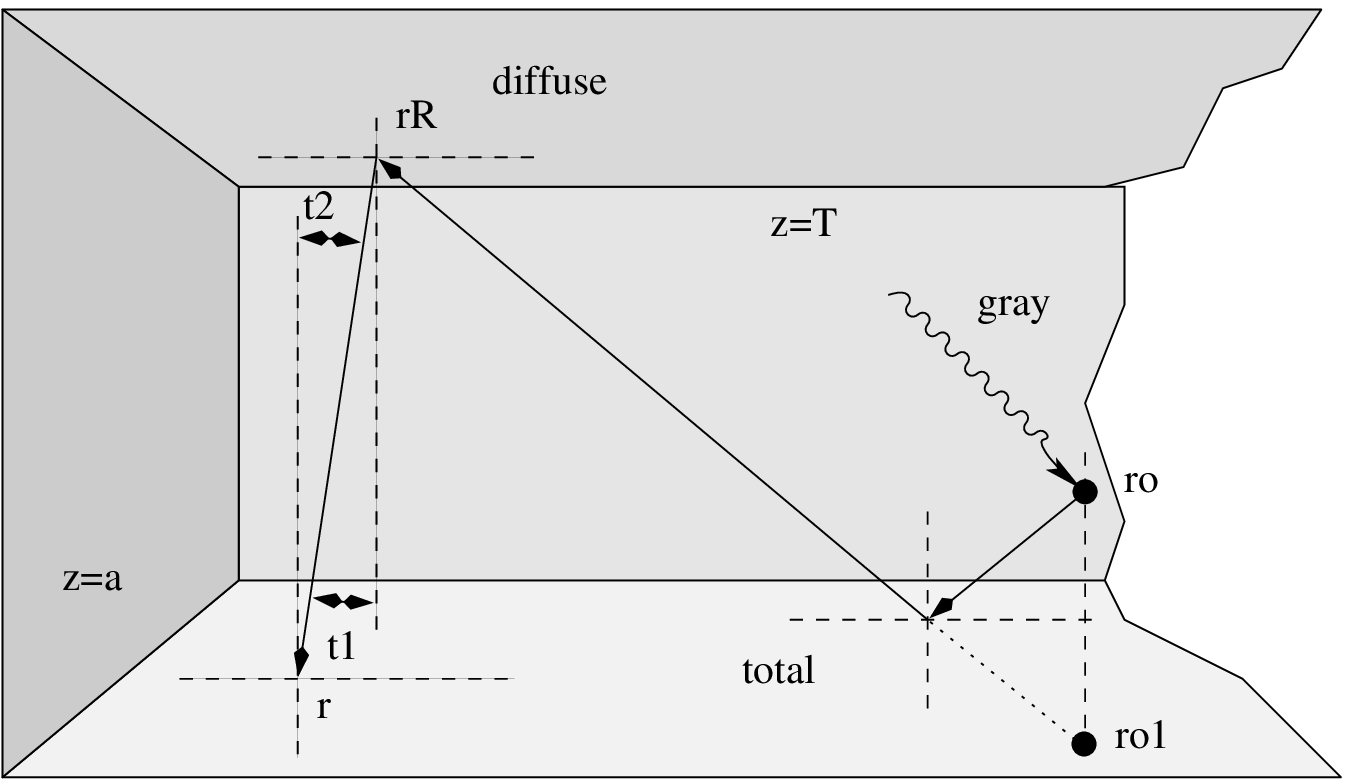}}\\
  \subfigure[][Diffuse reflection at black coatings normal to the
  entrance window.]{\label{subfig:border-reflect}
    \psfrag{ro}{$\mathbf{r}_\mathrm{c}$}
    \psfrag{r}{$\mathbf{r}$}
    \psfrag{z=a}{$z=0$}
    \psfrag{z=T}{$z=T+t_\mathit{eff}$}
    \psfrag{gray}{$\gamma$-ray}
    \psfrag{diffuse}{\rotatebox{90}{\parbox{4em}{{\scriptsize
          diffuse\vspace*{-1.6ex}\\reflection}}}}
    \psfrag{rR}{$\mathbf{r}_\tincaps{R}$}
    \psfrag{xR}{$x_\tincaps{R}$}
    \psfrag{yR}{$y_\tincaps{R}$}
    \psfrag{zR}{$z_\tincaps{R}$}
    \psfrag{t1}{$\vartheta_\tincaps{RE}$}
    \psfrag{t2}{$\vartheta_\tincaps{D}$}
  \includegraphics[width=0.48\textwidth]{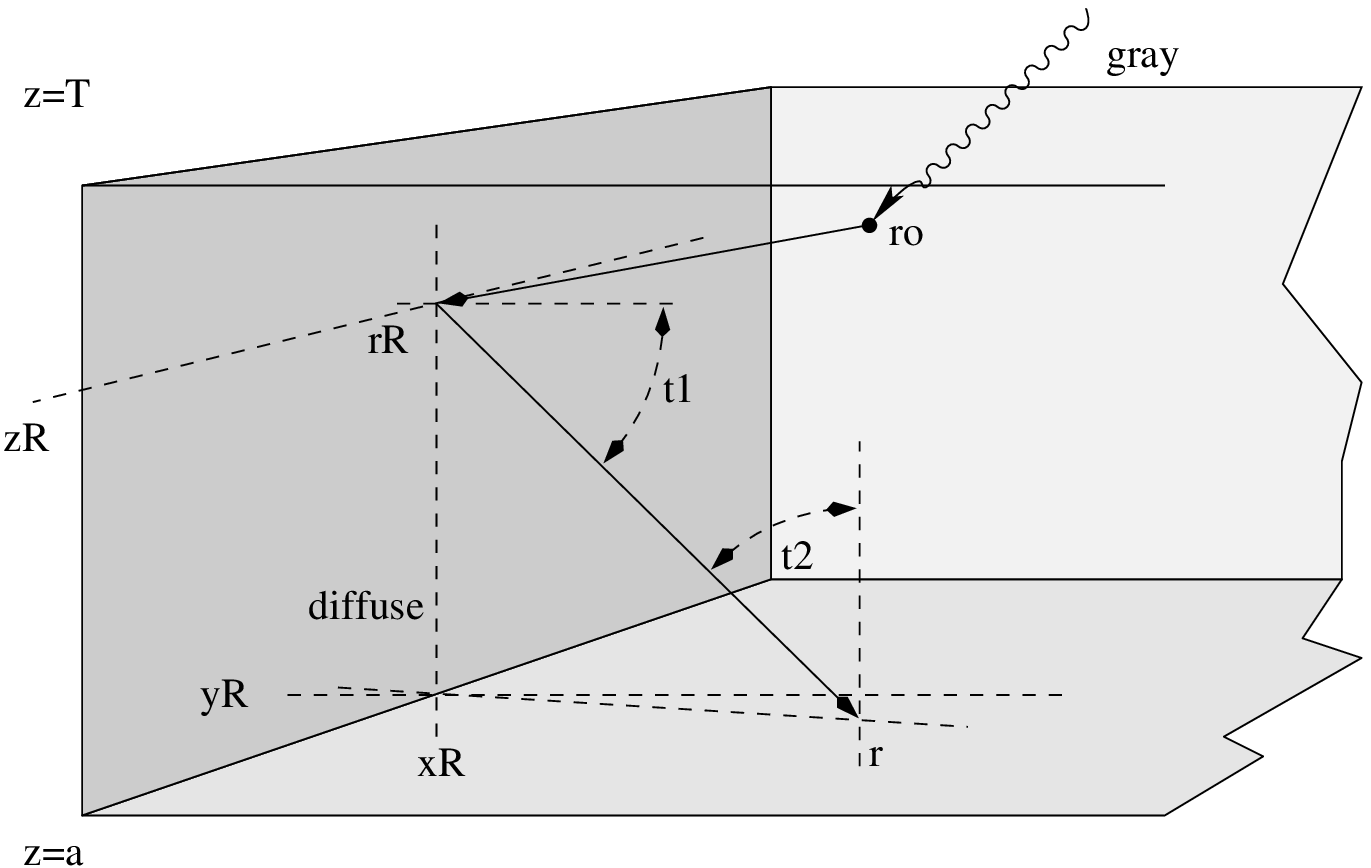}}
  \subfigure[][Total reflection at the scintillator-window interface
  and subsequent diffuse reflection at black coatings normal to the
  entrance window.]{\label{subfig:total-border-reflect}
    \psfrag{ro}{$\mathbf{r}_\mathrm{c}$}
    \psfrag{ro1}{${\mathbf{r}'}_{\!0}$}
    \psfrag{r}{$\mathbf{r}$}
    \psfrag{z=a}{$z=0$}
    \psfrag{z=T}{$z=T+t_\mathit{eff}$}
    \psfrag{gray}{$\gamma$-ray}
    \psfrag{diffuse}{\rotatebox{90}{\parbox{4em}{{\scriptsize
          diffuse\vspace*{-1.6ex}\\reflection}}}}
    \psfrag{total}{\parbox{4em}{{\scriptsize
          total\vspace*{-1.6ex}\\reflection}}}
    \psfrag{rR}{$\mathbf{r}_\tincaps{R}$}
    \psfrag{xR}{$x_\tincaps{R}$}
    \psfrag{yR}{$y_\tincaps{R}$}
    \psfrag{zR}{$z_\tincaps{R}$}
    \psfrag{t1}{$\vartheta_\tincaps{RE}$}
    \psfrag{t2}{$\vartheta_\tincaps{D}$}
  \includegraphics[width=0.48\textwidth]{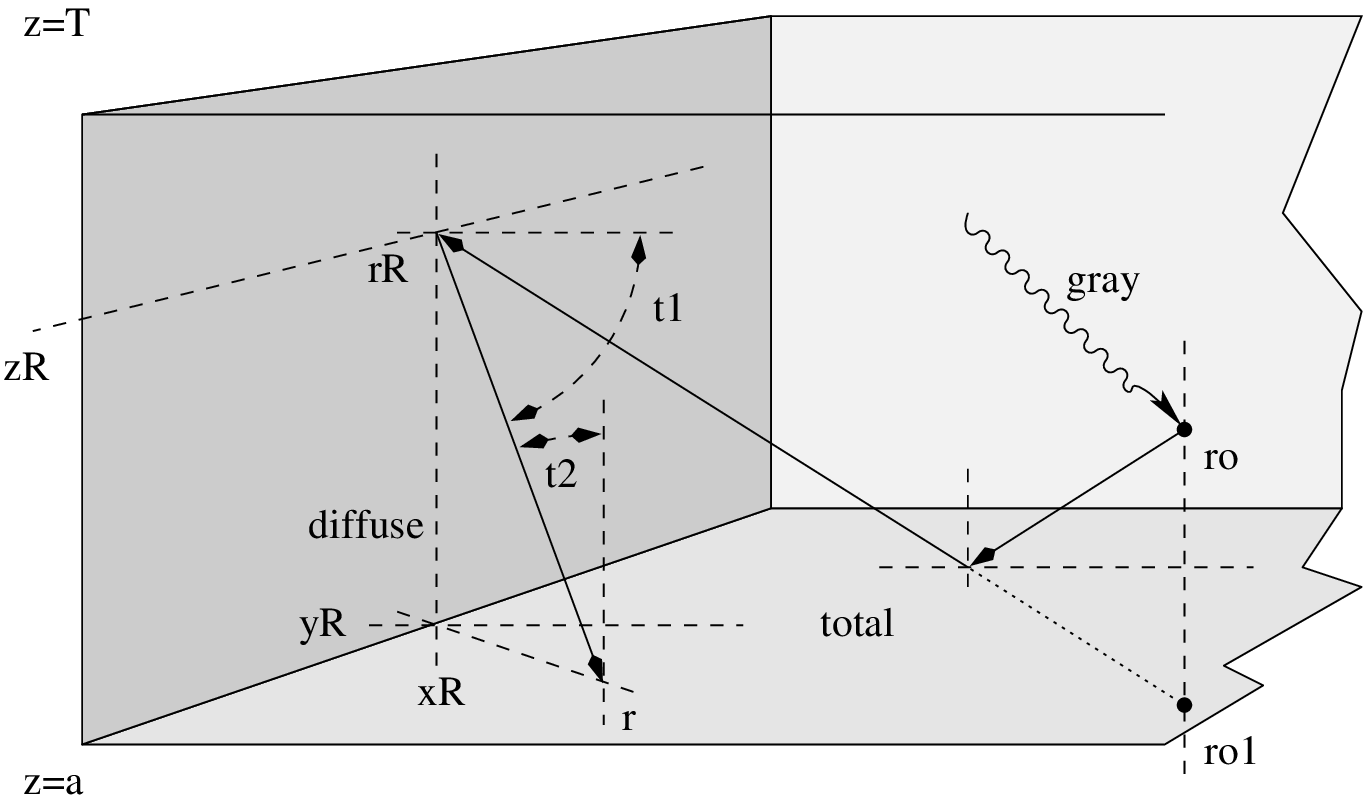}}
  \caption[Contributions to the background distribution caused by
    first-order diffusive reflections]{Contributions to the background distribution caused by
    first-order diffusive reflections.}
\label{fig:reflection-background}
\end{figure}

The application of the inverse square law of
equation~(\ref{eq:inv-square-law-2}) together with 
Lambert's cosine law (Born and Wolf, \cite{Born},
Xu {\em et al.}\ \cite{Xu:2003}) gives the four different contributions
shown in figure~\ref{fig:reflection-background}. One has 
diffuse reflection at the crystal border $\mathbf{r}_\tincaps{R}=(x_\tincaps{R},y_\tincaps{R},T+t_\mathit{eff})$
\begin{equation}
\label{eq:bg-dist-a}
  \mathcal{BG}_A(\mathbf{r},\mathbf{r}_\mathrm{c})=\xi\frac{J_\mathrm{c}}{4\pi}\iint_{\mathcal{S}_\tincaps{R}}
  \frac{(T+t_\mathit{eff})^2}
  {(\mathbf{r}_\tincaps{R}-\mathbf{r}_\mathrm{c})^2(\mathbf{r}-\mathbf{r}_\tincaps{R})^4}dx_\tincaps{R}dy_\tincaps{R},
\end{equation}
total reflection at the scintillator-window interface and subsequent diffuse
reflection at the crystal border
$\mathbf{r}_\tincaps{R}=(y_\tincaps{R},y_\tincaps{R},T+t_\mathit{eff})$
\begin{equation}
\label{eq:bg-dist-b}
  \mathcal{BG}_B(\mathbf{r},\mathbf{r}_\mathrm{c})=\xi\frac{J_\mathrm{c}}{4\pi}\iint_{\mathcal{S}_\tincaps{R}}
  \theta\left\{\cos^{-1}\left(\frac{z_0+T+t_\mathit{eff}}{|\mathbf{r}-\mathbf{r}_\mathrm{c}|}\right)-\vartheta_c\right\}
  \frac{(T+t_\mathit{eff})^2}
  {(\mathbf{r}_\tincaps{R}-\mathbf{r}_\mathrm{c})^2(\mathbf{r}-\mathbf{r}_\tincaps{R})^4}dx_\tincaps{R}dy_\tincaps{R},
\end{equation}
with $\mathbf{r}_\mathrm{c}=(x_c,y_c,-z_c)$;
diffuse reflection at any of the crystal borders
$\mathbf{r}_\tincaps{R}=(\pm L,y_\tincaps{R},z_\tincaps{R})$ and $\mathbf{r}_\tincaps{R}=(x_\tincaps{R},\pm L,z_\tincaps{R})$
\begin{equation}
\label{eq:bg-dist-c}
  \mathcal{BG}_C(\mathbf{r},\mathbf{r}_\mathrm{c})=\xi\frac{J_\mathrm{c}}{4\pi}\iint_{\mathcal{S}_\tincaps{R}}
  \frac{\sqrt{(x-x_\tincaps{R})^2+(y-y_\tincaps{R})^2}(z_\tincaps{R}-z)}
  {(\mathbf{r}_\tincaps{R}-\mathbf{r}_\mathrm{c})^2(\mathbf{r}-\mathbf{r}_\tincaps{R})^4}dv_\tincaps{R}dz_\tincaps{R},
\end{equation}
and total reflection at the scintillator-window interface and
subsequent diffuse reflection at any of the crystal borders
$\mathbf{r}_\tincaps{R}=(\pm L,y_\tincaps{R},z_\tincaps{R})$ and $\mathbf{r}_\tincaps{R}=(x_\tincaps{R},\pm L,z_\tincaps{R})$
\begin{equation}
\label{eq:bg-dist-d}
  \mathcal{BG}_D(\mathbf{r},\mathbf{r}_\mathrm{c})=\xi\frac{J_\mathrm{c}}{4\pi}\iint_{\mathcal{S}_\tincaps{R}}
  \theta\left\{\cos^{-1}\left(\frac{z_\tincaps{R}+z_0+2t}{|\mathbf{r}-\mathbf{r}_\mathrm{c}|}\right)-\vartheta_c\right\}
  \frac{\sqrt{(x-x_\tincaps{R})^2+(y-y_\tincaps{R})^2}(z_\tincaps{R}-z)}
  {(\mathbf{r}_\tincaps{R}-\mathbf{r}_\mathrm{c})^2(\mathbf{r}-\mathbf{r}_\tincaps{R})^4}dv_\tincaps{R}dz_\tincaps{R},
\end{equation}
with $\mathbf{r}_\mathrm{c}=(x_c,y_c,-z_c)$. $\theta\{x\}$ in
equations~(\ref{eq:bg-dist-b}) and (\ref{eq:bg-dist-d}) is the
unit-step distribution and $dv_\tincaps{R}$ in
equations~(\ref{eq:bg-dist-c}) and (\ref{eq:bg-dist-d})
has to be replaced by $dx_\tincaps{R}$ or $dy_\tincaps{R}$
depending on the surface $\mathcal{S}_\tincaps{R}$ over which the
integration is carried out. $T$ is the thickness of the
scintillation crystal and $2L$ its transverse dimension supposing a
square shape. The complete background light
distribution $\mathcal{BG}(\mathbf{r},\mathbf{r}_\mathrm{c})$ is given by the superposition of the contributions
(\ref{eq:bg-dist-a})-(\ref{eq:bg-dist-d}). 

\section{Complete Signal Distribution}
\label{sec:complete-signal-dist}

\begin{figure}[!htp]
  \centering
  \subfigure[][Total light density in arbitrary units for a
photoconversion at $\mathrm{\mathbf{r}_\mathrm{c}=(0,0,0)\,mm}$.]{\label{subfig:example-dist-1}
    \psfrag{x}{\hspace*{-1em}$x$ [mm]}
    \psfrag{y}{$y$ [mm]}
    \psfrag{dJ}{\raisebox{2.6ex}{$J_\mathit{tot}$}}
    \includegraphics[width=0.44\textwidth]{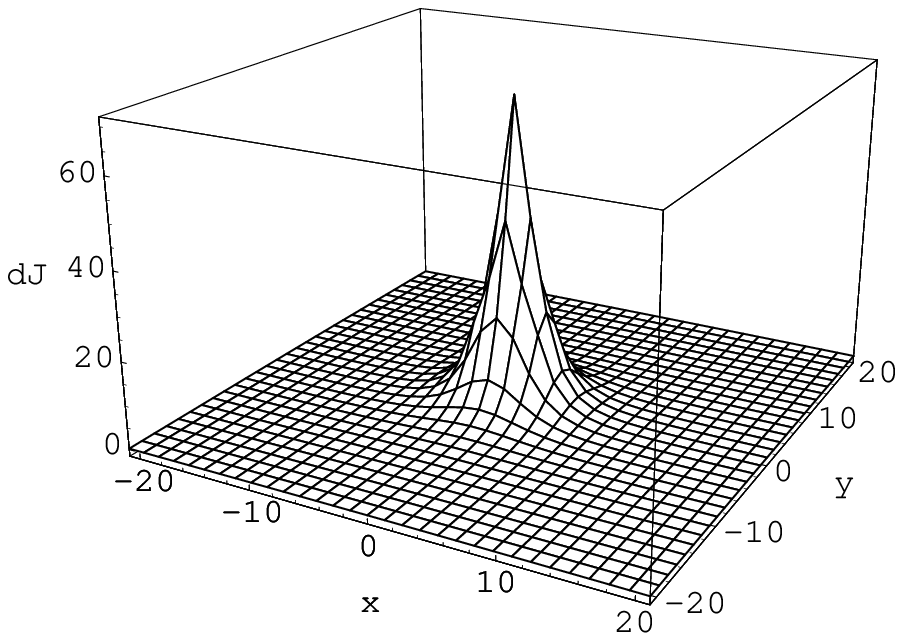}}
  \subfigure[][Background light density in arbitrary units for a
photoconversion at $\mathrm{\mathbf{r}_\mathrm{c}=(0,0,0)\,mm}$.]{\label{subfig:example-bg-1}
    \psfrag{x}{\hspace*{-1em}$x$ [mm]}
    \psfrag{y}{$y$ [mm]}
    \psfrag{dJ}{\raisebox{2ex}{$J_\tincaps{BG}$}}
    \includegraphics[width=0.44\textwidth]{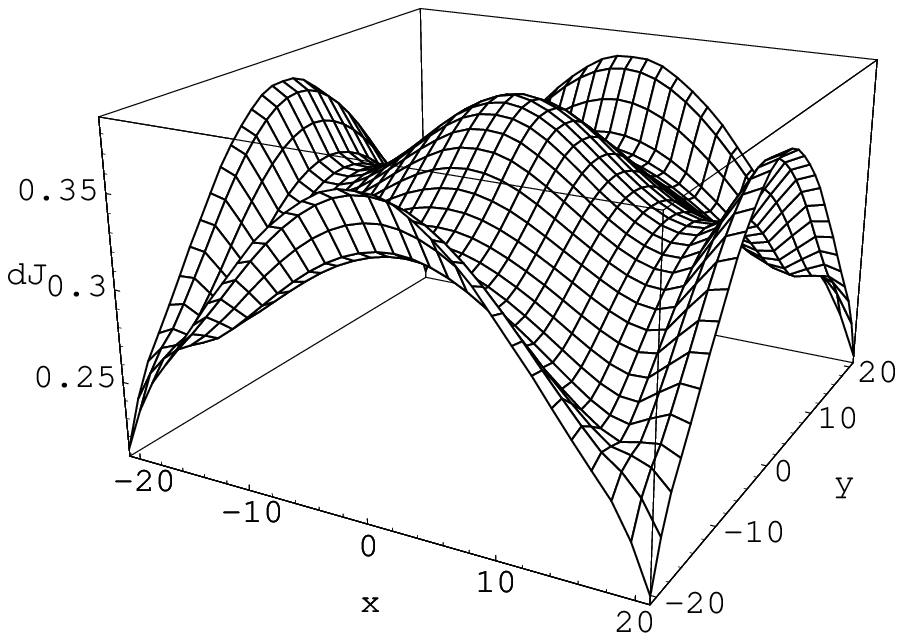}}\\
  \subfigure[][Total light density in arbitrary units for a
photoconversion at $\mathrm{\mathbf{r}_\mathrm{c}=(0,0,10)\,mm}$.]{\label{subfig:example-dist-2}
    \psfrag{x}{\hspace*{-1em}$x$ [mm]}
    \psfrag{y}{$y$ [mm]}
    \psfrag{dJ}{\raisebox{3ex}{$J_\mathit{tot}$}}
    \includegraphics[width=0.44\textwidth]{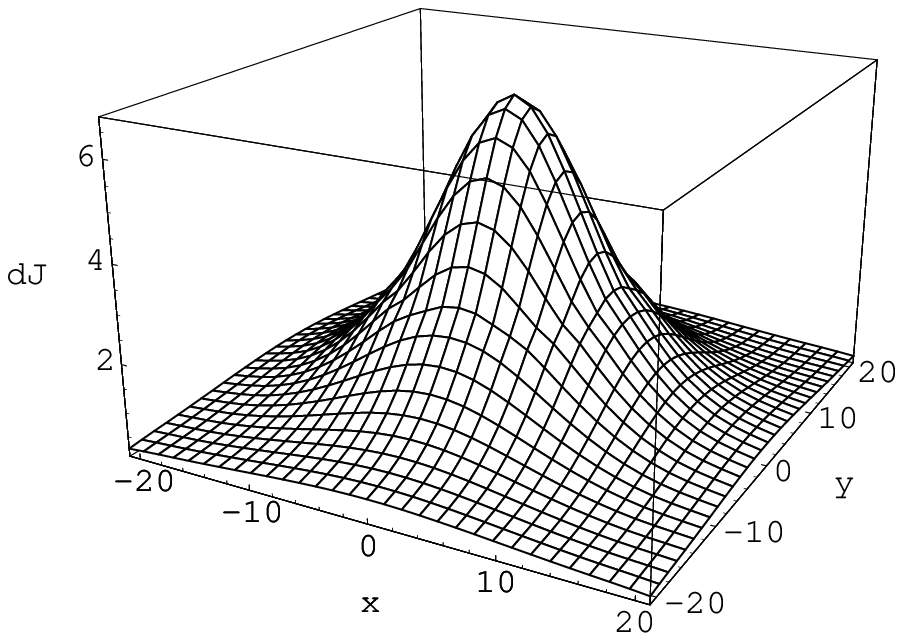}}
  \subfigure[][Background light density in arbitrary units for a
photoconversion at $\mathrm{\mathbf{r}_\mathrm{c}=(0,0,10)\,mm}$.]{\label{subfig:example-bg-2}
    \psfrag{x}{\hspace*{-1em}$x$ [mm]}
    \psfrag{y}{$y$ [mm]}
    \psfrag{dJ}{\raisebox{3ex}{$J_\tincaps{BG}$}}
    \includegraphics[width=0.44\textwidth]{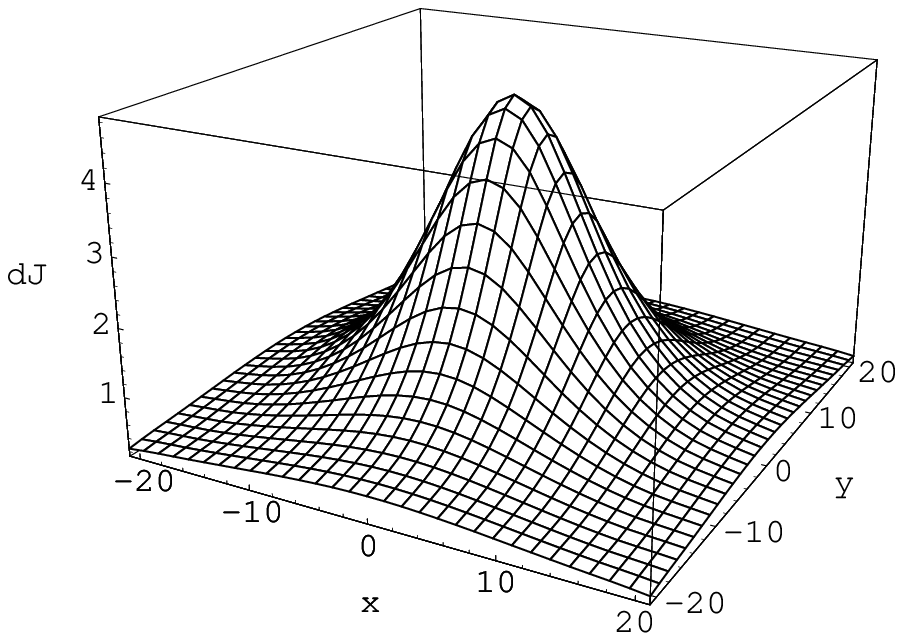}}\\
  \subfigure[][Total light density in arbitrary units for a
photoconversion at $\mathrm{\mathbf{r}_\mathrm{c}=(-16,16,5)\,mm}$.]{\label{subfig:example-dist-3}
    \psfrag{x}{\hspace*{-1em}$x$ [mm]}
    \psfrag{y}{$y$ [mm]}
    \psfrag{dJ}{\raisebox{0.6ex}{$J_\mathit{tot}$}}
    \includegraphics[width=0.44\textwidth]{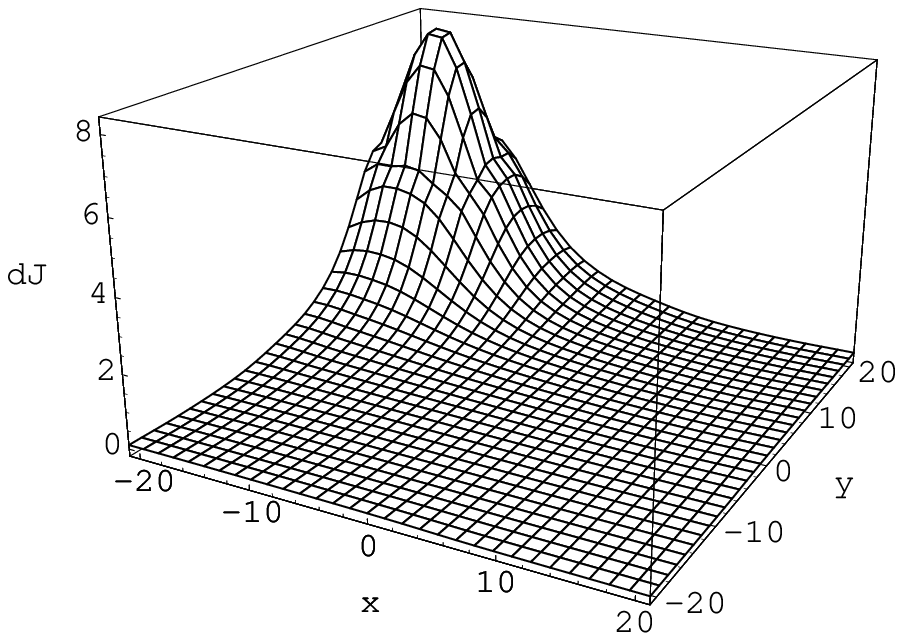}}
  \subfigure[][Background light density in arbitrary units for a
photoconversion at $\mathrm{\mathbf{r}_\mathrm{c}=(-16,16,5)\,mm}$.]{\label{subfig:example-bg-3}
    \psfrag{x}{\hspace*{-1em}$x$ [mm]}
    \psfrag{y}{$y$ [mm]}
    \psfrag{dJ}{\raisebox{1ex}{$J_\tincaps{BG}$}}
    \includegraphics[width=0.44\textwidth]{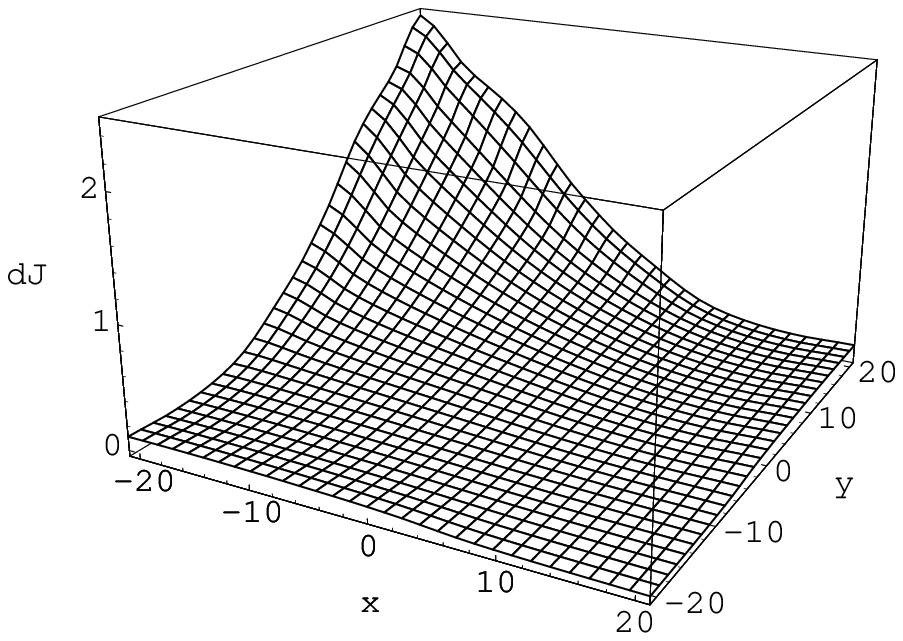}}
  \caption[Example distributions of the total light distribution and
    the assumed background light]{Example distributions of the total light distribution and
    the assumed background light at different possible $\gamma$-ray interaction positions.}
  \label{fig:light-dist-examples}
\end{figure}

\begin{figure}[!t]
  \centering
  \subfigure[Dependence of the combined second moment on the interaction distance
  at positions $\mathrm{(x_c,y_c)=(0,0)\,mm}$ (dashed line), $\mathrm{(x_c,y_c)=(21,0)\,mm}$
  (dot-dashed line) and $\mathrm{(x_c,y_c)=(21,21)\,mm}$ (solid
  line).]{\label{subfig:secmom-d-depth}
    \psfrag{x}{\hspace*{-1em}$z_c$ [mm]}
    \psfrag{y}{[a.u.]}
    \includegraphics[width=0.48\textwidth]{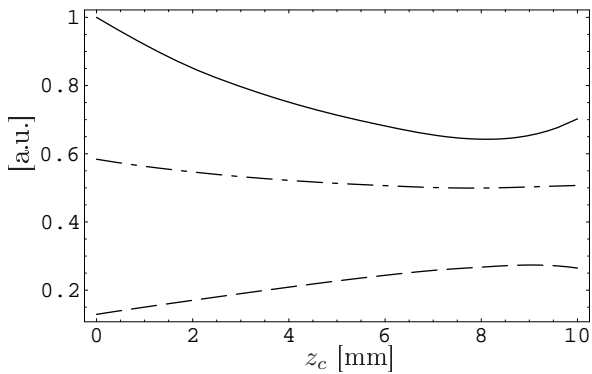}}
  \subfigure[Dependence of the standard deviation on the interaction distance
  at the positions $\mathrm{(x_c,y_c)=(0,0)\,mm}$ (dashed line), $\mathrm{(x_c,y_c)=(21,0)\,mm}$
  (dot-dashed line) and $\mathrm{(x_c,y_c)=(21,21)\,mm}$ (solid
  line).]{\label{subfig:sigma-d-depth}
    \psfrag{x}{\hspace*{-1em}$z_c$ [mm]}
    \psfrag{y}{[a.u.]}
    \includegraphics[width=0.48\textwidth]{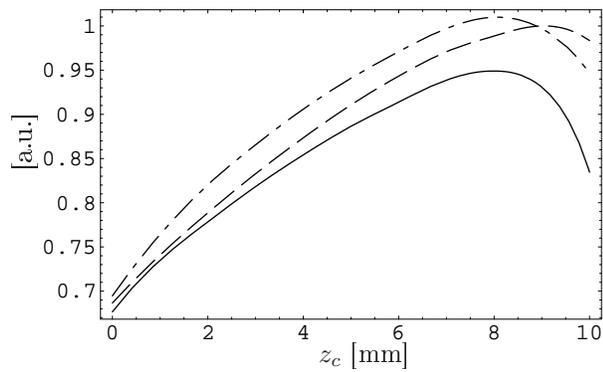}}
  \subfigure[Dependence of the combined second moment on the interaction distance
  at the positions $\mathrm{(x_c,y_c)=(0,0)\,mm}$ (dashed line), $\mathrm{(x_c,y_c)=(21,0)\,mm}$
  (dot-dashed line) and $\mathrm{(x_c,y_c)=(21,21)\,mm}$ (solid
  line) for a model distribution without reflected background.]{\label{subfig:secmom-d-depth-wo-bg}
    \psfrag{x}{\hspace*{-1em}$z_c$ [mm]}
    \psfrag{y}{[a.u.]}
    \includegraphics[width=0.48\textwidth]{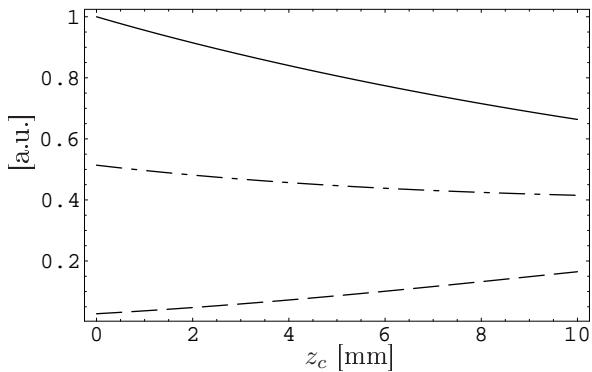}}
  \subfigure[Dependence of the standard deviation on the interaction distance
  at the positions $\mathrm{(x_c,y_c)=(0,0)\,mm}$ (dashed line), $\mathrm{(x_c,y_c)=(21,0)\,mm}$
  (dot-dashed line) and $\mathrm{(x_c,y_c)=(21,21)\,mm}$ (solid
  line) for a model distribution without reflected background.]{\label{subfig:sigma-d-depth-wo-bg}
    \psfrag{x}{\hspace*{-1em}$z_c$ [mm]}
    \psfrag{y}{[a.u.]}
    \includegraphics[width=0.48\textwidth]{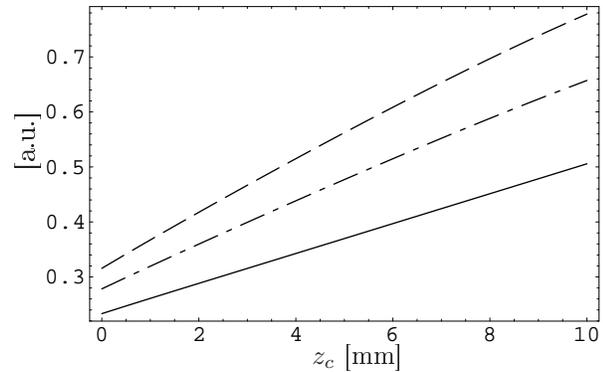}}
  \caption[Depth dependence of the normalized second moment and the standard deviation]{Dependence of the model distributions
    normalized second moment and the standard deviation on the
    interaction distance at three characteristic impact positions.}
  \label{fig:secmom-d-depth}
\end{figure}

The total scintillation light distribution is given as a superposition
 of the background light and the product of the contributions
discussed in
sections~\ref{sec:inv-square-law}-\ref{sec:angular-sens}. 
\begin{equation}
  \label{eq:total-lightdist}
  \mathcal{L}_\mathit{Detector}(\mathbf{r},\mathbf{r}_\mathrm{c}^\mathit{virt})=
  J(\mathbf{r},\mathbf{r}_\mathrm{c}^\mathit{virt})I(\mathbf{r},\mathbf{r}_\mathrm{c}^\mathit{virt})
  \mathcal{T}(\vartheta_2)\mathcal{A}(\vartheta_2)/I_0+\mathcal{BG}(\mathbf{r},\mathbf{r}_\mathrm{c})
\end{equation}
with
\begin{equation}
  \label{eq:virt-r-as-f-of-r}
  \mathbf{r}_\mathrm{c}^\mathit{virt}=\mathbf{r}_\mathrm{c}^\mathit{virt}(\mathbf{r},\mathbf{r}_\mathrm{c}).
\end{equation}
The subscript in (\ref{eq:total-lightdist}) indicates that the
light density does not only depend on the observation point
$\mathbf{r}$ and the $\gamma$-ray photoconversion position
$\mathbf{r}_\mathrm{c}^\mathit{virt}$, but also on the scintillation detector
configuration, {\em e.g.}\ the refraction indices, the entrance window
thickness, the crystals spatial dimensions, etc. Sample distributions
are plotted for different photoconversion positions in
figure~\ref{fig:light-dist-examples} with the following detector
design parameters: $\mathrm{t=2\,mm}$, $\mathrm{n_1=n_\tincaps{LSO}=1.82}$,
$\mathrm{n_2=n_\tincaps{BS}=1.51}$, $\mathrm{n_3=n_\tincaps{PC}=2.7+1.5i}$,
$\mathrm{L=21\,cm}$, $\mathrm{T=10\,mm}$, $\mathrm{\xi=2.5\%}$,
$\mathrm{\lambda=400\,nm}$ and $\mathrm{t_\tincaps{PC}=20\,nm}$. 
Also shown are the contributions of the background light due to
diffuse reflection at the crystal borders. It can be seen that this
contribution becomes important for large interaction distances $z_c$ and
near the crystal borders and edges, while it is very low for small
$z_c$. Also the expected variation of the distributions width with
the interaction distance can be observed.
If the second moment and the standard deviation are computed from the
distribution, the dependence can also be observed in this variables. 
However, both variables also depend on the planar impact
position to greater or lesser extent. 
Figure~\ref{fig:secmom-d-depth} shows the dependence of the standard
deviation and the second moment on the interaction distance for different
transverse photoconversion positions. The graphs shown in the
sub-figures~\ref{subfig:secmom-d-depth}
and \ref{subfig:sigma-d-depth} are obtained from a light distribution
with reflected background as discussed in
section~\ref{sec:background-light}. A strong variation in both
measures can be observed for large interaction distances shown in the
plots ($z_\mathrm{c}\gtrsim7\,\mathrm{mm}$).
This is not expected and not observed for a signal
distribution without any reflections
(sub-figures~\ref{subfig:secmom-d-depth-wo-bg} and
\ref{subfig:sigma-d-depth-wo-bg}) and may be an artefact caused by the
important approximations used for the derivation of the background
light model.

\chapterbib


   \cleardoublepage{}
\chapter{Enhanced Charge Dividing Circuits}
\label{ch:enhanced-charge-dividing-circuits}

 \chapterquote{%
   Character is like a tree and reputation like a shadow. 
   The shadow is what we think of it; the tree is the real thing.%
 }{%
   Abraham Lincoln, $\star$ 1809 -- $\dagger$ 1865
 }

\PARstart{T}{he} first time that the center of gravity algorithm (CoG) was used for
position estimation in nuclear medicine was in 1953, when Hal Anger
developed the first scintillation camera \mycite{Anger}{}{1958} (refer
to figure~\ref{fig:anger-patent} of chapter
\ref{ch:introduction}). As a consequence of its subtle and easy
electronic implementation, it has nowadays become  very  useful
for the position estimation of $\gamma$-ray imaging detectors
even though it introduces systematic errors.
The method is often called {\em centroiding} and the measured position
estimates for the $x$ and $y$ spatial directions are given the name
{\em centroid}. Actually, only in rare cases can one avoid using the
center of gravity algorithm for the analysis of the signal
distribution. High-energy physics experiments show a
dramatic increase in the detector segmentation and therefore a
tremendous amount of data have to be processed. As a consequence,
the use of the center of gravity
algorithm for position determination has become very widespread for
scientific and practical applications (Lauterjung {\em et al.}\
\cite{Lauterjung:1963}, Kuhlmann {\em et al.}\ \cite{Kuhlmann:1966a}, 
Bock {\em et al.}\ \cite{Bock:1966}, McDicken {\em et al.}\ \cite{McDicken:1967},
Doehring {\em et al.}\ \cite{Doehring:1969} and Landi \cite{Landi:2002}).

In this section, the mathematical properties of the
center of gravity
algorithm and especially the consequences that result from its
electronic implementation are reviewed in detail. However, the special
emphasis is put on the analysis of how the existing charge dividing
circuits can be enhanced to measure the second moment in addition
to the centroid and the total charge.

\section{Introduction, Conventions and General Considerations}

\subsection{Statistical Estimates}
\label{ch:stat-estimates}

By analogy with the geometric centroid, which
represents the center of gravity\footnote{Occasionally, the method is
  also called {\em center of mass} algorithm. In an uniform gravitational
  field, the center of mass and the center of gravity are identical.}
of a body, the generalized functional centroid is defined as
\begin{equation}
  \label{eq:func-centroid}
  \langle x\rangle = \frac{\int_\omega x \;\varphi(x) dx}{\int_\omega \varphi(x) dx}, 
\end{equation}
provided that the integrals exist. In equation~(\ref{eq:func-centroid}) 
$\varphi(x)$ is an integrable function and $\omega \subseteq
\Omega \subseteq \mathbb{R}$ is the support within the domain $\Omega$ of $\varphi(x)$. 
If $\varphi(x)$ is replaced by an arbitrary probability density function
$\mathcal{P}(x)$, equation~(\ref{eq:func-centroid}) reduces to 
\begin{equation}
  \label{eq:expec-val}
  \mathcal{E}(x) = \int_\omega x \mathcal{P}(x) dx, 
\end{equation}
with $\int_\omega \mathcal{P}(x) dx = 1$, since $\mathcal{P}(x)$
is a probability density function. $\mathcal{E}(x)$ is called the
{\em mean}, or more generally, the expectation value of $x$ respect to
$\mathcal{P}(x)$.
The mean (\ref{eq:expec-val}) is not to be confused with the {\em
  average probability} given by
\begin{equation}
  \label{eq:average-prob}
  \bar{\mathcal{P}} = \frac{\int_\omega \mathcal{P}(x) dx}{\int_\omega dx}.
\end{equation}
Equation~(\ref{eq:expec-val}) can be
generalized to yield the expectation value $\mathcal{E}[g(x)]$
for an arbitrary function $g(x)$ respect to the probability density
function $\mathcal{P}(x)$:
\begin{equation}
  \label{eq:expec-val-gen}
  \mathcal{E}[g(x)] = \int_\omega g(x) \mathcal{P}(x) dx, 
\end{equation}
whenever the integral is absolutely convergent. The same
generalizations can be made for equation~(\ref{eq:func-centroid}).
Of special statistical interest is the class of monomials $g(x)=x^k$
with $k\in\mathbb{N}$. Plugging in this class into
equation~(\ref{eq:func-centroid}) for the functional centroid, one
obtains
\begin{gather}
  \label{eq:func-moments}
  \mu_k=\frac{\int_\omega x^k\varphi(x)dx}{\int_\omega\varphi(x)
    dx}\quad\mbox{and}\\
  \mu_k'=\frac{\int_\omega (x-\mu_1)^k \varphi(x) dx}{\int_\omega \varphi(x) dx}.
\end{gather}
The sequences ${\mu_k}$ and ${\mu_k'}$ of real or complex numbers are called the
normalized {\em moments} and the normalized {\em central moments} of
$\varphi(x)$ respectively being $\mu_0,\mu_0'\equiv 1$ and $\mu_1'\equiv 0$ by
definition. Special names are given to the lowest order central moments of
probability distribution functions. As already mentioned, the first
moment $\mu_1$ is called the mean. $\mu_2'$ is given the name of {\em
  variance}. It is a measure of the spread or width of
$\mathcal{P}(x)$. The next two higher centered moments are called
skewness and kurtosis respectively. While the skewness is a measure
for the degree of antisymmetry of the function, the kurtosis
characterizes its relative flatness compared to a standard
distribution. Further details can be found in standard textbooks,
{\em e.g.}\ \cite{Lindley,Press:1992}. 
If the moments are not normalized, then $\mu_0$ and $\mu_0'$ give the norm of
$\varphi(x)$ and normally represent physical quantities. It is also possible 
(and straightforward) to define the {\em joint moments} and expectation values for 
multivariate functions $\varphi(\bar{x})$, $g(\bar{x})$ and
$\mathcal{P}(\bar{x})$ defined on $\omega \subseteq \Omega \subseteq
\mathbb{R}^n$:
\begin{gather}
  \label{eq:multivariate-moms-1}
  \mu_{k_1,\ldots,k_n} = \frac{\int_\omega \prod_i^n x_i^{k_i}
    \varphi(\bar{x}) d\bar{x}}{\int_\omega \varphi(\bar{x}) d\bar{x}},\\\label{eq:multivariate-moms-2}
  {\mu'}_{k_1,\ldots,k_n}  = \frac{\int_\omega \prod_i^n (x_i-\mu_{1_i})^{k_i}
    \varphi(\bar{x}) d\bar{x}}{\int_\omega \varphi(\bar{x}) d\bar{x}}\mbox{ and}\\\label{eq:multivariate-moms-3}
  \mathcal{E}[g(\bar{x})]=\int_\omega g(\bar{x}) \mathcal{P}(\bar{x}) d\bar{x}.
\end{gather}

\begin{figure}[!t]
  \centering
  \subfigure[][Median, mode and mean of the non-symmetric example
  distribution function
  $\mathcal{P}_a(x)=\frac{xe^{10-x/2}}{4e^{10}-44}$. In general, the
  three values are different.]{\label{subfig:stat-gen-func}%
    \psfrag{median}{\it median}
    \psfrag{C}{$\bar{\mathcal{P}}$}
    \psfrag{A}{\rotatebox{-20}{\it mode}}
    \psfrag{B}{\rotatebox{-20}{\it mean}}
    \includegraphics[width=0.488\textwidth]{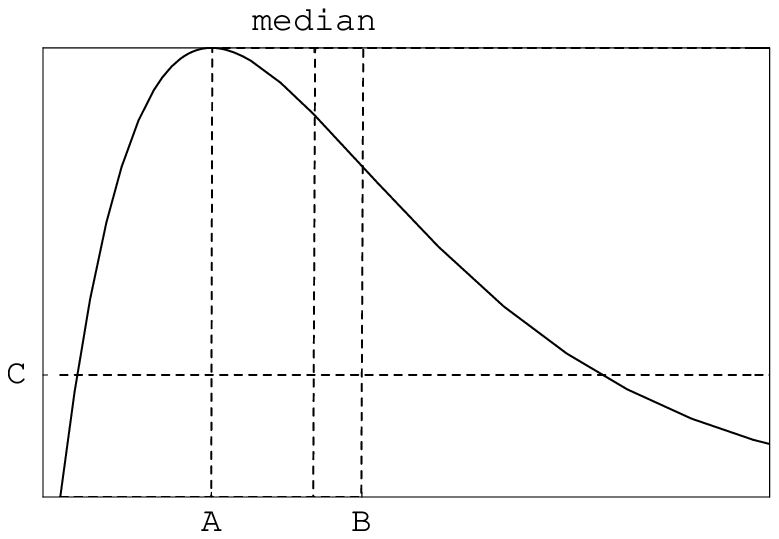}
  }  
  \subfigure[][Median, mode and mean of the axially symmetric example
  distribution function
  $\mathcal{P}_b(x)=\frac{e^{-(x-10)^2}}{\sqrt{\pi}\erf{10}}$. The
  three estimators give the same result for this special case.]{\label{subfig:stat-gauss-func}%
    \psfrag{median}{\it median}
    \psfrag{B}{$\bar{\mathcal{P}}$}
    \psfrag{A}{\rotatebox{-20}{\it mode}}
    \psfrag{C}{\rotatebox{-20}{\it mean}}
    \includegraphics[width=0.488\textwidth]{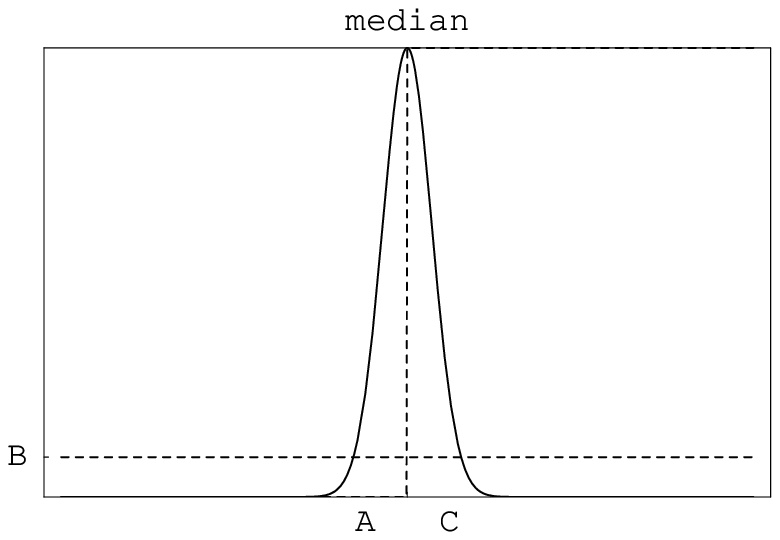}
  }  
  \caption[Mean, mode , median and average probability for
    two unimodal PDF]{Location of mean, mode, median and average probability for
    two special cases of unimodal probability distribution functions in
    a closed interval. The average probability of both distributions is also drawn.}
  \label{fig:stat-descript}
\end{figure}

Alternative estimators for probability distribution functions are the
{\em median} and the {\em mode}. The mode is the value $\bar{x}$,
where $\mathcal{P}(\bar{x})$ takes on a maximum. Clearly there exist
{\em multi-modal} probability distribution functions. The median of
$\mathcal{P}(x)$ is the value $x_\mathit{med}$ for which larger and
smaller values are equally probable, {\em i.e.}\ the value for which the
area below $\mathcal{P}(x)$ to the left from $x_\mathit{med}$ is equal
to the area below $\mathcal{P}(x)$ to the right from $x_\mathit{med}$.
Figures \ref{fig:stat-descript} show these values for the two
different probability distributions
$\mathcal{P}_a(x)=\frac{xe^{10-x/2}}{4e^{10}-44}$ and 
$\mathcal{P}_b(x)=\frac{e^{-(x-10)^2}}{\sqrt{\pi}\erf{10}}$ on the
support interval $\omega=[0,20]$. Note that for general probability
distribution functions, all mentioned estimators take on different
values, as can be seen for $\mathcal{P}_a(x)$ in figure
\ref{subfig:stat-gen-func}, while there are classes of distribution
functions for which some or even all of the mentioned estimators give the same results (figure
\ref{subfig:stat-gauss-func}). Actually the single condition of only
considering axially symmetric univariate functions obviously makes the
median coincide with the mean. For single mode axially symmetric
univariate functions these values also coincide with the mode. This
and the fact that numeric computation of the different estimators are
of different computational effort, are the reasons for occasionally taking one
of these estimators as the approximation for another. One has to take
special care with these approximations if one is dealing with asymmetric
distribution functions, since the introduced errors can become significant.

\subsection{Signal Characteristics of  Photomultiplier Tubes}
\label{subsec:pmt-as-ideal--current-source}

The anode current of a photomultiplier tube operated in the saturated
current region depends on the incident scintillation light flux but
not on the load that is connected to the anode \mycite{Flyckt}{{\em et al.}\
}{2002}. One can treat the anodes of the photomultiplier
tubes like ideal current sources, since they are more or less
sophisticated collection electrodes with necessarily low internal impedance
(Güttinger {\em et al.}\ \cite{Guettinger:1976}, McHose \cite{McHose:1989-tp133}).

\begin{figure}[!t]
  \centering
  \subfigure[][Anode circuit with load resistance $R_L$ and capacitance $C_L$]
  {\label{subfig:anode-connect}
    \psfrag{dn}{$\mathrm{D_n}$}
    \psfrag{dn1}{$\mathrm{D_{n-1}}$}
    \psfrag{dn2}{$\mathrm{D_{n-2}}$}
    \psfrag{A}{$\mathrm{A}$}
    \psfrag{Rl}{$\mathrm{R_L}$}
    \psfrag{Cl}{$\mathrm{C_L}$}
    \includegraphics[height=0.27\textwidth]{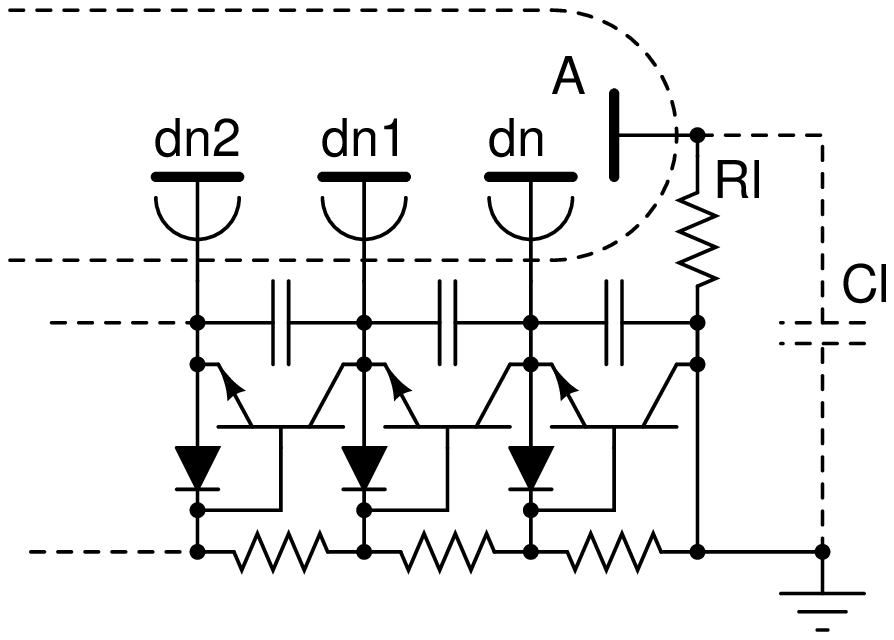}
  }\hspace*{0.04\textwidth}
  \subfigure[][Wave forms of the output voltages for different time
  constants of the anode load. The $\kappa$-values for the 5 different
  graphs are 0, 0.2, 0.4, 0.6, 0.8 and 1.]
  {\label{subfig:voltage-pulses}
    \psfrag{J}{\hspace*{-5em}$\mathcal{V}_\tincaps{PMT}(t)/\mathcal{V}_\tincaps{PMT}(0)\big|_{\kappa=0}$}
    \psfrag{t}{$t/\tau_\tincaps{Scint}$}
    \psfrag{k1}{\rotatebox{-70}{$\kappa=1$}}
    \psfrag{k0}{\rotatebox{75}{$\kappa=0$}}
    \includegraphics[height=0.32\textwidth]{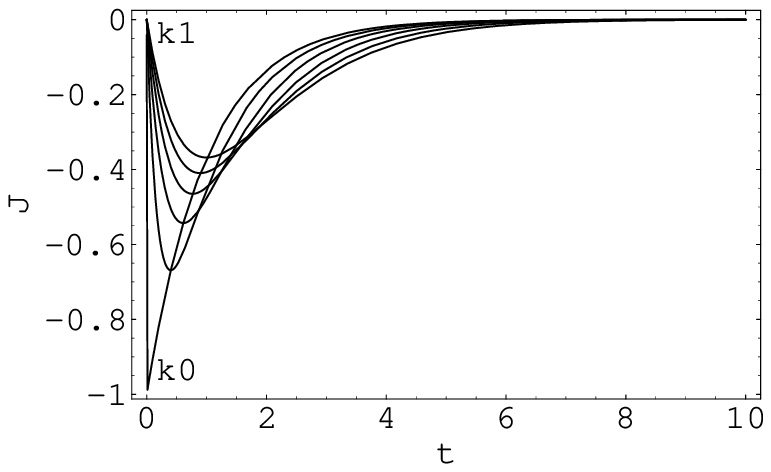}
  }
  \caption[Anode circuit and pulse shapes for anode
    loads]{Typical anode circuit (left) and pulse shapes for anode
    loads of different $\tau_\tincaps{RC}=R_L C_L$ (right).
  }
\end{figure}

The temporal dependence of the scintillation light emission
$\mathcal{L}(t)$ that was induced at time $t_0$ by a detected
$\gamma$-photon can be approximated by an exponential decay law 
\begin{equation}
  \label{eq:scint-time-depend}
  \mathcal{L}(t)=\frac{\bar{n}_\tincaps{Ph}(\mathcal{E})}{\tau_\tincaps{Scint}}e^{-\frac{t}{\tau_\tincaps{Scint}}},
\end{equation}
 $\tau_\tincaps{Scint}$ being the decay time constant of the scintillator
used and $\bar{n}_\tincaps{Ph}(\mathcal{E})$ the average number of scintillation
photons set free from the particle with energy $\mathcal{E}$.
The resulting number of photoelectrons is given by the convolution of
$\mathcal{L}(t)$ with the photomultiplier's pulse response
$\mathcal{R}_\tincaps{PMT}(t)$.  This response is mainly determined by the transit
time that requires the electron avalanche to pass the 
multiplying stages and to reach the anode. 
In particular, the difference in this time for each single
photoelectron leads to temporal impulse broadening \mycite{Kume}{}{1994}. If one assumes a
$\mathcal{R}_\tincaps{PMT}(t)$ of Gaussian shape, 
the current at the anodes becomes
\begin{equation}
  \label{eq:scint-pmt-response}
  \begin{split}
    \mathcal{J}_\tincaps{PMT}(t) & =  \mathcal{L}_\tincaps{Scint}(t)*\mathcal{R}_\tincaps{PMT}(t)\\
    & =  \frac{e^-G_\tincaps{PMT}\bar{n}_\tincaps{PE}(\mathcal{E})}{2\tau_\tincaps{Scint}}%
    \exp{\left[\frac{\tau_\tincaps{PMT}^{2}}{2\tau_\tincaps{Scint}^{2}}%
        -\frac{t}{\tau_\tincaps{Scint}}\right]}\left(1+\erf\left[\frac{t}{\sqrt{2}\tau_\tincaps{PMT}}%
        -\frac{\tau_\tincaps{PMT}}{\sqrt{2}\tau_\tincaps{Scint}}\right]\right),
  \end{split}
\end{equation}
where $e^-$ is the elementary charge, $G_\tincaps{PMT}$ the gain of the
photomultiplier tube, $\tau_\tincaps{PMT}$ the electron transit time
spread, $\bar{n}_\tincaps{PE}(\mathcal{E})$ the average number of
photoelectrons and $\erf$ the error function. Increasing the supply voltages
of the dynode system leads to a higher electric field strength between
the dynode stages and consequently leads to a higher electron speed and
shorter transit times. The geometric design type of the dynode
system and the total size of the photomultiplier are also of importance
for the timing resolution. After all, photomultiplier tubes are,
with rise times between $\mathrm{0.7\,ns}$ and $\mathrm{7\,ns}$, photodetectors of exceptionally
fast response (Kume {\em et al.}\ \cite{Kume:1988}, McHose \cite{McHose:1989-tp114}).
The response pulse width $\tau_\tincaps{PMT}$ is therefore normally
negligible compared to the decay time $\tau_\tincaps{Scint}$ of the
scintillator and equation \ref{eq:scint-pmt-response} can be approximated by
\begin{equation}
  \label{eq:approx-scint-pmt-response}
  \mathcal{J}_\tincaps{PMT}(t)\approx\frac{\bar{q}_a(\mathcal{E})}%
  {2\tau_\tincaps{Scint}}e^{-t/\tau_\tincaps{Scint}},
\end{equation}
where $e^-G_\tincaps{PMT}\bar{n}_\tincaps{PE}(\mathcal{E})$ has been
substituted by the average charge $\bar{q}_a(\mathcal{E})$ collected
at the anode. For proper signal collection, the anode is maintained at
ground potential via a load resistance $R_L$. This is shown in figure
\ref{subfig:anode-connect}. The parallel capacitance $C_L$
can be required, or is just a parasitic one and builds, together with
the load resistance $R_L$, an R-C network. As can be easily
verified by Fourier-analysis, the pulse response of this R-C network
is a decaying exponential with the time constant
$\tau_\tincaps{RC}=R_LC_L$. It has to be convolved with the
photomultiplier pulse given in equation~(\ref{eq:approx-scint-pmt-response})
leading to the following voltage pulse
\begin{equation}
  \label{eq:approx-voltage-pulse}
  \mathcal{V}_\tincaps{PMT}(t)\approx\frac{\bar{q}_a(\mathcal{E})}%
  {C_L}\frac{\tau_\tincaps{RC}}{\tau_\tincaps{Scint}-\tau_\tincaps{RC}}%
  \left(e^{-t/\tau_\tincaps{Scint}}-e^{-t/\tau_\tincaps{RC}}\right).
\end{equation}
The rise time of the voltage pulse is the time from the beginning of
the pulse at $t=0$ until it reaches its maximum value, and is 
given\footnote{In order to cope with unavoidable noise, the anode
  signal rise time is often defined as the elapsed time between 10\%
  and 90\% amplitude of the leading edge of the anode current pulse.} by
the position of the only extremum of (\ref{eq:approx-voltage-pulse}):
\begin{equation}
  \label{eq:tmax-value}
  t_\mathit{max}=\frac{\kappa\,\tau_\tincaps{Scint}}{\kappa-1}\log\kappa,
\end{equation}
being $\kappa=\tau_\tincaps{RC}/\tau\tincaps{Scint}$. For the ideal case of
$\tau_\tincaps{RC}=0$, the rise-time $t_\mathit{max}$ in equation~(\ref{eq:tmax-value}) becomes zero. However, this is due to the
approximation made in (\ref{eq:approx-scint-pmt-response}). The real
rise-time in this case would be given by the maximum position of (\ref{eq:scint-pmt-response}).
As $\kappa$ is switched on, the rise-time will be rapidly dominated by
the time constant $\tau_\tincaps{RC}$ of the anode circuit. This result is of
particular interest for the study of charge dividing circuits made in
this chapter. Since $\tau_\tincaps{RC}$ is proportional to $R_L$ as well as to
$C_L$, any change in their values leads to a variation in the pulse
shape of the anode signal. As will be seen, the resistor
$R_L$ especially is required to feature a large variation depending on the
position of the anode-segment of the position-sensitive
photomultiplier tubes used. This will lead to a strong variation of the
rise-time of the anode-signal depending on the impact position of the
$\gamma$-ray. If this signal were used to generate the trigger
signal for the coincidence detection, the varying pulse shape would
introduce a significant temporal jitter, lowering the temporal
resolution of the detector as a result. The use of the last dynode
instead avoids this problem completely and should be preferred when
a charge divider is used for position detection.

While these last considerations are of major importance for the timing
characteristics of $\gamma$-ray imaging detectors, especially in
coincidence mode, they are of sparse interest for the positioning
algorithm itself. All versions of the charge dividing circuits discussed here 
work without capacitors except for a few that are needed for
frequency and offset compensation of an operational amplifier. 
However, they do not change the pulse shape and do not affect the
computation of the first order moments of the signal distribution.
These are obtained exclusively by using resistances, which are supposed
to behave in an ideal way, {\em i.e.}\ any frequency dependence,
temperature dependency, etc.\ is neglected. Consequently, analyzing the currents that
are the temporal derivatives of the charges is equivalent to analyzing
the charges itself. In this work, except for a few exceptions, the
currents will be studied.

\subsection{General Preamplifier Design}
\label{ch:general-preamplifier-configuration}

\begin{figure}[t]
  \centering
  \subfigure[][Schematic for a current controlled voltage
  source.]{\label{subfig:i-u-converter}%
  \psfrag{R}{$R$}
  \psfrag{OP}{$\mathrm{OP}$}
  \psfrag{Ji}{$J_\mathit{In}$}
  \psfrag{Ui}{$U_\mathit{In}$}
  \psfrag{Ua}{$U_\mathit{Out}$}
    \includegraphics[width=0.28\textwidth]{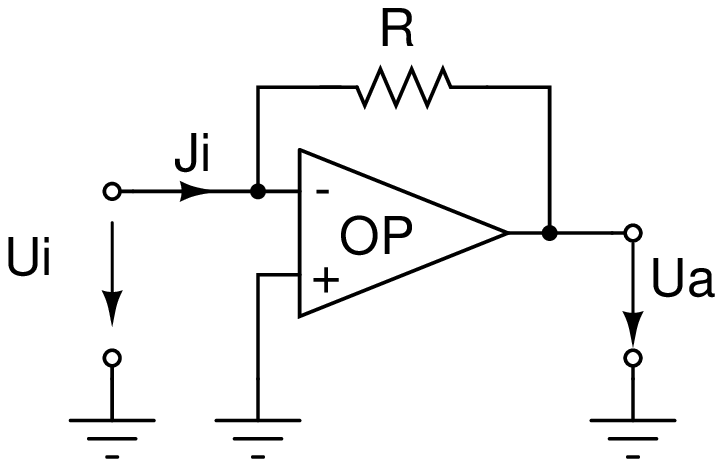}}
  \subfigure[][Schematic for a voltage controlled voltage
  source.]{\label{subfig:electrometer-v}%
  \psfrag{R1}{$R_1$}
  \psfrag{R2}{$R_2$}
  \psfrag{OP}{$\mathrm{OP}$}
  \psfrag{Ui}{$U_\mathit{In}$}
  \psfrag{Ua}{$U_\mathit{Out}$}
    \includegraphics[width=0.28\textwidth]{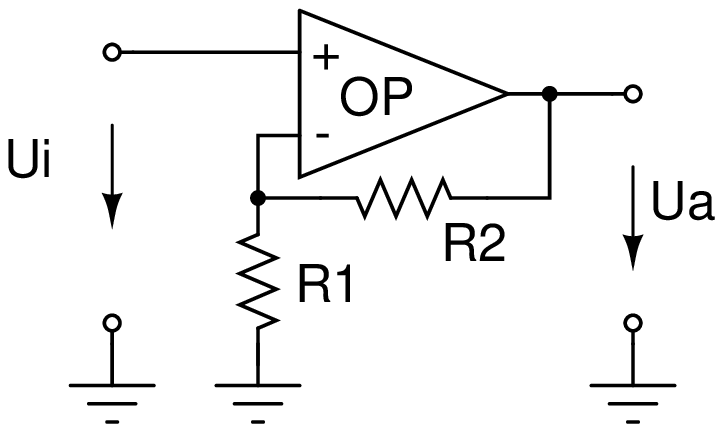}}
  \subfigure[][Schematic for the preamplifier configuration to be used
  with charge dividing circuits.]{\label{subfig:preamp-conf}%
  \psfrag{R}{$R$}
  \psfrag{OP1}{$\mathrm{OP1}$}
  \psfrag{OP2}{$\mathrm{OP1}$}
  \psfrag{Ji}{$J_\mathit{In}$}
  \psfrag{Ua}{$U_\mathit{Out}$}
    \includegraphics[width=0.42\textwidth]{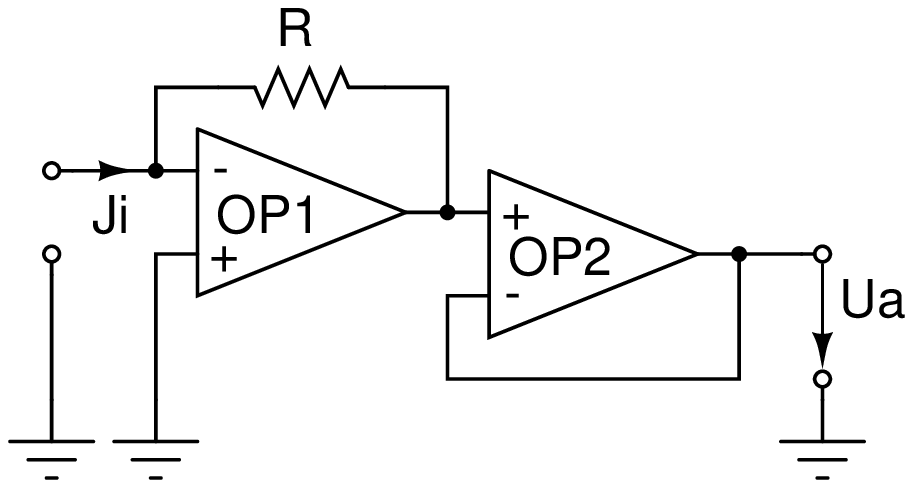}}
  \caption[Elementary operational amplifier
    configuration and current sensitive preamplifier]{Schematics of elementary operational amplifier
    configuration and a current sensitive preamplifier for charge
    dividing circuits.}
\end{figure}

In the following section, three possible electronic implementations of
the center of gravity algorithm are presented. All three versions
have in common that they provide currents which have to be measured
as exactly as possible. A particular property of all presented charge dividing
circuits is that the stages that read and amplify these currents 
necessarily need an input impedance $Z_\mathit{In}$ that is small
compared to the typical resistor values used for the charge
divider. Therefore, amplifier configurations with minimized 
$Z_\mathit{In}$ are required so that the described circuits work predictably.

As input stage, normally the current controlled voltage source
 shown in figure \ref{subfig:i-u-converter} is used. Expressions for its input
impedance $Z_\mathit{In}$, output voltage $U_\mathit{Out}$ and output
impedance $Z_\mathit{Out}$ can be found in standard textbooks, {\em e.g.}\
\cite{Tietze} and, when neglecting parasitic capacitances,  are given by 
\begin{equation}
  \label{eq:u-i-wandler-chars}
  \begin{array}{ccc}
    Z_\mathit{In}=\ds\frac{R}{A_D}\mbox{, }&U_\mathit{Out}=-RJ_\mathit{In}\mbox{, }&Z_\mathit{Out}=\ds\frac{Z_0}{g}.
  \end{array}
\end{equation}
Here, $A_D$ denotes the differential gain and $Z_0$ the output
impedance of the used operational amplifier OP. $g$ is the loop gain
of the circuit and for the inverting amplifier is given by $g=k_FA_D$,
with  $k_F$ a factor depending only on the external feedback
network. In the case of the current controlled voltage source, $k_F$
becomes very small, while the differential gain $A_D$ for modern
operational amplifiers easily exceeds $\mathrm{10^6}$. As one can see
from the middle equation in (\ref{eq:u-i-wandler-chars}), it is easy
to reduce the input impedance to a very small value. However, the output
resistance $Z_0$ of the same operational amplifier will typically be
of a few hundred Ohms, so that the output impedance of circuit
\ref{subfig:i-u-converter} can get very small. The problem of the low
output impedance of the $I$-$U$ converter stage can be solved by a
downstream voltage controlled voltage source, shown in
figure~\ref{subfig:electrometer-v}. Its input impedance, output
voltage and output impedance are given by
\begin{equation}
  \label{eq:electrometer-chars}
  \begin{array}{ccc}
    Z_\mathit{In}=R_\tincaps{CMR}\mbox{, }&U_\mathit{Out}=\ds\frac{U_\mathit{In}}{k_F}\mbox{, }&Z_\mathit{Out}=\ds\frac{Z_0}{g},
  \end{array}
\end{equation}
where $R_\tincaps{CMR}$ denotes the common mode resistance of the used
amplifier, which normally is of a few $M\Omega$ and can reach in
special devices up to $1T\Omega$. For the non-inverting amplifier, the
open loop gain is also given by $g=k_FA_D$, with $k_F$ taking the
value $R_1/(R_2+R_1)$. For the maximum value of $k_F=1$, the output
voltage follows the input voltage. It can be realized by removing
$R_1$ from the circuit and shorting $R_2$. In this case, one also
obtains the minimum possible output impedance. The final configuration
is shown in figure \ref{subfig:preamp-conf} and represents a standard
configuration for charge dividing circuit readout. It combines the
advantages of both sub-circuits discussed, and is still inexpensive and
easy to implement. There are further proposals to optimize
operational parameters of preamplifier circuits for photomultipliers,
which is outside the scope of this work (refer to Kume
\cite{Kume:1994}, Flyckt \cite{Flyckt:2002} and Fabris {\em et al.}\ \cite{Fabris:1999}).

\section{Charge Dividing Circuits for Position Determination}
\label{ch:charge-div-circuits}

The underlying principle of the positioning principle invented by
Anger for his scintillation camera \mycite{Anger}{}{1958} is based on
 Kirchhoff's rules. It uses the fact that a current $J$ that is
injected into a configuration consisting of two resistors, $R_l$ and
$R_r$ (refer to figure~\ref{subfig:simplest-cdr}), is divided into two
partial currents $J_l$ and $J_r$ depending only on the ratio
$R_l/R_r$. As a corollary, it is possible to deduce this ratio from
 both currents $J_l$ and $J_r$ for the notional  case that although the
sum of both resistor values $R_l$ and $R_r$ is known, their ratio is not.
\begin{equation}
  \renewcommand{\arraystretch}{1.3}
  \label{eq:kirchoff-rules}
  \begin{array}{c}
    J=J_l+J_r\\R_rJ_r=R_lJ_l
  \end{array}\Longrightarrow%
  \begin{array}{c}
    R_l=J_r\ds\frac{R_l+R_r}{J}\\
    R_r=J_l\ds\frac{R_l+R_r}{J}
  \end{array}
\end{equation}
This can be of special interest for position detection of the injected
current when there is a known correlation between the resistances  
$R_l$ and $R_r$ and the position. 
Consider the simplest case, when one
has the situation shown in figure \ref{subfig:wire-and-punctual-current}. 
A steady current $J$ is injected at the position
\mbox{$x_0\in\,\,\rbrack\frac{-L}{2},\frac{L}{2}\lbrack$} into a wire
of length $L$. An infinitely small wire segment has the resistance
$dR=g(x)dx$ with $dR>0$ and \mbox{$g(x)=\rho(x)/A(x)$}, where $\rho(x)$
is the specific resistance of the wire and $A(x)$ its
cross-section. The total resistance 
$R_\tincaps{W}$ of the wire from $-\frac{L}{2}$ to $\frac{L}{2}$
is obtained by integration of $g(x)$ over its length, while the
resistances to the left and to the right of the injection point $x_0$
are obtained by integrating over the corresponding wire-segment:
\begin{equation}
  \label{eq:r-values}
  R_\mathit{l}(x_0)=\int_{-\frac{L}{2}}^{x_0}g(x)\;dx\,,\quad%
  R_\mathit{r}(x_0)=\int^{\frac{L}{2}}_{x_0}g(x)\;dx\,,\quad%
  R_\tincaps{W}=R_\mathit{l}(x_0)+R_\mathit{r}(x_0)=\int_{-\frac{L}{2}}^{\frac{L}{2}}g(x)\;dx
\end{equation}

The current $J$, which is injected at position $x_0$ into the wire, will be divided into two fractions,
$J_\mathit{l}$ and $J_\mathit{r}$, corresponding to the ratio of 
$R_\mathit{l}$ and $R_\mathit{r}$ and thus on the injection position $x_0$.
Likewise, the parallel connection of $R_\mathit{l}$ and
$R_\mathit{r}$ in equation~(\ref{eq:r-values}) is given by  
\begin{equation}
  \label{eq:parallel-connection}
  R_\mathit{l}(x_0)\parallel
  R_\mathit{r}(x_0)=\frac{\int_{-\frac{L}{2}}^{x_0}%
    g(x)\;dx\,\,\int^{\frac{L}{2}}_{x_0}%
    g(x)\;dx}{\int_{-\frac{L}{2}}^{\frac{L}{2}}g(x)\;dx}.
\end{equation}
It is just the overall resistance that sees the current $J$.
Making use of Kirschoff's voltage law, 
$J_\mathit{r}(x_0)R_\mathit{r}(x_0)=R_\mathit{l}(x_0)J_\mathit{l}(x_0)$, 
the currents to the right and to the left of $x_0$, namely $J_\mathit{r}(x_0)$ and 
$J_\mathit{l}(x_0)$, can easily be computed:
\begin{gather}
  \label{eq:right-left-c-1}
  J_\mathit{r}(x_0)=\frac{R_\mathit{l}(x_0)}{R_\mathit{r}(x_0)}\big(J-J_\mathit{r}(x_0)\big)\;=\;\frac{R_\mathit{l}(x_0)
  }{R_\mathit{r}(x_0)+R_\mathit{l}(x_0)}J\\  
  \label{eq:right-left-c-2}
  J_\mathit{l}(x_0)=\frac{R_\mathit{r}(x_0)}{R_\mathit{l}(x_0)}\big(J-J_\mathit{l}(x_0)\big)\;=\;\frac{R_\mathit{r}(x_0)}
  {R_\mathit{r}(x_0)+R_\mathit{l}(x_0)}J  
\end{gather}
where $J_\mathit{r}(x_0)+J_\mathit{l}(x_0)=J$ was used. With either of
the two equations~(\ref{eq:right-left-c-1}) 
and~(\ref{eq:right-left-c-2}), it is now possible to reconstruct exactly the position $x_0$ on the wire 
where the current $J$ was injected. For this, the values of $J_\mathit{r}(x_0)$ and 
$J_\mathit{l}(x_0)$ and the distribution $g(x)$ of the resistance along the wire must be known. 
Normally one uses the difference of both currents for the computation:
\begin{figure}%
  \centering
  \subfigure[][Simplest case for a charge dividing
  configuration.]{\label{subfig:simplest-cdr}%
    \psfrag{J}{$J$}
    \psfrag{Jl}{$J_l$}
    \psfrag{Jr}{$J_r$}
    \psfrag{Rr}{\hspace*{-0.5em}$R_r$}
    \psfrag{Rl}{$R_l$}
    \psfrag{GND}{$\mathrm{GND}$}
    \includegraphics[width=0.19\textwidth]{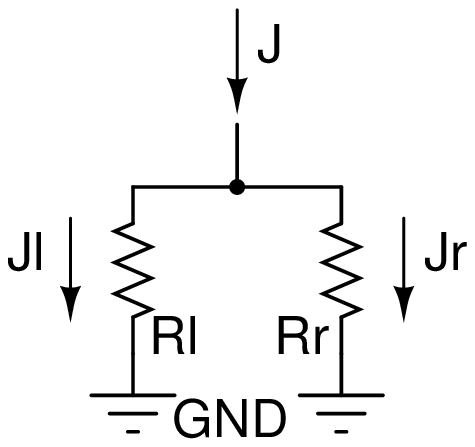}}\hspace*{1em}
  \subfigure[][A 'punctual' current $J\delta(x-x_0)$ is injected at the position $x_0$ 
  into the wire.]{\label{subfig:wire-and-punctual-current}%
    \psfrag{x}{$x_0$}
    \psfrag{L/2}{$\frac{L}{2}$}
    \psfrag{-L/2}{$\frac{-L}{2}$}
    \psfrag{x}{$x_0$}
    \psfrag{wire}{wire}
    \psfrag{Jl}{$J_\mathit{l}$}
    \psfrag{Jr}{$J_\mathit{r}$}
    \psfrag{Jd(x-xo)}{$J\delta(x-x_0)$}
    \psfrag{g(x)}{$g(x)$}
    \includegraphics[width=0.37\textwidth]{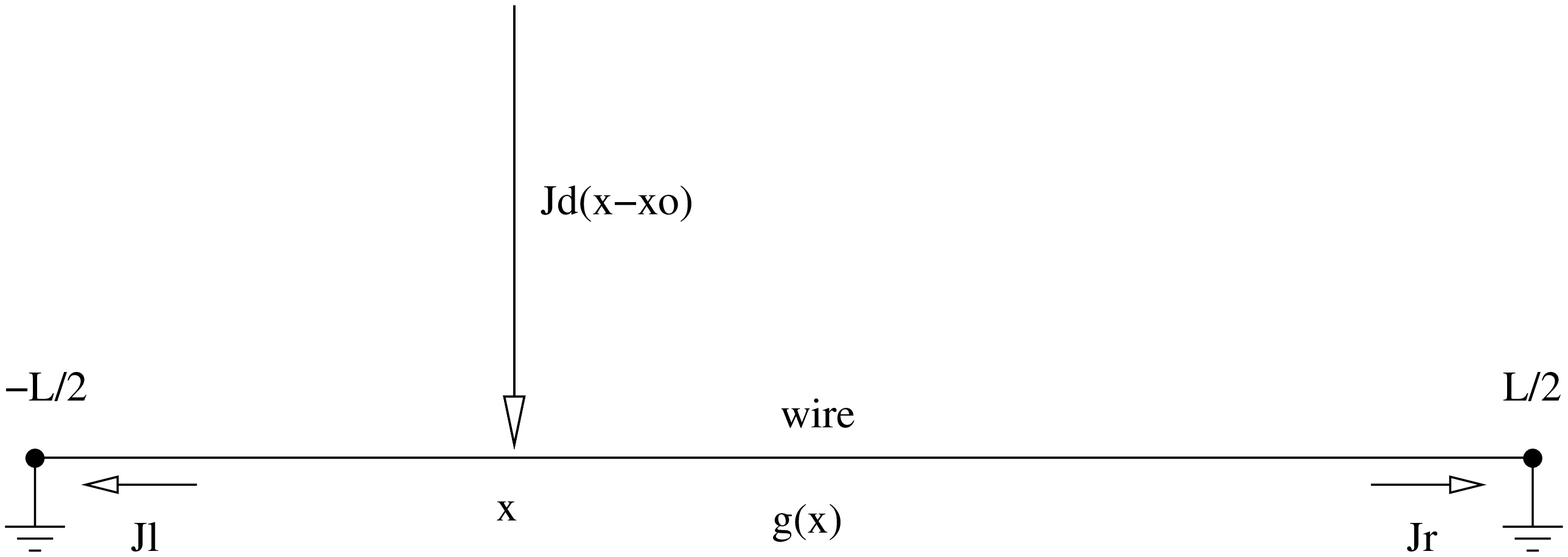}}\hspace*{1em}
  \subfigure[][A current distribution $Jg(x)$ is injected along the
  wire.]{\label{subfig:wire-and-extended-current}%
    \psfrag{L/2}{$\frac{L}{2}$}
    \psfrag{-L/2}{$\frac{-L}{2}$}
    \psfrag{wire}{wire}
    \psfrag{Jl}{$J_\mathit{l}$}
    \psfrag{Jr}{$J_\mathit{r}$}
    \psfrag{Jf(x)}{$J\varphi(x)$}
    \psfrag{Rg(x)}{$g(x)$}
    \includegraphics[width=0.37\textwidth]{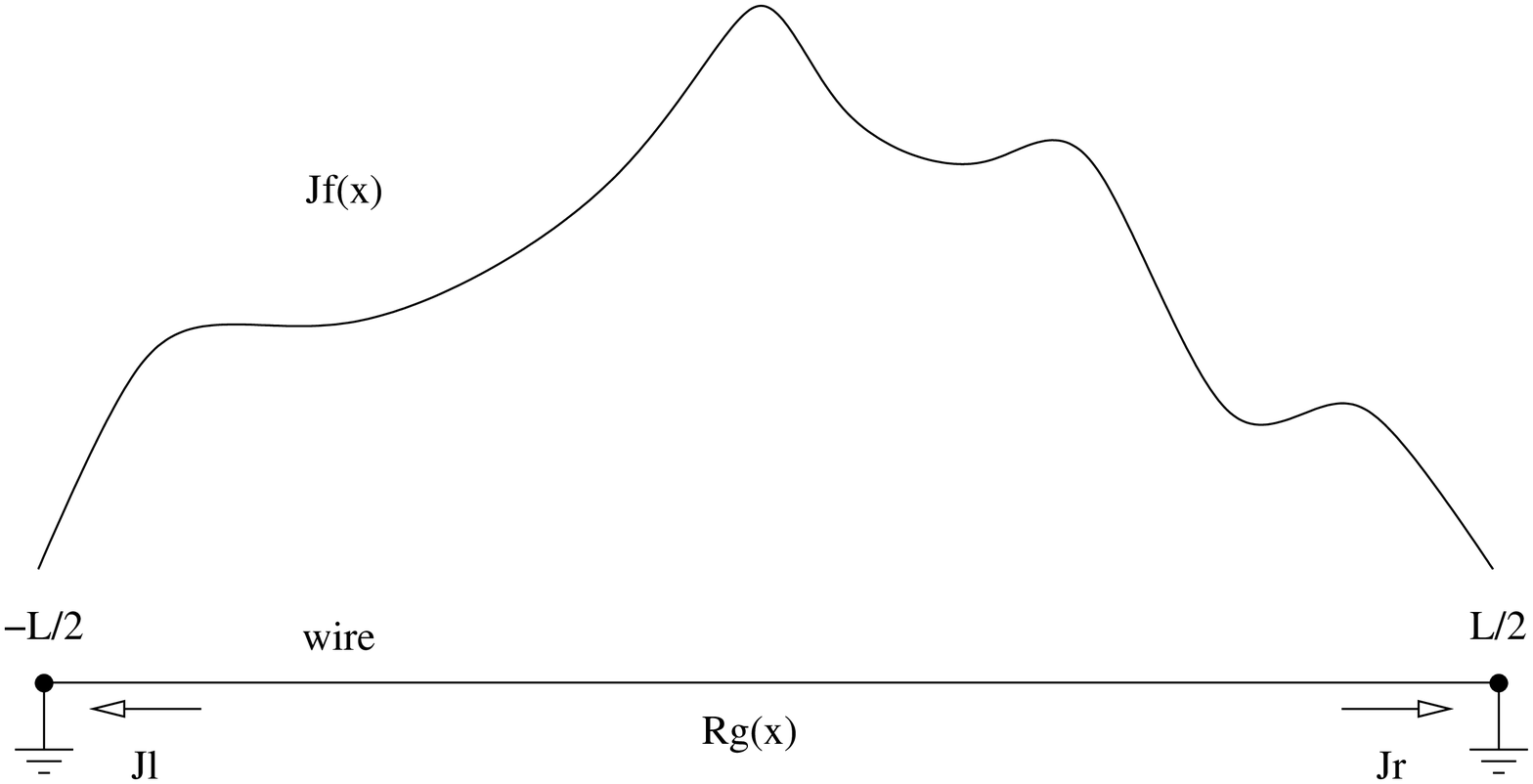}}
  \caption[Graphical representation for different injected current distributions
    along a conductor]{\label{fig:current-dists}%
    Graphical representation for different injected current distributions
    along a conductor. If the resistor values $R_l$ and $R_r$ in figure
    \ref{subfig:simplest-cdr} are chosen in such a way that they are
    proportional to a spatial direction, the
    currents $J_l$ and $J_r$ can be used for the determination of the
    position where $J$ is injected. }
\end{figure}
\begin{equation}
  \label{eq:position-general}
  \frac{J_\mathit{r}(x_0)-J_\mathit{l}(x_0)}{J_\mathit{r}(x_0)+J_\mathit{l}(x_0)}=%
  \frac{R_\mathit{r}(x_0)-R_\mathit{l}(x_0)}{R_\mathit{r}(x_0)+R_\mathit{l}(x_0)}
\end{equation}
If one supposes a constant resistance distribution $g(x)\equiv g$
along the wire, the resistance to the left and to the
right are then given by $R_l=(x_0+L/2)\;g$ and $R_r=(L/2-x_0)\;g$,
where $x_0\in\,\,]\frac{-L}{2},\frac{L}{2}[$. Thus, from
equations~(\ref{eq:kirchoff-rules}) and (\ref{eq:r-values}), one obtains  the
following expressions:
\begin{equation}
  \label{eq:example-consant-res-wire}
  \renewcommand{\arraystretch}{1.3}
  \begin{array}{c}
    (x_0+\frac{L}{2})\;g=J_r\frac{R_w}{J}\\
    (\frac{L}{2}-x_0)\;g=J_l\frac{R_w}{J}
  \end{array}\Longrightarrow%
  \begin{array}{c}
    x_0=(J_r-J_l)\frac{R_w}{2Jg}\\
    L=(J_r+J_l)\frac{R_w}{Jg}
  \end{array}\Longrightarrow%
  x_0=\frac{(J_r-J_l)}{(J_r+J_l)}\frac{L}{2}.
\end{equation}
For this reason, one can easily deduce the injection point of the current $J$ from
$J_l$ and $J_r$. 
Note that the equation on the far right of
(\ref{eq:example-consant-res-wire}) does not depend on the resistance
$g$ but only on the length of the wire. 

The statistical
uncertainties of this positioning method are obtained straightforwardly
by error propagation and are given by:
\begin{equation}
  \label{eq:anger-logic-errors}
  \delta x=\frac{L}{J^2}\sqrt{J_r^2\delta J_l^2+J_l^2\delta
    J_r^2}\quad\mbox{and}\quad\frac{\delta x}{x}=2\frac{\sqrt{J_r^2\delta
      J_l^2+J_l^2\delta J_r^2}}{J_r^2-J_l^2},
\end{equation}
with the limits
\begin{equation}
  \label{eq:a-log-error-lims}
  \frac{\delta x}{x}\Big|_{J_r\rightarrow0}=-\frac{2\sqrt{\delta J_{r}}}{J_l}\mbox{,
  }\frac{\delta x}{x}\Big|_{J_l\rightarrow0}=\frac{2\sqrt{\delta
    J_{l}}}{J_r}\quad\mbox{and}\quad\delta x\Big|_{J_{l,r}\rightarrow\frac{J}{2}}=\frac{L}{2J}\sqrt{\delta
    J_l^2+\delta J_r^2}=\frac{L}{4}\sqrt{\left(\frac{\delta
        J_l}{J_l}\right)^2+\left(\frac{\delta J_r}{J_r}\right)^2}
\end{equation}
Clearly, the relative error in the measurement of $x$ at the center
position $x=0$ diverges. The absolute error at this position is just
the geometric sum of the relative measurement errors in the currents
to the left and to the right, scaled by a quarter of the
wire-length. At both ends of the wire, the relative positioning error
depends only on the measurement error of the current extracted at the opposite
end, which does not necessarily equal zero, although the respective
current does. 

Now consider the case of a 
current distribution $J\varphi(x)$, that is injected simultaneously 
at all possible positions \mbox{$x_0\in\,\,]\frac{-L}{2},\frac{L}{2}[$}
along the wire as shown in figure \ref{subfig:wire-and-extended-current}. 
It is supposed that the injected current distribution $\varphi(x)$ is normalized.
\begin{equation}
  \label{eq:current-norm}
  \int_{-\frac{L}{2}}^{\frac{L}{2}}\varphi(x)\;dx=1
\end{equation}
Since the superposition principle applies to electromagnetic fields,
the impulse response of the wire (equations~\ref{eq:right-left-c-1}
and~\ref{eq:right-left-c-2}) must be summed up for all possible
injection points $x_0$, weighted by the value of the function at
this point $\varphi(x_0)$. This will transform the currents $J_\mathit{l}$
and $J_\mathit{r}$ into the following functionals of $\varphi(x)$:
\begin{equation}
  \label{eq:superposed}
  J_\mathit{r}[\varphi]=\frac{J}{R_\tincaps{W}}\int_{-\frac{L}{2}}^{\frac{L}{2}}R_\mathit{l}(x)\varphi(x)\;dx\,,\quad%
  J_\mathit{l}[\varphi]=\frac{J}{R_\tincaps{W}}\int_{-\frac{L}{2}}^{\frac{L}{2}}R_\mathit{r}(x)\varphi(x)\;dx,
\end{equation}
and leads to the equivalent of equation~(\ref{eq:position-general}),
\begin{equation}
  \label{eq:centroid}
  \frac{J_\mathit{r}[\varphi]-J_\mathit{l}[\varphi]}{J_\mathit{r}[\varphi]+J_\mathit{l}[\varphi]}=\frac{1}{R_\tincaps{W}}\int_{-\frac{L}{2}}^{\frac{L}{2}}%
  \big(R_\mathit{r}(x)-R_\mathit{l}(x)\big)\varphi(x)\;dx,          
\end{equation}
with\begin{equation}
  \label{eq:energy}
  J_\mathit{r}[\varphi]+J_\mathit{l}[\varphi]=\frac{J}{R_\tincaps{W}}\int_{-\frac{L}{2}}^{\frac{L}{2}}%
  \big(R_\mathit{r}(x)+R_\mathit{l}(x)\big)\varphi(x)\;dx=\frac{J}{R_\tincaps{W}}\int_{-\frac{L}{2}}^{\frac{L}{2}}%
  R_\tincaps{W}\varphi(x)\;dx=J.          
\end{equation}

Equations~(\ref{eq:centroid}) and (\ref{eq:energy}) are valid for any possible dependence
of the resistance-distribution $g(x)$ along the wire. Now
consider the simplest non-trivial case of a wire with a constant cross section $A(x)\equiv A$
and constant specific resistance $\rho(x)\equiv\rho$. One than has
$g(x)\equiv g=\mathit{const.}$
Setting $g(x)=g$ in equations~(\ref{eq:r-values}), the resistances to the 
left, to the right, of the whole wire (equation~\ref{eq:r-values}) and the parallel connection in 
equation~(\ref{eq:parallel-connection}) become
\begin{equation}
  \label{eq:left-r-special}
  R_\mathit{l}(x)=g\;\!L\left(\frac{1}{2}+\frac{x}{L}\right)\,,\quad%
  R_\mathit{r}(x)=g\;\!L\left(\frac{1}{2}-\frac{x}{L}\right)\,,\quad%
  R_\tincaps{W}(x)=g\;\!L,
\end{equation}
and
\begin{equation}
  \label{eq:parallel-spec}
  R_\mathit{r}(x)\parallel R_\mathit{l}(x)=g\;\!L\left(\frac{1}{4}-\frac{x^2}{L^2}\right).
\end{equation}
Equivalently, the currents to the left and to the right in
equation~(\ref{eq:superposed}) are given by the
simple relations
\begin{equation}
  \label{eq:superposed-spec}
  J_\mathit{r}[\varphi]=\frac{J}{2}+\frac{J}{L}\int_{-\frac{L}{2}}^{\frac{L}{2}}x\;\varphi(x)\;dx\quad\mbox{and}\quad%
  J_\mathit{l}[\varphi]=\frac{J}{2}-\frac{J}{L}\int_{-\frac{L}{2}}^{\frac{L}{2}}x\;\varphi(x)\;dx,
\end{equation}
that directly leads to the first moment of the arbitrary distribution
$\varphi(x)$ as discussed in the previous section~\ref{ch:stat-estimates}. 
\begin{equation}
  \label{eq:centroid-spec}
  \frac{J_\mathit{r}[\varphi]-J_\mathit{l}[\varphi]}{J_\mathit{r}[\varphi]+J_\mathit{l}[\varphi]}=\frac{2}{L}%
  \int_{-\frac{L}{2}}^{\frac{L}{2}}x\;\varphi(x)\;dx,
\end{equation}
when the currents (equations~\ref{eq:superposed-spec}) are plugged into equation~(\ref{eq:centroid}).
The discretized version of equation~(\ref{eq:centroid-spec}) is widely used in the field of nuclear 
medical imaging to compute the planar interaction position of the impinging $\gamma$-ray within the
scintillation crystal. Axial symmetry of the distribution $\varphi(x)$ and the existence of exactly 
one maximum is required for this method to work, since only for axial
symmetric distributions does the first 
moment coincide with the unique maximum of the distribution. If this
condition is not fulfilled, systematic errors
in real $\gamma$-ray detector setups are introduced. Since the length of the wire is always limited to a finite 
value, in the most normal case the distribution $\varphi(x)$ is truncated on its ends at the left and the 
right. This will destroy the symmetry of the distribution whenever the position of its maximum is 
different from the center of the wire. These and other errors of the
center of gravity algorithm will be discussed briefly in 
section~\ref{ch:errors-of-cog-and-cdr}.

\section{Anger's Approach}
\label{ch:anger-approach}

\begin{figure}[!t]
  \centering
  \psfrag{Ja}{$J^u$}    
  \psfrag{Jb}{$J^d$}    
  \psfrag{J1}{$J_1$}
  \psfrag{J2}{$J_2$}    
  \psfrag{Ji}{$J_i$}    
  \psfrag{Jn}{$J_n$}    
  \psfrag{Jn1}{$J_{n-1}$}    
  \psfrag{x1}{$x_1$}    
  \psfrag{x2}{$x_2$}    
  \psfrag{xi}{$x_i$}    
  \psfrag{xn1}{$x_{n-1}$}    
  \psfrag{xn}{$x_n$}    
  \psfrag{x}{$X$}    
  \psfrag{rin}{$R_\mathit{In}$}    
  \psfrag{ru1}{$R_1^u$}    
  \psfrag{rd1}{$R_1^d$}    
  \psfrag{rui}{$R_i^u$}    
  \psfrag{rdi}{$R_i^d$}    
  \psfrag{ru2}{$R_2^u$}    
  \psfrag{rd2}{$R_2^d$}    
  \psfrag{run}{$R_n^u$}    
  \psfrag{rdn}{$R_n^d$}    
  \psfrag{run1}{$R_{n-1}^u$}    
  \psfrag{rdn1}{$R_{n-1}^d$}    
  \psfrag{d}{$\Delta_x$}    
  \includegraphics[width=0.55\textwidth]{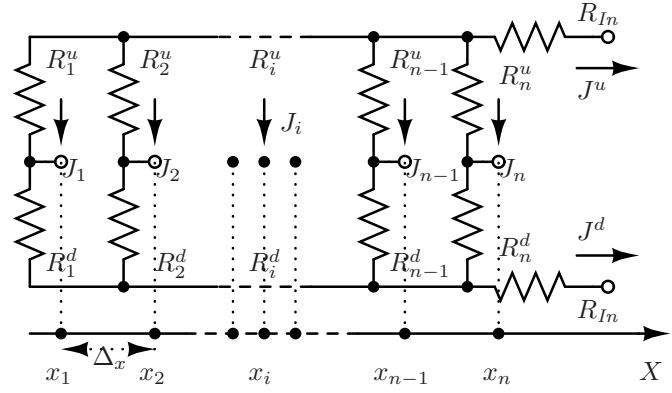}
  \caption[Circuit diagram showing the 1D-version of the anger
    positioning circuit]{Circuit diagram showing the 1D-version of the Anger
    positioning circuit. The resistances $R_{In}$ represent the input
    impedance of the following amplifier stage and are supposed to be
    connected to virtual ground because these impedances should be zero.}
  \label{fig:true-anger}
\end{figure}

For the Anger-type gamma camera one necessarily needs a discrete
positioning logic instead of wires with a constant resistivity since
commercially available photodetectors in general come with a finite
number of detector segments. The adaptation to that case can be achieved
by connecting a set of resistors to each photomultiplier or each
anode-segment with their values adjusted in such a way that there is a
linear correspondence between the currents and the position of the
photomultiplier tube. Figure~\ref{fig:true-anger} shows such a
configuration for one
dimension. For each position $x_i$, the injected current $J_i$ is
divided into $J^u_i$ and $J^d_i$ with a ratio that corresponds
uniquely to that position by choosing the values of resistors
$R^u_i$ and $R^d_i$ in an adequate way. Since electrical currents 
are subjected to the superposition principle, the currents $J^u_i$ and
$J^d_i$ are summed up on the upper and lower bus giving rise
to $J^u$ and $J^d$ respectively. The resistances $R_{In}$ represent the
input impedance of the attached amplifier stage, which have to be as low as possible for optimal linearity of
the positioning network. Rewriting equation~(\ref{eq:kirchoff-rules})
for this case, one obtains
\begin{equation}
  \label{eq:1D-anger-logik}
  \renewcommand{\arraystretch}{1.3}
  \begin{array}{c}
    J^u_i=\ds\frac{R^d_i}{R^d_i+R^u_i}J_i\\[0.9em]
    J^d_i=\ds\frac{R^u_i}{R^d_i+R^u_i}J_i
  \end{array}\Longrightarrow
  \begin{array}{c}
    J^u_i-J^d_i=\ds\frac{R^d_i-R^u_i}{R^d_i+R^u_i}J_i\\[0.9em]
    J^u_i+J^d_i=J_i
  \end{array}\Longrightarrow
  \frac{J^u_i-J^d_i}{J^u_i+J^d_i}=\frac{R^d_i-R^u_i}{R^d_i+R^u_i}.
\end{equation}
In the special case that there is only one single position $x_i$, with $J_i\ne0$
and all other $J_{j\ne i}=0$, one has $J^u\equiv J^u_i$ and $J^d\equiv
J^d_i$. For the general case, the superposition
principle has to be applied and all currents $J^u_i$ and $J^d_i$ must
be summed up to obtain $J^u$ and $J^d$ respectively. 
\begin{equation}
  \label{eq:sum-currents}
  \begin{array}{c}
    J^u = \sum_i^N J^u_i = \sum_i^N \frac{R^d_i}{R^d_i+R^u_i}J_i  \\[0.7em]
    J^d = \sum_i^N J^d_i = \sum_i^N \frac{R^u_i}{R^d_i+R^u_i}J_i,
  \end{array}\Longrightarrow
  \begin{array}{c}
    J^u-J^d = \sum_i^N \frac{R^d_i-R^u_i}{R^d_i+R^u_i}J_i  \\[0.7em]
    J^d+J^d = \sum_i^N J_i,
  \end{array}\Longrightarrow
  \frac{J^u-J^d}{J^u+J^d}=\frac{\sum_i^N \frac{R^d_i-R^u_i}{R^d_i+R^u_i}J_i}{\sum_i^N J_i}
\end{equation}
where $N$ is the total number of injection points $x_i$. If the input
impedance of the downstream current amplifier cannot be neglected, the
centroid would read
\begin{equation}
  \label{eq:sum-currents-with-input-impedance}
  \frac{J^u-J^d}{J^u+J^d}=\frac{\sum_i^N \frac{R^d_i-R^u_i}{R^d_i+R^u_i+2R_\mathit{In}}J_i}{\sum_i^N J_i}.
\end{equation}
In order to determine the required
resistor values $R^d_i$ and $R^u_i$ a current $J$ is considered that is
injected at all positions $x_i$, but only at one position at a time.
One can then compute the positions from the currents $J^u$ and $J^d$.
For any arbitrary linear position encoding one just need to choose the
resistor pairs to fulfill 
\begin{equation}
  \label{eq:1D-anger-logik-lin}
  \frac{R^d_i-R^u_i}{R^d_i+R^u_i}\stackrel{!}{=}ax_i+b.
\end{equation}

However, there are infinitely many possible functional dependences for
$R^u_i$ and $R^d_i$ which fulfill this requirement. $R^u_i+R^d_i=R_i$
is just the total resistance  and must not be the same for all
positions. Therefore, any pair of resistor values of the form 
\begin{equation}
  \label{eq:general-res-encoding}
  \renewcommand{\arraystretch}{1.3}
  \begin{array}{c}
    R^u_i=\frac{R_i}{2}(1-b-ax_i)\\[0.7em]
    R^d_i=\frac{R_i}{2}(1+b+ax_i)
  \end{array}
\end{equation}
fulfill equation~(\ref{eq:1D-anger-logik-lin}) and an
additional constraint for $R_i$ is necessary. Choosing a constant $R_i\equiv
R=\mathit{const.}$, yields
the trivial solution to (\ref{eq:1D-anger-logik-lin}). The parameter
$a$ is the scale and $b$ defines, which position $i$ within the resistor
array maps to $x=0$. A very common choice is $b=0, a=1$ for the center of
the resistor array and a unity scale. For these values, the right hand
side of equation~(\ref{eq:sum-currents}) transforms to the discrete
version for the computation of the centroid (equation~\ref{eq:func-centroid}).

The impedance seen by the source of the current $J_i$ is given by the
parallel connexion of each resistor pair,
\begin{equation}
  \label{eq:1D-anger-parallel}
  R^u_i\parallel R^d_i=\frac{R^u_i R^d_i}{R^u_i+R^d_i}=\frac{R_i}{4}\left(1-(ax_i+b)^2\right).
\end{equation}
As discussed in paragraph \ref{subsec:pmt-as-ideal--current-source},
photomultipliers act like an ideal current source and the
impedance~(\ref{eq:1D-anger-parallel}) does not affect the working of
the charge dividing circuit. The only restriction is given by the fact
that the anode currents $J_i$ lift the anode potential by the amount
$J_i(R^u_i\parallel R^d_i)$ with respect to the last dynode. When this
voltage becomes too large, the PMT will stop working correctly.

For situations where the impedance of the different inputs of the
charge divider is important, equation~(\ref{eq:1D-anger-parallel}) can be
used to adjust it to the required value for each $x_i$. Of special interest is the constraint 
$R^u_i\parallel R^d_i\equiv R^p=\mathit{const.}$\,, {\em i.e.}\ an input
impedance that is equal for all positions $x_i$. One then obtains for
the resistor values $R^u_i$, $R^d_i$ and $R_i$:
\begin{equation}
  \label{eq:const-para-encoding}
  R^u_i=\ds\frac{2R^p}{1+b+ax_i}\mbox{,}\quad R^d_i=\ds\frac{2R^p}{1-b-ax_i}\quad\mbox{and }\quad
  R_i=\ds\frac{4R^p}{1-(b+ax_i)^2},
\end{equation}
with $x_i,i\in\mathbb{N}$ equidistant points along the $x$-axis.

\begin{figure}[!t]
  \centering
  \subfigure[][The two-dimensional version of the positioning logic as
   proposed by Anger for the scintillation camera.]{\label{fig:2D-anger-logic}%
    \psfrag{j}{$J_{ij}$}
    \psfrag{ru}{$R^u_{ij}$}
    \psfrag{rl}{$R^l_{ij}$}
    \psfrag{rr}{$R^r_{ij}$}
    \psfrag{rd}{$R^d_{ij}$}
    \psfrag{jl}{$J^l$}
    \psfrag{jr}{$J^r$}
    \psfrag{ju}{$J^u$}
    \psfrag{jd}{$J^d$}
    \psfrag{jli}{$J^l_{ij}$}
    \psfrag{jl1}{$J^l_{ij-1}$}
    \psfrag{jl2}{$J^l_{ij+1}$}
    \psfrag{jri}{$J^r_{ij}$}
    \psfrag{jr1}{$J^r_{ij-1}$}
    \psfrag{jr2}{$J^r_{ij+1}$}
    \psfrag{jui}{$J^u_{ij}$}
    \psfrag{ju1}{$J^u_{i-1j}$}
    \psfrag{ju2}{$J^u_{i+1j}$}
    \psfrag{jdi}{$J^d_{ij}$}
    \psfrag{jd1}{$J^d_{i-1j}$}
    \psfrag{jd2}{$J^d_{i+1j}$}
    \includegraphics[height=0.41\textwidth]{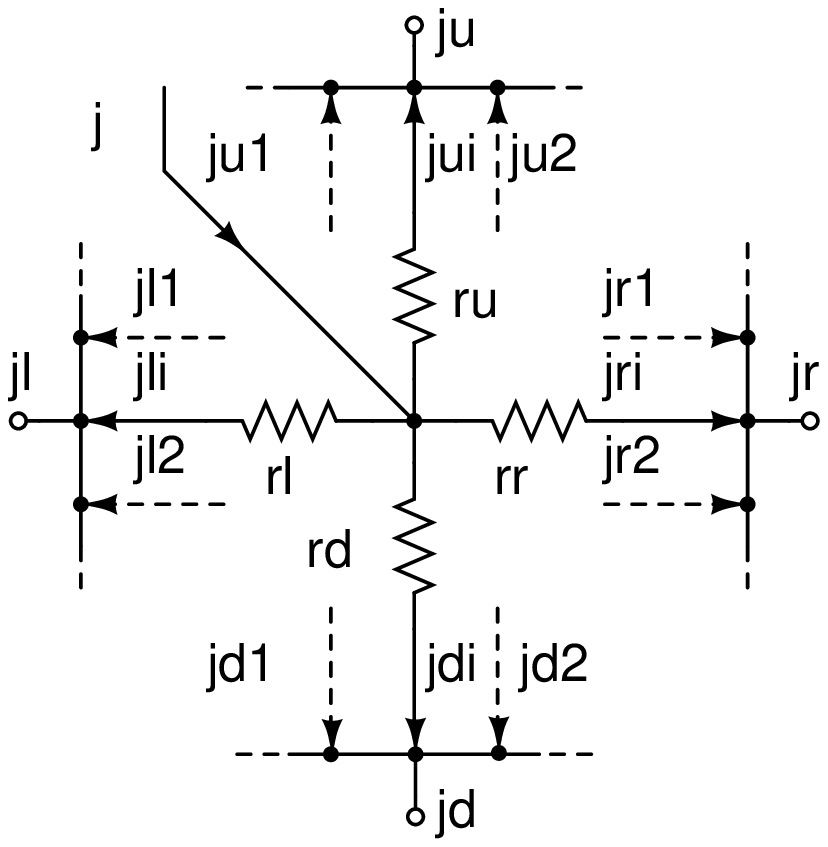}}
  \hspace*{0.03\textwidth}\subfigure[][General case of $k$
  dimensions. For each dimension an additional pair of resistors is
  required for every injection point
  $(x_{i_1},x_{i_2},\ldots,x_{i_k})$.]{\label{fig:k-D-anger-logic}%
    \psfrag{j}{$J_{i_1,i_2,\ldots,i_k}$}
    \psfrag{j1n1}{$J_a^{n_1}$}
    \psfrag{j2n1}{$J_b^{n_1}$}
    \psfrag{j1n2}{$J_a^{n_2}$}
    \psfrag{j2n2}{$J_b^{n_2}$}
    \psfrag{j1nk}{$J_a^{n_k}$}
    \psfrag{j2nk}{$J_b^{n_k}$}
    \psfrag{r1n1}{$R_{i_1i_2\ldots i_k}^{a,n_1}$}
    \psfrag{r2n1}{$R_{i_1i_2\ldots i_k}^{b,n_1}$}
    \psfrag{r1n2}{$R_{i_1i_2\ldots i_k}^{a,n_2}$}
    \psfrag{r2n2}{$R_{i_1i_2\ldots i_k}^{b,n_2}$}
    \psfrag{r1nk}{\hspace*{0.7em}$R_{i_1i_2\ldots i_k}^{a,n_k}$}
    \psfrag{r2nk}{\hspace*{0.7em}$R_{i_1i_2\ldots i_k}^{b,n_k}$}
    \includegraphics[height=0.41\textwidth]{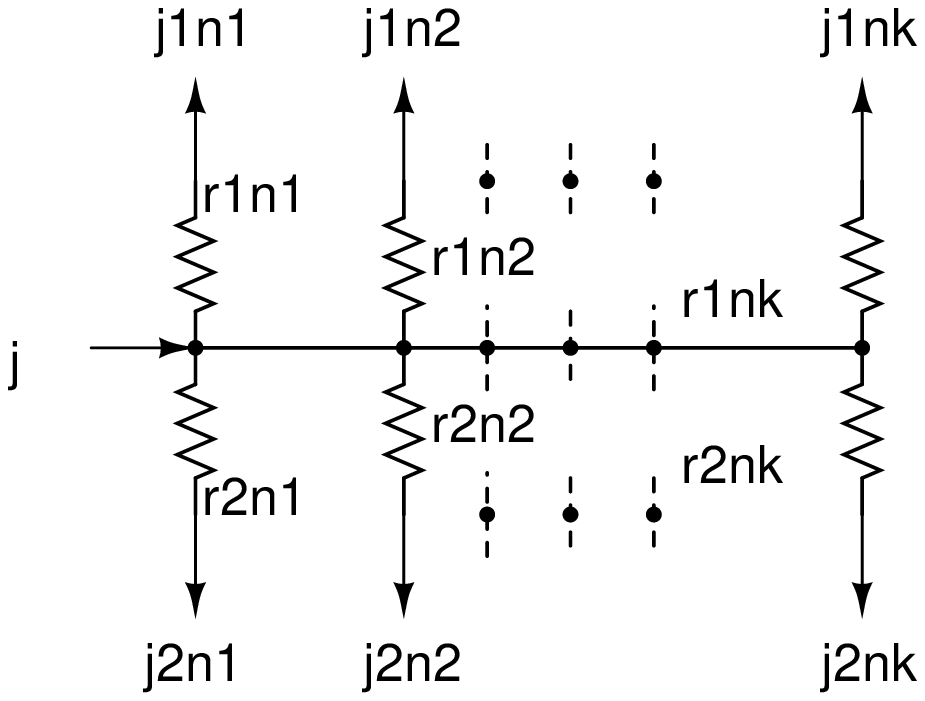}}
  \caption[Anger positioning logic for higher dimensions]{Anger positioning logic for higher dimensions.}
  \label{fig:multi-dim-anger}%
\end{figure}

The generalization to the higher dimensional case is straightforward,
however applications of the CoG using charge dividing circuits
normally do not exceed 3 dimensions. One only has to connect
one more charge divider with its corresponding busses for each
additional spatial dimension that has to be measured. 
This is shown in
figure \ref{fig:2D-anger-logic} for the two-dimensional case. On
obtains for the currents to the
left, right, up and down at each injection point $(x_i,y_j)$ 
\begin{equation}
  \label{eq:2D-anger-logic-elementar-current}
  \renewcommand{\arraystretch}{2.4}
  \begin{array}{cc}
    J^l_{ij}=\ds\frac{R^r_{ij}R^u_{ij}R^d_{ij}}{R_i}J_{ij}\,, &
    J^r_{ij}=\ds\frac{R^l_{ij}R^u_{ij}R^d_{ij}}{R_i}J_{ij} \\
    J^u_{ij}=\ds\frac{R^r_{ij}R^l_{ij}R^d_{ij}}{R_i}J_{ij}\,, &
    J^d_{ij}=\ds\frac{R^l_{ij}R^u_{ij}R^r_{ij}}{R_i}J_{ij}
  \end{array}
\end{equation}
and their equivalent bus currents $J^l$, $J^r$, $J^u$ and $J^d$
are obtained by summation over both position indexes $i$ and $j$
\begin{equation}
  \label{eq:2D-anger-logic-sum-current}
  \renewcommand{\arraystretch}{2.4}
  \begin{array}{cc}
    J^l=\ds\sum_{ij}\ts\frac{R^r_{ij}R^u_{ij}R^d_{ij}}{R_i}J_{ij}\,, &
    J^r=\ds\sum_{ij}\ts\frac{R^l_{ij}R^u_{ij}R^d_{ij}}{R_i}J_{ij} \\
    J^u=\ds\sum_{ij}\ts\frac{R^r_{ij}R^l_{ij}R^d_{ij}}{R_i}J_{ij}\,, &
    J^d=\ds\sum_{ij}\ts\frac{R^l_{ij}R^u_{ij}R^r_{ij}}{R_i}J_{ij}
  \end{array}
\end{equation}
with
$R_i=(R^d_{ij}+R^u_{ij})R^l_{ij}R^r_{ij}+(R^l_{ij}+R^r_{ij})R^d_{ij}R^u_{ij}$.

In analogy to equation~(\ref{eq:sum-currents}) one obtains
\begin{equation}
  \label{eq:2D-anger-positions}
  \begin{array}{ccc}
    \ds\frac{J^u-J^d}{J^u+J^d}=\frac{\ds\sum_{ij}^{N_x,N_y}\ds\frac{R^d_{ij}-R^u_{ij}}{R^d_{ij}+R^u_{ij}}J_{ij}}{\ds\sum_{ij}^{N_x,N_y}J_{ij}} &\mbox{and}&
    \ds\frac{J^r-J^l}{J^r+J^l}=\frac{\ds\sum_{ij}^{N_x,N_y}\ds\frac{R^r_{ij}-R^l_{ij}}{R^r_{ij}+R^l_{ij}}J_{ij}}{\ds\sum_{ij}^{N_x,N_y}J_{ij}}.
  \end{array}
\end{equation}
The impedance of the resistor network at point $(x_i,y_i)$ is given by
the parallel connexion of the four resistors $R^l_i$, $R^r_i$, $R^u_i$
and $R^d_i$, which is just the inverse of the sum of their inverse
values. Once again, there is no unique solution for the required
linear encoding for the current differences in
(\ref{eq:2D-anger-positions}) and  further constraints can be
implemented on the positioning circuit. Reasonable ones are the
requirement of a constant input impedance for all positions
$(x_i,y_i)$ or, as will be shown later, an input impedance  that
depends quadratically on the positions. It can be easily shown that
if the resistor values fulfill
\begin{equation}
  \label{eq:2D-anger-const-imp}
  \begin{array}{cccc}
    R^l_{ij} \propto \frac{1}{1-x_i}\,, & R^r_{ij} \propto
    \frac{1}{1+x_i}\,, & R^u_{ij} \propto \frac{1}{1+y_j}\mbox{ and } & R^d_{ij} \propto \frac{1}{1-y_j}
  \end{array}
\end{equation}
a constant input impedance will be achieved. Likewise, one obtains the above-mentioned
quadratic encoding of the impedance if the following
resistor-value position dependence instead of
(\ref{eq:2D-anger-const-imp}) are chosen.
\begin{equation}
  \label{eq:2D-anger-quad-imp}
  \begin{array}{cccc}
    R^l_{ij} \propto \frac{4(x_i^2+y_j^2)}{1-x_i}\,, & R^r_{ij} \propto
    \frac{4(x_i^2+y_j^2)}{1+x_i}\,, & R^u_{ij} \propto \frac{4(x_i^2+y_j^2)}{1+y_j}\mbox{ and } & R^d_{ij} \propto \frac{4(x_i^2+y_j^2)}{1-y_j}.
  \end{array}
\end{equation}
Equivalent results can be obtained for the $k$-dimensional case shown
in figure \ref{fig:k-D-anger-logic}.

\section{Proportional Resistor Chains}
\label{ch:prop-res-chains}

The second possible configuration of charge dividing circuits is based on
position-sensitive RC-line readouts for proportional gas chambers
\mycite{Borkowski}{{\em et al.}\ }{1970}. It was developed to determine
the position of ionizing events with large volume proportional
detectors. A discretized version designed for  use with multi-anode
position-sensitive photomultiplier tubes was presented by Siegel et
al.\ \cite{Siegel:1996}. This version needs significantly fewer
resistors than the original Anger logic and also reduces the wiring
effort, exposing a nearly equivalent positioning behavior. This is
clearly a great advantage with respect to its implementation.
As for the charge dividing circuit by Anger, it is
possible to extend the circuit to arbitrary many dimensions, as long
as the level of  electronic noise remains sufficiently low.

\begin{figure}[!t]
  \centering
  \psfrag{i}{$i$:}
  \psfrag{dx}{$\Delta x$}
  \psfrag{x}{$x$}
  \psfrag{x=0}{$x=0$}
  \psfrag{xi}{$x_i$}
  \psfrag{r1}{$R_1$}
  \psfrag{rnp}{$R_{n+1}$}
  \psfrag{Jr}{$J_\mathit{right}$}
  \psfrag{Jl}{$J_\mathit{left}$}
  \psfrag{Ji}{$J_i$}
  \psfrag{Ui}{$U_i$}
  \psfrag{1-np/2}{$\frac{1-n}{2}$}
  \psfrag{np-1/2}{$\frac{n-1}{2}$}
  \psfrag{3/2}{$\frac{3}{2}$}
  \psfrag{1/2}{$\frac{1}{2}$}
  \psfrag{-3/2}{$-\frac{3}{2}$}
  \psfrag{-1/2}{$-\frac{1}{2}$}
  \psfrag{3/2}{$\frac{3}{2}$}
  \psfrag{a1}{\hspace*{-2em}anode 1}
  \psfrag{anp}{anode $n$}
  \includegraphics[width=0.95\textwidth]{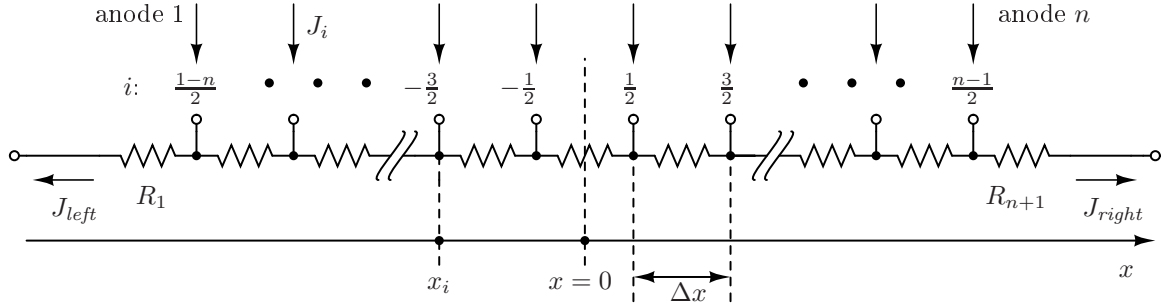}
  \caption[Normally used one-dimensional DPC-circuit for current PSPMTs]{\label{fig:prop-r-net}
    Normally used DPC-circuit for current PSPMTs in one dimension, especially multi-wire type PSPMTs, 
    with $n$ anode-wires. The injected currents $J_i$ will cause the voltages $U_i$ at the
    interconnection points.}
\end{figure}

The continuous case has already been treated in
section~\ref{ch:charge-div-circuits}. For the discrete case the wire
is replaced by a chain of a finite number of resistors $R_1\cdots
R_{n+1}$. Consider such a network (figure~\ref{fig:prop-r-net}) for
a number \mbox{$n\in\mathbb{N}$} of detector outputs. The anodes are
numbered by \mbox{$i\in\,[\frac{1-n}{2};\frac{n-1}{2}]$} and in
unit steps. In this way, one only has to multiply the index $i$ by 
the distance between the centers of two adjacent anode
segments $\Delta x$ to get the true position $x$.
If now a current
$J_i$ is injected at an arbitrary position $i$, it is grounded by the resistances,

\begin{gather}
  \label{eq:resistances_l}
  R_\mathit{l}(i)=\left(\frac{n+1}{2}+i\right)R_d=\left(\frac{R}{2R_d}+i\right)R_d\\
  \label{eq:resistances_r}
  R_\mathit{r}(i)=\left(\frac{n+1}{2}-i\right)R_d=\left(\frac{R}{2R_d}-i\right)R_d  
\end{gather}
to the left (equation~\ref{eq:resistances_l}) and to the right
(equation~\ref{eq:resistances_r}) from $i$ respectively. 
\mbox{$R=(n+1)R_d$} is the sum of all resistances used
in the network. Plugging equations~(\ref{eq:resistances_l}) and
(\ref{eq:resistances_r}) into the second theorem of Kirchoff 
\mbox{$R_\mathit{l}(i)J_\mathit{l}(i)=R_\mathit{r}(i)J_\mathit{r}(i)$}
and applying its first theorem
\mbox{$J(i)=J_\mathit{l}(i)+J_\mathit{r}(i)$}, 
one obtains the following currents at the two ends of the chain
\begin{gather}
  \label{eq:current-1}
  J_\mathit{l}(i)=\left(\frac{1}{2}-\frac{R_d}{R}i\right)J_i\quad\mbox{and} %
  \quad J_\mathit{r}(i)=\left(\frac{1}{2}+\frac{R_d}{R}i\right)J_i.  
\end{gather}
They depend linearly on the injection position $i$. In the case of more than one injected
current at two or more different positions, one obtains the resulting currents by superposition of all 
$n$ different currents $J_\mathit{l}(i)$ and $J_\mathit{r}(i)$:
\begin{gather}
  \label{eq:current-2}
  J_\mathit{l}=\frac{1}{2}\sum_iJ_i-\frac{R_d}{R}\sum_iiJ_i\quad\mbox{and} %
  \quad J_\mathit{r}=\frac{1}{2}\sum_iJ_i+\frac{R_d}{R}\sum_iiJ_i.  
\end{gather}
Equation~(\ref{eq:current-2}) directly leads to the relations normally used for the sum of 
currents and the centroid of the index:
\begin{gather}
  \label{eq:anger_result}
  J=\sum_iJ_i=J_\mathit{l}+J_\mathit{r}\quad\mbox{and}\quad%
  \frac{J_\mathit{r}-J_\mathit{l}}{J_\mathit{r}+J_\mathit{l}}=\frac{2R_d}{R}\frac{\sum_iiJ_i}{\sum_iJ_i}.  
\end{gather}
In order to get the centroid in the position space, one simply uses the
fact that $x_i=i \Delta x$ and thus $i=x_i/\Delta x.$
\begin{equation}
  \label{eq:1D-pos-centroid}
  \langle x\rangle = \frac{\sum_ix_iJ_i}{\sum_iJ_i} =
  \frac{R}{R_d}\frac{\Delta x}{2}\frac{J_\mathit{r}-J_\mathit{l}}{J_\mathit{r}+J_\mathit{l}}.    
\end{equation}
The current $J_i$ at the point $x_i$ will see the impedance
\begin{equation}
  \label{eq:dis-1D-r-chain-imp}
  R_l(i)\parallel R_r(i)=\frac{R_d}{n+1}\left(\frac{(n+1)^2}{4}-i^2\right).
\end{equation}

\subsection{2D Proportional Resistor Network}
\label{subsection:2D-prop-net}

\begin{figure}[!t]
  \centering
  \psfrag{Rd1}{${\Ss R_{v}}$}
  \psfrag{Rd2}{${\Ss R_{h}}$}
  \psfrag{Rv}{${R_{v}}$}
  \psfrag{Rh}{${R_{h}}$}
  \psfrag{Rh1}{${R_{h_1}}$}
  \psfrag{Rh2}{${R_{h_2}}$}
  \psfrag{Rhm}{${R_{h_m}}$}
  \psfrag{Rhm1}{${R_{h_{m-1}}}$}
  \psfrag{Ja}{${\ts J_A}$}
  \psfrag{Jb}{${\ts J_B}$}
  \psfrag{Jc}{${\ts J_C}$}
  \psfrag{Jd}{${\ts J_D}$}
  \psfrag{J11}{${J_{i_1j_1}}$}
  \psfrag{J12}{${J_{i_2j_1}}$}
  \psfrag{J21}{${J_{i_1j_2}}$}
  \psfrag{J22}{${J_{i_2j_2}}$}
  \psfrag{Jnm}{${J_{i_nj_m}}$}
  \psfrag{Jij}{${J_{ij}}$}
  \psfrag{Jl1}{${J^l_1}$}
  \psfrag{Jl2}{${J^l_2}$}
  \psfrag{Jlm}{${J^l_m}$}
  \psfrag{Jlm1}{${J^l_{m-1}}$}
  \psfrag{Jr1}{\hspace*{0.6em}${J^r_1}$}
  \psfrag{Jr2}{\hspace*{0.6em}${J^r_2}$}
  \psfrag{Jrm}{\hspace*{0.6em}${J^r_m}$}
  \psfrag{Jrm1}{\hspace*{0.3em}${J^r_{m-1}}$}
  \includegraphics[width=0.77\textwidth]{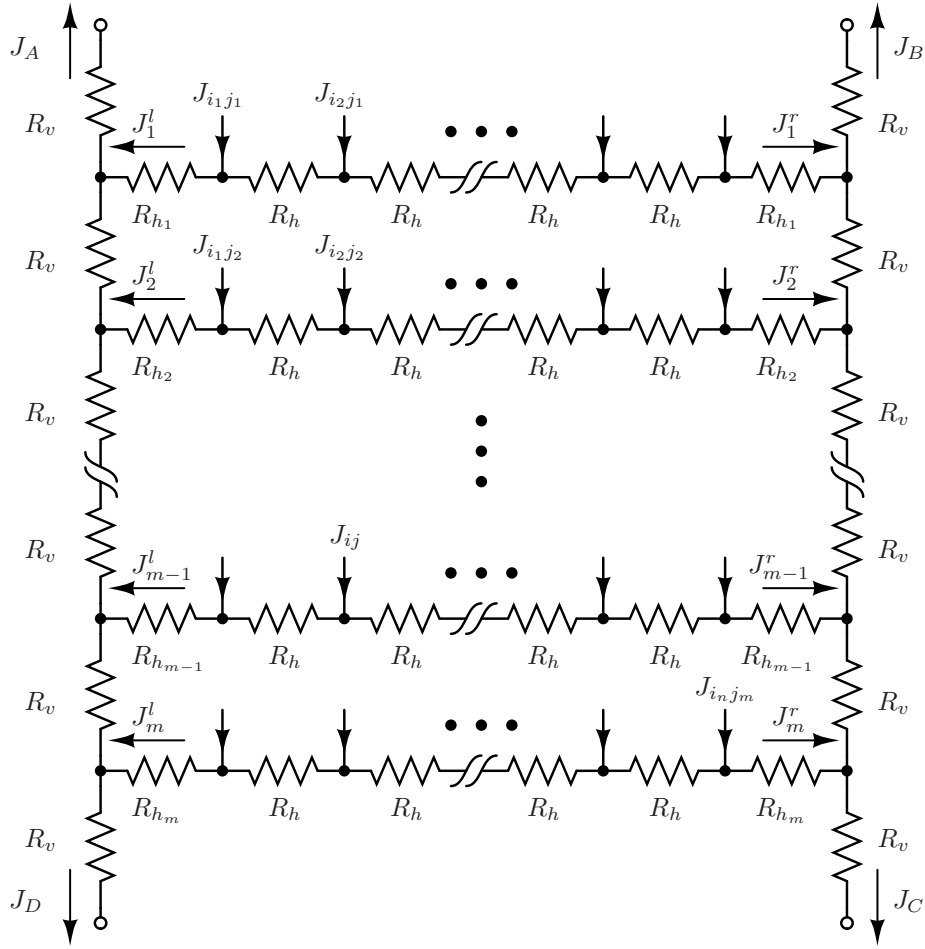}
  \caption[Two-dimensional proportional resistor network]{\label{fig:bidim-schem}%
    Two-dimensional proportional resistor network for a PSPMT
    with a $n\times m$ matrix of anode pads. The position indexes are
    given by $i_l=l-\frac{1+n}{2},\, l\in[1,2,\ldots,n]$ and
    $j_l=l-\frac{1+m}{2},\, l\in[1,2,\ldots,m]$.
  }
\end{figure}

It was shown that the generalization to higher dimensions in the case
of  Anger's solution is straightforward. For the proportional
resistor chain approach however, one faces a situation where the
electronic implementation of higher-dimensional detector-readouts is
easy, but it is very difficult to give explicit expressions for the
centroids as functions of the used resistor values and number of anode
segments. To achieve positional sensitivity for two dimensions
Borkowski {\em et al.}\ \cite{Borkowski:1970} proposed a circuit
configuration that was later discretized by Siegel {\em et al.}\
\cite{Siegel:1996} and which is shown in figure~\ref{fig:bidim-schem}. In this
configuration, the currents from the different sources are injected
into the interconnection points of $m$ (horizontal) 1D-proportional
resistor-chains, where $m$ is the number of anode-segments along the $y$
spatial direction. Equivalently, $n$ denotes the number of anode
segments along the $x$ spatial direction. The currents along one
horizontal resistor chain are divided and superposed according to
equations~(\ref{eq:current-1}-\ref{eq:current-2}) of section
\ref{ch:prop-res-chains}. Now one has $2m$ such currents $J^l_1,J^l_2,\ldots,
J^l_m$ and $J^r_1,J^r_2,\ldots,J^r_m$ that are injected into the two vertical
resistor chains where they are divided and superposed as in the 1D case.
However, contrary to this case, the horizontal sum currents $J^l_1,J^l_2,\ldots,
J^l_m$ and $J^r_1,J^r_2,\ldots,J^r_m$ do not see the same impedance to
ground. According to (\ref{eq:dis-1D-r-chain-imp}), they will see approximately 
\begin{equation}
  \label{eq:lateral-impedance}
  R_\mathit{Imp}(j)\approx R_u(j)\parallel R_d(j)=\frac{R_v}{m+1}\left(\frac{(m+1)^2}{4}-j^2\right)
\end{equation}
instead, if $R_v\ll R_h$. Therefore, the
equations~(\ref{eq:current-1}-\ref{eq:current-2}) have to be modified
in order to correctly describe the actual charge division that takes into
account the lateral impedance $R_u(j)\parallel R_d(j)$:
\begin{gather}
  \label{eq:imp-corrected-horiz-div}
  J_\mathit{l}(i,j)=\left(\frac{1}{2}-\frac{iR_d}{(n-1)R_d+2R_\mathit{Imp}(j)}\right)J_i\,\mbox{ and }\,
  J_\mathit{r}(i,j)=\left(\frac{1}{2}+\frac{iR_d}{(n-1)R_d+2R_\mathit{Imp}(j)}\right)J_i.
\end{gather}
The $j$ dependence in equation~(\ref{eq:imp-corrected-horiz-div}) will
propagate through the rest of the derivation for the centroids and
finally results in a nonlinear positioning behavior. Fortunately, one
can re-linearize the behavior of the network by varying the
value of the lateral horizontal resistors
$R_{h_1},R_{h_2},\ldots,R_{h_m}$ in figure~\ref{fig:bidim-schem}
\mycite{Siegel}{{\em et al.}\ }{1996}. However, the fact that there are
closed loops within this circuitry adds a new order of
complexity to the problem and makes a general description for $N$
dimension, by explicit expression for the centroids as a function of
the number of anode-segments per dimension and the position index
rather complicated. The solution can be found by network
analysis using the {\em branch current method}. In this method, one
establishes a set of equations that describes the relationship of the
currents and voltages to each other through Kirchhoff's current law and
Ohm's law. This set of algebraic equations can be solved, giving the
exact currents and voltages at each node of the network. 

\begin{figure}[t]
  \centering
  \psfrag{Rv}{$R_v$}
  \psfrag{Rh}{$7R_h$}
  \psfrag{R1}{$R_{h_1}$}
  \psfrag{R2}{$R_{h_2}$}
  \psfrag{R3}{$R_{h_3}$}
  \psfrag{R4}{$R_{h_4}$}
  \psfrag{Rl}{$R_l$}
  \psfrag{Rr}{$R_r$}
  \psfrag{Ja}{${\ts J_A}$}
  \psfrag{Jb}{${\ts J_B}$}
  \psfrag{Jc}{${\ts J_C}$}
  \psfrag{Jd}{${\ts J_D}$}
  \psfrag{J}{${\ts J}$}
  \psfrag{Ui}{${\ts U}$}
  \psfrag{Ur1}{${\ts U^r_1}$}
  \psfrag{Ur2}{${\ts U^r_2}$}
  \psfrag{Ur3}{${\ts U^r_3}$}
  \psfrag{Ur4}{${\ts U^r_4}$}
  \psfrag{Ur5}{${\ts U^r_5}$}
  \psfrag{Ur6}{${\ts U^r_6}$}
  \psfrag{Ur7}{${\ts U^r_7}$}
  \psfrag{Ur8}{${\ts U^r_8}$}
  \psfrag{Ul1}{${\ts U^l_1}$}
  \psfrag{Ul2}{${\ts U^l_2}$}
  \psfrag{Ul3}{${\ts U^l_3}$}
  \psfrag{Ul4}{${\ts U^l_4}$}
  \psfrag{Ul5}{${\ts U^l_5}$}
  \psfrag{Ul6}{${\ts U^l_6}$}
  \psfrag{Ul7}{${\ts U^l_7}$}
  \psfrag{Ul8}{${\ts U^l_8}$}
  \includegraphics[width=0.82\textwidth]{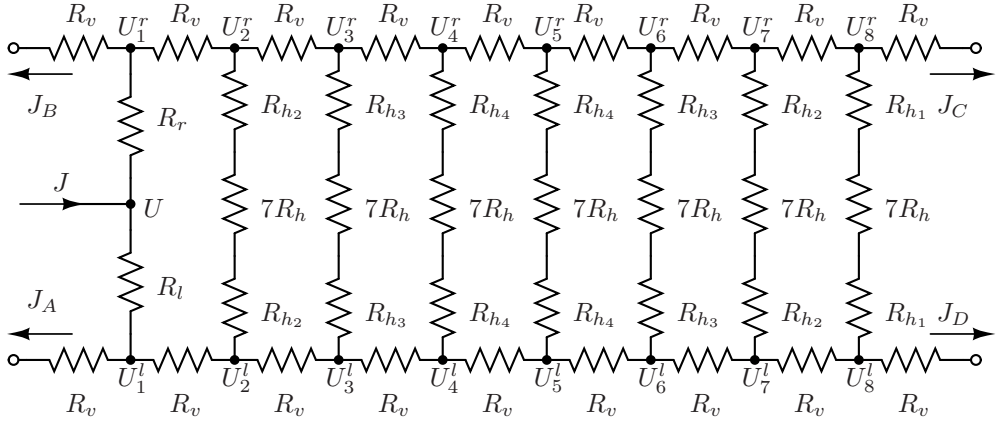}
  \caption[Electric circuit with naming convention for the computation
  of the lateral resistors]{Electric circuit with naming convention
    for the computation of the lateral resistors.}
  \label{fig:comp-lat-res}
\end{figure}

As a first step, one has to compute the lateral horizontal resistor values
$R_{h_1},R_{h_2},\ldots,R_{h_m}$ in order to recover the linearity of
the 2D-proportional positioning network. Since the position-sensitive
multi-anode photomultiplier tube H8500 \cite{data:H8500} from Hamamatsu Photonics Co.\
was used for the experimental
validation of the findings in this section, it is done here for the spacial case of
$n=m=8$ (refer to figure \ref{fig:comp-lat-res}).
Consider the case where, of all 64 anode-segments, only one is active,
and that the active is one of the first row in figure
\ref{fig:bidim-schem} (or the first column in figure
\ref{fig:comp-lat-res} respectively). Then the following set of equations is obtained:
\newcommand{\setsolv}{\stackrel{!}{=}}
\begin{equation}
  \label{eq:lateral-r-equations}
  \renewcommand{\arraystretch}{2.4}
  \begin{array}{ccc}
    \ds\frac{U-U^r_1}{R_r}+\frac{U-U^l_1}{R_l}&\setsolv&J \\
    \ds\frac{U^r_1}{R_v}+\frac{U^r_1-U^r_2}{R_v}-\frac{U-U^r_1}{R_r}&\setsolv&0 \\
    \ds\frac{U^l_1}{R_v}-\frac{U^l_1-U^l_2}{R_v}-\frac{U-U^l_1}{R_l}&\setsolv&0 \\
    \ds\frac{U^l_{j+1}-U^l_j}{R_v}+\frac{U^l_{j+1}-U^r_{j+1}}{7R_h+2R_{h_{j+1}}}+
    \frac{U^l_{j+1}-U^l_{j+2}}{R_v}&\setsolv&0 \\ 
    \ds\frac{U^r_{j+1}-U^r_j}{R_v}-\frac{U^l_{j+1}-U^r_{j+1}}{7R_h+2R_{h_{j+1}}}+
    \frac{U^r_{j+1}-U^r_{j+2}}{R_v}&\setsolv&0 \\ 
    \ds\frac{U^l_8-U^l_7}{R_v}+\frac{U^l_8-U^r_8}{7R_h+2R_{h_8}}+\frac{U^l_8}{R_v}&\setsolv&0 \\
    \ds\frac{U^r_8-U^r_7}{R_v}-\frac{U^l_8-U^r_8}{7R_h+2R_{h_8}}+\frac{U^r_8}{R_v}&\setsolv&0 
  \end{array}
  \qquad\mbox{with}\qquad 
  \begin{array}{c}
    j\in[1,2,\ldots,6]\\
    R_{h_8} =  R_{h_1}\\
    R_{h_7} =  R_{h_2}\\
    R_{h_6} =  R_{h_3}\\
    R_{h_5} =  R_{h_4}
  \end{array}
\end{equation}
The equations $R_{h_{9-j}} =  R_{h_j},\,j\in[1,2,3,4]$ are valid
due to the symmetry of the charge dividing circuit. If the
system~(\ref{eq:lateral-r-equations}) is solved, the
four currents $J_A$, $J_B$, $J_C$, and $J_D$ can be expressed as functions of the different
resistor values and $J$. The centroids are computed from these
currents using equations~(\ref{eq:anger_result}) that can be naturally
adapted to the 2D case as follows:
\begin{equation}
  \label{eq:bidim-cents}
  \langle
  j\rangle=c_j\frac{J_A+J_B-(J_C+J_D)}{J_A+J_B+J_C+J_D}\quad\mbox{and}\quad\langle i\rangle=c_i\frac{J_B+J_C-(J_A+J_D)}{J_A+J_B+J_C+J_D},
\end{equation}
where the constants $c_i$ and $c_j$ still have to be determined. If one
further constrains this set of equations with $R_r=R_{h_1}$ and
$R_l=7R_h+R_{h_1}$, one gets the constant value of 7/9 for $\langle
j\rangle$ independent of any resistor value. For the
expectation value $\langle i \rangle$, a quotient of
polynomials in $R_v$, $R_{h_1}-R_{h_4}$ and $R_{1}$ is obtained. However, this
value is supposed to give the same value as $\langle j\rangle$ owing
to the symmetry of the network and the requirement that it behaves in
the same manner for both spatial directions. Repeating this procedure
three times with currents only at the positions $(i,j)=(5/2,5/2)$,
$(i,j)=(3/2,3/2)$, and $(i,j)=(1/2,1/2)$, one gets a new set of 4
equations, whose solution gives the values for the lateral horizontal
resistors.
\begin{equation}
  \label{eq:solution-laterals}
  \begin{array}{cc}
    R_{h_1}=R_{h_8}=R_h-4R_v, & R_{h_2}=R_{h_7}=R_h-7R_v\\
    R_{h_3}=R_{h_6}=R_h-9R_v, & R_{h_4}=R_{h_5}=R_h-10R_v
  \end{array}
\end{equation}
They can be parameterized in the following way:
\begin{equation}
  \label{eq:solution-laterals-param-a}
  R^n_{h_l}=\frac{l}{2}\left(l-(n+1)\right)\,,\mbox{ with }l\in[1,2,\ldots,n]
\end{equation}
or
\begin{equation}
  \label{eq:solution-laterals-param-b}
  R^n_{h_l}=\frac{1}{2}\left(l^2-\frac{(n+1)^2}{4}\right)\,,\mbox{ with }l\in[\frac{1-np}{2},\ldots,\frac{n-1}{2}].
\end{equation}
Equations~(\ref{eq:solution-laterals-param-a}) and
(\ref{eq:solution-laterals-param-b}) have been verified for the cases
$n=m$ and $n,m\in[2,4,6,8,10]$. Once the 2D-proportional
resistor network has been linearized, one can determine the two proportionality constants 
$c_i$ and $c_j$. This can be done in exactly the same way as the
linearization of the network, but replacing the resistances seen by
the injected current to its left and right by
$R_l=(7/2+i)R_h+R_{h_j}$ and $R_r=(7/2-i)R_h+R_{h_j}$ in
equations~(\ref{eq:lateral-r-equations}) and using the
results~(\ref{eq:solution-laterals}). In analogy to
(\ref{eq:1D-pos-centroid}), for the expectation values 
of $x$ and $y$ one obtains
\begin{equation}
  \label{eq:2D-centroids}
  \renewcommand{\arraystretch}{1.4}
  \begin{array}{cc}
    \ds\langle x\rangle =\Delta
    x\frac{9}{2}\frac{J_B+J_C-(J_A+J_D)}{J_A+J_B+J_C+J_D}, &
    \ds\langle y\rangle =\Delta
    y\frac{9}{2}\frac{J_A+J_B-(J_C+J_D)}{J_A+J_B+J_C+J_D}.
  \end{array}  
\end{equation}
Note that the factors $9/2$ coincide with $(n+1)/2$ for
the case $n=8$. However, the general validity of this expression for
arbitrary $n$ has not been proved here.

The set of equations~(\ref{eq:lateral-r-equations}) can equally be used to
determine the impedance of the network of each of its 64 inputs. For
this, one solves equations (\ref{eq:lateral-r-equations}) and its three equivalents when
the current is injected in rows 2,3 and 4 for the voltage $U$ at the
injection point. By virtue of Ohm's law and together with the injected
current $J$, this is equivalent to the impedance. The other 4 rows are
given implicitly due to the symmetry of the network.

One obtains 4 equations for the impedances at the different
interconnection points along the horizontal resistor chains which are
given in Appendix \ref{app:exact_solutions}. An explicit
parameterization in the position indexes $i$ and $j$ can be given for
the case reported by \mycite{Siegel}{{\em et al.}\ }{1996} with $R_h=10
R_v$:
\begin{equation}
  \label{eq:exact-parametrization}
  R_\mathit{Imp}(i,j)=\frac{5}{18} \left(81-4 i^2\right) R_v+\left(a_2 j^2+a_0+i^2 \left(b_6 j^6+b_4 j^4+b_2 j^2+b_0\right)\right) R_v,
\end{equation}
with the parameter values
\begin{equation}
  \label{eq:exact-parametrization-constants}
  \begin{array}{llllll}
    a_0=-\frac{63}{16}, & a_2=-\frac{7}{36}, & b_0\approx0.24, & b_2\approx0.015,& b_4\approx0.19\cdot10^{-3},& b_6\approx0.19\cdot10^{-5}.
  \end{array}
\end{equation}
Equation~(\ref{eq:exact-parametrization}) shows that the quadratic dependence
of $R_\mathit{Imp}(i,j)$ is reproduced only for the $x$ spatial
direction, {\em i.e.}\ $j=\mathit{const.}$, while the $y$ spatial
direction includes $\mathcal{O}(j^4)$, manifesting in this way the
antisymmetry of the network in both spatial directions. Note that the
behavior is not due to the linearization of the position response by
varying the lateral horizontal resistor values $R_{h_j}$. If all $R_{h_j}$
are set to $R_h$, one will obtain the following parameter values  
\begin{equation}
  \label{eq:exact-parametrization-constants-b}
  \begin{array}{llllll}
    a_0=\frac{9}{8}, & a_2=-\frac{1}{18}, & b_0\approx0.048,& b_2\approx0.0021, & b_4\approx0.15\cdot10^{-4}, &  b_6\approx0.47\cdot10^{-7},
  \end{array}
\end{equation}
that still contain higher than quadratic orders in $j$. Note that
approximate values of the parameters $b_0,b_2,b_4$ and $b_6$ are given
here for clearity although they can be computed exactly. They are
given in Appendix~\ref{app:exact_solutions}.

\section{Hybrid Solution}

\begin{figure}[t]
  \centering
  \psfrag{Rd1}{${\Ss R^d_1}$}
  \psfrag{Rd2}{${\Ss R^d_2}$}
  \psfrag{Ru1}{${\Ss R^u_1}$}
  \psfrag{Ru2}{${\Ss R^u_2}$}
  \psfrag{Rh}{${R_{h}}$}
  \psfrag{Rum}{${R^u_m}$}
  \psfrag{Rdm}{${R^d_m}$}
  \psfrag{Rum1}{${R^u_{m-1}}$}
  \psfrag{Rdm1}{${R^d_{m-1}}$}
  \psfrag{Ja}{${\ts J_A}$}
  \psfrag{Jb}{${\ts J_B}$}
  \psfrag{Jc}{${\ts J_C}$}
  \psfrag{Jd}{${\ts J_D}$}
  \psfrag{J11}{${J_{i_1j_1}}$}
  \psfrag{J12}{${J_{i_2j_1}}$}
  \psfrag{J21}{${J_{i_1j_2}}$}
  \psfrag{J22}{${J_{i_2j_2}}$}
  \psfrag{Jnm}{${J_{i_nj_m}}$}
  \psfrag{Jij}{${J_{ij}}$}
  \psfrag{Jl1}{${J^l_1}$}
  \psfrag{Jl2}{${J^l_2}$}
  \psfrag{Jlm}{${J^l_m}$}
  \psfrag{Jlm1}{${J^l_{m-1}}$}
  \psfrag{Jr1}{${J^r_1}$}
  \psfrag{Jr2}{${J^r_2}$}
  \psfrag{Jrm}{${J^r_m}$}
  \psfrag{Jrm1}{${J^r_{m-1}}$}
  \includegraphics[width=0.8\textwidth]{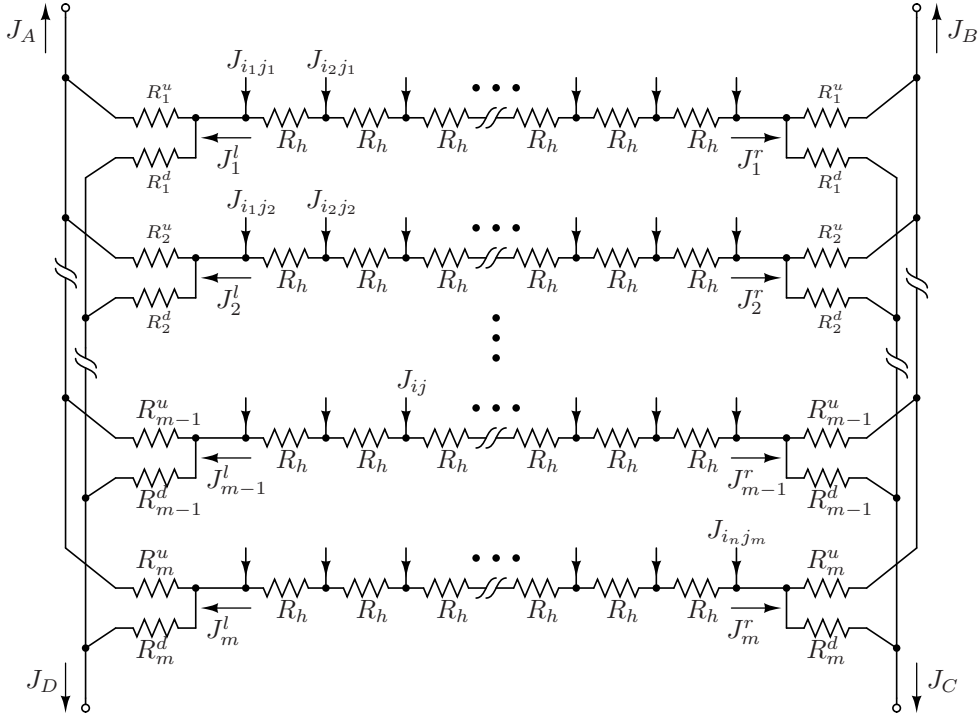}
  \caption[Circuit for a two-dimensional hybrid charge dividing readout]
  {Circuit for the 2D hybrid charge dividing readout proposed
    by \mycite{Siegel}{{\em et al.}\ }{1996}. It uses $m$ horizontal
    resistor chains for the $x$-centroid and Anger current scaling for
  the other spatial direction.}
  \label{fig:hybrid-solution}
\end{figure}

The third and last possible implementation of the CoG
algorithm using charge divider circuits is the combination of both
previous methods \mycite{Siegel}{{\em et al.}\ }{1996}. Clearly there can
be no hybrid version for one dimension. Instead, the different
previously mentioned approaches are applied to different spatial
dimensions in the circuit. This is shown in
figure~\ref{fig:hybrid-solution} for the 2D-case.

As discussed in section~\ref{ch:general-preamplifier-configuration}, 
it is possible to read the currents $J_A$, $J_B$,
$J_C$ and $J_D$ with shunt feedback transresistance amplifiers that 
have an input resistance $Z_\mathit{In}$ of virtually $\mathrm{0\,
  \Omega}$. In this case, the horizontal resistor chains in figure 
\ref{fig:hybrid-solution} decouple among each other and can be treated
separately according to section~\ref{ch:prop-res-chains}. It was also
shown that all resistors, including both at the chain
ends, need to have the same value in order to obtain correct results
from equations~(\ref{eq:anger_result}-\ref{eq:dis-1D-r-chain-imp}).
This can be assured using equations~(\ref{eq:const-para-encoding}) of
section~\ref{ch:anger-approach}, where the positioning
properties for the vertical component in
figure~\ref{fig:hybrid-solution} were also discussed.
Choosing the following values for $R^u_j$ and $R^d_j$ of
the $y$ spatial direction charge dividers:
\begin{equation}
  \label{eq:hybrid-lateral-values}
  R^u_j = \frac{R_h}{\frac{1}{2}-\frac{j}{m+1}}\quad\mbox{and}\quad
  R^d_j=\frac{R_h}{\frac{1}{2}+\frac{j}{m+1}},
\end{equation}
the expressions for the centroids along the $x$- and $y$-spatial
directions give
\begin{equation}
  \label{eq:2D-hybrid-centroids}
  \renewcommand{\arraystretch}{1.4}
  \begin{array}{cc}
    \ds\langle x\rangle =\Delta
    x\frac{n+1}{2}\frac{J_B+J_C-J_A-J_D}{J_A+J_B+J_C+J_D}, &
    \ds\langle y\rangle =\Delta
    y\frac{m+1}{2}\frac{J_A+J_B-J_C-J_D}{J_A+J_B+J_C+J_D}.
  \end{array}  
\end{equation}
Along the $x$-spatial direction, the impedance of the
hybrid-circuit inputs is the same as for the 1D proportional resistor
chain given by equation~(\ref{eq:dis-1D-r-chain-imp}). It does not
depend, however, on the index $j$ for the $y$-spatial direction, since this was
the constraint for finding
expression~(\ref{eq:hybrid-lateral-values}).
Thus one obtains for the input impedance of the hybrid solution 
\begin{equation}
  \label{eq:hybrid-impedance}
  R_\mathit{In}(i,j)=\frac{R_h}{n+1}\left(\frac{(n+1)^2}{4}-i^2\right).
\end{equation}

Siegel {\em et al.}\ state that the positioning performance of
the three circuits is quite similar. Therefore, the natural
consequence of choosing one of the three versions is to consider the 
complexity of implementation and their cost as the most important
criteria. In this case, the clear favorite is the multidimensional
version of the proportional resistor chain. First, it is the version
which can be managed with fewest resistors; only $m(n+1)+2(m+1)$
components are required. Second, there are no crossing wires in the
circuit which eases the layout considerably and allows for a compact
design. The most expensive solution is the proposal from Anger. It
requires $4nm$ resistors and a more complicated layout. The hybrid
solution gets by with $m(n+1)+4m$ resistors.

\section{Simultaneous Measurement of the Second Moment}
\label{sec:sim-measurement-of-sec-mom}

\begin{figure}[!t]
\centering
\subfigure[][Standard configuration of an analogue adder circuit.]{\label{subfig:simple-adder}
\psfrag{R1}{$R_1$}
\psfrag{Ri}{$R_i$}
\psfrag{Rn}{$R_n$}
\psfrag{Rc}{$R_c$}
\psfrag{Rf}{$R_f$}
\psfrag{Cc}{$C_c$}
\psfrag{Cf}{$C_f$}
\psfrag{U1}{$U_1$}
\psfrag{Ui}{$U_i$}
\psfrag{Un}{$U_n$}
\psfrag{Ua}{$U_\mathit{Out}$}
\psfrag{OP}{$\mathrm{OP}$}
\includegraphics[width=0.3\textwidth]{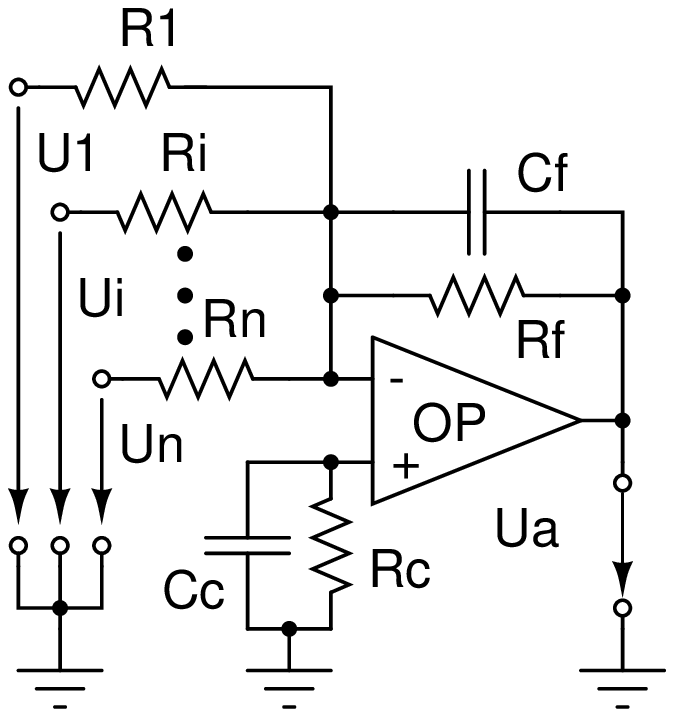}}
\subfigure[][Configuration with voltage follower for a high input
impedance of the individual branches.]{\label{subfig:adder-with-buffer}
\psfrag{R1}{$R_1$}
\psfrag{Ri}{$R_i$}
\psfrag{Rn}{$R_n$}
\psfrag{Rc}{$R_c$}
\psfrag{Rf}{$R_f$}
\psfrag{Cc}{$C_c$}
\psfrag{Cf}{$C_f$}
\psfrag{U1}{$U_1$}
\psfrag{Ui}{$U_i$}
\psfrag{Un}{$U_n$}
\psfrag{Ua}{$U_\mathit{Out}$}
\psfrag{OP}{$\mathrm{OP}$}
\includegraphics[width=0.38\textwidth]{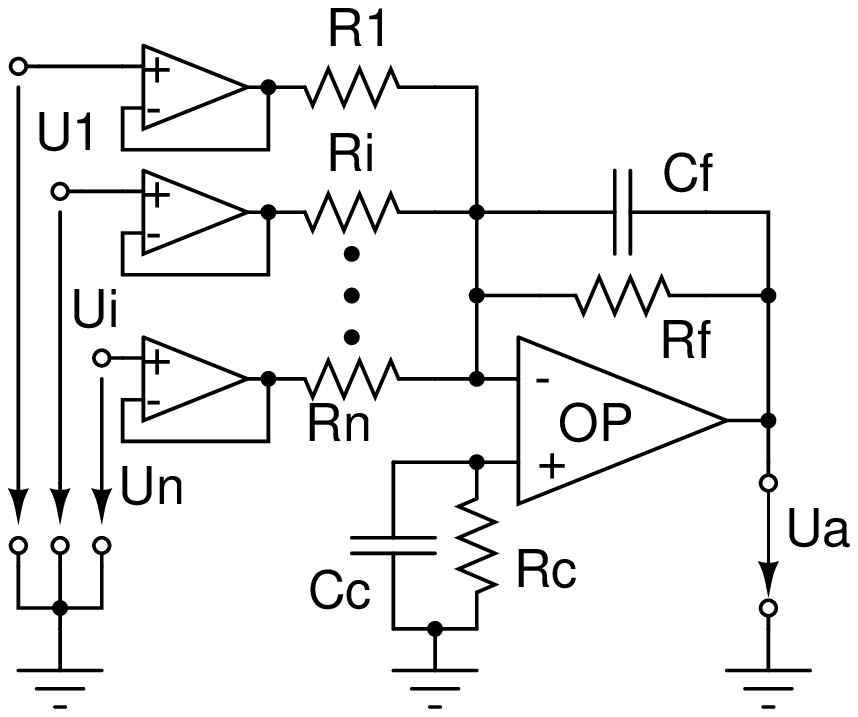}}
\subfigure[][Block-diagram of the combined summation and charge dividing
circuit.]{\label{subfig:block-sum-cdr}
  \psfrag{JA}{$J_A$}
  \psfrag{JB}{$J_B$}
  \psfrag{JC}{$J_C$}
  \psfrag{JD}{$J_D$}
  \psfrag{S}{$\sum$}
  \psfrag{J1}{$J_{1,1}$}
  \psfrag{Jn}{$J_{n,m}$}
  \psfrag{cdr}{CDR}
  \psfrag{PSPMT}{PSPMT}
  \psfrag{sumAmp}{adder}
  \includegraphics[width=0.28\textwidth]{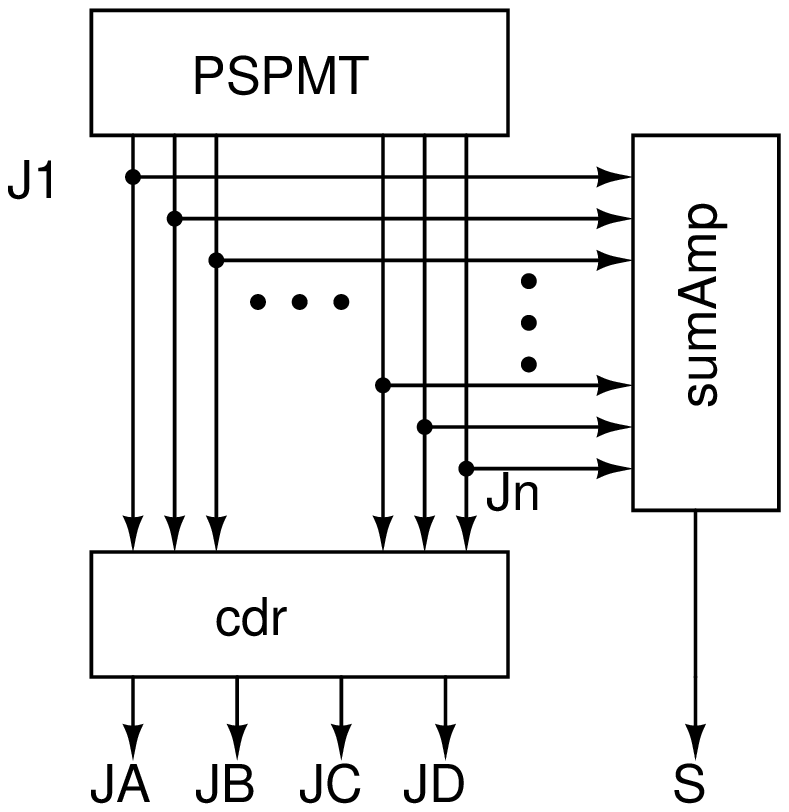}}
\caption[Inverting amplifier configuration for use as summation amplifier]{\label{fig:analogue-adder}
Inverting amplifier configuration used as summation amplifier. The
contributions of the voltages $U_i$ to the output voltages can be
adjusted with the resistors $R_i$ and $R_f$. $C_f$ is required for
phase correction and $C_c||R_c$ for offset correction.}
\end{figure}

After the discussion of how the centroids of a distribution can be
measured with different charge dividing circuits, the
emphasis is now put on the second moment. The motivation for measuring this
moment is justified by the fact that the square root of the centered second moment
is an excellent measure for the width of a distribution function.
It was discussed in detail in section \ref{ch:light-distribution} that
there is a strong correlation between the $\gamma$-ray's depth of
interaction within thick crystals and the induced scintillation light
distribution observed at the sensitive area of a photodetector.
Using this effect for depth of interaction determination has been
proposed at various times and by
different researchers (refer to section \ref{ch:doi-detectors}).
However, the problem consists of how to implement this measurement
in the detector electronics, while simultaneously meeting all typical requirements 
such as fast measurement, low computational effort, low costs
and sufficiently good measurement.
Observing the impedances seen by the currents fed into the charge
dividing circuits (equations~\ref{eq:1D-anger-parallel},
\ref{eq:dis-1D-r-chain-imp} and \ref{eq:hybrid-impedance}) a
possible solution becomes quite obvious. Since the impedances are
quadratically encoded with the injection position and owing to Ohm's
law one has $U=RJ$, one already disposes of the correct weighting for the sampled
distribution and the only step that remains is to do the summation
over these voltages. This can be done with a standard configuration of
a summation amplifier explained in electronics textbooks and shown in figure
\ref{subfig:simple-adder}. 

The output voltage of this circuit is the sum of the input voltages
$U_i$, weighted by $-R_f/R_i$:
\begin{equation}
  \label{eq:adder-output-voltage}
  U_\mathit{Out}=-R_f\sum_{i=1}^n\frac{U_i}{R_i}.
\end{equation}
Theoretically there is no upper limit on the number of
individual inputs $U_n$ when the electronic components behave as
ideal ones. In practice, however, this limit is imposed by the thermal
noise of the resistors $R_1,R_2,\ldots,R_n$ and the maximal output voltage
swing of the operational amplifier used. The spectral density for the
noise for each ohmic resistance $R_i$ in figure
\ref{fig:analogue-adder} is given by
$|U^\mathit{noise}_i(f)|^2=4k_BTR_i$, where $k_B$ is the Boltzmann constant
and $T$ the absolute temperature.
After integration over all frequencies, one obtains the measurable
effective noise. For this, the parasitic capacitance $C^p_{R_i}$ of each
resistance $R_i$ has to be taken into account which results in
$|U^\mathit{noise}_i(f)|\rightarrow 0$ for $f\rightarrow\infty$,
ensuring the convergence of the integral.
\begin{equation}
  \label{eq:effective-adder-noise}
  U^\mathit{noise,eff}_i=\sqrt{\frac{k_BT}{C^p_{R_i}}}.
\end{equation}
Equation~(\ref{eq:effective-adder-noise}) holds for each resistance
used in the configurations of figure \ref{fig:analogue-adder} and also
the impedances (equations~\ref{eq:1D-anger-parallel},
\ref{eq:dis-1D-r-chain-imp} and \ref{eq:hybrid-impedance}) of the
centroid networks. For ambient temperature and a typical capacitance of
$\mathrm{0.2 pF}$, equation~(\ref{eq:effective-adder-noise}) gives an effective
noise of approximately $\mathrm{150\mu V}$, which is three orders of magnitude
below the given signal level. A more important constraint is given by the
maximal output voltage swing that can be delivered by the
operational amplifier used. With an adequate choice of $R_f$, one has to
ensure that $U_\mathit{Out}$ is always smaller than this device
parameter, otherwise the operational amplifier saturates and distorts
the signal.

The impedance $R^\mathit{In}_i$ of the inputs of the summing amplifier
is approximately given by their input resistance $R_i$. Since
$R^\mathit{In}_i$ is connected in parallel to the impedances
(\ref{eq:1D-anger-parallel}), (\ref{eq:dis-1D-r-chain-imp})
and (\ref{eq:hybrid-impedance}) of the networks for the transverse
positioning, the desired quadratic dependence gets distorted.
In order to minimize this effect, one has to choose the individual 
$R^\mathit{In}_i$ very large compared to the impedances at the
different inputs of the charge dividing circuits. Alternatively one
can use the adder configuration of figure
\ref{subfig:adder-with-buffer}. In this version, each input of the
summation amplifier is buffered with a voltage follower described in
section \ref{ch:general-preamplifier-configuration}. In this way, one is
able to realize very high input impedances that are also independent of the weights
of the adder. However, the large number of required operational
amplifiers can pose thermal and power consumption problems on this solution. 
 
The resistance $R_c$ in figures \ref{fig:analogue-adder} is required
for offset compensation (Tietze and Schenk \cite{Tietze}). Since the inverting
terminal of the operational amplifier is connected to virtual ground
when using the inverting configuration, the input current causes
an offset voltage $U_\mathit{Off}=I_\mathit{In}R_f$ at the output of 
of the amplifier. This offset voltage can be compensated by a 
resistance $R_c$ of value
\begin{equation}
  \label{eq:com-r-value}
  R_c=\left(\frac{1}{R_f}+\sum_{i=1}^n\frac{1}{R_i}\right)^{-1}
\end{equation}
and connected as displayed in figures~\ref{fig:analogue-adder}.
In order not to introduce additional noise with this resistor,
the capacity $C_c$ is connected in parallel to  $R_c$, shortening by
this means all signal components with non-vanishing frequencies. Similarly, $C_f$
is not required for the correct working of the method of determining
the second moments, but is needed for compensation of the non-ideal behavior of
the amplifier configuration (Franco \cite{Franco}). It avoids
oscillations of the operational amplifier caused by stray capacities
at the inverting terminal.

The summation amplifiers of figure~\ref{fig:analogue-adder} can be
attached to any of the three discussed charge dividing circuits as
shown in figure~\ref{subfig:block-sum-cdr}. One can use the values of
the individual $R_i$ to adapt the summation amplifier to the
particular requirements of each of the three charge dividing circuits.
The value of $R_f$ can be used to adjust the amplification of the sum
signal for avoiding saturations of the operational amplifier.  
Throughout the remaining part of this work, only implementations of
the summation circuit shown in figure~\ref{subfig:simple-adder} will
be considered. This configuration was preferred to 
the buffered version shown in figure~\ref{subfig:adder-with-buffer},
which is complicated to realize in practice. The $n\times m$ voltage
follower requires space and impedes the
development of a simple circuit layout. On the other hand, the power
consumption of each of the required $n\times m$ operational amplifiers
imposes a serious design problem. A very low supply
current of $\mathrm{15\,mA}$  already leads to an overall consumption of approximately 
$1 A$ required for the buffers when a position sensitive photodetector
with $8\times8$ anode-segment is used. For this and other 
reasons, the solution proposed in figure \ref{subfig:adder-with-buffer}
is more suitable for ASICs ({\em application-specific integrated
  circuits}). 

As an important disadvantage, one introduces a systematic
error when computing the second moment using the circuit in figure 
\ref{subfig:simple-adder}, since the current coming from the
anode-segment of the photodetector will see the input impedance of the
summation amplifier in parallel to that of the charge dividing
circuit. This leads not only to improper weights for the signal
distribution but also extracts currents from the charge dividing
circuit for the transverse positioning. In order to minimize this
effect, the weighting resistors $R_i$ of the summation amplifier have
to be chosen as large as possible. Here, the fabrication tolerance of
1\% of the standard SMD-resistors used can serve for a reasonable design
criterion. If one chooses the weighting resistors $R_i$ at least 100
times larger than the smallest input impedance of all the inputs of
the charge dividing circuit, the errors introduced in the centroid and
the second moment will be smaller than 1\% for a 1D resistor chain.

\subsection{Anger Logic}

In the case of the true Anger Logic, the application of the analogue
summation circuit is straightforward. The specific design of this
circuit, which uses an individual charge divider for each current from
the anode-segments, produces a completely decoupled set of input
voltages for the summation circuit since the output currents of the
Anger logic are connected to virtual ground (refer to
figure~\ref{fig:2D-anger-logic}). One now has two possibilities to
obtain a quadratic weighting of the signal distribution. Choosing the
resistor values for the charge divider according to
equations~(\ref{eq:2D-anger-const-imp}), {\em i.e.}\ the same input impedance
for all CDR
inputs, implies the implementation of the quadratic dependence
using the resistors $R_i$ of the summation circuit. By virtue of
equation~(\ref{eq:adder-output-voltage}) and the constant input
impedance
$R_\mathit{In}^\mathit{CDR}$ of the Anger logic the relative values of the
summation weights $R_{ij}$ are given by
\begin{equation}
  \label{eq:anger-sum-const-input}
  R_{ij}\propto\frac{R_\mathit{In}^\mathit{CDR}R_f}{x_i^2+y_i^2}.
\end{equation}
Note that equation~(\ref{eq:anger-sum-const-input}) has a singularity
at $(x_i,y_i) = (0,0)$, {\em i.e.}\ for this position one would obtain
$R_{ij}=\infty$. In practice, this requirement can be implemented by
omitting this resistor, that is to say, one does not include this value
into the weighted sum in the same manner as this value is excluded
in the mathematical definitions (\ref{eq:multivariate-moms-1})-(\ref{eq:multivariate-moms-3}) for
$x_i = 0\,\forall\,i\in[1,n]$. In the majority of the cases,
however, there is no reason for this consideration, since most
position-sensitive photodetectors have an even number of detection
segments. Also one subsequently has to invert the signal since
summing amplifiers are realized by the inverting configuration of
operational amplifiers. For the determination of the absolute values
of the $R_{ij}$ one can use the above-mentioned design criterion 
of adjusting the minimal systematic error to the  resistor tolerance.

The second implementation configuration is given by equations
(\ref{eq:2D-anger-quad-imp}). Here, the quadratic impedance variation
seen by the anode-currents is inherently incorporated into the charge
divider. One therefore has to give all summands the same weights with
one single value $R_{ij}=R$. Once again the
error-criterion for the determination of the absolute values can be used.

\subsection{Proportional Resistor Chains}
\label{sec:proportional-resistor-chains}

In the case of the proportional resistor chains, the computation of
the second moment using the described summation amplifier is not as
trivial as in the case of the true Anger logic. The reason lies in the
fact that all resistors are used to generate the weights for each
position. Therefore, the charge dividing circuit does not decouple
the voltages produced by the input currents over the impedances of the
network from their neighborhood. Fortunately, this does not avert the
successful computation of the second moment but makes the mathematical
description more complicated. First the one-dimensional case is
studied, since for this it is possible to find an explicit and
exact expression for the summed voltage. Furthermore, it is of
interest for the hybrid case. Based on this result, approximate
solutions for the two-dimensional case can be found.

\subsubsection{One-Dimensional Case}
\label{subsection:dpc-with-2m-1D}

\begin{figure}[!t]
  \centering
  \psfrag{i}{$i$:}
  \psfrag{dx}{$\Delta x$}
  \psfrag{x}{$x$}
  \psfrag{x=0}{$x=0$}
  \psfrag{xi}{$x_i$}
  \psfrag{r1}{$R_1$}
  \psfrag{rnp}{$R_{n+1}$}
  \psfrag{Jr}{$J_\mathit{right}$}
  \psfrag{Jl}{$J_\mathit{left}$}
  \psfrag{Ji}{$J_i$}
  \psfrag{Rd}{$R_d$}
  \psfrag{Ui}{$U_i$}
  \psfrag{1-np/2}{$\frac{1-n}{2}$}
  \psfrag{np-1/2}{$\frac{n-1}{2}$}
  \psfrag{3/2}{$\frac{3}{2}$}
  \psfrag{1/2}{$\frac{1}{2}$}
  \psfrag{-3/2}{$-\frac{3}{2}$}
  \psfrag{-1/2}{$-\frac{1}{2}$}
  \psfrag{3/2}{$\frac{3}{2}$}
  \psfrag{a1}{\hspace*{-2em}anode 1}
  \psfrag{anp}{anode $n$}
  \psfrag{Uij}{$U_{ij}$}
  \psfrag{Uii}{$U_{i,i}$}
  \includegraphics[width=0.85\textwidth]{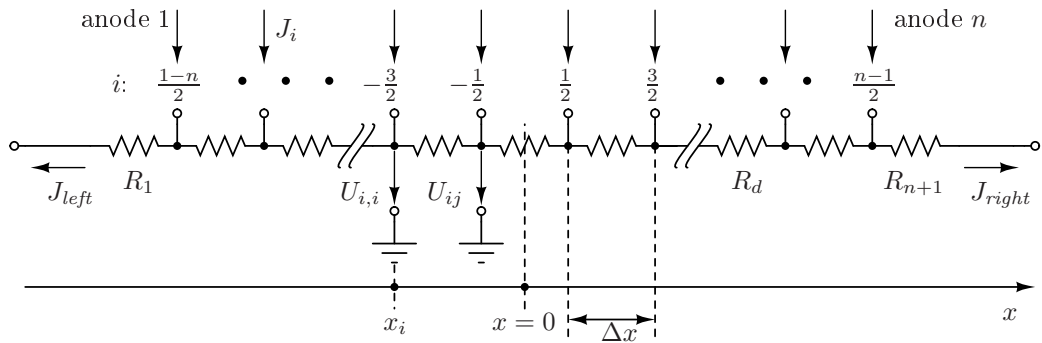}
  \caption[Schematic diagram of the resistor chain and naming conventions]{\label{fig:prop-r-net-with-voltage}
Schematic diagram of the resistor chain and naming conventions for the
computation of the sum of the voltage fraction induced by the injected
current $J_i$.}
\end{figure}

First, the situation that induces a current $J_i$ injected
at a single but arbitrary anode of index $i$ with
\mbox{$i\in\,[\frac{1-n}{2};\frac{n-1}{2}]$} is considered. Owing to Ohm's law,
the voltage $U_{i,i}$ seen at this same position is just the product of
the current $J_i$ and the impedance to ground, {\em i.e.}\ $R_l(i)\parallel
R_r(i)$ given in equation~(\ref{eq:dis-1D-r-chain-imp}):
\begin{equation}
  \label{eq:ij-voltage}
  U_{i,i} =\frac{J_iR_h}{n+1}\left(\frac{(n+1)^2}{4}-i^2\right)=J_i\,R\,\left(\frac{1}{4}-\left(\frac{R_h}{R}\right)^2i^2\right),
\end{equation}
where $R=R_h(n+1)$ was used for the last step. As in the second case
of the Anger logic, the required quadratic dependence is inherently incorporated
into the charge divider. However, an attached summation amplifier
would not give just the sum of all $U_{i,i}$ with
\mbox{$i\in\,[\frac{1-n}{2};\frac{n-1}{2}]$} since the resistor
chain in figure~\ref{fig:prop-r-net-with-voltage} would act as a voltage divider for
the $U_{i,i}$. Therefore the connected analogue adder will see and add
$U_{i,i}$ and all $U_{i,j}$ with $j\neq i;\,
i,j\in\,[\frac{1-n}{2};\frac{n-1}{2}]$ that appear due to the
current $J_i$ at all interconnection points $j\neq i$. Since all
resistors of the chain have the same value, the voltage fractions
$U_{j\leq i}$ and $U_{j>i}$ to the left and to the right of the
injection position $i$ are given by
\begin{equation}
  \label{eq:voltage-fracs}
  U_{j\leq i}=U_{i,i}\frac{j}{i}\qquad\mbox{and}\qquad U_{j>i}=U_{i,i}\frac{j}{n+1-i}.
\end{equation}
As a consequence of the superposition principle, the 
adder sums up all voltage fractions $U_{i,j}$ (including $U_{i,i}$) of
all $J_i$. Additionally each summand will be amplified by the (same)
weight $-R_f/R_s$. Therefore one has to sum over all $i$ and all $j$:
\begin{equation}
  \label{eq:voltage-sum-all}
  U_\Sigma\approx-\frac{R_f}{R_s}\sum_{i,j}^{n} U_{i,j}.
\end{equation}
Equation~(\ref{eq:voltage-sum-all}) can only be an approximation
since the finite impedance of the
summing amplifier's inputs was not taken into account. However, this approximation is sufficiently
good if the adder version of figure~\ref{subfig:adder-with-buffer}
is used. If one considers for a moment the situation where the current
 $J_i$ is only injected at one position, then the sum over $i$ collapses. The
remaining sum over $j$ reads
\begin{equation}
  \label{eq:fraction-sum}
  U_i\approx-\frac{R_f}{R_s}\sum_jU_{i,j}=-\frac{R_f}{R_s} U_{i,i}\left(\frac{1}{i}\sum_{j=1}^{i}j+\frac{1}{n-i+1}%
    \sum_{j=1}^{n-i}j\right)=-U_{i,i}\frac{R_f}{R_s}\frac{n+1}{2} = -U_{i,i}\frac{R_fR}{2R_sR_h},
\end{equation}
where the sum of the arithmetic series was computed in the next to
last step. 
Note that we did not impose any restriction on where the current is
injected, so the contribution of this partial sum to the complete sum
(\ref{eq:voltage-sum-all}) is the same constant factor for all
injection points $i$. It is this particular detail that allows the
computation of the second moment by attaching a summing amplifier to
the proportional resistor chain and hybrid charge divider. Besides, it leads to an
additional amplification of $(n+1)/2$ of the voltage signal $U_{i,i}$ since
$n$ fractions are fed  into the summing amplifier. For
a signal composed of various $J_i$ with different $i$, the sum over $i$
 gives
\begin{equation}
  \label{eq:voltage-sum-disc}
  U_\Sigma\approx\sum_iU_i=-\frac{R_fR^2}{2R_sR_h}\left(\frac{1}{4}\sum_iJ_i-\left(\frac{R_h}{R}\right)^2\sum_ii^2J_i\right).
\end{equation}
The second moment can be normalized by virtue of
equations~(\ref{eq:anger_result}). However, one has to take into
account that the electronic
channel and amplifiers for the voltage sum $U_\Sigma$ and the currents
$J_r$ and $J_l$ in general exhibit unequal amplifications $g_J$ and
$g_\Sigma$ due to different design requirements. One finally gets
\begin{equation}
  \label{eq:voltage-sum-disc-normal}
  \frac{U_\Sigma}{J}\approx\frac{|g_\Sigma|R_fR^2}{2|g_J|R_sR_h}\left(\frac{1}{4}-\left(\frac{R_h}{R}\right)^2\frac{\sum_ii^2J_i}{\sum_iJ_i}\right),
\end{equation}
which can be solved for the wanted second moment of the set of currents
$J_i$:
\begin{equation}
  \label{eq:1D-secmom-DPC}
  \frac{\sum_ii^2J_i}{\sum_iJ_i}\approx\frac{R^2}{4R_h^2}-\frac{2|g_J|R_s}{|g_\Sigma|R_fR_h}\,\frac{U_\Sigma}{J_r+J_l}
\end{equation}
and
\begin{equation}
  \label{eq:1D-secmom-DPC-spatial}
  \frac{\sum_ix_i^2J_i}{\sum_iJ_i}\approx\Delta x^2\left(\frac{R^2}{4R_h^2}-\frac{2|g_J|R_s}{|g_\Sigma|R_fR_h}\,\frac{U_\Sigma}{J_r+J_l}\right)
\end{equation}
respectively.

\begin{figure}[t]
  \centering
  \psfrag{Rh}{$R_h$}
  \psfrag{Rs}{$R_s$}
  \psfrag{Rf}{$R_f$}
  \psfrag{Rhl}{$R_h$}
  \psfrag{Ri1}{$R^\mathit{In}_1$}
  \psfrag{Ri2}{}
  \psfrag{Ri3}{$R^\mathit{In}_i$}
  \psfrag{Ri4}{}
  \psfrag{Ri5}{}
  \psfrag{Ri6}{}
  \psfrag{Ri7}{}
  \psfrag{Ri8}{$R^\mathit{In}_8$}
  \psfrag{OpAmp}{OpAmp}
  \includegraphics[width=0.85\textwidth]{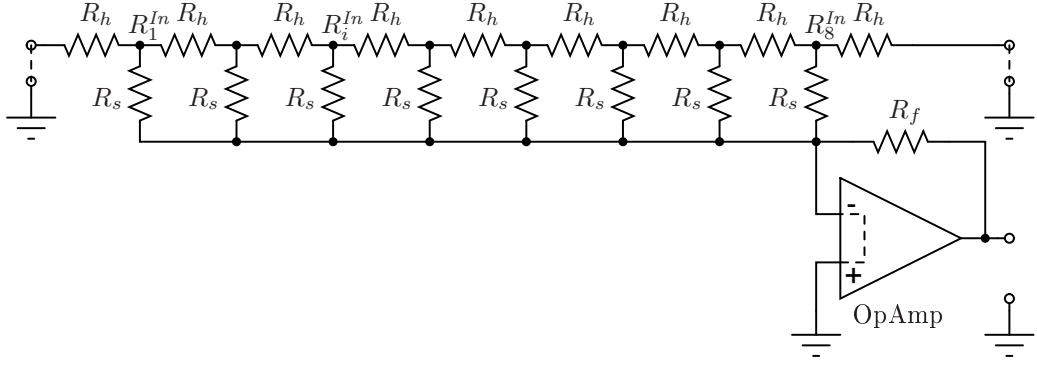}
  \caption[One dimensional proportional resistor chain for 8
    anodes]{One dimensional proportional resistor chain for 8
    anodes. For an ideal operational amplifier, signals that feed into
    its inverting terminal can be treated as if they were connected
    to ground (dashed line inside the amplifier). In this case one can
    compute the voltages at each connection of the resistor network.}
  \label{fig:r-chain-with-adder}
\end{figure}

For the simpler implementation version of the summing amplifier (refer
to figure~\ref{subfig:simple-adder}), an exact solution for the
voltage sum can be found, where {\em exact} means that all electronic
components are considered as ideal. That is to say, the input
impedance of each input is exactly given by $R_s$ (refer to
figure~\ref{fig:r-chain-with-adder}). With the branch current method
for analysis of the network (figure~\ref{fig:r-chain-with-adder}), an
explicit solution for the input impedance at the connection $i$ can be
found:
\begin{equation}
  \label{eq:dis-1D-r-chain-imp-exact}
  R^\mathit{In}_i=\frac{R_d}{n+1}\left(\frac{(n+1)^2}{4}-i^2\right)\kappa_\tincaps{I}(i,\epsilon);  
\end{equation}
\begin{equation}
  \label{eq:adder-imp-correction}
  \begin{split}
    \kappa_\tincaps{I}(i,\epsilon)&=a_{0}+a_{2}\;i^2+a_{4}\;i^4+a_{6}\;i^6;\\[1.4eX]
    a_0& =\frac{(\epsilon +2)
      (\epsilon  (\epsilon +4) (\epsilon  (\epsilon +4) (3637 \epsilon  (\epsilon +4)+25504)+54528)+36864)}{8192 (\epsilon +1) (\epsilon +3)
      \left(\epsilon  (\epsilon +3)^2+1\right) \left(\epsilon  (\epsilon
        +3)^2+3\right)},\\[1.4eX]
    a_2&=\frac{\epsilon  (\epsilon +2) (\epsilon +4) (\epsilon  (\epsilon +4) (1731 \epsilon  (\epsilon +4)+10304)+26880)}{71680
      (\epsilon +1) (\epsilon +3) \left(\epsilon  (\epsilon
        +3)^2+1\right) \left(\epsilon  (\epsilon +3)^2+3\right)},\\[1.4eX]
    a_4&= -\frac{\epsilon ^2 (\epsilon +2) (\epsilon +4)^2 (3
      \epsilon  (\epsilon +4)-224)}{17920 (\epsilon +1) (\epsilon +3) \left(\epsilon  (\epsilon +3)^2+1\right) \left(\epsilon  (\epsilon
        +3)^2+3\right)},\\[1.4eX]
    a_6&= \frac{\epsilon ^3 (\epsilon +2) (\epsilon +4)^3}{4480 (\epsilon +1) (\epsilon +3) \left(\epsilon  (\epsilon
        +3)^2+1\right) \left(\epsilon  (\epsilon +3)^2+3\right)},
  \end{split}
\end{equation}
with $\epsilon=R_h/R_s$. As expected, $\kappa_\tincaps{I}(i,\epsilon)$ tends to
unity for a vanishing $\epsilon$, and for $\epsilon\lesssim0.002$ one can
make $\kappa_\tincaps{I}(i,\epsilon)\gtrsim99/100$. Similarly, an implicit and exact
expression for the sum-voltage $U_i^\Sigma$ that is generated at the output of the
operational amplifier by a single current $J_i$ at the position $i$ 
can be computed.
\begin{gather}
  \label{eq:1D-r-chain-with-adder-exact-sum}
  \frac{U_i^\Sigma}{J_i}=-\frac{R_fR_d}{2R_s}\left(\frac{(n+1)^2}{4}-i^2\right)\kappa_\tincaps{II}(i,\epsilon)\\[1.4eX]
\kappa_\tincaps{II}(i,\epsilon)=\frac{\ds\frac{\left(64 i^6-48 i^4+6924 i^2+127295\right) \epsilon ^3}{1290240}+\ds\frac{\left(16 i^4+184
   i^2+3985\right) \epsilon ^2}{5760}+\ts\frac{1}{48}\ds\left(4 i^2+71\right) \epsilon +1}{(\epsilon +1)
   \left(\epsilon  (\epsilon +3)^2+1\right)}
\end{gather}
Also $\kappa_\tincaps{II}(i,\epsilon)$ in
equation~(\ref{eq:1D-r-chain-with-adder-exact-sum}) tends to unity if $\epsilon$
vanishes. However, the errors in each input impedance get summed up
and one now has to assure that $\epsilon\lesssim0.001$ if one requires a
deviation smaller than 1\% from the ideal sum-voltage that would have
been obtained with a summing amplifier with infinite input impedance.
If a compound signal consisting of various $J_i$ at different
positions is applied to the charge divider their contributions are
superposed. Thus, the sum over $i$ has to be performed as follows:
\begin{equation}
  \label{eq:1D-r-chain-with-adder-total-sum}
  U_\Sigma=-\frac{R_fR_d}{2R_s}\sum_i \left[J_i\left(\frac{(n+1)^2}{4}-i^2\right)\kappa_\tincaps{II}(i,\epsilon)\right].
\end{equation}

\subsubsection{Two-Dimensional Case}
\label{subsec:prop-2D-case}

For the two-dimensional case of the charge dividing circuit with
proportional resistor chains it becomes rather difficult to find an
explicit expression for the voltage sum. As for the one-dimensional
case, a current $J_{ij}$ injected at an arbitrary input $(i,j)$ of
the network causes not only a voltage $U_{ij}$\footnote{The notation for a `punctual'
  current $J_{ij}$ and of the voltage $U_{ij}$ for the two-dimensional cases
  of the charge divider are not to be confused with the voltages
  $U_{i,j}$ (equations~\ref{eq:ij-voltage}-\ref{eq:fraction-sum}) 
that will arise at positions $j$ when the single current $J_i$ is
  injected  into position $i$ of an one-dimensional resistor chain.}
 at this input but also
well-defined fractions at all the other inputs since they are coupled
among themselves through the weighting resistors for the first and
second moments. 
This leads to a distortion of the desired quadratic
weighting for the summing amplifier that is different for both spatial
directions and that can be corrected only partially. An additional
disadvantage is that the expressions that describe the dependency of
the quantity of interest on the configuration of the charge divider
are complex. Therefore it is a clear advantage of the true Anger logic that
each input has their charge dividing resistors for its own and that an exact
quadratic weighting can be implemented in this case. 

An important point is that the electronic configuration of the
proportional resistor network breaks the symmetry with respect to
rotations by 90\textdegree within the $x$-$y$-plane. It was shown
in section~\ref{subsection:2D-prop-net} that one is
able to correct this symmetry breaking for the determination of both
centroids. The input impedance however, which is used for the
determination of the second moment of the signal distribution, shows a
rather different functional dependence along both spatial
directions. Instead of the desired variation proportional to
$(i^2+j^2)$, the position dependent impedance
(\ref{eq:exact-parametrization}) was obtained. Unfortunately, one
cannot adjust the different resistor values of the network in order to
obtain the desired behavior as was done for the true anger
logic, since this would affect the computation of the
centroids. Clearly one also finds this unequal behavior in the
voltage sum that is needed for the measurement of the
second moment. Nevertheless, the summing amplifier itself gives us the
possibility to nearly regain symmetric behavior of the voltage sum. 
If an analogue adder, as shown in figures~\ref{subfig:simple-adder}
or \ref{subfig:adder-with-buffer}, is attached to the proportional
resistor network of figure~\ref{fig:bidim-schem}, one can give each row
of signals a different weight (see
equation~\ref{eq:adder-output-voltage}). But, as shown in the previous section,
along each row the weights of summing amplifier have to be the same
since this is the spatial direction for which the 2D-proportional
network intrinsically shows quadratic position encoding.

First consider the bare 2D-proportional resistor network and suppose
that an ideal summing amplifier is attached to it, {\em i.e.}\ with
$R_{i,j}^\mathit{In}=\infty$. A network analysis now gives a coupled
linear system of 76
equations and 76 unknowns. A general solution depending on the
different resistor values is very hard to find and the result would be
probably much more complex than the exact solutions for the 1D-resistor
chain (equations \ref{eq:dis-1D-r-chain-imp-exact} and
\ref{eq:1D-r-chain-with-adder-exact-sum} in section
\ref{subsection:dpc-with-2m-1D}). For this reason and within the scope
of this work, only a solution for the specific case of the charge
divider configuration reported in case B of Siegel {\em et al.}\
\cite{Siegel:1996} is given; (that is: $8\times8$ anode-segments,
$R_h=10 R_v$ and the corresponding values given in
equation~\ref{eq:solution-laterals}).
For other network configurations, a solution can be found in an
analogous way. 

\begin{figure}[t]
  \centering
    \psfrag{j}{\hspace*{2ex}$j$}
    \psfrag{i}{$i$}
    \psfrag{U}{\hspace*{-3ex}$\frac{U^\Sigma_{ij}}{J_{ij}}$}
    \includegraphics[width=0.47\textwidth]{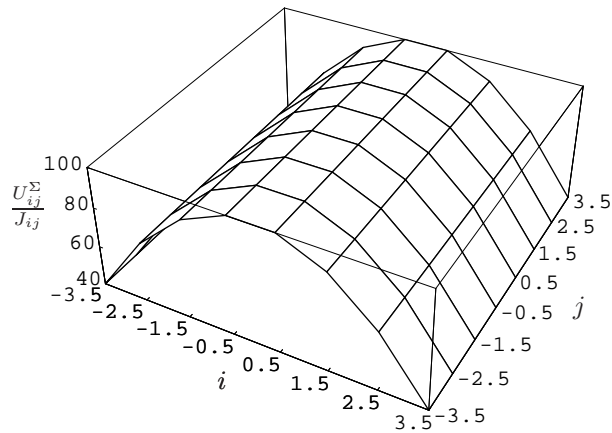}
  \caption[Dependence of the voltage sum on the position of the
  injected current]{Dependence of the voltage sum on the position of the
  injected current for the native 2D-proportional resistor
  network.}
  \label{fig:sumvoltage-native}
\end{figure}

The voltage that can be observed at the
output of the (ideal) summing amplifier of unity gain for all rows when the single current
$J_{ij}$ is injected into the charge divider at the position $(i,j)$
was found to be
\begin{equation}
  \label{eq:sum-voltage-ideal-2D-DPC}
  U_{ij}^\Sigma=-5 \left(\frac{81}{4}-i^2\right) R_v J_{ij}.
\end{equation}
While the column index has the desired quadratic behavior, the row
index $j$ does not even appear in equation
(\ref{eq:sum-voltage-ideal-2D-DPC}). Furthermore, a detailed
comparison between this result and equation
(\ref{eq:voltage-sum-disc}) brings out that the only difference is a
factor of ten. Figure \ref{fig:sumvoltage-native} gives a graphical
representation of $U_{ij}^\Sigma/R_vJ_{ij}$. Once again one has to
sum over $i$ and $j$ if instead of the single current a composite
signal is applied.

\begin{equation}
  \label{eq:SUM-voltage-ideal-2D-DPC}
  U^\Sigma=-5 R_v \sum_{i,j}\left(\frac{81}{4}-i^2\right) J_{ij}.
\end{equation}

Note that the last obtained result (\ref{eq:SUM-voltage-ideal-2D-DPC})
already allows for sensible measurement of the signal distribution's
second moment. This is a direct consequence of the rotational symmetry
about an axis normal to the $x$-$y$-plane and through the
photoconversion point of the signal distribution generated by the
detected $\gamma$-photon which was commented on in section
\ref{ch:light-distribution}. Owing to this symmetry, a supposed
resistor network with attached unity-gain summing amplifier is able to
successfully measure the second moment of the distribution since
they are equal along both spatial directions. The motivation for
searching for a symmetric equivalent to equation
(\ref{eq:SUM-voltage-ideal-2D-DPC}) is founded on the possible
maximization of the SNR of the measurement. An electronic
implementation of a quadratic variation for the $j$ direction would
lead to a highly increased signal while the statistical and
electronic errors remain the same. Actually, one already performs the sum over
all the 64 voltages, but because all rows have the same weight one will
get no benefit from the sum over the index $j$. The summing amplifier
itself allows us to implement different weights for each row of the
network by varying the values of the input resistors $R_1,R_2,\ldots R_n$
of the circuits shown in figures \ref{subfig:simple-adder} and
\ref{subfig:adder-with-buffer}. As a criterion for the determination of
these values, the symmetry behavior of the quadratic encoding of the
anode-segments was used. For this, the voltage sum of the 64 inputs
was computed as a function of the gains $g_1=-R_f/R_{S_1}$,
$g_2=-R_f/R_{S_2}$, $g_3=-R_f/R_{S_3}$ and $g_4=-R_f/R_{S_4}$.
Here, $R_{S_1},\ldots,R_{S_4}$ are the input resistances of the first
and eighth row, the second and seventh row, the third and sixth row and
the fourth and the fifth row. This choice already implements the
desired symmetry with respect to the center of the network. 
For better clarity, the summed voltage at the different inputs of
the enhanced charge divider network is shown in matrix form:
\begin{equation}
  \label{eq:voltage-sum-withgains}
  U^\Sigma_{ij}=-JR_vG_{ij},
\end{equation}
where the gain matrix $G_{ij}$ is given by 
{\scriptsize
\begin{equation}
    {\left(
        \begin{array}{c|c}
          \begin{array}{cccc}
            28 g_1+4 \left(g_2+g_3+g_4\right) &
            58 g_1+4 \left(g_2+g_3+g_4\right) &
            78 g_1+4 \left(g_2+g_3+g_4\right) &
            88 g_1+4 \left(g_2+g_3+g_4\right) \\
            20 g_2+4 \left(g_1+2 g_3+2 g_4\right) &
            50 g_2+4 \left(g_1+2 g_3+2 g_4\right) &
            70 g_2+4 \left(g_1+2 g_3+2 g_4\right) &
            80 g_2+4 \left(g_1+2 g_3+2 g_4\right) \\
            16 g_3+4 \left(g_1+2 g_2+3 g_4\right) &
            46 g_3+4 \left(g_1+2 g_2+3 g_4\right) &
            66 g_3+4 \left(g_1+2 g_2+3 g_4\right) &
            76 g_3+4 \left(g_1+2 g_2+3 g_4\right) \\
            4 \left(g_1+2 g_2+3 g_3\right)+16 g_4 &
            4 \left(g_1+2 g_2+3 g_3\right)+46 g_4 &
            4 \left(g_1+2 g_2+3 g_3\right)+66 g_4 &
            4 \left(g_1+2 g_2+3 g_3\right)+76 g_4
          \end{array} & \cdots \\&\\\hline \vdots & \ddots
        \end{array}\right).}\nonumber
\end{equation}}
\vspace*{-0.7eX}
\begin{equation}\label{eq:voltage-sum-gain-matrix}\end{equation}

The quadrants of the matrix which are not displayed have the same
elements as the one shown but in a different order and according to the
symmetry of the network. If one wants a second moment with same weighting
behavior along both directions $i$ and $j$, one has to make the matrix
$G_{ij}$ symmetric. Unfortunately, this is not
possible and one have to settle for an approximate symmetry between
the $i$ and the $j$ index. As possible Ansatz for finding an almost symmetric
weighting matrix, one can require that at least one row-column pair of 
(\ref{eq:voltage-sum-gain-matrix}) has the desired symmetry. This
gives a linear system of 4 equations which can be solved exactly and
as a function of $g_4$. One needs at least one degree of freedom left in
order to adjust the minimum input impedance of all 64 inputs to a
value that ensures the correct working of the centroid
determination. 

\begin{figure}[t]
  \subfigure[][Voltage sum for the 2D-proportional resistor
  network with maximized symmetry.]{\label{subfig:sumvoltage-symmetrized}
    \psfrag{j}{\hspace*{2ex}$j$}
    \psfrag{i}{$i$}
    \psfrag{U}{\hspace*{-4ex}$\frac{U^\Sigma_{ij}}{R_vJ_{ij}}$}
    \includegraphics[width=0.47\textwidth]{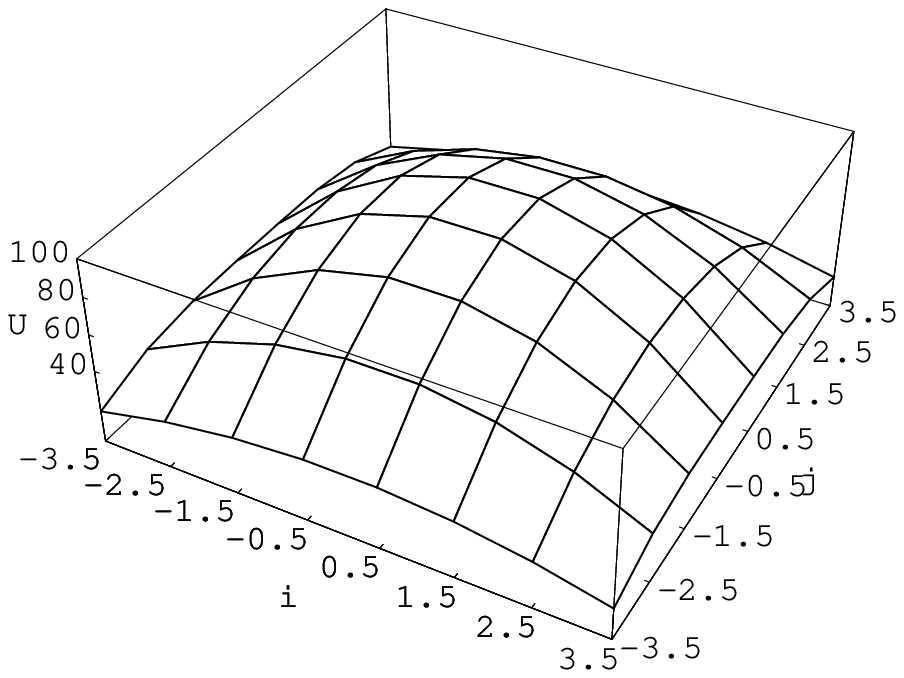}
  }\hspace*{0.5em}
  \subfigure[][Relative deviation in \% of the computed voltage sum from the
  results obtained by simulation.]{\label{subfig:sumvoltage-zymmetrized-errors}
    \psfrag{j}{\hspace*{2ex}$j$}
    \psfrag{i}{$i$}
    \psfrag{E}{\%}
    \includegraphics[width=0.47\textwidth]{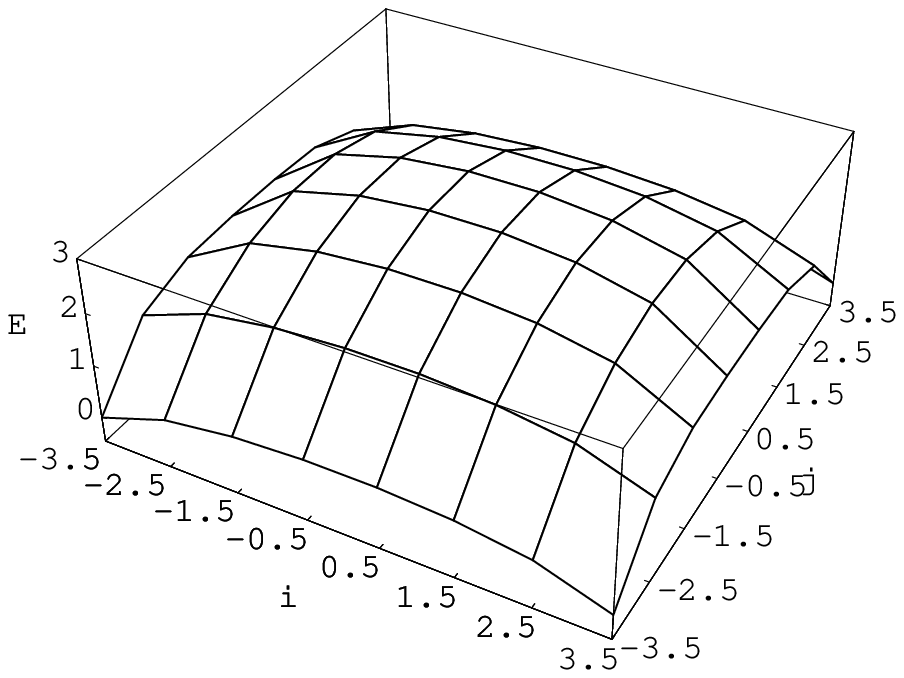}
  }
  \caption[Summed voltages and deviation from simulations for  the
  optimized case]{Summed voltages and deviation from simulations for
    the optimized case.}
  \label{fig:voltage-sums-sym}
\end{figure}

Since one is free in the choice of the column-row pair, there are 4
such sets of equations and for all exist unique but suboptimal
solutions that differ from each other. To find an optimum 
solution to the posed problem, one can measure the asymmetry
$\mathcal{S}[\mathbf{G}]$ of the matrix as follows:
\begin{equation}
  \label{eq:matrix-unsymmetry}
  \mathcal{S}[\mathbf{G}]=\sqrt{\sum_{i,j}\left[\frac{G_{ij}-G_{ji}}{G_{ij}}\right]^2}.
\end{equation}
The following values,
\begin{equation}
  \label{eq:opt-gain-vals}
  g_1 = 0.26972 g_4\mbox{,}\quad g_2=0.57368 g_4,\quad\mbox{and}\quad g_3=0.84035 g_4,
\end{equation}
will minimize expression (\ref{eq:matrix-unsymmetry}) and are the desired
optimum gains for the different resistor rows.  Since the gain for
each row of summing amplifier inputs is $g_i=-R_f/R_{S_i}$,
the corresponding resistor values are given by
\begin{equation}
  \label{eq:opt-r-vals}
  R_{S_1},R_{S_8}=3.7075 R_{S_4}\mbox{,}\quad R_{S_2},R_{S_7}=1.74314
  R_{S_4}\mbox{,}\quad R_{S_3},R_{S_6}= 1.18998 R_{S_4}\quad\mbox{and}\quad
  R_{S_5}=R_{S_4}
\end{equation}
for different weights of the input rows.
With this change in the summing amplifier, the voltage representing
the second moment of the signal distribution now becomes
\begin{equation}
  \label{eq:optimize-sum-voltage}
  U^\Sigma=-5 R_v R_f\sum_{i,j}\left(\frac{81}{4}-i^2\right) \frac{J_{ij}}{R_{S_j}}.
\end{equation}
A plot of the 64 values is shown in figure \ref{fig:voltage-sums-sym}
to the left. In the same figure but to the right, the
relative deviation between the approximate equation
(\ref{eq:optimize-sum-voltage}) and the results obtained with network
simulations is shown. These discrepancies are clearly due to the fact
that the current that is extracted by the analogue adder at each of
the 64 inputs distorts the ideal weighting as derived in section
\ref{subsection:dpc-with-2m-1D}. Furthermore, the network simulator
treats all elemental devices of the circuit as real as possible, while
for the derivation of equation~(\ref{eq:optimize-sum-voltage}) 
ideal components have been supposed.

\subsection{Hybrid Solution}

In this last case of possible implementations of charge division
circuits with second moment capability, one faces a similar situation as
just discussed. One once again has a circuit that is not symmetric with
respect to swapping the coordinates $x$ and $y$ and once again this
symmetry-breaking is not reflected in the measurement of the centroids
but can be observed very well in the measurement of the second
moment. It is clear from equation
(\ref{eq:1D-r-chain-with-adder-total-sum}) and the fact that the four
outputs $J_A,\ldots,J_D$ are connected through the downstream amplification
stage (see section \ref{ch:general-preamplifier-configuration}) to
virtual ground, that the voltage sum for the hybrid circuit has to be
\begin{equation}
  \label{eq:volt-sum-2D-hybrid}
  U_\Sigma=-\frac{R_d}{2}\sum_{i,j} \left[g_jJ_{ij}\left(\frac{(n+1)^2}{4}-i^2\right)\kappa_\tincaps{II}(i,\epsilon)\right].
\end{equation}
If one supposes a summing amplifier of the second type (figure
\ref{subfig:adder-with-buffer}), one can set
$\kappa_\tincaps{II}(i,\epsilon) = 1$. Alternatively one can equate
$\kappa_\tincaps{II}(i,\epsilon)\simeq1$ by choosing the input
resistors of the summing amplifier sufficiently
large. $\kappa_\tincaps{II}(i,\epsilon)$ can then be neglected for the
symmetrization of equation~(\ref{eq:volt-sum-2D-hybrid}).
This substantially eases the symmetrization
of equation~(\ref{eq:volt-sum-2D-hybrid}). Since the rows of the hybrid
circuit (figure~\ref{fig:hybrid-solution}) decouple from each other,
for even $n$ one only has to set
\begin{equation}
  \label{eq:hybrid-row-gain}
  g_j=\frac{4}{n
    \left(n+2\right)}\left(\frac{(n+1)^2}{4}-j^2\right),
\end{equation}
where the factor $4/(n(n+2))$ has been derived from the requirement
of unity gain for both central rows, {\em e.g.}\
$g_{n/2}=g_{n/2+1}=1$. For the case of odd $n$, equation
(\ref{eq:hybrid-row-gain}) has to be adjusted according to this
situation. With the aid of the feedback resistor $R_f$ of the summing
amplifier, the global gain $G=R_f/R_{S_{n/2}}$ is introduced and the
symmetrized version of the sum voltage (\ref{eq:volt-sum-2D-hybrid}) reads 
\begin{equation}
  \label{eq:volt-sum-2D-hybrid-final}
  U_\Sigma=-\frac{2 R_f
    R_d}{R_{S_{n/2}}n\left(n+2\right)}\sum_{i,j} 
\left[J_{ij}\left(\frac{(n+1)^2}{4}-i^2\right)\left(\frac{(n+1)^2}{4}-j^2\right)\right].
\end{equation}

\begin{figure}[t]
  \centering
  \subfigure[][Voltage encoding for an $8\times8$ array of anode
  segments when using the hybrid solution.]{\label{subfig:hybrid-sum-voltage}
    \psfrag{j}{\hspace*{2ex}$j$}
    \psfrag{i}{$i$}
    \psfrag{U}{\hspace*{-5ex}$\frac{U^\Sigma_{ij}}{R_d G J_{ij}}$}
    \includegraphics[width=0.47\textwidth]{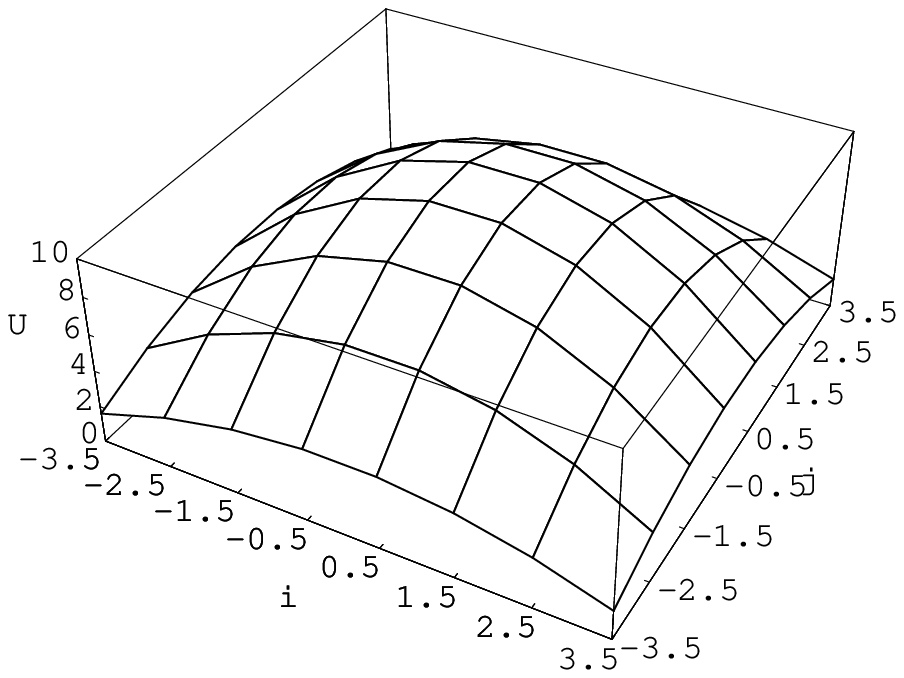}}
  \subfigure[][Relative difference in \% between theoretic model (equation \ref{eq:volt-sum-2D-hybrid-final})
  and simulation results for $n=8$.]{\label{subfig:hybrid-sum-voltage-errors}
    \psfrag{j}{\hspace*{2ex}$j$}
    \psfrag{i}{$i$}
    \psfrag{E}{\%}
    \includegraphics[width=0.47\textwidth]{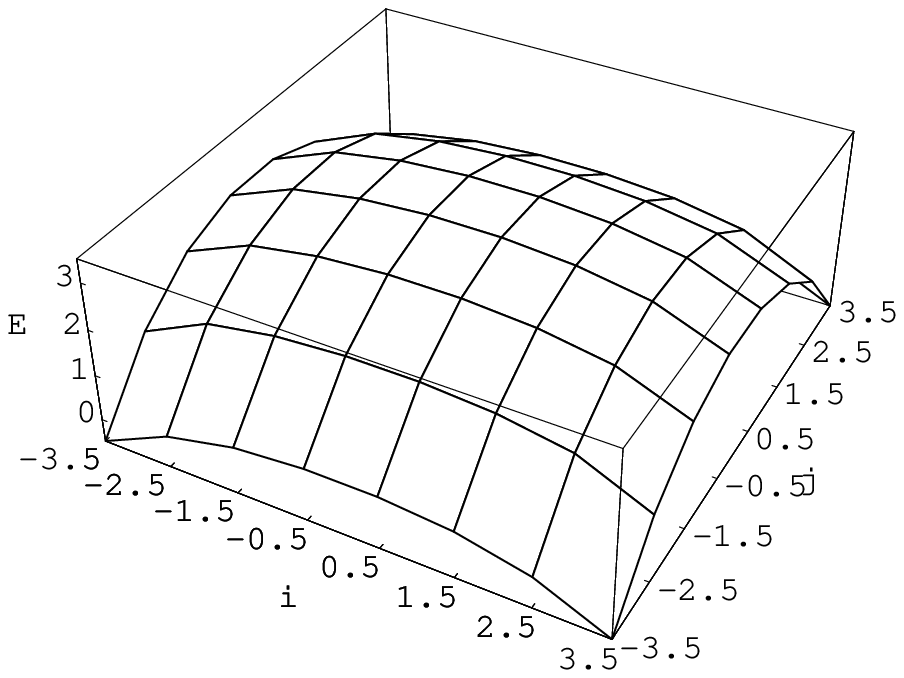}}
  \caption{Corrected sum voltage for the hybrid charge divider configuration.}
\end{figure}
The voltage values are plotted in figure
\ref{subfig:hybrid-sum-voltage} for the case of $n=8$. As for the
other results, expression~(\ref{eq:volt-sum-2D-hybrid-final}) for the
voltage sum of the hybrid solution was verified with {\sc Spice} simulations
({\em Simulation Program with Integrated Circuits Emphasis}, see for
instance Tietze and Schenk \cite{Tietze}). The small differences between the model and the
simulations are shown in figure
\ref{subfig:hybrid-sum-voltage-errors}. They are mainly due to the
approximation made with $\kappa(i,\epsilon)=1$, that is, by the finite
input impedances of the summing amplifier.

\section{Anode Inhomogeneity Compensation}

One point that has not been considered so far are the differences in the
response of the different anode-segments. For real devices one clearly
cannot expect all the detection segments to behave in exactly the same
manner. The deviation in the total sensitivity $S_\mathit{tot}$,
that defines the ratio of the anode current and the incident light
flux can alter significantly from one segment to another, especially
for large size position sensitive PMTs. $S_\mathit{tot}$ is generally
an accumulation of various
effects like the photocathode sensitivity $S_\tincaps{PC}$, the
quantum efficiency $\mathit{QE}_\tincaps{PC}$, the photoelectron
collection efficiency $\eta$, the gain $G_\tincaps{Dyn}$ of the
electron multiplier system and the anode sensitivity
$S_\tincaps{Anode}$ \cite{Flyckt:2002}. The variation over the
detector's spatial extension of each of these factors is finally reflected
in the signal distribution at the output of the detector. Due to the
large number of contributions involved, the inhomogeneity of the
sensitivity is mostly determined experimentally and provided by the
manufacturer in the form of an anode uniformity map for each single device. 

As shown in chapter~\ref{ch:experiment}, the normalized nontrivial
moments that are intended to be measured with the presented
enhanced charge division circuits are nearly unaffected by these
variations. This however does not apply to the 0th moment which
represents the energy. In favor of a high intrinsic energy resolution,
one is tempted to correct for this detector inhomogeneity. In systems
that digitize all channels of the used photodetector independently, the
way to go would be a software correction of these aberrations from
their optimal behavior. In the present case, many computation steps are
performed analogically and the detector-caused inhomogeneity of the
signal results in a systematic error of these computations.
The problem can be solved satisfyingly with additional active
electronic components or, in the case of the Anger logic, with a
passive compensation of the anode inhomogeneity.

\subsection{Passive Compensation}

\begin{figure}[t]
  \centering
  \begin{minipage}[c]{0.52\textwidth}
    \centering
    \subfigure[][Passive compensation method proposed by
    Tornai {\em et al.}\ \cite{Tornai:1996} for use with the Anger
    positioning logic.]{\label{subfig:2D-anger-logic-with-comp}%
      \psfrag{j}{$J_{ij}$}\psfrag{ru}{$R^u_{ij}$}\psfrag{rl}{$R^l_{ij}$}
      \psfrag{rr}{$R^r_{ij}$}\psfrag{rd}{$R^d_{ij}$}\psfrag{rc}{$R^c_{ij}$}
      \psfrag{jl}{$J^l$}\psfrag{jr}{$J^r$}\psfrag{ju}{$J^u$}\psfrag{jd}{$J^d$}
      \psfrag{jli}{$J^l_{ij}$}\psfrag{jl1}{$J^l_{ij-1}$}\psfrag{jl2}{$J^l_{ij+1}$}
      \psfrag{jri}{$J^r_{ij}$}\psfrag{jr1}{$J^r_{ij-1}$}\psfrag{jr2}{$J^r_{ij+1}$}
      \psfrag{jui}{$J^u_{ij}$}\psfrag{ju1}{$J^u_{i-1j}$}\psfrag{ju2}{$J^u_{i+1j}$}
      \psfrag{jdi}{$J^d_{ij}$}\psfrag{jd1}{$J^d_{i-1j}$}\psfrag{jd2}{$J^d_{i+1j}$}
      \includegraphics[width=0.94\textwidth]{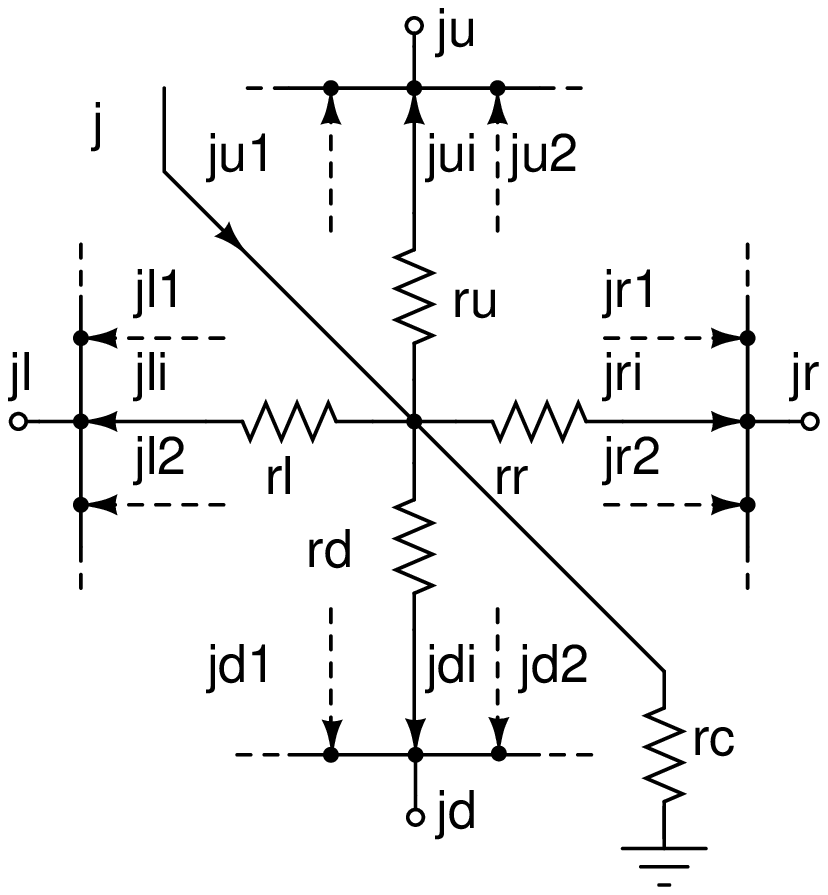}}
  \end{minipage}\hfill
  \begin{minipage}[c]{0.48\textwidth}
    \centering
    \subfigure[][Simplest non-trivial case of proportional charge
    dividing resistor chains with intended passive compensation.]{\label{subfig:1D-r-chain-with-comp}%
      \psfrag{R1}{$R_1$}\psfrag{R2}{$R_2$}\psfrag{R3}{$R_3$}
      \psfrag{Rs1}{$R_{S_1}$}\psfrag{Rs2}{$R_{S_2}$}\psfrag{aJ}{$\alpha J$}
      \psfrag{bJ}{$\beta J$}\psfrag{(1-a)bJ}{\hspace*{-0.6ex}$(1\!-\!\alpha)\,\beta J$}
      \psfrag{(1-b)aJ}{\hspace*{-1ex}$(1\!-\!\beta)\;\alpha J$}
      \includegraphics[width=0.94\textwidth]{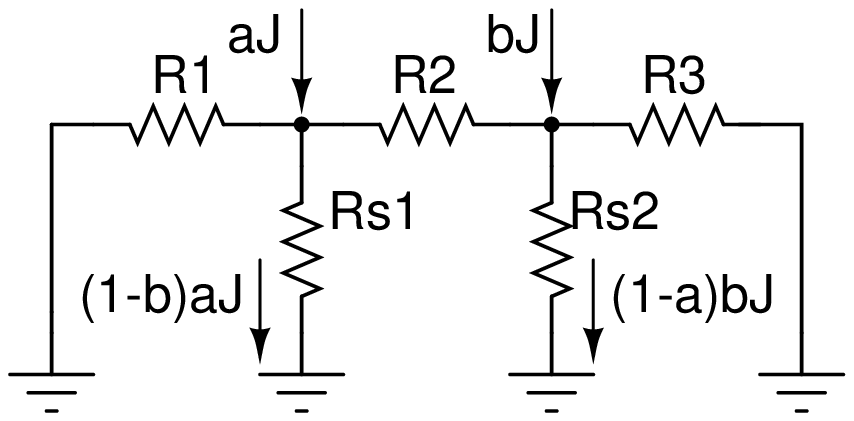}}
    \subfigure[][Special case $0<\beta<\alpha=1$,
    $R_{S_1}=R_S$ and $R_{S_2}=\infty$.]{\label{subfig:1D-r-chain-with-comp-beta}%
      \psfrag{R1}{$R_1$}\psfrag{R2}{$R_2$}\psfrag{R3}{$R_3$}\psfrag{Jl}{$J_l$}
      \psfrag{Rs}{$R_{S}$}\psfrag{J}{$J$}\psfrag{bJ}{$\beta J$}
      \psfrag{(1-b)J}{\hspace*{-1ex}$(1-\beta)J$}\psfrag{Jr}{$J_r$}
      \includegraphics[width=0.94\textwidth]{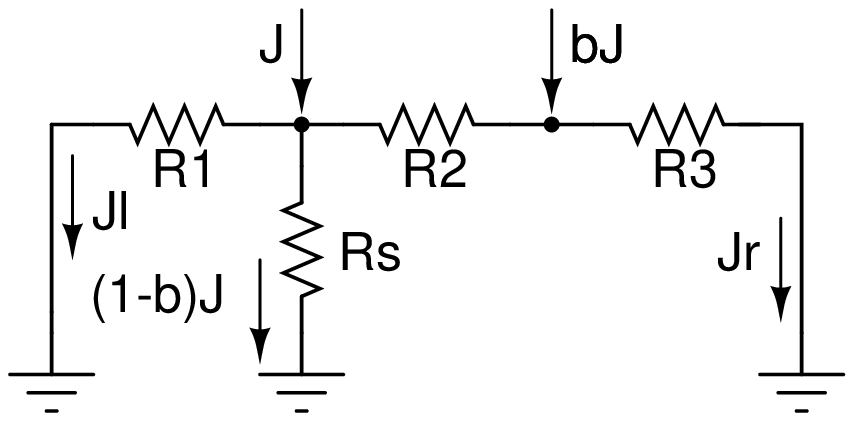}}
  \end{minipage}
  \caption[Electronic configurations for passive anode inhomogeneity
  compensation]
{\label{fig:multi-dim-anger-compensated}%
    Electronic configurations for passive compensation of the spatial
    sensitivity inhomogeneity of photodetectors.}
\end{figure}

The true Anger logic described in section \ref{ch:anger-approach} has
the already mentioned advantage that each anode-segment has its own
 independent charge divider. Therefore it is possible to connect
a (different) correction shunt $R^c_{ij}$ as shown in figure
\ref{subfig:2D-anger-logic-with-comp} at each input of the Anger logic
\mycite{Tornai}{{\em et al.}\ }{1996}. The value of $R^c_{ij}$ that is
required for the compensation of the anode inhomogeneity can be easily
computed from the manufacturer's anode uniformity map and the input
impedance of the Anger logic at the respective position according to
\begin{equation}
  \label{eq:input-imp-anger-logic}
  R_\mathit{In}=\left(\frac{1}{R_{ij}^u}+\frac{1}{R_{ij}^l}+\frac{1}{R_{ij}^d}+\frac{1}{R_{ij}^r}\right)^{-1}.
\end{equation}

For the proportional resistor chain and its derived positioning
circuits, this compensation method cannot be used. This can be
demonstrated for the simple case of a 1D-resistor chain of two inputs
shown in figure \ref{subfig:1D-r-chain-with-comp}. Suppose that the
two anode-segments whose signals $\alpha J$ and $\beta J$ are fed into
this charge divider are of different strength, {\em e.g.}\ $\alpha\neq\beta$,
and are wanted to be equalized. Without loss of generality one can
further assume that $0<\beta<\alpha=1$. The passive compensation works
by reducing all signals of the photodetector to the strength of the
lowest one. Therefore one encounters the situation shown in figure
\ref{subfig:1D-r-chain-with-comp-beta}. A first condition on the
values of the resistors $R_1,R_2,R_3$ and $R_S$ is given by the fact
that the computation of the first moment requires linear weights for
the currents $J$ and $\beta J$. Therefore, the condition 
\begin{equation}
  \label{eq:1-codition-linear-compensated-DPC}
  R_2=R_3=\frac{R_1 R_S}{R_1+R_S}
\end{equation}
has to be fulfilled.
In order to equalize both currents $J$ and $\beta J$, the fraction
$(1-\beta)J$ has to be bled to ground by the resistor $R_S$ and one
obtains (neglecting the trivial solution of $R_S=R_1=R_2=R_3=0$)
\begin{equation}
  \label{eq:2-codition-linear-compensated-DPC}
  R_S=\frac{R_1}{3}\left(\frac{1+\beta}{1-\beta}\right).
\end{equation}
With this configuration, the circuit behaves as desired for the
current $J$ fed into the left input of the circuit. Unfortunately,
 the current $\beta J$ will also be divided at the left node and 
the computation of the centroid gives
\begin{equation}
  \label{eq:erroneous-centroid}
  \langle x\rangle\big|_{\beta\neq1} = \frac{J_r-J_l}{J_r+J_l}=\frac{5}{3}+\frac{8}{3
    (\beta -3)}\neq \langle x\rangle\big|_{\beta\to1}=\frac{1}{3}. 
\end{equation}
This leads to unacceptable systematic errors in the centroid even for
$\beta$ near to unity. As a consequence, one obtains for a composed
signal a positioning error, which is comparable in its size to the
mispositioning, that would exist if the network were not compensated
at all.

\subsection{Active Compensation}

\begin{figure}[!t]
  \centering
  \psfrag{vga}{{\small FGA}}
  \psfrag{Jr}{$J_\mathit{r}$}\psfrag{Jl}{$J_\mathit{l}$}
  \psfrag{Ji}{$J_i$}\psfrag{a1}{anode 1}\psfrag{anp}{anode $n$}
  \psfrag{rnp}{}\psfrag{r1}{}
  \includegraphics[width=0.95\textwidth]{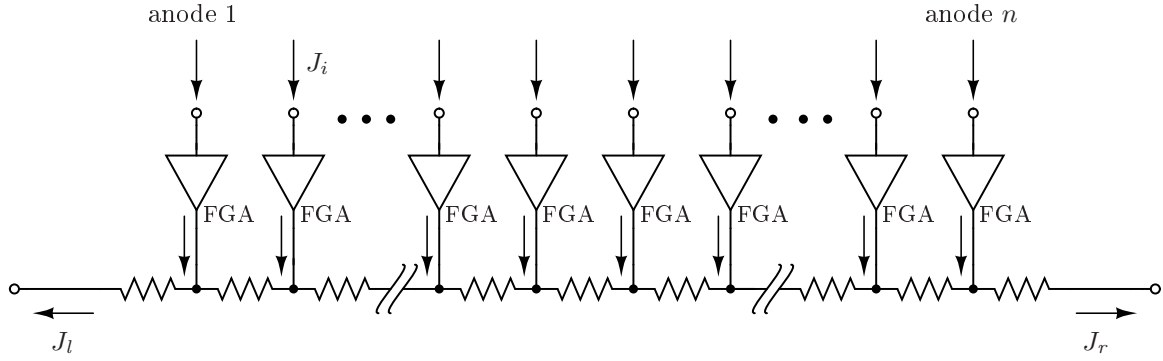}
  \caption[Active anode inhomogeneity compensation using current amplifiers]{\label{fig:active-comp}
    Anode inhomogeneity compensation using current amplifiers. Each of
    these amplifiers will have a different amplification according to
    the anode-segment that it has to correct. It is important that
    the input impedance of the FGA is very low. Otherwise the
    currents $J_i$ will generate voltages $J_i
    R_\mathit{In}^\tincaps{FGA}$ that are opposite in sign to the bias
    voltages of the photodetector and lead to a decrease or even an
    complete breakdown of electron collection or multiplication.}
\end{figure}

A second obvious possibility which works with each of the described
enhanced charge dividers is to use an independent current amplifier for
each of the inputs of the positioning circuit. 
Figure~\ref{fig:active-comp} shows this for a 1D-resistor chain. One
only has to adjust the gain of each single amplifier corresponding to
the anode uniformity map. Besides the universality of this circuit,
another important advantage is that it does not bleed parts of the
signal to ground but amplifies it. For the correct working of this
charge divider circuit, the fixed gain amplifiers (FGAs) have to deliver
currents at their outputs. Furthermore, a very small input impedance
is needed, since  photodetectors for photon-counting applications
normally act as current sources. The main drawback of this method is
its complicated implementation owing to power consumption and required
space on the printed circuit board.

\section{Errors of The Center of Gravity Algorithm}
\label{ch:errors-of-cog-and-cdr}

At the end of the present section, the systematic and statistic errors
that are introduced by the center of gravity algorithm as well as
the imperfection of the photodetectors and readout electronics are
discussed. Important sources for systematic errors are discretization effects 
and border effects. The first are a consequence of the sampling of an
arbitrary signal distribution that enforcedly includes its
discretization. Almost always, this destroys existing symmetries of
the distribution. Border effects arise because it is impossible to realize
detectors of infinite spatial extension and lead also to breaking of
the symmetries of the signal distribution. On the other hand, one faces
signal fluctuations that are caused by the quantization of the signal
and a quantum detection efficiency which in no real case can reach 100\%. 
The thermal noise of the electronic components falls far short of the
other mentioned error sources and it will be shown that one must not 
worry too much about this error.

\subsection{Signal Fluctuations}

Fluctuations in the signal distribution are caused by various
underlying processes. The signal generation starts with the
photoconversion of the $\gamma$-ray into one or more electrons within
the scintillation crystal. Depending on the underlying process,
{\em e.g.}\ the Compton effect, photoelectric effect or pair production, one
or more primary electrons are generated. These primary electrons
produce the scintillation light via the decay of the secondarily ionized
atoms. Together with these random processes, inhomogeneities of the
material and non-proportional energy dependency lead to important
fluctuations in the total light output of the scintillator. While this
affects all anode-segments by the same amount, the distribution of the
finite number of these scintillation photons over the sensitive area of the
detector leads to fluctuation that differs for all
anode-segments. In the following step, the visible
light photons will be converted independently and one by one into
photoelectrons at the cathode of the detector. This is a Poisson
process and probably the most important cause for segment-dependent signal
fluctuations for photomultiplier tubes. Finally, the fluctuation will
lead to an uncertainty $\delta_\tincaps{P}\mu_1$ in the centroid
measurement $\mu_1$, which
can be estimated using error propagation. For this, a set
$\bar{f}^{(\!x_o\!)}=\{f^{(\!x_o\!)}_j\},\,j=[1,2,\ldots,n]$ of $n\in\mathbb{N}$ values that
represent the signals of the anode-segments with center-positions
$x_j$ is considered.
The superscript $(x_0)$ indicates that the signals are generated by
the photoconversion of a $\gamma$-photon at position $x_0$.
If the errors of the anode positions are assumed to be negligible, one
obtains for the error $\delta_\tincaps{P}\mu_1(\bar{f}^{(\!x_o\!)})$ in the centroid
$\mu_1(\bar{f}^{(\!x_o\!)})$ the expression
\begin{equation}
  \label{eq:stat-centroide-error}
  \delta_\tincaps{P}\mu_1(\bar{f}^{(\!x_o\!)})=\sqrt{\sum_j\left[\frac{\partial}{\partial
        f_j}\left(\frac{\sum_i x_i f_i^{(\!x_o\!)}}{\sum_i f_i^{(\!x_o\!)}}\right)\delta
      f_j^{(\!x_o\!)}\right]^2}=\frac{1}{\sqrt{\sum_if_i^{(\!x_o\!)}}}\sqrt{\frac{\sum_j
      (x_j-\mu_1(\bar{f}^{(\!x_o\!)}))^2\delta f_j^2}{\sum_i f_i^{(\!x_o\!)}}},
\end{equation}
where the centroid is given by 
\begin{equation}
  \label{eq:centroide-reminder}
  \mu_1(\bar{f}^{(\!x_o\!)})=\frac{\sum_i x_if_i^{(\!x_o\!)}}{\sum_i f_i^{(\!x_o\!)}}.
\end{equation}
Setting $\delta f_j\propto\sqrt{f_j}$ for the fluctuations of Poisson
dominated process, equation~(\ref{eq:stat-centroide-error})
reduces to 
\begin{equation}
  \label{eq:stat-centroide-error-final}
  \delta_\tincaps{P}\mu_1(\bar{f}^{(\!x_o\!)})\propto\frac{\sigma_\mathit{\mbox{\tiny SD}}}{\sqrt{\sum_i f_i^{(\!x_o\!)}}},
\end{equation}
where $\sigma_\mathit{\mbox{\tiny SD}}$ denotes the standard deviation
for the set $\bar{f}^{(\!x_o\!)}$.
This result is not unexpected. Since the standard deviation gives an
idea of the dispersion of the set of variables $\{f_j^{(\!x_o\!)}\}$,
equation~(\ref{eq:stat-centroide-error-final}) states that the error
of the
centroid is smaller the narrower the signal distribution is.
Furthermore, the result  also scales with the square root of the 
sum of all signals $\{f_j^{(\!x_o\!)}\}$, {\em e.g.}\ the total amount of photons collected 
by all the anode-segments. One therefore expects a better position estimate 
from the center of gravity algorithm the more light is released by 
the $\gamma$-ray and the narrower the distribution of this light is.

\begin{figure}[t]
  \centering
  \subfigure[][Interaction depth
  $2\,\mathrm{mm}$.]{\label{subfig:sampling-broadening-2mm}
    \psfrag{J}{}
    \psfrag{x}{$x$}
    \includegraphics[width=0.45\textwidth]{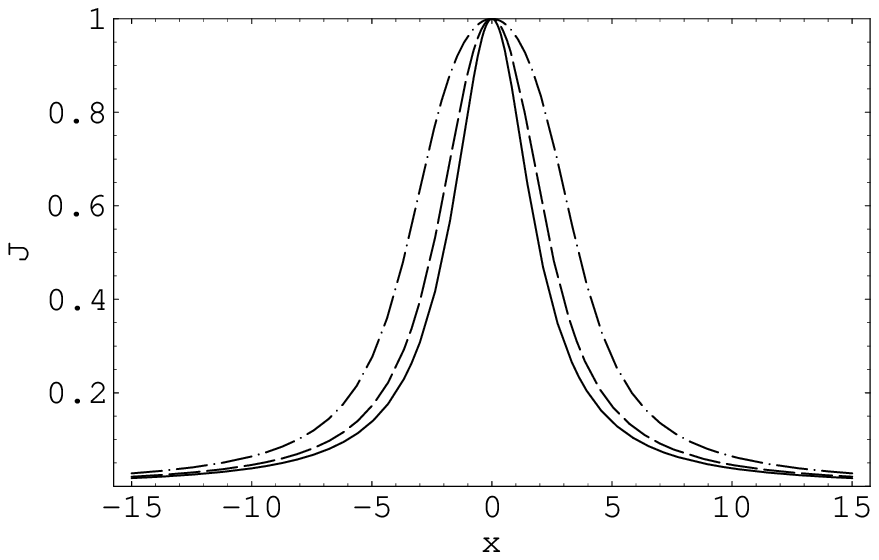}}\hspace*{0.05\textwidth}
  \subfigure[][Interaction depth $6\,\mathrm{mm}$.]{\label{subfig:sampling-broadening-6mm}
    \psfrag{J}{}
    \psfrag{x}{$x$}
    \includegraphics[width=0.45\textwidth]{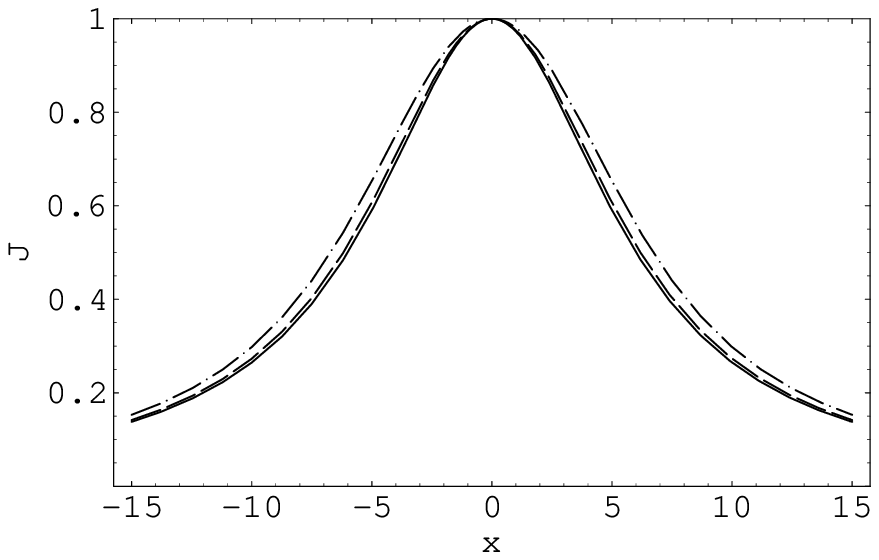}}
  \caption[Broadening of the signal distribution owing to the pixel size]{Broadening of the original distribution (solid line)
    introduced by a sampling with an anode-segment size of $3\,\mathrm{mm}$
    (dashed line) and $6\,\mathrm{mm}$ (dot-dashed line) for 2 different
    interaction depths. All graphs are normalized to
    $f_{x_o}(x)\big|_{x=0}=1$ for easier comparison of
    the width.}
  \label{fig:sampling-broadening}
\end{figure}

As already mentioned, any measurement of the (characteristic) parameter
of the distribution observed with the photodetector has to be sampled 
and therefore will be only available in the form of the finite set of 
values $\{f_j^{(\!x_o\!)}\}$. Normally, this sampling is done by integrating the 
distribution of interest piecewise over many small and equidistant 
intervals. Mathematically, this amounts to a convolution \mycite{Landi}{}{2002}
of the real distribution $\varphi(x)$ with the interval function\footnote{The 
interval function is defined through $\Pi(x)=\left\{
    \begin{array}{ccc}
      1 & & |x|< 1/2 \\ 1/2 & \mbox{for} & |x|=1/2 \\ 0  & & |x|> 1/2 
    \end{array}\right.$.} $\Pi(x)$ 
of width $\tau$: 
\begin{equation}
  \label{eq:sampling-convolution}
  f_{x_o}(x)=\int_{-\infty}^{\infty}\Pi\left(\frac{x-x'}{\tau}\right)\varphi(x'-x_0)dx'.
\end{equation}
The set of measured values $\{f_j^{(\!x_o\!)}\}$ is just given by the functional value of 
$f_{x_o}(x)$ at the positions $x=x_j$:
\begin{equation}
  \label{eq:sampling-values}
  f_j^{(\!x_o\!)}=f_{x_o}(x)\big|_{x_j}.
\end{equation}
It is clear from equation~(\ref{eq:sampling-convolution}),
that the resulting distribution $f_{x_o}(x)$ has to be wider than 
the original distribution $\varphi(x)$, since one always has $\tau>0$ 
for any real photodetector. This means together with
equation~(\ref{eq:stat-centroide-error-final}),
that when the CoG algorithm is used together with a continuous
light distribution, the pixel size of the detector used is of great
importance. As an example, the projection of the normalized
one-dimensional inverse square law onto the abscissa $x$,
\begin{equation}
  \label{eq:test-func-for-stat-err-cog}
  \varphi(x)=\frac{d J}{\pi(x^2+d^2)},
\end{equation}
is used to study the influence of the detector pixel size $\tau$ on
the width of the sampled distribution $f_{x_o}(x)$. For the
computation one can set $x_0=0$ w.l.o.g. While the undisturbed
distribution~(\ref{eq:test-func-for-stat-err-cog}) has a FWHM equal to
$2d$, the sampled distribution
$f_{x_o}(x)$ will have the width $\sqrt{\tau^2+4d^2}$. The results are
plotted in figure~\ref{fig:sampling-broadening} for two different
depths $d$ and different values of $\tau$. Obviously, this result is
not directly applicable to
equation~(\ref{eq:stat-centroide-error-final}) since there the standard
deviation was computed. Nevertheless, the qualitative result that
one gets a larger statistical error of the CoG for a
coarser sampling of the distribution still holds since the standard
deviations as well as the FWHMs of the distributions $f_{x_o}(x)$ and
$\varphi(x)$ are strictly increasing functions.

The statistical error made in the second moment is not related to the
center of gravity algorithm but will be given for completeness. It can be computed
equivalently to (\ref{eq:stat-centroide-error}) and yields
\begin{equation}
  \label{eq:stat-secmom-error}
  \delta_\tincaps{P}\mu_2(\bar{f}^{(\!x_o\!)})=\sqrt{\sum_j\left[\frac{\partial}{\partial
        f_j}\left(\frac{\sum_i x^2_i f_i^{(\!x_o\!)}}{\sum_i f_i^{(\!x_o\!)}}\right)\delta
      f_j^{(\!x_o\!)}\right]^2}\,\,\stackrel{\delta f_j=\sqrt{f_j}}{=}\,\,\frac{1}{\sqrt{\sum_i
      f_i^{(\!x_o\!)}}}\sqrt{\sum_j\left[\left(x^2_j-\frac{\sum_ix_i^2f_i^{(\!x_o\!)}}{\sum_if_i^{(\!x_o\!)}}\right)^2 f_j^{(\!x_o\!)}\right]}.
\end{equation}
Note that $\sum_i(x_i^2f_i^{(\!x_o\!)})/(\sum_if_i^{(\!x_o\!)})$ is just the central second
moment for $x_0=0$. 

\subsection{Discretization Errors}

The necessary discretization of the signal distribution leads
to additional systematic errors \mycite{Landi}{}{2002}.
The definition of the moments for a distribution $\varphi(x)$ was given by (see
equation~(\ref{eq:func-moments}) in section \ref{ch:stat-estimates})
\begin{equation}
  \label{eq:gen-moment-once-again}
  \mu_k(x)=\int_{\omega}x^k\varphi(x)dx.
\end{equation}
At this level and for normalized $\varphi(x)$, the method gives a
perfect position measurement for $k=1$ and all other moments for
$k\neq1, k\in\mathbb{N}$ whenever the integration interval $\omega$
covers the support of the distribution. Unfortunately, $\varphi(x)$ is not
accessible to experiments, $\varphi(x)$ is not normalized, nor is it
possible to build detectors of infinite spatial extension.
Each real detector performs a discrete reduction that essentially modifies the
properties of the method. In its simplest form, 
the sampling is a set of finite disjoint integrations of the
distribution. Therefore the detector readout is the set of the
following $n$ numbers
\begin{equation}
  \label{eq:sampling-values-alt}
  f^{(\!x_o\!)}_j=\int_{\tau(n-\half\!)}^{\tau(n+\half\!)}\!\!\!\varphi(x-x_0)\,dx,
\end{equation}
which is just another definition of $\{f^{(\!x_o\!)}_j\}$  equivalent
to the one given with equation~(\ref{eq:sampling-convolution}). Except for the case of
$\tau\to0$ and $n\to\infty$, the centroid~(\ref{eq:centroide-reminder})
differs from $x_0$ almost everywhere by a systematic error due to the
discretization. This error was studied in detail by G.\ Landi
\cite{Landi:2002}. It was found to be 
\begin{equation}
  \label{eq:discretization-cent-error}
  \mu_1(\bar{f}^{(\!x_o\!)})=x_0+\frac{\tau}{\pi}\sum_{k=1}^\infty\frac{(-1)^k}{k}
  \sin{(2\pi kx_0/\tau)}\Phi(2\pi k/\tau),
\end{equation}
where $\Phi(\omega)$ is the Fourier transform of $\varphi(x)$.
The $x_0$ dependency of the discretization error has the form of a
Fourier Series with the amplitudes $(-1)^k\Phi(2\pi k/\tau)/k$ that
scales with the sampling interval $\tau$. Evidently the discretization
error completely vanishes for all $x_0=\tau n/2,\, n\in\mathbb{Z}$,
that is, for all those cases when $x_0$ is located exactly over the
center of one interval or exactly between two intervals. At these
special points, the symmetry of the distribution will be correctly
reproduced by the detector. If $\varphi(x)$ converges to the Dirac
$\delta$-function, the discretization error reaches its maximum
values. This is intuitively expected because the exact position of the
Dirac $\delta$-function within one interval cannot be determined.

One obtains the following simple form of the Fourier transform for the
special signal distribution~(\ref{eq:test-func-for-stat-err-cog}) with
frequencies $\omega=2\pi k/\tau$:
\begin{equation}
  \label{eq:foureier-weigths-test-func}
  \Phi(2\pi k/\tau)=\frac{e^{-\frac{2 d k \pi }{\tau }} J}{\sqrt{2 \pi }}.
\end{equation}
With this result, one can easily find an upper limit for the
discretization error (\ref{eq:discretization-cent-error}), since
$e^{-2d\pi/\tau}/\sqrt{2\pi}$ majorizes\footnote{One also has to pay
attention to the sine factor within the
series~(\ref{eq:discretization-cent-error}). Generally it is possible
that this factor cancels exactly the alternate sign
$(-1)^k$ for all $k\in\mathbb{N}$ inside the sum. This happens for
$x_0=\tau/2$ and $k=(2n+1)/2,\,n\in\mathbb{N}$ which is
just the previously discussed case where the center of gravity
algorithm does not break the symmetry of the distribution and does not
contribute any error at all.} the series
(\ref{eq:discretization-cent-error}) and by computing the alternating
harmonic series one gets the following expression for the upper limit
of the discretization error:
\begin{equation}
  \label{eq:upp-lim-discrete-error}
  \delta_\tincaps{D}\mu_1(\bar{f}^{(\!x_o\!)})=\mu_1(\bar{f}^{(\!x_o\!)})-x_0=\pm\frac{e^{-\frac{2d\pi}{\tau}}\tau\ln(2)}{\sqrt{2}\pi^{3/2}}.
\end{equation}
From the previous section it is known that for the present example
distribution $d$ is proportional to its width. Therefore
equation~(\ref{eq:upp-lim-discrete-error}) states that the discretization
error becomes significant for detector setups with $\tau$
much larger than the width of the distribution to be measured.

\subsection{Symmetry Breaking of the Current Distribution}
\label{sec:symmetry-breaking}

Another systematic error which is of special interest within the scope of the
present work is introduced by the fact that all practicable
detectors can never be of infinite size. Owing to this intrinsic
property of the devices, the sampled signal distribution has a finite
support. This error is well known and because of its severity it has
been subjected to detailed studies by many investigation groups 
({\em e.g.}\ Freifelder {\em et al.}\ \cite{Freifelder:1993}, Clancy {\em et al.}\
\cite{Clancy:1997}, Siegel {\em et al.}\ \cite{Siegel:1995} and Joung et
al.\ \cite{Joung:2002}) in order to avoid or combat this degradation
often referred to as {\em image compression} or {\em edge effects}.
Actually, this error is a simple consequence of applying the CoG 
algorithm with an integration interval smaller than the support for
the signal distribution. It can be observed for all moments whose
distribution is (necessarily) 
truncated by the detector design. Being one of the most important of
all errors considered here, it often forces the designer of the
detector to restrict the usable area to a central region and render
inoperative a large peripheral area. There are many recipes to
extract the best results from each detector technology and geometry,
{\em e.g.}\ scintillator or semiconductor, pixels or continuous. The error
discussed in this section depends crucially on the detector
configuration. Here, only the special configuration of a scintillation
detector with continuous crystal is considered.
Even in this case, the response performance can be varied by different
approaches in order to optimize total light yield and other important
parameters. Some well-known design enhancements, for example
diffusive reflective coating, are, however, excluded for an optimal performance of
the depth of interaction measurement. Since reflection on the crystal
edges would destroy the correlation between the distribution width and
the depth of interaction (not the second moment), all crystal
surfaces that are not read out by a photodetector have to be coated 
anti-reflective.

First, the 0th moment is considered since it is required for the
energy measurement and for the normalization of the center of gravity
algorithm according to equation~(\ref{eq:func-centroid}). The error
that is made by truncating the integration becomes clear from
splitting the integral into the integration over the support $L$ of
the detector and two residual parts $]\!-\infty,\frac{-L}{2}]$ and
$[\frac{L}{2},\infty\;[$ to the left and to the right respectively:
\begin{equation}
  \label{eq:sym-error-energy}
  \mu_0^{(\!x_0\!)}=J\int_{-\frac{L}{2}}^{\frac{L}{2}}\varphi(x-x_0)\;dx=
  J-J\int_{\frac{L}{2}}^{\infty}\left\{\varphi(x-x_0)+\varphi(-x-x_0)\right\}dx  .
\end{equation}
Note that the signal distribution will have its symmetry center
at $x_0$ and not $x=0$. Therefore, except for the case $x_0=0$, the
integration of $\varphi(x-x_0)$ over the intervals $]\!-\infty,\frac{-L}{2}]$ and
$[\frac{L}{2},\infty\;[$ give different results. The generalization of equation
(\ref{eq:sym-error-energy}) to any $k$-th moment is straightforward and
given by
\begin{equation}
  \label{eq:sym-error-k-mom}
  \mu_k^{(\!x_0\!)}=J\int_{-\frac{L}{2}}^{\frac{L}{2}}x^k\varphi(x-x_0)\;dx=
  J\mu_k-J\int_{\frac{L}{2}}^{\infty}x^k\left\{\varphi(x-x_0)+(-1)^k\varphi(-x-x_0)\right\}dx,
\end{equation}
where the $\mu_k$ are the normalized distribution moments according to
its definition~(\ref{eq:func-moments}), and the normalization of the
moments $\mu_k^{(\!x_0\!)}$ has to be done with
$\mu_0^{(\!x_0\!)}$. From equations~(\ref{eq:sym-error-energy}) and
(\ref{eq:sym-error-k-mom}) it becomes clear that the introduced error
depends on the signal distribution $\varphi(x)$ in a non-trivial way.
Nevertheless, some general properties can be extracted from them.
Consider the case of a distribution that is symmetric around the
impact point $x_0$, {\em i.e.}\ $\varphi(x-x_0)=\varphi(x_0-x)$. Actually,
this is a property of many signal distributions of interest in the
field of experimental high energy physics and/or when using
scintillation detectors. Result (\ref{eq:sym-error-k-mom}) splits then
into two cases depending on whether $k$ is odd or even. The error in the
odd moments vanishes completely for $x_0=0$.
Contrariwise, the error in the even moments will never vanish for
the class of non-trivial distributions with $\varphi(x)\neq0,\,x\in[L/2,\infty\;[$.  
Obviously this is the symmetry behavior of the CoG
algorithm and the error that distinguishes the real impact position
$x_0$ from the centroid $\mu_1^{(\!x_0\!)}$ comes along with the
breaking of the symmetry of the distribution by the integration over
the finite interval.

\begin{figure}[t]
  \centering
  \subfigure[][Behavior of positioning for different depth of
  interactions ($d=0,2,4,\ldots,10\,\mathrm{mm}$) if the center of gravity
  algorithm is used. $L$ was set to $20\,\mathrm{cm}$.]{\label{subfig:theo-sample-centroid}
    \psfrag{x}{\hspace*{-8ex}true position [mm]}
    \psfrag{y}{\hspace*{-6ex}centroid [mm]}
    \includegraphics[width=0.45\textwidth]{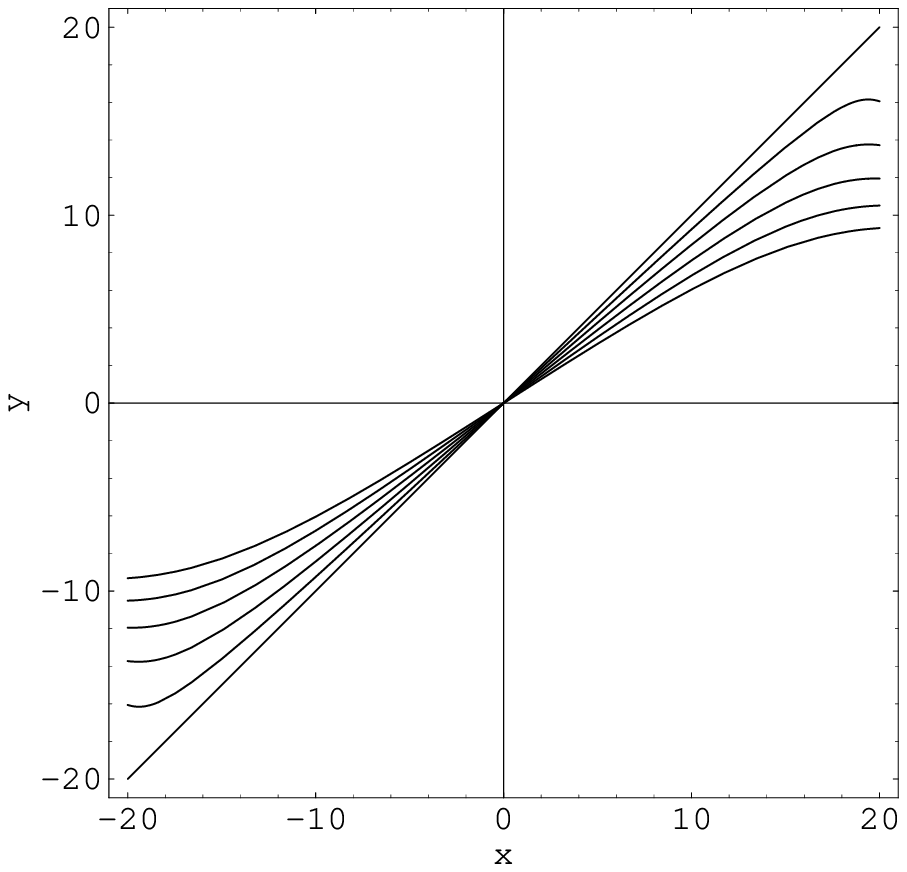}}\hspace*{0.05\textwidth}
  \subfigure[][Broadening of the point spread function due to the
  nonlinearity in the position response of the truncated center of
  gravity algorithm. The identity $x=y$ is plotted as dashed line.]{\label{subfig:inversion-broadening}
    \psfrag{x}{\hspace*{-9ex}measured centroid [mm]}
    \psfrag{y}{\hspace*{-9ex}inverted position [mm]}
    \includegraphics[width=0.45\textwidth]{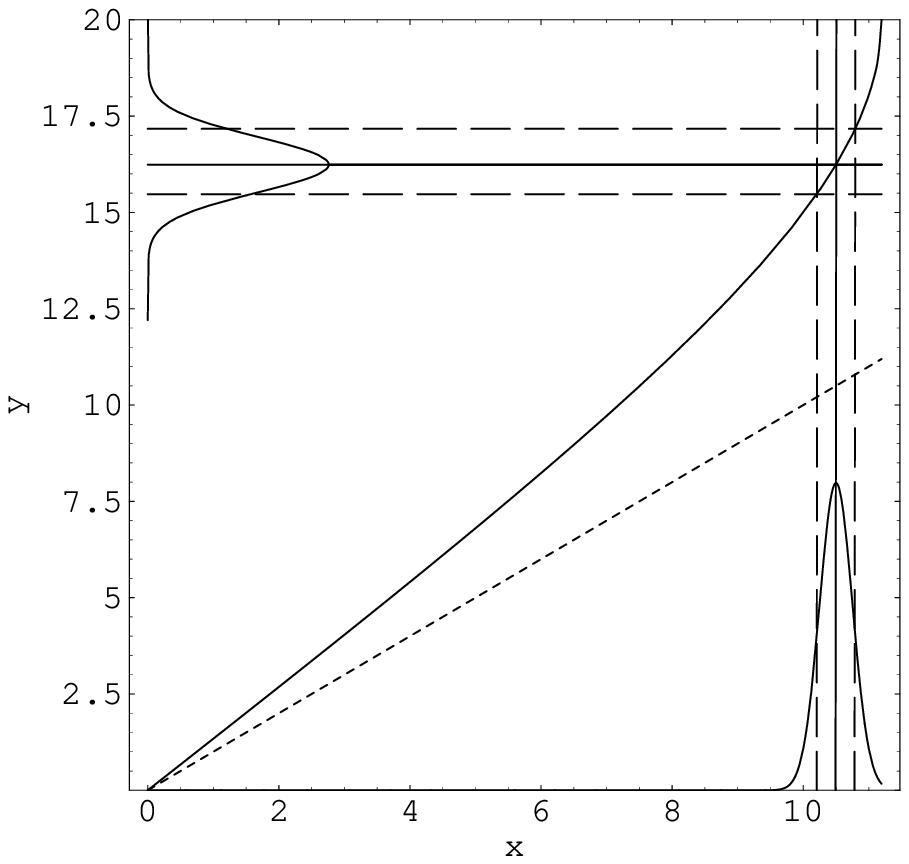}}
  \caption[Behavior of positioning of the center of gravity
  algorithm]{Behavior of positioning of the center of gravity
    algorithm and consequences for the linearization.}
  \label{fig:theo-sample-centroid}
\end{figure}

For the test distribution (\ref{eq:test-func-for-stat-err-cog}) the
centroid actually measured is given by:
\begin{equation}
  \label{eq:moments-for-sample-dist}
  \mu_1^{(\!x_0\!)}=x+\frac{d \log \left(\frac{4 d^2+(L-2 x)^2}{4 d^2+(L+2 x)^2}\right)}{2 \left(\cot ^{-1}\left(\frac{2 d}{L-2 x}\right)+\cot
   ^{-1}\left(\frac{2 d}{L+2 x}\right)\right)}.
\end{equation}
The other finite support moments $\mu_k^{(\!x_0\!)}$ can be obtained similarly by straightforward
application of the definition (\ref{eq:sym-error-k-mom}). Result
(\ref{eq:moments-for-sample-dist}) is plotted for various impact
depths $d$ and a typical detector size $L$ in figure
\ref{subfig:theo-sample-centroid}. From equation
(\ref{eq:moments-for-sample-dist}) and its graphical representation in
figure \ref{subfig:theo-sample-centroid}, it is seen that the truncation
of the signal distribution at the detector's limits compresses the
identity $y=x$ at the detector's edges. The closer the impact position
is to the edge, the stronger the compression. The degree of
compression varies also with the depth of interaction. This compression not
only results in a characteristic mapping of the impact position, but
also leads to a resolution loss at the detector's edges. Since it will
be attempting to find the inverse mapping of the dependences in figure 
\ref{subfig:theo-sample-centroid}, the point spread function of the
detector will be broadened as shown in figure
\ref{subfig:inversion-broadening}. A detector response that
compresses the images at its edges is corrected by the application of an
expanding reverse mapping. However, not only will the center-points of the point
spread function be dispersed, but also the point
spread function itself. Together with the depth dependence of the
detector's positioning characteristic, the distortion at peripheral
zones of the sensitive area are subjected to a significant resolution
degradation. The different image compressions for different impact
depths easily lead to superposition of impact positions that makes the
bare center of gravity algorithm useless except for a small region at
the center, especially for applications that require thick
scintillators. 

\subsection{Electronic Noise}

Finally, the influence of thermal noise generated by the passive
and active electronic devices is considered. In general, the
spectral density of the noise power $\hat{P}$ at the frequency $f$ is given by 
$\hat{P}=k_BTf$, where $k_B$ denotes the Boltzmann constant and
$T$ the absolute temperature of the device (Tietze and Schenk \cite{Tietze}). Since the
power is also given by $P=U^2/R$ and by virtue of Ohms law, the
spectral density of the noise voltage and current can be derived to yield
\begin{equation}
  \label{eq:noise-voltage}
  \hat{U}=\sqrt{4Rk_BTf}\quad\mbox{ and }\quad\hat{I}=\sqrt{4R^{-1}k_BTf},
\end{equation}
where the factor 4 arises from averaging the $\sin^2(t)$ function over a
full period. At room temperature, $4k_BT$ amounts to $1.6\cdot10^{-20}\,\mathrm{Ws}$
and therefore the noise voltage of a resistor is given by 
$\hat{U}_R\simeq0.13\,\mathrm{nV}\times\sqrt{Rf}$. Clearly the noise voltages of
any two resistors are not correlated at all and the $n\times n$
noise voltages of the summing amplifier have to be added
quadratically, while one has to take into account that they will be
amplified along with the desired signal voltage.
\begin{equation}
  \label{eq:sum-noise-voltage}
  \hat{U}_\mathit{noise}^\Sigma\simeq\sqrt{\hat{U}^2_\mathit{noise_{OP}}+\hat{U}^2_\mathit{R_C}+4k_BTf\sum_{i,j}^{n,n}g^2_{ij}R_{ij}}
\end{equation}
Here, $\hat{U}_\mathit{noise_{OP}}$ is the noise contribution to the output
signal of the operational amplifier itself and
$\hat{U}^2_\mathit{R_C}$ the contribution caused by the compensation
resistor $R_c$ (refer to figure \ref{fig:analogue-adder} in section
\ref{sec:sim-measurement-of-sec-mom}). Using the network configuration
of the two-dimensional proportional charge divider (figure
\ref{fig:bidim-schem}) together with the optimum values for the
summation weights computed in
section~\ref{sec:proportional-resistor-chains}, one obtains a typical
total noise voltage of $U_\mathit{noise}^\Sigma\approx160\,\mathrm{\mu V}$.
The spectral noise $\hat{U}_\mathit{noise_{OP}}$ of the
operational amplifier was assumed to be
of $10\,\mathrm{nV\sqrt{Hz}}$, which is a typical value for
modern operational amplifier. Therefore, the strongest contribution in the sum
(\ref{eq:sum-noise-voltage}) arises, with $\approx140\,\mathrm{\mu V}$, from the
operational amplifier itself. The reason for the rather low contribution
from the summation network ($\approx25\,\mathrm{\mu V}$) lies in the posterior
attenuation of the signal. Actually, the presented summation amplifier,
with a mean gain of factor $\approx0.003$, is more likely to be called
a summation attenuator. This very small gain is needed in order
to avoid saturations of the operational amplifier but also effectively
suppresses the  thermal noise of the resistor network. 

Typical  signal amplitudes observed with our experimental
setup described in chapter \ref{ch:experiment} are of $50\,\mathrm{mV}$
and more. Thus, the electronic noise of the summing amplifier can
be neglected compared with the other error sources discussed in this
section. The same holds for the currents $J_A$, $J_B$, $J_C$ and $J_D$.

\chapterbib


  \cleardoublepage{}
\chapter{Compton Scattered Events}
\label{ch:compton}

\chapterquote{%
We did not inherit this earth from our parents,
we are borrowing it from our children.}{
old native American proverb
}

\PARstart{A}{n} important issue that has not been addressed so far is the influence
of Compton scattered events on the presented method for the determination
of the interaction depth. Compton scattering is the process in which
a $\gamma$-photon and an electron scatter off one another. Owing to
its small rest energy $E_e=m_ec^2=511\,\mathrm{keV}$, a
significant amount of the $\gamma$-ray's incidence energy is
transferred to the target electron. It is therefore referred to as
{\em incoherent scattering} in contrast to the {\em coherent} Rayleigh
scattering that describes the process in which a $\gamma$-photon and
an atom scatter off of one another. Coherent scattering entails
usually only a small change in the photon's original direction and
there is virtually no loss of energy. Compton and Rayleigh scattering
are only two of a number of individual processes by which photons
interact with matter, {\em e.g.}\ photoelectric absorption, pair production
and the nuclear photoelectric effect.\footnote{The nuclear photoelectric effect
  describes the process of liberating an electron from the atomic
  nucleus or one of its nucleons.} The probability that the incident photon
undergoes one of these processes when traversing the target matter can
be expressed in terms of the mass attenuation coefficient
$\mu/\rho\,\mathrm{[cm^2/g]}$, where $\mu\,\mathrm{[cm^{-1}]}$ is the (density dependent)
linear attenuation coefficient and $\rho\,\mathrm{[g/cm^3]}$ the material's density. 
The mass attenuation coefficient is related to the total atomic
cross section $\sigma_\mathit{tot}\,\mathrm{[cm^2/atom]}$ by the atomic
weight. This quantity can also be obtained as the sum of cross
sections for the mentioned different types of possible interactions of
the photons with the target material:
\begin{equation}
  \label{eq:cross-section-sum}
  \sigma_\mathit{tot}=\sigma_\mathit{pe}+\sigma_\mathit{incoh}+\sigma_\mathit{coh}+\sigma_\mathit{pair}+\sigma_\mathit{phn},
\end{equation}
in which $\sigma_\mathit{pe}$ is the atomic photoelectric effect cross section,
$\sigma_\mathit{incoh}$ and $\sigma_\mathit{coh}$ are the Compton and
the Rayleigh cross sections respectively, $\sigma_\mathit{pair}$ is
the cross section for pair production and $\sigma_\mathit{phn}$ is the
nuclear photoelectric effect cross section (refer to Hubbell
\cite{Hubbell:1999}, Johns and Cunningham \cite{Johns:1983}).
All contributions in equation~\ref{eq:cross-section-sum} show a strong
energy dependency. For photon energies below $2m_ec^2$ pair production
is not allowed. The major interaction process in the energy range
between $\mathrm{15\,keV}$ and $\mathrm{511\,keV}$ are
coherent and incoherent scattering and photoelectric absorption with a
cross section for Rayleigh scattering that is one or two orders of
magnitudes smaller than $\sigma_\mathit{tot}$ in
equation~\ref{eq:cross-section-sum}. It can therefore be neglected
without grave inconvenience with the remaining major processes Compton
scattering and photoelectric absorption.

When discussing the consequences of these two processes, one has to
distinguish between interaction inside (detector scatter) and outside
(object scatter) the detector's sensitive volume, {\em i.e.}\ inside and
outside the scintillation crystal. Object scatter can cause serious
problems at the moment of image reconstructions. Its impact and
possible correction has been addressed by numerous studies (see Zaidi
and Koral \cite{Zaidi:2004}) but will not be treated here. In object
Compton scattering, the photon leaves the interaction point with a new
direction and a new energy. The error will be introduced when
constructing the line of response (in coincidence imaging) or the
origin of the $\gamma$-ray (in single photon tomography) and not when
determining the impact position inside the detector, especially if the
detector has poor energy resolution.

\begin{figure}[t]
  \centering
  \psfrag{E}{\hspace*{-5em}Photon Energy [MeV]}
  \psfrag{C}{\hspace*{-6em}Attenuation Coefficients [$\mathrm{cm^2}/g$]}
  \includegraphics[width=0.8\textwidth]{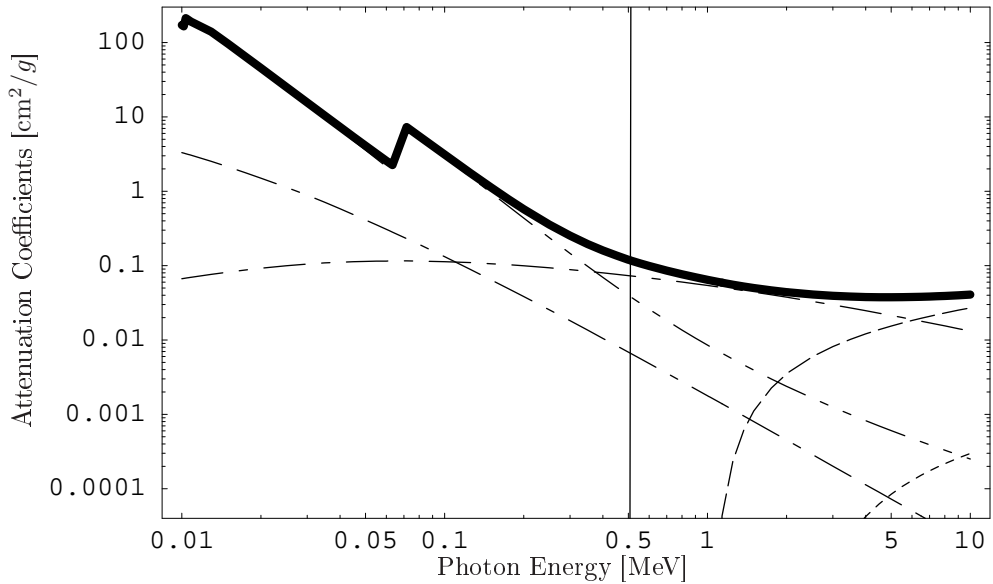}
  \caption[Total $\gamma$-ray attenuation for LSO from
  $10\,\mathrm{keV}$ to $10\,\mathrm{MeV}$]{%
    Total $\gamma$-ray attenuation for LSO  from $10\,\mathrm{keV}$ to $10\,\mathrm{MeV}$
    (thick solid line). Also shown are the attenuation coefficient for
  coherent scattering (dot-dashed-dashed line), the attenuation coefficient for
  incoherent scattering (dot-dashed line), the attenuation coefficient for
  photoelectric absorption (dot--dot-dashed line), the attenuation coefficient for
  pair production in the nuclear field (long-dashed line), and  the attenuation coefficient for
  pair production in the electron field (short-dashed line);
  Berger and Hubbell \cite{Hubbell:1987}. The vertical solid line marks the energy of $511\,\mathrm{keV}$.}
  \label{fig:LSO-crossections}
\end{figure}

Compton scattering inside the scintillation crystal instead affects the
position determination. The $\gamma$-photon can undergo photoelectric
absorption or can have been Compton scattered multiple times within the
crystal, and at each interaction position it will deposit a fraction
of its energy. At best, the scattered $\gamma$-photon escapes the
sensitive volume after one or more interactions taking away a
sufficiently large energy fraction and allowing a discrimination
against photoelectric events. Obviously, photoelectric absorption does not
introduce any position blurring and is therefore the preferred
interaction processes. While it is not possible to control which
elementary process a single photon undergoes, one can optimize the
fractions (or probability) of Compton-scattered events and
photoelectric events by choosing an adequate scintillator. For
the detection of 
annihilation radiation with $511\,\mathrm{keV}$ energy, materials of high
density and high atomic number are required (cf.\
 section~\ref{subsec:scintillators}). LSO matches all requirements
for positron emission tomography quite well and is one of the  scintillators of
choice for this imaging modality. Nevertheless, its Compton
attenuation coefficient at $511\,\mathrm{keV}$ is, at  $\mathrm{0.07\,cm^2/g}$, nearly twice
as large as the photoelectric attenuation coefficient of $\mathrm{0.04\,cm^2/g}$
(Berger and Hubbell \cite{Hubbell:1987}). In
figure~\ref{fig:LSO-crossections}, all LSO coefficients are plotted
for the energy range $10\,\mathrm{keV}$ to $10\,\mathrm{MeV}$.

\begin{figure}[t]
  \centering
  \subfigure[][Polar plot for the energy $E_1$ in
  equation~\ref{eq:annihil-compton-kinematic} (solid line, in units of
  $E_0$) and the
  energy $E_0-E_1$ that is transferred to the electron (dashed line, in units of
  $E_0$).]{\label{subfig:final-photon-energy}
    \includegraphics[width=0.46\textwidth]{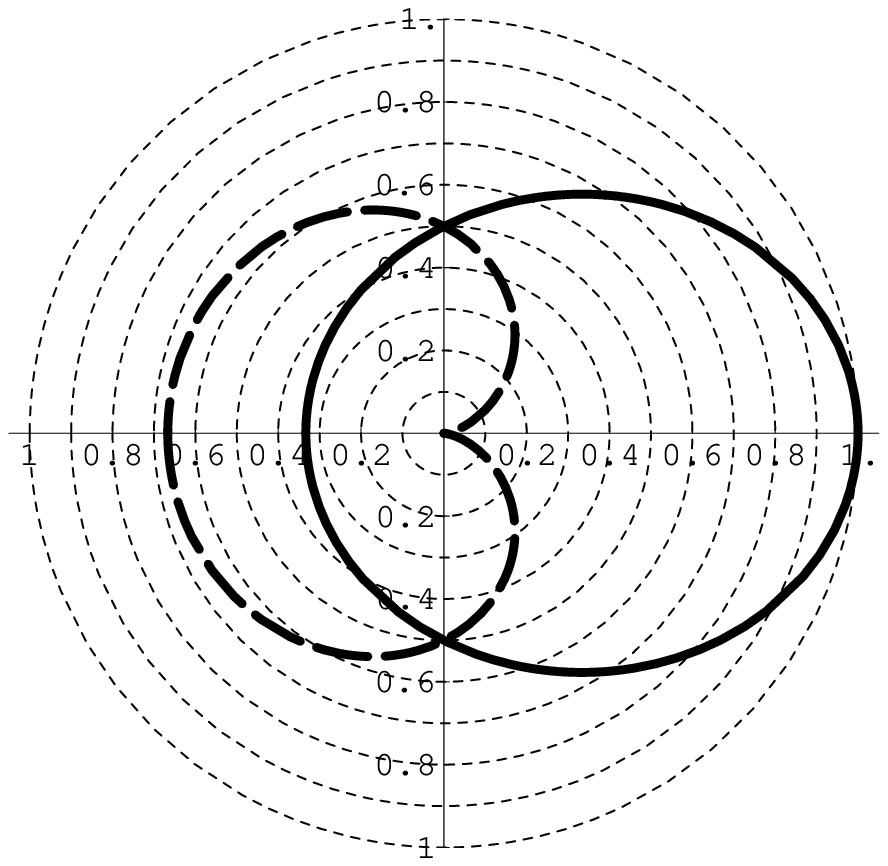}}
  \subfigure[][Polar plot of the Klein-Nishina
  formula~\ref{eq:klein-nishina} for $\gamma=1$ and normalized to its
  maximum at $\vartheta=0$.]{\label{subfig:diff-cross-section}
    \includegraphics[width=0.46\textwidth]{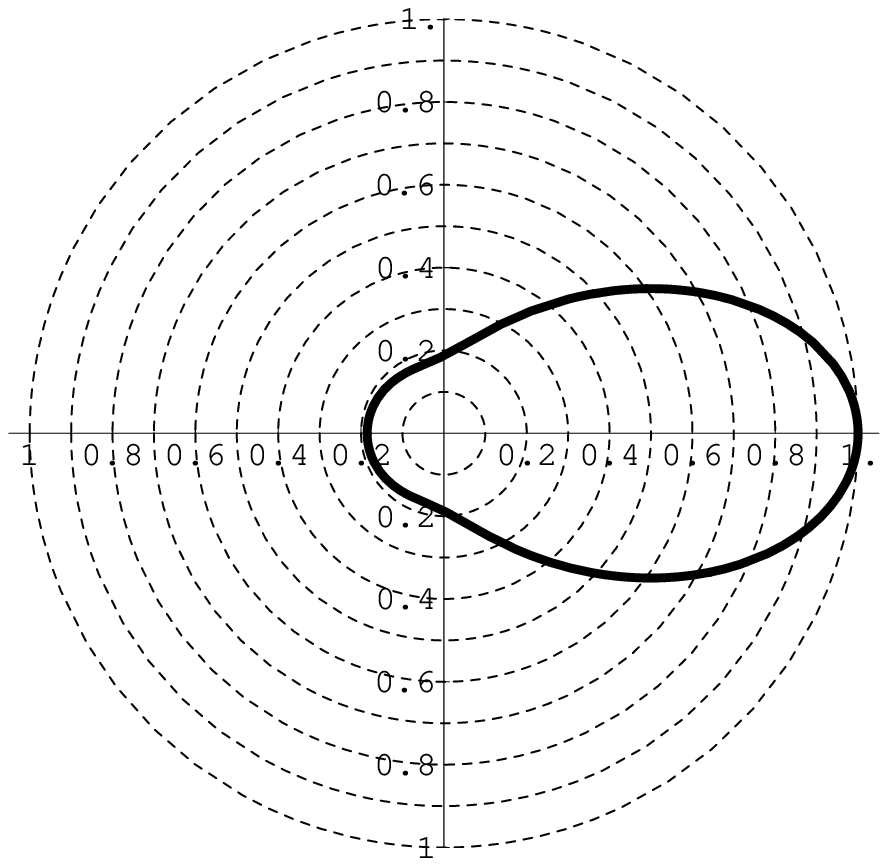}}
  \caption[Angular dependency of the photon's final energy, energy
  transfer and diff.\ cross section]{Angular dependency of the photon's final energy $E_1$, the
    energy transfer $E_0-E_1$ and the differential cross section $\frac{d\sigma}{d\Omega}$.}
  \label{fig:compton-kienmatics}
\end{figure}

The physics behind Compton scattering is very well understood. From
relativistic energy and momentum conservation the equation for its
kinematics can be derived:
\begin{equation}
  \label{eq:compton-kinematic}
  E_1=\frac{E_0}{1+\gamma(1-\cos\vartheta)}, 
\end{equation}
with $\gamma=E_0/E_e$ the photon's relativistic $\gamma$-factor,
$E_0$ the photon's initial energy, $E_1$ the photon's final energy and
$\vartheta$ the change in the photon's direction. For annihilation
radiation with $E_0=E_e$
equation~\ref{eq:compton-kinematic} transforms into the simpler expression
\begin{equation}
  \label{eq:annihil-compton-kinematic}
  E_1=\frac{E_e}{2-\cos\vartheta}.
\end{equation}
From equation~(\ref{eq:annihil-compton-kinematic}) it can be seen that
the maximum energy transfer to the electron is given by
$2E_e/3$ for backscattering and that the photon no loses
energy at all if it is scattered forward. This is the most likely
direction in which the photon is expected to be scattered as can be
seen from the Klein-Nishina (see for instance Leo \cite{Leo:1994}) differential
cross section for Compton scattering:
\begin{equation}
  \label{eq:klein-nishina}
  \frac{d\sigma}{d\Omega}=\frac{r_e^2}{2}\frac{1}{\left[1+\gamma(1-\cos{\vartheta})\right]^2}\
  \left(1+\cos^2{\vartheta}+\frac{\gamma^2(1-\cos{\vartheta})^2}{1+\gamma(1-\cos{\vartheta})}\right),
\end{equation}
in which $r_e=e^2/E_e$ is the classical electron radius
($r_e\approx 2.83\cdot10^{-13}\,\mathrm{cm}$). The photon's final energy, the
energy transfer to the electron and the (normalized) differential
cross section are shown for $\gamma=1$ in figure~\ref{fig:compton-kienmatics}.

\section{Inner Crystal Compton Scattering}

From the plot of the attenuation coefficients for LSO in
figure~\ref{fig:LSO-crossections}, it is clear that a large fraction
of the impinging $511\,\mathrm{keV}$ $\gamma$-rays is expected to undergo single
or multiple Compton scattering in the scintillation crystal.
For a detected $\gamma$-photon, there exist different possibilities of how
they have deposited their energy. If the very first interaction is a
photoelectric effect, a single scintillation light distribution as
described in chapter~\ref{ch:light-distribution} arises from this
impact position and no blurring due to Compton scatter will occur.
This is just the case that was considered throughout this work up to
now. Another possibility is that the scattered $\gamma$-photon
escapes from the scintillation crystal. If $\Delta E$ is the energy
resolution of the detector and less than approximately
100\%-$\Delta E/2$ of the $511\,\mathrm{keV}$ are deposited in the scintillator, this event can
be discriminated using the energy window. The last possibility is the
total absorption after one or more Compton interaction(s). These events
will deposit the whole photon energy within the scintillator and cannot be
discriminated by virtue of the energy. These are of the kind of events
studied in this chapter. Various light distributions of the
form~(\ref{eq:general-inv-square-law}) originating from the
different interaction points are then superposed at the
photodetector and the moments will be computed from this
superposed signal distribution. However, only the first $\gamma$-ray's
interaction point is connected to its origin by a single straight
line segment (line of flight). Although the energy is determined for each possible
scatter angle (and vice versa) by
equation~(\ref{eq:compton-kinematic}) and therefore allows, at least
theoretically, the reconstruction of the scattered path, the information
that is necessary for doing so is lost because the moments are
computed from  the superposed light distribution. Even with much more
sophisticated detectors (Sánchez {\em et al.} \cite{Sanchez:1996}) it is rather difficult to
exactly distinguish the several consecutive Compton interaction
position.

Monte Carlo simulations have been carried out in order to investigate
the impact of this effect on the spatial resolution and the depth of
interaction resolution in $\gamma$-ray imaging detectors using
continuous LSO crystals (Sánchez \cite{sanchez:privcom}). In a similar
study, the same effect was examined for different scintillators
($\mathrm{NaI\doped Tl}$ and $\mathrm{BGO}$) and detector geometries, {\em e.g.}\
individual crystal pixel readout, different Block detector designs
and large-sized continuous
crystals together with PSPMTs (Thompson \cite{Thompson:1990}).
He found that the strength of the caused resolution degradation strongly
depends on the detector's geometry and components. For this reason, and
because the results also depend strongly on the applied scintillation
material, their findings are not directly applicable to our detector
setup.

Instead of setting up from-scratch programs for Monte Carlo
simulations as done by Thompson,
the \mbox{GEANT 3} library (Brun and Carminati, \cite{Geant3:1994})
was used.  
A large LSO crystal of rectangular shape and with $x$, $y$ and $z$ dimensions
$\mathrm{40\times40\times20\,mm^3}$ was simulated with the $511\,\mathrm{keV}$
$\gamma$-rays impinging on the center of the $x$-$y$
plane and normal to the same. A total of 20000 events were simulated and the results were
classified by their number of interactions within the scintillator.
An event is registered as detected when it deposits all its
energy ($511\,\mathrm{keV}$) within the crystal. Table \ref{tab:geant_results}
shows the obtained probabilities.
\begin{table}[t]
  \renewcommand{\arraystretch}{1.2}
  \centering
  \begin{tabular}{{cccccc}}
    \hline\hline &&&&&\\
    \parbox{0.12\textwidth}{Number of\\ Interactions} &
    \parbox{0.07\textwidth}{Counts} & \parbox{0.10\textwidth}{Fraction\\ of
      Events} & \parbox{0.09\textwidth}{Detected\\ Events} &
    \parbox{0.17\textwidth}{Total Fraction of\\ Detected Events} &
    \parbox{0.19\textwidth}{Relative Fraction of\\ Detected Events} \\ &&&&&\\\hline\hline
    0 & 3807 & 19\% & 0 & 0\% & 0\% \\
    1 & 7389 & 37\% & 5499 & 27\% & 40\% \\
    2 & 5752 & 29\% & 5273 & 26\% & 38\% \\
    3 & 2302 & 12\% & 2218 & 11\% & 16\%\\
    4 & 614  & 3\%  & 600  & 3\%  & 4\%\\
    5 & 114  & 0.6\% & 114 & 0.6\% & 0.8\%\\
    6 & 18   & 0.1\% & 18  & 0.1\% & 0.1\%\\
    7 & 4   & 0.02\% & 4  & 0.02\% & 0.03\%\\
    \hline\hline
  \end{tabular}
  \caption[Counts, total fractions and relative fractions of possible
    events]{Counts, total fractions and relative fractions of possible
    events obtained with the Monte Carlo simulation.}
  \label{tab:geant_results}  
\end{table}

3807 of the $\gamma$-rays pass through the crystal without any
interaction. 7389 events occur with exactly one interaction within the
simulated crystal volume. However, only 5499 of these were counted as
detected events. These interact via the photoelectric effect and do not
cause any spatial resolution degradation. The remaining 1890 $\gamma$-photons
were scattered coherently or incoherently and escaped
from the LSO block. With growing number of interactions, the
probability of being detected rises significantly. The total of all
detected events gives the efficiency at the center of the detector and reaches
approximately 68\%. 

\begin{figure}[!tp]
  \centering\vspace*{-1eX}
  \subfigure[][Displacement histogram for the transverse component
  for all the events registered as detected.]{\label{subfig:compton-r-allhist}
    \psfrag{r}{\hspace*{-1.3em}$\bar{r}$ [mm]}
    \psfrag{C}{\hspace*{-3.5em}detected events}
    \includegraphics[width=0.45\textwidth]{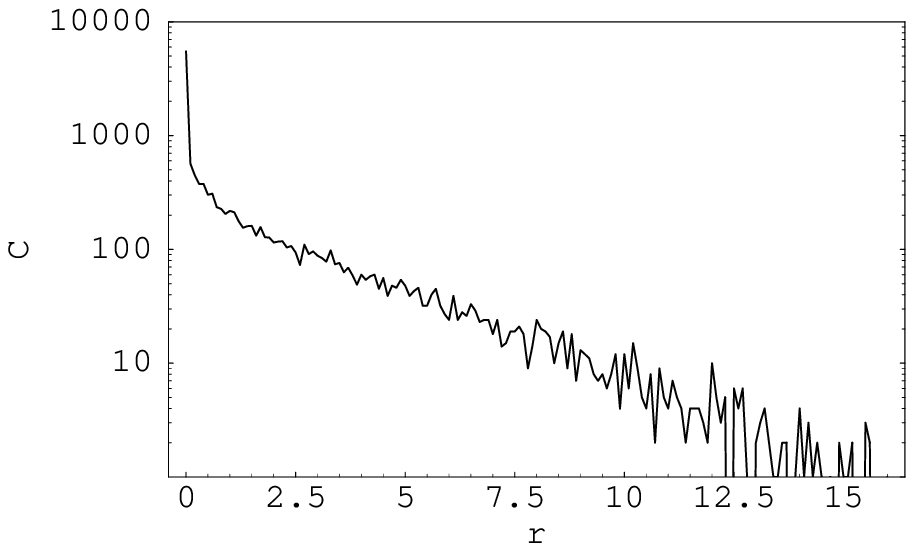}
  }
  \subfigure[][Displacement histogram for the normal component for all
  the events registered as detected.]{\label{subfig:compton-z-allhist}
    \psfrag{r}{\hspace*{-1.3em}$\bar{z}$ [mm]}
    \psfrag{C}{\hspace*{-3.5em}detected events}
    \includegraphics[width=0.45\textwidth]{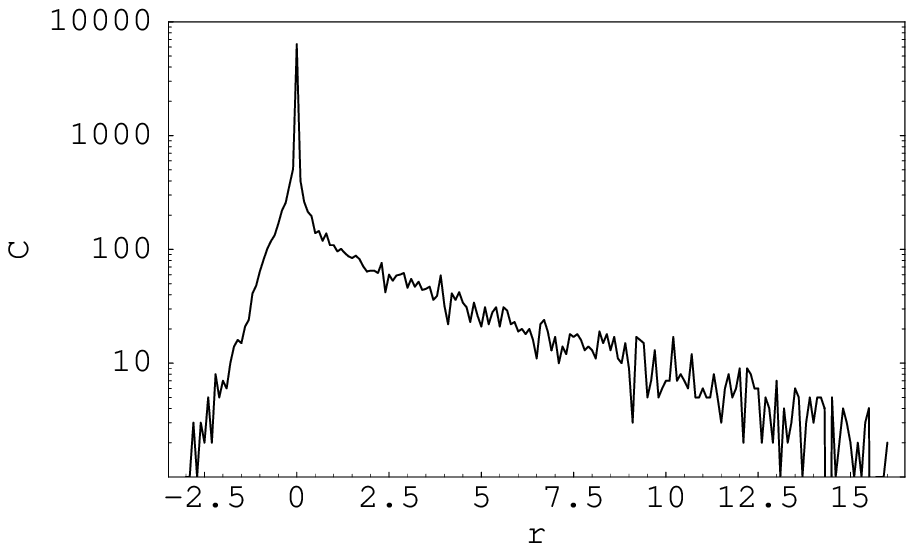}
  }\\\vspace*{-3eX}
  \subfigure[][Displacement histograms for the transverse component
  and classified by the number of interactions.]{\label{subfig:compton-r-hists}
    \psfrag{0}{$\mathrm{0}$}
    \psfrag{1}{$\mathrm{1}$}
    \psfrag{2}{$\mathrm{2}$}
    \psfrag{3}{$\mathrm{3}$}
    \psfrag{4}{$\mathrm{4}$}
    \psfrag{5}{$\mathrm{5}$}
    \psfrag{6}{$\mathrm{6}$}
    \psfrag{7}{$\mathrm{7}$}
    \psfrag{8}{$\mathrm{8}$}
    \psfrag{12}{$\mathrm{12}$}
    \psfrag{14}{$\mathrm{14}$}
    \psfrag{16}{$\mathrm{16}$}
    \psfrag{18}{$\mathrm{18}$}
    \psfrag{10}{$\mathrm{10}$}
    \psfrag{20}{$\mathrm{10^2}$}
    \psfrag{30}{$\mathrm{10^3}$}
    \psfrag{40}{$\mathrm{10^4}$}
    \psfrag{x}{\hspace*{-1em}\rotatebox{66}{\hspace*{-4em}\# of interactions}}
    \psfrag{y}{\hspace*{-1.3em}$\bar{r}$ [mm]}
    \psfrag{z}{\hspace*{0.3em}\rotatebox{90}{\hspace*{-3em}detected events}}
    \includegraphics[width=0.72\textwidth]{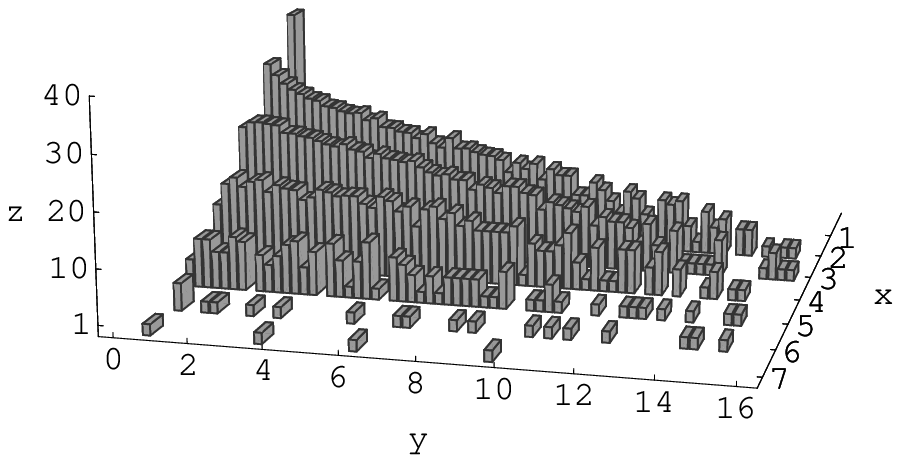}
  }\vspace*{-5eX}
  \subfigure[][Displacement histograms for the normal component
  and classified by the number of interactions.]{\label{subfig:compton-z-hists}
    \psfrag{-2}{$\mathrm{-2}$}
    \psfrag{-4}{$\mathrm{-4}$}
    \psfrag{0}{$\mathrm{0}$}
    \psfrag{1}{$\mathrm{1}$}
    \psfrag{2}{$\mathrm{2}$}
    \psfrag{3}{$\mathrm{3}$}
    \psfrag{4}{$\mathrm{4}$}
    \psfrag{5}{$\mathrm{5}$}
    \psfrag{6}{$\mathrm{6}$}
    \psfrag{7}{$\mathrm{7}$}
    \psfrag{8}{$\mathrm{8}$}
    \psfrag{10}{$\mathrm{10}$}
    \psfrag{12}{$\mathrm{12}$}
    \psfrag{14}{$\mathrm{14}$}
    \psfrag{16}{$\mathrm{16}$}
    \psfrag{18}{$\mathrm{18}$}
    \psfrag{20}{$\mathrm{10^2}$}
    \psfrag{30}{$\mathrm{10^3}$}
    \psfrag{40}{$\mathrm{10^4}$}
    \psfrag{x}{\hspace*{-1em}\rotatebox{66}{\hspace*{-4em}\# of interactions}}
    \psfrag{y}{\hspace*{-1.3em}$\bar{z}$ [mm]}
    \psfrag{z}{\hspace*{0.3em}\rotatebox{90}{\hspace*{-3em}detected events}}
    \includegraphics[width=0.72\textwidth]{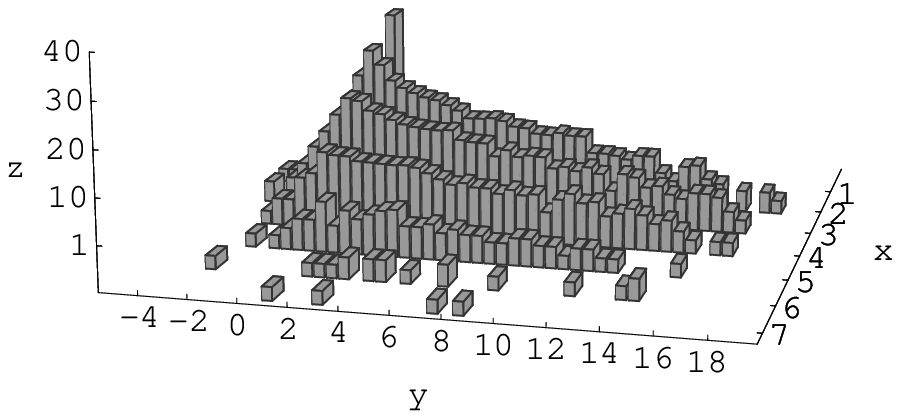}
  }\vfill
  \caption[Histograms for the displacements of the centroids due to
    Compton scattering]{Histograms for the displacements of the centroids due to
    Compton scattering. Figures~\ref{subfig:compton-r-allhist} and
    \ref{subfig:compton-z-allhist} show the displacement of both
    projections for the total of detected
    events. In figures~\ref{subfig:compton-r-hists} and
    \ref{subfig:compton-z-hists} separate histograms are shown for the
    seven subsets of events with same number of interactions.  
  }
  \label{fig:compton-histograms}
\end{figure}

For estimating the position blurring caused, the displacement vectors
$\mathbf{r}_{1i}=\mathbf{r}_i-\mathbf{r}_1$ were
computed. Here, $\mathbf{r}_1$ is the position of the first
interaction of the $\gamma$-ray and $\mathbf{r}_i$ with
$i=\{2,3,\ldots,7\}$ the positions of all following interactions.
Obviously, $\mathbf{r}_1$ has the same $x$-$y$ coordinates as the
$\gamma$-ray source. The displacement vectors were then projected onto
the $x$-$y$ plane ($\hat{\mathbf{r}}_{1i}$) and onto the $z$-axis
($\hat{\mathbf{z}}_{1i}$) and their centroids $\bar{r}$ and $\bar{z}$
were computed:
\begin{gather}
  \label{eq:transversal-cent}
  \bar{r}=\frac{\sum_{i=1}^7\hat{\mathbf{r}}_{1i}E_i}{\sum_{i=1}^7E_i}\\
  \label{eq:normal-cent}
  \bar{z}=\frac{\sum_{i=1}^7\hat{\mathbf{z}}_{1i}E_i}{\sum_{i=1}^7E_i}.
\end{gather}
Figure~\ref{fig:compton-histograms} shows the
histograms of the displacements of the centroids of all detected events
for the projections onto the $x$-$y$ plane and onto the $z$-axis.

The two distributions of $\bar{r}$ and $\bar{z}$ show a long tail
and a very sharp maximum which is caused by the photoelectric events.
These events have displacement vectors $\bar{r}$ and $\bar{z}$ equal to
zero and cause no blurring at all. From
figures~\ref{subfig:compton-r-hists} and \ref{subfig:compton-z-hists} 
it can be seen that the slope of the tail is flatter the higher
the number of interactions is.

For want of a suitable model distribution, the data has been
interpolated in order to compute the FHWM and FWTM. The results have
been summarized together with additional statistical estimates, {\em e.g.}\ mean,
standard deviation and median, in
table~\ref{tab:statistical-characterization}. 
The same computations have been repeated for the total of all
those detected events that undergo at least one Compton
scattering. Rather large values can be observed for the mean values
and standard deviations of all four statistics. In contrast, one
obtains small values for the full widths at half maximum. This
combination is characteristic for distributions with a narrow maximum
and a long tail of low intensity. Except for the case of transverse
displacement without photoelectric absorptions, the computed medians
are also very small.

\section{Screening of Forward Scattered Events}

Another remarkable feature is that
the transverse displacement gives a broader statistics than the
normal displacement. This is obviously not expected, since the
Klein-Nishina formula strongly favors forward scattering (see
figure~\ref{eq:klein-nishina}). One instead would expect a much
stronger blurring along the direction parallel to the $\gamma$-ray
beam. The reason lies in the geometry of the scintillator block. 
As already observed by Thompson \cite{Thompson:1990}, the degree of spatial
blurring that results from inner crystal Compton scattering depends
strongly on the configuration of the $\gamma$-ray imaging detector.
In the case discussed here, the scintillator has a spatial extension
of $\mathrm{40\times40\times20\,mm^3}$. That is, its transverse
extension is twice as large as its normal extension. 

\begin{figure}[t]
  \centering
  \subfigure[][Histogram of the scattering angles $\vartheta_1$ of the very first
  interaction inside the crystal for detected events.]{\label{subfig:first-scatter-angle}
    \psfrag{A}{\raisebox{-0.5eX}{\hspace*{-2em}$\vartheta_1$ [degree]}}
    \psfrag{C}{\hspace*{-3.4em}detected events}
    \includegraphics[width=0.4\textwidth]{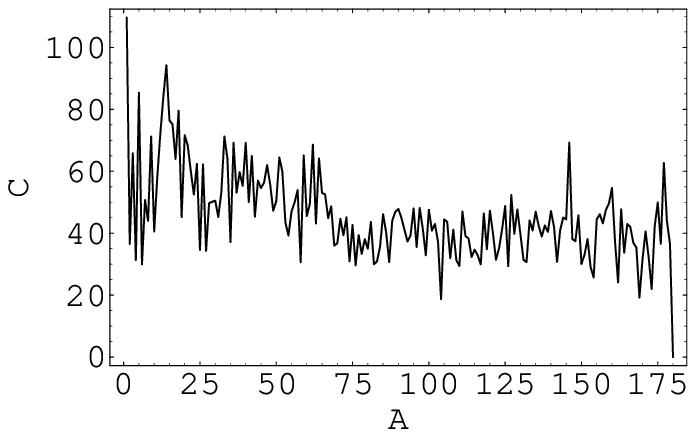}}\hspace*{0.08\textwidth}
  \subfigure[][Polar plot of the scattering angles $\vartheta_1$ of the very first
  interaction inside the crystal (solid black line). The Klein-Nishina angular
  dependencies (solid gray line: simulation, dashed gray line: theory)
  are also plotted for comparison. Both
  distributions are normalized to the maximum values of the
  theoretical dependency.]{\label{subfig:first-scatter-angle-polar}
    \includegraphics[width=0.32\textwidth]{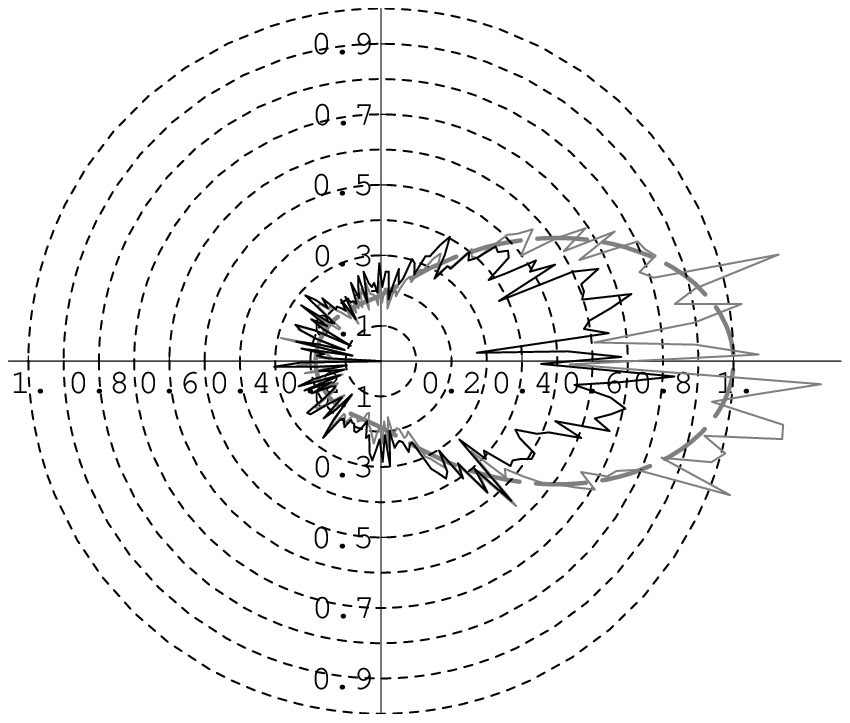}}
  \subfigure[][$z$ component of the very first interaction position of
  all detected events.]{\label{subfig:first-z-pos}
    \psfrag{Z}{\hspace*{-2em}$z_1\,\mathrm{[mm]}$}
    \psfrag{C}{\hspace*{-3.4em}detected events}
    \includegraphics[width=0.4\textwidth]{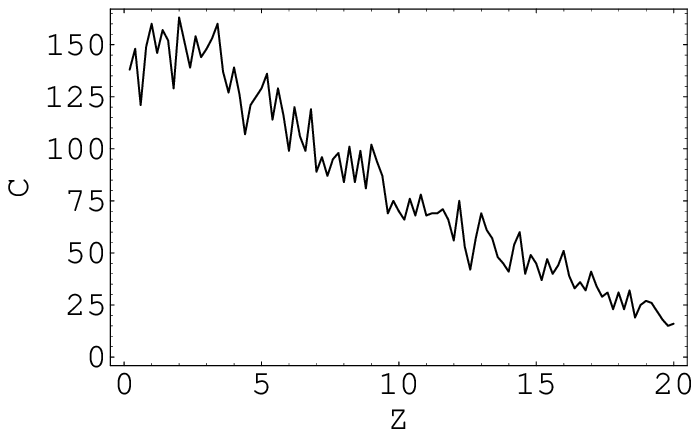}}
  \subfigure[][Energy deposition at the very first interaction position of
  all detected events.]{\label{subfig:first-energy}
    \psfrag{E}{\hspace*{-2em}$E_1\,\mathrm{[keV]}$}
    \psfrag{C}{\hspace*{-3.4em}detected events}
    \includegraphics[width=0.4\textwidth]{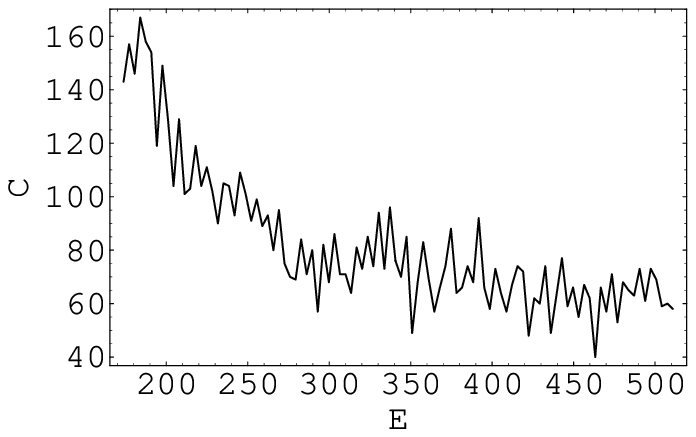}}
  \caption[Distribution of the scattering angle obtained from the subset of all
    detected events]{Distribution of the scattering angle obtained from the subset of all
    detected events and for the very first interaction of the
    $\gamma$-ray. The $z$-component and the energy of the interaction position of
    these events is shown in the figures below.}
  \label{fig:detected-angles}
\end{figure}

Suppose an incident $\gamma$-ray is forward scattered. Then, the
energy $E_1$ of the scattered photon is approximately $E_e$ by virtue
of equation~(\ref{eq:annihil-compton-kinematic}). This scattered photon
now sees a remaining crystal thickness of $\mathrm{20\,mm}-z_1$ being $z_1$ the
$z$-component of this first interaction position. The distribution of
$z_1$ is shown in figure~\ref{subfig:first-z-pos} for the subset of
all these events that are detected (68\% of all simulated events).
As is expected for the $511\,\mathrm{keV}$ annihilation radiation, a large
fraction of $\gamma$-rays travel a large distance along the
$z$-direction before interacting, due to the lower cross section for
higher energies. The mean value is approximately $7\,\mathrm{mm}$
for the present simulation. Therefore, all those events that are
forward scattered in the first interaction most likely escape from the
crystal without any further interaction because their energy is still
high and the remaining crystal thickness is small. This is not the
case for events whose first interaction results in a large scattering
angle. First, the spatial extension of the crystal along the $x$- and
$y$-direction is larger and the scattered photons need to travel a
larger distance through the scintillator in order to escape. Secondly,
their energy is significantly reduced and therefore their probability
for interaction increased (refer to
figure~\ref{fig:LSO-crossections}). As a consequence, the distribution
of the first scattering angles of all detected events is 
biased towards large scattering angles because the detector described here
 is nearly ``blind'' to the forward scattered events. The
distribution is plotted in figures~\ref{subfig:first-scatter-angle}
and \ref{subfig:first-scatter-angle-polar}. From the
histogram~\ref{subfig:first-scatter-angle} the mean scattering angle can
be computed to be about 63\textdegree. If all events are
considered, the mean value of the first scattering angle is, at
approximately 57\textdegree, noticeably lower.
For an angle of 63\textdegree, the energy of the
outgoing photon is $\approx\frac{2}{3}E_e$ and its probability of
interaction almost twice as large. The distribution of the energy
deposition at the first interaction position of all detected events is
shown in figure~\ref{subfig:first-energy}. It is also biased
towards lower energy and the mean value is 315keV.

\begin{table}[t]
  \centering
  \begin{tabular}{cccccc}
  \hline\hline &&&&&\\
    Component & FWHM & FWTM & Mean & Standard deviation & Median
    \\ &&&&& \\\hline\hline
  \parbox[][7eX][c]{10em}{transverse (all\\ detected events)} & 0.2 &
  1.33 & 1.46 & 2.36 & 0.3 \\
  \parbox[][9eX][c]{10em}{transverse (detected\\ events with nr. of\\
    interactions $\mathrm{>1}$)} & 1.08 &
  8.51 & 2.44 & 2.64 & 1.49 \\
  \parbox[][7eX][c]{10em}{normal (all\\ detected events)} & 0.22 &
  0.4 & 1.04 & 2.56 &  0.11 \\
  \parbox[][9eX][c]{10em}{normal (detected\\ events with nr. of\\
    interactions $\mathrm{>1}$)} & 0.38 &
  2.86 & 1.04 & 2.56 & 0.11 \\\hline\hline
  \end{tabular}
  \caption[Different statistical estimates in $\mathrm{mm}$ for the displacement
    distributions]{Different statistical estimates for the displacement
    distributions that were found by Monte Carlo simulations.}
  \label{tab:statistical-characterization}
\end{table}

\chapterbib


  \cleardoublepage
\chapter{Experimental Verification}
\label{ch:experiment}

\chapterquote{%
Die gefährlichste Weltanschauung ist die Weltanschaung
derjenigen, die die Welt nicht angeschaut haben.
}{%
Alexander von Humboldt, $\star$ 1769 -- $\dagger$ 1859 
}

\PARstart{T}{he} experimental verification of the derived theoretical
models of chapters~\ref{ch:light-distribution} and
\ref{ch:enhanced-charge-dividing-circuits} is hampered
by the fact that it is nearly impossible to design a measurement
setup that allows a sufficiently good definition of the
$\gamma$-rays interaction depth. The same statement holds for the evaluation of
the detector performance concerning the energy resolution, spatial
resolution and depth of interaction resolution. For a $\gamma$-ray
impinging normally to the plane of the photodetector's sensitive area, 
the depth at which the photon interacts is subjected
to a random process and it is not accessible by experimental preparation. The
possibility of using a sandwich-like configuration consisting of several 
inactive layers and only one scintillation layer, but all having the
same optical properties, has been discarded for economic
reasons. Moreover, it is questionable if these set of sandwiches
really reproduce the properties of a single thick scintillator slab.
Lateral irradiation with a collimated $\gamma$-ray, as was explained by
Moses and Derenzo \cite{Moses:1994} or Huber {\em et al.}\ \cite{Huber:1999} for the use with small
scintillation crystal needles, has to be discarded for another
reason: due to  the large transverse spatial extension of the 
crystal slab, the intensity of the $\gamma$-ray would decrease quickly
from the edge 
and lead to very high counting statistics at the side of the detector
that faces the radioactive source and very low statistics at the
opposite side. 

Another important problem is caused by the collimation of the
$\gamma$-ray beam. When using collimators, one has to keep in mind
that, for the $\mathrm{511\,keV}$ energy of the annihilation photons, the probability for
scattering is already rather high.
For lead or tungsten, which are the most suitable materials for
collimators, the photon cross-sections for photoelectric absorption
and incoherent Compton scattering actually have very similar values.
This leads to a high
fraction of Compton scattered photons in the $\gamma$-ray beam and fans
out the collimated beam. A second possibility is given by
electronic collimation using a second coincidence detector,
which is required anyway for Positron Emission Tomography.
In this case, the spatial extension of the radioactive source has to be taken 
into account. Commercially available positron emitters normally have a
diameter of one millimeter or more. Modern $\gamma$-ray imaging
detectors already reach intrinsic spatial resolutions in the range of
$\mathrm{1\,mm}$ to a few $\mathrm{mm}$ (Correia {\em et al.}\ \cite{Correia:1999}, Joung {\em et al.}\
\cite{Joung:2002} and Tavernier {\em et al.}\ \cite{Tavernier:2005})
and one has to correct the obtained results for the spatial extension of
the $\gamma$-ray source. The same arguments hold for beams collimated
with a tungsten or lead collimator.
  
In this section a method is explained, detailing how information about the
depth of interaction can be obtained without the necessity of
preparing several experiments for different DOIs. This method is
subsequently used to verify the analytic model for the signal
distribution derived in chapter~\ref{ch:light-distribution}.
Also, the four first-order moments, namely $\mu_0$, $\mu^x_1$,
$\mu^y_1$ and $\mu_2$ (refer to
chapter~\ref{ch:enhanced-charge-dividing-circuits}),
are measured at 81 uniformly distributed test positions within the
sensitive area of the photodetector.

\section{Experimental Setup}
\label{sec:exp-setup}

For the measurement of the moments, two identical and opposite
$\gamma$-ray detectors were used. Each module consisted of a single large-sized
scintillator-block with spatial dimensions $\mathrm{42\times42\times10\,mm^3}$.
As scintillation material, cerium-doped lutetium oxyorthosilicate
($\mathrm{Lu_2(SiO_4)O:Ce^{3+}}$, also LSO, invented by Melcher and
Schweitzer, \cite{Melcher:1992}) was chosen for the excellent
matching of its properties with the requirements for Positron
Emission Tomography (cf.\ section~\ref{subsec:scintillators}). 
According to the arguments given in
section~\ref{sec:included-contribs} and the findings of
section~\ref{sec:complete-signal-dist}, the five side-surfaces that
were not coupled to the photodetector, had to be covered with a
highly anti-reflective layer. Furthermore, these sides were not
polished but only fine ground. The reason for this decision was not only
economical. A polished crystal surface together with the glue
of the painting or the epoxy possibly builds an interface of two
optical media with different refractive indices. Most probably, this would
lead to total reflections because the refraction index
$\mathrm{n_\tincaps{LSO}}$ of LSO is,
at 1.82, rather high. Black epoxy resins can be used to realize high
quality anti-reflective coatings that generally out-perform most
other resin types in terms of mechanical properties and resistance to
environmental degradation. We also observed that the reflections
caused by simply black painted side-surfaces result in a high
fraction of reflected scintillation light that renders the determination
of the distribution's second moment with the method presented here  almost
impossible.

\begin{figure}[t]
  \centering
  \psfrag{electronics}{electronics}
  \includegraphics[angle=270,width=0.8\textwidth]{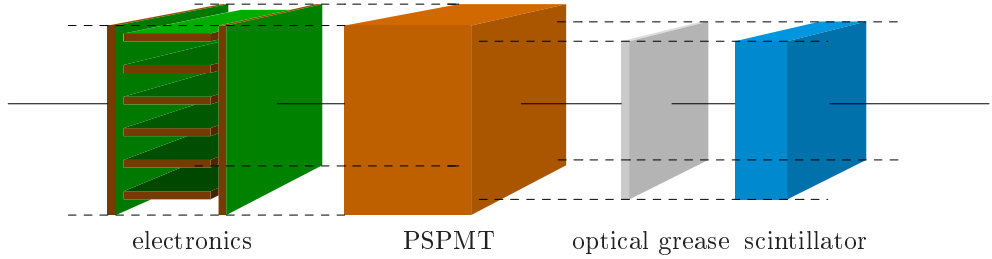}\\~\\
  electronics\hspace*{2cm}PSPMT\hspace*{1cm}optical grease\hspace*{0.2cm}scintillator
  \caption[Explosion view of a single $\gamma$-ray detector module]
  {Explosion view of a single $\gamma$-ray detector module
    (without housing) consisting of the large-sized continuous
    scintillation crystal, the photodetector, the printed circuit
    boards for the amplifiers and high-voltage supply.}
  \label{fig:detector-explode}
\end{figure}

As position sensitive photodetector, the Flat-Panel type multi-anode
PMT H8500 from Hamamatsu Photonics K.K.\ \cite{data:H8500} was
chosen. Photomultipliers generally have outstanding signal
characteristics, namely very high gains together with low noise, fast
response, high stability and a long life. The price per unit
sensitive area is also very low compared to other photodetectors
making it a very  versatile light detector. The H8500 is characterized
by a $\mathrm{49\times49\,mm^2}$ sensitive area with a dimensional
outline of only $\mathrm{52\times52\times28\,mm^3}$. That is to say, it
is an extremely compact PMT with a very low dead area at the borders,
allowing a high packing density and very compact $\gamma$-ray imaging
detectors. The small thickness of the H8500 indicates that the 12-stage
metal channel dynode system is also very compact, leading to a low
electron transit time and electron transit time spread. The H8500 has
a borosilicate window of refractive index
$\mathrm{n_\tincaps{W}}=\mathrm{1.51}$ at $\mathrm{400\,nm}$ incident
wavelength and a thickness of $\mathrm{2\,mm}$. It
 has a standard bi-alkali photocathode of thickness $\mathrm{200}$ \r{A}
\cite{hamamatsu:privcom} with a complex refractive index of
$\mathrm{2.54+1.59i}$ (Motta and Schönert \cite{Motta:2005}). The
anode consists of an array of $\mathrm{8\times8}$ segments of size
$\mathrm{5.8\times5.8\,mm^2}$ and an inter-segment pitch of $\mathrm{6.08\,mm}$.

A non-curing optical grease (Rhodorsil P\^{a}te 7, Rhodia Siliconi
Italia) was used to attach the LSO crystal to the entrance window of the
H8500. The refractive index of this gel
($\mathrm{n_\tincaps{Gel}\approx1.6}$) lies in between the refractive indices of the
scintillator and the entrance window. It avoids air-gaps between the
two media and optimizes the scintillation light
collection. Figure~\ref{fig:detector-explode} shows schematically the
 internal configuration of the detector module.

The electronics contained in each detector module includes the
proportional resistor network displayed in figure~\ref{fig:bidim-schem}
of section~\ref{subsection:2D-prop-net} together with the summation
amplifier (figures~\ref{fig:summing-amp} and
figures~\ref{subfig:simple-adder}, \ref{subfig:block-sum-cdr} of
section~\ref{subsection:2D-prop-net}), the $\mathrm{1000\,V}$ high-voltage power
supply (Hamamatsu, \cite{data:C4900}) and the
preamplifiers and line drivers displayed in
figures~\ref{fig:inverting-preamp} and \ref{fig:noninverting-preamp}.
of Appendix~\ref{app:elec-config}. The preamplifier and line-driver
modules perform the current-to-voltage conversion, restore the baseline
of the signal and amplify it for an optimal matching of the input voltage
range with the
analog-to-digital converter module. The basic configuration from
Siegel {\em et al.} (\cite{Siegel:1996}) was adopted for the 2D
proportional resistance network and the values
$\mathrm{806\,k\Omega}$, $\mathrm{402\,k\Omega}$, $\mathrm{270\,k\Omega}$
and $\mathrm{243\,k\Omega}$ were taken for the summation resistors (refer to
section~\ref{subsec:prop-2D-case}).

Although the method for depth of interaction detection presented in
this work is not confined to imaging modalities based on coincidence
detection, the experiments have to be carried out in temporal
coincidence for two reasons. First, approximately $\mathrm{2.6\,\%}$ of the natural
lutetium contained in LSO is radioactive ($\mathrm{^{176}Lu}$,
$\mathrm{t_{1/2}\approx4\cdot10^{10}}$ years) causing a background of
roughly 280 detected single events per second and $\mathrm{cm^3}$ of
scintillation material. $\mathrm{^{176}Lu}$
decays via $\beta^-$-decay with a maximum energy of
$\mathrm{1.192\,MeV}$ and with subsequent decays of the
$\mathrm{597\,keV}$ and $\mathrm{998\,keV}$ levels of
$\mathrm{^{176}Hf}$ through $\gamma$-decay with transition energies of $\mathrm{88\,keV}$,
$\mathrm{202\,keV}$, $\mathrm{307\,keV}$ and $\mathrm{401\,keV}$
(Lauckner {\em et al.}\ \cite{Lauckner:2001}, Melcher and Schweitzer \cite{Melcher:1992},
Huber {\em et al.}\ \cite{Huber:2002}). Therefore, a $\gamma$-ray source of
high activity is required for the use of LSO with single photon
imaging modalities, whereas the $\mathrm{^{176}Lu}$ isotope
produces only a negligible background event rate  in temporal
coincidence mode. A second important advantage of coincidence
detection is that the $\gamma$-ray beam can be collimated
electronically as long as the coincidence detector is also sensitive
to the $\gamma$-ray's impact position. Such a collimation should
be preferred for the reasons explained above and because it allows
simple off-line manipulation of the data set as well as a variation of the
collimation diameter.

\begin{figure}[!t]
  \centering
  \psfrag{LED}{\renewcommand{\baselinestretch}{0.7}\hspace*{-0.6em}\parbox[t]{7em}{\small
        leading edge\\discriminator}}
  \psfrag{AND}{\renewcommand{\baselinestretch}{0.7}\parbox[t]{5em}{\small
        coincidence\\unit}}
  \psfrag{DELAY}{\renewcommand{\baselinestretch}{0.7}\parbox[t]{5em}{\small
        gate\\generator}}
  \psfrag{Switch}{\raisebox{1.4ex}{\renewcommand{\baselinestretch}{0.7}\parbox[t]{5em}{\small
        Ethernet\\switch}}}
  \psfrag{ADC}{\renewcommand{\baselinestretch}{0.8}\hspace*{-0.3em}\parbox[t]{5em}{\small
        12 channel\\ADC-card}}
  \psfrag{Workstation}{\renewcommand{\baselinestretch}{0.7}\parbox[t]{5em}{\small
        personal\\computer}}
  \psfrag{MODII}{MOD II}
  \psfrag{MODI}{MOD I}
  \psfrag{Dynodes}{\raisebox{1.4ex}{\renewcommand{\baselinestretch}{0.7}\parbox[t]{5em}{\small
        from last\\dynodes}}}
  \psfrag{2.5us Inhibit}{\renewcommand{\baselinestretch}{0.7}\parbox[t]{5em}{\small
        $\mathrm{2.5\mu s}$\\Inhibit}}
  \psfrag{200ns Delay}{\renewcommand{\baselinestretch}{0.7}\parbox[t]{5em}{\small
        $\mathrm{200ns}$\\delay}}
  \psfrag{400ns IntWin}{\renewcommand{\baselinestretch}{0.7}\parbox[t]{5em}{\small
        $\mathrm{400ns}$\\integration\\window}}
  \psfrag{S1}{$\scriptstyle\mathrm{\sum_I}$}
  \psfrag{S2}{$\scriptstyle\mathrm{\sum_{II}}$}
  \psfrag{A1}{$\scriptstyle\mathrm{A_I}$}
  \psfrag{A2}{$\scriptstyle\mathrm{A_{II}}$}
  \psfrag{B1}{$\scriptstyle\mathrm{B_I}$}
  \psfrag{B2}{$\scriptstyle\mathrm{B_{II}}$}
  \psfrag{C1}{$\scriptstyle\mathrm{C_I}$}
  \psfrag{C2}{$\scriptstyle\mathrm{C_{II}}$}
  \psfrag{D1}{$\scriptstyle\mathrm{D_I}$}
  \psfrag{D2}{$\scriptstyle\mathrm{D_{II}}$}
  \psfrag{MA}{\hspace*{-0.5em}\raisebox{0.2eX}{$\mathrm{\genfrac{}{}{0pt}{}{LeCroy}{623B}}$}}
  \psfrag{MB}{\hspace*{-0.5em}\raisebox{0.2eX}{$\mathrm{\genfrac{}{}{0pt}{}{Ortec}{CO4020}}$}}
  \psfrag{MC}{\hspace*{-0.5em}\raisebox{0.2eX}{$\mathrm{\genfrac{}{}{0pt}{}{Ortec}{GG8020}}$}}
  \psfrag{MD}{}
  \includegraphics[width=1\textwidth]{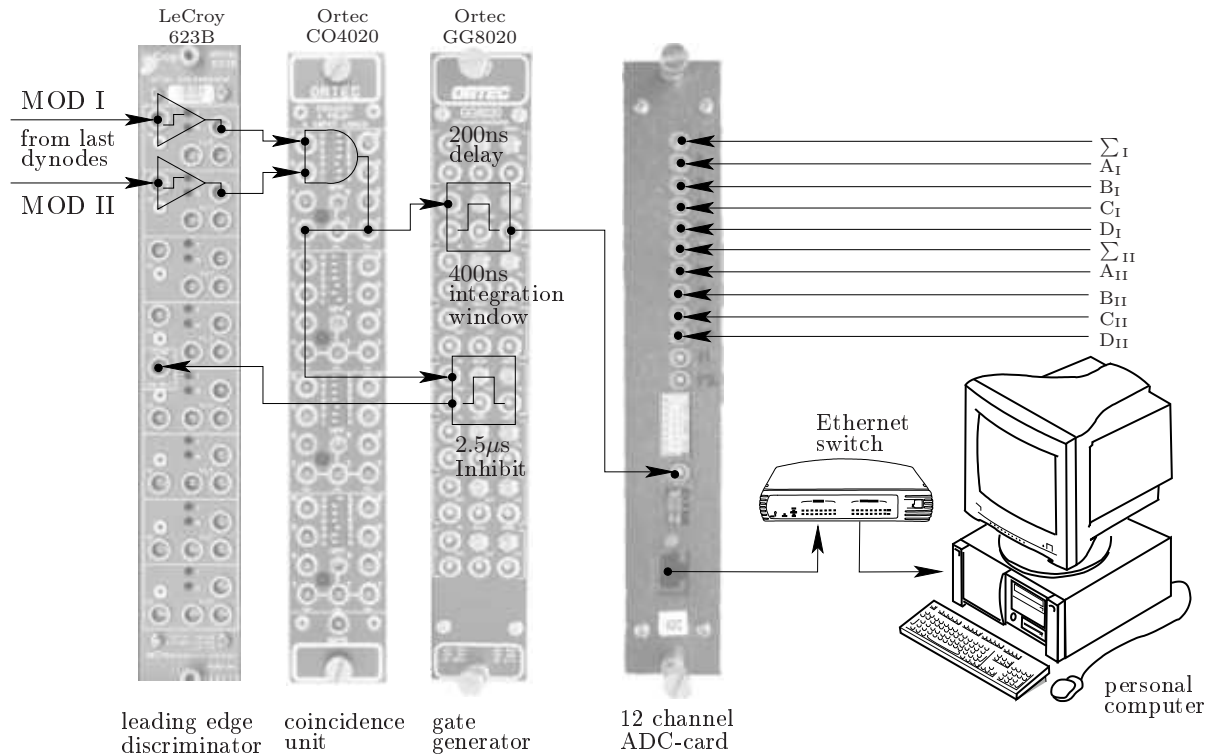}
  \vspace*{1eX}
  \caption[Layout of the configuration for data-acquisition and
    coincidence-trigger generation]{Layout of the configuration for data-acquisition and
    coincidence-trigger generation. The (buffered) signals from the
    last dynodes of both modules are fed into the leading-edge
    discriminator and evaluated by a logical AND-gate. This signal
    was conditioned for its use as a trigger for the ADC module and
    as common veto for the discriminators. The digital
    data was transferred to a personal computer via Ethernet.}
\label{fig:coincidence-and-data-aquisition}
\end{figure}

The electronic configuration of the generation of the coincidence
trigger and the data-acquisition system is shown in
figure~\ref{fig:coincidence-and-data-aquisition}. For the derivation
of the trigger impulse one preferably uses the additional last dynode
signal provided by the H8500 PSPMT. The signal from the last dynode is
synchronous to those from the $\mathrm{64}$ anodes and its pulse height
is proportional to the total amount of detected scintillation light.
Therefore, we can use this signal instead of a fraction of the
anode-signals for energy discrimination and detection of temporally
coincident events. Two channels of the octal leading edge
discriminator 623B from LeCroy were used to establish the low energy
threshold for both dynode signals. The sum of the width of both
resulting logic pulses defines the time coincidence window which
was set to its lower limit of approximately $\mathrm{9\,ns}$, {\em i.e.}\
using the minimum pulse width of $\mathrm{4.5\,ns}$ provided by each discriminator
module. These signals are fed
directly into a logic unit (Ortec CO4020) configured for coincidence
mode operation. From this signal two pulses are generated. The first
pulse has a duration of $\mathrm{400\,ns}$ and is delayed by
$\mathrm{200\,ns}$. This signal is used for triggering the 12-channel charge integrating analog-to-digital
converter module (Zavarzin and Earle, \cite{Zavarzin:1999}). Each of the 12 channels is
internally equipped with a $\mathrm{200\,ns}$ delay line, baseline restoration,
signal shaping and an analogue integrator whose integration time is
defined by the width of the trigger pulse. The internal delay of the
input signals by $\mathrm{200\,ns}$ comes in very handy for the derivation of the
coincidence trigger. The second pulse of $\mathrm{2.5\,\mu s}$
width from the logic unit and without delay is fed into the common
veto input of the discriminator unit in order to avoid re-triggering of
the ADC module while a conversion is in progress. Once the digital data
is available, it is transfered to a PC using a 100MB Ethernet
connection.

\begin{figure}[!t]
  \centering
  \includegraphics[angle=270,width=\textwidth]{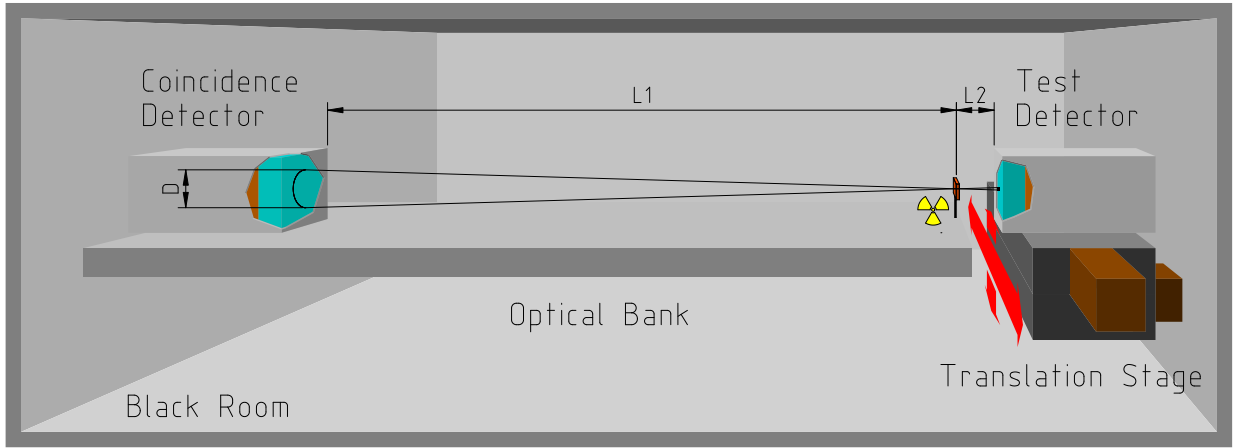}
  \vspace*{1eX}
  \caption[Mechanical setup of the experiment]{Mechanical setup of
    the experiment. All components required for the measurements are
    mounted inside a black room in order to avoid detection of exterior
    stray light. The distance $\mathit{L_1}$ from the radioactive 
    source to the coincidence module was $\mathrm{24\,cm}$ and the
    distance $\mathit{L_2}$ from the radioactive source to the test detector was
    approximately $\mathrm{3\,mm}$.}
  \label{fig:coincidence-setup}
\end{figure}

The complete experimental setup was mounted inside a black room
in order to minimize detection of stray light (refer to
figure~\ref{fig:coincidence-setup}). While the coincidence
module was mounted stationary, the test detector was mounted on top
of a $x$-$y$ translation stage with a precision of $\mathrm{10\,\mu m}$
and driven by PC-controlled stepper motors. 
A $\mathrm{^{22}Na}$ source of $\mathrm{10\,\mu Ci}$ nominal activity
was aligned with the center of the coincidence detector. Furthermore, it
was placed at the large distance $\mathrm{L_1=24\,cm}$ from this detector but
as close as possible to the test detector (distance
$\mathrm{L_2\approx3\,mm}$). 
Due to geometric arguments, the positions of valid annihilation events detected
at the test detector have to lie within a small circular spot of
diameter $\mathrm{d=DL_2/L_1\approx D/80}$ whenever they are detected at the
coincidence detector within the circle of diameter $\mathrm{D}$. 
In the experiment, $\mathrm{D}$ was set to approximately $\mathrm{1.2\,cm}$.

After the alignment of the radioactive source and both modules, the
test detector was calibrated for offsets introduced by the
preamplifier and the analogue-to-digital converters. For this purpose,
the charge integrating digitizing card provides an internal clock
allowing a fast simultaneous offset measurement of all 12 electronic
channels with the high voltage of the Photomultiplier tubes turned
off. The gain of the different electronic channels was supposed to be
nearly the same and was not calibrated.

\begin{figure}[!t]
  \centering
  \includegraphics[height=0.18\textheight]{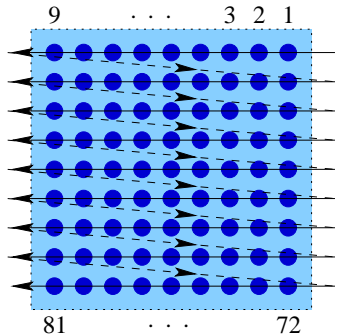}
  \caption[Graphic illustration of the 81 test positions]{Graphic
    illustration of the 81 test positions. The starting point of the
    measurement sequence is the upper left corner; then the test
    detector is moved to the right until the last position of the
    upper row in the matrix of positions is reached. Thereafter, the
    detector is moved to the left of the next row and begins measuring the
    next row and continues in this way till the end position with
    number 81.}
  \label{fig:sample-pos}
\end{figure}

The energy, both centroids and the second moment of the signal distributions of valid
coincidence events were measured at
81 different positions $\mathrm{(x,y), \mbox{with }
  x,y\in[\pm19,\pm14.25,\pm9.5,\pm4.75,0]}$. With the aid of the 
translational stage, the test-detector was automatically moved to each
of these positions as indicated in figure~\ref{fig:sample-pos} and
 192000 coincidence events were registered before moving to the next
position. The data was stored in raw format (all 12 electronic
channels for each event) for subsequent processing, {\em i.e.}\ computation
of the moments, electronic collimation and energy and position filtering.
Approximately $\mathrm{6-8\,\%}$ of all events produced an ADC
overflow or underflow. These events have been removed from the
data sets.

\subsection{Spatial Extension of the Radioactive Test-Source}
\label{sec:spat-extend-source}

\begin{figure}[!t]
  \centering
  \subfigure[][Normalized density plot of the simulated positron endpoint positions.]{\label{subfig:pos-range-density}
    \psfrag{x}{\hspace*{-1em}x [mm]}
    \psfrag{y}{\hspace*{-1em}y [mm]}
  \includegraphics[height=0.19\textheight]{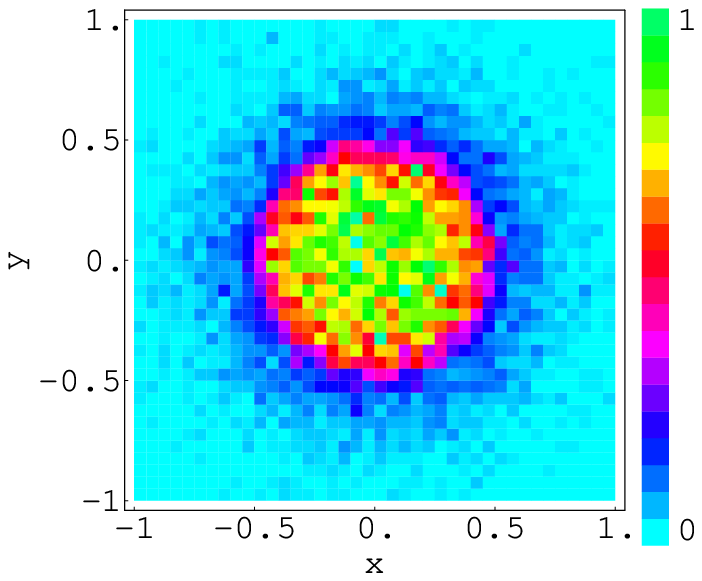}}
  \subfigure[][Histogram of the distances $|\mathbf{r}|$ from the
  center of the source to the positron endpoint positions. 
  The larger relative errors for small $|\mathbf{r}|$ are due to
  the normalization of the bin contents.]{\label{subfig:pos-range-in-source}
    \psfrag{counts}{\hspace*{-0.7em}counts [a.u.]}
    \psfrag{range}{\hspace*{-1em}$|\mathbf{r}|$ [mm]}
  \includegraphics[height=0.19\textheight]{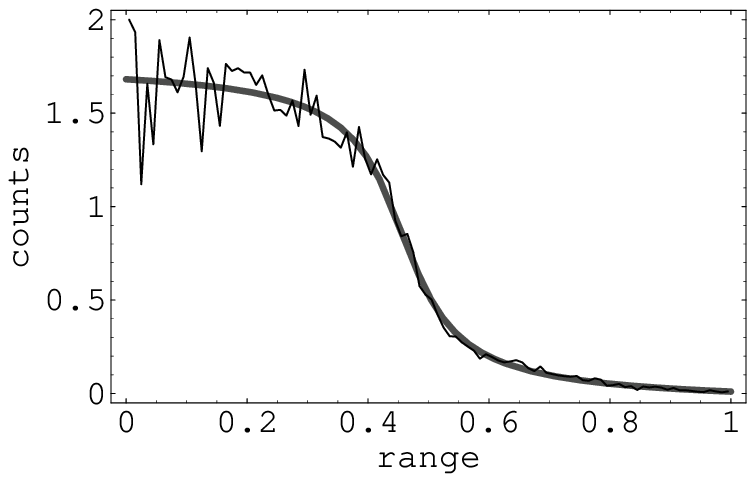}}
  \caption[Simulated positron endpoint distribution of the 
     test source]{Simulated positron endpoint distribution of the
     test source.}
  \label{fig:positron-range-source}
\end{figure}

A $\mathrm{^{22}Na}$ point source of $\mathrm{370\,kBq}$ nominal
activity was used for these measurements. The radioactive isotope
is absorbed in an ion-exchange bead and fixed at the geometric center
of an epoxy resin capsule. However, this source with a diameter of
$\mathrm{1\,mm}$ cannot be considered as a point-like source for the
present experiment, since the spatial resolution is expected to be of the same
order and therefore one has to correct the results for this finite size effect.
Furthermore, one has to take into account that the source is a
$\mathrm{\beta^+}$-emitter and that the positrons have a finite range
before giving rise to the annihilation radiation (refer to
section~\ref{subsec:radio-source-errors}). This
leads to a penetration of $\mathrm{\beta^+}$-particles into
the epoxy resin and also to an expected higher activity at the center
of the source pellet. The resulting effective source diameter can
hardly be measured and a description with analytic models would be too
cumbersome. Instead, it was estimated using Monte Carlo
simulations (Sanchez, \cite{sanchez:privcom}) with \mbox{GEANT 3}
(Brun and Carminati \cite{Geant3:1994}).

Figure~\ref{subfig:pos-range-density} shows a normalized density plot
of the positron endpoint positions of $\mathrm{20000}$ simulated events projected
onto the $x$-$y$-plane. However, for the estimation of the effective diameter
only the radius $|\mathbf{r}|=\mathrm{\sqrt{x^2+y^2}}$ is of
interest. One has to normalize the bin-content of the histogram for
$|\mathbf{r}|$  with $|\mathbf{r}|$ because
the circumference and therefore the counts in each bin grow with the radius. The
simulation results are displayed in
figure~\ref{subfig:pos-range-in-source}. 
In order to compute the FWHM from the histograms of
the radii, the following empirical model distribution 
\begin{equation}
  \label{eq:source-extension-fitmodel}
  N(|\mathbf{r}|)=a + b\,\arctan (d + c\;|\mathbf{r}|)
\end{equation}
was least-square fitted to the data
(figure~\ref{subfig:pos-range-in-source}). The obtained best-fit
parameters are:
\begin{align}
  \label{eq:bet-fit-params-pos-range}
  a&=830\pm26, & b&=580\pm18, & c&=-13.9\pm0.8\,\mathrm{mm}^{-1},& d&=6.3\pm0.4.
\end{align}
Equation~\ref{eq:source-extension-fitmodel} can be easily inverted for
finding the FWHM of the distribution:
\begin{equation}
  \label{eq:fwhm-pos-range}
  |\mathbf{r}|_\tincaps{FWHM}=-\left( \frac{d + \tan (\frac{a - b\,\arctan (d)}{2\,b})}{c} \right),
\end{equation}
and leads to the final result for the effective source diameter
\begin{equation}
  \label{eq:effective-source-diameter}
  d_\mathit{eff}=2|\mathbf{r}|_\tincaps{FWHM}=0.92\pm0.15\,\mathrm{mm}.
\end{equation}

\subsection{Model Distribution for Event Statistics}
\label{sec:model_dist_for_event_stat}

As mentioned in the introduction of this chapter, another experimental difficulty in
the setup is imposed by the fact that one cannot easily prepare 
a measurement for the distribution moments at well defined interaction
depths. Unfortunately, one is only able to define the impact position
along the two
spatial directions that are normal to the $\gamma$-ray beam. The
position of interaction along the remaining direction parallel to the
beam is subjected to the random processes of 
photoabsorption and Compton scattering. For thick scintillation crystals one therefore faces
a large uncertainty in this third position component while thinner
crystals produce a lower uncertainty. However, the attempt to reduce
the crystal thickness would lead to lower interaction probability
(efficiency) and would clearly change the optical behavior of the
crystal. 

\begin{figure}[t]
  \centering
  \subfigure[][Sketch of the derivation of the model
  distribution for the second moment of the detected
  events.]{\label{subfig:fitmodel-derive-a}%
    \psfrag{e}{$e^{-\alpha z}$}
    \psfrag{a}{$a$}
    \psfrag{b}{$b$}
    \psfrag{D}{\hspace*{-1em}$z$ [a.u.]}
    \psfrag{C}{\hspace*{-2.4em}Counts [a.u.]}
    \includegraphics[width=0.485\textwidth]{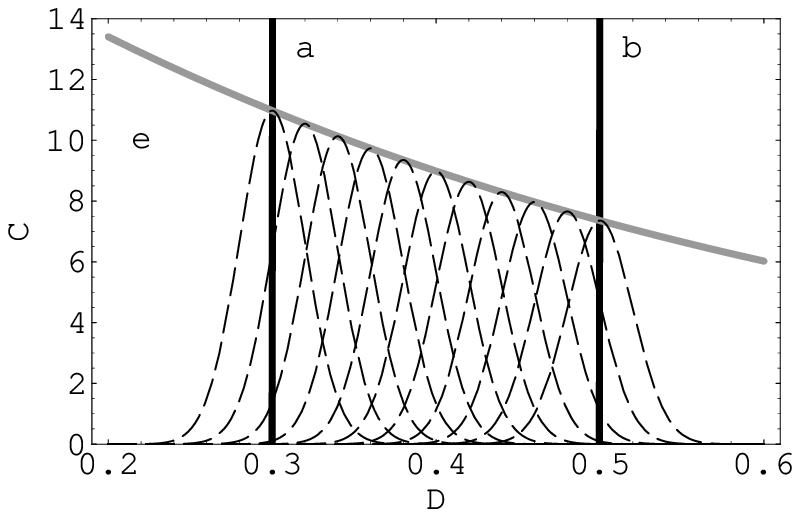}}
  \subfigure[][Resulting model distribution.]{\label{subfig:fitmodel-derive-b}%
    \psfrag{e}{$e^{-\alpha z}$}
    \psfrag{a}{$a$}
    \psfrag{b}{$b$}
    \psfrag{D}{\hspace*{-1em}$z$ [a.u.]}
    \psfrag{C}{\hspace*{-2.4em}Counts [a.u.]}
    \includegraphics[width=0.485\textwidth]{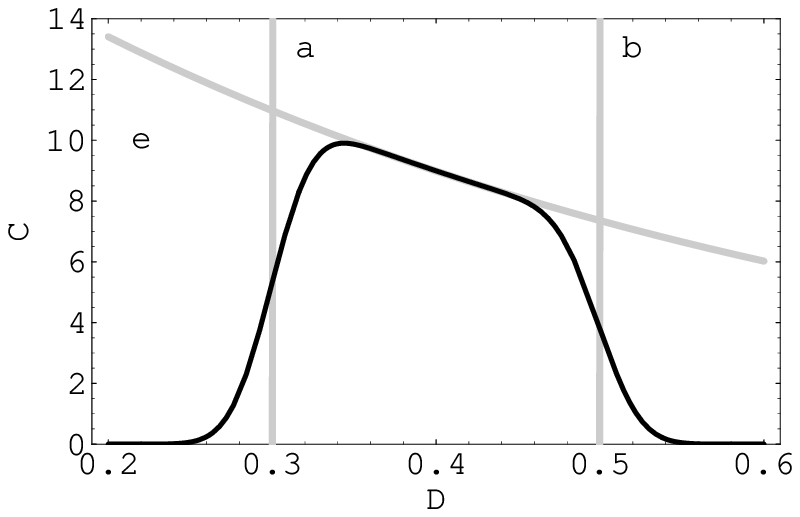}}
  \subfigure[][Sample distributions for $A=10$, $a=2$, $b=6$,
  $\alpha=0.1$ and $\hat{\sigma}=\{\simeq0,0.2,0.4,\ldots,1.4\}$.]{\label{subfig:fit-model-schar-sigma}%
    \psfrag{x}{\scriptsize\hspace*{-1em}$z$ [a.u.]}
    \psfrag{N}{\scriptsize\hspace*{-2.4em}Counts [a.u.]}
    \includegraphics[width=0.32\textwidth]{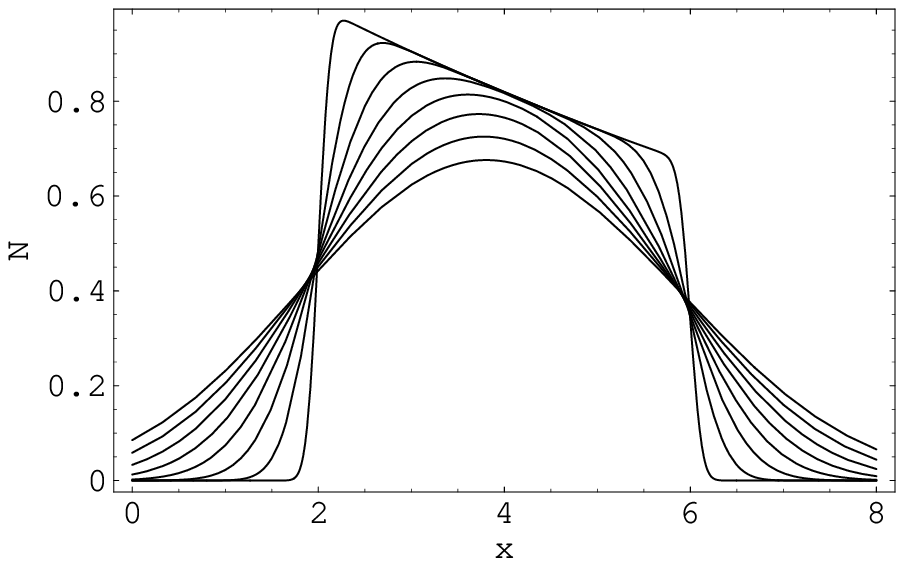}}
  \subfigure[][Sample distributions for $A=10$, $a=2$, $b=6$,
  $\hat{\sigma}=0.2$ and $\alpha=\{\simeq0,0.1,0.2,\ldots,1.1\}$.]{\label{subfig:fit-model-schar-alpha}%
    \psfrag{x}{\scriptsize\hspace*{-1em}$z$ [a.u.]}
    \psfrag{N}{\scriptsize\hspace*{-2.4em}Counts [a.u.]}
    \includegraphics[width=0.32\textwidth]{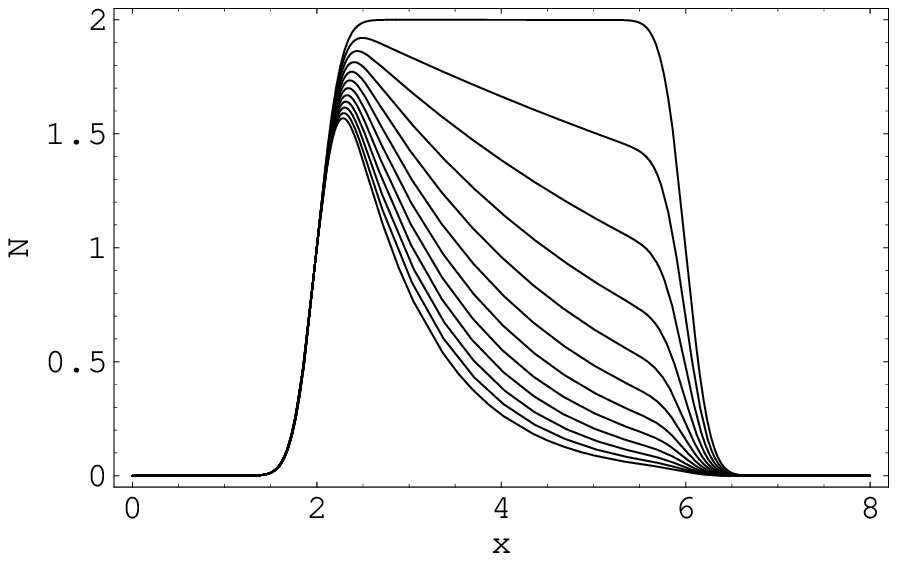}}
  \subfigure[][Sample distributions for $A=10$, $a=2$, $\hat{\sigma}=0.4$
  and $\alpha=0.1$ $b=\{2.5,3,3.5,\ldots,6\}$.]{\label{subfig:fit-model-schar-b}%
    \psfrag{x}{\scriptsize\hspace*{-1em}$z$ [a.u.]}
    \psfrag{N}{\scriptsize\hspace*{-2.4em}Counts [a.u.]}
    \includegraphics[width=0.32\textwidth]{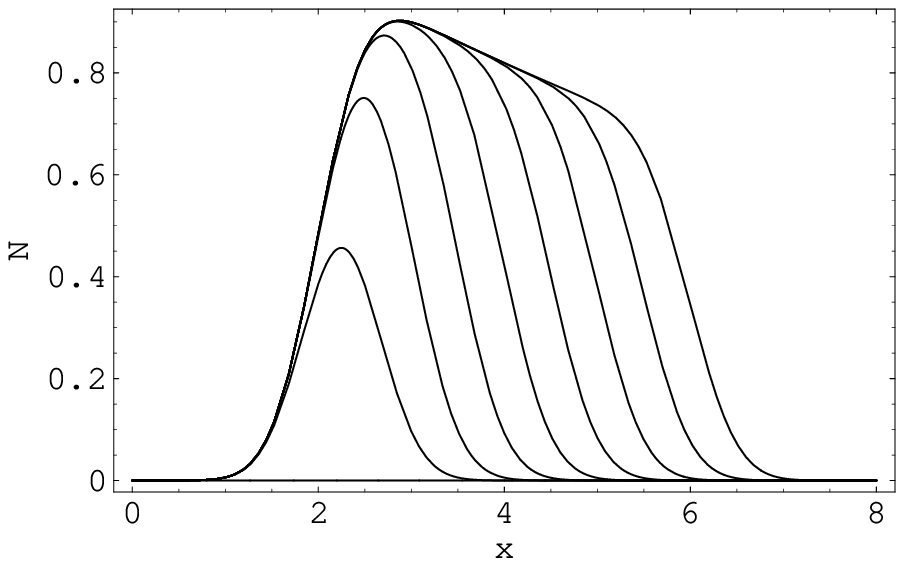}}
  \caption[Graphical illustration for the derivation of the fit model]{Graphical illustration for the derivation of the fit model. The lower figures illustrate the change of
    the shape for variations in some of the parameters. Note that the
    parameters do not have subscripts referring to the position or
    moment space because this general behavior does not depend on the
    choice of representation space.}
  \label{fig:fit-model-dist}
\end{figure}

Another possibility consists of properly defining the depth of interaction
and one of the spatial dimensions $x$ or $y$. This can be achieved 
experimentally when a $\gamma$-ray beam parallel either to
the $x$-axis or the $y$-axis enters the crystal at one of the small
sides normal to the sensitive area. The remaining component
of the impact position can be measured sufficiently well by means of
the center of gravity algorithm, since no superposition from different
depths occurs. An important drawback of this method is the attenuation
of the $\gamma$-ray beam along the indeterminate spatial direction. 
Especially for large-sized crystals this leads to low statistics at
the side opposite to the $\gamma$-ray source. Furthermore, it does not
allow fast and handy calibrations of the detector.

The depth enhanced charge dividing circuits presented in
chapter~\ref{ch:enhanced-charge-dividing-circuits} allow for a
more suitable approach since all three spatial components of the
impact position can be measured simultaneously, provided that the
spatial resolution is sufficiently good and the dependency of the
second moment on the depth of interaction is known and sufficiently
pronounced. In this case, one expects a characteristic distribution of
the second moment which has to be a superposition of the detector
response at all possible interaction depths. By making the following
assumptions, an adequate model for the distribution of the second
moment was derived. First, it is supposed that there is a constant intrinsic 
resolution for the spatial direction parallel to the
$\gamma$-ray beam. This is clearly only an approximation since there
are various possible effects that can make this intrinsic resolution
depend on the position. However, the superposition requires
integration over the crystal's spatial extension along $z$-axis and,
therefore, does not admit very sophisticated integrands. Furthermore, the detector's
response function for an exactly defined depth of interaction is
expected to have Gaussian shape and width $2\hat{\sigma}$.
The second
assumption is that the second moment at all points $(x,y)$ depends linearly on the depth of
interaction. This allows for linear transformations of the integration
variable and legitimates the use of the model for the second moment
instead of the true depth of interaction. Evidence for this second
assumption has been reported by Antich {\em et al.}\ \cite{Antich:2002} 
and was also predicted by the model derived in
chapter~\ref{ch:light-distribution} (refer to
figures~\ref{subfig:secmom-d-depth}-\ref{subfig:sigma-d-depth-wo-bg} in
section~\ref{sec:complete-signal-dist}).

The $\gamma$-ray beam's intensity
decreases exponentially due to the attenuation within the
crystal. Therefore, the amplitudes of the Gaussians decrease with
increasing depth of interaction and one obtains the convolution
\begin{equation}
  \label{eq:fit-model-synthese}
  \mathcal{D}_\mathit{m}(z_\mathit{m})=\frac{A_\mathit{m}\alpha_\mathit{m}}{{\sqrt{2\,\pi
      }}\,\hat{\sigma}_\mathit{m}}\int_{a_\mathit{m}}^{b_\mathit{m}}
  \exp\left\{-\alpha_\mathit{m}({z_\mathit{m}}'-a_\mathit{m})-
  \frac{{\left( z_\mathit{m} - {z_\mathit{m}}'\right)}^2}{2\,{\hat{\sigma}_\mathit{m} }^2}\right\}d{z_\mathit{m}}',
\end{equation}
where $\mathrm{\alpha_m}$ is the attenuation coefficient of the
scintillation crystal for the $\gamma$-ray energy of interest, and
$\mathrm{a_m}$ and $\mathrm{b_m}$ are the upper and lower limits of the
crystal respectively. Outside of these limits one does not expect any events and
the distribution is equal to zero. The subscript $m$ indicates that all
parameters in equation~\ref{eq:fit-model-synthese} are understood to
refer to the moment space. Figure~\ref{subfig:fitmodel-derive-a}
illustrates the integral in equation~\ref{eq:fit-model-synthese}. 
An explicit form is obtained by resolving the previous integral:
\begin{equation}
  \label{eq:fit-model-dist}
  \mathcal{D}_\mathit{m}(z_\mathit{m})=\frac{A_\mathit{m}\alpha_\mathit{m}}{2}%
  \exp\left\{\alpha_\mathit{m}\left(a_\mathit{m}-z_\mathit{m}+\ts\frac{\alpha_\mathit{m}{\hat{\sigma}_\mathit{m}}^2}{2}\ds\right)\right\}
  \left[\erf\left\{\frac{b_\mathit{m}-z_\mathit{m}+\alpha_\mathit{m}{\hat{\sigma}_\mathit{m}}^2}{{\sqrt{2}\;}\hat{\sigma}_\mathit{m}}\right\}%
    -\erf\left\{\frac{a_\mathit{m}-z_\mathit{m}+\alpha_\mathit{m}{\hat{\sigma}_\mathit{m}}^2}{{\sqrt{2}\;}\hat{\sigma}_\mathit{m}}\right\}\right]\!,
\end{equation}
where $\erf$ stands for the error function.
Since there is a linear dependence between the second moment and the depth of
interaction, the corresponding intrinsic depth resolution
$\hat{\sigma}_\mathit{p}$ in the position space is given by scaling
$\hat{\sigma}_\mathit{m}$. The integration limits $a_\mathit{m}$ and
$b_\mathit{m}$ in equation~\ref{eq:fit-model-synthese} define with
$|a_\mathit{m}-b_\mathit{m}|$ a measure for the true
crystal thickness $\mathrm{T}$ in the moment space. Therefore, in position space
one obtains the resolution
\begin{equation}
  \label{eq:doi-in-position-space}
  \hat{\sigma}_\mathit{p}=\hat{\sigma}_\mathit{m}\frac{T}{|a_\mathit{m}-b_\mathit{m}|}.
\end{equation}

In figures~\ref{subfig:fit-model-schar-sigma} -
\ref{subfig:fit-model-schar-b}, plots of the model distribution for
different parameters are shown. One can observe that the distribution
adopts a Gaussian shape for the case
$\mathrm{|a_m-b_m|\ll\hat{\sigma}_m}$. Actually, it
can be seen directly from the definition~\ref{eq:fit-model-synthese}
that the distribution approaches a Gaussian-like shape in the limit $\mathrm{b_m\to a_m}$.
If $\mathrm{\alpha_m}$ is large compared to $\mathrm{\hat{\sigma}_m}$ and
$\mathrm{\alpha_m^{-1}\ll|a_m-b_m|}$, the
distribution approaches an exponential decay distribution that is
horizontally shifted by $\mathrm{a_m}$. For both these limits, fits
with this special model distribution cannot produce reliable
estimates for the parameters $\mathrm{a_m}$ and $\mathrm{b_m}$.
Clearly, these limits have to be avoided with a proper experimental
setup. In the experiment, the majority of histograms did not approach
these limits and the fit routine gave acceptable results. 

Note that all assumptions that were made for the derivation of the
model also hold for the zeroth moment, {\em i.e.}\ the energy, and both first
moments (the centroids). As shown in
section~\ref{ch:errors-of-cog-and-cdr}, these moments also exhibit
depth of interaction dependence. In particular,
near the crystal borders it
shows the characteristic shape displayed in the examples in
figures~\ref{subfig:fitmodel-derive-a}-\ref{subfig:fit-model-schar-b}
and, therefore, has to be preferred to the Gaussian model. The
model~(\ref{eq:fit-model-dist}) was modified in some cases by adding
a constant or linear background.

\begin{figure}[!t]
  \centering
  \subfigure[][Second moment at the
  center of the PSPMT, \mbox{$\nu=123$}, \mbox{$\chi^2=1.15$},
  \mbox{$Q=0.87$}.]{\label{subfig:fit-example-1}
    \psfrag{x}{\hspace*{-1.3em}$z$ [a.u.]}
    \psfrag{y}{\hspace*{-1.3em}counts}
  \includegraphics[width=0.45\textwidth]{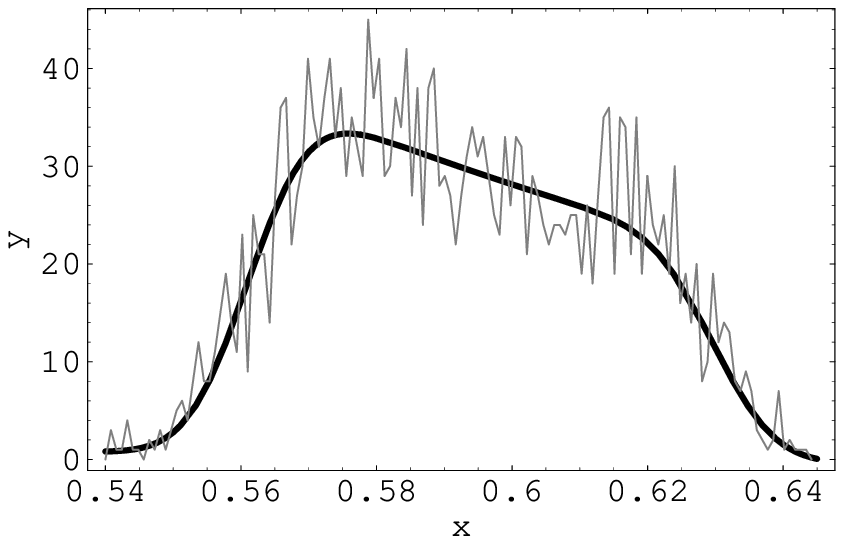}}
  \subfigure[][Second moment at a
  corner, \mbox{$\nu=145$}, \mbox{$\chi^2=1.37$}, 
  \mbox{$Q=1$}.]{\label{subfig:fit-example-2}
    \psfrag{x}{\hspace*{-1.3em}$z$ [a.u.]}
    \psfrag{y}{\hspace*{-1.3em}counts}
  \includegraphics[width=0.45\textwidth]{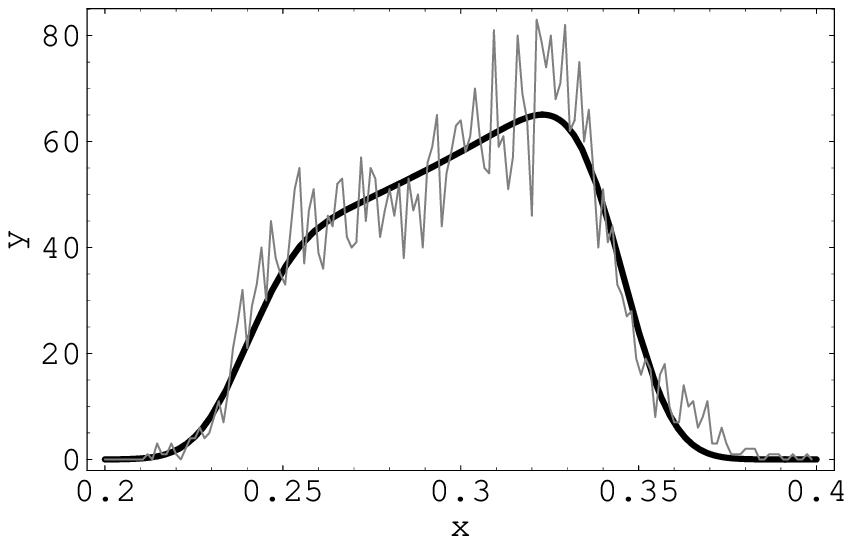}}\\
  \vspace*{1eX}
  \subfigure[][Dispersion of the $x$-coordinate at the center, \mbox{$\nu=144$}, \mbox{$\chi^2=1.2$}, 
  \mbox{$Q=0.95$}.]{\label{subfig:fit-example-3}
    \psfrag{x}{\hspace*{-1.3em}$x$ [a.u.]}
    \psfrag{y}{\hspace*{-1.3em}counts}
  \includegraphics[width=0.45\textwidth]{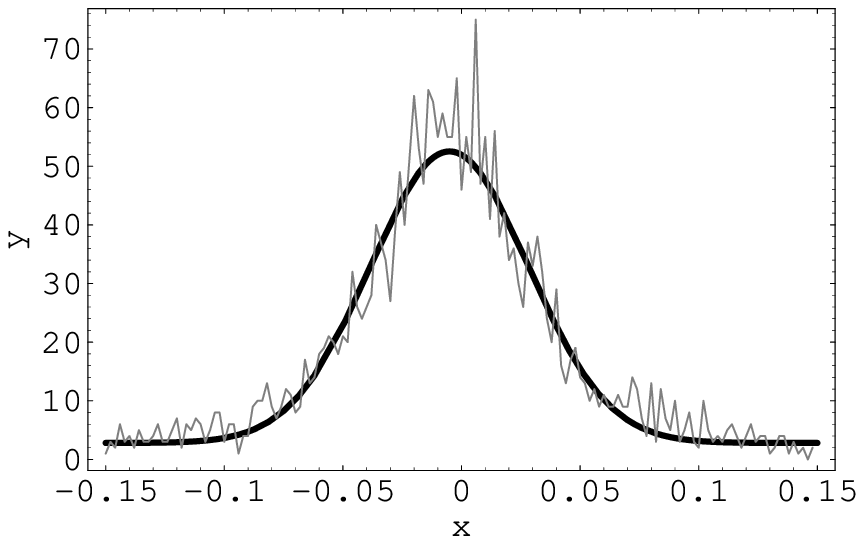}}
  \subfigure[][Dispersion of the $x$-coordinate at a corner, \mbox{$\nu=143$}, \mbox{$\chi^2=1.13$}, 
  \mbox{$Q=0.86$}.]{\label{subfig:fit-example-4}
    \psfrag{x}{\hspace*{-1.3em}$x$ [a.u.]}
    \psfrag{y}{\hspace*{-1.3em}counts}
  \includegraphics[width=0.45\textwidth]{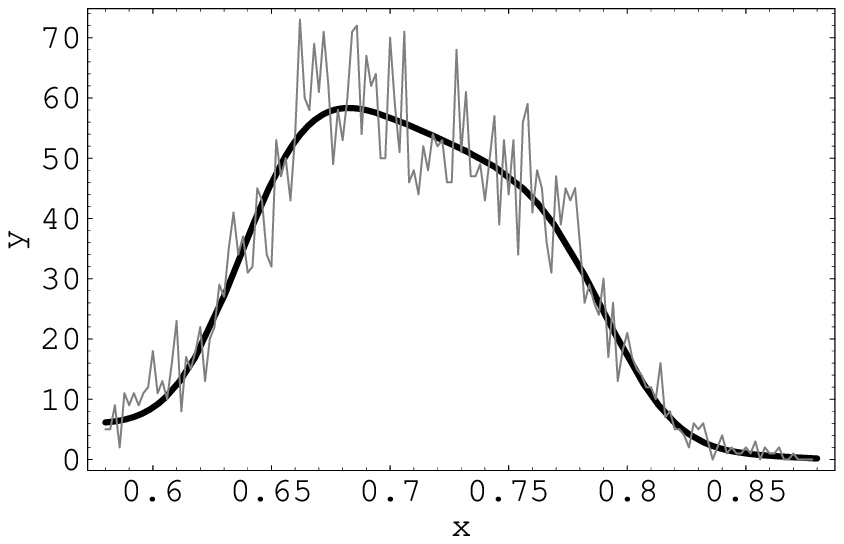}}\\
  \vspace*{1eX}
  \subfigure[][Dispersion of the energy at the center, \mbox{$\nu=145$}, \mbox{$\chi^2=1.16$}, 
  \mbox{$Q=0.9$}.]{\label{subfig:fit-example-5}
    \psfrag{x}{\hspace*{-3em}energy [a.u.]}
    \psfrag{y}{\hspace*{-1.3em}counts}
  \includegraphics[width=0.45\textwidth]{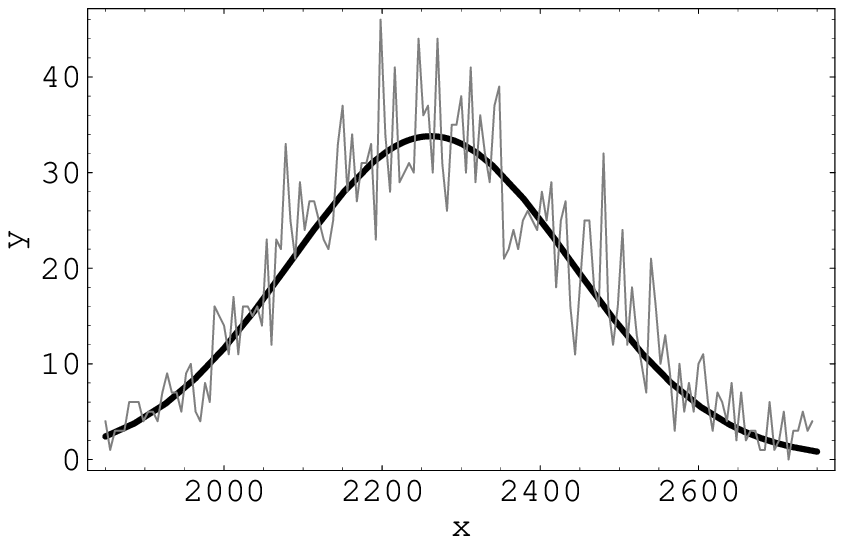}}
  \subfigure[][Dispersion of the energy at a corner, \mbox{$\nu=113$}, \mbox{$\chi^2=1.24$}, 
  \mbox{$Q=0.96$}.]{\label{subfig:fit-example-6}
    \psfrag{x}{\hspace*{-3em}energy [a.u.]}
    \psfrag{y}{\hspace*{-1.3em}counts}
  \includegraphics[width=0.45\textwidth]{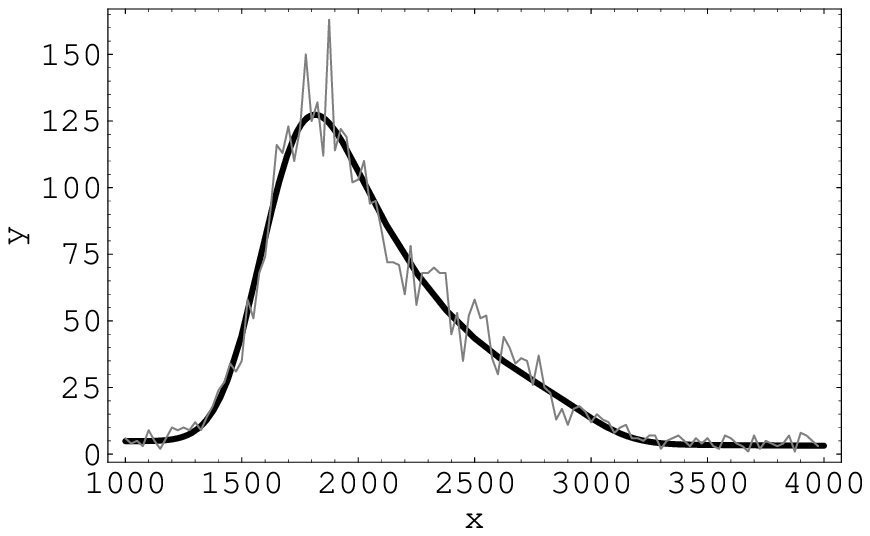}}\\
  \vspace*{1eX}
  \caption[Examples of distributions from moment
  measurements]{Distributions of the second
    moment, $x$-coordinate of the centroid and energy at two
    different positions over the photocathode. True data (thin
    light-gray lines) are plotted together with the best fits (black lines).}
  \label{fig:fit-examples}
\end{figure}

Figures~\ref{subfig:fit-example-1}-\ref{subfig:fit-example-6} show
some examples of the model fitted to distributions measured with the
experimental setup that was described above. The left column shows
best fits for the case when the $\gamma$-ray beam entered the
detector at the center of the PSPMTs sensitive area ($\mathrm{x=y=0}$). The
right column shows the results obtained when locating the beam
at one of the corner positions, {\em e.g.}\ $\mathrm{x,y=\pm19}$. It can be
seen that the model reproduces well the measured distributions
for both cases. All other positions will result in
distributions of a form in between these two extreme cases.  

A more reliable method to estimate the goodness of the fit results
is given in Press {\em et al.}\ \cite{Press:1992} and is reasoned as
follows. The probability $Q$ that a value for $\chi^2$ obtained by
fitting the model to experimental data occurs for a determinate number
$\nu$ of degrees of freedom is given by 
\begin{equation}
  \label{eq:good-of-fit}
  Q=1-\Gamma\left(\frac{\nu}{2},\frac{\chi^2}{2}\right)/\Gamma\left(\frac{\nu}{2}\right),
\end{equation}
where $\Gamma(z)$ is the Euler gamma function and $\Gamma(a,z)$ is the
incomplete gamma function. The values of $Q$, $\nu$ and $\chi^2$ are
given together with the graphs in
figures~\ref{subfig:fit-example-1}-\ref{subfig:fit-example-6}.
Note that in this estimation nothing is known about the achieved
precision of the best fit parameters. It is possible to obtain 
good values for $Q$ $(\mathrm{Q\simeq1})$ and $\chi^2$ while the errors of the adjusted
parameters could be rather large. Actually, this is the case when the
measured distributions becomes similar to the Gaussian or exponential case. These limits
contain only poor information on $\mathrm{a}$, $\mathrm{b}$ and $\sigma$ 
 and, thus, their errors are quite large.

\section{Results}

First of all, the method of depth of interaction determination is
tested using an inclined $\gamma$-ray beam. This inclination must be
reflected somehow in the histograms for the second moment and the
centroids. Likewise, the broken symmetry of the light distribution
that was explained in section~\ref{sec:symmetry-breaking} must become
apparent from the same histograms if the $\gamma$-ray interacts with the
crystal near one of its absorbing borders. Once this has been tested,
it will be shown that the quality of the second moment is sufficiently good
in order to filter subsets with similar DOI parameters out of
the data. 

After these qualitative crosschecks, the moments measured at the 81 test positions
were compared to the predictions of the model for the
signal distribution established in chapter~\ref{ch:light-distribution}.
In addition, the possible resolution which can be
obtained for each of the lower-order moments is measured.

\subsection{Qualitative Verification of the Method}

\begin{figure}[!t]
  \centering
  \subfigure[][Density plot of $\mu_{x_2,y_0}+\mu_{x_0,y_2}$ vs.\ $\mu_{x_1,y_0}$.]{\label{subfig:inclined_sig-vs-x}
    \includegraphics[width=0.48\textwidth]{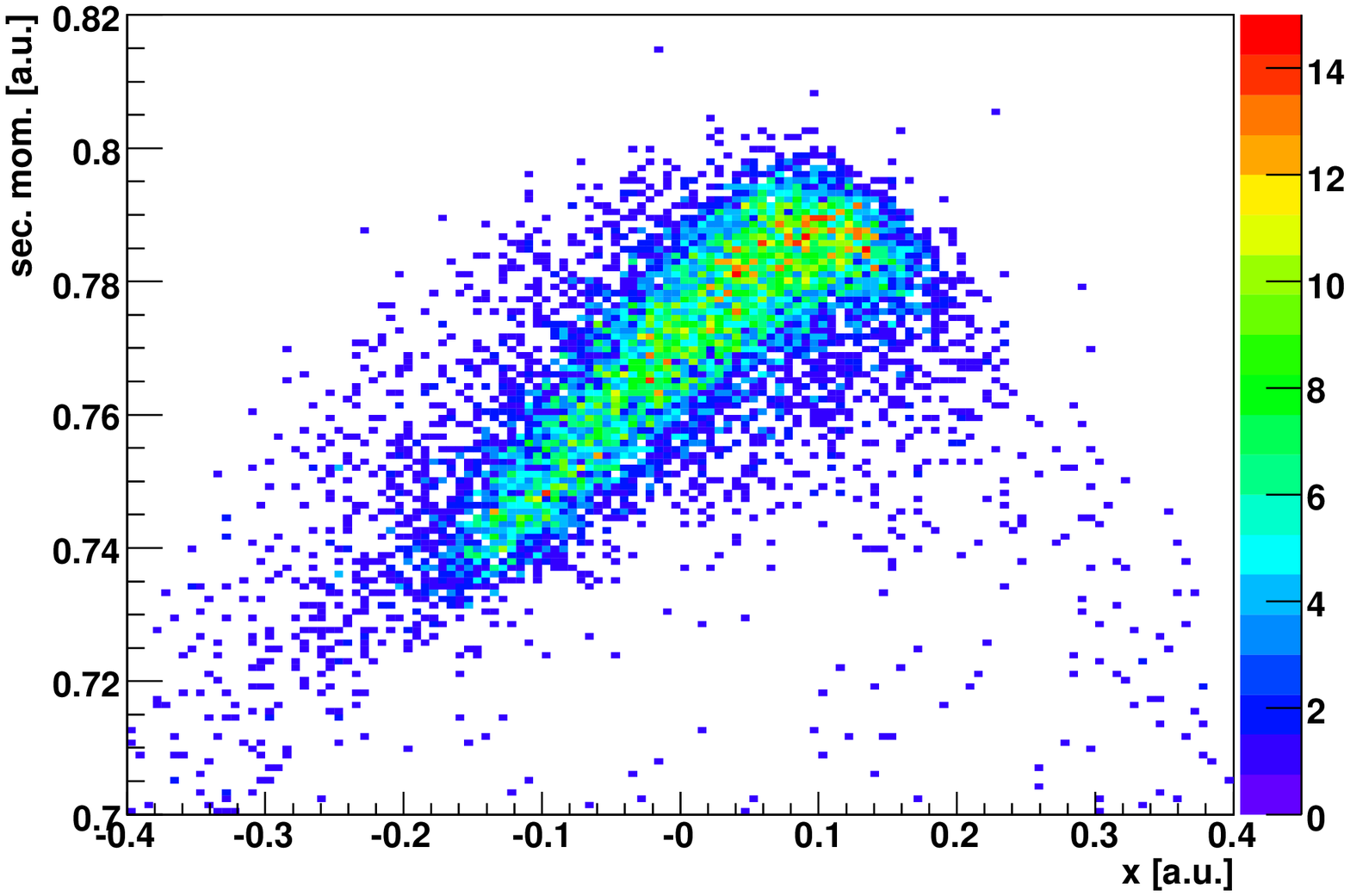}}
  \subfigure[][Density plot of $\mu_{x_2,y_0}+\mu_{x_0,y_2}$ vs.\ $\mu_{x_0,y_1}$.]{\label{subfig:inclined_sig-vs-y}
    \includegraphics[width=0.48\textwidth]{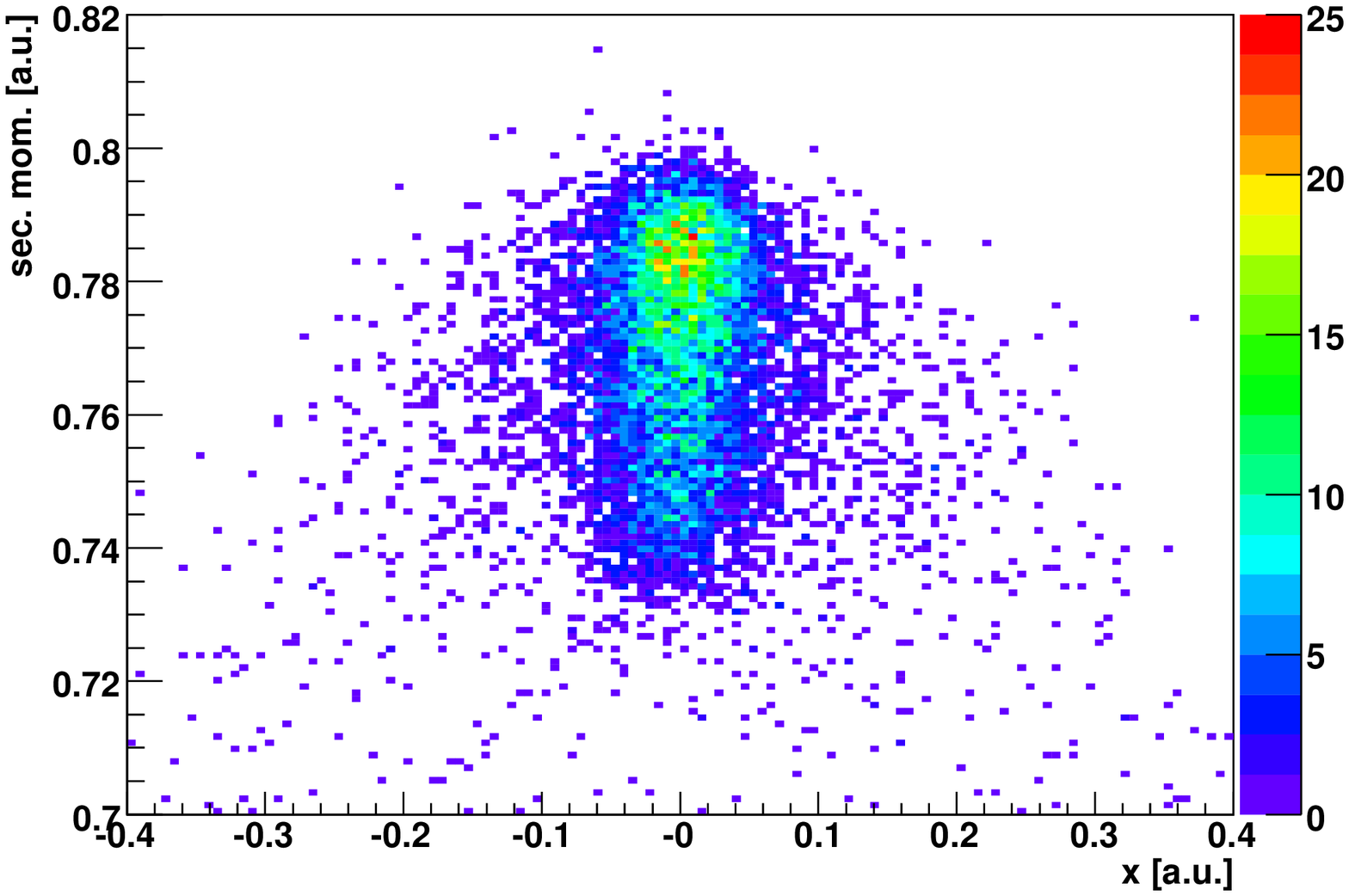}}
  \caption[Correlation between $x$, $y$ and $\mu_2$ for an inclined
  $\gamma$-ray beam]{Measured values for the second moment vs.\ the
    $x$ centroid (left figure) and the $y$ centroid
    (right figure) when the test detector is inclined by 
    $\approx 45\mathdegree$ against the $\gamma$-ray beam.}
  \label{fig:inclined_g_ray}
\end{figure}

In order to crosscheck the presented method, 
the test detector was irradiated with an electronically 
collimated $\gamma$-ray beam inclined by $\mathrm{\approx
  45\mathdegree}$ with
respect to the $z$-axis.  The beam was adjusted in such a way that its
axis crossed the geometric center of the crystal and laid in the
$x$-$z$-plane. In figure~\ref{subfig:inclined_sig-vs-x}, $\hat{\sigma}_\mathit{m}$
is plotted versus the $x$-centroid and a correlation between
these two measures is apparent. In contrast to this, one cannot observe this
correlation for the $y$-centroid (figure~\ref{subfig:inclined_sig-vs-y})
since the $\gamma$-ray beam is perpendicular to this direction.
Antich {\em et al.}\ \cite{Antich:2002} already reported this result. They
studied the possibility of using crossed-wire photomultiplier tubes
together with large-sized $\mathrm{NaI\doped Tl}$ scintillator crystals 
for three-dimensional position readout for SPECT/PET.  However, they
needed to digitize all 32 channels from the Hamamatsu R2486 PSPMT in
order to obtain this impact parameter. Using the enhanced charge
dividing circuits described in
chapter~\ref{ch:enhanced-charge-dividing-circuits}, only 5 electronic
channels were necessary and the moments were computed analogically and
online.

\begin{figure}[!t]
  \begin{center}
    \subfigure[][$x$- $y$-centroid distribution for inclined
    $\gamma$-ray beam.]{\label{subfig:inclined_g_ray_all}
      \includegraphics[width=0.48\textwidth,hiresbb=true]{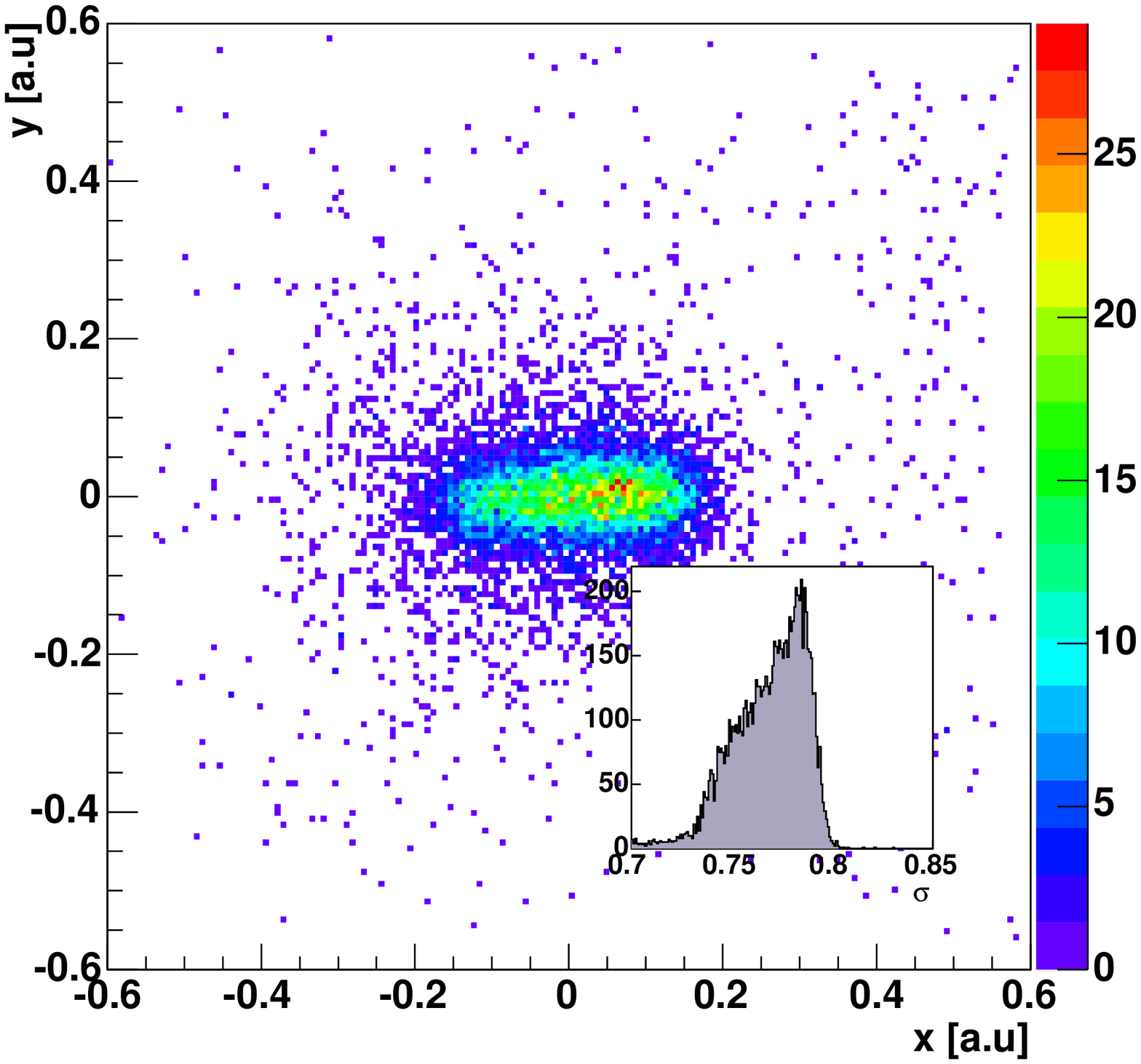}}
    \subfigure[][$x$- $y$-centroid distribution for 
    $\gamma$-ray beam impinging on a corner.]{\label{subfig:corner_DOI_all}
      \includegraphics[width=0.48\textwidth,height=0.44\textwidth,hiresbb=true]{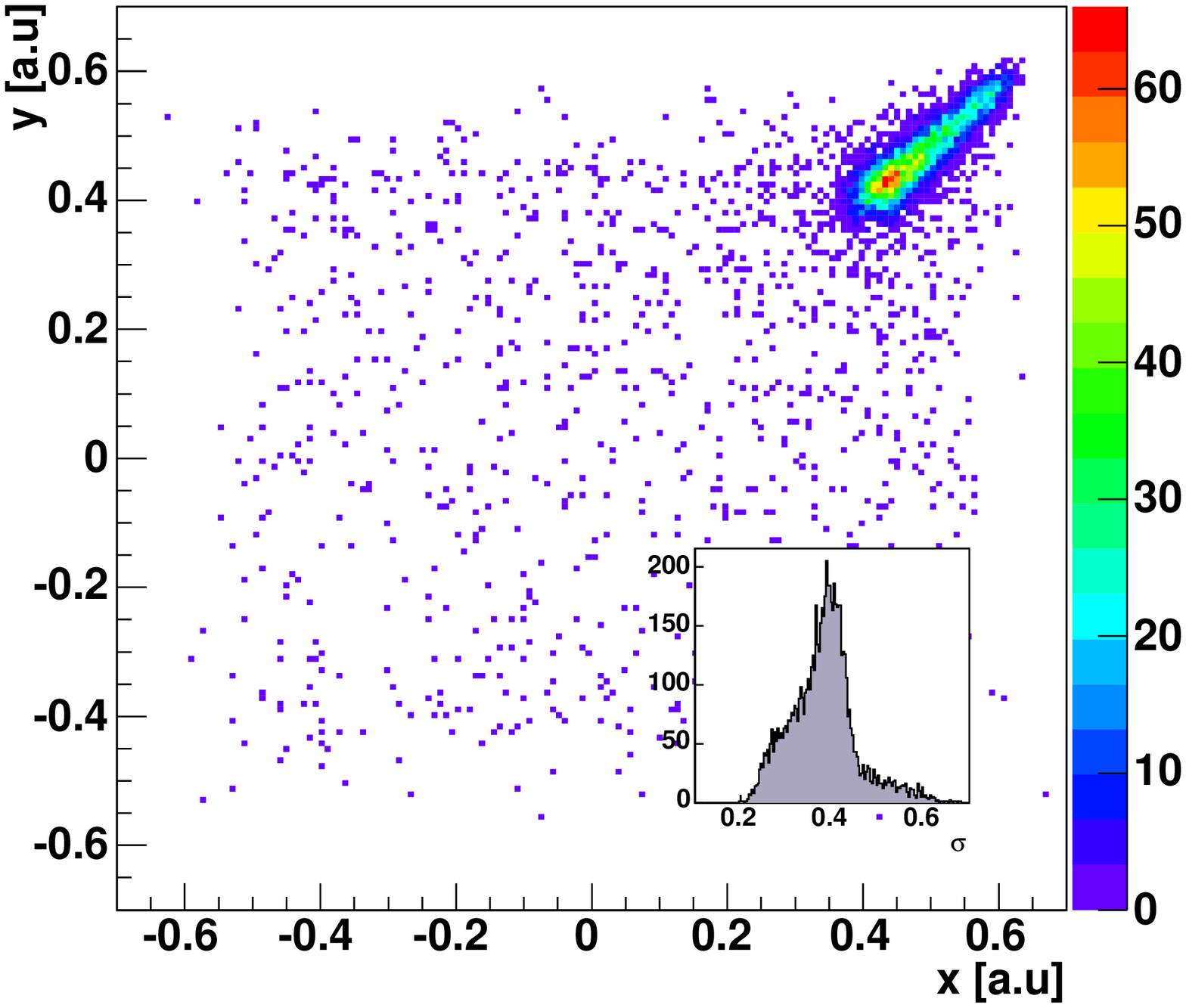}}
  \end{center}
  \caption[Density plot of the centroids for an inclined beam]{Density
    plot of the centroids that were registered when
    the test detector was inclined by an angle of $\approx
    45\mathdegree$ against the $\gamma$-ray beam (l.h.s.) and 
    when the collimated $\gamma$-ray was positioned at
    $(x,y)=(19,19)\;\mathrm{mm}$, {\em i.e.}\ at only 2 mm from the
    crystals borders (r.h.s.). The small insets show the
    corresponding $\hat{\sigma}_\mathit{m}$ distributions.}
  \label{fig:center-inclined-corner}
\end{figure}

The distribution of the measured centroids within the $x$-$y$-plane is shown
in the density plot~\ref{subfig:inclined_g_ray_all}. There is an
apparent blurring along the $x$-direction and almost none along the
$y$-direction. The small inset shows the histogram for the standard
deviation of the signal distribution. In
the same figure to the right,
the distribution for the $x$- and $y$-centroids is shown for a
$\gamma$-ray beam hitting the test detector at an outer
corner of the photocathode. The blurring of both centroids is now
caused by the effect described in section~\ref{sec:symmetry-breaking}
and not by the $\gamma$-ray beam itself, because this time it is
normal to the photocathode. The small inset shows the corresponding distribution of
the second moment. Note that due to
mechanical limitations of the experimental setup, the radioactive
source could not be placed very close to the detector when the
detector was inclined by $\mathrm{\approx 45\mathdegree}$. Therefore, a more
important blurring in all directions is observed in
figure~\ref{subfig:inclined_g_ray_all}.

\begin{figure}[!tp]
  \centering
  \vspace*{-2eX}
  \subfigure[][Events (inclined beam) with low $\hat{\sigma}_m$.]{\label{subfig:inclined_high}
    \includegraphics[width=0.39\textwidth,keepaspectratio,hiresbb=true]{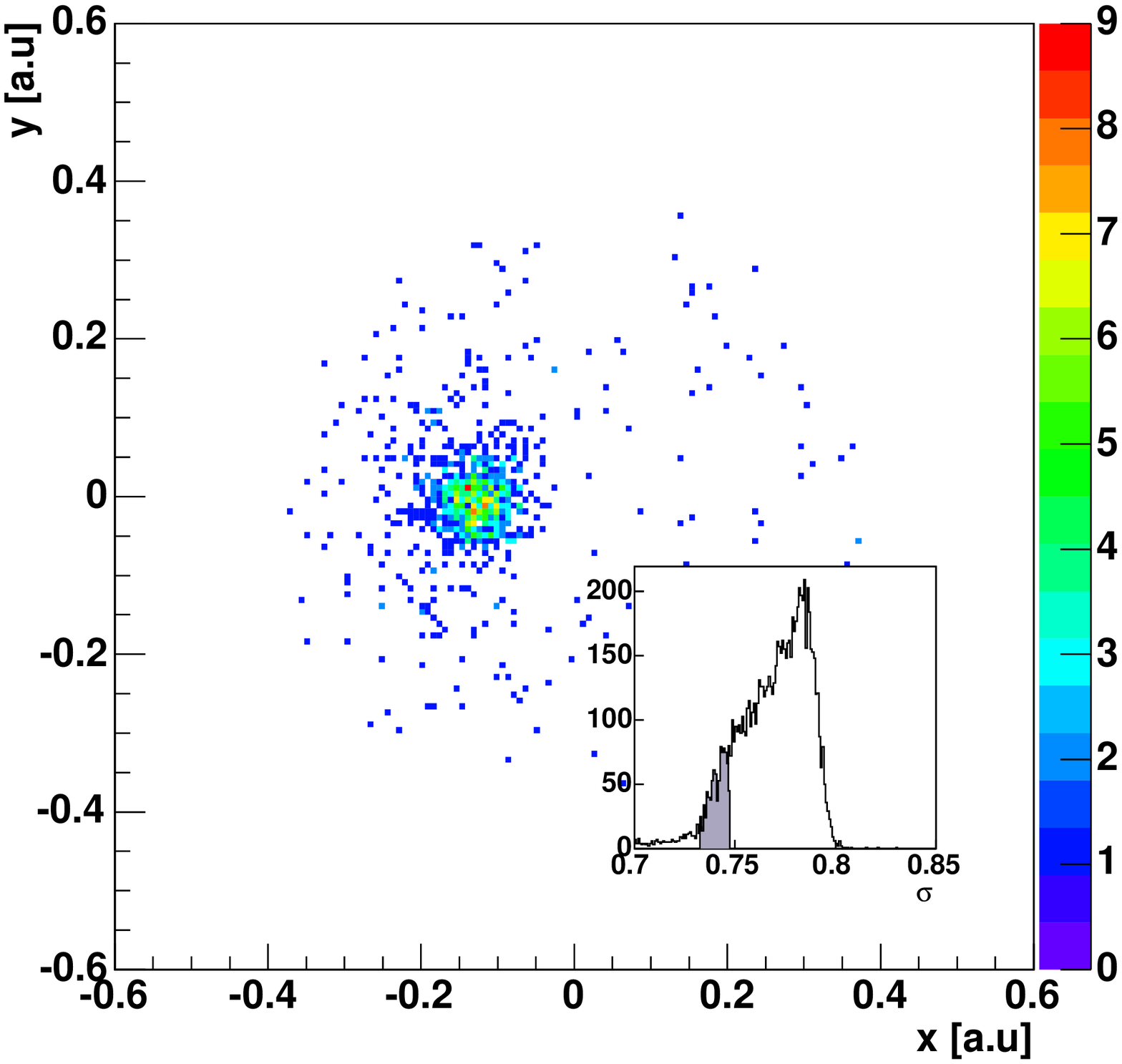}}\hspace*{4em}
  \subfigure[][Events (beam at corner) with low $\hat{\sigma}_m$.]{\label{subfig:corner_high}
    \includegraphics[width=0.39\textwidth,keepaspectratio,hiresbb=true]{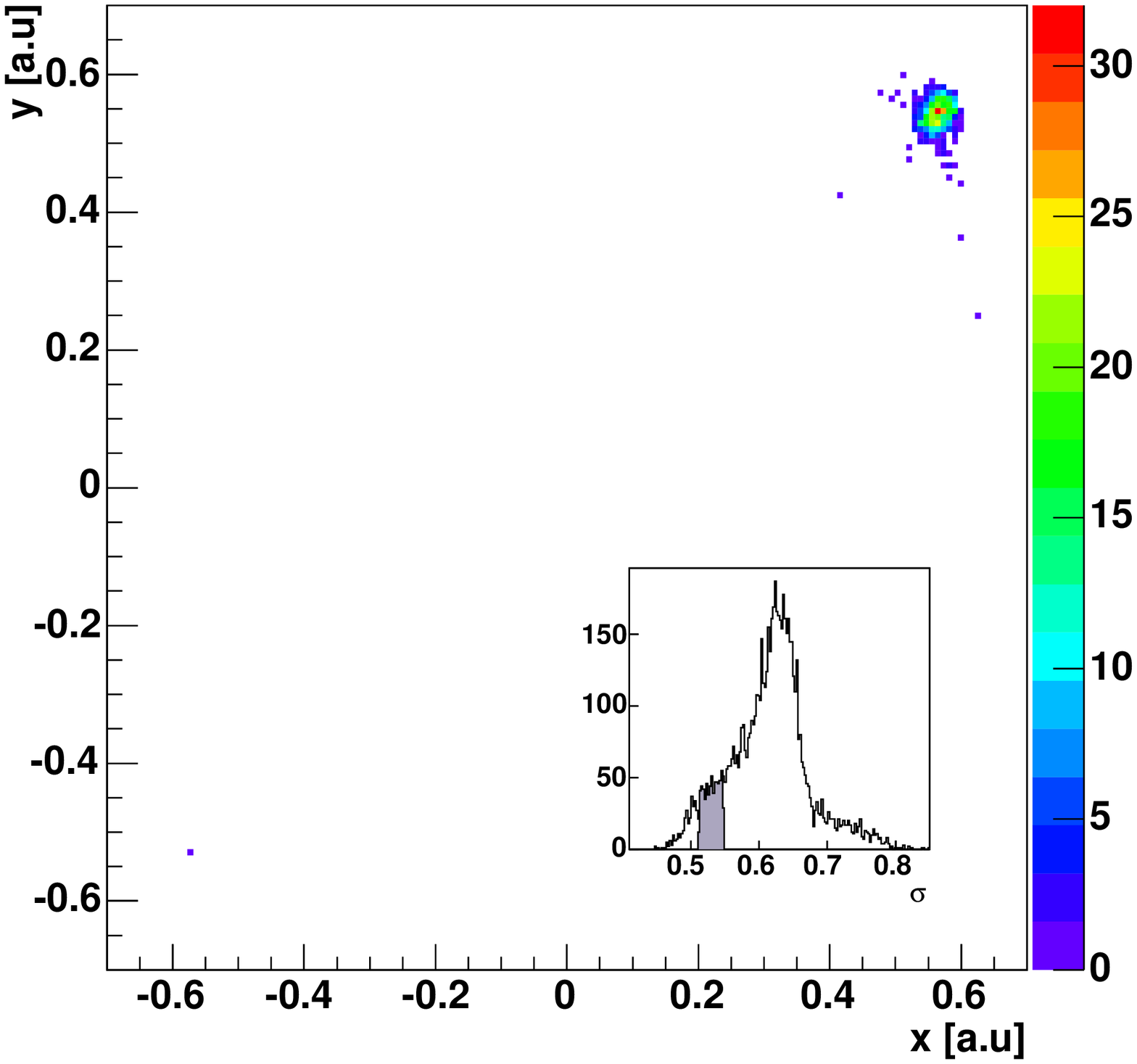}}\\\vspace*{-2eX}
  \subfigure[][Events (inclined beam) with medium $\hat{\sigma}_m$.]{\label{subfig:inclined_medium}
    \includegraphics[width=0.39\textwidth,keepaspectratio,hiresbb=true]{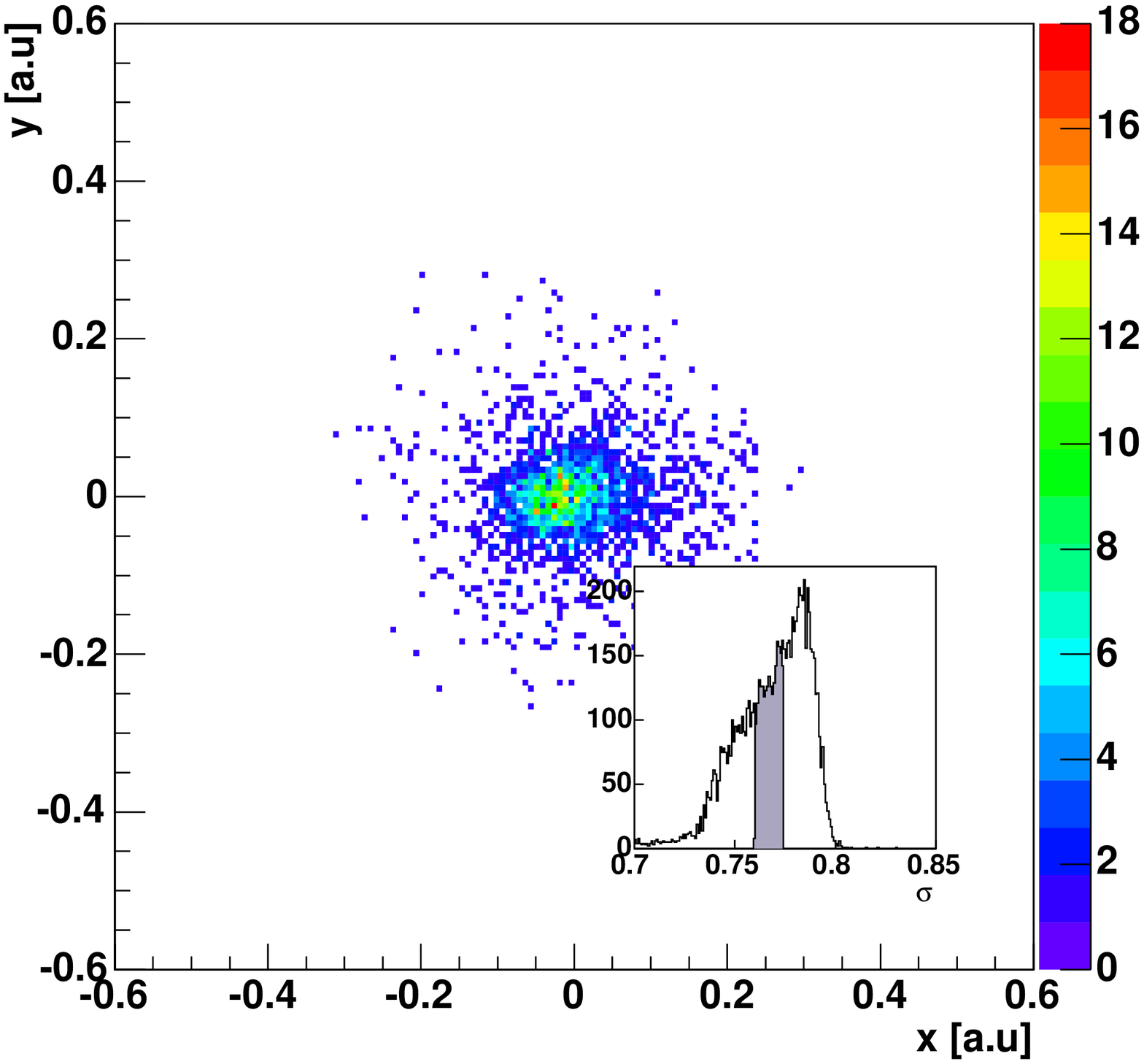}}\hspace*{4em}
  \subfigure[][Events (beam at corner) with medium
  $\hat{\sigma}_m$.]{\label{subfig:corner_medium}
    \includegraphics[width=0.39\textwidth,keepaspectratio,hiresbb=true]{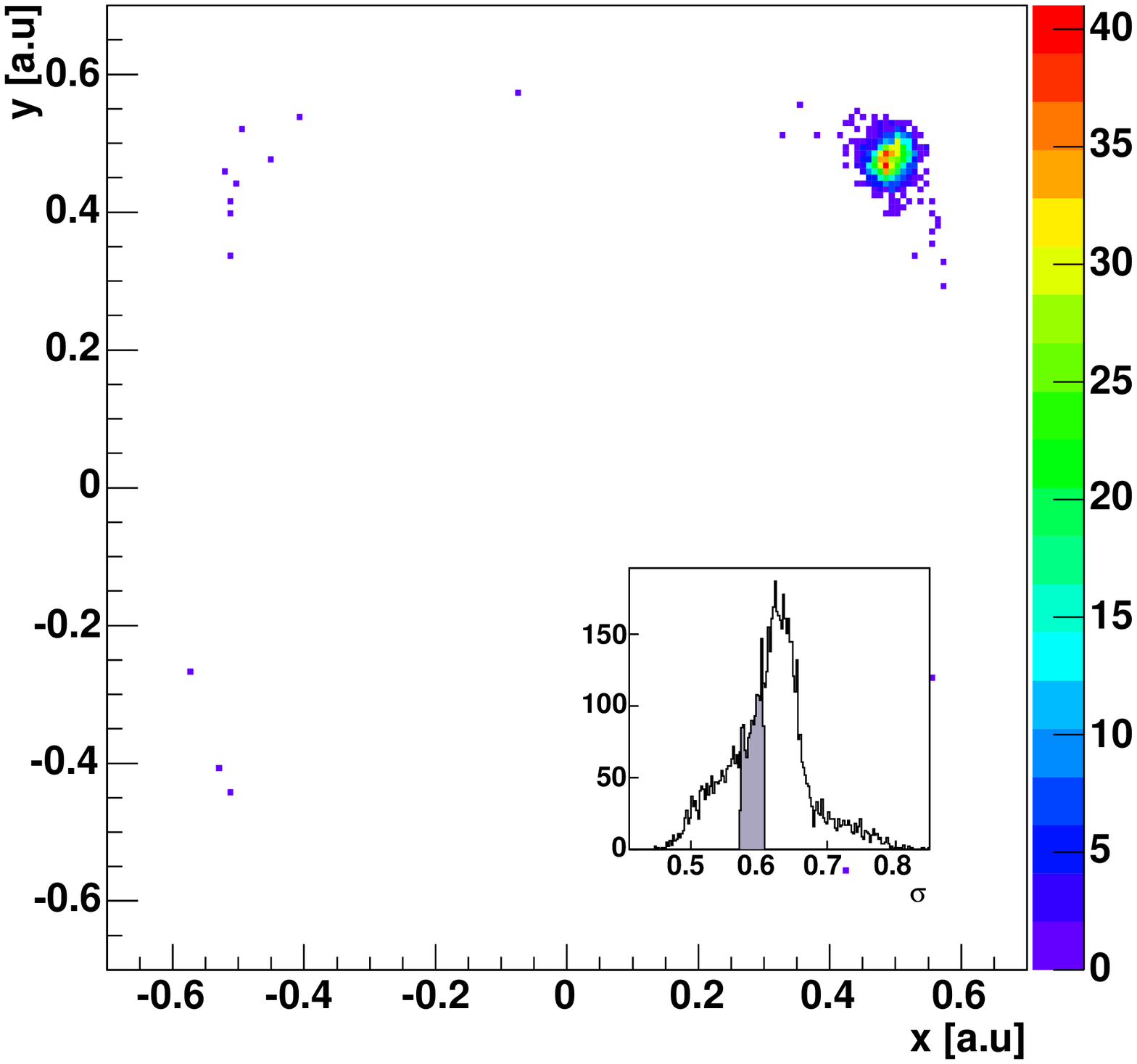}}\\\vspace*{-2eX}
  \subfigure[][Events (inclined beam) with high $\hat{\sigma}_m$.]{\label{subfig:inclined_low}
    \includegraphics[width=0.39\textwidth,keepaspectratio,hiresbb=true]{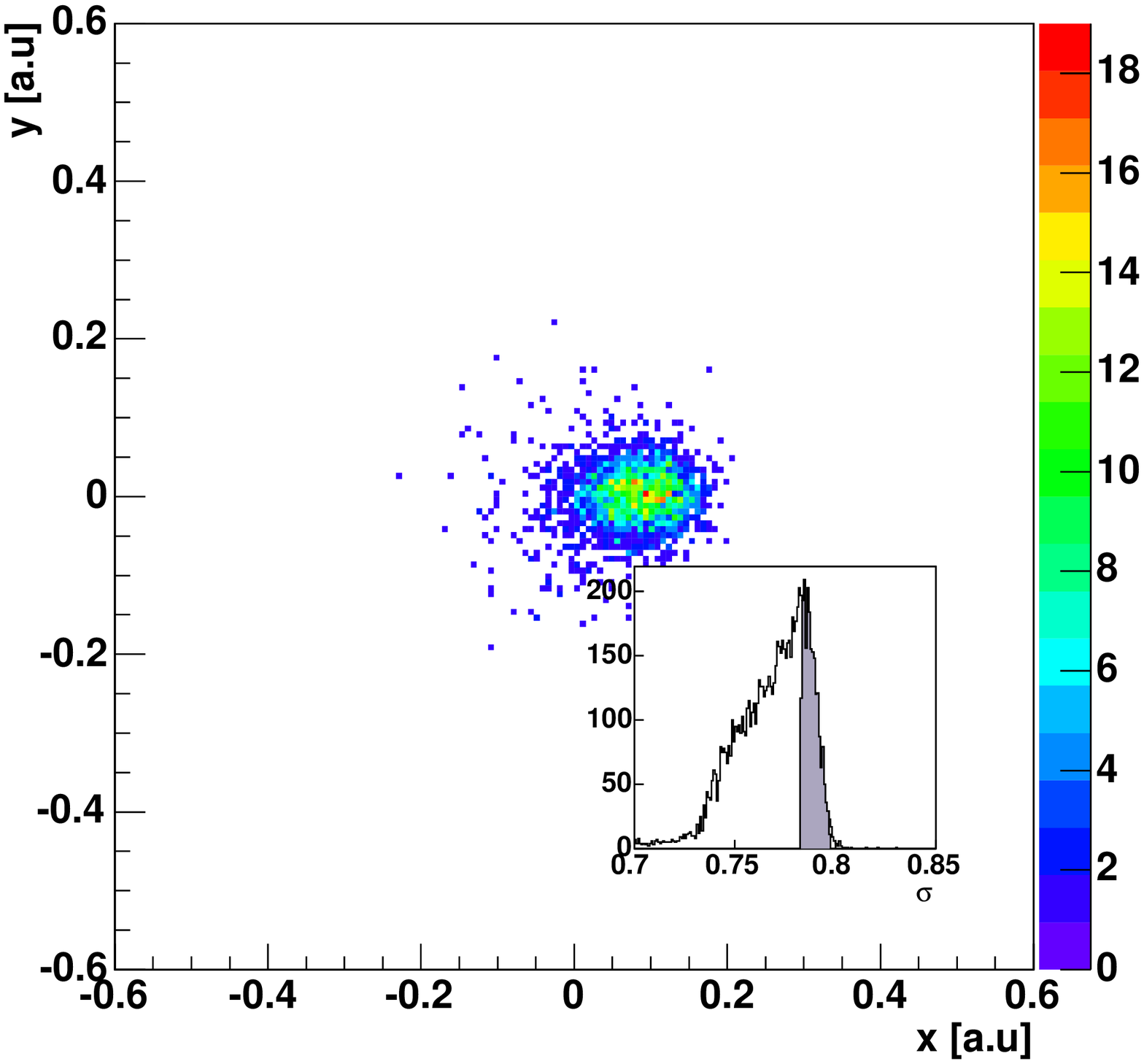}}\hspace*{4em}
  \subfigure[][Events (beam at corner) with high $\hat{\sigma}_m$.]{\label{subfig:corner_low}
    \includegraphics[width=0.39\textwidth,keepaspectratio,hiresbb=true]{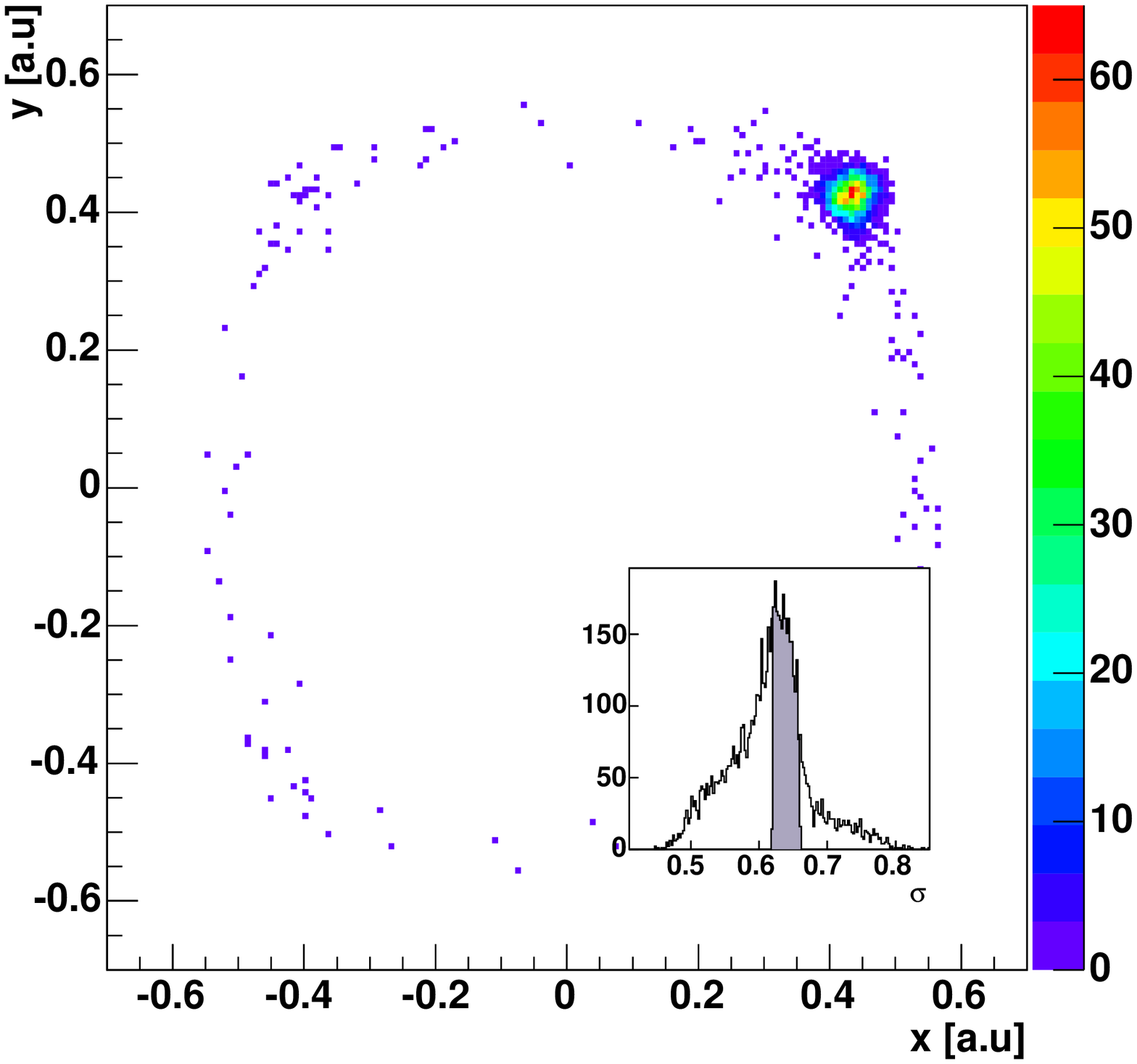}}\vspace*{-2eX}
  \caption[Density plot of the centroids for an inclined beam
  (filtered)]{Density plots for the same data displayed in
    figure~\ref{fig:center-inclined-corner} but now the additional
    moment is used to separate the events into subsets of similar
    $\hat{\sigma}_m$. The gray shaded areas in the small insets mark the
    events that are selected for display in the large density plots.}
  \label{fig:inclined_separated}
\end{figure}

The information that is given for each detected event with the second
moment (or alternatively the standard deviation) can be used
for filtering out events with a specific second moment. This is illustrated in
figures~\ref{subfig:inclined_high}-\ref{subfig:corner_low}.
In the left column, three density plots for the $x$- and $y$-centroids
are displayed. The shaded regions of the histograms in the small
insets indicate those events that are selected for display in large density
plots. This was repeated for three characteristic subsets of
$\hat{\sigma}_m$-values (low, medium
and high). By this means,
the long spot in figure~\ref{subfig:inclined_g_ray_all} could be
fragmented into three smaller spots of almost circular shape. The same
procedure was repeated for the normal $\gamma$-ray beam impinging on
an outer corner of the photocathode. Once again, a clear
discrimination of subsets of similar $x$- and $y$-centroids is
possible. All observed effects in these preliminary experiments can be
easily explained assuming that: 
\begin{enumerate}
\item  the signal distribution behaves like
the analytic model derived in chapter~\ref{ch:light-distribution},
\item the enhanced charge divider circuit presented in
  chapter~\ref{ch:enhanced-charge-dividing-circuits} measures indeed
  the second moment, and
\item the arguments given for deriving the fit model
  (expression~\ref{eq:fit-model-dist}) in fact describe the
  steps of the formation of this distribution.
\end{enumerate}
Throughout the rest of the present work it is therefore assumed that
the second moment (or alternatively the standard deviation) is
appropriate for estimating the depth of interaction.

\subsection{Validity of the Model for the Signal Distribution}
\label{subsec:mod-val}

In order to test the accuracy of the analytic model for the signal
distribution, some further effects have to be taken into account. The
model includes the most important physical effects until the
scintillation light is absorbed, with or without its detection. In addition, the
photomultiplier is assumed to behave as ideal. That is, neither
photoemission from the dynode-system, nor reflections inside the PSPMT,
nor effects like dark-current are considered. Actually,
photomultipliers behave almost ideally in many aspects. 

After the photoemission of an electron, it is attracted by the strong
electric potential of the first dynode where its impact causes various
secondary electrons. In the case of the H8500 from Hamamatsu, this
process is repeated twelve times because it has twelve dynode
stages. The special design of these dynodes provides for low
transverse dispersion of the electron avalanches. These are finally
collected by the $\mathrm{8\times8}$ anode pads. The size and pitch of the
anode pixels are found to be $\mathrm{5.8\times5.8\,mm^2}$ and $\mathrm{6.08\,mm}$
respectively \cite{data:H8500}. However, the effective sensitive area
of the device ($\mathrm{49\times49\,mm^2}$) is larger than
$\mathrm{64\times5.8\times5.8\,mm^2}$ or
$\mathrm{64\times6.08\times6.08\,mm^2}$. This is
because the electron avalanches can be focused towards the anode pads
and the small gaps between them are not critical. Dividing the effective
area by the total number of pixels, one can define the effective pixel
size $\mathrm{L_\mathit{eff}^\tincaps{AS}}$, area
$\mathrm{A_\mathit{eff}^\tincaps{AS}}$ and pitch $\mathrm{P_\mathit{eff}^\tincaps{AS}}$
to be $\mathrm{6.125\,mm}$, $\mathrm{6.125\times6.125\,mm^2}$ and $\mathrm{6.125\,mm}$,
respectively. 

The particular combination of photocathode, focusing dynodes and
anode matrix configuration performs an integration by intervals of the
scintillation light. The intervals are defined by the effective pixel
size $\mathrm{L_\mathit{eff}^\tincaps{AS}}$
and its position. The integration gives a set of 64 
positive numbers that correspond to the amounts of scintillation light
detected at each of the 64 equal-sized segments of the photocathode.
It was already mentioned that the light sensitivity can vary
significantly from one anode segment to another. The typical value
for this uniformity is given by the manufacturer and is of the order
of 1:3. Eventually ratios as high as 1:6 can be reached. Therefore, an anode
uniformity map is provided with each delivered PSPMT. In order to
obtain reasonable predictions of the detector energy response behavior, one
has to multiply each of the 64 values obtained by integration with the
relative anode sensitivity $\epsilon_{i,j}$ of the segment that
corresponds to this integration interval.

From this set of corrected numbers, the moments can now be
computed. The energy $\mu_{x_0,y_0}(\mathbf{r}_c)$ is obtained
by simply summing up all 64 corrected numbers. For computing the centroids 
and the second moment, the numbers have to be multiplied with linear
and quadratic weights respectively before building the sum. Since it was
shown there that all configurations of charge dividers can be configured to
reproduce an exact linear weighting along both transverse directions,
the weights for the centroids (normalized first order moments) are
nothing but the center positions of the anode segments. 
The situation for the second moment is more complex due to several
reasons. First of all, one has
to decide which type of second moment is appropriate. By
definition~(\ref{eq:multivariate-moms-1}), many second moments
are possible in the multivariate case, {\em e.g.}\ $\mu_{x_2,y_0}$,
$\mu_{x_0,y_2}$, $\mu_{x_2,y_2}$ {\em etc.}. Due to 
limitation by electronic design, minimum cost requirements and the desire to maximize the
SNR of the measured moment, the {\em composite second moment}
$(\mu_{x_2,y_0}+\mu_{x_0,y_2})(\mathbf{r}_c)$ is chosen. In the ideal case, the
weights would then be given by the sum of squares of the
anode center positions: $(x^2_{i,j}+y^2_{i,j})$. However, this
behavior can only be approximated with the charge divider configuration
that uses proportional resistor chains. Therefore, higher orders and
mixed terms of the center position also occur. Furthermore, the
electronic signals have to be amplified and prepared for
integration, introducing corrections that would be far too complex to
be treated analytically. For these reasons, the impulse response of the circuitry for
the second moment computation for each of the 64 segments was
simulated using the circuit simulation program {\sc Spice}.
The response was fitted using a polynomial Ansatz $\mathcal{W}(x,y)$ consisting of
even orders of $x$ and $y$ as well as their mixed contributions, and
with the following result:
\begin{equation}
  \label{eq:sec-mom-fit-ansatz}
  \mathcal{W}(x,y)\approx906-30x^2-37y^2+0.09y^2x^2
-0.008x^4+0.001y^4-8\cdot10^{-8}y^4x^4.
\end{equation}

\begin{figure}[!t]
  \centering
  \vspace*{-1eX}
  \subfigure[][Weights for the
  $x$-centroid.]{\label{subfig:xcent-weights}
    \psfrag{x}{$i$}
    \psfrag{y}{$j$}
    \psfrag{z}{\hspace*{-0.6em}$x_{i,j}$}    
    \includegraphics[width=0.39\textwidth]{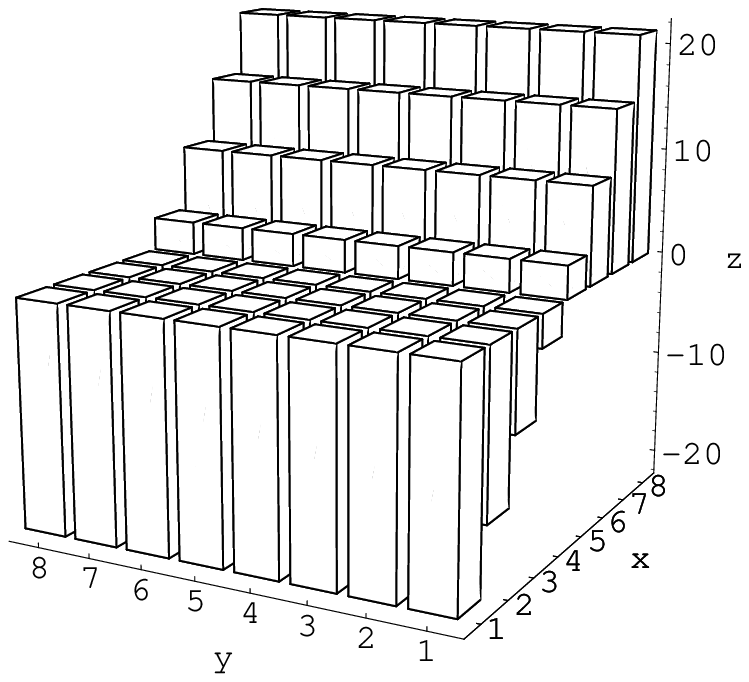}}\hspace*{3em}
  \subfigure[][Weights for the
  $y$-centroid.]{\label{subfig:ycent-weights}
    \psfrag{x}{$i$}
    \psfrag{y}{$j$}
    \psfrag{z}{\hspace*{-0.4em}$y_{i,j}$}    
    \includegraphics[width=0.39\textwidth]{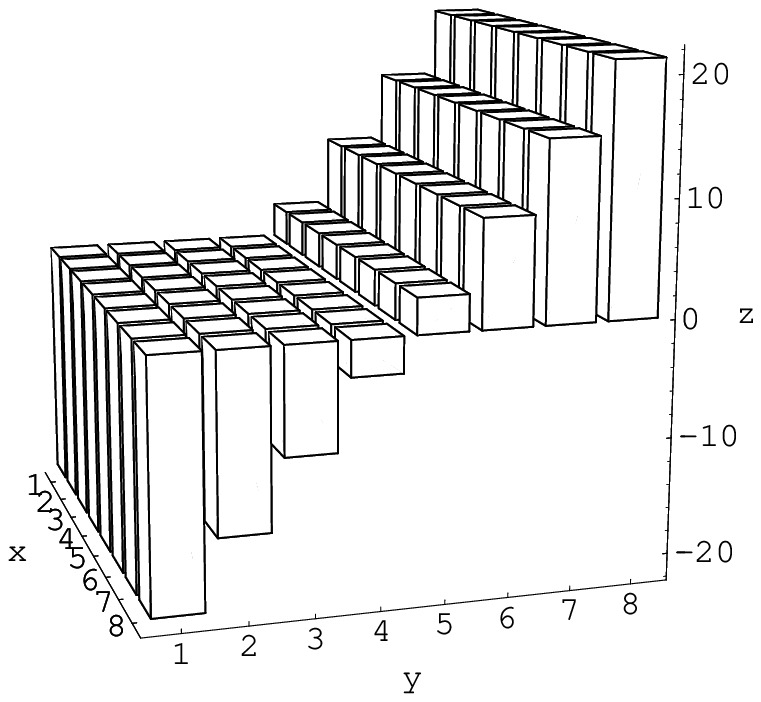}}\\
  \vspace*{-1eX}
  \subfigure[][Weights for the composite second
  moment (divided by $\mathrm{10^3}$).]{\label{subfig:sigma-weights}
    \psfrag{x}{$i$}
    \psfrag{y}{$j$}
    \psfrag{z}{\rotatebox{90}{\hspace*{-1.5em}$\mathcal{W}(x_{i,j},y_{i,j})$}} 
    \includegraphics[width=0.39\textwidth]{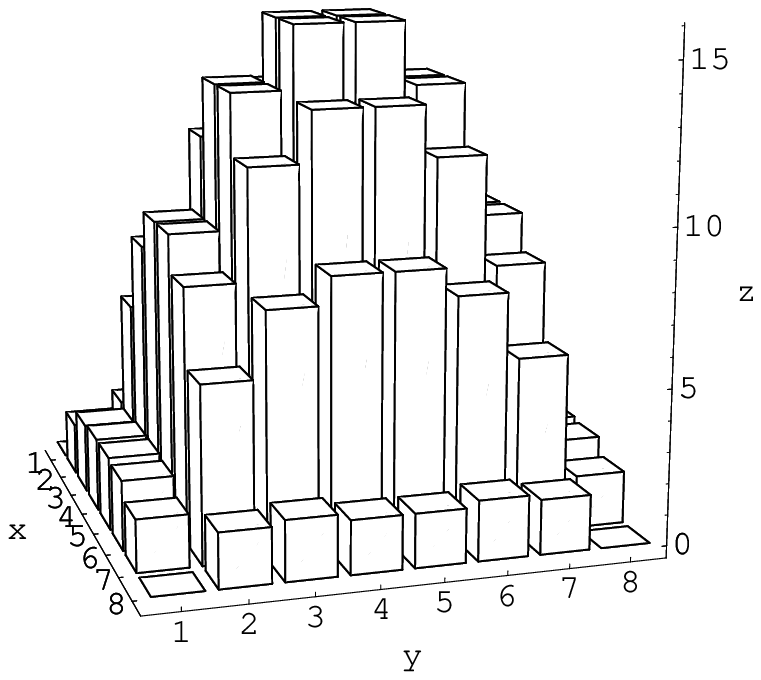}}\hspace*{3em}
  \subfigure[][Relative sensitivity of the anode
  segments.]{\label{subfig:anode-weights}
    \psfrag{x}{$i$}
    \psfrag{y}{$j$}
    \psfrag{z}{\hspace*{0.9em}$\epsilon_{i,j}$} 
    \includegraphics[width=0.39\textwidth]{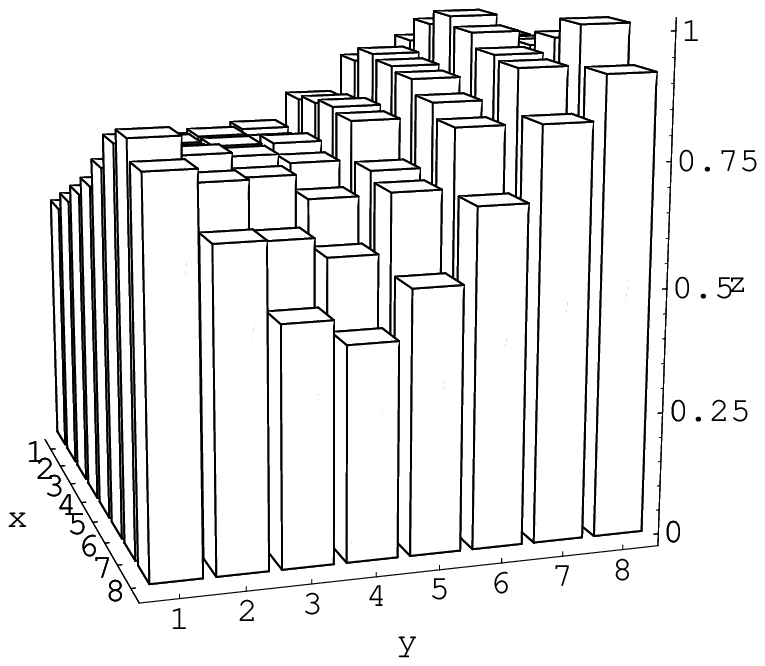}}\\
  \vspace*{-1eX}
  \caption[Weights for the centroids and the composite second
  moment]{Weights for the centroids and the composite second moment
    (\ref{subfig:xcent-weights}-\ref{subfig:sigma-weights}). 
    Figure~\ref{subfig:anode-weights} shows the 
    relative anode sensitivity of the particular PSPMT H8500
    from Hamamatsu.}
  \label{fig:weights-plots}
\end{figure}

All mathematical operations performed by the PSPMT as part 
 of the present $\gamma$-ray imaging detector can now be
written in the form
\begin{gather}
  \label{eq:exp-0-mom}
  \mu_{x_0,y_0}(\mathbf{r}_c)=\sum_{i,j}\epsilon_{i,j}\iint_{\omega_{i,j}}\mathcal{L}_\mathit{Detector}(\mathbf{r},\mathbf{r}_c)\,dxdy\\
  \label{eq:HMP-1x-mom}
  \mu_{x_1,y_0}(\mathbf{r}_c)=\frac{1}{\mu_{x_0,y_0}(\mathbf{r}_c)}\sum_{i,j}x_{i,j}\epsilon_{i,j}\iint_{\omega_{i,j}}\mathcal{L}_\mathit{Detector}(\mathbf{r},\mathbf{r}_c)\,dxdy\\
  \label{eq:exp-1y-mom}
  \mu_{x_0,y_1}(\mathbf{r}_c)=\frac{1}{\mu_{x_0,y_0}(\mathbf{r}_c)}\sum_{i,j}y_{i,j}\epsilon_{i,j}\iint_{\omega_{i,j}}\mathcal{L}_\mathit{Detector}(\mathbf{r},\mathbf{r}_c)\,dxdy\\
  \label{eq:exp-2-mom}
  (\mu_{x_2,y_0}+\mu_{x_0,y_2})(\mathbf{r}_c)=\frac{1}{\mu_{x_0,y_0}(\mathbf{r}_c)}\sum_{i,j}\epsilon_{i,j}\mathcal{W}(x_{i,j},y_{i,j})\iint_{\omega_{i,j}}\mathcal{L}_\mathit{Detector}(\mathbf{r},\mathbf{r}_c)\,dxdy.
\end{gather}
The values for all parameters of the signal distribution
$\mathcal{L}_\mathit{Detector}(\mathbf{r},\mathbf{r}_c)$ together with
the configuration of the enhanced charge divider circuit was described in
section~\ref{sec:exp-setup}. The electronic amplifier configuration is
sketched in appendix~\ref{app:elec-config}.
Figures~\ref{subfig:xcent-weights}-\ref{subfig:anode-weights} 
graphically represent the weights for the centroids, the second moment and the anode
uniformity.

The moments were then computed numerically for the 
81 different transverse positions $(x,y)$, with 
$x,y\in[\pm19,\pm14.25,\pm9.5,\pm4.75,0]\,\mathrm{mm}$ and at the two limit
values for the depth of interaction $z_c=0$ and $z_c=10$.
The data obtained from the measurement of the different moments 
at these positions are, however, distributions of all possible values
of $z_c$ between these two limits. One can obtain these limits
(parameters $a$ and $b$) from the experimental data by means of the
fit model (expression~\ref{eq:fit-model-dist}) described in
section~\ref{sec:model_dist_for_event_stat}.

In order to compare the parameters ${a}$ and ${b}$ to the computed moments for
the limits ${z_c=0}$ and ${z_c=10}$ one has to take into account that the
analog-to-digital conversion introduces a global and linear shift
of the measured moments which is very difficult to predict
analytically. The ADC module internally shapes, delays and integrates
the current pulses fed into their inputs before the analog-to-digital
converter maps the area to an integer of the interval ${]0,4096[}$ 
(${0}$ and ${4096}$ are reserved for under- and overflows,
respectively). This sequence of transformations produces both a global
offset ${t_\tincaps{ADC}}$ and a global proportionality constant
${m_\tincaps{ADC}}$. The determination of these parameters
is best done by minimizing the sum of differences since the measurements are subjected to
statistical errors. Hence, the following sum is defined:
\begin{equation}
  \label{eq:theo-meas-adjust}
  \zeta_\mu\mdef\sum_i^{9}\sum_j^{9}\left[\left(a_{i,j}^\mathit{meas}-m_\tincaps{ADC}\cdot\mu_{i,j}|_{(z_c=10)}-t_\tincaps{ADC}\right)^2+\left(b_{i,j}^\mathit{meas}-m_\tincaps{ADC}\cdot\mu_{i,j}|_{(z_c=0)}-t_\tincaps{ADC}\right)^2\right].
\end{equation}
Here, $\mu$ stands for one of the four moments
${\mu_{x_0,y_0}(\mathbf{r}_c)}$, ${\mu_{x_1,y_0}(\mathbf{r}_c)}$,
${\mu_{x_0,y_1}(\mathbf{r}_c)}$ and
${(\mu_{x_2,y_0}(\mathbf{r}_c)+\mu_{x_0,y_2})(\mathbf{r}_c)}$, each 
having its own pair of transformation parameters
${m_\tincaps{ADC}}$ and ${t_\tincaps{ADC}}$. That is, the sum in
equation~(\ref{eq:theo-meas-adjust}) has to be computed four
times. The optimum values for ${m_\tincaps{ADC}}$ and
${t_\tincaps{ADC}}$ of each moment are given by those that minimize ${\zeta_\mu}$.
Table~\ref{tab:scaling-signal} summarizes the results for a signal
model without diffuse reflected background light and another including
the latter contribution.

\begin{table}[!ht]
  \centering\renewcommand{\arraystretch}{1.4}
  \begin{tabular}{ccc}\hline\hline
    $\mu$ & $m_\tincaps{ADC}$ & $t_\tincaps{ADC}$\\\hline
    $\mu_{x_0,y_0}(\mathbf{r}_c)$ & $1.124$ & $875$\\
    $\mu_{x_1,y_0}(\mathbf{r}_c)$ & $0.954$ & $-0.434$\\
    $\mu_{x_0,y_1}(\mathbf{r}_c)$ & $0.947$ & $0.225$\\
    $(\mu_{x_2,y_0}+\mu_{x_0,y_2})(\mathbf{r}_c)$ & $2.774\cdot10^{-5}$ & $0.204$\\\hline\hline
  \end{tabular}\hspace*{2.5em}
  \begin{tabular}{ccc}\hline\hline
    $\mu$ & $m_\tincaps{ADC}$ & $t_\tincaps{ADC}$\\\hline
    $\mu_{x_0,y_0}(\mathbf{r}_c)$ & $0.866$ & $608$\\
    $\mu_{x_1,y_0}(\mathbf{r}_c)$ & $1.053$ & $-0.78$\\
    $\mu_{x_0,y_1}(\mathbf{r}_c)$ & $1.06$ & $0.196$\\
    $(\mu_{x_2,y_0}+\mu_{x_0,y_2})(\mathbf{r}_c)$ & $3.471\cdot10^{-5}$ & $0.165$\\\hline\hline
  \end{tabular}
  \caption[Scaling of the theoretical moment predictions]{Scaling of the theoretical moment predictions for a signal
    distribution without reflected background (l.h.s.) and a signal
    distribution with reflected background (r.h.s.).}
  \label{tab:scaling-signal}
\end{table}

\begin{figure}[!htp]
  \centering
  \vspace*{-1.5eX}
  \subfigure[][Parameter ${a}$ of ${\mu_{x_1,y_0}}$
  (values).]{\label{subfig:parameter-a-xcent}
    \psfrag{x}{\hspace*{-2em}{\small position \#}}
    \psfrag{y}{\hspace*{-2em}{\small value [mm]}}
    \includegraphics[width=0.40\textwidth]{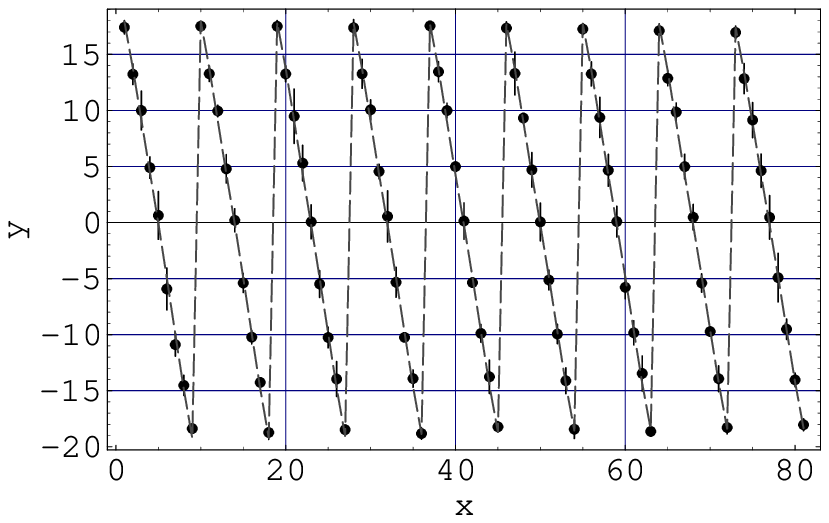}}\hspace*{2.2em}
  \subfigure[][Parameter ${a}$ of ${\mu_{x_1,y_0}}$
  (errors).]{\label{subfig:parameter-a-xcent-err}
    \psfrag{x}{\hspace*{-2em}{\small position \#}}
    \psfrag{y}{\hspace*{-2.8em}{\small abs.\ error [mm]}}
    \includegraphics[width=0.40\textwidth]{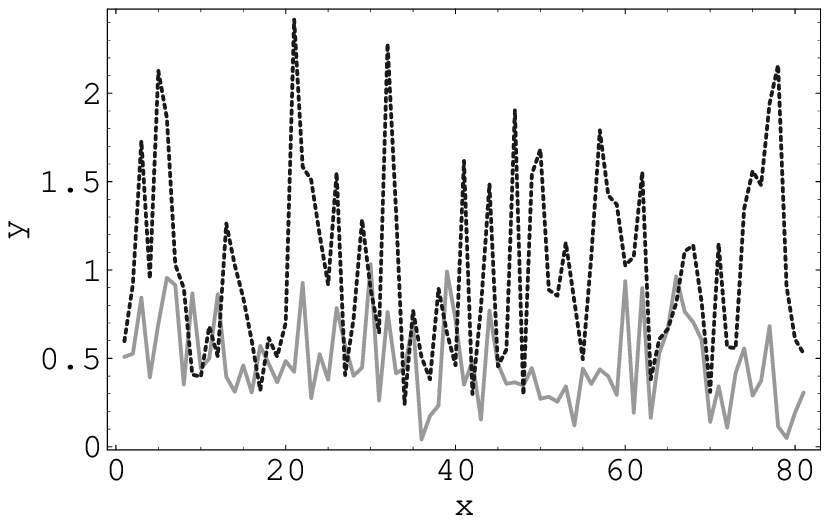}}\\
  \vspace*{-1.5eX}
  \subfigure[][Parameter ${b}$ of ${\mu_{x_1,y_0}}$
  (values).]{\label{subfig:parameter-b-xcent}
    \psfrag{x}{\hspace*{-2em}{\small position \#}}
    \psfrag{y}{\hspace*{-2em}{\small value [mm]}}
    \includegraphics[width=0.40\textwidth]{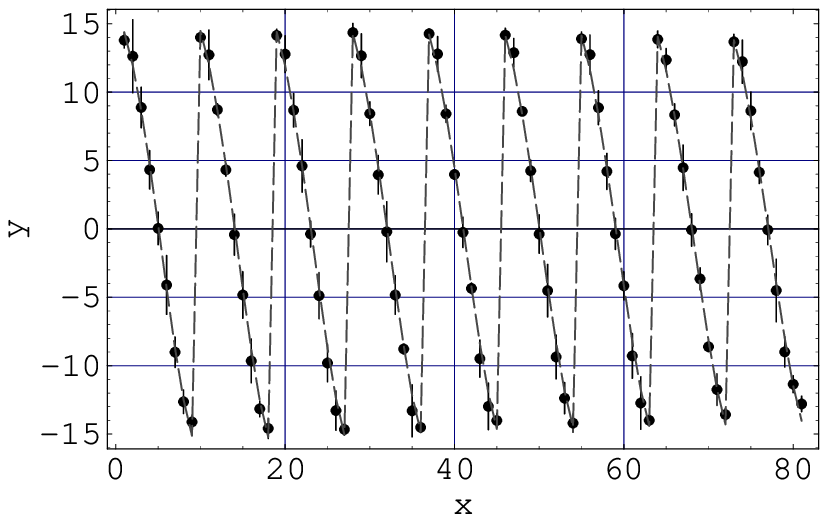}}\hspace*{2.2em}
  \subfigure[][Parameter ${b}$ of ${\mu_{x_1,y_0}}$
  (errors).]{\label{subfig:parameter-b-xcent-err}
    \psfrag{x}{\hspace*{-2em}{\small position \#}}
    \psfrag{y}{\hspace*{-2.8em}{\small abs.\ error [mm]}}
    \includegraphics[width=0.40\textwidth]{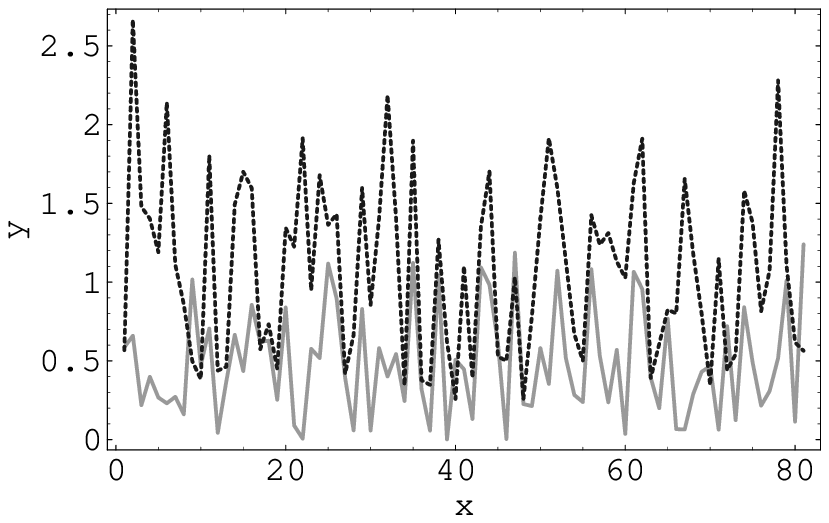}}\\
  \vspace*{-1.5eX}
  \subfigure[][Parameter ${a}$ of ${\mu_{x_0,y_1}}$
  (values).]{\label{subfig:parameter-a-ycent}
    \psfrag{x}{\hspace*{-2em}{\small position \#}}
    \psfrag{y}{\hspace*{-2em}{\small value [mm]}}
    \includegraphics[width=0.40\textwidth]{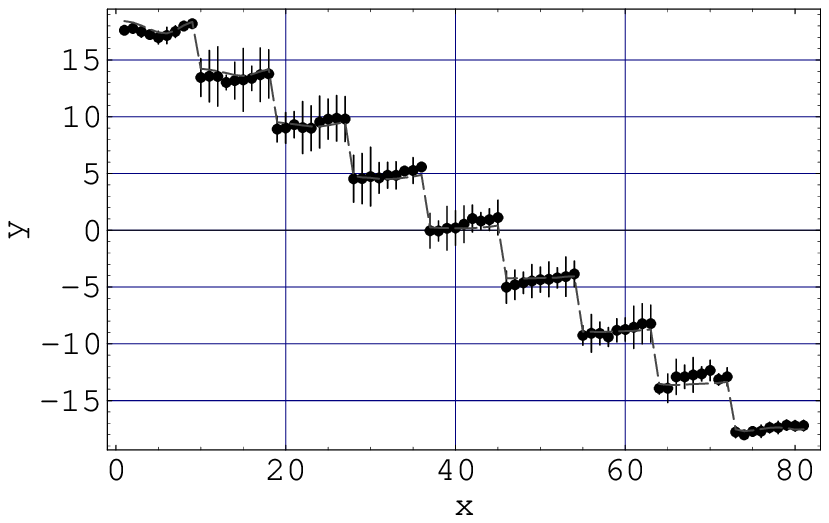}}\hspace*{2.2em}
  \subfigure[][Parameter ${a}$ of ${\mu_{x_0,y_1}}$
  (errors).]{\label{subfig:parameter-a-ycent-err}
    \psfrag{x}{\hspace*{-2em}{\small position \#}}
    \psfrag{y}{\hspace*{-2.8em}{\small abs.\ error [mm]}}
    \includegraphics[width=0.40\textwidth]{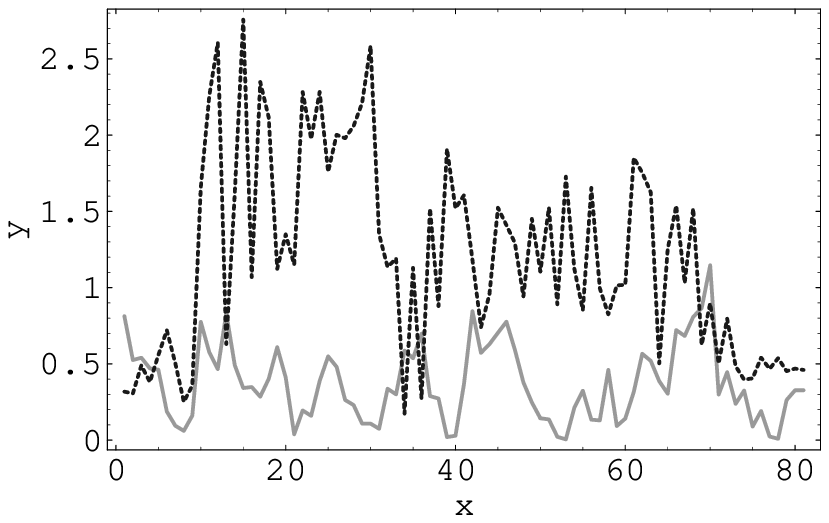}}\\
  \vspace*{-1.5eX}
  \subfigure[][Parameter ${b}$ of ${\mu_{x_0,y_1}}$
  (values).]{\label{subfig:parameter-b-ycent}
    \psfrag{x}{\hspace*{-2em}{\small position \#}}
    \psfrag{y}{\hspace*{-2em}{\small value [mm]}}
    \includegraphics[width=0.40\textwidth]{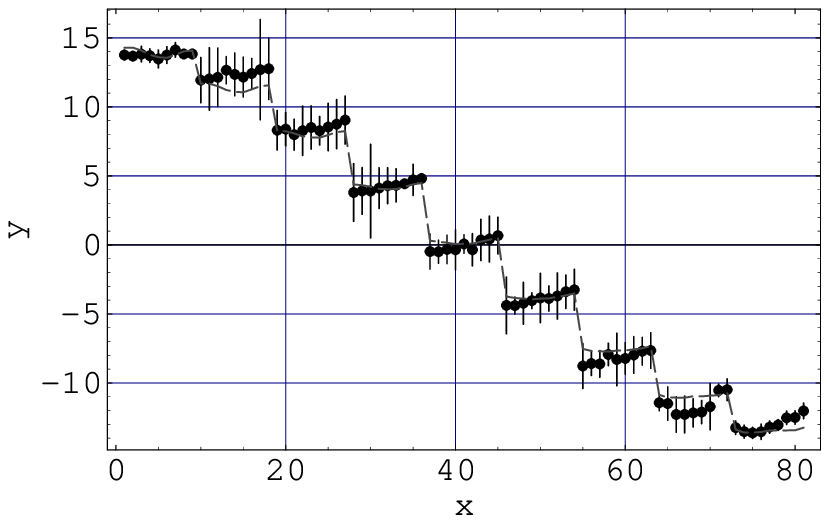}}\hspace*{2.2em}
  \subfigure[][Parameter ${b}$ of ${\mu_{x_0,y_1}}$
  (errors).]{\label{subfig:parameter-b-ycent-err}
    \psfrag{x}{\hspace*{-2em}{\small position \#}}
    \psfrag{y}{\hspace*{-2.8em}{\small abs.\ error [mm]}}
    \includegraphics[width=0.40\textwidth]{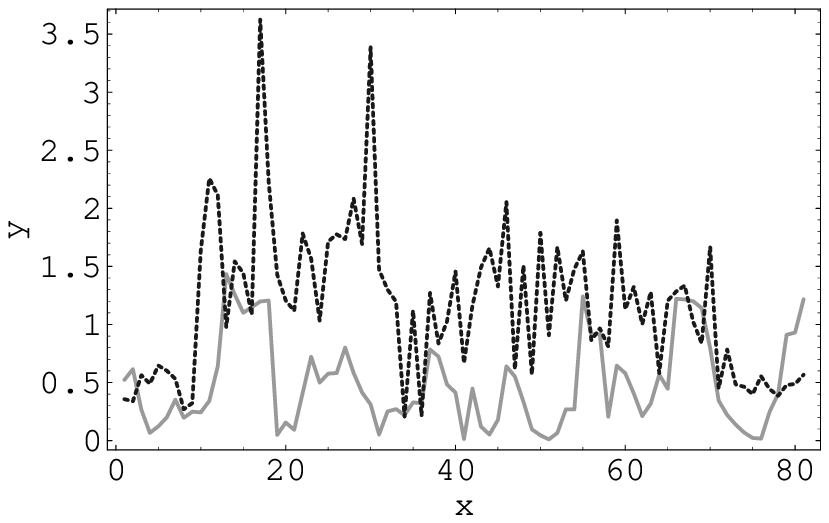}}\\
  \vspace*{-1.5eX}
  \caption[Plots of measured values, errors and
    theoretical predictions for both centroids]{Comparison
      between measurements (errors only at the left hand side) and
    theoretical predictions for both centroids in the limits ${z_c=0}$
    and ${z_c=10}$. The model signal distributions used include
    residual reflections at the absorbing surfaces.}
\end{figure}

\begin{figure}[!htp]
  \centering
  \vspace*{-1.5eX}
  \subfigure[][Parameter ${a}$ of ${\mu_{x_2,y_0}+\mu_{x_0,y_2}}$ (values).]{\label{subfig:parameter-a-secmom}
    \psfrag{x}{\hspace*{-2em}{\small position \#}}
    \psfrag{y}{\hspace*{-2em}{\small value [a.u.]}}
    \includegraphics[width=0.40\textwidth]{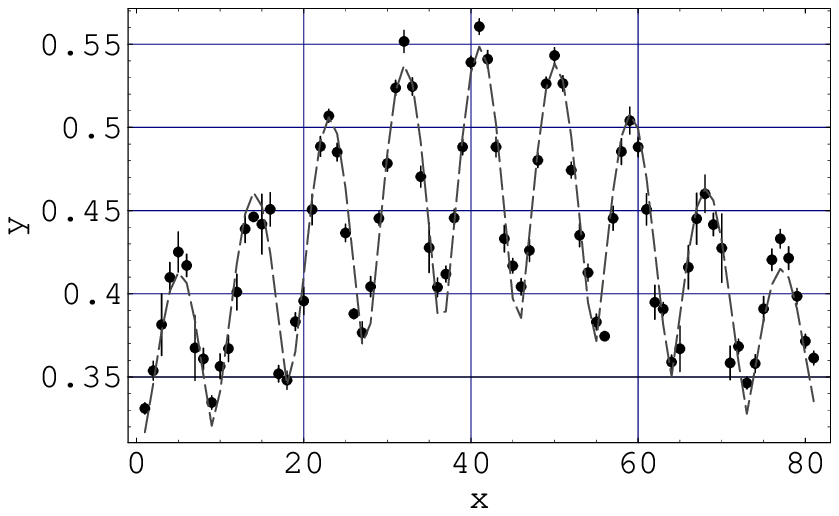}}\hspace*{2.2em}
  \subfigure[][Parameter ${a}$ of ${\mu_{x_2,y_0}+\mu_{x_0,y_2}}$ (errors).]{\label{subfig:parameter-a-secmom-err}
    \psfrag{x}{\hspace*{-2em}{\small position \#}}
    \psfrag{y}{\hspace*{-2.4em}{\small rel.\ error [\%]}}
    \includegraphics[width=0.40\textwidth]{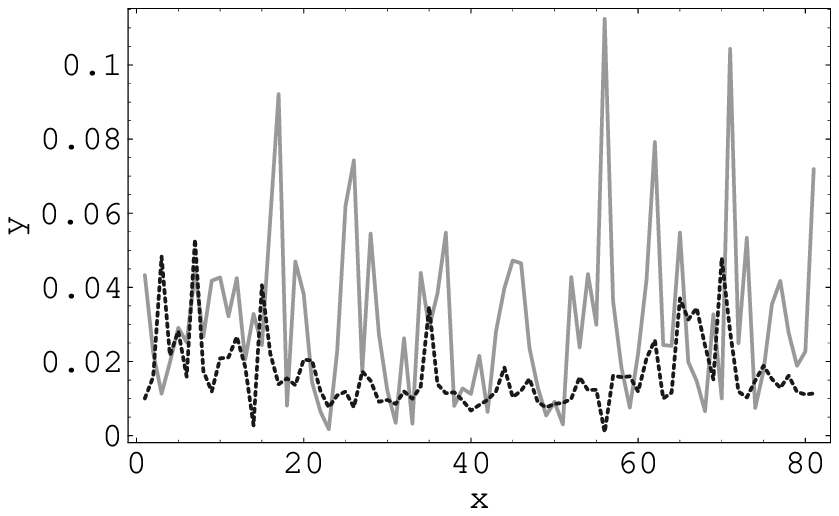}}\\
  \vspace*{-1.5eX}
  \subfigure[][Parameter ${b}$ of ${\mu_{x_2,y_0}+\mu_{x_0,y_2}}$
  (values).]{\label{subfig:parameter-b-secmom}
    \psfrag{x}{\hspace*{-2em}{\small position \#}}
    \psfrag{y}{\hspace*{-2em}{\small value [a.u.]}}
    \includegraphics[width=0.40\textwidth]{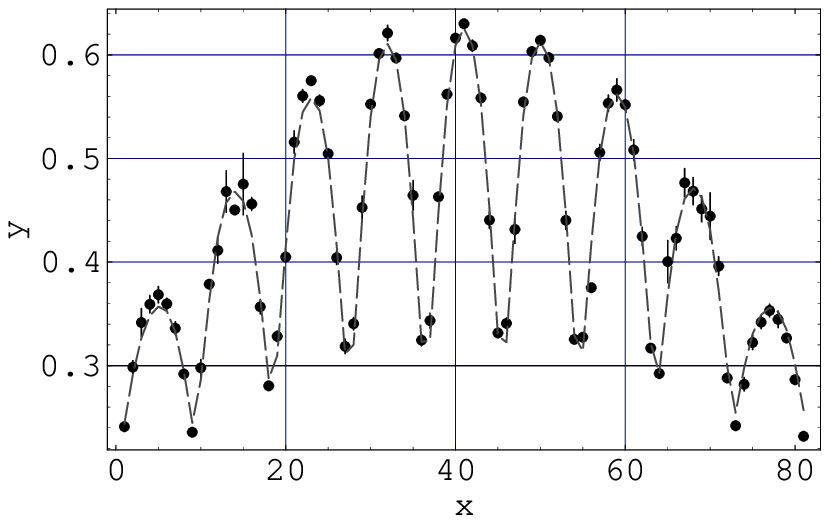}}\hspace*{2.2em}
  \subfigure[][Parameter ${b}$ of ${\mu_{x_2,y_0}+\mu_{x_0,y_2}}$
  (errors).]{\label{subfig:parameter-b-secmom-err}
    \psfrag{x}{\hspace*{-2em}{\small position \#}}
    \psfrag{y}{\hspace*{-2.4em}{\small rel.\ error [\%]}}
    \includegraphics[width=0.40\textwidth]{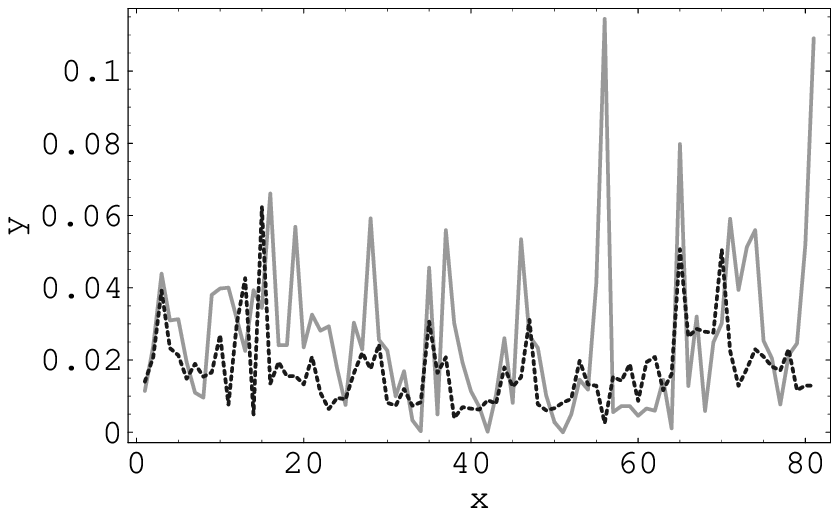}}\\
  \vspace*{-1.5eX}
  \subfigure[][Parameter ${a}$ of ${\mu_{x_0,y_0}}$
  (values).]{\label{subfig:parameter-a-energ}
    \psfrag{x}{\hspace*{-2em}{\small position \#}}
    \psfrag{y}{\hspace*{-2em}{\small value [ch.]}}
    \includegraphics[width=0.40\textwidth]{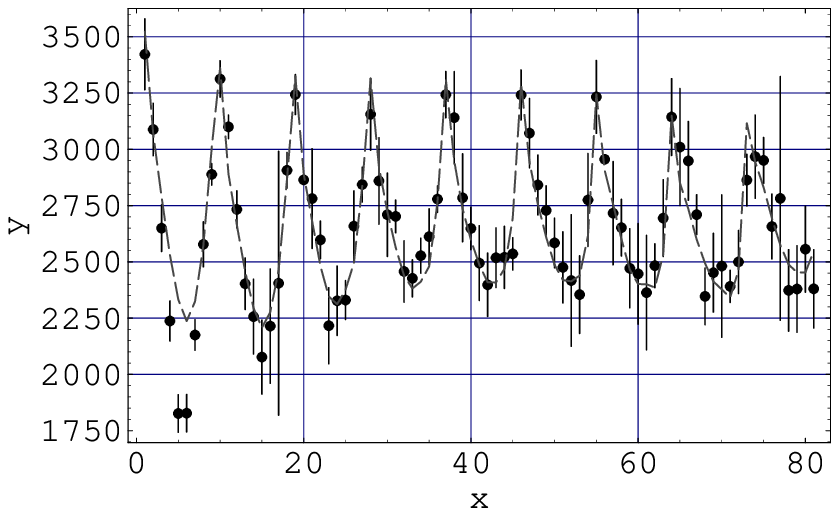}}\hspace*{2.2em}
  \subfigure[][Parameter ${a}$ of ${\mu_{x_0,y_0}}$
  (errors).]{\label{subfig:parameter-a-energ-err}
    \psfrag{x}{\hspace*{-2em}{\small position \#}}
    \psfrag{y}{\hspace*{-2.4em}{\small rel.\ error [\%]}}
    \includegraphics[width=0.40\textwidth]{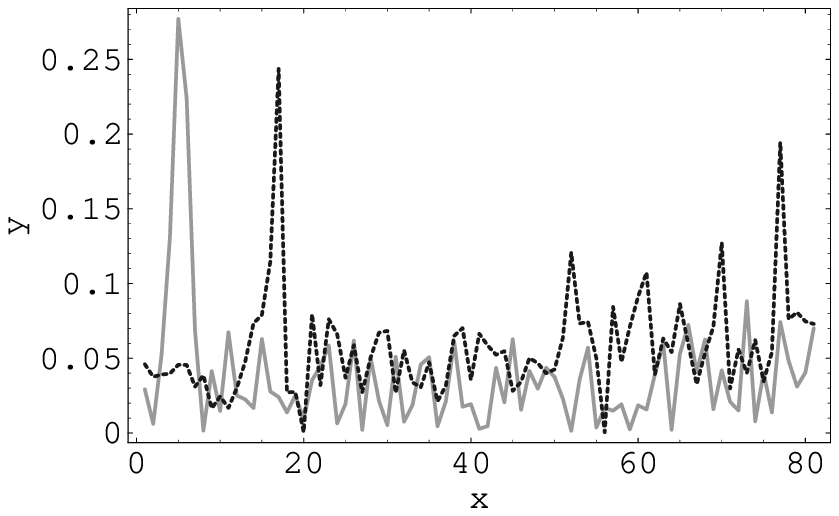}}\\
  \vspace*{-1.5eX}
  \subfigure[][Parameter ${b}$ of ${\mu_{x_0,y_0}}$
  (values).]{\label{subfig:parameter-b-energ}
    \psfrag{x}{\hspace*{-2em}{\small position \#}}
    \psfrag{y}{\hspace*{-2em}{\small value [ch.]}}
    \includegraphics[width=0.40\textwidth]{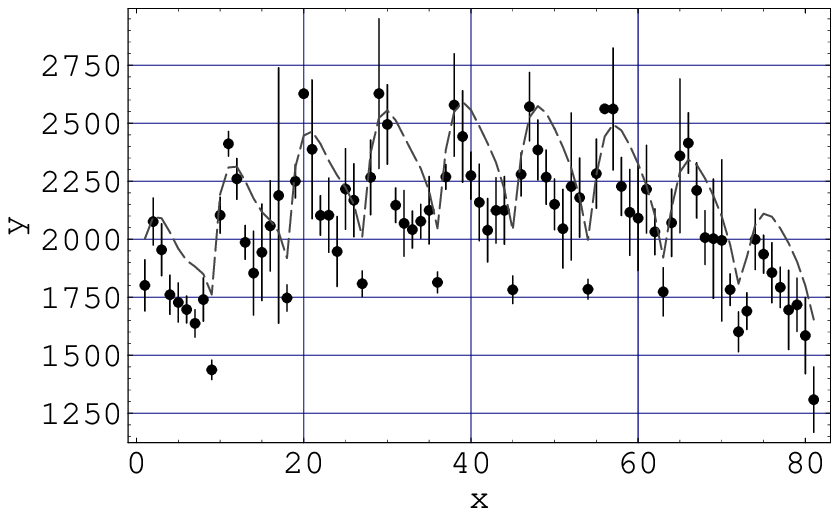}}\hspace*{2.2em}
  \subfigure[][Parameter ${b}$ of ${\mu_{x_0,y_0}}$
  (errors).]{\label{subfig:parameter-b-energ-err}
    \psfrag{x}{\hspace*{-2em}{\small position \#}}
    \psfrag{y}{\hspace*{-2.4em}{\small rel.\ error [\%]}}
    \includegraphics[width=0.40\textwidth]{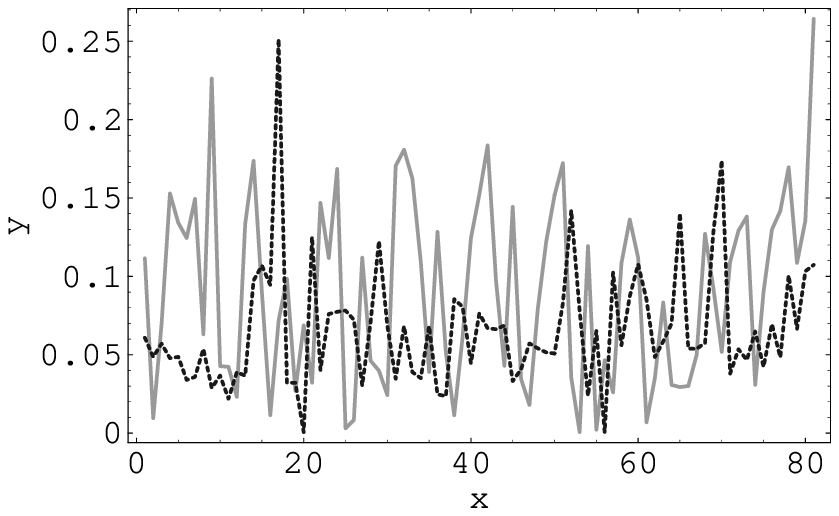}}\\
  \vspace*{-1.5eX}
  \caption[Plots of measured values, errors and
    theoretical predictions for energy and second moment]{Comparison
      between measurements (errors only at the left hand side) and
    theoretical predictions for energy and composite second moment in the limits ${z_c=0}$
    and ${z_c=10}$. The model signal distributions used include
    residual reflections at the absorbing surfaces.}
\end{figure}

In figures~\ref{subfig:parameter-a-xcent}-\ref{subfig:parameter-b-energ-err},
the measured and predicted values are displayed for a model that includes the
reflective background discussed in section~\ref{sec:background-light}.
For clarity, the predicted values (gray dashed line) are displayed in
2D-plots together with the measured data (black data points) 
and in the order illustrated in figure~\ref{fig:sample-pos}. The right
column of plots shows the deviation between the model prediction
and the measured values (light-gray solid lines), displayed together with the
measurement errors (dashed black lines).  

\begin{figure}[!t]
  \centering
  \subfigure[][Parameter $a$ of $\mu_{x_0,y_0}$
  (values).]{\label{subfig:parameter-a-energ-wo-bg}
    \psfrag{x}{\hspace*{-2em}{\small position \#}}
    \psfrag{y}{\hspace*{-2em}{\small value [ch.]}}
    \includegraphics[width=0.40\textwidth]{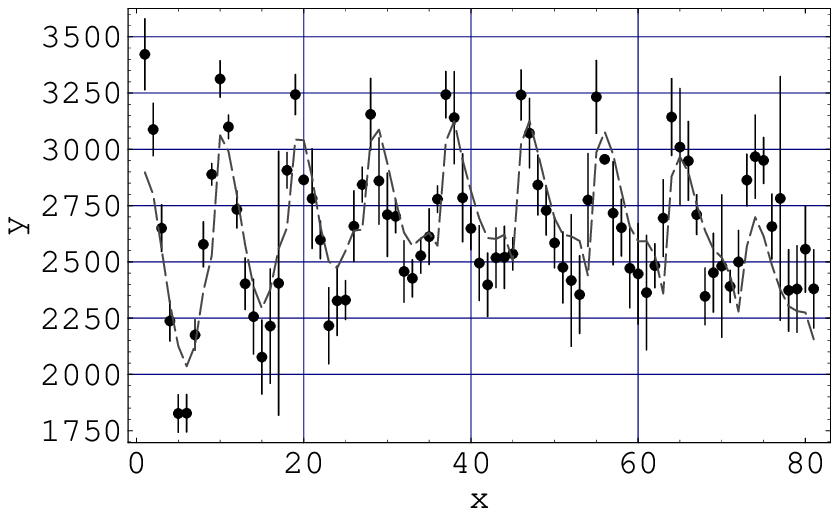}}\hspace*{2.2em}
  \subfigure[][Parameter $a$ of $\mu_{x_0,y_0}$
  (errors).]{\label{subfig:parameter-a-energ-err-wo-bg}
    \psfrag{x}{\hspace*{-2em}{\small position \#}}
    \psfrag{y}{\hspace*{-2.4em}{\small rel.\ error [\%]}}
    \includegraphics[width=0.40\textwidth]{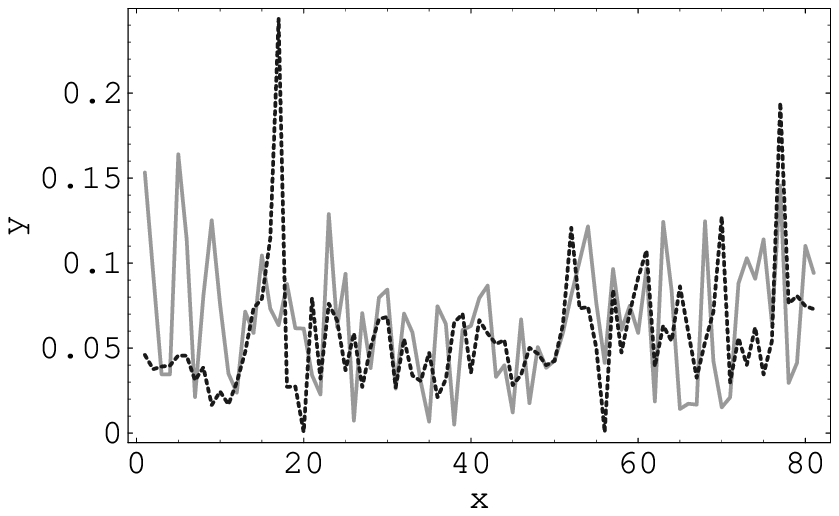}}\\
  \vspace*{-1.5eX}
  \subfigure[][Parameter $b$ of $\mu_{x_0,y_0}$
  (values).]{\label{subfig:parameter-b-energ-wo-bg}
    \psfrag{x}{\hspace*{-2em}{\small position \#}}
    \psfrag{y}{\hspace*{-2em}{\small value [ch.]}}
    \includegraphics[width=0.40\textwidth]{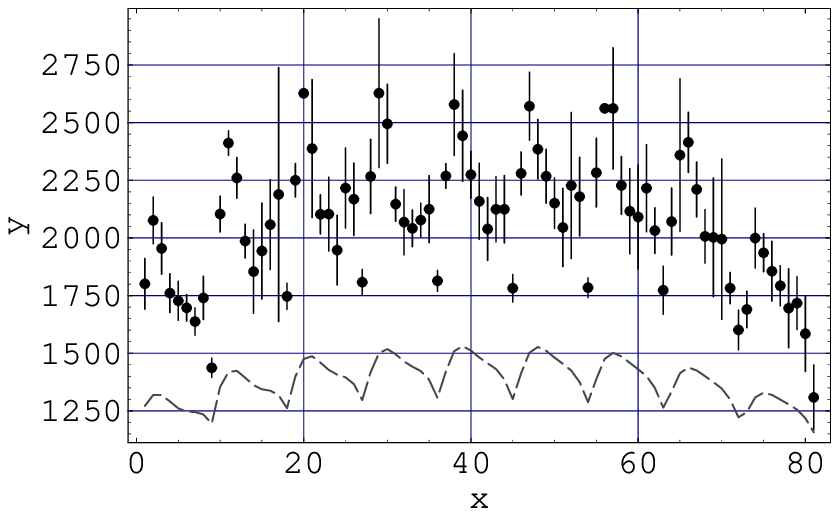}}\hspace*{2.2em}
  \subfigure[][Parameter $b$ of $\mu_{x_0,y_0}$
  (errors).]{\label{subfig:parameter-b-energ-err-wo-bg}
    \psfrag{x}{\hspace*{-2em}{\small position \#}}
    \psfrag{y}{\hspace*{-2.4em}{\small rel.\ error [\%]}}
    \includegraphics[width=0.40\textwidth]{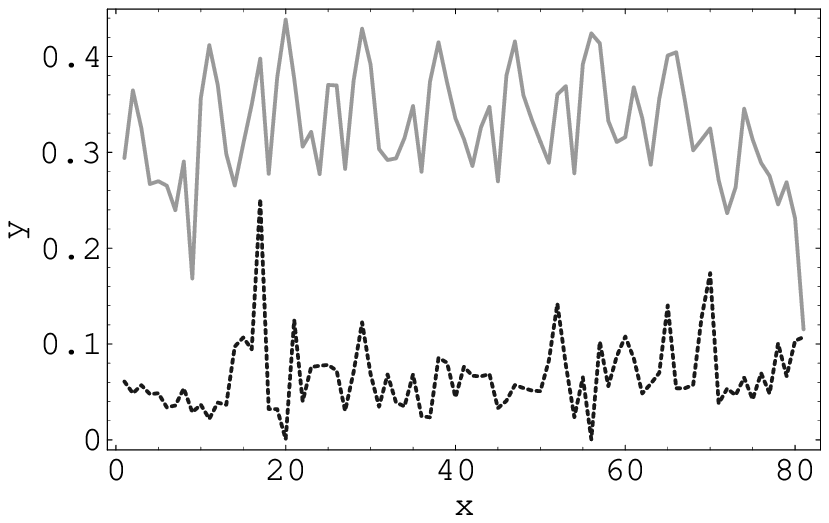}}\\
  \vspace*{-1.5eX}
  \caption[Plots of measured values, errors and
    theoretical predictions for energy without reflective bg.]{Comparison
      between measurements (errors only at the left hand side) and theoretical predictions for energy in the limits $z_c=0$
    and $z_c=10$. The model signal distributions used do not include
    residual reflections at the absorbing surfaces.}
\end{figure}

An important observation is that the model distribution derived in
chapter~\ref{ch:light-distribution} reproduces very well the spatial
dependence of the three non-trivial moments. In the case of the zeroth
moment, {\em i.e.}\ the energy, the agreement was, however, poorer. 
The model fails to reproduce correctly the details of the energy
variation over the sensitive area especially for the upper limit in
the interaction distance (${z_c=10}$). Note that this affects the
non-trivial moments only marginally because the aim of normalization is
to suppress the dependency on the trivial moment. The measured
centroids
(figures~\ref{subfig:parameter-a-xcent}-\ref{subfig:parameter-b-ycent-err})
exhibit errors at the central positions that are much
larger than the discrepancy between prediction and
measurement. Most likely this is caused by the fact that the
parameters ${a}$ and ${b}$ cannot be estimated very well with the
fit model~(\ref{eq:fit-model-dist}) when the distribution are nearly
of Gaussian shape. In the case of the composite second moment
(figures~\ref{subfig:parameter-a-secmom}-\ref{subfig:parameter-b-secmom-err}),
it is observed that the agreement between model and measurement
is much better in the limit ${z_c=0}$ (parameter ${b}$) than in the limit
${z_c=10}$ (parameter ${a}$). Since this also occurs for the zero-order
moment
(figures~\ref{subfig:parameter-a-energ}-\ref{subfig:parameter-b-energ-err}),
it has to be assumed that the model for the background light does not
reproduce well the distribution of the additional light. However,
 a model with the background light switched off, fails
completely to predict the trivial moment 
(figures~\ref{subfig:parameter-a-energ-wo-bg}-
\ref{subfig:parameter-b-energ-err-wo-bg}). 
Therefore, a detailed study of the reflection properties is necessary in order to
find a more reasonable background model.


Other reasons for the remaining deviations between measurements and
those obtained from the theoretical model are
Compton scattering as described in chapter~\ref{ch:compton}, 
lack of mechanical precision and possible light reflections inside the
PSPMT, {\em e.g.}\ at the dynode system and the housing. 
In particular, the mechanical tolerance may be of relevance. While the
precision of the motorized translation stage used 
($10\,\mathrm{\mu m}$) is 
sufficiently good for our purposes, the remaining parts of the
mechanical setup like housing, source holder and test detector
mounting, not only miss this precision but are estimated to introduce
errors in the transverse directions up to $\mathrm{1\,mm}$. This precision
has to be significantly increased if a model with higher accuracy is
required. Table~\ref{tab:error-stats} summarizes some statistic estimators
of all observed measurement errors and deviations between model
(with residual reflections) and
measurement. Except for the centroids, relative errors are
used. Absolute errors have been preferred for the centroids because at
nominal positions around $0$, the relative error of the centroid will
diverge even for very small absolute errors. All relative errors are
of the order of 10\%, while the absolute errors of the centroid are
about $\mathrm{1\,mm}$. Note that this corresponds to the estimated mechanical
uncertainty of the test detector assembly.

\begin{table}[!t]
  \centering\renewcommand{\arraystretch}{1.4}
{\small 
  \begin{tabular}{cc|c|c|c}\hline\hline
    moment & p. & model-measure deviation & measurement error&unit \\
    &&    \begin{tabularx}{17em}{>{\hsize=0.9\hsize}X>{\hsize=1.2\hsize}X>{\hsize=1.1\hsize}X>{\hsize=0.8\hsize}X}
      Mean & StdDev & Min & Max\\ 
    \end{tabularx}&
    \begin{tabularx}{16em}{>{\hsize=0.9\hsize}X>{\hsize=1.2\hsize}X>{\hsize=1.1\hsize}X>{\hsize=0.8\hsize}X}
      Mean & StdDev & Min & Max\\    \end{tabularx}&\\\hline
    $\mu_{x_0,y_0}$ & $a$ &
    \begin{tabularx}{17em}{>{\hsize=0.9\hsize}X>{\hsize=1.2\hsize}X>{\hsize=1.1\hsize}X>{\hsize=0.8\hsize}X}
      3.8 & 4.2 & 0.1 & 27.8
    \end{tabularx}&
    \begin{tabularx}{16em}{>{\hsize=0.9\hsize}X>{\hsize=1.2\hsize}X>{\hsize=1.1\hsize}X>{\hsize=0.8\hsize}X}
      5.7 & 3.6 & <0.1 & 24.4
    \end{tabularx}&\%\\
    $\mu_{x_0,y_0}$ & $b$ &
    \begin{tabularx}{17em}{>{\hsize=0.9\hsize}X>{\hsize=1.2\hsize}X>{\hsize=1.1\hsize}X>{\hsize=0.8\hsize}X}
      9.0 & 5.9 & <0.1 & 26.4
    \end{tabularx}&
    \begin{tabularx}{16em}{>{\hsize=0.9\hsize}X>{\hsize=1.2\hsize}X>{\hsize=1.1\hsize}X>{\hsize=0.8\hsize}X}
      6.6 & 3.8 & <0.1 & 25.1
    \end{tabularx}&\%\\
    $\mu_{x_1,y_0}$ & $a$ &
    \begin{tabularx}{17em}{>{\hsize=0.9\hsize}X>{\hsize=1.2\hsize}X>{\hsize=1.1\hsize}X>{\hsize=0.8\hsize}X}
      0.48 & 0.25 & 0.04 & 1.03
    \end{tabularx}&
    \begin{tabularx}{16em}{>{\hsize=0.9\hsize}X>{\hsize=1.2\hsize}X>{\hsize=1.1\hsize}X>{\hsize=0.8\hsize}X}
      1.01 & 0.53 & 0.24 & 2.42
    \end{tabularx}&mm\\
    $\mu_{x_1,y_0}$ & $b$ &
    \begin{tabularx}{17em}{>{\hsize=0.9\hsize}X>{\hsize=1.2\hsize}X>{\hsize=1.1\hsize}X>{\hsize=0.8\hsize}X}
      0.49 & 0.34 & <0.01 & 1.24
    \end{tabularx}&
    \begin{tabularx}{16em}{>{\hsize=0.9\hsize}X>{\hsize=1.2\hsize}X>{\hsize=1.1\hsize}X>{\hsize=0.8\hsize}X}
      1.08 & 0.56 & 0.26 & 2.67
    \end{tabularx}&mm\\
    $\mu_{x_0,y_1}$ & $a$ &
    \begin{tabularx}{17em}{>{\hsize=0.9\hsize}X>{\hsize=1.2\hsize}X>{\hsize=1.1\hsize}X>{\hsize=0.8\hsize}X}
      0.37 & 0.25 & <0.01 & 1.15
    \end{tabularx}&
    \begin{tabularx}{16em}{>{\hsize=0.9\hsize}X>{\hsize=1.2\hsize}X>{\hsize=1.1\hsize}X>{\hsize=0.8\hsize}X}
      1.19 & 0.65 & 0.17 & 2.76
    \end{tabularx}&mm\\
    $\mu_{x_0,y_1}$ & $b$ &
    \begin{tabularx}{17em}{>{\hsize=0.9\hsize}X>{\hsize=1.2\hsize}X>{\hsize=1.1\hsize}X>{\hsize=0.8\hsize}X}
      0.49 & 0.38 & 0.01 & 1.43
    \end{tabularx}&
    \begin{tabularx}{16em}{>{\hsize=0.9\hsize}X>{\hsize=1.2\hsize}X>{\hsize=1.1\hsize}X>{\hsize=0.8\hsize}X}
      1.16 & 0.65 & 0.21 & 3.62
    \end{tabularx}&mm\\
    $\mu_{x_2,y_0}+\mu_{x_0,y_2}$ & $a$ &
    \begin{tabularx}{17em}{>{\hsize=0.9\hsize}X>{\hsize=1.2\hsize}X>{\hsize=1.1\hsize}X>{\hsize=0.8\hsize}X}
      3.1 & 2.3 & 0.2 & 11.2
    \end{tabularx}&
    \begin{tabularx}{16em}{>{\hsize=0.9\hsize}X>{\hsize=1.2\hsize}X>{\hsize=1.1\hsize}X>{\hsize=0.8\hsize}X}
      1.7 & 1 & 0.1 & 5.3
    \end{tabularx}&\%\\
    $\mu_{x_2,y_0}+\mu_{x_0,y_2}$ & $b$ &
    \begin{tabularx}{17em}{>{\hsize=0.9\hsize}X>{\hsize=1.2\hsize}X>{\hsize=1.1\hsize}X>{\hsize=0.8\hsize}X}
      2.7 & 2.2 & <0.1 & 11.4
    \end{tabularx}&
    \begin{tabularx}{16em}{>{\hsize=0.9\hsize}X>{\hsize=1.2\hsize}X>{\hsize=1.1\hsize}X>{\hsize=0.8\hsize}X}
      1.8 & 1.1 & 0.2 & 6.2
    \end{tabularx}&\%\\\hline\hline
  \end{tabular}
}
  \caption[Statistic estimators for the errors of the limit parameter
    $a$ and $b$ of all four moments]{Statistic estimators for the errors of the limit parameters
    $a$ and $b$ of all four moments. Relative errors are given except
    for the centroids, where absolute errors have been preferred.}
  \label{tab:error-stats}
\end{table}

For the case of a model without diffuse reflections, only the results
for the zeroth order moment have been plotted
(figure~\ref{subfig:parameter-b-energ-err-wo-bg}). 
The results for the
other moments are very similar to the case with reflected background and their errors of the same
order. Actually, this model was only applied to demonstrate that
even for very low reflectanceivity of the absorbing coatings, residual
reflections on these surfaces cannot be neglected if the model is
also required to yield good agreement for the trivial moment. 

\subsection{Moments as 3D Position Estimate}
\label{sec:moments-as-pos-estimate}

At the end of this chapter, an estimation is given of what 3D-spatial
resolution can be expected from the presented detector setup if the
raw moments were used as position estimates. While it has been
reported elsewhere (Clancy {\em et al.}\ \cite{Clancy:1997}, Joung {\em et al.}\
\cite{Joung:2002}, Tavernier {\em et al.}\ \cite{Tavernier:2005}) that this
is not satisfactory for many reasons, the results are given here not only for
completeness. Instead, the availability of an analogically computed
second moment that can be used to estimate the depth of interaction 
via the standard deviation is an interesting extension of the
conventional Anger logic. Unfortunately, this new estimate suffers
from the same problems as the centroids: as the $\gamma$-ray's impact
position gets closer to one of the borders, the light
distribution is distorted and the moment is subjected to an
additional variation not caused by the interaction depth. 
The following chapter will treat this problem by presenting an
algorithm for recovering the true three-dimensional impact position
from the set of moments.

\begin{figure}[!t]
  \centering
  \subfigure[][Measured energy resolution at the 81 positions.]{\label{subfig:zero-moment-measure2D}
    \psfrag{x}{\hspace*{-2em}{\small position \#}}
    \psfrag{y}{\hspace*{-1.5em}{\small $\Delta E$ [\%]}}
    \includegraphics[width=0.52\textwidth]{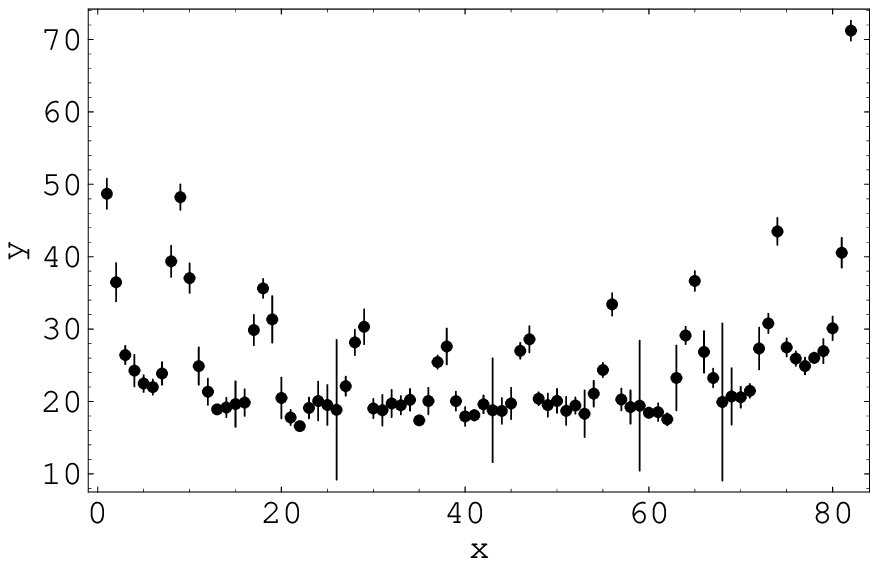}
  }\hspace*{1.2em}
  \subfigure[][2D-dependency of the energy resolution.]{\label{subfig:zero-moment-measure3D}
    \psfrag{x}{{\small $x$}}
    \psfrag{y}{{\small $y$}}
    \psfrag{z}{\hspace*{-0.4em}\rotatebox{90}{\hspace*{-1.5em}{\small $\Delta E$ [\%]}}}
    \includegraphics[width=0.40\textwidth]{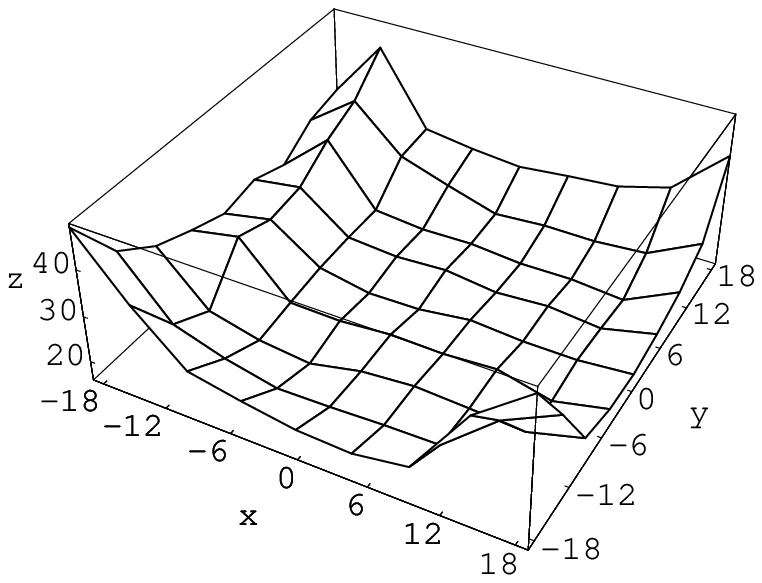}
  }\\
  \subfigure[][Measured $X$ resolution at the 81 positions.]{\label{subfig:firstX-moment-measure2D}
    \psfrag{x}{\hspace*{-2em}{\small position \#}}
    \psfrag{y}{\hspace*{-1.5em}{\small $\Delta X$ [mm]}}
    \includegraphics[width=0.52\textwidth]{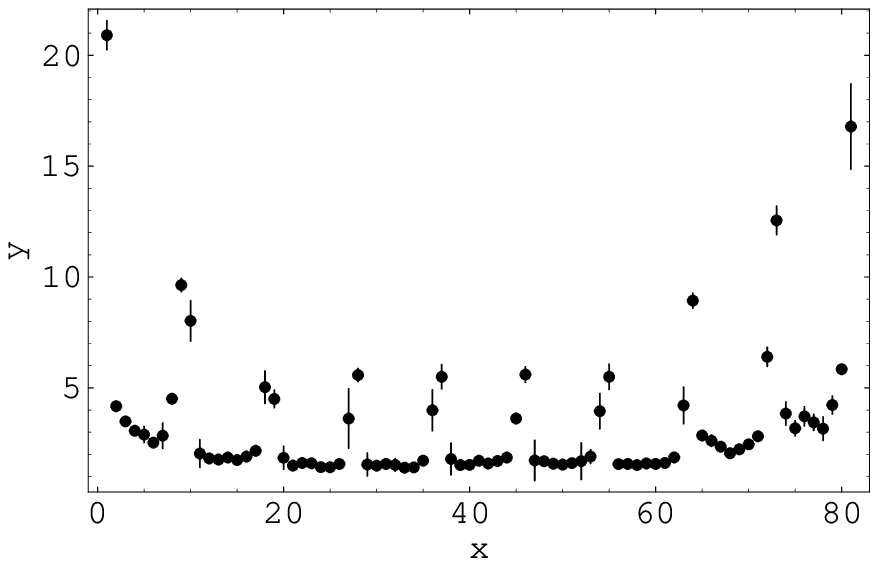}
  }\hspace*{1.2em}
  \subfigure[][2D-dependency of $X$ resolution.]{\label{subfig:firstX-moment-measure3D}
    \psfrag{x}{{\small $x$}}
    \psfrag{y}{{\small $y$}}
    \psfrag{z}{\hspace*{-0.4em}\rotatebox{90}{\hspace*{-1.5em}{\small $\Delta X$ [mm]}}}
    \includegraphics[width=0.40\textwidth]{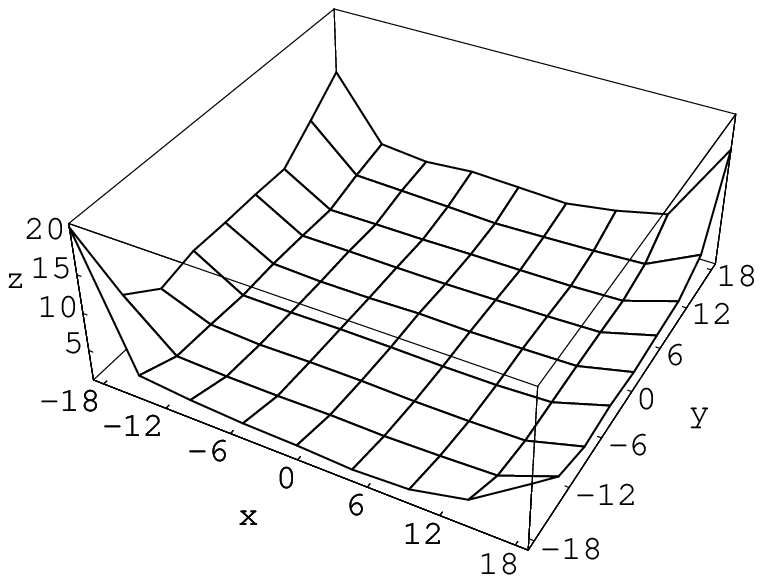}
  }\\
  \subfigure[][Measured $Y$ resolution at the 81 positions.]{\label{subfig:firstY-moment-measure2D}
    \psfrag{x}{\hspace*{-2em}{\small position \#}}
    \psfrag{y}{\hspace*{-1.5em}{\small $\Delta Y$ [mm]}}
    \includegraphics[width=0.52\textwidth]{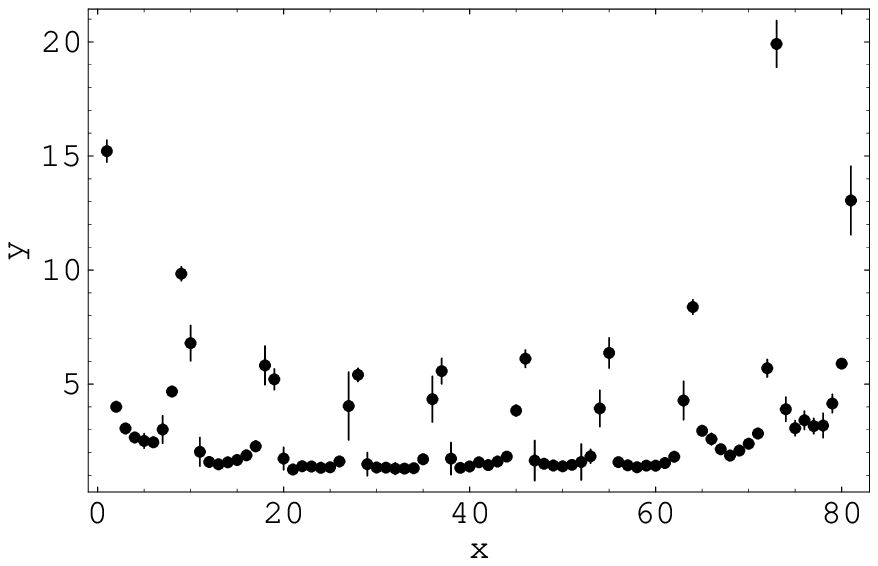}
  }\hspace*{1.2em}
  \subfigure[][2D-dependency of $Y$ resolution.]{\label{subfig:firstY-moment-measure3D}
    \psfrag{x}{{\small $x$}}
    \psfrag{y}{{\small $y$}}
    \psfrag{z}{\hspace*{-0.4em}\rotatebox{90}{\hspace*{-1.5em}{\small $\Delta Y$ [mm]}}}
    \includegraphics[width=0.40\textwidth]{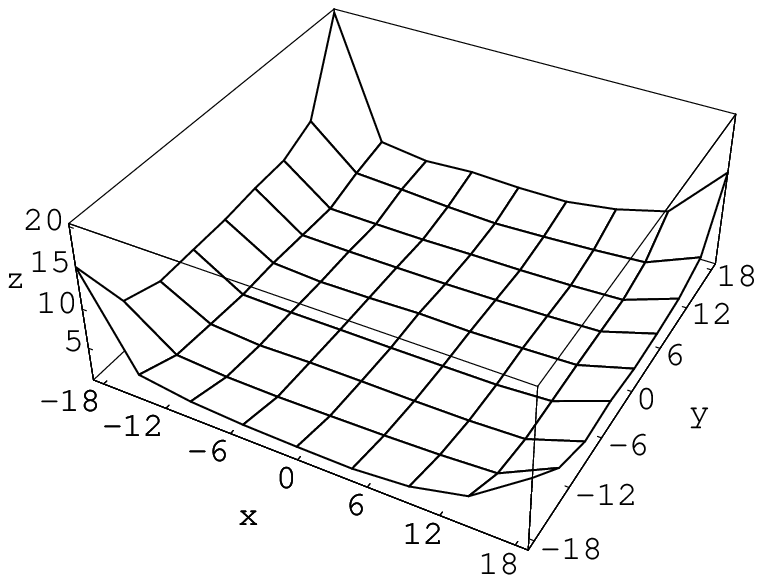}
  }
  \caption[Measured resolutions for the $x$- and $y$- centroid and the
    energy]{Measured resolutions for the $x$- and $y$- centroid and the
    energy. To the left; the values are displayed with error bar. To
    the right, 2D-plots are shown for better recognition of the
    functional dependence on the transverse coordinates.}
  \label{fig:all-normal-mom-res}
\end{figure}

In order to obtain the standard deviation $\sigma_\tincaps{ID}$ from
the three non-trivial and normalized moments $\mu_{x_1,y_0}$,
$\mu_{x_2,y_1}$ and $\mu_{x_2,y_0}+\mu_{x_0,y_2}$, the definition of
the variance for two dimensions can be applied only with some modifications:
\begin{equation}
  \label{eq:definition-variance}
\begin{split}
  \var(X+Y)&=\var(X)+\var(Y)+2\cov(X,Y)\\
  &=\mathcal{E}(X^2)-\mathcal{E}(X)^2+\mathcal{E}(Y^2)-\mathcal{E}(Y)^2+2\mathcal{E}\left[(X-\mathcal{E}(X))(Y-\mathcal{E}(Y))\right]\\
  &=\mathcal{E}(X^2)+\mathcal{E}(Y^2)-\mathcal{E}(X)^2-\mathcal{E}(Y)^2+2\mathcal{E}(XY)-\mathcal{E}(X)\mathcal{E}(Y),
\end{split}
\end{equation}
where $\mathcal{E}(A)$ denotes the expectation value of $A$. Note that
$\mathcal{E}(X^2)+\mathcal{E}(Y^2)$ is just the additional moment 
$\mu_{x_2,y_0}+\mu_{x_0,y_2}$ provided by the enhanced charge
divider. $\mathcal{E}(X)$ and $\mathcal{E}(Y)$ are the centroids
$\mu_{x_1,y_0}$ and $\mu_{x_0,y_1}$, respectively. However, one has to
take into account that different electronic amplifier designs (refer
to Appendix~\ref{app:elec-config}) are used for the centroids and for the
composite second moment. Therefore, they arrive with different
gain-factors at the ADC-module. Moreover, as it was discussed before
in this section and also in section~\ref{ch:prop-res-chains},
proportional resistor chains do not provide exactly the second moment,
but also include higher than quadratic orders. We define, therefore, the
ID-estimator $\sigma_\tincaps{ID}$ as the square root of the composite
second moment $\mu_{x_2,y_0}+\mu_{x_0,y_2}$, reduced by the transverse
dependency $\mathcal{W}(\mu_{x_1,y_0},\mu_{x_0,y_1})$ defined in
equation~(\ref{eq:sec-mom-fit-ansatz}):
\begin{equation}
  \label{eq:my-doi-sigma}
  \sigma_\tincaps{ID}\mdef\sqrt{(\mu_{x_2,y_0}+\mu_{x_0,y_2})-\mathcal{W}(\mu_{x_1,y_0},\mu_{x_0,y_1})}.
\end{equation}

\begin{figure}[!t]
  \centering
  \subfigure[][Measured resolution of the bare second moment at the 81 positions.]{\label{subfig:second-moment-measure2D}
    \psfrag{x}{\hspace*{-2em}{\small position \#}}
    \psfrag{y}{\hspace*{-4em}{\small $\mu_{x_2,y_0}+\mu_{x_0,y_2}$ [mm]}}
    \includegraphics[width=0.52\textwidth]{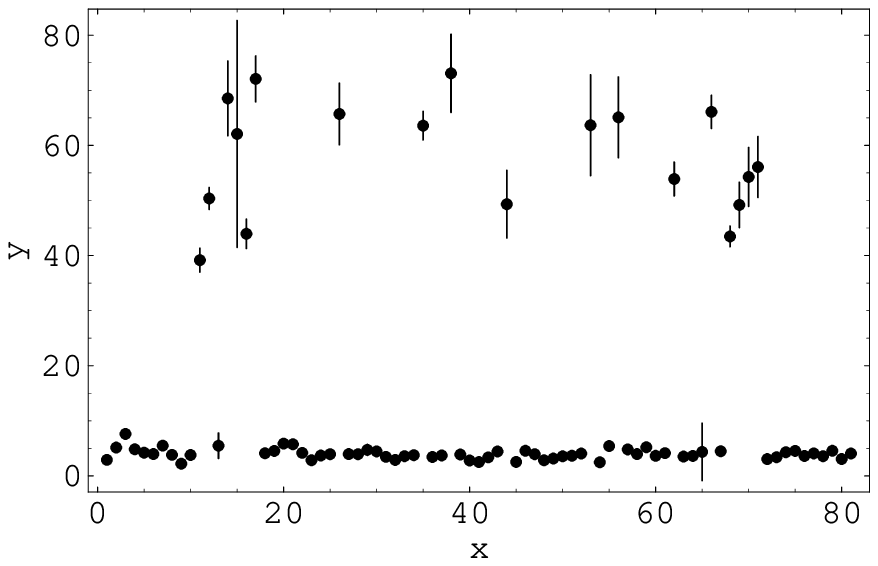}
  }\hspace*{1.4em}
  \subfigure[][2D-dependency of the resolution of the bare second moment.]{\label{subfig:second-moment-measure3D}
    \psfrag{x}{{\small $x$}}
    \psfrag{y}{{\small $y$}}
    \psfrag{z}{\hspace*{-1.2em}\rotatebox{90}{\hspace*{-3.5em}{\small $\mu_{x_2,y_0}+\mu_{x_0,y_2}$ [mm]}}}
    \includegraphics[width=0.40\textwidth]{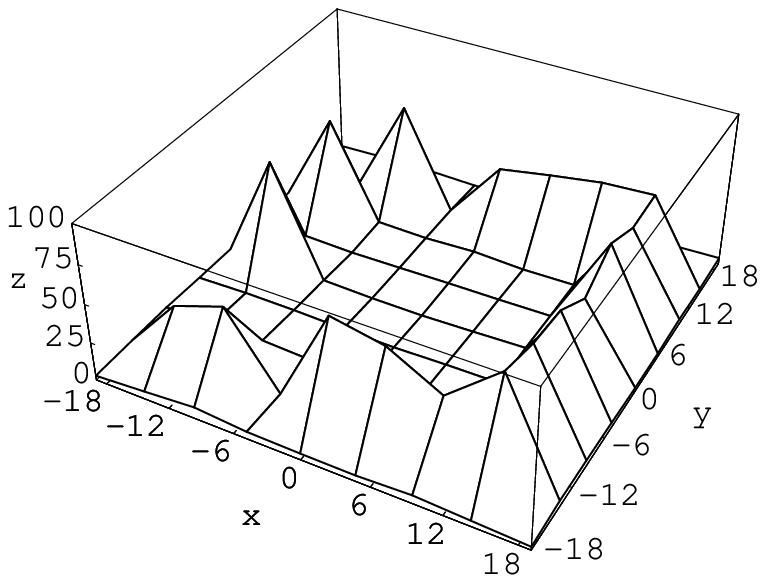}
  }\\
  \subfigure[][Measured the resolution of $\sigma_\tincaps{ID}$ at the 81 positions.]{\label{subfig:sigma-measure2D}
    \psfrag{x}{\hspace*{-2em}{\small position \#}}
    \psfrag{y}{\hspace*{-1.5em}{\small $\sigma_\tincaps{ID}$ [mm]}}
    \includegraphics[width=0.52\textwidth]{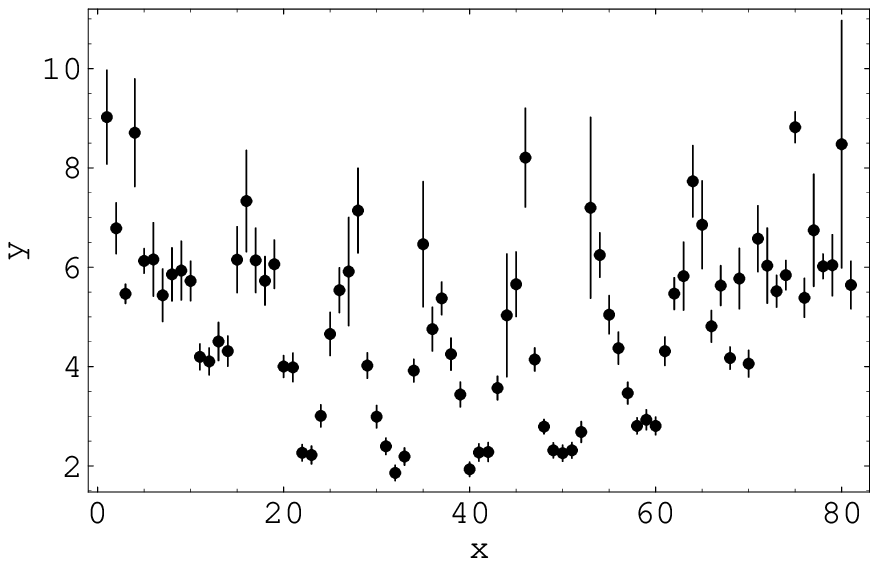}
  }\hspace*{1.4em}
  \subfigure[][2D-dependency of the resolution of $\sigma_\tincaps{ID}$.]{\label{subfig:sigma-measure3D}
    \psfrag{x}{{\small $x$}}
    \psfrag{y}{{\small $y$}}
    \psfrag{z}{\hspace*{-0.4em}\rotatebox{90}{\hspace*{-1.5em}{\small $\sigma_\tincaps{ID}$ [mm]}}}
    \includegraphics[width=0.40\textwidth]{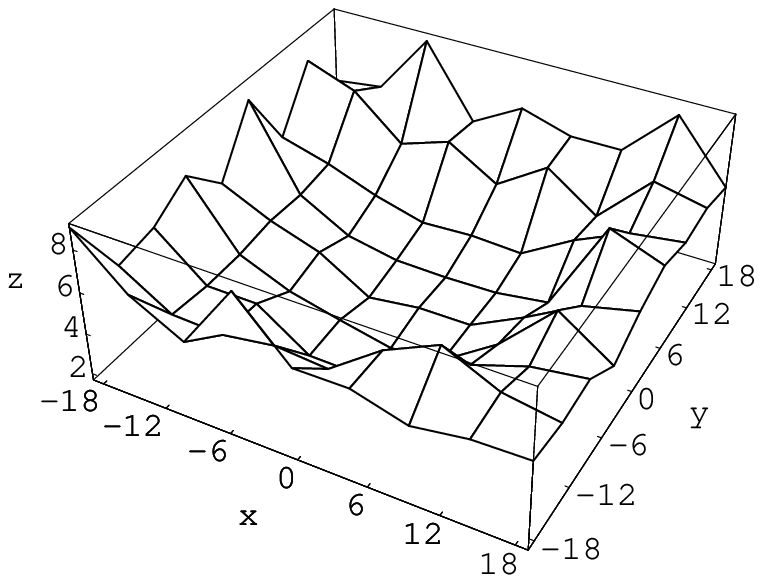}
  }
  \caption[Measured resolutions for $\mu_{x_2,y_0}+\mu_{x_0,y_2}$ and
  $\sigma_\tincaps{ID}$]{Measured resolutions for the
    $\mu_{x_2,y_0}+\mu_{x_0,y_2}$ and $\sigma_\tincaps{ID}$. To the
    left; the values are displayed with error bar. To
    the right, 2D-plots are shown for better recognition of the
    functional dependence on the transverse coordinates.}
  \label{fig:all-higher-mom-res}
\end{figure}

For measuring  the three-dimensional resolution of the detector,
model~(\ref{eq:fit-model-dist}) is fitted to the histograms of the
centroids and the energy for all 81 test positions. The FWHM is computed
from the best fit curves and its error is estimated by
applying standard error
propagation. Equation~(\ref{eq:doi-in-position-space}) with the
corresponding error propagation was used for the bare second moment
$\mu_{x_2,y_0}+\mu_{x_0,y_2}$ and $\sigma_\tincaps{ID}$.
Since the centroids always reproduce a compressed image of the
radioactive sources, they have to be expanded by applying the
inverse of the position-to-centroid mapping. However, not only the
center-position is shifted in this way but the point-spread function is
also proportionally blurred. This expansion has to be applied to the
FWHMs of the centroid distributions and the FWHMs also have to be corrected for
the finite source diameter~(\ref{eq:effective-source-diameter}). The
results are displayed in
figures~\ref{subfig:zero-moment-measure2D}-\ref{subfig:sigma-measure3D}.
As expected, the energy and centroid resolutions degrade considerably towards the
edges and borders of the crystal. The values for
$\mu_{x_2,y_0}+\mu_{x_0,y_2}$ and $\sigma_\tincaps{ID}$ likewise
depend strongly on the transverse coordinates. Nevertheless, a good
depth of interaction estimate is given by $\sigma_\tincaps{ID}$,
reaching a mean DOI resolution of $\approx5\,\mathrm{mm}$ and its best value of 
$1.8\,\mathrm{mm}$ at the center. This a high resolution in 
comparison with the methods discussed in section~\ref{ch:doi-detectors}.
The resolution of the bare second moment becomes very poor at certain test
positions. These positions are part of a connected region of the
PSPMT's sensitive area, at which the resolution of this
moment is subjected to strong variations. This effect can be easily explained
by the fact that the parabolas of the bare second moments for different
DOIs intersect at this particular region. As a consequence, even large variations
of the interaction depth lead only to very small variation for this
moment, although the intrinsic width of the point-spread
function for the DOI remains the same. Then, the denominator of
equation~(\ref{eq:doi-in-position-space}) becomes very small and this
explains the observed large values (see also Lerche {\em et al.}\ \cite{Lerche:2005b}).

Mean values, standard deviations, maximum and minimum values for the
resolutions of the five moments are summarized in table~\ref{tab:resolution-moms}.

\begin{table}[!ht]
  \centering\renewcommand{\arraystretch}{1.4}
  \begin{tabular}{cccccc}\hline\hline
    Moment&Mean&StdDev&Min&Max&Unit\\\hline
    $\mu_{x_0,y_0}$&24.9&8.8&16.6&71.2&\%\\
    $\mu_{x_1,y_0}$&3.4&3.2&1.4&20.9&mm\\
    $\mu_{x_0,y_1}$&3.3&3.1&1.3&19.9&mm\\
$\mu_{x_2,y_0}+\mu_{x_0,y_2}$&15.9&23&2.2&73.1&mm\\
$\sigma_\tincaps{ID}$&4.9&1.8&1.9&9.0&mm\\\hline\hline
  \end{tabular}
  \caption[Mean, std.\ dev., max.\ and min.\ values for the
  resolutions of the four moments]{Mean, standard deviation, maximum
    and minimum values for the resolutions of the different moments.}
  \label{tab:resolution-moms}
\end{table}

\chapterbib


\cleardoublepage{}
\chapter{3D-Impact Position Reconstruction}
\label{ch:position-reconstruction}

\chapterquote{%
Imagination is the real and eternal world of which this vegetable
universe is but a faint shadow.}{%
William Blake, $\star$ 1757 -- $\dagger$ 1827 
}

\PARstart{A}{s} mentioned in chapter~\ref{ch:motivation}, the reason
for using large-sized continuous scintillation
crystals\footnote{In the sense of
  section~\ref{sec:included-contribs}.} is mainly founded in the
aim to reduce the cost of $\gamma$-ray imaging detectors.
Clearly this can only be justified if
there is no significant loss in the performance of the
different characterizing parameters. With the development of the block
detector by Nutt and Casey \cite{Casey:1986}, a promising technique for
building PET scanners appeared. Since the intrinsic spatial resolution of the
$\gamma$-ray imaging detector can be obtained by selecting an adequate
size of the crystal element, this design has been adopted by
several research groups for use with small animal positron
emission tomography. There have been some alternative 
efforts to use continuous crystals for $\gamma$-ray imaging 
for coincidence and single photon imaging
(Siegel {\em et al.}\ \cite{Siegel:1995} and Seidel {\em et al.}\
\cite{Seidel:1996}), relying on the Center of Gravity algorithm
implemented as described in section~\ref{ch:charge-div-circuits}. 
They report strong edge artefacts, especially
when thick crystals for the detection of high energy photons are
required. As shown in section~\ref{ch:errors-of-cog-and-cdr}, these
errors are inherent in the CoG algorithm for crystals of finite size
and thickness. Other groups abandoned the conventional Anger
positioning scheme in favor of statistic based positioning (Joung et
al.\ \cite{Joung:2002}) or the use of neural networks (Tavernier {\em et al.}\
\cite{Tavernier:2005}), but at the expense of independently digitizing all
photodetector segments. Despite the fact that a better spatial resolution can
be achieved by using these methods, they are obtained by drastically
increasing the number of electronic channels. 
Due to the fact that
the data acquisition and analysis system has to be more complex,
the final cost of the device is significantly increased.

\section{The Truncated Moment Problem}

In chapter~\ref{ch:enhanced-charge-dividing-circuits}, DOI enhanced
designs of charge dividing networks
were presented. These networks have been developed primarily in order
to allow for an efficient estimation of the depth of interaction of the 
$\gamma$-ray. However, the experimental results of chapter~\ref{ch:experiment}
revealed that using the standard deviation as DOI estimator yields a
good resolution at the central region of the sensitive area but
degrades towards the crystal edges. The reason for this effect is that
near the edges, the detected signal
distribution is distorted because a large fraction of the
scintillation light is absorbed by the black painted surfaces.
This causes a change in all moments and their use as 
three-dimensional position and energy estimators becomes a poor
approximation for a large fraction of the sensitive area.

Using directly the second moment instead of the square root of the
central second moment ({\em i.e.}\ the standard deviation) is not possible
without problems because it is subjected to strong quadratic
variations along the transverse spatial directions. Moreover, the
second moments for different DOI intersect at an annular
region of the sensitive area. That is, there is a region of the
sensitive area where the measured second moments of events with
different depth but same $x$-$y$-position are almost equal to each
other. For points within this region, no
DOI information can be obtained directly from the second moment since
all DOIs are mapped to a similar value. However, this applies only to
the non-centered second moment. For the standard deviation no
intersections are observed (refer to chapter~\ref{ch:experiment} and
also Lerche {\em et al.}\ \cite{Lerche:2005b}). 

As a consequence, one is tempted to recover the
true impact positions from the available moments given by equations
(\ref{eq:exp-0-mom})-(\ref{eq:exp-2-mom}). This is a typical inverse
problem and for its solution one needs to find the following three
functions, or at least good approximations for them.
\begin{gather}
  \label{eq:inverse-func-1}
  x=\mathcal{X}\left(\frac{\mu_{x_1,y_0}}{\mu_{x_0,y_0}},\frac{\mu_{x_2,y_1}}{\mu_{x_0,y_0}},\frac{\mu_{x_2,y_0}+\mu_{x_0,y_2}}{\mu_{x_0,y_0}}\right)\\
  \label{eq:inverse-func-2}
  y=\mathcal{Y}\left(\frac{\mu_{x_1,y_0}}{\mu_{x_0,y_0}},\frac{\mu_{x_2,y_1}}{\mu_{x_0,y_0}},\frac{\mu_{x_2,y_0}+\mu_{x_0,y_2}}{\mu_{x_0,y_0}}\right)\\
  \label{eq:inverse-func-3}
  z=\mathcal{Z}\left(\frac{\mu_{x_1,y_0}}{\mu_{x_0,y_0}},\frac{\mu_{x_2,y_1}}{\mu_{x_0,y_0}},\frac{\mu_{x_2,y_0}+\mu_{x_0,y_2}}{\mu_{x_0,y_0}}\right).
\end{gather}
No inverse function is required for the energy since $\mu_{x_0,y_0}$ is used for
normalization of the moments involved and the true energy of the detected
event, {\em i.e.}\ the amplitude $J_0$ of the signal distribution
\ref{eq:total-lightdist}, acts as a multiplicative constant.

A special feature of the present inverse problem is that one has to
reconstruct distribution parameters from some known moments of the
distribution. Thus, it is very similar to the {\em truncated moment
problem} (Kre\u{\i}n and Nudel'man \cite{Krein}, Tkachenko {\em et al.}\
\cite{Tkachenko:1996}). There are three essentially different types of
moment problems depending on which type of intervals they are defined
on. For historical reasons, the moment problem on the semi-infinite
interval $[0,\infty[$ is called the {\em Stieltjes} moment problem. On
$\mathbb{R}$ it is called the {\em Hamburger} moment problem and on
the bounded interval $[0,1]$ it is referred to as the {\em Hausdorff}
moment problem.

\section{Polynomial Interpolation}

A promising technique for inverting the system of
equations~(\ref{eq:exp-0-mom})-(\ref{eq:exp-2-mom}) is given with
standard polynomial interpolation. The approximation of functions by
interpolation embraces a large variety of methods including Spline
interpolation, Taylor series, continued fractions, etc., which are well
understood and widely used (see for instance Kincaid and Cheney
\cite{Kincaid} and Press {\em et al.}\ \cite{Press:1992}). 

Polynomial interpolation has been recently applied to a rather similar
problem (Olcott {\em et al.}\ \cite{Olcott:2005}, \cite{Olcott:privcom}). 
They studied the spatial
response of a scintillation crystal array coupled to a position-sensitive
avalanche photodiode (PSAPD). These photodetecting devices
present a promising technology for $\gamma$-ray imaging detectors but
suffer from a strong {\em pincushion}-like distortion. Olcott {\em et al.}\
used finite element methods for solving Laplace's equation on the
resistive layer of the PSAPD in order to predict its spatial response
function. Finally, they expanded the supposed detected positions using a
two-dimensional polynomial basis, thereby obtaining a matrix of coefficients
that could be inverted and applied for linearity correction with
encouraging results. 

The differences compared to the inverse problem posed by
equations~(\ref{eq:exp-0-mom})-(\ref{eq:exp-2-mom}) and the detector
setup discussed in the present work are minor. Instead, the
fact that an additional measure of the distribution, {\em i.e.}\
the second moment, is made available, probably increases the performance of the method.

\subsection{Polynomial Interpolation in One Dimension}
\label{sec:poly-expand-1D}

Before applying this method to the inverse problem, some useful
properties are summarized (refer to Kincaid and Cheney
\cite{Kincaid}). The interpolation problem consists in finding a
polynomial $p(x)$ of lowest possible degree $k$ for a given set of
$n+1$ known data points ${(x_i,y_i)},\;{i=0,1,\ldots,n}$. If the interpolation
points $x_0,x_1,\ldots,x_n$ are distinct real numbers, then there is a
unique polynomial $p_n(x)$ of degree $k$ at most $n$ such that 
\begin{equation}
  \label{eq:1D-poly-interpol}
  p_n(x_i)=y_i\mbox{, with } (0\leq i\leq n).
\end{equation}
This theorem holds for arbitrary values $y_0,y_1,\ldots,y_n$ and a
proof can be found in Kincaid and Cheney \cite{Kincaid}.
The interpolation polynomial can be written in terms of powers of $x$
\begin{equation}
  \label{eq:i-pol-repres}
  p_n(x)=\sum_{j=0}^na_jx^j
\end{equation}
which, together with the interpolation
conditions~(\ref{eq:1D-poly-interpol}), leads to a system of $n+1$ linear
equations that can be used for determining the coefficients
$a_0,a_1,\ldots,a_n$. This system has the following form:
\begin{equation}
  \label{eq:interpol-lineareq-sys}
  y_i=p_n(x_i)=\sum_{j=0}^na_jx^j_i.
\end{equation}
The coefficient matrix $\mathbf{X}:=x_i^j$ is known as the {\em Vandermonde
  Matrix} (also called an
alternant matrix), and has the form
\begin{equation}
  \label{eq:vandermonde-matrix-form}
  \mathbf{V}=\left[
  \begin{array}{ccccc}
    1&x_0&x_0^2&\cdots&x_0^n\\
    1&x_1&x_1^2&\cdots&x_1^n\\
    1&x_2&x_2^2&\cdots&x_2^n\\
    \vdots&\vdots&\vdots&\ddots&\vdots\\
    1&x_n&x_n^2&\cdots&x_n^n
  \end{array}
  \right].
\end{equation}
While the Vandermonde matrix is non-singular because the system
has an unique solution for any choice of $y_0,y_1,\ldots,y_n$, it is
often ill-conditioned and does not allow an accurate determination of
the coefficients $a_i$. There are still other algorithms for
polynomial interpolation that all produce the same result, since the
solution is unique as stated above. Depending on the posed problem,
they have advantages and disadvantages. The amount of work involved to
obtain $p_n(x)$ in equation~(\ref{eq:i-pol-repres}) seems excessive
and algorithms that intrinsically implement Horner's
scheme\footnote{Horner's scheme rearranges a polynomial into the
  recursive form $p_n(x)=a_0+x(a_1+x(a_2+\cdots
  x(a_{n-1}+a_nx)\cdots))$ and therefore requires only 
$n$ additions and $n$ multiplications for evaluating the polynomial.} may be
preferred. However, as will be shown later, the expansion into the
polynomial $p_n(x)$ never has to be computed because there is no
need to know the functional form of the distribution. Only the
three-dimensional impact position has to be estimated.

The discrepancy between the true function $f(x)$ that produced the
data points ${(x_i,y_i)}_{i=0,1,\ldots,n}$ and the interpolating
polynomial at the position $x$ is given by
$\epsilon_n(x)=f(x)-p_n(x)$. This error can be expressed as a function
of the ($n+1$)th derivative of $f(x)$ at the corresponding position
$\xi_x$ as follows: 
\begin{equation}
  \label{eq:interpol-error}
  \epsilon_n(x)=\frac{1}{(n+1)!}f^{(n+1)}(\xi_x)\prod_{i=0}^n(x-x_i),
\end{equation}
where it is supposed that all distinct interpolation nodes
$x_0,x_1,\ldots,x_n$ lie in the interval $[a,b]$, and that to each 
$x$ in this interval corresponds a point $\xi_x$ in $]a,b[$ that satisfies
equation~(\ref{eq:interpol-error}). The error
$\epsilon_n(x)$ of the  polynomial $p_n(x)$ can be optimized by
choosing adequate interpolation nodes. When the roots of the Chebyshev
polynomials defined by 
\begin{equation}
  \label{eq:chebychev-pols}
  \mathcal{T}_n(x)=\cos\left[n\cos^{-1}(x)\right]\mbox{, with }n\geq0
\end{equation}
on the interval $[-1,1]$ are used as nodes,
equation~(\ref{eq:interpol-error}) transforms into 
\begin{equation}
  \label{eq:minimized-error}
  |\epsilon_n(x)|\leq\frac{1}{2^n(n+1)!}\max_{|t|\leq1}\left|f^{(n+1)}(t)\right|.
\end{equation}
The roots of the Chebyshev polynomials are given by the following
closed-form expression:
\begin{equation}
  \label{eq:tcheby-i-points}
  x_i=\cos\left(\frac{2i+1}{2n+1}\pi\right).
\end{equation}
Again, proofs for both error estimates~(\ref{eq:interpol-error}) and
(\ref{eq:minimized-error}) can be found in Kincaid and Cheney
\cite{Kincaid}.

\subsection{Polynomial Interpolation in Higher Dimensions}
\label{pol-interpol-hd}

For clarity, higher dimensional interpolation is explained
considering the two-dimensional case. Generalizations to higher
dimensions are straightforward using the results derived in this paragraph.
In the two dimensional case, the interpolation problem consists in
finding a smooth interpolant $p$ for the set of $n+1$ distinct interpolation
points $(x_0,y_0),(x_1,y_1),\ldots,(x_n,y_n)$. With each point
$(x_i,y_i)$ there is an associated real function value $f_i$. Clearly,
the interpolation function has to reproduce these values at the
interpolation points such that
\begin{equation}
  \label{eq:interpol-2d}
  p(x_i,y_i)=f_i\mbox{, with } 0\leq i\leq n.
\end{equation}

Within the scope of this work, the higher dimensional interpolation
can be reduced to univariate cases using a tensor product of
the interpolation described above. For this, the
interpolation nodes have to form a Cartesian grid such that
$\mathcal{N}_\mathit{ip}=\{(x_l,y_m):0\leq l\leq r, 0\leq m\leq q;rq=n\}$
with
$x_0,x_1,\ldots,x_r$ being the interpolation points in the $x$-spatial
direction and $y_0,y_1,\ldots,y_q$ the interpolation points in the $y$-spatial
direction. As in the one-dimensional case, the interpolation
polynomial can be written in the form
\begin{equation}
  \label{eq:poly-interpol-hd}
  p_{k_x,k_y}(x,y)=\sum_{j_x=0}^{k_x}\sum_{j_y=0}^{k_y}c_{j_xj_y}x^{j_x}y^{j_y}.
\end{equation}
with $0\leq k_x\leq r$ and $0\leq k_y\leq q$
Equations~(\ref{eq:interpol-2d}) and (\ref{eq:poly-interpol-hd}) again
result in a set of linear equations that can be used to find the
coefficients $c_{j_xj_y}$ of the interpolating polynomial:
\begin{equation}
  \label{eq:poly-interpol-hd-coeff}
  f_i=p_{k_x,k_y}(x_i,y_i)=\sum_{j_x=0}^{k_x}\sum_{j_y=0}^{k_y}c_{j_xj_y}x_i^{j_x}y_i^{j_y}.
\end{equation}
The Vandermonde matrix now has the form
\begin{equation}
  \label{eq:vandermond-3D}
  \renewcommand{\arraystretch}{1.6}
  \mathbf{V}=\left[
  \begin{array}{ccccccccccccc}
    1&x_1&y_1&\cdots&x_1^{k_x}&y_1^{k_y}&x_1y_1&\cdots&x_1y_1^{k_y}&\cdots&x_1^{k_x}y_1&\cdots&x_1^{k_x}y_1^{k_y}\\
    1&x_2&y_2&\cdots&x_2^{k_x}&y_2^{k_y}&x_2y_2&\cdots&x_2y_2^{k_y}&\cdots&x_2^{k_x}y_2&\cdots&x_2^{k_x}y_2^{k_y}\\
    \vdots&\vdots&\vdots&\ddots&\vdots&\vdots&\vdots&\ddots&\vdots&\ddots&\vdots&\ddots&\vdots\\
    1&x_n&y_n&\cdots&x_n^{k_x}&y_n^{k_y}&x_ny_n&\cdots&x_ny_n^{k_y}&\cdots&x_n^{k_x}y_n&\cdots&x_n^{k_x}y_n^{k_y}
  \end{array}
  \right].
\end{equation}

So far, only functions that map a region into the one dimensional
space have been considered. Let now the unknown function that has to
interpolated be a mapping from the two-dimensional space into the
two-dimensional space, {\em e.g.}\ $f:U\subset\mathbb{R}^2\mapsto B\subset\mathbb{R}^2$.
Then, each interpolation node $(x_i,y_i)$ has a pair of  associated real
numbers $(v_i,u_i)$. Obviously, a single polynomial cannot
reproduce this two-dimensional mapping. However, one can define a
combined function using one polynomial for each dimension of the image
as follows:
\begin{equation}
  \label{eq:two-dimensional-poly}
  P(x,y)\mdef(p_u(x,y),p_v(x,y)),
\end{equation}
where $p_u(x,y)$ and $p_v(x,y)$ are ordinary one-dimensional
polynomials that satisfy the following relations:
\begin{gather}
  \label{eq:1d-pols-for-2d-mapping-a}
  p_u(x_i,y_i)=u_i\mbox{, with } 0\leq i\leq n\quad\mbox{and}\\
  \label{eq:1d-pols-for-2d-mapping-b}
  p_v(x_i,y_i)=v_i\mbox{, with } 0\leq i\leq n.
\end{gather}
As in the previous cases, one can write both polynomials $p_u(x,y)$ and $p_v(x,y)$ 
as a sum of the different orders weighted with the coefficients
$a_{j_xj_y}$ and $b_{j_xj_y}$. Together with
equations~(\ref{eq:1d-pols-for-2d-mapping-a}) and
(\ref{eq:1d-pols-for-2d-mapping-b}) one now obtains two systems of
linear equations, one for the coefficients $a_{j_xj_y}$ and another
for the coefficients $b_{j_xj_y}$:
\begin{gather}
  \label{eq:poly-interpol-2dto2d-a}
  u_i=p_u(x_i,y_i)=\sum_{j_x=0}^{k_x}\sum_{j_y=0}^{k_y}a_{j_xj_y}x_i^{j_x}y_i^{j_y}\\
  \label{eq:poly-interpol-2dto2d-b}
  v_i=p_v(x_i,y_i)=\sum_{j_x=0}^{k_x}\sum_{j_y=0}^{k_y}b_{j_xj_y}x_i^{j_x}y_i^{j_y}.
\end{gather}
Note that the Vandermonde matrix is exactly the same in both
systems~(\ref{eq:poly-interpol-2dto2d-a}),
(\ref{eq:poly-interpol-2dto2d-b}) and they can be written in
matrix form as follows: 
\begin{equation}
  \label{eq:fwd-matrix-equations}
  \mathbf{u}=\mathbf{V}\mathbf{a}\quad\mbox{ and }\quad\mathbf{v}=\mathbf{V}\mathbf{b}.
\end{equation}
Equation~(\ref{eq:fwd-matrix-equations}) can be rewritten in one
single equation by gathering all $u$ and $v$ values in the matrix
$\mathbf{Y}\mdef [\mathbf{u}\,\mathbf{v}]$ and all $a$ and $b$ values
in the coefficient matrix $\mathbf{C}\mdef
[\mathbf{a}\,\mathbf{b}]$. In this way, one obtains the single matrix equation
\begin{equation}
  \label{eq:final-interpol-m-eq}
  \mathbf{Y}=\mathbf{V}\mathbf{C},
\end{equation}
where the Vandermonde matrix $\mathbf{V}$ is obtained from the
interpolation nodes $\mathcal{N}_\mathit{ip}$. 
As stated formerly, $\mathbf{V}$ is non-singular
and its determinant is therefore nonzero if the interpolation nodes 
$\mathcal{N}_\mathit{ip}$ are distinct. Therefore,
equation~(\ref{eq:final-interpol-m-eq}) can be inverted in order to
obtain the coefficients $\mathbf{C}$ as a function of the function-values 
$\mathbf{Y}$ at the interpolation nodes. Thus one has
\begin{equation}
  \label{eq:inverted-interpol-m-eq}
  \mathbf{C}=\mathbf{V}^{^+}\mathbf{Y},
\end{equation}
were, $\mathbf{V}^{^+}$ denotes the pseudo-inverse (also called Moore-Penrose
matrix inverse) of $\mathbf{V}$ described in the next section.

\subsection{Moore-Penrose Matrix Inverse}

The matrix that has to be inverted in order to obtain
(\ref{eq:inverted-interpol-m-eq}) from (\ref{eq:final-interpol-m-eq})
is the Vandermonde matrix (\ref{eq:vandermond-3D}). In general, this
matrix is not a square matrix but a $n\times(k_x+1)(k_y+1)$ matrix, where
$n$ is the total number of interpolation points and $k_x$ and $k_y$
the interpolation orders along the $x$- and $y$-spatial directions
respectively. By definition, the inverse $\mathbf{A}^{-1}$ of an
matrix $\mathbf{A}$ has to satisfy the relation
\begin{equation}
  \label{eq:matrix-inverse-def}
  \mathbf{A}^{-1}\mathbf{A}=\mathbf{A}\mathbf{A}^{-1}=\mathbf{I},
\end{equation}
where $\mathbf{I}$ denotes the identity matrix. Clearly, this is only
possible, if the matrix $\mathbf{A}$ is a square matrix. For general
$l\times m$ matrices with $l\neq m$, the matrix product does not commute
and the first identity in expression~(\ref{eq:matrix-inverse-def})
does not hold. A generalization of the matrix inverse for non-square
matrices was independently given by E.H.\ Penrose and R.\
Moore. Whenever the inverse of $\mathbf{A}^{T}\mathbf{A}$ exist, one
can define a new matrix $\mathbf{A}^{^+}$ as
\begin{equation}
  \label{eq:moore-penrose-inverse}
  \mathbf{A}^{^+}={\left(\mathbf{A}^{T}\mathbf{A}\right)}^{-1}\mathbf{A}^{T}.
\end{equation}
This matrix is either called the pseudo-inverse or the
Moore-Penrose inverse. $\mathbf{A}^{T}$ is the transpose of
the matrix $\mathbf{A}$. In the present case, the pseudo-inverse of the
Vandermonde matrix can be computed, since $\mathbf{V}$ has a
non-vanishing determinant. In addition, the inverse of
$\mathbf{V}^{T}\mathbf{V}$ exists as a consequence of distributivity of
determinants and because $\det \mathbf{A}^{T}=\det\mathbf{A}$.

\section{Inverse Mapping of the Gamma-Ray Impact Positions}
\label{sec:inverse-mapping}

At this stage, an algorithm for computing the true three-dimensional
photoconversion position of the $\gamma$-photon from the moments of
the produced signal distribution can be given. For this purpose, the method
given by Olcott {\em et al.}\ \cite{Olcott:2005} was adapted to the
three-dimensional case.

As a first step, the 
detector's response is computed by using the signal distribution
$\mathcal{L}_\mathit{Detector}(\mathbf{r},\mathbf{r}_c)$
of section~\ref{sec:complete-signal-dist} at a three-dimensional
grid of interpolation nodes
$\mathbf{X}\mdef[\mathbf{x}\,\mathbf{y}\,\mathbf{z}]$ that are the
supposed $x$-, $y$- and $z$-coordinates of the photoconversion. At
each position, a set of
three non-trivial moments, representing the centroids and the second
moment are computed from the signal distribution as was described in
section~\ref{subsec:mod-val}. This is the response that one expects from
the detector by virtue of chapters~\ref{ch:light-distribution} and
\ref{ch:experiment} and forms the matrix
$\mathbf{Y}\mdef[\mathbf{x}_\mu\,\mathbf{y}_\mu\,\mathbf{z}_\mu]$, where
the subscript indicates that these values are the moments extracted
using an enhanced charge divider configuration. These moments are
used as interpolation nodes and the Vandermonde $\mathbf{V}$ matrix is
built from them.
The coefficients that map the polynomial expansion of the detector
response back into the space of impact positions can be found using the
Moore-Penrose inverse of the previous paragraph.
\begin{gather}
  \label{eq:system-mapping}
  \mathbf{X}=\mathbf{S}\mathbf{V}\mbox{ with }\\
  \label{eq:inv-system-mapping}
  \mathbf{S}=\mathbf{X}\mathbf{V}^{^+}
\end{gather}
Throughout the remaining work, $\mathbf{S}$ is called the {\em system}
matrix of the used $\gamma$-ray imaging detector.

In order to apply the inverse mapping to a measured set of moments
$(\mu_x,\mu_y,\mu_z)$, this vector is expanded over the same
three-dimensional polynomial basis as the Vandermonde matrix, giving a
$1\times(k_x+1)(k_y+1)(k_z+1)$ matrix $\mathbf{Y}_\mathit{meas}$ that
has to be contracted with the system matrix to give an estimate for
the true impact position:
\begin{equation}
  \label{eq:true-impact}
  \mathbf{x}\simeq\mathbf{S}\mathbf{Y}_\mathit{meas}. 
\end{equation}

\section{Results}

\begin{figure}[!t]
  \centering
  \includegraphics[width=0.4\textwidth]{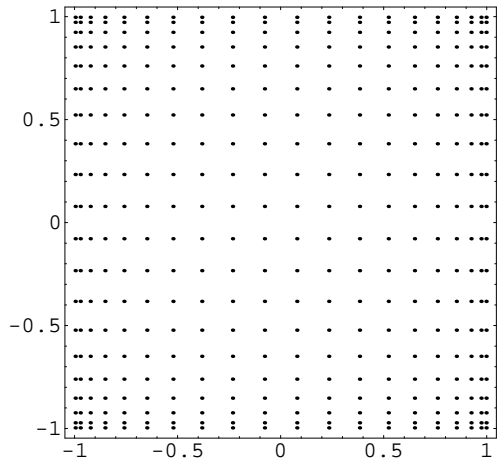}
  \caption[Grid of transverse test-positions]{Grid of transverse test-positions obtained by using the roots of the 20th Chebyshev polynomial.}
  \label{fig:cheby-sample-grid}
\end{figure}

Before the three dimensional resolution of the detector using the
reconstructed photoconversion position is measured at the same 81
positions as in chapter~\ref{ch:experiment}, the position
reconstruction from the moments is tested qualitatively.
The generalization from section~\ref{pol-interpol-hd} to the
three-dimensional case is straightforward.
For this, a three-dimensional array for the detector's response at
$n_\tincaps{TP}=n_x\times n_y\times n_z=50\times50\times16$ different impact positions
has been created. Along the transverse directions, the
roots of the 50th order Chebyshev polynomial have been taken as coordinates
for the test-points. An important remark has to be made
here. In section~\ref{sec:poly-expand-1D} it was mentioned
that the error of the approximating polynomial can be minimized
using the roots of the Chebyshev polynomials. However, the chosen
test-points are {\em not} the interpolation nodes, because they are
the  photoconversion positions from which the moments are
computed. Instead, one has to use the moments that correspond to
these photoconversion positions as interpolation nodes.
Due to the nonlinear response function
of the CoG algorithm (refer to
section~\ref{ch:errors-of-cog-and-cdr}), the moments will not coincide
with the roots. Nevertheless, the roots of the Chebyshev Polynomials
have been used for generating the test-points. In this way, the density
of the sample points increases towards the edges of the photocathode
where the maximum distortion is caused by the CoG. A example for a
$20\times20$ grid is shown in figure~\ref{fig:cheby-sample-grid}.
Along the $z$-direction, 16 equidistant depths have
been used. Note that it is highly recommendable to use interpolation
nodes inside the intervals $]-1,1[$ and $]0,1[$ for the transverse and
normal directions respectively. For these intervals, all monomials in
the representations~(\ref{eq:i-pol-repres}) and
(\ref{eq:poly-interpol-hd}) are confined to the image intervals
$]-1,1[$ and $]0,1[$ respectively. The same argument is valid for the non-unity
elements of the Vandermonde matrices by definition. Outside these
intervals, the monomials can become rather large even for values only
slighter larger than 1. This constraint is easily achieved by the linear
(and therefore bijective) mappings $[a,b]\to[-1,1]$ and
$[a,b]\to[0,1]$. The limits of the preimage interval $a$ and $b$ are
obviously given by the limits $L$, $-L$ and $T$ of the scintillation
crystal. This also
holds for both centroids because the signal distribution is non-negative
everywhere. In the case of the composite second moment nothing has to
be done because it is confined to an interval $I_{\mu_2}\subset]0,1[$
as a consequence of the experimental configuration (see
figures~\ref{subfig:parameter-a-secmom} and
\ref{subfig:parameter-b-secmom}). It is important that none of the
$50\times50\times16$ test-points contain any crystal limit $L$, $-L$
and $T$ as its coordinate since the signal distribution model of chapter~
\ref{ch:light-distribution} is not defined at these
points. Furthermore, it has been observed that the use of test-points
with the particular transverse coordinates $x=0$, $y=0$ or $x=y=0$
lead to large reconstruction errors in the near neighborhood of these
points. The reason for this behavior is not known but it can be
easily avoided by choosing an even number of test-positions along the
$x$- and $y$-spatial directions. Then, the mis-positioned events
disappear completely. 

Once the three-dimensional grid of test-positions has been created, the
detector response is computed at each of them as described in
section~\ref{subsec:mod-val}. In particular, the parameter values for
the signal distribution and the optimum scalings resumed in
table~\ref{tab:scaling-signal} have been used. This gives two matrices
$\mathbf{X}$ and $\mathbf{Y}$ of dimension $n_\tincaps{TP}\times3$
representing the true impact position and the corresponding three
non-trivial moments respectively. Moments and positions were then
scaled down to the intervals $]-1,1[$ and $]0,1[$. The system matrix
$\mathbf{Y}$ of the test detector was then computed by virtue of
equation~(\ref{eq:inv-system-mapping}). Different interpolation orders
have been tested.

\subsection{Qualitative Validation of the Method}

First of all, the impact positions of all detected events were
reconstructed from the measured moments at all 81 test-positions. 
All two-dimensional position plots were inspected to make sure that the
algorithm works in a qualitatively correct way, {\em i.e.}, that no artefacts were
present and that the reconstructed events were grouped around a reasonably
small region around the nominal $\gamma$-ray beam position. At a
first glance, it could be observed that the method works as
expected. For impact positions within the central region of the
photocathode, transverse positions and centroids are nearly the
same because the positioning error of the CoG algorithm is small for
these points. The second moment is, however, subjected to a strong
quadratic variation with the impact position which is successfully
removed to a great extent by the interpolation algorithm. As a result,
the majority of the reconstructed interaction depths lie
within the interval $[0,10]$ of possible values of the DOI. A low
fraction of events lead to reconstructed DOI values outside this
interval. There are two possible reasons for this. First, there are
always statistical errors in the measurements of the moments and
actually the intrinsic resolution for the DOI parameter is finite. On the
other hand, there is a significant fraction of Compton scattered events, as
was explained in chapter~\ref{ch:compton}, which are not
included in the response prediction of
chapters~\ref{ch:light-distribution} and \ref{ch:experiment}. 
While the centroids will always be smaller than the true impact
positions, (there can be no event outside the interval
$[-1,1]$, neither for the $x$ nor the $y$ spatial directions) there is
no well defined upper limit for the second moment. Consider an incident
$\gamma$-ray perpendicular to the plane of the photocathode
that is scattered off at 90\textdegree\, to the normal
direction. It will deposit $E_e/2$ at this position and the scattered
photon will travel parallel to the photocathode. If the scattered photon
covers a large distance and is photoelectrically absorbed at its
second interactions, one will get an event with a very large second moment
that does not correspond to the the interaction distance because the
signal distribution is a superposition of two equal distributions like
in expression~(\ref{eq:total-lightdist}) but displaced horizontally.
Such events cannot be reconstructed because they are not included in
the predicted position response. While the centroids remain inside the
interval ${[-1,1]}$ even for this special event, the second moment seems to
originate from an event of very high ID. Therefore, the polynomial
interpolation will project it outside the interval $[0,10]$. In
chapter~\ref{ch:compton}, the frequency and expected range of events
with large scattering angles was discussed. Due to the screening of forward scattered events, many
of the undiscriminated Compton scattered events will behave similarly to
the discussed example. Nevertheless, it was found that half of all
Compton scattered events have a transverse range
of the scattered photons of the order of $\mathrm{300\mu m}$. This is
significantly less than the obtained spatial resolution for both 
transverse and normal directions (see
section~\ref{sec:moments-as-pos-estimate}) and therefore will not lead
to large positioning errors for the majority of the cases. The other
half of the Compton scattered events are part of the long tails of the
histograms~\ref{subfig:compton-r-allhist}-\ref{subfig:compton-z-hists}.
They will produce strongly misplaced positions and large
reconstructed interaction depths. Note that this may provide
an excellent possibility for filtering inner-crystal Compton scattered
events. Unfortunately, the errors in the reconstructed positions are
still too large and an improvement of the method's accuracy is
required. This possibility has been mentioned here for completeness
but has not been studied in detail.

\begin{figure}[!tp]
  \centering
  \subfigure[][2D-density plot of the centroids for two $\gamma$-ray beams
  impinging on the positions ${(x,y)}$ = $(14.25,0)\mathrm{\,mm}$ and
  ${(x,y)}$ = $(19,0)\mathrm{\,mm}$.]{\label{subfig:mom-ovlerlay}
    \psfrag{x}{\hspace*{-2em}{\small $x$ centroid}}
    \psfrag{y}{\hspace*{-2em}{\small $y$ centroid}}
    \includegraphics[width=0.44\textwidth]{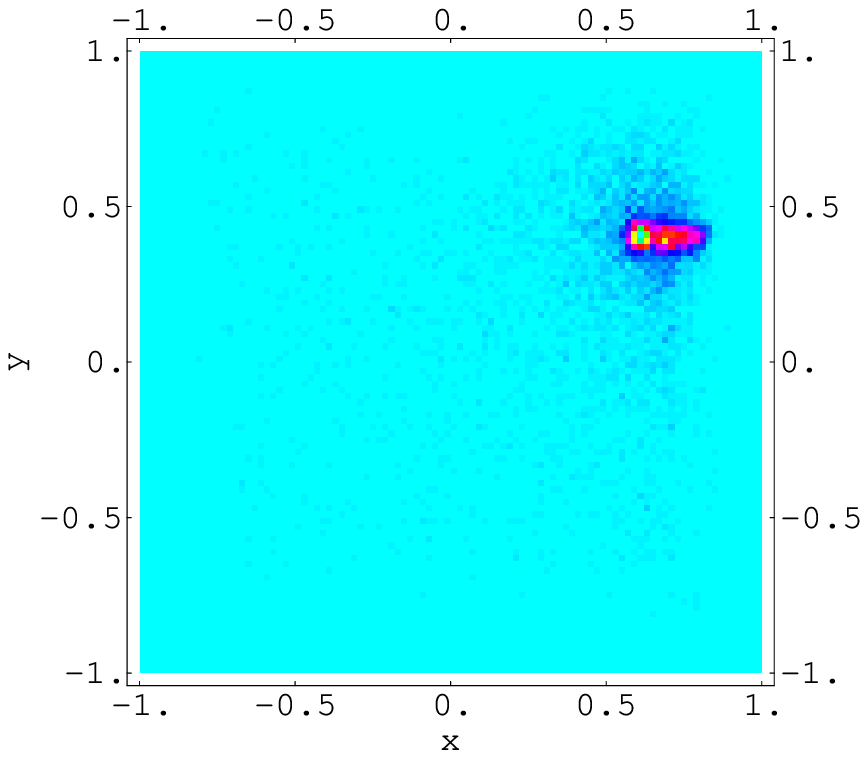}}
  \subfigure[][2D-density plot of the reconstructed positions for two
  $\gamma$-ray beams
  impinging on the positions ${(x,y)}$ = $(14.25,0)\mathrm{\,mm}$ and
  ${(x,y)}$ = $(19,0)\mathrm{\,mm}$.]{\label{subfig:pos-ovlerlay}
    \psfrag{x}{\hspace*{-2em}{\small $x$ position}}
    \psfrag{y}{\hspace*{-2em}{\small $y$ position}}
    \includegraphics[width=0.44\textwidth]{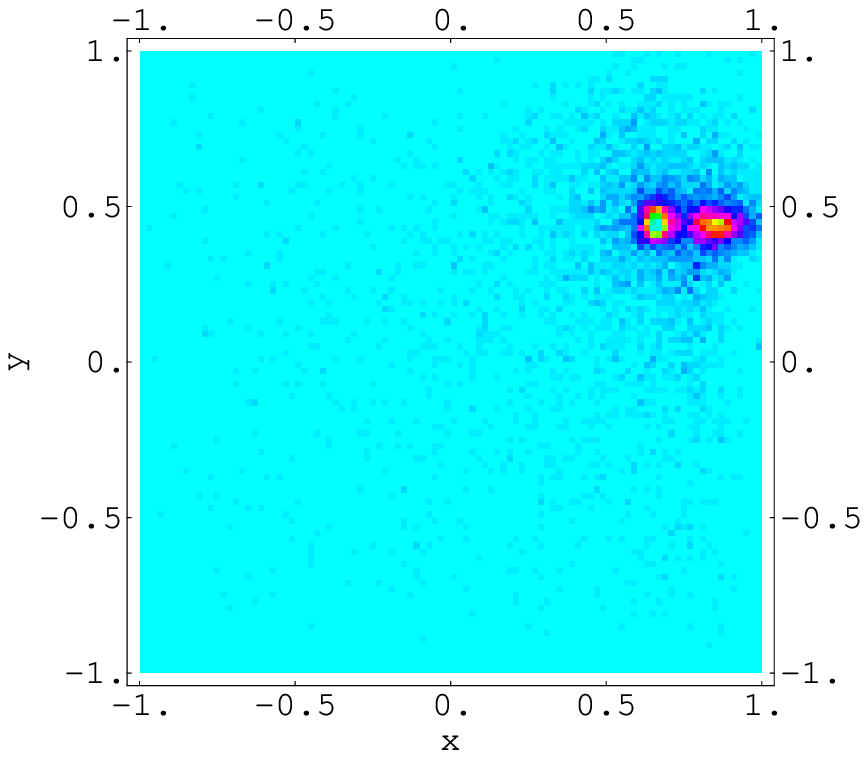}}\\
  \vspace*{-1eX}
  \subfigure[][Surface plot of the centroids for two $\gamma$-ray beams
  impinging on the positions ${(x,y)}$ = $(14.25,0)\mathrm{\,mm}$ and
  ${(x,y)}$ = $(19,0)\mathrm{\,mm}$.]{\label{subfig:mom-ovlerlay-3D}
    \psfrag{x}{\hspace*{-2em}{\small $x$ centroid}}
    \psfrag{y}{\hspace*{0.5em}\rotatebox{70}{\hspace*{-2em}{\small $y$ centroid}}}
    \psfrag{z}{\hspace*{-1.5em}{\small counts}}
    \includegraphics[width=0.44\textwidth]{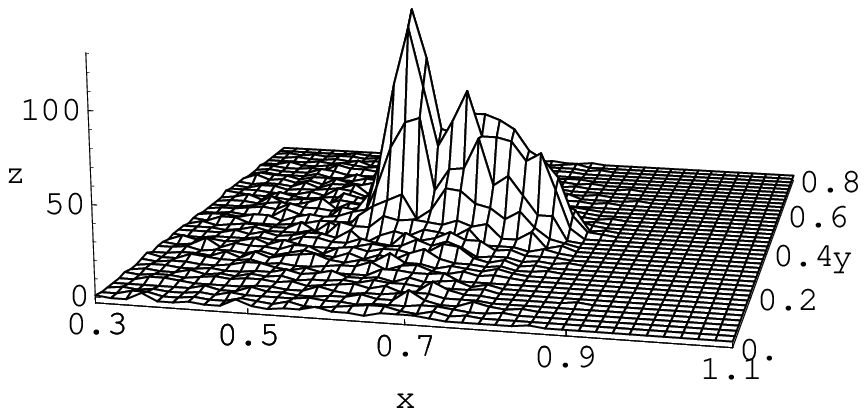}}
  \subfigure[][Surface plot of the reconstructed positions for two
  $\gamma$-ray beams
  impinging on the positions ${(x,y)}$ = $(14.25,0)\mathrm{\,mm}$ and
  ${(x,y)}$ = $(19,0)\mathrm{\,mm}$.]{\label{subfig:pos-ovlerlay-3D}
    \psfrag{x}{\hspace*{-2em}{\small $x$ position}}
    \psfrag{y}{\hspace*{0.5em}\rotatebox{70}{\hspace*{-2em}{\small $y$ position}}}
    \psfrag{z}{\hspace*{-1.5em}\raisebox{5eX}{{\small counts}}}
    \includegraphics[width=0.44\textwidth]{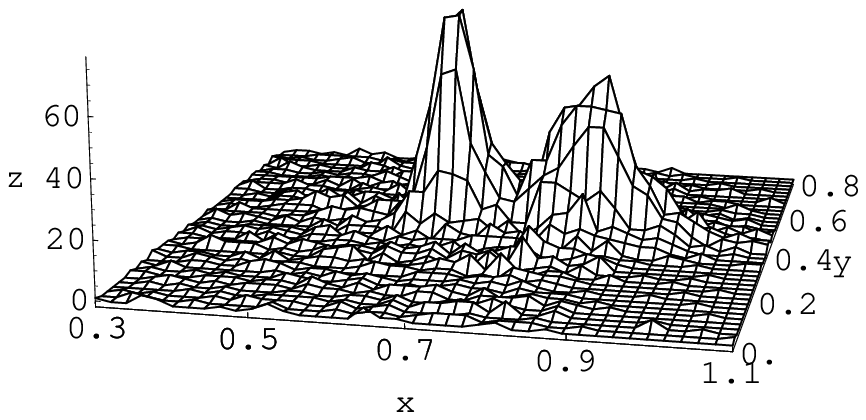}}\\
  \vspace*{1eX}
  \subfigure[][$x$-centroid for the same beam positions. Measured data
  (light-gray lines) and best fits (black lines) are plotted together.]{\label{subfig:mom-ovlerlay-xproj}
    \psfrag{x}{\hspace*{-2em}{\small $x$ centroid}}
    \psfrag{y}{\hspace*{-1.5em}{\small counts}}
    \includegraphics[width=0.44\textwidth]{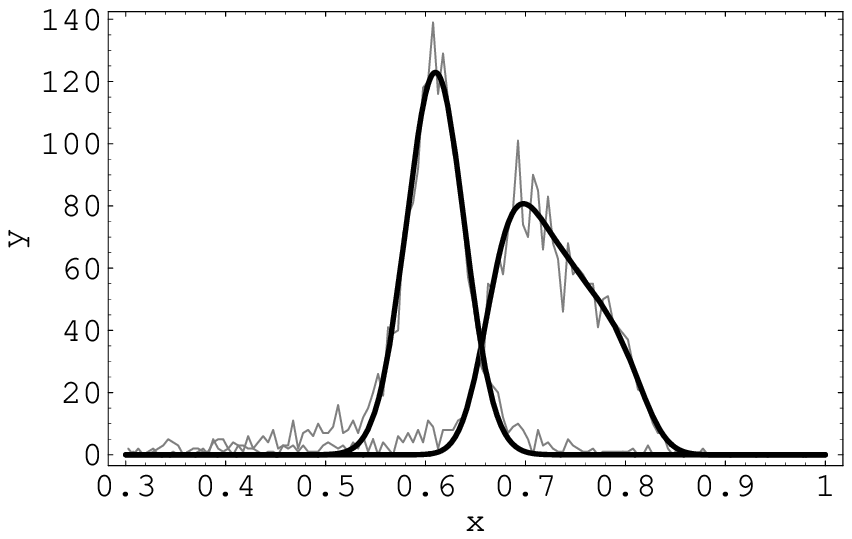}}
  \subfigure[][Reconstructed $x$-positions for the same beam positions. Measured data
  (light-gray lines) and best fits (black lines) are plotted together.]{\label{subfig:pos-ovlerlay-xproj}
    \psfrag{x}{\hspace*{-2em}{\small $x$ centroid}}
    \psfrag{y}{\hspace*{-1.5em}{\small counts}}
    \includegraphics[width=0.44\textwidth]{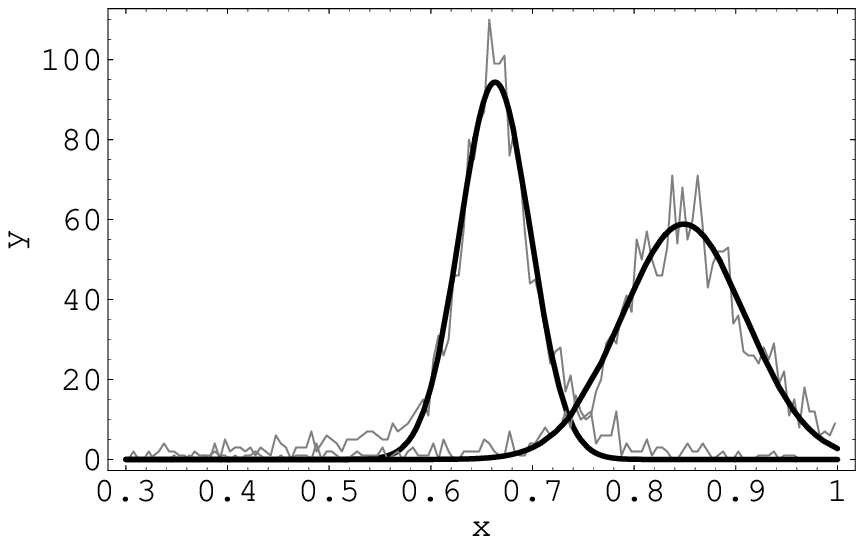}}
  \caption[Qualitative comparison of the quality of moments and reconstructed
    positions]{Comparison of the quality of moments and reconstructed
    positions for a pair of two representative beam positions.}
  \label{fig:mom-to-pos-comp}
\end{figure}

Figure~\ref{fig:mom-to-pos-comp} shows various graphical
representations of the centroids for the adjacent beam positions
$\mathrm{(x,y)}$ = $(14.25,9.5)\mathrm{\,mm}$ and $(x,y)=$
$(19,9.5)\mathrm{\,mm}$. 
A total of 192000 temporal coincidence
events have been registered at each beam position. The electronic
collimation was adjusted to filter out all events that are confined to
the central circular region of diameter $\diameter\approx12\mathrm{\,mm}$
at the coincidence detector corresponding to a circular region of diameter
$\diameter\lesssim0.2\mathrm{\,mm}$ at the test detector (refer to
section~\ref{sec:exp-setup}). Only
photopeak events have been considered. The moments have been computed
and were used for position reconstruction as described above with the
results plotted in
figures~\ref{subfig:mom-ovlerlay}-\ref{subfig:pos-ovlerlay-3D}.
When using the centroids as impact position estimate, it is hardly
possible to distinguish each position. The typical elongation of the
centroid along the $x$ direction can be clearly observed for the
distribution caused by the beam at position
$(x,y)$ = $(19,9.5)\mathrm{\,mm}$. The image is also
compressed giving maximum positions of approximately $(0.6,0.4)$ and $(0.7,0.4)$ instead of the true
beam positions $(0.68,0.45)$ and $(0.9,0.45)$\footnote{These positions
  have been scaled to the reference interval $[-1,1]$.}. The use of the
reconstructed positions improves the quality of the image in all
aspects. Both peaks are now well separated and can be easily
distinguished. The distribution of the penultimate beam
position is centered very close to the true impact position and the
Gaussian shape of the other distribution is restored. However, the
center position is slightly underestimated and the width is
considerably larger than the width for the penultimate position.

A quantitative evaluation of the quality of the algorithm can be
given in the following way. One-dimensional Gaussians are fitted to the measured $x$-position
distributions and the distribution of ${x}$-centroids that corresponds
to the beam position ${(0.68,0.45)}$. For the remaining distribution,
model~(\ref{eq:fit-model-dist}) is used instead, because it exhibits
strong CoG distortion. The results are shown in
figures~\ref{subfig:mom-ovlerlay-xproj}  and
\ref{subfig:pos-ovlerlay-xproj}. With the aid of the best fits one can
compute the fraction of mis-positioned events by determining the
intersection of both centroid distributions and both position
distributions and computing the area-fractions below the curves. In the
case of the centroids one obtains that a total of 5.3\% of the events
correspond to the overlapping area and cannot be unequivocally assigned to
one position or the other. In the case of the reconstructed positions,
this fraction drops to 2.5\%. For the ideal case when both
reconstructed positions have the same width as the inner one and are
perfectly centered at the nominal beam positions, one expects an
overlapping fraction smaller than 0.1\%. This shows that the method
improves the positioning but also that it is far from being perfect, and further
optimization and improvement must be carried out. Equal evaluations have been
made for other pairs of points and similar results have been obtained.

\begin{figure}[!tp]
  \centering
  \vspace*{-3.5eX}
  \subfigure[][Moments for the beam at ${(x,y)}$
  = $(19,9.5)\mathrm{\,mm}$.]{\label{subfig:pos-3D-2}
    \psfrag{x}{$\mu_{x_1,y_0}$}
    \psfrag{y}{\hspace*{0.7em}$\mu_{x_0,y_1}$}
    \psfrag{z}{\hspace*{-1.5em}\rotatebox{90}{\hspace*{-2.5em}$\mu_{x_2,y_0}+\mu_{x_0,y_2}$}}
    \includegraphics[width=0.34\textwidth]{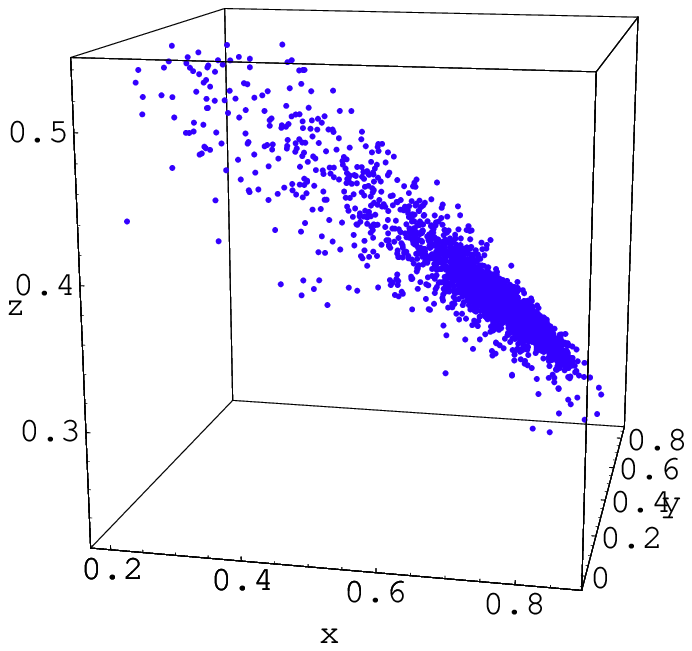}}\hspace*{3em}
  \subfigure[][Reconstructed positions for the beam at ${(x,y)}$
  = $(19,9.5)\mathrm{\,mm}$.]{\label{subfig:mom-3D-2}
    \psfrag{x}{$\bar{x}$}
    \psfrag{y}{\hspace*{0.7em}$\bar{y}$}
    \psfrag{z}{$\bar{z}$}
    \includegraphics[width=0.34\textwidth]{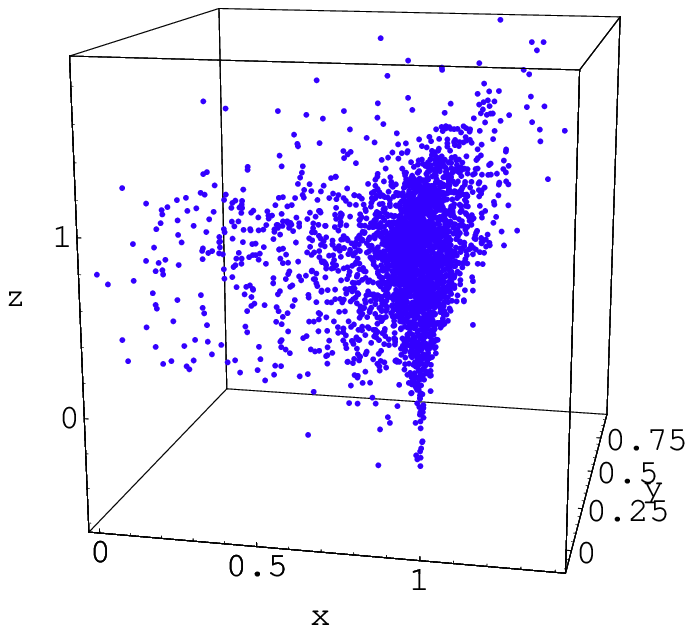}}\\
  \vspace*{-2.5eX}
  \subfigure[][Moments for the beam at ${(x,y)}$
  = $(14.25,9.5)\mathrm{\,mm}$.]{\label{subfig:pos-3D-3}
    \psfrag{x}{$\mu_{x_1,y_0}$}
    \psfrag{y}{\hspace*{0.7em}$\mu_{x_0,y_1}$}
    \psfrag{z}{\hspace*{-1.5em}\rotatebox{90}{\hspace*{-2.5em}$\mu_{x_2,y_0}+\mu_{x_0,y_2}$}}
    \includegraphics[width=0.34\textwidth]{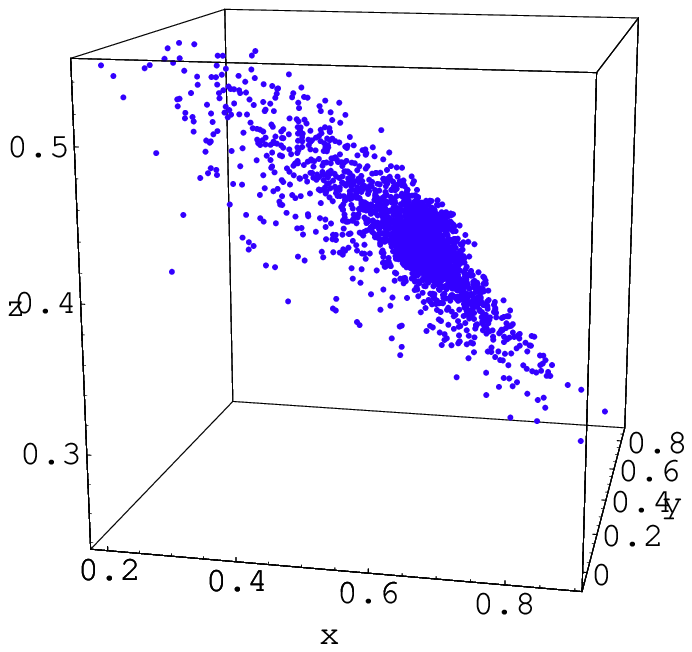}}\hspace*{3em}
  \subfigure[][Reconstructed positions for the beam at ${(x,y)}$
  = $(14.25,9.5)\mathrm{\,mm}$.]{\label{subfig:mom-3D-3}
    \psfrag{x}{$\bar{x}$}
    \psfrag{y}{\hspace*{0.7em}$\bar{y}$}
    \psfrag{z}{\raisebox{1ex}{$\bar{z}$}}
    \includegraphics[width=0.34\textwidth]{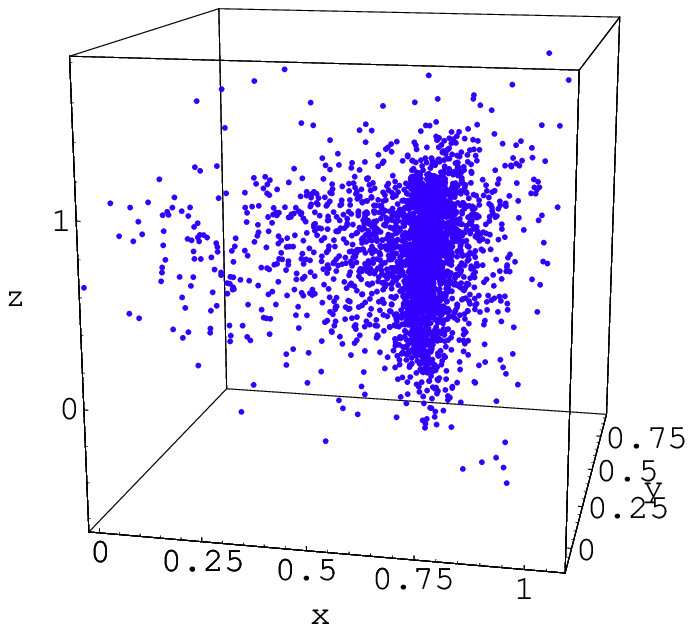}}\\
  \vspace*{-2.5eX}
  \subfigure[][Moments for both $\gamma$-ray beams at ${(x,y)}$
  = $(14.25,9.5)\mathrm{\,mm}$ and ${(x,y)}$
  = $(19,9.5)\mathrm{\,mm}$.]{\label{subfig:pos-3D-1}
    \psfrag{x}{$\mu_{x_1,y_0}$}
    \psfrag{y}{\hspace*{0.7em}$\mu_{x_0,y_1}$}
    \psfrag{z}{\hspace*{-1.5em}\rotatebox{90}{\hspace*{-2.5em}$\mu_{x_2,y_0}+\mu_{x_0,y_2}$}}
    \includegraphics[width=0.34\textwidth]{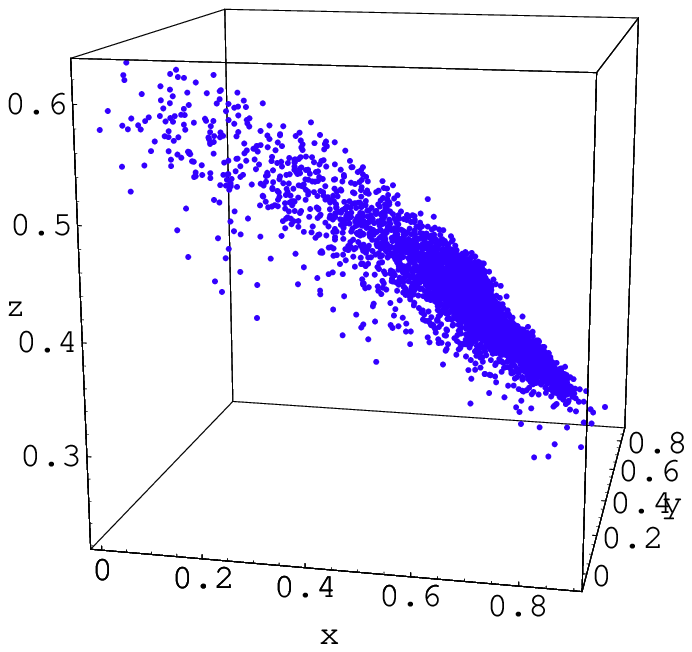}}\hspace*{3em}
  \subfigure[][Reconstructed positions for both beams at ${(x,y)}$
  = $(14.25,9.5)\mathrm{\,mm}$ and ${(x,y)}$
  = $(19,9.5)\mathrm{\,mm}$.]{\label{subfig:mom-3D-1}
    \psfrag{x}{$\bar{x}$}
    \psfrag{y}{\hspace*{0.7em}$\bar{y}$}
    \psfrag{z}{$\bar{z}$}
    \includegraphics[width=0.34\textwidth]{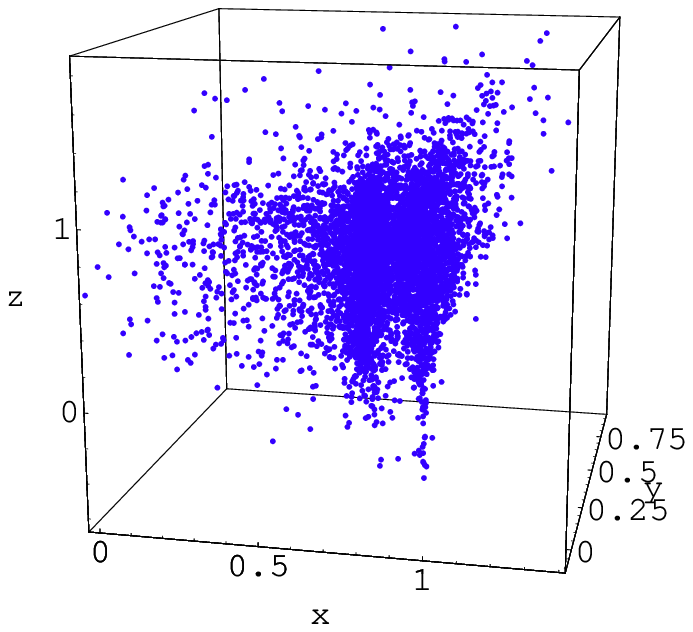}}
  \vspace*{-2eX}
  \caption[Three-dimensional representation of the selected coincidence
  events]{Three-dimensional representation of the selected coincidence events. The
    figures to the right hand side show the reconstructed positions
    for both beam positions separately and together. To the left,
    the moments of the corresponding events are shown.}
  \label{fig:3d-represent-of-pos-and-mom}
\end{figure}

Figures~\ref{subfig:mom-3D-1}-\ref{subfig:pos-3D-3} show the same
selection of data-points displayed as three-dimensional scatter plots
for the case of two $\gamma$-rays impinging on the crystal at
different positions. 
The right column shows the reconstructed impact positions of the events
and the left column the measured moments. The comparison of the six
cases reveals the deficiency of using the bare moments as
position estimate. In the case of a detector without distortion by the
CoG algorithm, the events would
appear as two thin vertical line-like distributions.
It can be seen that the distributions of the reconstructed
positions are indeed line-like and vertical. The thickness represents
just the spatial resolution for the different interaction distances.
However, the distributions of
the bare moments are neither vertical nor line-like and therefore
introduce a large systematic error. Note that the interaction depths
are rather well confined to the expected interval $[0,1]$. 

\subsection{3D Spatial Resolution}

\begin{figure}[!t]
  \centering
  \subfigure[][Measured resolutions in the ${x}$-direction of the
  reconstructed events at all 81
  positions.]{\label{subfig:xpos-errlistplot}
    \psfrag{x}{\hspace*{-2em}{\small position \#}}
    \psfrag{y}{\hspace*{-1.5em}{\small $\Delta x$ [mm]}}
    \includegraphics[width=0.52\textwidth]{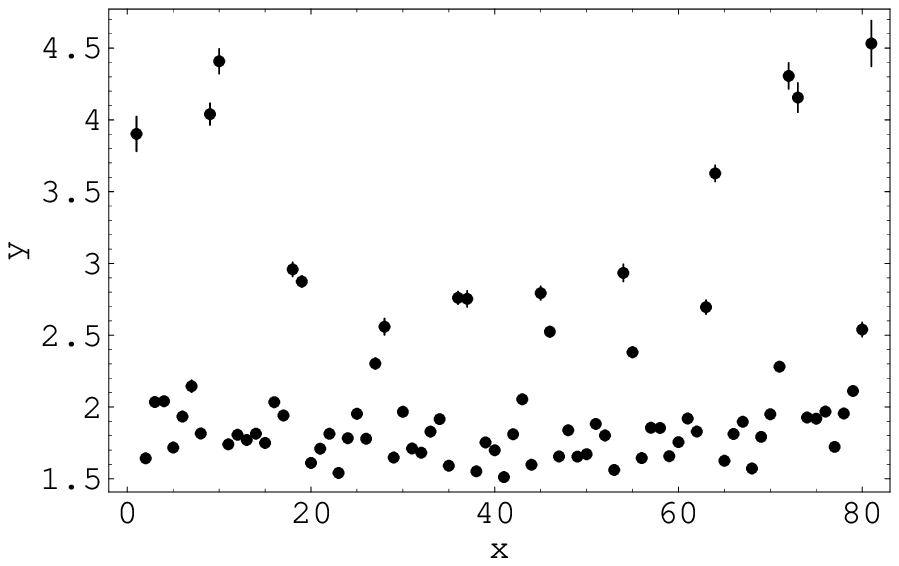}}\hspace*{1.2em}
  \subfigure[][2D-dependency of resolution in the $x$-direction of the
  reconstructed events.]{\label{subfig:xpos-listplot3d}
    \psfrag{x}{{\small $x$}}
    \psfrag{y}{{\small $y$}}
    \psfrag{z}{\hspace*{-0.9em}\rotatebox{90}{\hspace*{-1.5em}{\small $\Delta x$ [mm]}}}
    \includegraphics[width=0.40\textwidth]{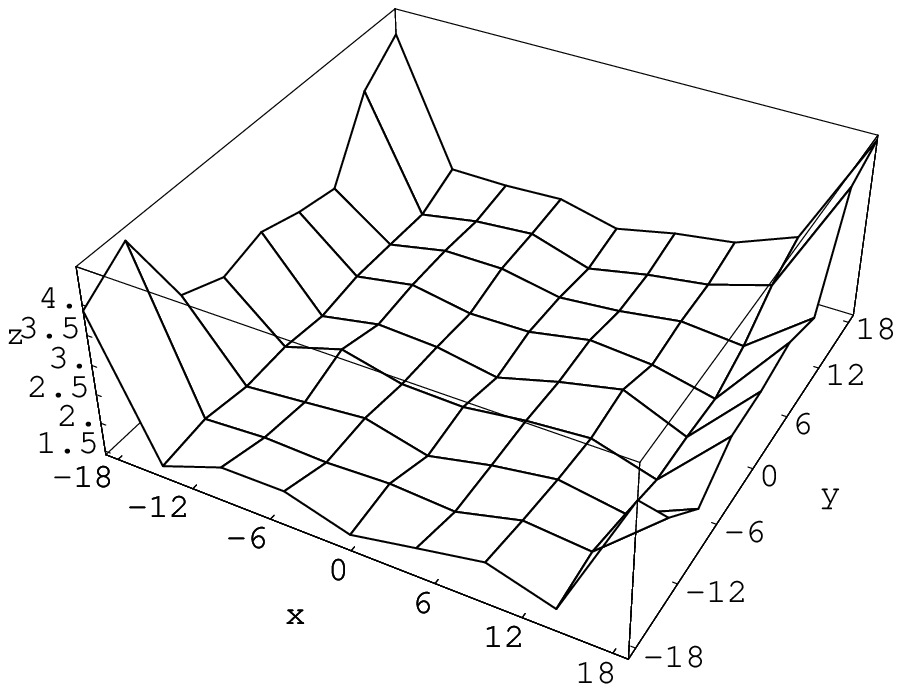}}
  \subfigure[][Measured resolutions in the $y$-direction of the
  reconstructed events at all 81
  positions.]{\label{subfig:ypos-errlistplot}
    \psfrag{x}{\hspace*{-2em}{\small position \#}}
    \psfrag{y}{\hspace*{-1.5em}{\small $\Delta y$ [mm]}}
    \includegraphics[width=0.52\textwidth]{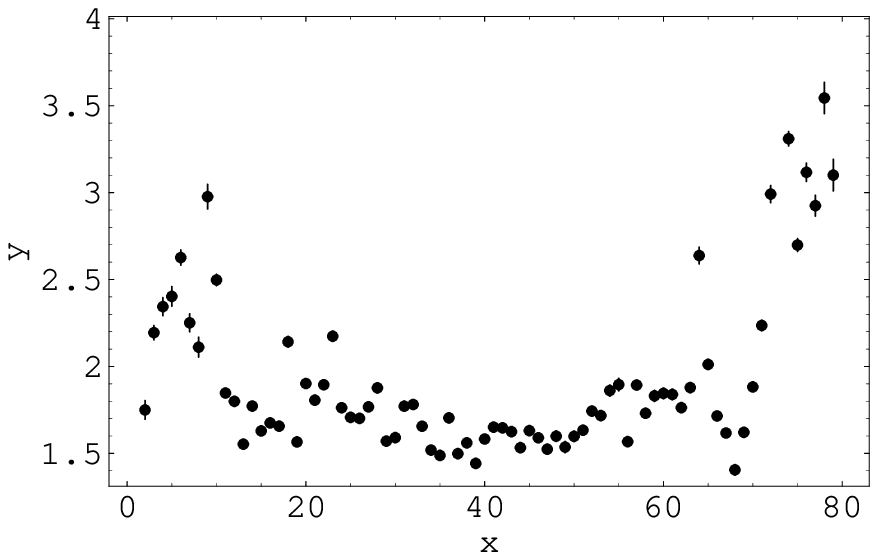}}\hspace*{1.2em}
  \subfigure[][2D-dependency of resolutions in the $y$-direction of
  the reconstructed events.]{\label{subfig:ypos-listplot3d}
    \psfrag{x}{{\small $x$}}
    \psfrag{y}{{\small $y$}}
    \psfrag{z}{\hspace*{-0.9em}\rotatebox{90}{\hspace*{-1.5em}{\small $\Delta y$ [mm]}}}
    \includegraphics[width=0.40\textwidth]{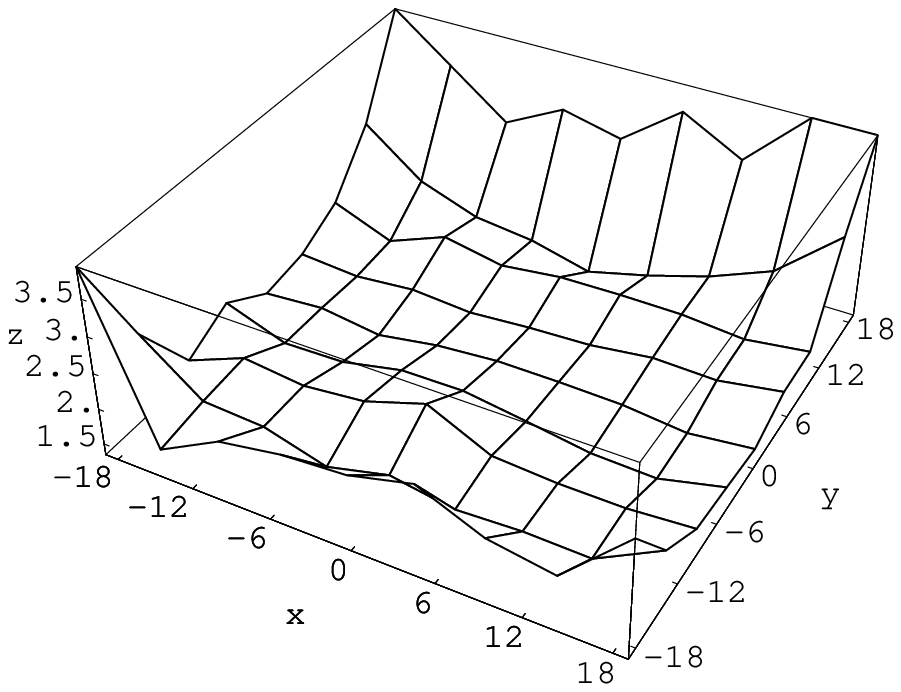}}
  \caption[Measured resolutions for the reconstructed $x$- and
  $y$-positions]%
  {Measured resolutions for the reconstructed $x$- and
  $y$-positions. To the left: the resolutions are displayed with their error bars. To
    the right:  3D-plots are shown for better recognition of the
    functional dependence on the transverse coordinates.}
  \label{fig:recons-pos}
\end{figure}

\begin{figure}[!tp]
  \centering
  \subfigure[][Measured resolutions in the $z$-direction of the reconstructed events at all
  81 positions.]{\label{subfig:z-errlistplot}
    \psfrag{x}{\hspace*{-2em}{\small position \#}}
    \psfrag{y}{\hspace*{-1.5em}{\small $\Delta z$ [mm]}}
    \includegraphics[width=0.5\textwidth]{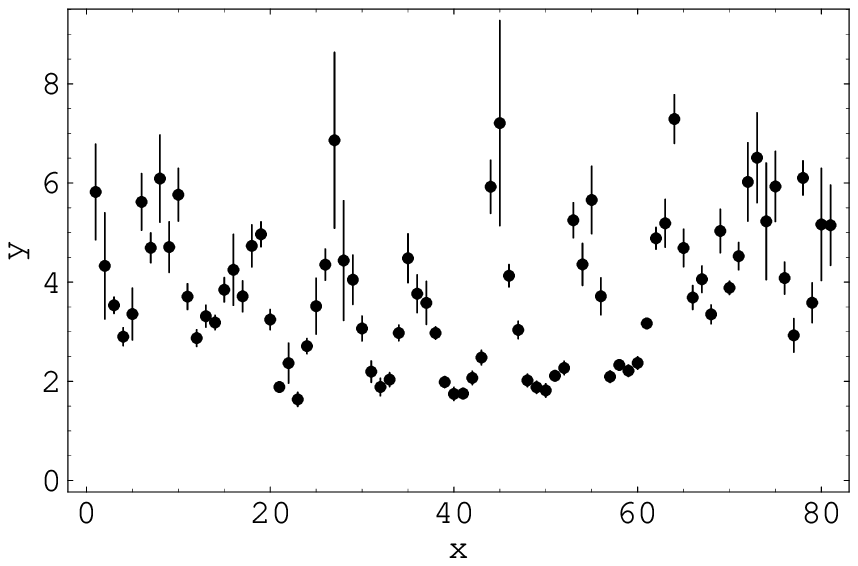}}\hspace*{1.2em}
  \subfigure[][2D-dependency of resolutions in the $z$-direction of the
  reconstructed events.]{\label{subfig:z-listplot3d}
    \psfrag{x}{{\small $x$}}
    \psfrag{y}{{\small $y$}}
    \psfrag{z}{\hspace*{-0.9em}\rotatebox{90}{\hspace*{-1.5em}{\small $\Delta z$ [mm]}}}
    \includegraphics[width=0.38\textwidth]{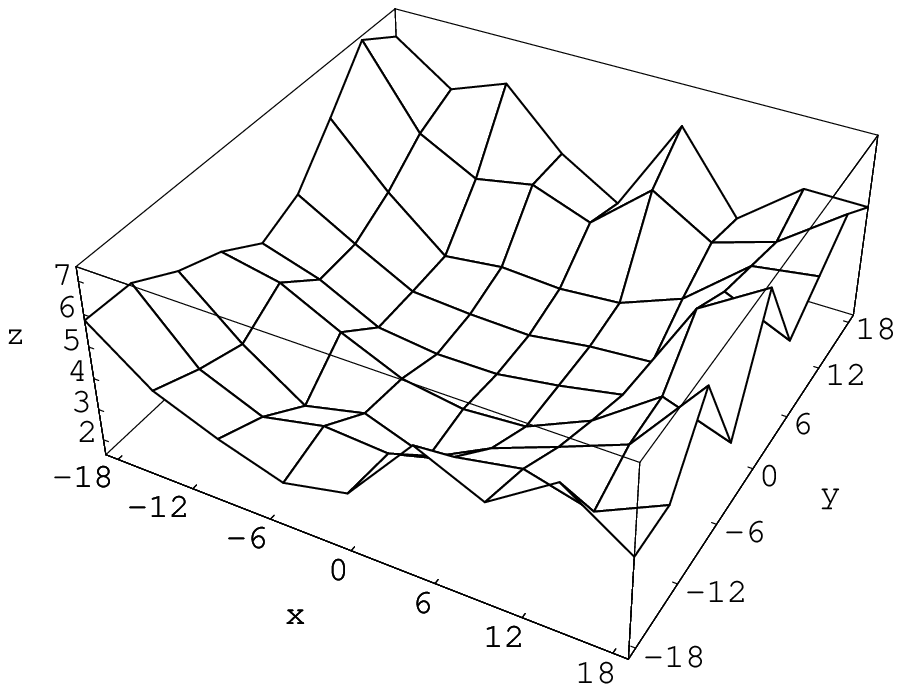}}\\
  \vspace*{2eX}
  \subfigure[][Measured lower limits for the interaction depth at all
  81 positions.]{\label{subfig:a-errlistplot}
    \psfrag{x}{\hspace*{-2em}{\small position \#}}
    \psfrag{y}{\hspace*{-1.0em}{\small $z$ [mm]}}
    \includegraphics[width=0.5\textwidth]{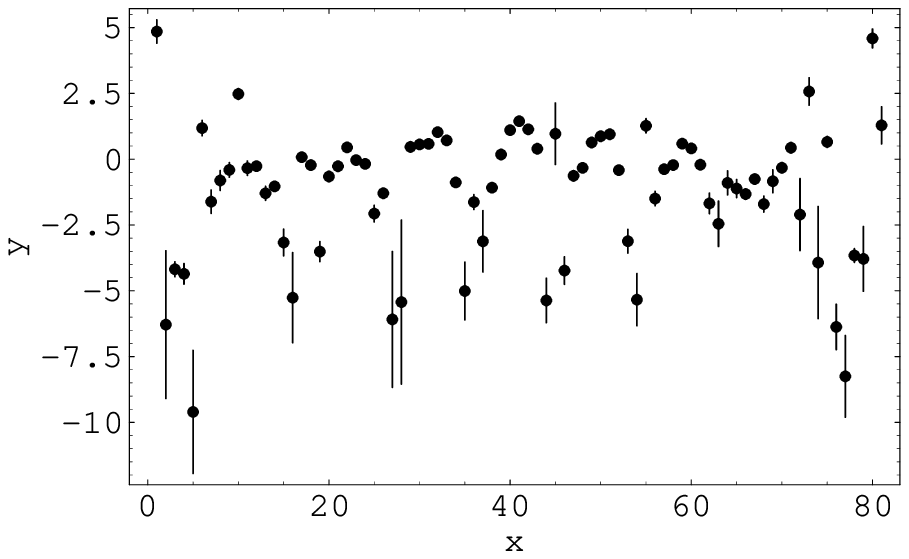}}\hspace*{1.2em}
  \subfigure[][2D-dependency of the measured lower limits of the
  reconstructed events.]{\label{subfig:a-listplot3d}
    \psfrag{x}{{\small $x$}}
    \psfrag{y}{{\small $y$}}
    \psfrag{z}{\hspace*{-0.9em}\rotatebox{90}{\hspace*{-1.5em}{\small $z$ [mm]}}}
    \includegraphics[width=0.38\textwidth]{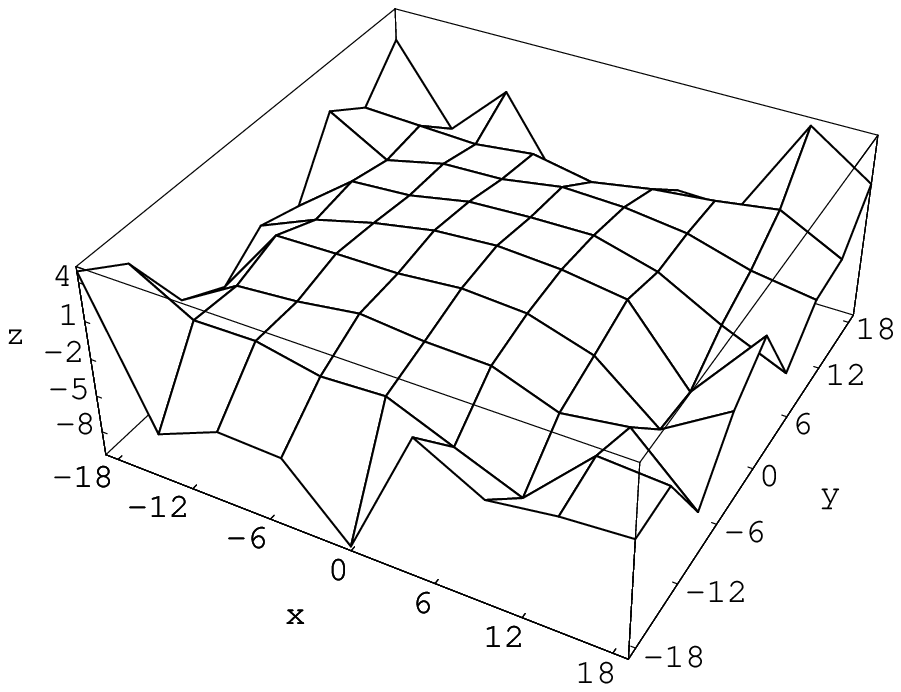}}\\
  \vspace*{2eX}
  \subfigure[][Measured upper limits for the interaction depth at all
  81 positions.]{\label{subfig:b-errlistplot}
    \psfrag{x}{\hspace*{-2em}{\small position \#}}
    \psfrag{y}{\hspace*{-1.0em}{\small $z$ [mm]}}
    \includegraphics[width=0.5\textwidth]{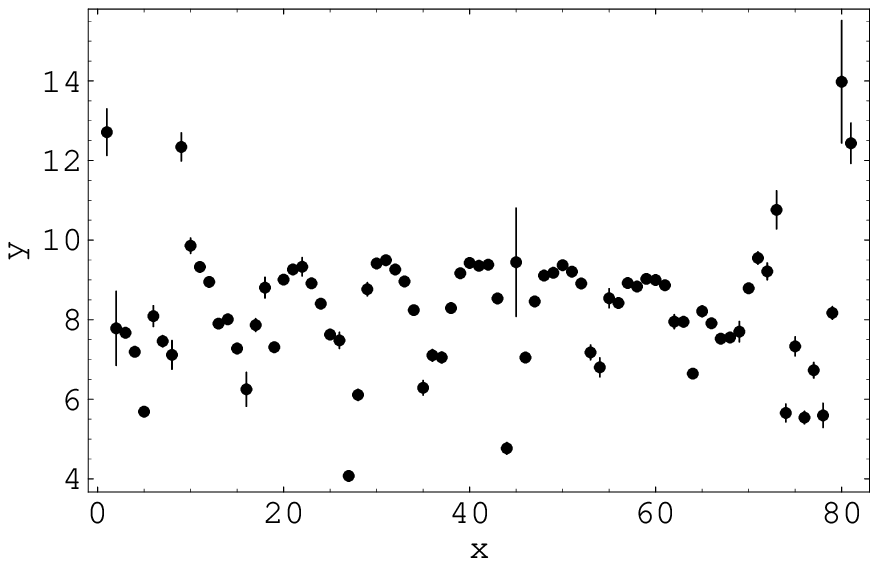}}\hspace*{1.2em}
  \subfigure[][2D-dependency of the measured upper limits of the
  reconstructed events.]{\label{subfig:b-listplot3d}
    \psfrag{x}{{\small $x$}}
    \psfrag{y}{{\small $y$}}
    \psfrag{z}{\hspace*{-0.9em}\rotatebox{90}{\hspace*{-1.5em}{\small $z$ [mm]}}}
    \includegraphics[width=0.38\textwidth]{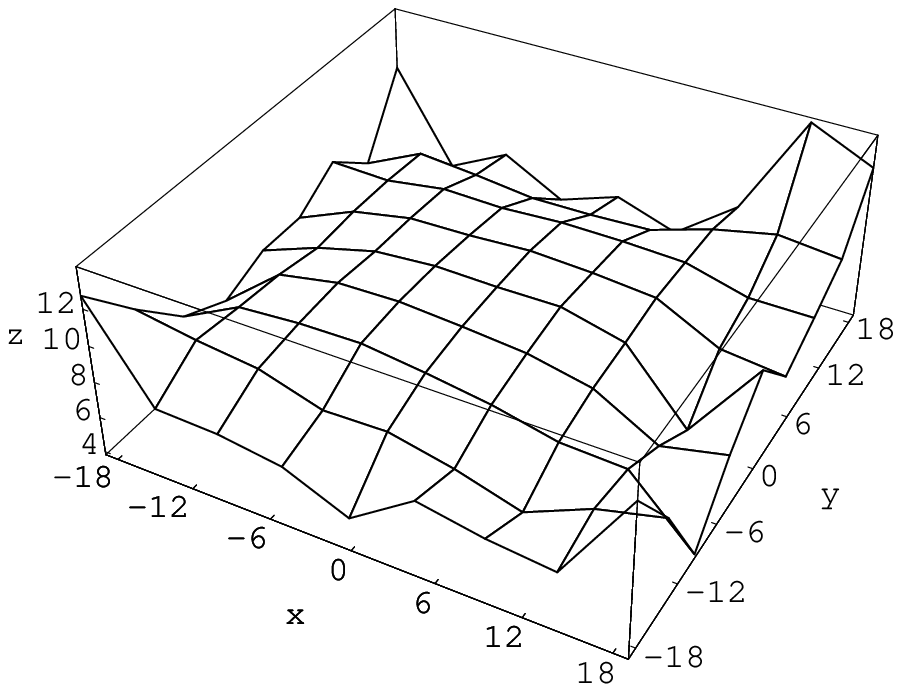}}
  \caption[Measured resolution for the reconstructed $z$-positions]{Measured resolutions for the reconstructed $z$-positions and
  maximum and minimum measured interaction depths. To the left: the values are displayed with error bars. To
    the right: 2D-plots are shown for better recognition of the
    functional dependence on the transverse coordinates.}
  \label{fig:recons-z}
\end{figure}

The intrinsic spatial resolution of the test detector after impact
position reconstruction is measured in the same way as it was done in
section~\ref{sec:moments-as-pos-estimate}. Since the Gaussian shape of
the detector point spread function along the transverse directions is
recovered by the inversion algorithm to a large extent, simple Gaussian distributions with
linear backgrounds are used instead of the fit model derived in
section~\ref{sec:model_dist_for_event_stat}. The spatial resolutions are
then given by
\begin{equation}
  \label{eq:spatial-res-gauss-case}
  \Delta
  x=\sqrt{(2.35\sigma^X_\tincaps{Gauss})^2-d_\mathit{eff}^2}\quad\mbox{and}\quad  \Delta y=\sqrt{(2.35\sigma^Y_\tincaps{Gauss})^2-d_\mathit{eff}^2},
\end{equation}
where $\sigma^X_\tincaps{Gauss}$ and $\sigma^Y_\tincaps{Gauss}$ are
the widths of the Gaussian distribution for the $x$- and
$y$-coordinate respectively that best fits to the measured data, and
$d_\mathit{eff}$ is the effective
source diameter (\ref{eq:effective-source-diameter}) discussed in
section~\ref{sec:spat-extend-source}. Finally, a residual linearity
correction has to be applied. This is necessary because the
reconstruction algorithm in its actual form fails to completely
recover the linear behavior and thus, the spatial resolution would be
underestimated.
Results for the spatial resolutions of both
transverse directions are plotted in
figures~\ref{subfig:xpos-errlistplot}-\ref{subfig:ypos-listplot3d}.
Clear improvements compared to the spatial resolutions in the case of
using the bare moments as position estimate (refer to
section~\ref{sec:moments-as-pos-estimate}) can be observed.

The resolution of the depth of interaction is obtained in the same way
as for the resolution of the
composite second moment in section~\ref{sec:moments-as-pos-estimate}.
The results are displayed in figures~\ref{subfig:z-errlistplot} and
\ref{subfig:z-errlistplot}. For this impact parameter the best results
were obtained with interpolation orders of 13 for the transverse
components and 2 for the normal component. The reason is not
understood and will be subject of  further
investigation. Only a slight
improvement compared to using the standard deviation is achieved in
the spatial resolution of the $z$ position.  An important advantage is,
however, the more accurate mapping of the reconstructed $z$ to the
correct interval of values ($[0,1]$ or $[0,10]$ after scaling).
This is shown in figures~\ref{subfig:a-errlistplot}-\ref{subfig:b-listplot3d}.
Mean, standard deviation, minimum and maximum values for the achieved
resolutions of all three impact coordinates are summarized in
table~\ref{tab:resolution-3Dpos}, where the results for  the spatial resolution obtained in
section~\ref{sec:moments-as-pos-estimate} are also shown for better 
comparison. The improvement of the detector performance is
unfortunately not sufficient for the correction of the energy.
Two different methods for energy correction have been tested.
The first included the energy as a fourth dimension in the matrices of the detector response $\mathbf{Y}$
and the test-position $\mathbf{X}$ for polynomial interpolation. This
method failed completely and rendered the position information
unusable. The second method consists in computing multiplicative correction factors. For
this, the energy prediction obtained from the model in
chapter~\ref{ch:experiment} was interpolated at the three-dimensional grid of
$n_\tincaps{TP}$ test-points. Since the energy
is proportional to the amount of released scintillation light $J_0$ in
expression~(\ref{eq:total-lightdist}), the correction factor is
obtained as the inverse of the predicted energy multiplied by the 
detected zeroth moment. This method does not lead to any improvement even
though the theoretical model for the signal distribution predicts a
variation of the energy with the impact position (see also Gagnon et
al.\ \cite{Gagnon:1993}). 
A possible reason is that the accuracy of the position reconstruction
is still low at the outer regions where exact position information
is most required for the energy reconstruction. Also, the strong
variation of the anode sensitivity may make higher interpolation orders
necessary. With the personal computer used, orders higher than 12 for the
transverse components and higher than 5 for the normal component were not
possible to use. 

\begin{table}[!t]
  \centering\renewcommand{\arraystretch}{1.4}
  \begin{tabular}{cccccc}\hline\hline
    Coordinate&Mean&StdDev&Min&Max&Unit\\\hline
    $x$&1.9&0.9&1.2&4.8&mm\\
    $y$&1.9&1&1.1&6.1&mm\\
    $z$&3.9&1.5&1.6&7.3&mm\\\hline
    $\mu_{x_1,y_0}$&3.4&3.2&1.4&20.9&mm\\
    $\mu_{x_0,y_1}$&3.3&3.1&1.3&19.9&mm\\
$\sigma_\tincaps{DOI}$&4.9&1.8&1.9&9.0&mm\\\hline\hline
  \end{tabular}
  \caption[Mean, std.\ dev., max.\ and min.\ values for
  resolutions of the 3D positions]{Mean, standard deviation, maximum
    and minimum values for resolutions of the reconstructed 3D
    positions. For comparison, the results from
    section~\ref{sec:moments-as-pos-estimate} have been included in
    the same table.}
  \label{tab:resolution-3Dpos}
\end{table}

\subsection{Linearity of the Positioning Scheme}
\label{sec:lin-pos-scheme}

\begin{figure}[!t]
  \centering
  \subfigure[][Reconstructed $x$-position as a function of the true
  impact position.]{\label{subfig:xpos-linearityplot}
    \psfrag{x}{$x$}
    \psfrag{y}{$y$}
    \psfrag{z}{\hspace*{-0.8em}\rotatebox{90}{\hspace*{-2em}{\small $x$-position}}}
    \includegraphics[width=0.45\textwidth]{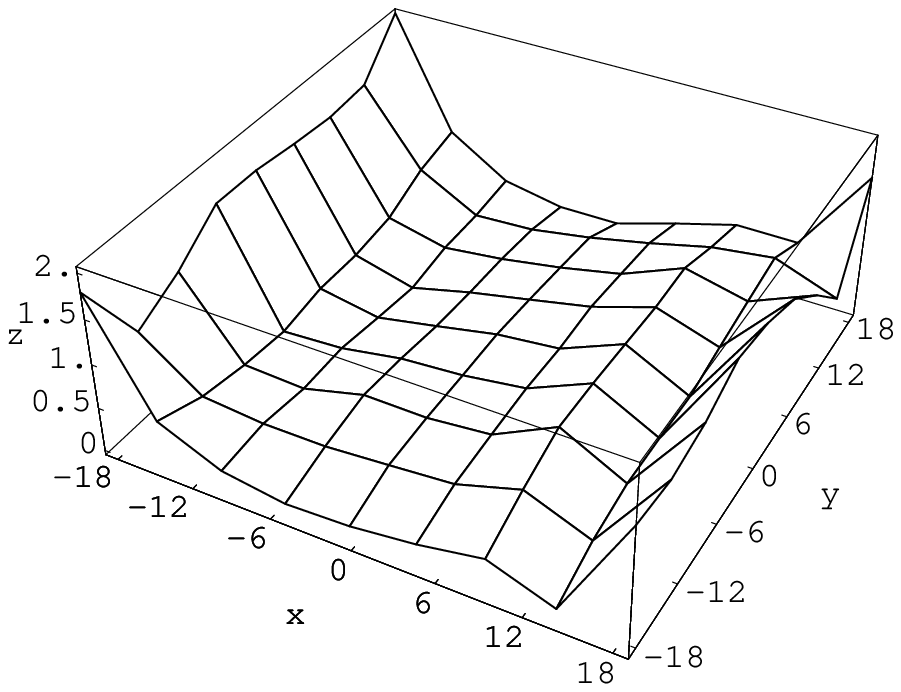}}
  \subfigure[][$x$-centroid as a function of the true
  impact position.]{\label{subfig:xcent-linearityplot}
    \psfrag{x}{$x$}
    \psfrag{y}{$y$}
    \psfrag{z}{\hspace*{-0.0em}\rotatebox{90}{\hspace*{-2em}{\small $x$-centroid}}}
    \includegraphics[width=0.45\textwidth]{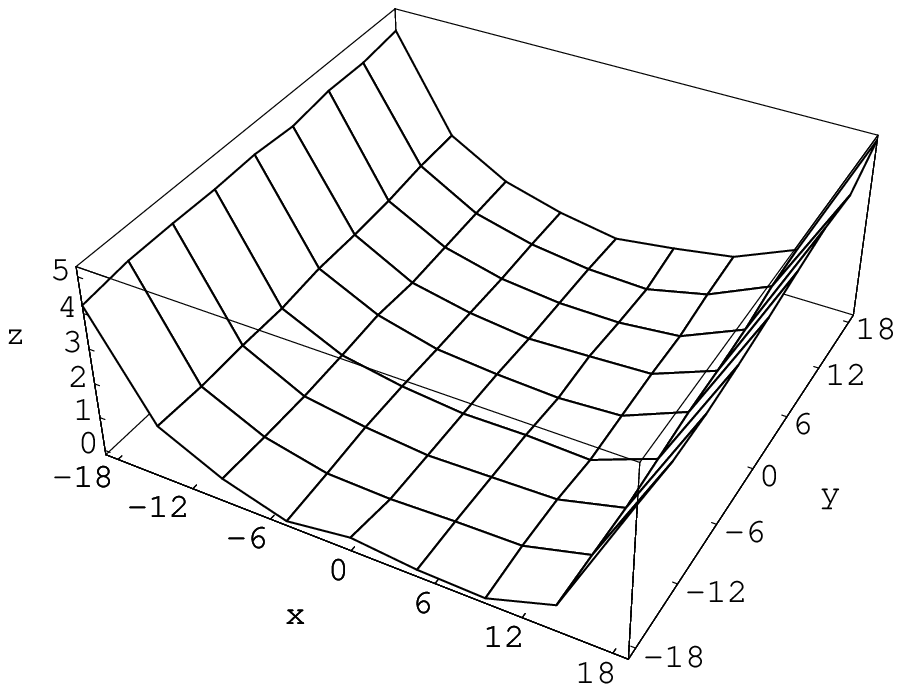}}\\
  \subfigure[][Reconstructed $y$-position as a function of the true
  impact position.]{\label{subfig:ypos-linearityplot}
    \psfrag{x}{$x$}
    \psfrag{y}{$y$}
    \psfrag{z}{\hspace*{-0.8em}\rotatebox{90}{\hspace*{-2em}{\small $x$-position}}}
    \includegraphics[width=0.45\textwidth]{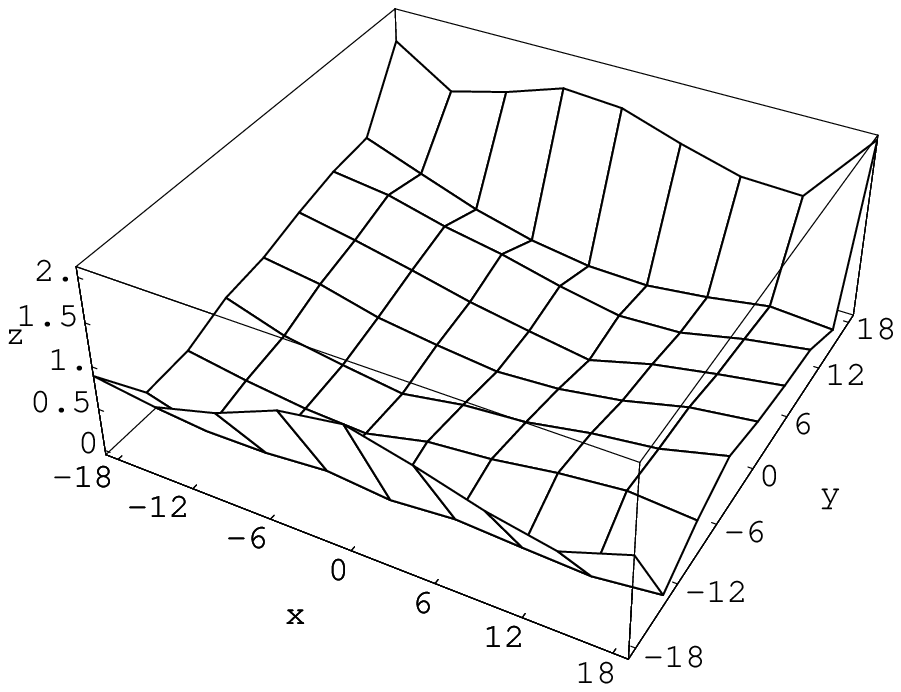}}
  \subfigure[][$y$-centroid as a function of the true
  impact position.]{\label{subfig:ycent-linearityplot}
    \psfrag{x}{$x$}
    \psfrag{y}{$y$}
    \psfrag{z}{\hspace*{-0.0em}\rotatebox{90}{\hspace*{-2em}{\small $x$-centroid}}}
    \includegraphics[width=0.45\textwidth]{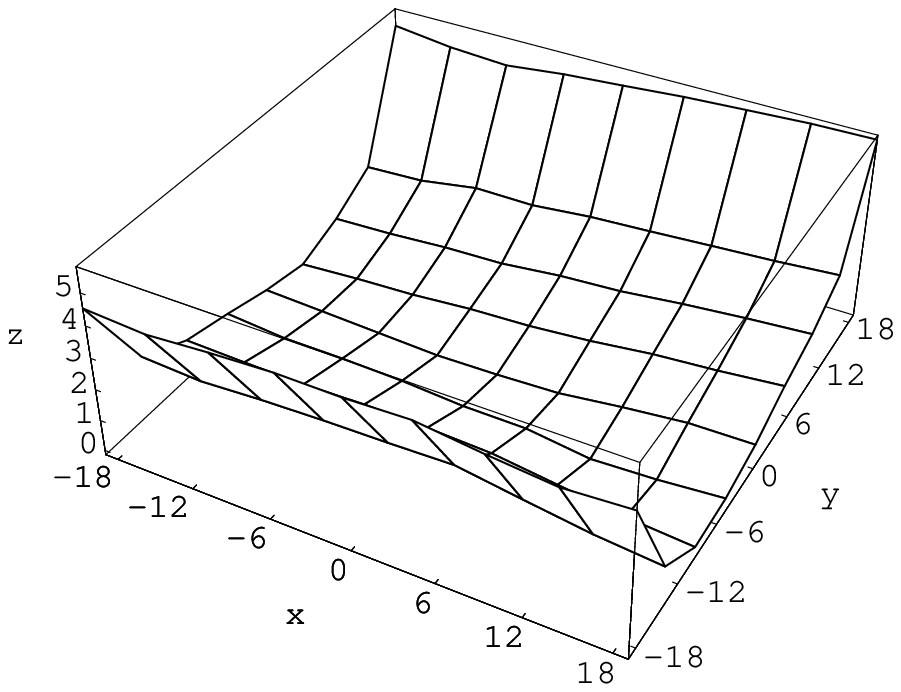}}\\
  \caption[Linearity behavior of the reconstructed positions
    and the centroids]{Linearity behavior of both the reconstructed positions
    (l.h.s.) and the centroids (r.h.s.). Note that the scales for the
    function values are not the same.}
  \label{fig:linearity-compare}
\end{figure}

\begin{table}[!ht]
  \centering\renewcommand{\arraystretch}{1.4}
  \begin{tabular}{cccccc}\hline\hline
    Coordinate&Mean&StdDev&Min&Max&Unit\\\hline
    $x$&0.7&0.5&0.2&2.0&mm\\
    $y$&0.5&0.6&<0.1&2.1&mm\\\hline\hline
  \end{tabular}
  \caption[Mean, std.\ dev., max.\ and min.\ values of the
  linearity error for the 3D positions]{Mean, standard deviation, maximum
    and minimum values of the absolute linearity error for the 3D positions} 
  \label{tab:linearity-3Dpos}
\end{table}

The linearity of the position response has
been tested for the reconstructed impact position. As mentioned above,
the method fails to completely restore the three-dimensional impact
position and a residual non-linearity of the position mapping
remains. This is shown in
figures~\ref{subfig:xpos-linearityplot}-\ref{subfig:ycent-linearityplot}
together with the linearity behavior of the centroids. Once again, a
clear improvement is observed. Only the last row (or column) 
of the beam positions shows a significant deviation from the linear
behavior of the reconstructed impact positions.
In the case of using the centroids as position estimates, the non-linearity
is much more pronounced.   
Mean, standard deviation, minimum and maximum values of the absolute
systematic positioning errors are summarized in
table~\ref{tab:linearity-3Dpos}. Note that the maximum values
for the positioning error correspond without exception to the beam
positions ${(x,y)}$ = $(\pm19,y)\mathrm{\,mm}$ and ${(x,y)}$
= $(x,\pm19)\mathrm{\,mm}$. Consequently, the
maximum relative positioning error is of about 10\%, which was just the
achieved accuracy for the signal distribution model. A further
improvement in the position correction probably requires a higher
accuracy of the model signal distribution. Note that the maximum
linearity error in the case of using the centroids is of the order of
25\%.

\subsection{Execution Time}
\label{sec:exec-time}

The inversion algorithm was implemented on a standard single CPU personal
computer with a 2.4GHz Pentium IV CPU. Execution times of $\mathrm{\sim 20\,\mu s}$
per event were achieved. This corresponds to processing rates of about
$\mathrm{50\cdot10^3}$ events per second thus being a factor of 10
slower than the data aquisition card (Zawarzin \cite{Zavarzin:1999}).

\chapterbib


 \cleardoublepage{}
\chapter{Conclusions \& Outlook}
\label{ch:conclusiones-and-outlook}

\chapterquote{%
  \hspace*{0.23\linewidth}\begin{minipage}{0.54\linewidth}
    \begin{flushleft}
      Still round the corner there may wait\newline
      A new road or a secret gate;\newline
      And though I oft have passed them by,\newline
      A day will come at last when I\newline
      Shall take the hidden paths that run\newline
      West of the Moon, East of the Sun.      
    \end{flushleft}
  \end{minipage}\hfill

  \vspace*{1eX}
}{John Ronald Reuel Tolkien, $\star$ 1892 -- $\dagger$ 1973}

\PARstart{I}{n} all nuclear imaging modalities, \g-photons in the energy
range between $\mathrm{15keV}$ and $\mathrm{511keV}$ must be
detected. Simultaneously, measurements of their energies and 
interaction positions have to be performed. From these measured values, 
the distribution of a function-specific radiopharmaceutical in the
organism can be
reconstructed and this allows one to make conclusions about the explored
metabolic function. High accuracy of this computed distribution
is equivalent to high quality of the imaging process and can be
quantified by means of the spatial resolution, the SNR and the image contrast. High spatial resolution, high SNR and 
high contrast require \g-ray detectors that provide full
three-dimensional information about the position of the 
photoconversion inside the crystal, good energy resolution and high efficiency.
It was repeatedly stressed that, above all, the requirement for high
intrinsic efficiency and high spatial resolution constitute an
important design conflict for conventional Anger-type \g-ray imaging
detectors. These detectors cannot provide depth of interaction
information without modifications and approximate the transverse
components of the impact position by centroids, {\em i.e.}\ the normalized
first moments of the scintillation light distribution. 
However, for the PET modality in particular, thick scintillation
crystals are required for effectively
stopping the incident $\mathrm{511\,keV}$ $\gamma$-ray.
It is well known that the center of gravity algorithm results in a
non-linear and depth-dependent position mapping and leads to a
significantly lowered spatial resolution near the edges and corners of
the scintillation crystal, especially for cases when the thickness of
the crystal is no longer significantly smaller than its transverse
extension.

A detailed study in section~\ref{ch:errors-of-cog-and-cdr}
revealed that this is a consequence of the breaking of the
signal distribution's symmetry caused by the finite crystal dimension.
Therefore, it is not a deficiency of the charge dividing circuits but
an indication that one of the conditions for using the center of gravity 
algorithm is not fulfilled. This condition is clearly the geometric
shape of the scintillation crystal. 
In other words: although the centroids can be measured with high
precision, they do not represent suitable impact position
estimates for \g-ray imaging detectors. Actually, it was shown in
chapter~\ref{ch:experiment} that the moments can be measured with
rather high accuracy. The detailed study of charge dividing circuits
also revealed that the input impedance of these circuits together with
the charge signals from photodetectors produce voltages that are
inherently quadratically position encoded. The connection of a
standard analog adder presented in
chapter~\ref{ch:enhanced-charge-dividing-circuits} allows the
addition of these voltages and provide by this means an additional
signal, which depends linearly on the second moment of the signal
distribution. 

Previous studies by other researchers showed that the width of the
signal distribution caused by photoconversion of the impinging
\g-ray and the free propagation of the liberated scintillation light
is strongly correlated with the interaction depth of the \g-ray
(Rogers {\em et al.}\ \cite{Rogers:1986}, Matthews {\em et al.}\
\cite{Kenneth:2001}, Antich {\em et al.}\ \cite{Antich:2002}).
As a consequence, the successful and precise measurement of the
second moment provides good depth of interaction information. 

In
chapter~\ref{ch:enhanced-charge-dividing-circuits}, it was demonstrated
that all known unidimensional and bidimensional charge dividing
circuits can be easily modified for
computing the second moment analogically. 
Approximated but explicit expressions were given for the dependence of
the quadratic voltage encoding on the position of the divider
nodes. Comparison with simulations were carried out with the {\sc
  Spice} program and showed a
a very good agreement with low systematic deviations of the order of
$\mathrm{3\%}$. Further results concern the symmetry of the three
configurations. The true Anger logic is inherently symmetric with respect to
interchanging the $x$- and the $y$-coordinate. The configurations based on
proportional resistor chains and the hybrid network are not. Adequate
dimensioning of the resistor values can, however, completely restore this
symmetry for the centroid measurements in both cases. In the case of
the second moment, only for the hybrid configuration  can an appropriate choice of
resistor values be found in order to make its response in all four
moments symmetric. For the two-dimensional resistor chain network,
the symmetric behavior of the second moment can, however, be optimized and
residual asymmetries $\mathrm{<1\%}$ can be achieved. This choice of
optimum values for the resistances leads also to a functional
behavior slightly different from an exact parabola. Terms of
$\mathcal{O}(x^4)$, $\mathcal{O}(y^4)$ and $\mathcal{O}(x^2y^2)$ are
introduced. However, no difficulty is expected from this result if the
polynomial interpolation of chapter~\ref{ch:position-reconstruction}
is used, because the interpolated points must not fulfill special
requisites apart from building a Cartesian grid on the volume
$\mathrm{]-1,1[^2\times]0,1[}$. It may, however, result in additional
errors if the standard deviation is used for estimating the depth of
interaction. 
In addition, the values of the resistors for the summation amplifier have to be chosen in such a way that a
tradeoff between electronic noise and accuracy of the centroid is achieved. As
a design criterion, the following requirement was established: at any node
less than $\mathrm{1\%}$ of the total current should be extracted for
the computation of the second moment. Higher accuracy does
not make sense because the tolerance of
commercial resistors is of order $\mathrm{1\%}$. 
It was found that the DOI enhanced true Anger logic is the least practical
configuration due to the large number of required resistors.
The two remaining configurations have very similar analytic properties when compared
to each other. Their real
performance still has to be tested by experiment in further
investigations. For the present work, the two-dimensional configuration of
the proportional resistor chain based charge divider has been chosen to
set up a DOI enhanced small animal PET detector module because it
is very easy to implement.

In chapter~\ref{ch:experiment}, measurements of all four lower moments
for a real \g-ray imaging detector have been presented. The detector
is based on a $\mathrm{42\times42\times10\,mm^3}$ LSO parallelepiped
together with the large area PSPMT H8500 from Hamamatsu Photonics
Inc. and is suitable for small animal PET. The design is modular and a
full ring made from 8 such modules would have a central aperture of
$\mathrm{\approx10\,cm}$. The experimental verification demonstrated that the
measurement of the centroids is not affected at all by the
simultaneous measurement of the compound second moment. The mean
resolution is  in all non-trivial moments
rather high ($\mathrm{\lesssim5\%}$). However, the direct use of these
moments as an estimate for
the three-dimensional photoconversion position leads to a very low
spatial resolution in all components for \g-ray impact positions near
the edges and corners of the scintillation crystal. This is no
surprise as it was reported by many research groups and by virtue of
the arguments given above.

The trivial moment represents the energy of the detected \g-ray. This
moment is affected by additional effects and requirements and its
achieved resolution falls behind those of the non-trivial
moments. One of these effects is the
anode inhomogeneity of the position sensitive photomultiplier
tube used. Since the difference in efficiency and gain from one segment to
another can achieve ratios up to 1:3, a strong position dependence of
the zero order moment has to be expected and is also observed. A
passive compensation as proposed by Tornai {\em et al.}\ \cite{Tornai:1996}
is neither possible for the proportional resistor chain based charge
divider nor for the hybrid configuration. For the same reasons mentioned in
conjunction with the active impedance converters, an active
compensation is only possible if a dedicated ASIC is developed. This
possibility is the subject of further development. The second reason for
lower quality of this moment is the intrinsic energy resolution of the
scintillator used. The finite number of scintillation photons together with
the low quantum efficiency of the photomultiplier leads to an energy
resolution of about $\mathrm{14\%}$ for LSO and $\mathrm{511keV}$
photons. This resolution is further lowered because the second moment
measurement does not tolerate diffuse or specular reflections at any
inner crystal surface. Reflections would lead to superposed light
distributions with widths that differ from those of the
distribution of direct scintillation light and thus destroy the
correlation between DOI and distribution width. 
Therefore, all surfaces except for the one that
is coupled to the PSPMT have been covered with highly absorptive black
epoxy resin coatings. This clearly lowers the light collection
efficiency and leads to poorer energy resolution especially at the
regions near the crystal borders. A mean energy resolution of
$\mathrm{25\%}$ together with the best value of $\mathrm{17\%}$ at the
center and the poorest value of $\mathrm{70\%}$ at one corner have been observed. The very poor
energy resolution at the crystal corners is due to a strong dependence
of the light yield on the DOI. For events that undergo photoelectric effect
near the photocathode, the total amount of detected light is much
larger than for events that occur far from the photocathode. Therefore,
the photopeak is shifted towards lower values of the energy spectrum
when the photoconversion position gets closer to the absorbing surfaces.
As a result, one detects a lower $\gamma$-ray energy although the true
energy of the photon remains the same.
However, it is possible to correct this shift if a high quality
measurement of the true impact position is achieved. A first attempt
was made with the reconstructed positions obtained in
chapter~\ref{ch:position-reconstruction}, but no improvement could be
achieved. This is probably due to the still low spatial resolution.
Apart from this variation in the total amount of detected
scintillation light, there is, of course, also a higher statistical
error. Probably, a large amount of the light lost by absorption can be
recovered with the aid of a micro-machined retroreflector. The
perspectives of using such micro-optical components have been
investigated several times by different research groups (Karp and Muehllehner \cite{Karp:1985},
Rogers {\em et al.}\ \cite{Rogers:1986} and McElroy {\em et al.}\
\cite{McElroy:2002}), but emphasis was put mainly on position
resolution. 
Retroreflectors reflect the incoming light back onto a direction that is
parallel to that one of the incident light. Therefore, in the ideal case, the light
distribution is expected to be exactly the same as for an absorbing
coating but with twice the intensity. Initial tests have been carried
out, giving encouraging results, and further investigation will
probably end up using these (cheap) devices for better detector performance.

As stated above, the bare moments are rather poor estimates for
the true \g-ray photoconversion position but they can be used to
reconstruct the true photoconversion position. This is a typical
inverse problem and is also known as the truncated moment problem. 
In chapter~\ref{ch:position-reconstruction}, a method for position
reconstruction from the three non-trivial moments is presented.
The complete understanding of all aspects of the formation of the
signal distribution and the moments computed from it is a {\em sine
  qua non} condition 
for the successful position reconstruction. A parameterization for the
signal distribution has been derived in
chapter~\ref{ch:light-distribution}. The starting point is the inverse
square law, but other important effects have been included:
geometric refraction at the crystal-window interface, Fresnel
transition at the same, angular response of the photocathode,
exponential attenuation of the scintillation light and background from
residual diffuse reflections at the black epoxy resin surfaces.
The signal distribution was derived for the photoelectric effect.
Compton scattering has been treated as an independent effect and its
impact on the moment determination method has been studied in detail
in chapter~\ref{ch:compton}.

The model has been verified by experiment and the results are summarized
in chapter~\ref{ch:experiment}. For the three non-trivial moments, a
very good agreement with measurements  
 was observed (errors $\mathrm{\lesssim11\%}$). Once again, the trivial moment
is an exception. Good agreement was achieved only for the lower limit
of possible DOI values. For the upper limit, the model with background
fails to reproduce all details of the measured
dependency. Nevertheless, background light is present in the true
detector and is an important contribution especially for large DOI
values. This has been verified with an alternative model where the
background light was switched off. In this case, the model fails
completely to reproduce the measurement values for this moment. The
non-trivial moments are only slightly affected by this variation because they
are normalized. The above mentioned dependency of the distribution
width on the interaction depth was better reproduced by the model
without background. Reasons for these disagreements between model
behavior and measured moments may be mainly due to the rough
approximation made during the derivation of the distribution of the
residual diffuse reflections at the epoxy resin surfaces. It is
especially this contribution of the model that has to be improved if
higher accuracy is required. Other possible reasons are the blurring
caused by Compton scattering and, above all, poor mechanical precision.
The absolute error of the source position is
 too high ($\mathrm{\approx1\,mm}$) for achieving better agreement between
model and measurement. In chapter~\ref{ch:compton} it was found that
Compton scattered effects result in a position blurring that is very
low for the vast majority of events and can be very high for a small
fraction of events. The FWHMs and the medians of the caused point
spread function are rather low, being only a few hundred $\mathrm{\mu m}$, and
the main error has to be attributed to the low mechanical precision.

In the last chapter, standard polynomial interpolation in higher
dimensions has been adopted to solve the inverse problem of position
reconstruction from the moments. For this, the moment response of the
detector was predicted by virtue of the model derived in
chapter~\ref{ch:light-distribution} at a total of 40000 possible
impact positions. This response has then been interpolated with
polynomials of order 13 for the transverse spatial directions and a
polynomial of order 6 for the normal component. The coefficients that
map from the expansion onto the grid of interpolation nodes can be
solved by using the pseudo-inverse. Measured events where then expanded
onto the same polynomial basis and can be corrected by multiplication
by a system matrix obtained from the pseudo-inverse and the true 40000
interaction points. Approximately $\mathrm{13^2\cdot 6\cdot 3}$ multiplications and 
the same number of additions are required for the three-dimensional impact position
reconstruction of one single event. Therefore, the method will be fast
enough for its practical use. The intrinsic mean spatial resolution of
the detector, when using these reconstructed impact positions, was found to be $\mathrm{1.9\,mm}$ for the transverse
components and $\mathrm{3.9\,mm}$ for the depth of interaction. This
is a significant improvement compared to the bare moments, where 
$\mathrm{3.4\,mm}$ and  $\mathrm{4.9\,mm}$ respectively have been measured for the
corresponding resolutions. Also the maximum and minimum values show a
significant improvement. In particular, the achieved DOI
resolution is an exceptional result which is achieved by very few of
the alternative measurement methods presented in
chapter~\ref{ch:introduction}. The cost for the necessary
detector improvements are, however, for the present method essentially
negligible. A residual non-linearity of about 10\% remains present in
the position information after reconstruction and is probably due to 
insufficient prediction accuracy of the analytic model. An increased
accuracy here will very likely lead to even better results for the
position reconstruction and presumably make possible effective
correction of the trivial moment. The spatial resolution achieved at
the moment turned out to be insufficient for that purpose.

In this work, a novel technique for measuring the DOI in $\gamma$-ray
imaging detectors was presented. The high quality of the DOI
measurement makes fast three-dimensional impact position computation
from the centroids possible and therefore allows the construction of
cheap detectors that provide a performance comparable to $\gamma$-ray
imaging detectors based on pixelated scintillators. Future
investigation should put their emphasis on higher mechanical precision
of the experimental setup and the radioactive
source, better calibration and an essential improvement in the model
for the residual reflections at the crystal borders. Other possible
improvements of the method are the mentioned retroreflector and a
PSPMT with smaller detector segments. The H9500 from the same company
is an interesting alternative to the type used for the present study.
It has 256 anode segments of $\mathrm{\approx3\times3\,mm^3}$ and a
thinner entrance window. All other parameters are exactly the same and
the depth enhanced charge divider circuits are easily adapted to this
device without making necessary additional electronic channels.

\chapterbib


  \begin{appendix}
    \renewcommand{\theequation}{\mbox{\Alph{chapter}.\arabic{equation}}}

    \cleardoublepage{}
\chapter{Common Radiopharmaceuticals}
\label{app:common-radiotracer}

The following table summarizes important tests with its corresponding
radiopharmaceutical. The list was not intended to be complete and the
data was taken from the indicated sources.

\newcommand{\footcite}[1]{\cite{#1}}

\renewcommand{\baselinestretch}{0.9}{\small\renewcommand{\arraystretch}{1.6}
\begin{longtable}{>{\raggedright\arraybackslash}p{0.2\textwidth}>{\raggedright\arraybackslash}p{0.32\textwidth}>{\raggedright\arraybackslash}X>{\raggedright\arraybackslash}p{0.3\textwidth}}
  \hline\hline  Test & Agent & Isotope & Purpose \& Indications\\\hline\hline\endfirsthead
  \hline  Test & Agent & Isotope & Purpose \& Indications\\\hline\endhead
  \\\hline\endfoot
  \\\hline\hline\endlastfoot
  Plasma volume & Serum albumin & $\mathrm{^{121}I}$ & Determination of plasma
  volume  excessive blood loss, burns, cardiovascular or renal
  disease; \footcite{Johns:1983,UIC:2004}\\
  Red cell volume & Erythrocytes & $\mathrm{^{51}Cr}$ &
  Determination of red cell volume and their half life
  gastro-intestinal protein loss; \footcite{Johns:1983}\\
  Vitamin $\mathrm{B_{12}}$ absorption & Vitamin $\mathrm{B_{12}}$
  with intrinsic factor (IF) and Vitamin $\mathrm{B_{12}}$ &
  $\mathrm{^{57}Co}$, $\mathrm{^{58}Co}$ & Determination of
  Vitamin $\mathrm{B_{12}}$ absorption/excretion,  Pernicious anemia; \footcite{Johns:1983}\\
  Thyroid & Na salt & $\mathrm{^{131}I}$ & Determination of thyorid
  function, thyorid imaging, also diagnosis of abnormal liver
  function, renal blood flow, urinary tract obstruction; \footcite{Johns:1983,UIC:2004}\\
  Thyroid (Imaging) & $\mathrm{NaTcO_4^-}$ & $\mathrm{^{99m}Tc}$ &
  Determination of areas with decreased/increased uptake,
  hyperthyoridism, tumor; \footcite{Johns:1983}\\
  Brain imaging & Glucoheptonate & $\mathrm{^{99m}Tc}$ &
  Study of vascularity of brain  physical injury, tumor; \footcite{Johns:1983}\\
  liver/spleen imaging & sulphur colloid & $\mathrm{^{99m}Tc}$ &
  Study of uptake of colloid, abnormal liver/spleen; \footcite{Johns:1983}\\
  Kidney & Glucoheptonate & $\mathrm{^{99m}Tc}$ &
  Study of structure and filtration rate, renal trauma, tumors or cysts; \footcite{Johns:1983}\\
  Lung & Albumin macroaggregates & $\mathrm{^{99m}Tc}$ &
  Macroaggregates trapped in pulmonary capillary bed, defective
  perfusion; \footcite{Johns:1983}\\
  Internal, mammary and iliac lymph nodes & Antimony sulphide colloid & $\mathrm{^{99m}Tc}$ &
  Localization of internal mammary and pelvic lymph nodes, 
  localization of sentinel nodes; \footcite{Johns:1983}\\
  Skeleton & Phosphates,HDP & $\mathrm{^{99m}Tc}$ &
  Uptake/hlaf-life of tracer (phosphate tracer behaves like calcium),
  metastatic disease, regions of trauma, arthritic regions; \footcite{Johns:1983}\\
  Tumors & $\mathrm{^{67}Ga}$-Citrate & $\mathrm{^{67}Ga}$ &
  Regional tracer uptake,
  recurrence of lymphomas, regions of inflammation; \footcite{Johns:1983}\\
  Heart & Sestamibi, Tetrofosmin & $\mathrm{^{99m}Tc}$ &
  Myocardial perfusion imaging (MPI), diagnosis of heart conditions,
  coronary artery disease (CAD), acute ischemic syndrome, also location of low-grade
  lymphomas; \footcite{Johns:1983,UIC:2004,Acampa:2000}\\
  Heart & TlCl (via the ATP-pump for Ka) & $\mathrm{^{201}Tl}$ &
  Myocardial viability, regional bloodflow, CAD, prediction of
  future cardiac events; \footcite{Johns:1983,UIC:2004,Acampa:2000}\\
  Heart & Ru (via the ATP-pump for Ka) & $\mathrm{^{82}Ru}$ &
  Myocardial perfusion imaging, stress/rest protocols; \footcite{Acampa:2000}\\
  Lung & Krypton-gas & $\mathrm{^{81m}Kr}$ &
  Functional imaging of pulmonary ventilation,
  asthma, early diagnosis of lung disease; \footcite{UIC:2004}\\
  Breast & Sestamibi & $\mathrm{^{99m}Tc}$ &
  Imaging of dense breast tissue,
  detection of breast cancer; \footcite{Johns:1983}\\
  Enzymes & Seleno-methionine & $\mathrm{^{75}Se}$ &
  Study of production of digestive enzymes; \footcite{UIC:2004}\\
    Tumors/Heart & Fluoro-dioxy glucose (FDG) & $\mathrm{^{18}F}$ &
    Distribution of FDG (glycolysis), detection of tumors, infection,
post-traumatic inflammation, glycolysis of myocardium; \footcite{Johns:1983,UIC:2004,Acampa:2000}\\
    Heart & Ammonia (via the Na/K ATPase pump) & $\mathrm{^{13}N}$ &
    Quality images of the myocardium; \footcite{Acampa:2000}\\
    Heart & Radioiodinated fatty acid & $\mathrm{^{123}I}$ &
    Metabolic imaging of the myocardium; \footcite{Acampa:2000}\\
    Heart & 15-(p-$\mathrm{^{123}I}$ iodophenyl-)-3-R,S-methyls
pentadecanoic acid (BMIPP) & $\mathrm{^{123}I}$ &
    Cardiac metabolic activity, myocardial viability; \footcite{Acampa:2000}\\
    Heart & Palmitate & $\mathrm{^{11}C}$ &
    Myocardial blood flow (tracer uptake), metabolic metabolism
(tracer clearance); \footcite{Acampa:2000}\\
    Somatostatin receptor & $\mathrm{^{111}In}$-D-Phe-DTPA-octreotide (Octreoscan) & $\mathrm{^{111}In}$ &
    Clinical imaging of primary and metastatic breast
carcinoma, lymphoma and neuroendocrine tumors; \footcite{Blankenberg:2002}\\
    Somatostatin receptor & $\mathrm{^{99m}Tc}$-depreotide (Neotect) & $\mathrm{^{99m}Tc}$ &
    Clinical imaging of primary and metastatic breast
carcinoma, lymphoma and neuroendocrine tumors; \footcite{Blankenberg:2002}\\
    $\mathrm{(GABA)_A}$ receptor (Gamma-Aminobutyric Acid) &
[$\mathrm{^{123}I}$]iomazenil & $\mathrm{^{123}I}$ &
    Imaging of neurologic disorders, Alzheimer presenile dementia,
    pre/intra-operative definition of cerebral tumor margins (SPECT); \footcite{Blankenberg:2002}\\
    $\mathrm{(GABA)_A}$ receptor & [$\mathrm{^{11}C}$]flumazenil & $\mathrm{^{11}C}$ &
    Imaging of neurologic disorders, epilepsy (PET); \footcite{Blankenberg:2002,Fowler:2003}\\
    Dopamine Transporter (DAT) & [$\mathrm{^{123}I}$]$\beta$-CIT,[$\mathrm{^{123}I}$]FP-CIT & $\mathrm{^{123}I}$ &
    Accelerated presynaptic dopaminergic degeneratin in Parkinson disease; \footcite{Blankenberg:2002}\\
    $\mathrm{D_2}$ Dopamine Receptor & $\mathrm{^{123}I}$-iodo-benzamide,
$\mathrm{^{123}I}$-iodolisuride & $\mathrm{^{123}I}$ &
    Abnormal increase in $\mathrm{D_2}$ receptor binding found in schizophrenia; \footcite{Blankenberg:2002}\\
    Dopamine Metabolism & [$\mathrm{^{18}F}$]FluoroDOPA & $\mathrm{^{18}F}$ &
    Loss of dopaminergic neurons (dopamine synthesis); \footcite{Fowler:2003}\\
    $\mathrm{D_2}$ Dopamine Receptor & N-methylspiroperidol &
$\mathrm{^{18}F}$, $\mathrm{^{11}C}$ & Measurement of dopamine
receptor availibility, absolute concentration in the normal/diseased
brain, probe of dopamine receptor occupancy by antipsychotic drugs; \footcite{Fowler:2003}\\
    Dopamine Transporters & [$\mathrm{^{11}C}$]nomifensine, [$\mathrm{^{11}C}$]cocaine & $\mathrm{^{11}C}$ &
    Study of neurodegeneration, aging, dopamine transporter occupancy
by drugs; \footcite{Fowler:2003}\\
    Vesicular amine transporters & $\alpha(+)-[\mathrm{^{11}C}]$-di\-hydro\-tetra\-benazine & $\mathrm{^{11}C}$ &
    Normal aging, Parkinson's disease; \footcite{Fowler:2003}\\
    Catechol-O-Methyl\-transferase (estrogen metabolism) & [$\mathrm{^{18}F}$]Ro41-0960 & $\mathrm{^{18}F}$ &
    Promising tracer for Breast cancer (preclinical studies); \footcite{Fowler:2003}\\
    Brain serotonine system (5-HT receptors) &
[$\mathrm{^{18}F}$]altanserin, [$\mathrm{^{18}F}$]setoperone & $\mathrm{^{18}F}$ &
    Anxiety, depression, sleep/eating disorders, violence; \footcite{Fowler:2003}\\
    Brain opiate system & [$\mathrm{^{11}C}$]Carfentanil & $\mathrm{^{11}C}$ &
    Respiratory depression, analgesia, reward, sedation; \footcite{Fowler:2003}\\
    Benzo\-dia\-zepine system & isoquinoline [$\mathrm{^{11}C}$]PK 11195 & $\mathrm{^{11}C}$ &
    High, saturable uptake in glioma relative to normal brain tissue,
multiple sclerosis; \footcite{Fowler:2003}\\
    Muscarinic-Cholinergic Receptors &
[$\mathrm{^{11}C}$]N-methyl-4-piperidinyl\-benzilate &
$\mathrm{^{11}C}$ & Muscarinic receptor concentration in normal aging
and Alzheimer's disease; \footcite{Fowler:2003}\\
    Amino-Acid transport and Protein synthesis &
[$\mathrm{^{11}C}$]D/L-methionine & $\mathrm{^{11}C}$ & Delineation of
tumors; \footcite{Fowler:2003}\\
    Amino-Acid transport and Protein synthesis &
[$\mathrm{^{11}C}$-carboxyl]L-leucine & $\mathrm{^{11}C}$ & Protein
synthesis rate (brain development, regeneration, repair); \footcite{Fowler:2003}\\
    Amino-Acid transport and Protein synthesis &
O-(2-[$\mathrm{^{18}F}$]fluoroethyl)-L-tyrosine & $\mathrm{^{18}F}$ &
Mammary carcinoma in mice, recurrent astrocytoma in humans; \footcite{Fowler:2003}\\
    DNA synthesis & $\mathrm{^{14}C}$/$\mathrm{^{3}H}$-labeled
thymidine & $\mathrm{^{14}C}$, $\mathrm{^{3}H}$ &
Measurement of tissue proliferation and growth kinetics; \footcite{Fowler:2003}\\
    DNA synthesis & [$\mathrm{^{18}F}$]Fluorouridine, [$\mathrm{^{11}F}$]thymidine &
$\mathrm{^{18}F}$, $\mathrm{^{11}C}$ & Measurement of tissue proliferation and growth kinetics; \footcite{Fowler:2003}\\
    DNA synthesis & [$\mathrm{^{18}F}$]FLT &
$\mathrm{^{18}F}$ & High-contrast images of normal bone marrow and
tumors in canine and human subjects; \footcite{Fowler:2003}
\end{longtable}
}

\chapterbib

    \cleardoublepage{}
\chapter{Common Inorganic Scintillators}
\label{app:com-inorg-scints}

\renewcommand{\arraystretch}{2.6}
\renewcommand{\baselinestretch}{0.4}
The following table resumes the physical properties of various
common scintillators. The data has been taken mainly from 
Novotny \cite{Novotny:2005}, Melcher \cite{Melcher:2005}, 
van Eijk \cite{vanEijk:2002,Eijk:2003}, Pidol et al.\
\cite{Pidol:2003}, Eriksson et al.\ \cite{Eriksson:2004} and Derenzo
and Moses \cite{Derenzo:1992bp}.

\renewcommand{\baselinestretch}{0.9}{\small\renewcommand{\arraystretch}{1.6}
\begin{longtable}{>{\raggedright\arraybackslash}p{0.205\textwidth}>{\raggedright\arraybackslash}p{0.04\textwidth}>{\raggedright\arraybackslash}X>{\raggedright\arraybackslash}X>{\raggedright\arraybackslash}X>{\raggedright\arraybackslash}X>{\raggedright\arraybackslash}X>{\raggedright\arraybackslash}p{0.1\textwidth}>{\raggedright\arraybackslash}X}
  \hline\hline Name/Dopand & $\mathrm{Z_{eff}}$ & Density [$\mathrm{g/cm^2}$] &
  $\mathrm{\lambda}$ [nm] & photons [$\mathrm{keV^{-1}}$] & decay
  time [ns] & refrac.\ index & radiation length [cm] \& photo\-fract. &
  Hygro\-scopic \\\hline\hline\endfirsthead
  \hline\hline Name/Dopand & $\mathrm{Z_{eff}}$ & Density [$\mathrm{g/cm^2}$] &
  $\mathrm{\lambda}$ [nm] & photons [$\mathrm{keV^{-1}}$] & decay
  time [ns] & refrac.\ index & radiation length [cm] \& photo\-fract. &
  Hygro\-scopic \\\hline\hline\endhead
  \hline\endfoot
  \hline\hline\endlastfoot
  \chemform{NaI\doped Tl} & 50.8 & 3.67 & 415 & 38 & 230 & 1.85 & 2.6/17 & yes \\
  \chemform{CsI} & 54 & 4.51 & {305 450} & 2 &
  {2 20} & 1.8 & 1.86/21 & slight \\
  \chemform{CsI\doped Na} & 54 & 4.51 & 420 & 39 & {460
    4180} & 1.84 & 1.86/21 & slight \\
  \chemform{CsI\doped Tl} & 54 & 4.51 & {550 565} & 65 &
  {680 3340} & 1.8 & 1.86/21 & slight \\
  {\chemform{Bi_4Ge_3O_{12}} (BGO)} & 75 & 7.13 & 480 &
  8.2 & 300 & 2.15 & 1.12/40 & no \\
  \chemform{BaF_2} & 51 & 4.88 & {220 310} &
  {1.4 9.5} & {0.6 630} & 1.56 & 2.03/19 & no \\
  \chemform{CaF_2\doped Eu} & 16.5 & 3.18 & 435 & 24 & 940 & 1.44 & 3.05
  & no \\
  \chemform{CdWO_4} & 66 & 8 & 470 & 15 & {1.1$\mathrm{\mu
      s}$ 14.0$\mathrm{\mu s}$} & 2.3 & 1.06/29 & no \\
  \chemform{CaWO_4} &  & 6.1 & 430 & 6 & 940 & 1.92 & 1.5 & no \\
  \chemform{LiF\doped W} & 8.31 & 2.64 & 430 & 1.5 & 4000 & 1.4 & & no \\
  \chemform{LiI\doped Eu} &  & 4.08 & {470 485} & 11 &
  1400 & 1.96 & 2.73 & yes \\
  {\chemform{LuAlO_3\doped Ce} (LuAP)} & 65 & 8.4 & 365 &
  17 & 17 & 1.94 & 1/30 & no \\
  {\chemform{Lu_{1-x}Y_{x}AlO_3\doped Ce}  (LuYAP)} & 61.3
  & 7.1 & 367 & 6-8 & {20-26 200} & 1.4 & 1.3/27 & no \\
  {\chemform{Lu_2(SiO_4)O\doped Ce^+} (LSO)} & 66.4 & 7.4
  & 420 & 25 & 40 & 1.81 & 1.14/32 & no \\
  {\chemform{Lu_{2-x}Y_{x}(SiO_4)O\doped Ce^+} (LYSO)} & 66 & 7.4
  & 428 & 32 & 41 & 1.81 & 1.1/30 & no \\
  {\chemform{Lu_{2-x}Gd_{x}(SiO_4)O\doped Ce^+} (LGSO)} & 57 & 6.7
  & {410 500} & 18 & 65 & &  & no \\
  {\chemform{Lu_2Si_2O_7\doped Ce} (LPS)} & 64 & 6.2 &
  380 & 20 & 30 & 1.74 & 1.4/29 & no \\
  {\chemform{Lu_2S_3\doped Ce}} &  & 6.2 & 590 & 28 &
  32 &  & 1.4 & no \\
  {\chemform{Gd_2SiO_5\doped Ce^+} (GSO)} & 59.4 & 6.71 & 440
  & 9 & {56 400} & 1.85 & 1.4/25 & no \\
  {\chemform{Gd_2O_2S\doped Pr} (GOS/UFC)} & 61.3 & 7.34 &
  510 & 50 & 3000 &  & 1.27/27 & no \\
  {\chemform{Y_{1.34}Gd_{0.6}O_3S\doped Eu_{0.06}}
    \chemform{Y_{1.34}Gd_{0.6}O_3S\doped Pr_{0.06}} (Hilight)} & 52.2 &
  5.9 & 610 & 44 & 1000 &  & 1.78/16 & no \\
  {\chemform{Gd_3Ga_5O_{12}\doped Cr,Ce}} & 53.5 & 7.1 & 730 &
  40 & 140$\mathrm{\mu s}$ &  & 1.48/18 & no \\
  \chemform{PbSO_4} & 70.4 & 6.20 & 350 & 3.8 & 100 & 1.88 & 1.28 & no \\
  \chemform{CsF} & 53 & 4.64 & 390 & 1.5 & 5 & 1.48 & 2.69 & yes \\
  \chemform{CeF_3} & & 6.16 & {310 340} & 4.4 &
  {5 27} & 1.68 & 1.68 & no \\
  {\chemform{Y_3AlO_3\doped Ce^+} (YAP)} & 36 & 5.37 & 370
  & 18 & 27 & 1.95 & 2.7 & no \\
  {\chemform{Y_3Al_5O_{12}\doped Ce^+} (YAG)} & 14 & 4.56
  & 550 & 17 & {88 302} & 1.82 & 3.5 & no \\
  \chemform{LiF/ZnS\doped Ag} & 26.1 & 4.09 & 450 & 75 & 200 & 2.36 & & no \\
  \chemform{Lu_2O_3\doped Eu,Tb} & 68.7 & 9.4 & 610 & 30 & $>$1000 &
   &8.7/35  & no \\
  \chemform{PbWO_4} & 76 & 8.28 & {440 $\sim$500} & 0.2 &
  5 $\sim$ 15 & 2.16 & 0.89 & no \\
  \chemform{PbWO_4\doped Mo,Y} &  & 8.3 & $\sim$ 500 & 0.6 & $\sim$
  15 &  & 0.87 & no \\
  \chemform{PbWO_4\doped Mo,Nb} &  & 8.3 & $\sim$ 500 & 0.4 & $10
  \sim 10^3$ &  & 0.87 & no \\
  \chemform{LaF_3\doped Ce} & 52 & 5.94 & {290 340} & 2.2 & 26 & 1.7 & 1.85 & no \\
  \chemform{LaCl_3\doped Ce} & 50 & 3.86 & 330 & 46 & 25 &  & 27.8 & yes \\
  \chemform{LaBr_3\doped Ce} & 47 & 5.3 & 360 & 61 & 25 & 1.88 & 2.13/15 & yes \\
  \chemform{LuI_3\doped Ce} & 61 & 5.6 & 474 & 33 & 34,360 &  & 1.7 & yes \\
  \chemform{LuPo_4\doped Ce} & 62.5 & 6.53 & 360 & 17 & 24 &  & 1.43/30 & no \\
  \chemform{ZnSe\doped Te} & 33 & 5.42 & {610 $\sim$640} & 80 & {1000 $\sim$30000} & 2.6 & & no \\
  \chemform{ZnS\doped Ag} &  & 4.1 & 450 & 49 & 200 & 2.4 & 2.94 & no \\
  \chemform{ZnO\doped Ga} &  & 5.6 & 385 & 15 & 1.48 & 2.0 & 2.16 & no \\
  \chemform{ZnWO_4} &  & 7.87 & 480 & 10 & 5000 & 2.2 & 1.2 & no \\
  anthracene &  & 1.25 & 450 & 16 & 30 & 1.62 & 8.79 & no \\
\end{longtable}
}

\chapterbib


    \cleardoublepage{}
\begin{landscape}

\chapter{Supplementary Analytic Results}
\label{app:exact_solutions}

  \section{Impedances of the 2D Proportional Resistor Network}

The parameterization~(\ref{eq:exact-parametrization}) for the 
impedance at each input of the 2D proportional resistor network
described in section~\ref{subsection:2D-prop-net} has to be understood
as an empirical ad-hoc parameterization. Except for the symmetry of the
used charge divider, no physical argument was used to motivate a
polynomial of 6th order as model behavior. For reasons of symmetry,
odd monomials have been discarded. By this means one obtains a linear
system of 4 unknowns and 4 equations with exactly one real
solution. Another point is that it was not possible to find this
parameterization since it is still a function of the used resistor values.
For this reason, the partial solution of the network analysis
mentioned in section~\ref{subsection:2D-prop-net} is given here in
its most general form and for a supposed $8\times8$ anode-segment matrix.
Equations~(\ref{eq:exact-imp-solution-a1})-(\ref{eq:exact-imp-solution-a4})
are obtained for a proportional resistor network with linearized
$y$-centroid behavior. That is to say,
solutions~(\ref{eq:solution-laterals}) have been plugged into the node
equations~(\ref{eq:lateral-r-equations}). The second set of
equations~(\ref{eq:exact-imp-solution-b1})-(\ref{eq:exact-imp-solution-b4})
is obtained when all horizontal resistors in
figure~(\ref{fig:bidim-schem}) have the same value $R_h$ instead.

  \begin{gather}
    \label{eq:exact-imp-solution-a1}
    U_{i1}=A \left(i^2 R_v^2\left(\frac{32 \left(-17739 R_h^3+90990 R_v
            R_h^2-151428 R_v^2 R_h+81200 R_v^3\right)
        }{D}+\frac{864}{8 R_v-9 R_h}\right)-81 B-56
      C\right)\\\label{eq:exact-imp-solution-a2}
    U_{i2}=A \left(i^2R_v^2 \left(\frac{16 \left(-116397 R_h^3+530604 R_v
            R_h^2-741192 R_v^2 R_h+296800 R_v^3\right)
        }{D}+\frac{96}{8 R_v-9 R_h}+\frac{2352}{14 R_v-9 R_h}\right)-81
      B-98 C\right)\\\label{eq:exact-imp-solution-a3}
    U_{i3}=A \left(i^2 R_v^2\left(\frac{144 \left(-23571 R_h^3+97047 R_v
            R_h^2-120912 R_v^2 R_h+44800 R_v^3\right)}{D}-\frac{324}{R_h-2
          R_v}+\frac{96}{8 R_v-9 R_h}+\frac{588}{14 R_v-9 R_h}\right)-81 B-126
      C\right)\\\label{eq:exact-imp-solution-a4}
    U_{i4}=A \left(i^2 R_v^2\left(\frac{16\left(-300105 R_h^3+1139157 R_v
            R_h^2-1343904 R_v^2 R_h+481600 R_v^3\right)}{D}
        -\frac{324}{R_h-2 R_v}+\frac{96}{8 R_v-9 R_h}+\frac{588}{14 R_v-9
          R_h}\right)-81 B-140 C\right)
  \end{gather}

  \begin{gather}
    \label{eq:exact-imp-solution-b1}
    U_{i1}=A \left(i^2 R_v^2\left(-\frac{24}{9 R_h+2 R_v}-\frac{8}{27 R_h+2 R_v}-\frac{24\left(891 R_h^2+180 R_v
            R_h+8 R_v^2\right)}{E}-\frac{24\left(729 R_h^2+180 R_v R_h+8 R_v^2\right)}{F}\right)-81 B+16 C\right)\\\label{eq:exact-imp-solution-b2}
    U_{i2}= A \left(i^2 R_v^2\left(-\frac{24}{9 R_h+2 R_v}-\frac{8}{27 R_h+2 R_v}-\frac{48\left(1539 R_h^2+252 R_v
            R_h+10 R_v^2\right)}{E}-\frac{144\left(243 R_h^2+48 R_v R_h+2 R_v^2\right)}{F}\right)-81B+28 C\right)\\\label{eq:exact-imp-solution-b3}
    U_{i3}=A \left(i^2R_v^2\left(-\frac{72 \left(405 R_h^2+90 R_v R_h+4 R_v^2\right)}{E}-\frac{216 \left(621 R_h^2+102 R_v R_h+4
            R_v^2\right)}{F}\right)-81B+36 C\right)\\\label{eq:exact-imp-solution-b4}
    U_{i4}= A \left(i^2 R_v^2\left(-\frac{24}{9 R_h+2 R_v}-\frac{8}{27 R_h+2 R_v}-\frac{24\left(7209 R_h^2+1152 R_v R_h+44 R_v^2\right)
        }{E}-\frac{24\left(9 R_h+2 R_v\right) \left(27 R_h+2 R_v\right)}{F}\right)-81B+40 C\right)
  \end{gather}
  with
  \begin{gather}
    \label{eq:exact-imp-solution-substs}
    A=2916^{-1}\nn\\B=\left(4 i^2-81\right) R_h\nn\\
    C=\left(4i^2+81\right) R_v\nn\\
    D=6561 R_h^4-38880 R_v R_h^3+82404 R_v^2 R_h^2-73056 R_v^3 R_h+22400R_v^4\nn\\
    E=729 R_h^3+1458 R_v R_h^2+216 R_v^2 R_h+8 R_v^3\nn\\
    F=2187 R_h^3+1458 R_v R_h^2+216 R_v^2 R_h+8 R_v^3\nn
  \end{gather}

With these expressions, an explicit parameterization in the position indexes $i$ and $j$ can be given for
the case explained in section~\ref{subsection:2D-prop-net} with $R_h=10
R_v$:
\begin{equation}
  \label{eq:exact-parametrization-2}
  R_\mathit{Imp}(i,j)=\frac{5}{18} \left(81-4 i^2\right) R_v+\left(a_2 j^2+a_0+i^2 \left(b_6 j^6+b_4 j^4+b_2 j^2+b_0\right)\right) R_v,
\end{equation}
with the exact parameter values
\begin{equation}
  \label{eq:exact-parametrization-constants-2}
  \renewcommand{\arraystretch}{1.4}
  \begin{array}{llll}
    a_0=-\frac{63}{16}, & a_2=-\frac{7}{36} &  &  \\
    b_0=-\frac{18348064865}{74993190912}, & b_2=\frac{105010791245}{6918121861632},& b_4=-\frac{327412375}{1729530465408},& b_6=\frac{816155}{432382616352}.
  \end{array}
\end{equation}
If all $R_{h_j}$
are set to $R_h$, one will obtain the following exact parameter values
\begin{equation}
  \label{eq:exact-parametrization-constants-b-2}
  \renewcommand{\arraystretch}{1.4}
  \begin{array}{llll}
    a_0=\frac{9}{8}, & a_2=-\frac{1}{18} & &  \\
    b_0=\frac{44570871214009595}{922346100698554368},& b_2=-\frac{52753280350225}{25620725019404288}, & b_4=-\frac{97452730375}{6405181254851072}, &
    b_6=-\frac{74531275}{1601295313712768}.
  \end{array}
\end{equation}

\clearpage
\end{landscape}


    \cleardoublepage{}
\chapter{Electronic Amplifier Configuration for the Experiment}
\label{app:elec-config}

\section*{Summing Amplifier}

\begin{figure}[ht]
  \centering
  \psfrag{R1}{\small $R_1$}
  \psfrag{R2}{\small $R_2$}
  \psfrag{R3}{\small $R_3$}
  \psfrag{R4}{\small $R_4$}
  \psfrag{R5}{\small $R_5$}
  \psfrag{R6}{\small $R_6$}
  \psfrag{C1}{\small $C_1$}
  \psfrag{C2}{\small $C_2$}
  \psfrag{C3}{\small $C_3$}
  \psfrag{C4}{\small $C_4$}
  \psfrag{C5}{\small $C_5$}
  \psfrag{C6}{\small $C_6$}
  \psfrag{C7}{\small $C_7$}
  \psfrag{D1}{\small $D_1$}
  \psfrag{D2}{\small $D_2$}
  \psfrag{U1}{\small $U_1$}
  \psfrag{Vout}{\small $V_\tincaps{Out}$}
  \psfrag{S}{\small $\sum$}
  \includegraphics[width=0.4\textwidth]{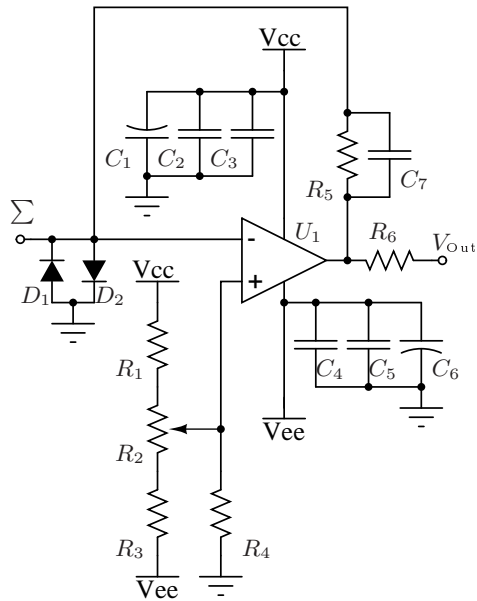}
  \caption[Schematics of the summation amplifier]{Schematics of the summation amplifier.}
  \label{fig:summing-amp}
\end{figure}

For the computation of the sum of the 64 voltages sensed at the inputs
of the charge divider circuits, the standard analogue adder
configuration of figure~\ref{fig:summing-amp} was used. The diodes
$D_1$ and $D_2$ have the function of protecting the inputs of the
operational amplifier. $R_1$, $R_2$ and $R_3$ are used for offset
corrections and $C_7$ prevents the circuit from oscillations.

\begin{table}[ht]
  \centering
  \begin{tabular}{|ll|ll|ll|}\hline
    Component & Value & Component & Value & Component & Value \\\hline
    $R_1$, $R_3$ & $1k\Omega$ & $R_6$ & $50\Omega$ & $D_1$, $D_2$ & $1SS335$ \\
    $R_2$ & $10k\Omega$ & $C_1$, $C_6$ & $10\mu F$ & $U_1$ & $AD8055$\\
    $R_4$ & $1.5k\Omega$ & $C_2$, $C_5$ & $100n F$ & $C_7$ & $3.9pF$ \\
    $R_5$ & $2.7k\Omega$ & $C_3$, $C_4$  & $1nF$ &  & \\\hline
  \end{tabular}
  \caption[Electronic configuration for the
    summation amplifier]{Electronic configuration (component values) for the
    summation amplifier shown in figure~\ref{fig:summing-amp}.}
  \label{tab:summing-component-values}
\end{table}

\section*{Inverting/Non-inverting Amplifier and Line-driver}

\begin{figure}[ht]
  \centering
  \psfrag{R1}{\small $R_1$}
  \psfrag{R2}{\small $R_2$}
  \psfrag{R3}{\small $R_3$}
  \psfrag{R4}{\small $R_4$}
  \psfrag{R5}{\small $R_5$}
  \psfrag{R6}{\small $R_6$}
  \psfrag{R7}{\small $R_7$}
  \psfrag{R8}{\small $R_8$}
  \psfrag{R9}{\small $R_9$}
  \psfrag{R10}{\small $R_{10}$}
  \psfrag{C1}{\small $C_1$}
  \psfrag{C2}{\small $C_2$}
  \psfrag{C3}{\small $C_3$}
  \psfrag{C4}{\small $C_4$}
  \psfrag{C5}{\small $C_5$}
  \psfrag{C6}{\small $C_6$}
  \psfrag{C7}{\small $C_7$}
  \psfrag{C8}{\small $C_8$}
  \psfrag{C9}{\small $C_9$}
  \psfrag{C10}{\small $C_{10}$}
  \psfrag{C11}{\small $C_{11}$}
  \psfrag{C12}{\small $C_{12}$}
  \psfrag{C13}{\small $C_{13}$}
  \psfrag{C14}{\small $C_{14}$}
  \psfrag{C15}{\small $C_{15}$}
  \psfrag{C16}{\small $C_{16}$}
  \psfrag{C17}{\small $C_{17}$}
  \psfrag{C18}{\small $C_{18}$}
  \psfrag{C19}{\small $C_{19}$}
  \psfrag{D1}{\small $D_1$}
  \psfrag{D2}{\small $D_2$}
  \psfrag{U1}{\small $U_1$}
  \psfrag{U2}{\small $U_2$}
  \psfrag{U3}{\small $U_3$}
  \psfrag{U4}{\small $U_4$}
  \psfrag{U1}{\small $U_1$}
  \psfrag{Vcc}{\small \raisebox{0.4ex}{$V_\tincaps{CC}$}}
  \psfrag{Vee}{\small $V_\tincaps{EE}$}
  \psfrag{Iin}{\small $I_\tincaps{In}$}
  \psfrag{Iout}{\small $I_\tincaps{Out}$}
  \includegraphics[width=\textwidth]{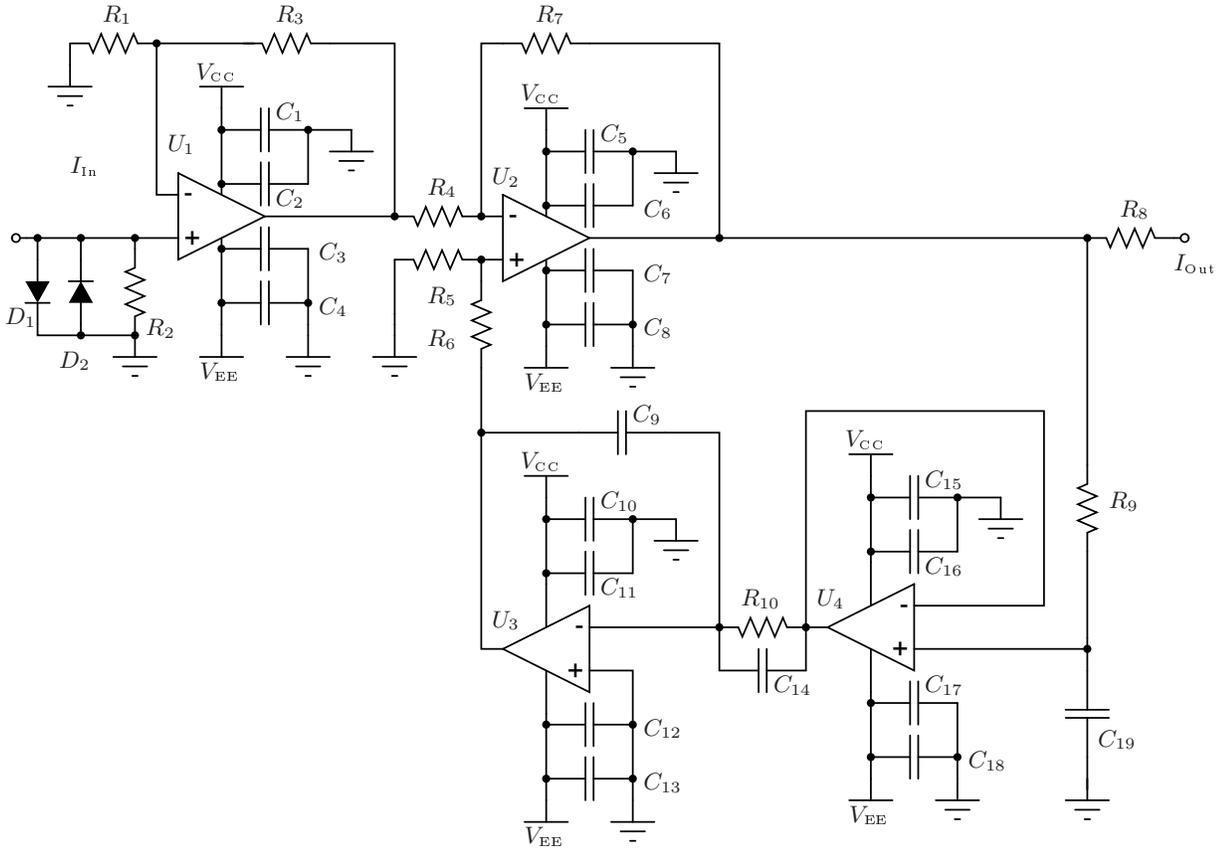}
  \caption[Schematic circuit diagram of the non-inverting
    preamplifier/line driver]{Schematic circuit diagram of the inverting
    preamplifier/line driver. The values of the different components are
  given in table~\ref{tab:inverting-component-values}.}
  \label{fig:inverting-preamp}
\end{figure}

\begin{figure}[ht]
  \centering
  \psfrag{R1}{\small $R_1$}
  \psfrag{R2}{\small $R_2$}
  \psfrag{R3}{\small $R_3$}
  \psfrag{R4}{\small $R_4$}
  \psfrag{R5}{\small $R_5$}
  \psfrag{R6}{\small $R_6$}
  \psfrag{R7}{\small $R_7$}
  \psfrag{R8}{\small $R_8$}
  \psfrag{R9}{\small $R_9$}
  \psfrag{R10}{\small $R_{10}$}
  \psfrag{C1}{\small $C_1$}
  \psfrag{C2}{\small $C_2$}
  \psfrag{C3}{\small $C_3$}
  \psfrag{C4}{\small $C_4$}
  \psfrag{C5}{\small $C_5$}
  \psfrag{C6}{\small $C_6$}
  \psfrag{C7}{\small $C_7$}
  \psfrag{C8}{\small $C_8$}
  \psfrag{C9}{\small $C_9$}
  \psfrag{C10}{\small $C_{10}$}
  \psfrag{C11}{\small $C_{11}$}
  \psfrag{C12}{\small $C_{12}$}
  \psfrag{C13}{\small $C_{13}$}
  \psfrag{C14}{\small $C_{14}$}
  \psfrag{C15}{\small $C_{15}$}
  \psfrag{C16}{\small $C_{16}$}
  \psfrag{C17}{\small $C_{17}$}
  \psfrag{C18}{\small $C_{18}$}
  \psfrag{C19}{\small $C_{19}$}
  \psfrag{D1}{\small $D_1$}
  \psfrag{D2}{\small $D_2$}
  \psfrag{U1}{\small $U_1$}
  \psfrag{U2}{\small $U_2$}
  \psfrag{U3}{\small $U_3$}
  \psfrag{U4}{\small $U_4$}
  \psfrag{U1}{\small $U_1$}
  \psfrag{Vcc}{\small \raisebox{0.4ex}{$V_\tincaps{CC}$}}
  \psfrag{Vee}{\small $V_\tincaps{EE}$}
  \psfrag{Iin}{\small $I_\tincaps{In}$}
  \psfrag{Iout}{\small $I_\tincaps{Out}$}
  \includegraphics[width=\textwidth]{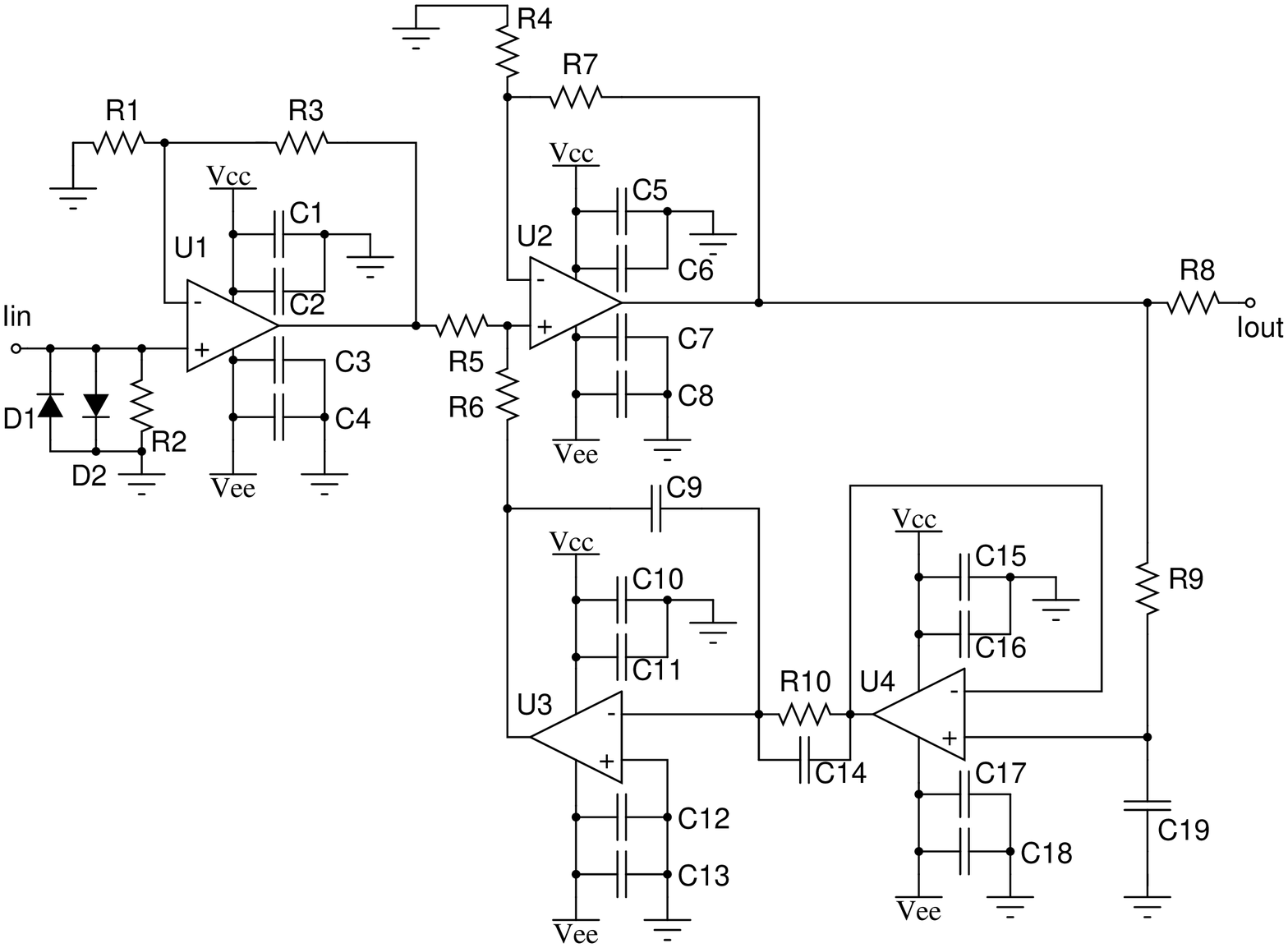}
  \caption[Circuit diagram of the non-inverting
    preamplifier/line driver]{Circuit diagram of the non-inverting
    preamplifier/line driver. The values of the different components are
  given in table~\ref{tab:noninverting-component-values}.}
  \label{fig:noninverting-preamp}
\end{figure}

The configuration for the inverting (figure~\ref{fig:inverting-preamp})
and the non-inverting (figure~\ref{fig:noninverting-preamp}) amplifier
are the
same except for the detail that the operational amplifier $U_2$
inverts the signal from $U_1$ in the first case and does not invert
the corresponding signal in the latter case. $R_2$ converts the input
current $I_\tincaps{In}$ into a voltage which enters the first stage
build by $U_1$. $U_2$ builds a differential amplifier which
subtracts the output of the integrator $U_3$. $U_4$ senses the output
voltage of the differential amplifier and feeds it into the
integrator. In this way, the baseline of the fast PMT signals are
restored. The diodes protect the inputs of the first stages.

\begin{table}[ht]
  \centering
  \begin{tabular}{|ll|ll|ll|}\hline
    Component & Value & Component & Value & Component & Value \\\hline
    $R_1$ & $270\Omega$ & $R_9$ & $10k\Omega$ & $D_1$, $D_2$ & $1SS335$ \\
    $R_2$, $R_8$ & $50\Omega$ & $R_{10}$ & $4.7k\Omega$ & $U_1$ & $AD8009$\\
    $R_3$, $R_6$, $R_7$ & $270\Omega$ & $C_1$, $C_4$, $C_5$, $C_8$,
    $C_{10}$, $C_{13}$, $C_{15}$, $C_{18}$ & $1\mu F$ & $U_2$, $U_4$ &
    $AD8055$ \\
    $R_4$, $R_5$ & $100\Omega$ & $C_2$, $C_3$, $C_6$, $C_7$,
    $C_{11}$, $C_{12}$, $C_{14}$, $C_{16}$, $C_{17}$, $C_{19}$ & $100nF$ & $U_3$ &
    $OP97FS$ \\\hline
  \end{tabular}
  \caption[Electronic configuration (component values) for the inverting
    preamplifier/line driver]{Electronic configuration (component values) for the inverting
    preamplifier/line driver.}
  \label{tab:inverting-component-values}
\end{table}

\begin{table}[ht]
  \centering
  \begin{tabular}{|ll|ll|ll|}\hline
    Component & Value & Component & Value & Component & Value \\\hline
    $R_1$ & $47\Omega$ & $R_9$ & $10k\Omega$ & $D_1$, $D_2$ & $1SS335$ \\
    $R_2$, $R_8$ & $50\Omega$ & $R_{10}$ & $4.7k\Omega$ & $U_1$ & $AD8009$\\
    $R_3$, $R_6$, $R_7$ & $270\Omega$ & $C_1$, $C_4$, $C_5$, $C_8$,
    $C_{10}$, $C_{13}$, $C_{15}$, $C_{18}$ & $1\mu F$ & $U_2$, $U_4$ &
    $AD8055$ \\
    $R_4$, $R_5$ & $100\Omega$ & $C_2$, $C_3$, $C_6$, $C_7$,
    $C_{11}$, $C_{12}$, $C_{14}$, $C_{16}$, $C_{17}$, $C_{19}$ & $100nF$ & $U_3$ &
    $OP97FS$ \\\hline
  \end{tabular}
  \caption[Electronic configuration (component values) for the non-inverting
    preamplifier/line driver]{Electronic configuration (component values) for the non-inverting
    preamplifier/line driver.}
  \label{tab:noninverting-component-values}
\end{table}

\cleardoublepage{}


  \end{appendix}

 \listoffigures

 \listoftables

\cleardoublepage{}

\pagestyle{empty}
\cleardoublepage{}

\vspace*{5eX}

\centerline{\LARGE\bf Acknowledgments}

\vspace*{2eX}

\begin{center}
  \hspace*{0.09\linewidth} \begin{minipage}[t]{0.8\linewidth}
    I want to thank everyone who helped me finish this
    dissertation research and writing. First of all, I'd like to thank
    José-M.~Benlloch for being my supervisor during the last 4 years.
    Special thanks go to Filomeno Sánchez and José Luis Taín for their continuous support
    and many clarifying discussions. I also would like to thank Michael Döring, Igor
    Tkachenko, Adoración Abellan, Magdalena
    Rafecas, Noriel Pavón, Francisco García-de-Quirós and Jose Antonio
    Palazón for helpful suggestions and comments.

    \vspace*{1.8eX}
    I am also indebted to Luis Caballero, Antonio Gonzalez, Ana Ros,
    Mike Ramage and all others who helped me revise my dissertation.

    \vspace*{1.8eX}
    Further thanks go to María Fernández, Cibeles Mora, Eva Nerina Giménez
    and Marcos Giménez and others. Their help on various occasions is gratefully
    acknowledged.

    \vspace*{1.8eX}
    I would specially thank Daniel Bornkessel and Thomas Übermeier 
    who helped me with several computer issues.

    \vspace*{1.8eX}
    My thanks and appreciation also go to the German Academic Exchange
    Service (DAAD) and the Spanish Ministry of Science and Education for
    the funding of this doctoral thesis.

    \vspace*{1.8eX}
    At last, I want to thank my parents Sieglinde and Karl-Heinz 
    and my sister Sabine for their endless support. Without their help and courage 
    this work would not have been possible.

    \vspace*{3eX}
    Thank You everyone
  \end{minipage} \hspace*{0.09\linewidth}
\end{center}
\cleardoublepage{}


\end{document}